\documentclass[aps,pra,10pt,amsmath,amssymb,notitlepage,onecolumn,nofootinbib,superscriptaddress,longbibliography]{revtex4-1}

\pdfoutput=1

\usepackage{amsmath}
\usepackage{tabularx,graphicx}
\usepackage{epstopdf}
\usepackage{makecell}
\usepackage{graphicx}
\usepackage{latexsym}
\usepackage{amssymb}
\usepackage{amsthm}
\usepackage{color, colortbl}
\usepackage{bbm}
\usepackage{bm}
\usepackage{dsfont}
\usepackage{feynmp}
\usepackage{longtable}
\usepackage{slashed}
\usepackage{multirow}
\usepackage{url}
\usepackage{cancel}
\usepackage{mathtools}
\usepackage{tikz}
\usepackage{tikz-3dplot} 
\usepackage[letterpaper,margin=0.6in]{geometry}
\usetikzlibrary {shapes.geometric,arrows.meta,bending}

\newtheorem{theorem}{Theorem}
\newtheorem{definition}[theorem]{Definition}
\newtheorem{conj}[theorem]{Conjecture}
\newtheorem{cor}[theorem]{Corollary}

\newtheorem{statement}[theorem]{Statement}

\newtheorem{remark}[theorem]{Remark}
\newtheorem{verification}[theorem]{Verification}

\newcommand{\mc}[1]{\mathcal{#1}}

\newcommand{\beq}{\begin{eqnarray}}
	\newcommand{\eeq}{\end{eqnarray}}

\newcommand{\Tr}{{\rm Tr}}

\DeclareMathOperator{\im}{im}
\newcommand{\bsp}{\begin{aligned}}
	\newcommand{\esp}{\end{aligned}}

\usepackage{xcolor}
\definecolor{darkblue}{rgb}{0.,0.,0.4}
\definecolor{darkred}{rgb}{0.5,0.,0.}
\definecolor{BlueViolet}{RGB}{138,43,226}
\definecolor{SkyBlue}{RGB}{30,144,255}
\definecolor{DarkGreen}{RGB}{0,100,0}

\newcommand{\customyinyang}{%
    \begin{tikzpicture}[scale=0.58,rotate=90]
      \draw[line width = 0.05ex,transform canvas={yshift=0.12ex}] (0,0) circle (1ex);
      \path[fill=black,transform canvas={yshift=0.12ex}] (90:1ex) arc (90:-90:0.5ex)
                        (0,0)    arc (90:270:0.5ex)
                        (0,-1ex) arc (-90:-270:1ex);
    \end{tikzpicture}}

\tdplotsetmaincoords{80}{20} %
\tdplotsetrotatedcoords{0}{0}{0} %
\newcommand{\translatepoint}[1]%
{   \coordinate (mytranslation) at (#1);
}
\newcommand{\cornersharingtetrahedraI}%
{

\draw[color = red] (0,0+1/4,0+1/4) -- (0+1/4,0+1/4,0);
\draw[color = red] (0,0,0) -- (0-1/4,0,0-1/4);
\draw[color = green] (1/8,1/8,1/8) -- (0,0,0);
\draw[color = green] (1/8,1/8,1/8) -- (1/4,1/4,0);
\draw[color = green] (1/8,1/8,1/8) -- (0,1/4,1/4);
\draw[color = green] (1/8,1/8,1/8) -- (1/4,0,1/4);
\draw[color = green] (-1/8,-1/8,-1/8) -- (0,0,0);
\draw[color = green] (-1/8,-1/8,-1/8) -- (-1/4,-1/4,0);
\draw[color = green] (-1/8,-1/8,-1/8) -- (-1/4,0,-1/4);
\draw[color = green] (-1/8,-1/8,-1/8) -- (0,-1/4,-1/4);
\shade[rotated axis,greenBall] (-1/8,-1/8,-1/8) circle (0.05cm);
\draw[color = red] (0,0,0) -- (0,0+1/4,0+1/4);
\shade[rotated axis,greenBall] (1/8,1/8,1/8) circle (0.05cm); 
\draw[color = red] (0,0,0) -- (0+1/4,0,0+1/4);
\draw[color = red] (0,0,0) -- (0+1/4,0+1/4,0);
\draw[color = red] (0+1/4,0,0+1/4) -- (0,0+1/4,0+1/4);
\draw[color = red] (0+1/4,0+1/4,0) -- (0+1/4,0,0+1/4);
\draw[color = red][color = red] (0,0,0) -- (0-1/4,0-1/4,0);
\draw[color = red] (0,0,0) -- (0,0-1/4,0-1/4);
\draw[color = red] (0-1/4,0-1/4,0) -- (0-1/4,0,0-1/4);
\draw[color = red] (0-1/4,0,0-1/4) -- (0,0-1/4,0-1/4);
\draw[color = red] (0,0-1/4,0-1/4) -- (0-1/4,0-1/4,0);
\shade[rotated axis,redBall] (0,0,0) circle (0.035cm); 
\shade[rotated axis,redBall] (0,1/4,1/4) circle (0.035cm); 
\shade[rotated axis,redBall] (1/4,0,1/4) circle (0.035cm); 
\shade[rotated axis,redBall] (1/4,1/4,0) circle (0.035cm); 
\shade[rotated axis,redBall] (-1/4,-1/4,0) circle (0.035cm); 
\shade[rotated axis,redBall] (-1/4,0,-1/4) circle (0.035cm); 
\shade[rotated axis,redBall] (0,-1/4,-1/4) circle (0.035cm); 
}

\newcommand{\cornersharingtetrahedraII}%
{
\draw[color = blue] (0,0+1/4,0+1/4) -- (0+1/4,0+1/4,0);
\draw[color = blue] (0,0,0) -- (0-1/4,0,0-1/4);
\draw[color = yellow] (1/8,1/8,1/8) -- (0,0,0);
\draw[color = yellow] (1/8,1/8,1/8) -- (1/4,1/4,0);
\draw[color = yellow] (1/8,1/8,1/8) -- (0,1/4,1/4);
\draw[color = yellow] (1/8,1/8,1/8) -- (1/4,0,1/4);
\draw[color = yellow] (-1/8,-1/8,-1/8) -- (0,0,0);
\draw[color = yellow] (-1/8,-1/8,-1/8) -- (-1/4,-1/4,0);
\draw[color = yellow] (-1/8,-1/8,-1/8) -- (-1/4,0,-1/4);
\draw[color = yellow] (-1/8,-1/8,-1/8) -- (0,-1/4,-1/4);
\shade[rotated axis,yellowBall] (-1/8,-1/8,-1/8) circle (0.05cm);
\draw[color = blue] (0,0,0) -- (0,0+1/4,0+1/4);
\shade[rotated axis,yellowBall] (1/8,1/8,1/8) circle (0.05cm); 
\draw[color = blue] (0,0,0) -- (0+1/4,0,0+1/4);
\draw[color = blue] (0,0,0) -- (0+1/4,0+1/4,0);
\draw[color = blue] (0+1/4,0,0+1/4) -- (0,0+1/4,0+1/4);
\draw[color = blue] (0+1/4,0+1/4,0) -- (0+1/4,0,0+1/4);
\draw[color = blue][color = blue] (0,0,0) -- (0-1/4,0-1/4,0);
\draw[color = blue] (0,0,0) -- (0,0-1/4,0-1/4);
\draw[color = blue] (0-1/4,0-1/4,0) -- (0-1/4,0,0-1/4);
\draw[color = blue] (0-1/4,0,0-1/4) -- (0,0-1/4,0-1/4);
\draw[color = blue] (0,0-1/4,0-1/4) -- (0-1/4,0-1/4,0);
\shade[rotated axis,blueBall] (0,0,0) circle (0.035cm); 
\shade[rotated axis,blueBall] (0,1/4,1/4) circle (0.035cm); 
\shade[rotated axis,blueBall] (1/4,0,1/4) circle (0.035cm); 
\shade[rotated axis,blueBall] (1/4,1/4,0) circle (0.035cm); 
\shade[rotated axis,blueBall] (-1/4,-1/4,0) circle (0.035cm); 
\shade[rotated axis,blueBall] (-1/4,0,-1/4) circle (0.035cm); 
\shade[rotated axis,blueBall] (0,-1/4,-1/4) circle (0.035cm); 
}

\makeatletter

\pgfkeys{
  /pgf/arrow keys/.cd,
  pitch/.code={%
    \pgfmathsetmacro\pgfarrowpitch{#1}
    \pgfmathsetmacro\pgfarrowcospitch{abs(cos(\pgfarrowpitch))}
    \pgfmathsetmacro\pgfarrowsinpitch{    sin(\pgfarrowpitch)}
  }
}

\pgfdeclarearrow{
  name = Cone1,
  defaults = {       %
    length     = +3.6pt +5.4,
    width'     = +0pt +0.5,
    line width = +0pt 1 1,
    pitch      = +0, %
  },
  cache = false,     %
  setup code = {},   %
  drawing code = {   %
    \pgfmathsetmacro\pgfarrowhalfwidth{.5\pgfarrowwidth}
    \pgfmathsetmacro\pgfarrowhalfwidthsin{\pgfarrowhalfwidth*abs(\pgfarrowsinpitch)}
    \pgfpathellipse{\pgfpointorigin}{\pgfqpoint{\pgfarrowhalfwidthsin pt}{0pt}}{\pgfqpoint{0pt}{\pgfarrowhalfwidth pt}}
    \pgfusepath{fill}
    \pgfmathsetmacro\pgfarrowlengthcos{\pgfarrowlength*\pgfarrowcospitch}
    \ifdim\pgfarrowlengthcos pt>\pgfarrowhalfwidthsin pt
      \pgfmathsetmacro\pgfarrowlengthtemp{\pgfarrowhalfwidthsin*\pgfarrowhalfwidthsin/\pgfarrowlengthcos}
      \pgfmathsetmacro\pgfarrowwidthtemp{\pgfarrowhalfwidth/\pgfarrowlengthcos*sqrt(\pgfarrowlengthcos*\pgfarrowlengthcos-\pgfarrowhalfwidthsin*\pgfarrowhalfwidthsin)}
      \pgfpathmoveto{\pgfqpoint{\pgfarrowlengthcos pt}{0pt}}
      \pgfpathlineto{\pgfqpoint{\pgfarrowlengthtemp pt}{ \pgfarrowwidthtemp pt}}
      \pgfpathlineto{\pgfqpoint{\pgfarrowlengthtemp pt}{-\pgfarrowwidthtemp pt}}
      \pgfpathclose
      \pgfusepath{fill}
    \else
      \ifdim\pgfarrowsinpitch pt>0pt
      \pgfpathcircle{\pgfqpoint{\pgfarrowlengthcos pt}{0pt}}{.5\pgfarrowlinewidth}
      \pgfsetcolor{white}
      \pgfusepath{fill}
      \fi
    \fi
    \pgfpathmoveto{\pgfpointorigin}
  }
}

\def\pgfarrowtangenttosincos#1{
    #1
    \tdplotcrossprod(\pgf@xx,\pgf@yx,\pgf@zx)(\pgf@xy,\pgf@yy,\pgf@zy)
    \pgfmathsetmacro\pgfarrowtangentxxyy{\pgf@x*\pgf@x+\pgf@y*\pgf@y}
    \pgfmathsetmacro\pgfarrowtangentxy{sqrt(\pgfarrowtangentxxyy)}
    \pgfmathsetmacro\pgfarrowtangentz{(\pgftemp@x*\tdplotresx+\pgftemp@y*\tdplotresy+\pgftemp@z*\tdplotresz)/72.27*2.54}
    \pgfmathsetmacro\pgfarrowtangentxyz{sqrt(\pgfarrowtangentxxyy+\pgfarrowtangentz*\pgfarrowtangentz)}
    \pgfmathsetmacro\pgfarrowcospitch{\pgfarrowtangentxy/\pgfarrowtangentxyz}
    \pgfmathsetmacro\pgfarrowsinpitch{\pgfarrowtangentz/\pgfarrowtangentxyz}
}
\pgfkeys{
  /pgf/arrow keys/.cd,
  tangent/.code={%
    \tikz@scan@one@point\pgfarrowtangenttosincos#1
  }
}
\pgfdeclarearrow{
  name = Cone2,
  defaults = {             %
    length     = +3.6pt +5.4,
    width'     = +0pt +0.5,
    line width = +0pt 1 1,
    tangent    = {(1,0,0)} %
  },
  cache = false,           %
  setup code = {},         %
  drawing code = {         %
    \pgfmathsetmacro\pgfarrowhalfwidth{.5\pgfarrowwidth}
    \pgfmathsetmacro\pgfarrowhalfwidthsin{\pgfarrowhalfwidth*abs(\pgfarrowsinpitch)}
    \pgfpathellipse{\pgfpointorigin}{\pgfqpoint{\pgfarrowhalfwidthsin pt}{0pt}}{\pgfqpoint{0pt}{\pgfarrowhalfwidth pt}}
    \pgfusepath{fill}
    \pgfmathsetmacro\pgfarrowlengthcos{\pgfarrowlength*\pgfarrowcospitch}
    \ifdim\pgfarrowlengthcos pt>\pgfarrowhalfwidthsin pt
      \pgfmathsetmacro\pgfarrowlengthtemp{\pgfarrowhalfwidthsin*\pgfarrowhalfwidthsin/\pgfarrowlengthcos}
      \pgfmathsetmacro\pgfarrowwidthtemp{\pgfarrowhalfwidth/\pgfarrowlengthcos*sqrt(\pgfarrowlengthcos*\pgfarrowlengthcos-\pgfarrowhalfwidthsin*\pgfarrowhalfwidthsin)}
      \pgfpathmoveto{\pgfqpoint{\pgfarrowlengthcos pt}{0pt}}
      \pgfpathlineto{\pgfqpoint{\pgfarrowlengthtemp pt}{ \pgfarrowwidthtemp pt}}
      \pgfpathlineto{\pgfqpoint{\pgfarrowlengthtemp pt}{-\pgfarrowwidthtemp pt}}
      \pgfpathclose
      \pgfusepath{fill}
    \else
      \ifdim\pgfarrowsinpitch pt>0pt
      \pgfpathcircle{\pgfqpoint{\pgfarrowlengthcos pt}{0pt}}{.5\pgfarrowlinewidth}
      \pgfsetcolor{white}
      \pgfusepath{fill}
      \fi
    \fi
    \pgfpathmoveto{\pgfpointorigin}
  }
}

\usepackage{silence}
\usepackage[colorlinks=true,linkcolor=darkblue,citecolor=blue,urlcolor=darkred,bookmarksdepth=1]{hyperref}
\makeatletter
\def\@bibdataout@aps{%
 \immediate\write\@bibdataout{%
  @CONTROL{%
   apsrev41Control%
   \longbibliography@sw{%
    ,author="60",editor="1",pages="1",title="0",year="0"%
   }{%
    ,author="08",editor="1",pages="0",title="",year="1"%
   }%
  }%
 }%
 \if@filesw
  \immediate\write\@auxout{\string\citation{apsrev41Control}}%
 \fi
}%
\let\SGC@toc@subsection\l@subsection
\newcommand{\SGCHideSubsectionsInTOC}{\let\l@subsection\@gobbletwo}
\newcommand{\SGCRestoreSubsectionsInTOC}{\let\l@subsection\SGC@toc@subsection}
\makeatother
\usepackage[normalem]{ulem}

\usepackage{mathrsfs}
\usepackage{dutchcal}
\usepackage[title]{appendix}

\usepackage{etoolbox}
\patchcmd{\appendices}{\quad}{: }{}{}
\usepackage{comment}

\newcommand{\Z}{\mathbb{Z}}

\def\U{\mathrm{U}(1)}
\newcommand{\SO}{\mathrm{SO}}

\usepackage{booktabs}
\usepackage{yfonts}
\makeatletter

\usepackage[caption=false]{subfig}

\begin{document}
\raggedbottom

\title{Crystallography, Group Cohomology, and Lieb--Schultz--Mattis Constraints}

\author{Chunxiao Liu}
\affiliation{Department of Physics, University of California, Berkeley, California, USA 94720}
\email{chunxiaoliu@berkeley.edu}

\author{Weicheng Ye}
\affiliation{Department of Physics and Astronomy, and Stewart Blusson Quantum Matter Institute, University of British Columbia, Vancouver, BC, Canada V6T 1Z1}
\email{victoryeofphysics@gmail.com}

\begin{abstract}

We present a computational study of the mod-2 cohomology of three-dimensional (3D) space groups, with an eye toward their applications in Lieb--Schultz--Mattis constraints. We prove finite-generation results for the cohomology of crystallographic groups and give ring presentations for $H^*(G,\mathbb Z_2)$ for \emph{all} 230 3D space groups, together with explicit inhomogeneous representatives for the degree-$\leq 3$ cocycles used in the lattice applications. The all-degree interpretation of the ring presentations is organized through finite LHS-spectral-sequence and Hilbert-series verification checks. We then associate distinguished classes in $H^3(G,\mathbb Z_2)$ to irreducible Wyckoff positions and use these classes as cohomological representatives of Lieb--Schultz--Mattis anomaly candidates when the on-site projective representations are classified by powers of $\mathbb Z_2$. Finally, we apply the resulting anomaly data to $\mathrm U(1)$ quantum spin liquids on the 3D pyrochlore lattice and compare the symmetry-fractionalization constraints with projective-symmetry-group calculations.
\end{abstract}

\maketitle

\tableofcontents

\section{Introduction}

The $k$-dimensional crystallographic groups constitute an important family of infinite groups within group theory. In both mathematical and crystallographic literature, the two-dimensional and three-dimensional crystallographic groups are commonly referred to as wallpaper groups and space groups, respectively. These groups have played a fundamental role in understanding crystal properties and classifying phases of matter. Beyond condensed matter physics, the concept of $k$-dimensional crystallographic groups also appears in various areas of theoretical physics \cite{dixon1986strings,deBoer:2001wca,doi:10.1142/S0129167X21500786} and mathematics \cite{scott1983geometries,charlap2012bieberbach,HILD2007208}.

While the classification of wallpaper and space groups was completed over a century ago \footnote{For an interesting classification of 3D space groups using orbifolds, see \cite{CDHT}.}, and their representation-theoretic properties have been extensively documented in crystallographic references, the calculation of group cohomology and other homological-algebraic properties for these crystallographic groups remains largely incomplete. The main challenge in deriving these properties arises from the infinite order and complex group structures of crystallographic groups, which make it difficult to implement the methods designed for finite group calculations in an efficient way \cite{GAP4,ellis2019invitation,Ouyang_2021}.

From a physical perspective, the homological-algebraic properties of crystallographic groups can greatly help understand the symmetry properties of phases of matter in crystals. Of particular importance is the concept of \emph{quantum anomalies} \cite{Hooft1980}. Roughly speaking, quantum anomaly refers to the obstruction to having a unique, symmetric, gapped ground state, and different anomalies are captured by distinct elements of the cohomology of the symmetry group \cite{DijkgraafWitten}. For symmetries acting on a lattice system, a prototypical quantum anomaly is called Lieb--Schultz--Mattis (LSM) anomaly \cite{lieb1961two,Cheng2015}, which exists in a lattice system with on-site internal symmetry such that the microscopic, on-site degrees of freedom carry projective representations of the internal symmetry. For example, in 1D, LSM anomaly is present in a spin-1/2 chain with translation and $\SO(3)$ rotation symmetry. 

Quantum anomaly is important because of the principle of \emph{anomaly matching}: 
in a lattice system, consider an infrared (IR) theory that emerges on the lattice with some emergent low-energy degrees of freedom, the quantum anomaly of this IR theory must match
the anomaly present in the original, microscopic (UV) theory. 
As such, one can ask how the UV symmetries (including crystalline symmetry $G$ and internal symmetry) act on the emergent IR degrees of freedom, and this action is heavily constrained by anomaly matching \cite{10.21468/SciPostPhys.13.3.066,Ye2024}. 
Such data of symmetry actions (referred to as ``quantum numbers" in physics) can then be numerically tested, providing additional information regarding phases on the lattice \cite{Song2018a,ferrari2019,Wietek2024}. This makes anomaly matching a quite powerful theoretical tool besides conventional methods of detecting phases of matter. 

To implement the program of anomaly matching, we need to acquire three pieces of data: (1) the UV anomaly of the lattice system; (2) the IR anomaly; and (3) the check that the UV and IR anomalies match. In the context of a spin-1/2 lattice magnet, where the UV anomaly is of the LSM type, the anomaly information is encoded in the mod-2 cohomology of crystallographic groups. Additionally, knowing the complete ring structure of mod-2 cohomology is essential in performing anomaly matching.

A related matter is the classification of symmetry-protected topological phases (SPTs) protected by crystalline symmetries (crystalline SPTs), where the homological-algebraic aspects of crystallographic groups also play a crucial role. Bosonic crystalline SPTs are classified by the group cohomology of crystalline groups \cite{PhysRevX.8.011040,10.1093/ptep/ptad086}. On the fermionic side, while fermionic crystalline SPTs are classified by bordism groups \cite{Freed2019,Debray2021,Zhang2022,zhang2022construction}, the mod-2 cohomology ring (as a module of the Steenrod algebra) still serves as input to the Adams spectral sequence calculation. Therefore, from the perspectives of both quantum anomalies and crystalline SPTs, a deep understanding of the mod-2 cohomology is of paramount importance.

In this work, we conduct a comprehensive study of the homological-algebraic properties of all 3D space groups. Taking advantage of recent progress in computational homotopy \cite{ellis2019invitation}, we systematically obtain presentations for the mod-2 cohomology rings of these groups. Furthermore, we explicitly give standard inhomogeneous cochain expressions for the degree-$\leq 3$ or lower-degree cocycles needed in the lattice applications. As the mathematical foundation of the computation, we analyze the Lyndon--Hochschild--Serre (LHS) spectral sequence, prove that the $\mathbb{F}$-cohomology ring of any crystallographic group $G$ is finitely generated, and isolate the finite verification steps used to certify the degree bounds for generators and relations in the 3D tables.

Next, we turn to an important physical application of these mod-2 cohomology results: LSM constraints and LSM anomaly data. We construct, for every IWP of every 3D space group, a distinguished class in $H^3(G,\mathbb Z_2)$ detected by explicit topological invariants. The conversion of this cohomological data into a full many-body no-go theorem is stated separately as a physical conjecture/conditional statement, in keeping with the present status of general 3D LSM theorems. Finally, we apply these data to the symmetry actions on the emergent IR degrees of freedom in $\U$ quantum spin liquids on the 3D pyrochlore lattice and compare the result with projective symmetry group (PSG) calculations.

\subsection{Statement of main results}

We now present a summary of the main results and outline the organization of the paper.

\begin{enumerate}

\item We prove that the $\mathbb{F}$ cohomology ring of any crystallographic group $G$ must be finitely generated with a finite number of generators and relations.

\item We obtain presentations for the mod-2 cohomology rings of all 230 3D space groups. These cohomology rings are collected in Appendix~\ref{collection230}. The all-degree completeness of the presentations is tied to the finite verification described in Sec.~\ref{subsec:degree_bound}. We give an explicit expression for each of the 1-, 2-, and 3-cocycle functions of the mod-2 cohomology used in the LSM tables (except for the 3-cocycles of groups No.~\hyperref[subsub:sg225]{225}, \hyperref[subsub:sg227]{227}, and \hyperref[subsub:sg229]{229}, where the current labels denote deterministic GAP basis completions). These data are collected in our online GitHub files \cite{github}.

\item To organize the LSM anomaly candidates, to each Irreducible Wyckoff Position (IWP, defined in Sec.~\ref{subsec:IWP}) of every 3D space group we associate a distinguished element in the third cohomology of the space group with $\Z_2$ coefficients. These data are listed in the IWP table for each of the 230 groups, collected in Appendix~\ref{collection230}. A cohomological-operational characterization of these elements is given in Statement~\ref{Thm:LSM3dmath}.

\item The IWP data allow us to state a 3D LSM no-go criterion for lattice magnets whose on-site degrees of freedom carry projective representations classified by powers of $\mathbb{Z}_2$. Because a fully general many-body proof of this 3D criterion is not presently part of the paper, we state it as Conjecture~\ref{Thm:LSM3dphys} and use it as the physical input for the anomaly interpretation. 

\item In Sec.~\ref{sec:anomaly_matching}, we demonstrate the physical significance of our result by performing anomaly matching for $U(1)$ quantum spin liquids on the pyrochlore lattice.
We obtain the symmetry fractionalization of electric and magnetic charges in Eq.~\eqref{eq:e_SF} and Eq.~\eqref{eq:mag_SF}, respectively, for a type of symmetry actions. The result of the symmetry fractionalization of fermionic electric charges is consistent with the PSG calculation in Ref.~\cite{Liu2021pyrochlore}.

\end{enumerate}

To make the logical status of the results transparent, we explicitly state formal theorems, finite computational verifications, and physical inputs of our results.

\begin{enumerate}
\item Finite generation of $H^*(G,\mathbb F)$ for crystallographic groups is a theorem, proved in Theorem~\ref{thm:finite}.
\item The exhaustive all-page calculation in Verification~\ref{ver:degree-certificate} and Theorem~\ref{thm:finite3D_ref} prove uniform all-degree bounds for generators and defining relations of the mod-2 cohomology rings of all 230 space groups.
\item The finite-quotient method using $P_3=G/T^3$ is used to produce low-degree inhomogeneous cocycle representatives; its all-degree surjectivity is reduced to a finite low-degree check in Theorem~\ref{thm:p4_low_degree}.
\item The 3D LSM no-go theorem for all IWP classes is stated as the physical Conjecture~\ref{Thm:LSM3dphys}. The IWP--$H^3(G,\mathbb Z_2)$ correspondence is an algorithmic cohomological construction using the topological invariants in Sec.~\ref{sec:LSM} and concisely summarized in Statement~\ref{Thm:LSM3dmath}.
\item The pyrochlore anomaly matching is an application of the cohomological anomaly data under the usual anomaly-matching assumptions.
\end{enumerate}

The paper proceeds as follows. In Sec.~\ref{sec:basics}, we give a brief introduction of crystallographic groups and the concept of IWP, and set up the notation that will be used for the rest of the paper.  In Sec.~\ref{sec:structure}, we state and prove several theorems for cohomology of crystallographic groups, which serve as the mathematical foundation of our code. In Sec.~\ref{sec:cohomology}, we outline our methods to obtain the mod-2 cohomology of 3D space groups, in particular the ring structure, and discuss some features of our results. In Sec.~\ref{sec:LSM}, we review existing LSM constraints in 2D and their connection to cohomology. Then we move on to 3D, state the various LSM constraints from IWPs and spell out the algorithm we use to associate each IWP of space groups $G$ with an element in $H^3(G, \Z_2)$, with several notable examples included in Sec.~\ref{subsec:examples}. In Sec.~\ref{sec:anomaly_matching}, we apply LSM anomaly and anomaly matching to the understanding of $U(1)$ quantum spin liquids on the pyrochlore lattice and compare our results with existing PSG calculations. We close the paper with open questions and outlooks. Several appendices are included to provide the necessary information for 3D space groups and their cohomology, with the complete enumeration of results for all 230 space groups given in Appendix~\ref{collection230}. 

Many of the results are obtained with the help of the software packages GAP \cite{GAP4}, SageMath \cite{sagemath}, and Mathematica \cite{Mathematica}. The code is available on GitHub \cite{github}.

\section{Basics of crystallographic groups}\label{sec:basics}

In this section, we introduce the basic concepts of crystallographic groups, including the notion of Irreducible Wyckoff Positions (IWPs), which will be central to the description of Lieb--Schultz--Mattis constraints. We also set up the notation that will be used throughout this paper.

\subsection{Crystallographic groups in 3D}\label{subsec:crystal}

Crystallographic groups are groups consisting of symmetry operations of crystals that leave the crystal structure invariant \cite{brown1978crystallographic}. We first give its formal definition:
\begin{definition}
(Crystallographic groups of dimension $k$. \cite{ellis2019invitation})
Let $G$ be a group of invertible affine transformations of $\mathbb{R}^k$, such that any element $g\in G$ has the form $g\colon \mathbb{R}^k\rightarrow \mathbb{R}^k,\bm{x}\mapsto A\bm{x}+\bm{b}$, where the matrix $A\in GL_k(\mathbb{R})$ and the vector $\bm{b}\in \mathbb{R}^k$. $g$ is called pure translation if $A$ is the identity matrix. Define $T$ as the normal subgroup consisting of pure translations in $G$. The group $G$ is a crystallographic group of dimension $k$ if $T$ is a free abelian group of rank $k$ and the quotient group $P:=G/T$ is finite.
\end{definition}
The translation group $T$ is, by definition, isomorphic to $\mathbb{Z}^k$. The quotient group $P$ is called \emph{point group} associated with $G$. The three groups fit into the short exact sequence 
\begin{equation}\label{TGP}
1\rightarrow T\rightarrow G\rightarrow P\rightarrow 1.
\end{equation}
The short exact sequence Eq.~\eqref{TGP} offers a complementary view for the three groups $T$, $G$, and $P$. Given $P$ as a finite group, $T$ is an integral representation of $P$ whose action is defined by conjugation
\begin{equation}\label{intrep}
\rho\colon P\times T\rightarrow T,~~ (p,t)\mapsto s(p)\cdot t \cdot s(p)^{-1},
\end{equation}
with $s(p)$ any pre-image of $p$ in $G$. $G$ is then thought of as the group extension of $P$ by $T$, characterized by the integral representation $\rho$ and an element $\omega$ of the second cohomology group $H^2_\rho(P,T)$.

Unless otherwise stated, we will always use $G$ to denote a crystallographic group, $P$ to denote its associated point group, and $T$ to denote the group of pure translations $\Z^k$. Sometimes we will also use $\mathcal{G}$ to denote the full symmetry group of a physical system, including lattice symmetries and internal symmetries. 

Going to 3D, the classification of space groups was done separately by Fedorov and Schönflies in 1891. Treating $P$ as abstract groups, we know there are 18 possibilities in 3D
\begin{equation}
\begin{aligned}
&\mathbb{Z}_1, \mathbb{Z}_2, \mathbb{Z}^2_2, \mathbb{Z}^3_2, \mathbb{Z}_4, \mathbb{Z}_4\times \mathbb{Z}_2, Dih_4, Dih_4\times \mathbb{Z}_2,\\
&\mathbb{Z}_3, \mathbb{Z}_3\times \mathbb{Z}_2,\mathbb{Z}_3\times \mathbb{Z}^2_2, Dih_3, Dih_3\times \mathbb{Z}_2,  Dih_3\times \mathbb{Z}^2_2,\\
&A_4, A_4\times \mathbb{Z}_2, S_4, S_4\times \mathbb{Z}_2, 
\end{aligned}
\end{equation}
where $Dih_n,A_4,S_4$ stands for dihedral group with $2n$ elements, the alternating group on 4 letters, and the symmetric group on 4 letters. 

The abstract classification of point groups do not capture all the features of crystal symmetry. Depending on the symmetry operations (rotation, mirror reflection, inversion, etc.) on $\mathbb{R}^3$, the abstract point groups are further distinguished into 32 point groups. The detailed information about the 18 abstract point groups and the 32 point groups are listed in Table \ref{tablePT1}. 

To name the generators of the point groups, we use the following convention:
\begin{itemize}
\item $C_n$: $n$-fold rotation $C_n$ 
\item $M$: mirror reflection
\item $I$: inversion
\item $G$: glide
\item $S_n$: $n$-fold screw 
\item $\overline{C}_n$: $n$-fold rotoinversion (only $\overline{C}_4$ will explicitly appear). 
\end{itemize}

The 32 point groups, combined with the 14 Bravais lattice types, determine a total of 73 \emph{arithmetic crystal classes}, in one-to-one correspondence with the distinct integral representations $\rho$ (see Eq.~\eqref{intrep}).
For each arithmetic class, the 2nd cohomology $H^2_\rho(P,T)$ classifies the distinct group extensions. The arithmetic classes together with $H^2_\rho(P,T)$ determines a total number of 219 (non-isomorphic) abstract space groups as possible extensions of $P$ by $T$.
The detailed information about the arithmetic classes and the extension classes is given in Table \ref{tableExt}. Each arithmetic class contains a split extension, for which the space group is a semidirect product $G=T\rtimes P$. In crystallography literature, the split crystallographic groups are called \emph{symmorphic} groups. Therefore, there are in total 73 symmorphic groups, one for each arithmetic crystal classes. These symmorphic groups are labeled with a ``$\rtimes$" in the third column of Table \ref{tablemod2230}.

Finally, out of the 219 abstract space groups, 11 of them describe crystal structures with a designated chirality \footnote{A left-handed space group is related to the right-handed one by the conjugation of certain affine transformation $\bm{x}\mapsto A\bm{x}+\bm{b}$, such that $A\in GL_n(\mathbb{R})$ has determinant $-1$.}. 
The following pairs of isomorphic space groups are introduced to further distinguish the two chiralities
\begin{equation}\label{isomG}
\begin{aligned}
&\hyperref[subsub:sg76]{76~(P4_1)}\cong \hyperref[subsub:sg78]{78~(P4_3)},&&
\hyperref[subsub:sg91]{91~(P4_122)}\cong \hyperref[subsub:sg95]{95~(P4_322)},&&
\hyperref[subsub:sg92]{92~(P4_12_12)}\cong \hyperref[subsub:sg96]{96~(P4_32_12)},\\
&\hyperref[subsub:sg144]{144~(P3_1)}\cong \hyperref[subsub:sg145]{145~(P3_2)},&&
\hyperref[subsub:sg151]{151~(P3_112)}\cong \hyperref[subsub:sg153]{153~(P3_212)},&&
\hyperref[subsub:sg152]{152~(P3_121)}\cong \hyperref[subsub:sg154]{154~(P3_221)},\\
&\hyperref[subsub:sg169]{169~(P6_1)}\cong \hyperref[subsub:sg170]{170~(P6_5)},&&
\hyperref[subsub:sg171]{171~(P6_2)}\cong \hyperref[subsub:sg172]{172~(P6_4)},&&
\hyperref[subsub:sg178]{178~(P6_122)}\cong \hyperref[subsub:sg179]{179~(P6_522)},\\
&\hyperref[subsub:sg180]{180~(P6_222)}\cong \hyperref[subsub:sg181]{181~(P6_422)},&&
\hyperref[subsub:sg212]{212~(P4_332)}\cong \hyperref[subsub:sg213]{213~(P4_132)},
\end{aligned}    
\end{equation}
and this brings the final number of space groups to 230. %

In Appendix \ref{collection230}, for each group, we list the generators and their actions on the coordinate systems. Our choice of the coordinate systems is the same as the ``Standard/Default Setting" on Bilbao Crystallographic Server \cite{Bilbao} (these ``Standard/Default Settings" are always one of the coordinate setups given in the International Tables for Crystallography (ITC) \cite{aroyo2013international}). The numbering of the 230 space groups follows the standard crystallography numbering \cite{aroyo2013international}.

\subsection{(Irreducible) Wyckoff position and lattice homotopy}\label{subsec:IWP}

In this subsection, we introduce the concept of Wyckoff positions and irreducible Wyckoff positions (IWPs).

Even within a given crystallographic symmetry setting, arranging degrees of freedom (spins) in different configurations can yield distinct lattices (which share the same crystalline symmetry). A crystallographic lattice $\Lambda$ is a set of discrete points in $\mathbb{R}^k$ invariant under the crystallographic group $G$. We will assume that there are some physical degrees of freedom (e.g. spins) located at each site of the lattice, and they are acted upon by some internal symmetry \footnote{For most of our discussion, we assume that crystalline symmetry operations only permute the physical degrees of freedom without extra effects, and that the internal symmetry is $\SO(3)$ whose projective representation is classified by $H^2(\SO(3), \U)\cong \Z_2$. We will address possible generalizations as they become relevant.}. To fully characterize all possible lattice structures resulting from these arrangements, we introduce the concept of the \emph{Wyckoff position}. 
Suppose $\Lambda$ contains a point $s\in \mathbb{R}^k$, we need to include  all points $g.s$ that are related to $s$ by a crystalline symmetry $g\in G$. Still, there are some elements $g\in G$ such that they keep $s$ invariant, and they form a subgroup of $G$ that must be one of the point groups. We call this subgroup the \emph{little group} of the site $s$ (which is also called the \emph{stabilizer} of the site $s$ in the math literature). 

\begin{definition} (Wyckoff position \cite{eick1997computing})
 A Wyckoff position of a crystallographic group $G$ is an equivalence class of points in Euclidean space, such that their associated little groups are conjugate subgroups of $G$. 
\end{definition}

A Wyckoff position can be understood as a set of points (or orbits) on a crystal lattice that are transformed in the same way under the crystallographic group. For a given crystallographic group, placing identical physical degrees of freedom on different Wyckoff positions—or even on multiple Wyckoff positions simultaneously—will result in distinct lattice structures. The little group of a Wyckoff position does not fully specify the Wyckoff position: there can be distinct Wyckoff positions with isomorphic little groups. Still, the little group of a Wyckoff position is its most important property.

In addition, to analyze symmetry actions on the lattice structure and extract anomaly from these symmetry actions, we are free to symmetrically move degrees of freedom around in a continuous way. This operation is part of so-called \emph{lattice homotopy} \cite{Po2017latticehomotopy,Else2020}. This gives rise to the concept of \emph{irreducible} Wyckoff positions.

\begin{definition} (Irreducible Wyckoff Position, IWP)
Given a Wyckoff position, consider the closure of all points belonging to the Wyckoff position. If the little group of the Wyckoff position is not a proper subgroup of any other points in the closure, we will call the Wyckoff position irreducible.
\end{definition}

Colloquially, a Wyckoff position is reducible if we can \emph{symmetrically} tune the points in the Wyckoff position to a nearby ``high symmetry points'', and thereby enhance the little group to a bigger group. To analyze the anomaly of symmetry actions, we are free to symmetrically move all degrees of freedom to these IWPs, and hence simplify the analysis significantly.

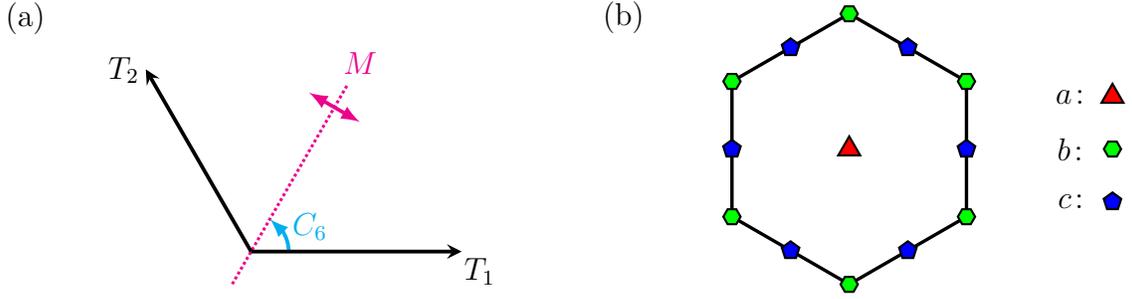
\begin{figure}
    \centering
\begin{tikzpicture}[thick]
\node[] at (-3cm,3.11cm) {\large (a)};
\draw [densely dotted, line width=0.4mm,,draw = magenta, fill=magenta](-120:0.5cm) -- (60:2.55cm);
\draw[>=latex,<->,line width=0.5mm,draw = magenta, fill=magenta] (50:2.25cm) -- (70:2.25cm);
\node[] at (-5:3.05cm) {\large $T_1$};
\node[] at (125:2.95cm) {\large $T_2$};
\node[] at (60:2.9cm) {\large $\textcolor{magenta}{M}$};
\node[] at (23:0.85cm) {\large $\textcolor{cyan}{C_6}$};
\draw[thick,>=latex, ->,line width=0.5mm,,draw = cyan, fill=cyan] (0.5cm,0) arc (0:60:0.5cm);
\draw [->,>=stealth,line width=0.5mm] (0,0) -- (0:2.8cm);
\draw [->,>=stealth,line width=0.5mm] (0,0) -- (120:2.8cm);
\end{tikzpicture}
\hspace{1cm}
\begin{tikzpicture}[thick]
\node[] at (-3cm,1.75cm) {\large (b)};
\newdimen\R
   \R=1.8cm
   \draw [line width=0.45mm] (30:\R) \foreach \x in {30,90,...,390} {  -- (\x:\R) };
   \foreach \x/\l/\p in
     { 30/{}/above,
      90/{}/above,
      150/{}/left,
      210/{}/below,
      270/{}/below,
      330/{}/right
     }
     \node[inner sep=2pt,regular polygon,regular polygon sides=6,draw,fill=green] at (\x:\R) {};
\newdimen\S
   \S=1.5588cm
   \foreach \x/\l/\p in
     { 0/{}/above,
      60/{}/above,
      120/{}/left,
      180/{}/below,
      240/{}/below,
      300/{}/right
     }
     \node[inner sep=2pt,regular polygon,regular polygon sides=5,draw,fill=blue] at (\x:\S) {};
\node[inner sep=1.7pt,regular polygon,regular polygon sides=3,draw,fill=red] at (0:0) {};
\node[inner sep=1.7pt,regular polygon,regular polygon sides=3,draw,fill=red,label={left: {\large $a\colon $}}] at (3.5cm,0.7cm) {};
\node[inner sep=2pt,regular polygon,regular polygon sides=6,draw,fill=green,label={left: {\large $b\colon $}}] at (3.5cm,0cm) {};
\node[inner sep=2pt,regular polygon,regular polygon sides=5,draw,fill=blue,label={left: {\large $c\colon $}}] at (3.5cm,-0.7cm) {};
\end{tikzpicture}

    \caption{Panel (a) shows the generators of the wallpaper group $p6m$. In panel (b), the hexagon is a translation unit cell of the wallpaper group $p6m$. It has three IWPs, conventionally labeled by $a$, $b$ and $c$ in crystallography, and they form the sites of the triangular, honeycomb and kagome lattices, respectively.}
    \label{fig:p6m}
\end{figure}

For instance, in 2D, consider the wallpaper group $p6m$, generated by two translations $T_{1,2}$, a six-fold rotation $C_6$ and a mirror reflection $M$. This group has three distinct IWPs in total, corresponding to the center of six-fold rotation $C_6$ (a), the center of three-fold rotation $T_1C_6^2$ (b) and the center of two-fold rotation $T_1C_6^3$ (c), respectively, as illustrated in Figure~\ref{fig:p6m}. When spins are placed on these three IWPs, they form triangular, honeycomb, and kagome lattices. While all three lattices share the same crystalline symmetry $p6m$, the little group at each site differentiates them.

\begin{figure}
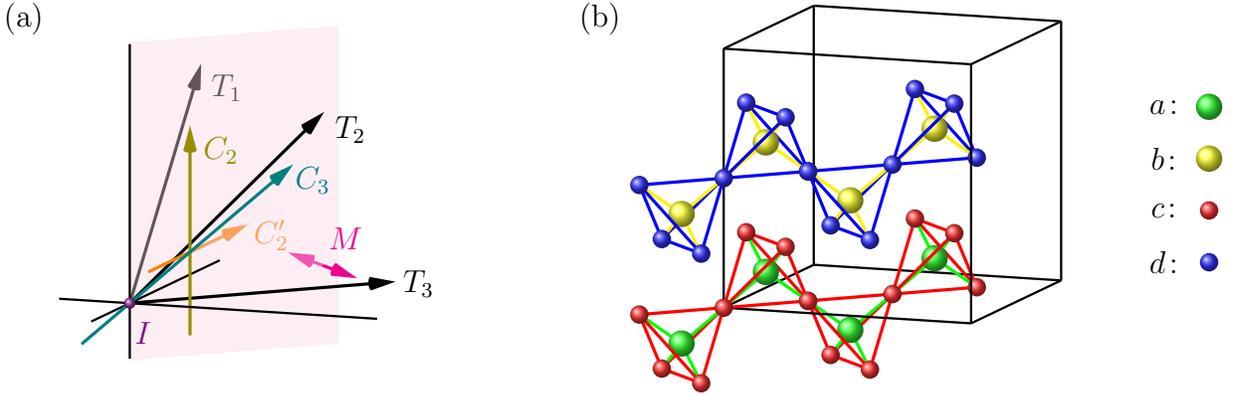



    \caption{Panel (a) shows the generators of the space group \hyperref[subsub:sg227]{No. 227~($Fd\bar{3}m$)}. In panel (b), the cube is a conventional unit cell of the $Fd\bar{3}m$. It has four IWPs, conventionally labeled by $a$, $b$, $c$, and $d$ in crystallography, where $a$ and $b$ form two sets of diamond lattices and $c$ and $d$ form two sets of pyrochlore lattices.}
    \label{fig:fd-3m}
\end{figure}

As a 3D example, consider the space group \hyperref[subsub:sg227]{No. 227~ ($Fd\bar{3}m$)}, generated by three translations $T_{1,2,3}$, two-fold rotations $C_2$ and $C'_2$, a three-fold rotation $C_3$, a mirror $M$, and an inversion $I$. This group has four distinct IWPs in total, corresponding to centers of inversions $I$ (c) and $T_1T_2I$ (d), and the intersection points of two pairs of orthogonal two-fold rotations $(C_2,C'_2)$ (a), and $(T_3C_2,T_2C'_2)$ (b). When spins are placed on one of the inversion centers one gets a pyrochlore lattice, and when spins are placed on one of the intersection points of a pair of orthogonal two-fold rotations one gets a diamond lattice. The symmetry operations and the four IWPs are illustrated in Figure \ref{fig:fd-3m}.

Besides moving degrees of freedom around, for the analysis of anomaly, we are also allowed to fuse these degrees of freedom together in a symmetric way. This includes the following two possibilities \cite{Else2020,chen2021lieb}: (1) remove sites that carry linear representation under all symmetry actions (especially the internal symmetry) and (2) combine two sites at the same location into a single site which carries the tensor product representation of the two original sites. For example, when we have on-site internal $\SO(3)$ symmetry, we can fuse two spin-1/2 local moments and discard them, since the outcome is a linear representation of $\SO(3)$ that does not contribute to the analysis of anomaly. 

\emph{Lattice homotopy} is the operation of moving and fusing degrees of freedom according to the rules above. For a given lattice symmetry $G$, (local) physical degrees of freedom in $\mathbb{R}^k$ form equivalence classes under the operation of lattice homotopy, which we call \emph{lattice homotopy classes} \cite{Po2017latticehomotopy,10.21468/SciPostPhys.13.3.066,chen2021lieb} and carry the structure of abelian group under local fusion. When the internal symmetries only act on-site and do not mix with crystalline symmetries, the lattice homotopy classes should carry the structure of $PR^n$, where $n$ is the number of IWPs and $PR=H^2(G_{\text{int}}, \U)$ labels the projective representation of internal symmetry $G_{\text{int}}$ at each site.

\section{Structure theorems for cohomology of crystallographic groups}\label{sec:structure}

In this section, we state and prove several structure theorems for cohomology of crystallographic groups. In Sec.~\ref{subsec:SS}, we analyze the LHS spectral sequence for mod-2 cohomology of crystallographic groups and illustrate where a generator and relation can appear from the point of view of LHS spectral sequence through specific examples. Building on this, in Sec.~\ref{subsec:finite}, we prove that the $\mathbb{F}$-cohomology ring of any crystallographic groups must be finitely generated as an $\mathbb{F}$-algebra for some field $\mathbb{F}$. %
This proof is adapted from the proofs for finite groups in e.g. Refs.~\cite{Evens1991,adem2013cohomology,ellis2019invitation}. The two subsections serve as the mathematical foundation for our code for calculating the mod-2 cohomology rings. 

To collect additional miscellaneous results of cohomology of crystallographic groups, in Sec.~\ref{subsec:integral}, we connect the results of mod-2 cohomology with integral cohomology. Lastly, in Sec.~\ref{subsec:periodic}, we discuss when the resolution of crystallographic groups admits a periodic resolution.

\subsection{Spectral sequence}\label{subsec:SS}

Although most calculations are carried out without using Lyndon–Hochschild–Serre (LHS) spectral sequence, the LHS spectral sequence remains an invaluable tool in our analysis, since it gives an upper bound on the degree at which new generators and relations can appear. 
In this subsection, we summarize key facts about the LHS spectral sequence, emphasizing its application to the cohomology of crystallographic groups. For further reading, there are several standard textbooks on spectral sequences, such as Refs.~\cite{Evens1991,neukirch2013cohomology,brown2012cohomology}. Many of the formal discussions can be applied to any field $\mathbb{F}$ where $G$ acts trivially, hence we will keep $\mathbb{F}$ and specialize to $\mathbb{Z}_2$ when discussing specific examples.

Given an arbitrary (discrete) group $G$, with a normal subgroup $N\triangleleft G$ and the quotient group $P=G/N$, and a field $\mathbb{F}$ which $G$ acts trivially, we have a short exact sequence of groups,
\begin{equation}
    1\rightarrow N \rightarrow G \rightarrow P \rightarrow 1.
\end{equation}
Then we have the LHS spectral sequence, which can be written as 
\begin{equation}\label{lhse}
E^{p,q}_2 = H^p(P,H^q(N,\mathbb{F}))\Rightarrow H^{p+q}(G,\mathbb{F}).
\end{equation}
An important property is the structure of bilinear product in LHS spectral sequence, which we summarize below.

\begin{theorem} (Bilinear product of LHS spectral sequence \cite{brown2012cohomology,ellis2019invitation})
The LHS spectral sequence Eq.~\eqref{lhse} admits bilinear products
\begin{equation}
    E_r^{p, q} \times E_r^{s, t} \rightarrow E_r^{p+s, q+t}
\end{equation}
for $r \geq 1$ which satisfy the following properties:
\begin{itemize}

\item Each differential $d_r \colon E_r \rightarrow E_r$ satisfies the Leibniz rule
\begin{equation}\label{eq:Leibniz}
d_r(x y)=\left(d_r x\right) y+(-1)^{p+q} x\left(d_r y\right).
\end{equation}

\item The Leibniz rule ensures that the product on $E_r$ induces a product on $\ker d_r/\im d_r$, which is the same (ring isomorphism) as the bilinear product on $\ker d_r/\im d_r \cong E_{r+1}$.

\item The bilinear product on the $E_2$ page $E_2^{p, q} \times E_2^{s, t} \rightarrow E_2^{p+s, q+t}$ is the composition
\begin{equation}
\begin{gathered}
H^p\left(Q, H^q(N, \mathbb{F})\right) \times H^s\left(Q, H^t(N, \mathbb{F})\right) \xrightarrow{\cup_1} H^{p+s}\left(Q, H^q(N, \mathbb{F}) \otimes_{\mathbb{F}} H^t(N, \mathbb{F})\right) \\
\xrightarrow{\cup_2} H^{p+s}\left(Q, H^{q+t}(N,\mathbb{F})\right) \xrightarrow{\times(-1)^{q s}} H^{p+s}\left(Q, H^{q+t}(N,\mathbb{F})\right),
\end{gathered}
\end{equation}
where $\cup_2$ is induced by the cup product
$H^q(N, \mathbb{F}) \times H^t(N, \mathbb{F}) \xrightarrow{\cup} H^{q+t}(N, \mathbb{F})$, 
and $\cup_1$ is the cup product induced by a diagonal map $\Delta_*: R_*^Q \rightarrow R_*^Q \otimes R_*^Q$ on a free $\Z Q$-resolution $R_*^Q$.

\item The bilinear product on the $E_\infty$ page $E_\infty^{p, q} \times E_\infty^{s, t} \rightarrow E_\infty^{p+s, q+t}$ is the same as the one induced from the cup product 
$H^q(G, \mathbb{F}) \times H^t(G, \mathbb{F}) \xrightarrow{\cup} H^{q+t}(G, \mathbb{F})$ on the filtration corresponding to the $E_\infty$ page.

\end{itemize}
\end{theorem}

This makes $E_r^{*, *}$ a double-graded ring on top of the structure of $\mathbb{F}$-algebra. To connect to the cohomology ring $H^*(G, \mathbb{F})$ of $G$, recall that $H^*(G, \mathbb{F})$ can be written as
\begin{equation}
    \mathbb{F}[x,y,\dots]/\text{Relations},
\end{equation}
where $x,y,\dots$ are called \emph{generators}, and ``Relations'' generate an ideal in the ring that are quotiented out and treated as zero in the ring. This is called a \emph{free presentation} of $H^*(G, \mathbb{F})$ treated as an $\mathbb{F}$-algebra. For $E_r^{*, *}$, we also have such a free presentation, except that now $x,y,\dots$ are also double-graded. We have the following theorem for the generators and relations of the double-graded ring $E_\infty^{*, *}$ and those for the cohomology ring $H^*(G, \mathbb{F})$ itself.
\begin{theorem}\label{thm:filtered-presentation}
    If the $E_\infty$ page $E_\infty^{*, *}$ of the LHS spectral sequence admits a free presentation with generators of total degree $\leq d$ and relations of total degree $\leq n$, then the cohomology ring $H^*(G,\mathbb{F})$
    also admits a free presentation with generators of degree $\leq d$ and relations of degree $\leq n$. 
\end{theorem}
\begin{proof}
Let $A=H^*(G,\mathbb F)$ and let $F^\bullet A$ be the finite multiplicative filtration associated with the LHS spectral sequence, so that $\operatorname{gr} A\cong E_\infty$ as graded $\mathbb F$-algebras. Choose homogeneous generators $\bar x_i$ of $\operatorname{gr} A$ of degrees at most $d$, and choose filtered lifts $x_i\in A$ of the same total degrees. These lifts generate $A$: if not, choose a homogeneous class $a\in A$ of minimal total degree not generated by the $x_i$, and among such classes choose one with maximal filtration. Its initial term in $\operatorname{gr}A$ is a polynomial in the $\bar x_i$. Subtracting the corresponding polynomial in the $x_i$ either kills $a$ or lowers its filtration, and induction on the finite filtration gives a contradiction.

It remains to control the relations. Let $S=\mathbb F[X_i]$ with $|X_i|=|x_i|$, and give $S$ the filtration induced from the chosen filtrations of the $x_i$. Let $I=\ker(S\to A)$ and $J=\ker(S\to\operatorname{gr}A)$. By assumption $J$ is generated by homogeneous elements of total degree at most $n$. Choose filtered lifts in $S$ of these generators. Their evaluations in $A$ have strictly lower filtration, and the same descending-filtration argument used above expresses the lower-filtration errors using relations of no larger total degree. Thus $I$ is generated in degrees at most $n$. Equivalently, a bounded presentation of the associated graded algebra lifts to a bounded presentation of the filtered algebra without increasing total degree.
\end{proof}
Hence, even though we cannot obtain the full ring structure of $H^*(G, \mathbb{F})$ from the LHS spectral sequence alone, we can still obtain a lot of information about generators and relations. Moreover, when computing the cohomology ring of $G$, there is an algorithm using LHS spectral sequences to see whether we obtain a complete set of generators and relations \cite{ellis2019invitation}. Here we will also use the LHS spectral sequence to deduce similar results for crystallographic groups. 

We call the \emph{standard} LHS spectral sequence for crystallographic groups the one associated with 
the short exact sequence \eqref{TGP},
and we have
\begin{equation}\label{lhse2}
E^{p,q}_2 = H^p(P,H^q(T,\mathbb{F}))\Rightarrow H^{p+q}(G,\mathbb{F}).
\end{equation}
In $k$-dimensions, the translation group $T\cong \Z^k$ has cohomology
\begin{equation}
H^q(T,\mathbb{F}) = \left\{\begin{array}{ll} \mathbb{F}^{\binom{k}{q}}, &0\leq q \leq k,\\
0, & \text{other }q.\end{array}\right.
\end{equation}
Hence, for the standard LHS spectral sequence, starting from the $E_2$ page, only the first $(k+1)$-rows have nonzero entries. Since this is a finite number, we have an easy corollary.
\begin{cor}\label{cor:end}
    For $k$-dimensional crystallographic groups, the standard LHS spectral sequence collapses at the $E_{k+2}$ page, i.e. $E_{k+2}\cong E_\infty$.
\end{cor}
Specializing to $\mathbb{F}=\Z_2$, the information for the $E_2$ page of the standard LHS spectral sequence for all 230 space groups is listed in the last three columns of Table \ref{tableExt}, written in terms of the $\Z_2$ ranks for each entry. 

For our analysis of the cohomology ring, we do not attempt to obtain the full bilinear product structure for the standard LHS spectral sequences, since usually it suffices to write down each row as a module of the ring $H^*(P, \mathbb{F})$. The top and bottom row is always isomorphic to the ring $H^*(P, \mathbb{F})$ itself, while the middle rows can be a nontrivial module. Especially, we focus on the degree where a nontrivial generator (of the module) appears. 

To make our discussion more concrete, we provide two examples, one for the 2D wallpaper group $p4g$ and the other for the 3D space group \hyperref[subsub:sg42]{$Fmm2$}. Although these results will not be explicitly used in our automated calculations in GAP, these examples nicely illustrate how generators and relations appear from a spectral sequence point of view. 

\subsubsection{Example: \texorpdfstring{$p4g$}{p4g}}

For 2D wallpaper groups, consider $p4g$, which has the most complicated form of mod-2 cohomology ring among all wallpaper groups. The point group of $p4g$ is $D_4$, which as an abstract group is the dihedral group $Dih_4$. $p4g$ is generated by two translations $T_{1,2}$ along two perpendicular directions, a four-fold rotation $C_4$, and a glide reflection $G$ whose reflection axis passes through the rotation center of $C_4$ and bisects the two translation vectors. Acting on the 2D Euclidean space, we have
\begin{subequations}
 \begin{align}
T_1 &\colon (x,y)\rightarrow (x+1, y),\\ 
T_2 &\colon (x,y)\rightarrow (x, y+1),\\
C_4 &\colon (x,y)\rightarrow (-y, x),\\ 
G &\colon (x,y)\rightarrow (y+1/2, x+1/2).
\end{align}
\end{subequations}

To write down the standard LHS spectral sequence for $p4g$, first we write down the mod-2 cohomology ring for $D_4$. $D_4\cong Dih_4$ is generated by a four-fold rotation $C_4$ and a reflection $M$ such that 
\begin{equation}
    C_4^4 = 1, \quad M^2 = 1,\quad M C_4 M = C_4^{-1},
\end{equation}
and every element in $D_4$ can be written as $C_4^c M^m$ with $c\in \{0,1,2,3\},m\in \{0,1\}$. The mod-2 cohomology ring of $D_4$ is
\begin{equation}\label{D4cohoring}
    H^*(D_4, \Z_2) = \Z_2[A_c, A_m, B_\alpha]/(A_c^2+A_c A_m),
\end{equation}
where the generators $A_c$, $A_m$ and $B_\alpha$ have the following explicit cochain representative in the bar resolution,
\begin{equation}
    A_c(C_4^c M^m) = c,\quad A_m(C_4^c M^m) = m, \quad B_\alpha(C_4^{c_1} M^{m_1}, C_4^{c_2} M^{m_2}) = \frac{c_1+(-1)^{m_1}c_2 - (c_1+(-1)^{m_1}c_2 \mod 4)}{4}.
\end{equation}
We can say that $A_c$ and $A_m$ are $\Z_2$ characters for $C_4$ and $M$, respectively. 

The $q=0$ and $q=2$ rows of the LHS spectral sequence at the $E_2$ page can be directly obtained from \eqref{D4cohoring}. The $q=1$ row gives $H^*(D_4, (\Z_2)^2)$, which is a module over the point group cohomology ring $H^*(D_4, \Z_2)$, where both $C_4$ and $M$ permute the two $\Z_2$ factors of $(\Z_2)^2$. It turns out to be
\begin{equation}
    H^*(D_4, (\Z_2)^2) = H^*(D_4, \Z_2).[\omega_{01}, \omega_{11}]/((A_c + A_m) \omega_{01}, (A_c + A_m) \omega_{11}),
\end{equation}
where $\omega_{01}$ and $\omega_{11}$ are two generators (of the module) at degree 0 and 1, respectively. Thus, we can write down the standard LHS spectral sequence for $p4g$ in terms of these generators,
\begin{equation}\label{p4gLHSSS}
\begin{array}{c|ccccc}
q=2&\omega_{02}&A_c \omega_{02}, A_m \omega_{02}&A_c^2 \omega_{02}, A_m^2 \omega_{02}, B_\alpha \omega_{02}&A_c^3 \omega_{02}, A_m^3 \omega_{02}, B_\alpha A_c \omega_{02}, B_\alpha A_m \omega_{02}&\cdots\\
q=1&\omega_{01}&A_m \omega_{01}, \omega_{11}&A_m^2 \omega_{01}, B_\alpha \omega_{01}, A_m \omega_{11} &A_m^3 \omega_{01}, B_\alpha A_m \omega_{01}, A_m^2 \omega_{11}, B_\alpha \omega_{11}&\cdots\\
q=0&1&A_c, A_m&A_c^2, A_m^2, B_\alpha&A_c^3, A_m^3, B_\alpha A_c, B_\alpha A_m&\cdots\\
\hline
E^{p,q}_2&p\!=\!0&p\!=\!1&p\!=\!2&p\!=\!3&\cdots
\end{array}
\end{equation}
subject to the relations $(A_c + A_m) \omega_{01} = 0$ and $(A_c + A_m) \omega_{11} = 0$. Also, to account for elements that do not explicitly appear in the spectral sequence, several relations must exist:  $\omega_{01}^2 +\dots = 0$, $\omega_{11}^2+\dots = 0$, $\omega_{01} \omega_{11} +\dots =0$, $\omega_{02}^2+\dots = 0$, $\omega_{02} \omega_{01} + \dots =0$, and $\omega_{02} \omega_{11} + \dots =0$. %

For $p4g$, we have a nontrivial differential $d_2$. From the Leibniz rule Eq.~\eqref{eq:Leibniz}, it suffices to write down the image of the generators $\omega_{01},\omega_{11}$ and $\omega_{02}$ under $d_2$, which are
\begin{equation}
    d_2(\omega_{01}) = A_c^2,\quad d_2(\omega_{11}) = 0,\quad d_2(\omega_{02}) = A_m \omega_{11}.
\end{equation}
Thus, at the $E_3$ page, we have
\begin{equation}\label{p4gLHSSSE3}
\begin{array}{c|ccccc}
q=2&0&\widetilde{\omega_{12}} & A_m \widetilde{\omega_{12}} & A_m^2 \widetilde{\omega_{12}}, B_\alpha \widetilde{\omega_{12}}&\cdots\\
q=1&0&\omega_{11}&0 &B_\alpha \omega_{11}&\cdots\\
q=0&1&A_c, A_m&A_m^2, B_\alpha&A_m^3, B_\alpha A_c, B_\alpha A_m&\cdots\\
\hline
E^{p,q}_3&p\!=\!0&p\!=\!1&p\!=\!2&p\!=\!3&\cdots
\end{array}
\end{equation}
Note that even though $\omega_{01}$ and $\omega_{02}$ have nontrivial image under $d_2$ and are hence killed at the $E_3$ page, not everything involving $\omega_{01}$ or $\omega_{02}$ is killed. Specifically, $d_2((A_c+A_m)\omega_{02}) = (A_c + A_m)A_m\cdot \omega_{11} = 0$ hence $(A_c+A_m)\omega_{02}$ survives and descends to $\widetilde{\omega_{12}}$. It turns out that $E_3$ collapses to $E_\infty$ and the analysis is done. From this, we can directly see that there are five generators in total, i.e., $A_c,A_m$ at degree $p+q=1$, $\omega_{11},B_\alpha$ at degree $p+q=2$ and $\widetilde{\omega_{12}}$ at degree $p+q=3$. These five generators exactly correspond to the five generators of the mod-2 cohomology ring of $p4g$. Moreover, there are relations involving $A_c^2$, $A_c A_m$, $A_c \omega_{11}$, $A_m \omega_{11}$, $\omega_{11}^2$, $A_c\widetilde{\omega_{12}}$, $\omega_{11} \widetilde{\omega_{12}}$, $\widetilde{\omega_{12}}^2$ in $E^{*,*}_\infty$, all of which should correspond to an independent relation of the mod-2 cohomology ring of $p4g$. 

Indeed, from our code, the mod-2 cohomology ring of $p4g$ is given by
\begin{equation}
 \mathbb{Z}_2\left[A_c, A_m, B_{\alpha}, B_{\beta}, C_{\gamma}\right]/\langle\mathcal{R}_2,\mathcal{R}_3,\mathcal{R}_4,\mathcal{R}_5,\mathcal{R}_6\rangle,
 \end{equation}
where the relations $\mathcal{R}_{2,3,4,5,6}$ at degree $2,3,4,5,6$ are 
\begin{subequations} 
 \begin{align}
\mathcal{R}_2\colon & ~~A_c^2,~~A_c A_m,\\
\mathcal{R}_3\colon & ~~A_c B_\beta,~~A_m (B_\alpha + B_{\beta}),\\
\mathcal{R}_4\colon & ~~B_\beta(B_\alpha + B_\beta),~~A_c C_{\gamma},\\
\mathcal{R}_5\colon & ~~(B_\alpha + B_\beta) C_{\gamma},\\
\mathcal{R}_6\colon & ~~C_\gamma^2+B_\beta^3 + A_m B_\alpha C_\gamma.
\end{align} 
 \end{subequations}
 We see that all the generators and relations match our analysis from the LHS spectral sequence. In particular, there are indeed five generators in total, with $A_c,A_m,B_\alpha,B_\beta,C_\gamma$ \footnote{In Ref.~\cite{10.21468/SciPostPhys.13.3.066}, they are denoted as $A_c,A_s,B_{c^2},B_{c(x+y)},C_{c^2(x+y)}$, respectively. Here we follow the naming convention of 3D space groups in Sec.~\ref{subsec:3D}.} descending to $A_c,A_m,B_\alpha,\omega_{11},\widetilde{\omega_{12}}$ . The relations all match our analysis of the LHS spectral sequence, even though we cannot fix their explicit form solely from this analysis. Also, we see how a degree-3 generator of the mod-2 cohomology ring of wallpaper groups can appear. 
 
According to the analysis of all 17 wallpaper groups in Ref.~\cite{10.21468/SciPostPhys.13.3.066}, it turns out that $p4g$ is the \emph{only} wallpaper group which has a degree-3 generator, while other wallpaper groups have at most degree-2 generators. We note that the group $p4g$ is a subgroup of the 3D space group \hyperref[subsub:sg100]{No. 100 ($P4bm$)} upon removing the generator $T_z$; furthermore, the mod-2 cohomology ring of $p4g$ is isomorphic to that of \hyperref[subsub:sg100]{No. 100} upon removing the generator $A_z$.

\subsubsection{Example: No. 42 \texorpdfstring{($Fmm2$)}{(Fmm2)} and collapse at \texorpdfstring{$E_2$}{E2} page}

The analysis for 3D space groups is much more complicated. Still, we manage to perform the analysis for several nontrivial groups. As a side quest, we ask the following question: \emph{when does the standard LHS spectral sequence collapse at $E_2$ page?} This question was raised for general crystallographic groups in any dimension at the end of Ref.~\cite{adem2008compatible}. A non-collapsing $E_2$, or equivalently the existence of nonzero differentials $d_2$, can have important physical implications \cite{PhysRevB.101.245160,Wang2021}.

For this purpose, let us study the group \hyperref[subsub:sg42]{No. 42 ($Fmm2$)} in more detail. The point group of \hyperref[subsub:sg42]{No. 42} is $C_{2v}$, which as an abstract group is $\Z_2\times\Z_2$. This group is generated by three translations $T_{1,2,3}$, a two-fold rotation $C_2$, and a mirror reflection $M$:
\begin{subequations}
 \begin{align}
T_1 &\colon (x,y,z)\rightarrow (x, y+1/2, z+1/2),\\ 
T_2 &\colon (x,y,z)\rightarrow (x+1/2, y, z+1/2),\\
T_3 &\colon (x,y,z)\rightarrow (x+1/2, y+1/2, z),\\
C_2 &\colon (x,y,z)\rightarrow (-x, -y, z),\\ 
M &\colon (x,y,z)\rightarrow (x, -y, z).
\end{align}
\end{subequations}
This group is a symmorphic space group which splits, i.e., $G=T\rtimes P$. Yet, we will see that the $E_2$ page of $G$ does not collapse, i.e. $E_2$ page is not isomorphic to $E_\infty$.

The mod-2 cohomology ring for $C_{2v}\cong\Z_2\times \Z_2$ is
\begin{equation}\label{C2vcohoring}
    H^*(C_{2v}, \Z_2) = \Z_2[A_c, A_m],
\end{equation}
where the generators $A_c$ and $A_m$ have the following explicit cochain representative in the bar resolution,
\begin{equation}
    A_c(C_2^c M^m) = c,\quad A_m(C_2^c M^m) = m,\quad c,m\in\{0,1\}.
\end{equation}
We can say that $A_c$ and $A_m$ are $\Z_2$ characters for $C_2$ and $M$, respectively. 

The $q=0$ and $q=3$ rows of the $E_2$ page can be directly obtained from Eq.~\eqref{C2vcohoring}. For the $q=1$ row, we have
\begin{equation}
    H^*(C_{2v}, H^1(\Z^3,\Z_2)) = H^*(C_{2v}, \Z_2).[\omega_{01}, \omega_{01}']/(A_c \omega_{01} + A_m \omega_{01}'),
\end{equation}
where $\omega_{01}$ and $\omega_{01}'$ are two generators (of the module) at degree 0;
For the $q=2$ row, we have
\begin{equation}
    H^*(C_{2v}, H^2(\Z^3,\Z_2)) = H^*(C_{2v}, \Z_2).[\omega_{02}, \omega_{12}]/(A_c \omega_{02}, A_m \omega_{02}).
\end{equation}
where $\omega_{02}$ and $\omega_{12}$ are two generators  at degree 0 and 1, respectively. Therefore, the LHS spectral sequence at the $E_2$ page has the following form
\begin{equation}\label{Fmm2LHSSSE2}
\begin{array}{c|cccccc}
q=3&\omega_{03}&A_c \omega_{03}, A_m \omega_{03}&A_c^2 \omega_{03},A_c A_m \omega_{03}, A_m^2 \omega_{03}&A_c^3 \omega_{03}, A_c^2 A_m \omega_{03}, A_c A_m^2 \omega_{03}, A_m^3 \omega_{03}&\cdots\\
q=2&\omega_{02}&\omega_{12}&A_c \omega_{12}, A_m \omega_{12}&A_c^2 \omega_{12},A_c A_m \omega_{12}, A_m^2 \omega_{12}&\cdots \\
q=1&\omega_{01},\omega_{01}'&A_c \omega_{01}, A_m \omega_{01}, A_c \omega_{01}'&A_c^2 \omega_{01}, A_c A_m \omega_{01}, A_m^2 \omega_{01}, A_c^2 \omega_{01}'&A_c^3 \omega_{01}, A_c^2 A_m \omega_{01}, A_c A_m^2 \omega_{01}, A_m^3 \omega_{01}, A_c^3 \omega_{01}'&\cdots\\
q=0&1&A_c, A_m&A_c^2,A_c A_m, A_m^2&A_c^3, A_c^2 A_m, A_c A_m^2, A_m^3&\cdots\\
\hline
E^{p,q}_2&p\!=\!0&p\!=\!1&p\!=\!2&p\!=\!3&\cdots
\end{array}
\end{equation}
It turns out that $d_2$ maps $\omega_{01}$, $\omega_{01}'$, $\omega_{02}$, and $\omega_{03}$ to zero while we calculate in Appendix~\ref{app:calculation_d2} that it maps $\omega_{12}$ to
\begin{equation}
    d_2(\omega_{12}) = (A_c^3 + A_c A_m^2) \omega_{01} .
\end{equation}
Hence, the $E_2$ page is non-collapsing for the symmorphic group \hyperref[subsub:sg42]{No. 42~($Fmm2$)}. At the $E_3$ page, we have
\begin{equation}\label{Fmm2LHSSSE3}
\begin{array}{c|cccccc}
q=3&\omega_{03}&A_c \omega_{03}, A_m \omega_{03}&A_c^2 \omega_{03},A_c A_m \omega_{03}, A_m^2 \omega_{03}&A_c^3 \omega_{03}, A_c^2 A_m \omega_{03}, A_c A_m^2 \omega_{03}, A_m^3 \omega_{03}&\cdots\\
q=2&\omega_{02}&0&0&0&\cdots \\
q=1&\omega_{01},\omega_{01}'&A_c \omega_{01}, A_m \omega_{01}, A_c \omega_{01}'&A_c^2 \omega_{01}, A_c A_m \omega_{01}, A_m^2 \omega_{01}, A_c^2 \omega_{01}'&A_c^3 \omega_{01}, A_c^2 A_m \omega_{01}, A_m^3 \omega_{01}, A_c^3 \omega_{01}'&\cdots\\
q=0&1&A_c, A_m&A_c^2,A_c A_m, A_m^2&A_c^3, A_c^2 A_m, A_c A_m^2, A_m^3&\cdots\\
\hline
E^{p,q}_2&p\!=\!0&p\!=\!1&p\!=\!2&p\!=\!3&\cdots
\end{array}
\end{equation}
$E_3$ collapses to $E_\infty$ and the analysis is done. Also, we have $\omega_{02} = \omega_{01} \omega_{01}'$. From this, we can directly see that there are five generators in total, i.e., $A_c,A_m,\omega_{01},\omega_{01}'$ at degree $p+q=1$ and $\omega_{03}$ at degree $p+q=3$, and we should have relations involving $A_c \omega_{01} + A_m \omega_{01}'$, $\omega_{01}^2$, $\omega_{01}'^2$, $(A_c^3+A_c A_m^2)\omega_{01}$, $\omega_{01} \omega_{03}$, $\omega_{01}' \omega_{03}$ and $\omega_{03}^2$.

The $\mathbb{Z}_2$ cohomology ring of \hyperref[subsub:sg42]{No. 42~($Fmm2$)} is given by
\begin{equation}
\mathbb{Z}_2[A_c,A_m,A_{x+z},A_{y+z},C_{xyz}]/\langle\mathcal{R}_2,\mathcal{R}_4,\mathcal{R}_6\rangle
 \end{equation}
where the relations are 
\begin{subequations} 
 \begin{align}
\mathcal{R}_2\colon & ~~
A_c A_{x+z} + A_m A_{x+z} + A_c A_{y+z},~~A_{x+z}^2 + A_c A_{y+z},~~A_{y+z} (A_c + A_{y+z}),\\
\mathcal{R}_4\colon & ~~
A_{x+z} C_{xyz},~~A_{y+z} C_{xyz},\\
\mathcal{R}_6\colon & ~~
C_{xyz}^2.
\end{align} 
 \end{subequations}
There are indeed five generators in total, with $A_c,A_m,A_{x+z}+A_{y+z},A_{x+z},C_{xyz}$ descending to $A_c,A_m,\omega_{01},\omega_{01}',\omega_{03}$. Note that the relation $(A_c^3+A_c A_m^2)(A_{x+z} + A_{y+z})$ is not an independent relation and can be derived from the relations $\mathcal{R}_2$. Other relations match the relations in $E^{*,*}_\infty$.

A number of examples of split crystallographic groups whose LHS spectral sequence does not collapse at $E_2$ were given in Ref.~\cite{totaro1996cohomology}, and later the obstruction characteristic classes were studied in Ref.~\cite{PETROSYAN20092916}. Here we provide a complete list for this phenomenon to occur in 3D space groups. Out of the 73 symmorphic space groups, the following 10 symmorphic groups do not have a collapsing $E_2$ page: 
\begin{equation}
\begin{aligned}
&\hyperref[subsub:sg42]{42~(Fmm2)},&&\hyperref[subsub:sg69]{69~(Fmmm)},&&\hyperref[subsub:sg87]{87~(I4/m)},&&\hyperref[subsub:sg107]{107~(I4mm)},&&\hyperref[subsub:sg121]{121~(I\bar{4}2m)},\\
&\hyperref[subsub:sg139]{139~(I4/mmm)},&&\hyperref[subsub:sg202]{202~(Fm\bar{3})},&&\hyperref[subsub:sg217]{217~(I\bar{4}3m)},&&\hyperref[subsub:sg225]{225~(Fm\bar{3}m)},&&\hyperref[subsub:sg229]{229~(Im\bar{3}m)}.
\end{aligned}
\end{equation}
Note the subgroup relations
\begin{equation}
\begin{aligned}
&\hyperref[subsub:sg42]{42}\subset \hyperref[subsub:sg69]{69},&&
\hyperref[subsub:sg69]{69},\hyperref[subsub:sg87]{87},\hyperref[subsub:sg107]{107},\hyperref[subsub:sg121]{121}\subset \hyperref[subsub:sg139]{139},&&\hyperref[subsub:sg121]{121}\subset \hyperref[subsub:sg217]{217},\\
&\hyperref[subsub:sg69]{69}\subset \hyperref[subsub:sg202]{202},&& \hyperref[subsub:sg139]{139},\hyperref[subsub:sg202]{202}\subset \hyperref[subsub:sg225]{225},&&\hyperref[subsub:sg139]{139},\hyperref[subsub:sg217]{217}\subset \hyperref[subsub:sg229]{229},
\end{aligned}
\end{equation}
therefore, the issue of a nonzero differential $d_2$ starts to appear at \hyperref[subsub:sg42]{No. 42~($Fmm2$)} \footnote{Note that this group is an example for Corollary 1 of Ref.~\cite{totaro1996cohomology}.}, and this propagates to the larger symmorphic groups containing \hyperref[subsub:sg42]{No. 42} as a subgroup. Moreover, the following 13 nonsymmorphic groups have a collapsing $E_2$ page: 
\begin{equation}\label{nonsymcolElapse2}
\begin{aligned}
\hyperref[subsub:sg77]{77~(P4_2)},
~~\hyperref[subsub:sg93]{93~(P4_222)},
~~\hyperref[subsub:sg144]{144~(P3_1)},
~~\hyperref[subsub:sg145]{145~(P3_2)},
~~\hyperref[subsub:sg151]{151~(P3_112)},
~~\hyperref[subsub:sg152]{152~(P3_121)},
~~\hyperref[subsub:sg153]{153~(P3_212)},\\
~~\hyperref[subsub:sg154]{154~(P3_221)},~~\hyperref[subsub:sg171]{171~(P6_2)},~~\hyperref[subsub:sg172]{172~(P6_4)},~~\hyperref[subsub:sg180]{180~(P6_222)},~~\hyperref[subsub:sg181]{181~(P6_422)},~~\hyperref[subsub:sg208]{208~(P4_232)}.
\end{aligned}
\end{equation}
These groups and the 63 (of the 73) symmorphic groups that have a collapsing $E_2$ page are labeled with a ``\,\customyinyang\," in the third column of Table \ref{tablemod2230}. The 13 nonsymmorphic groups in Eq.~\eqref{nonsymcolElapse2} and the 73 symmorphic groups constitute the 86 groups having $E_\infty^{0,3} = \mathbb{Z}_2$ in the standard LHS spectral sequence. All other 144 groups have $E^{0,3}_2=\mathbb{Z}_2$ and $E^{0,3}_3=0$.

Finally, a refined conjecture was raised by Adem, Ge, Pan, and Petrosyan \cite{adem2008compatible}: the $E_2$ page collapses for $k$-dimensional crystallographic groups $G=\mathbb{Z}^k\rtimes P$ when $P$ is a cyclic group. Here we find that this conjecture holds for all the 3D ($k=3$) symmorphic space groups of this form. A counterexample has been found in the 6D ($k=6$) crystallographic groups \cite{LANGER20121318}.

\subsection{Finiteness theorem}\label{subsec:finite}

Building on the discussion regarding the LHS spectral sequence, in this subsection, we prove the following theorem, which states that the mod-2 cohomology ring of any crystallographic group $G$ is finitely generated.
\begin{theorem} \label{thm:finite} (Finiteness condition)
    The $\mathbb{F}$ cohomology ring of any crystallographic group $G$ must be finitely generated with a finite number of generators and relations. 
\end{theorem}

The proof follows the corresponding proof for finite groups in Ref.~\cite{Evens1991}, also based on the analysis of the LHS spectral sequence. This theorem guarantees that our code always ends at a finite degree.

\begin{proof}

Consider the standard LHS spectral sequence. At the $E_2$ page, every row is of the form $H^*(P, \mathbb{F}^m)$ where $\mathbb{F}^m$ is a finitely generated $\mathbb{F}$-module. The following theorem dictates that every row is a finitely generated module, i.e., a Noetherian module, over $H^*(P, \mathbb{F})$ \footnote{In general, a module is said to be \emph{Noetherian} if every submodule is finitely generated. Still, we know that $H^*(P, \mathbb{F})$ is a Noetherian ring, and a finitely generated module over a Noetherian ring is a Noetherian module.}.

\begin{theorem} (Evens \cite{Evens1991})
Let $P$ be a finite group, $k$ a commutative ring on which $P$ acts trivially, and $M$ a $kP$-module. If $M$ is Noetherian as a $k$-module, then $H^*(P, M)$ is Noetherian over $H^*(P, k)$.    
\end{theorem}

Together with the bilinear product at the $E_2$ page, we see that the $E_2$ page of the standard LHS spectral sequence is a Noetherian module over $H^*(P, \mathbb{F})$. Starting from the $E_2$ page, the $E_{r+1}$ page is a quotient module of a submodule, i.e., $\ker d_r/\im d_r$, of the $E_r$ page. Since the submodule and the quotient module of a Noetherian module are all Noetherian, the $E_r$ page for any $r$ is Noetherian. Corollary \ref{cor:end} says that $E_{k+2}=E_\infty$, hence the $E_\infty$ page is also a Noetherian module over $H^*(P, \mathbb{F})$. The same proof as in Lemma 7.4.5 of Ref.~\cite{Evens1991} then dictates that $H^*(G, \mathbb{F})$ is a Noetherian module over $H^*(P, \mathbb{F})$. Therefore, the Hilbert basis theorem dictates that $H^*(G, \mathbb{F})$ is a Noetherian ring itself. By the usual argument, the cohomology in positive degrees, denoted as $H^+(G, \mathbb{F})$, is a finitely generated ideal, and a set of ideal generators is also a set of
generators for $H^*(G, \mathbb{F})$ as an algebra over $\mathbb{F}$. The Noetherian condition also guarantees that we only need a finite number of relations.

\end{proof}

\subsection{On the integral cohomology of crystallographic groups}\label{subsec:integral}

In this subsection, we discuss the relevant maps to relate mod-2 cohomology with integral cohomology. This connects our results of mod-2 cohomology with cohomology with other coefficients. The complete results for the integral cohomology of the 230 space groups are collected in Table~\ref{tableZZor230}.

Let $\mathbb Z^\rho$ denote the abelian group $\mathbb Z$ equipped with a sign action $\rho\colon G\to\{\pm1\}$. In the applications below, $\rho$ is either the trivial action or the orientation action, where orientation-reversing crystalline symmetries act by $-1$. The coefficient sequence
\begin{equation}\label{eq:twisted-short-exact}
0\longrightarrow \mathbb Z^\rho
\xrightarrow{\times 2}
\mathbb Z^\rho
\xrightarrow{\mathrm{mod}\,2}
\mathbb Z_2
\longrightarrow 0
\end{equation}
is $G$-equivariant because the sign action becomes trivial after reduction mod $2$. It gives the long exact sequence
\begin{equation}\label{eq:LES}
    \cdots \rightarrow H^{n-1}(G, \Z_2) \xrightarrow{\beta_{n-1}^{\rho}} H^n(G, \Z^\rho)\xrightarrow{\times 2}H^n(G,\mathbb{Z}^\rho)
\xrightarrow{\mathrm{mod}\,2}
H^{n}(G,\mathbb{Z}_2)\xrightarrow{\beta_n^\rho}\cdots .
\end{equation}
Here $\beta_n^\rho$ is the Bockstein homomorphism associated with Eq.~\eqref{eq:twisted-short-exact}.

Equivalently, for each $n$ there is a natural short exact sequence
\begin{equation}\label{Z2relatedtoU1}
0\longrightarrow H^n(G,\mathbb Z^\rho)\otimes\mathbb Z_2
\longrightarrow H^n(G,\mathbb Z_2)
\xrightarrow{\beta_n^\rho}
\operatorname{Tor}^{\mathbb Z}_1\!\left(H^{n+1}(G,\mathbb Z^\rho),\mathbb Z_2\right)
\longrightarrow 0,
\end{equation}
where the last term may also be written as the subgroup of elements of order dividing $2$ in $H^{n+1}(G,\mathbb Z^\rho)$. After choosing a splitting, Eq.~\eqref{Z2relatedtoU1} gives the noncanonical decomposition
\begin{equation}
H^n(G,\mathbb{Z}_2) \cong (H^n(G,\mathbb{Z}^\rho)\otimes \mathbb{Z}_2) \oplus 
\mathrm{Tor}^{\mathbb{Z}}_1(H^{n+1}(G,\mathbb{Z}^\rho),\mathbb{Z}_2).
\end{equation}
Thus the first summand is intrinsically characterized as $\ker \beta_n^\rho$.

The LSM classes discussed later lie in $\ker \beta_n^{\mathrm{or}}$, where $\mathbb Z^{\mathrm{or}}$ is the orientation local system dictated by the Crystalline Equivalence Principle \cite{PhysRevX.8.011040}. Reducing the twisted Bockstein modulo $2$ defines the twisted Steenrod operation
\begin{equation}\label{eq:twisted_SQ1}
    \mathcal{SQ}^1_\rho := (\mathrm{mod}\,2)\circ \beta_n^\rho\colon H^n(G, \Z_2) \rightarrow H^{n+1}(G, \Z_2).
\end{equation}
When $\rho$ is the trivial action, $\mathcal{SQ}^1_\rho$ is the ordinary Steenrod square $Sq^1$. For a sign local system, the standard cochain calculation gives
\begin{equation}
\mathcal{SQ}^1_\rho(\lambda) = Sq^1(\lambda) + w_\rho\cup \lambda,
\end{equation}
where $w_\rho\in H^1(G,\mathbb Z_2)$ is the mod-$2$ character detecting the elements acting by $-1$ on $\mathbb Z^\rho$, and in the orientated case $w_\rho$ is just trivial. We will return to this operation in Sec.~\ref{sec:LSM} when we discuss LSM constraints.

\subsection{Periodicity of mod-2 cohomology}\label{subsec:periodic}

In this subsection, we collect some interesting theorems about the periodicity of $\Z_2$ rank for mod-2 cohomology of crystallographic groups. Our results for 2D wallpaper groups and 3D space groups indeed align with these general theorems, which apply to all crystallographic groups.

We have the following theorem regarding the periodicity of the resolution for a group $G$ \footnote{We always assume that the resolution is with respect to $\Z$.}.
\begin{theorem}\label{thm:periodic}
(Brown \cite{brown2012cohomology}) A group $G$ with a finite-index torsion-free nilpotent subgroup admits a resolution which is periodic in sufficiently high degrees if and only if all of its finite-index subgroups admit periodic resolutions.
\end{theorem}
Therefore, if we want the $\Z_2$ ranks for mod-2 cohomology of crystallographic groups to be periodic in sufficiently high degrees, we need to check all of its finite index subgroups. 

An interesting class of crystallographic groups is called Bieberbach group \cite{charlap2012bieberbach}, which is the extreme limit of crystallographic groups which satisfy the criterion of Theorem~\ref{thm:periodic}.
\begin{definition} (Bieberbach group)
    A crystallographic group is a Bieberbach group if it is torsion-free.
\end{definition}
In fact we can say more about these Bieberbach groups \cite{charlap2012bieberbach}. For a $k$-dimensional crystallographic group $G$, consider the orbit space $\mathbb{R}^k/G$. For a general crystallographic group $G$, $\mathbb{R}^k/G$ may contain orbifold singularities. However, when $G$ is a Bieberbach group, $\mathbb{R}^k/G$ is a smooth compact flat $k$-manifold. Conversely, every compact flat $k$-manifold arises in this way. Moreover, $G$ is the fundamental group of $\mathbb{R}^k/G$, and $\mathbb{R}^k/G$ is a $K(G,1)$.

This correspondence identifies the group cohomology of $G$ with the ordinary cohomology of the associated flat manifold. The correct top-degree statement must remember orientability.
\begin{theorem}\label{thm:Bieberbach}
Let $G$ be a $k$-dimensional Bieberbach group and let $M=\mathbb R^k/G$. Then $G$ is a Poincar\'e duality group of dimension $k$. In particular,
\begin{equation}
H^n(G,\mathbb Z^\rho)=0\qquad (n>k)
\end{equation}
for the trivial and orientation local systems. Moreover,
\begin{equation}
H^k(G,\mathbb Z)\cong
\begin{cases}
\mathbb Z, & M\text{ orientable},\\
0, & M\text{ nonorientable},
\end{cases}
\qquad
H^k(G,\mathbb Z^{\mathrm{or}})\cong \mathbb Z,
\end{equation}
and
\begin{equation}
H^k(G,\mathbb Z_2)\cong\mathbb Z_2,
\qquad
H^n(G,\mathbb Z_2)=0\quad(n>k).
\end{equation}
\end{theorem}
\begin{proof}
Because $G$ is torsion-free and acts freely, properly discontinuously, and cocompactly on the contractible manifold $\mathbb R^k$, the quotient $M=\mathbb R^k/G$ is a closed aspherical $k$-manifold. Hence $BG\simeq M$. The assertions are precisely Poincar\'e duality for the closed manifold $M$, with the orientation local system in the nonorientable case, together with the fact that a $k$-manifold has no cohomology in degrees greater than $k$.
\end{proof}

From the point of view of a $k$-dimensional lattice with symmetry $G$, $\mathbb{R}^k/G = K(G, 1)$ can be thought of as the \emph{fundamental domain} associated with crystallographic symmetry $G$, which also corresponds to the unique IWP for the Bieberbach group $G$.

The criterion of Theorem~\ref{thm:periodic} can also be satisfied if the point group $P$ of $G$ admits periodic resolution. Hence, we have the following easy corollary (see also Ref.~\cite{ellis2019invitation} page 247--248), 
\begin{cor}\label{cor:periodic}
For an $k$-dimensional crystallographic group $G$ with the associated point group $P$, suppose that the point group $P$ admits a periodic resolution of period $d$. Then, the crystallographic group $G$ admits a periodic resolution of period $d$ in degrees greater than $n$. In particular, $H^k(G, \Z_2) = H^{k+d}(G, \Z_2)$ for all integers $k>n$. 
\end{cor}

\subsubsection{Example: 2D wallpaper groups}

To illustrate these theorems, let us check the criterion for 2D wallpaper groups and we will see that these theorems indeed match with the explicit calculation for individual groups in Ref.~\cite{10.21468/SciPostPhys.13.3.066}.

\begin{enumerate}

\item There are exactly two Bieberbach groups in 2D, $p1$ and $pg$. For both groups $G$, we have $H^2(G, \Z_2) = \Z_2$ and $H^k(G, \Z_2)=0,k\geq 3$. The flat manifolds $\mathbb{R}^2/G=K(G,1)$ for $p1$ and $pg$ are in fact torus and the Klein bottle, respectively.

\item For 2D wallpaper groups, there are six point groups satisfying the condition of having a periodic resolution:
\begin{equation}
C_2,D_1 \cong \mathbb{Z}_2, ~~~ C_4\cong \mathbb{Z}_4,~~~C_3\cong \mathbb{Z}_3,~~~ D_3\cong Dih_3,~~~ C_6 \cong \mathbb{Z}_6. 
\end{equation}
The period is one for $\Z_2$, two for $\Z_4$, $\Z_3$, $\Z_6$, and four for $Dih_3$, and for each of these groups, the mod-2
cohomology groups in all positive degrees are the same \cite{brown2012cohomology,swan1960periodic}. The corresponding wallpaper groups are $p2$ ($C_2$), $pm$, $pg$, $cm$ ($D_1$), $p4$ ($C_4$), $p3$ ($C_3$), $p3m1$, $p31m$ ($D_3$) and $p6$ ($C_6$). We see that indeed for each of these wallpaper groups, the mod-2 cohomology groups in degrees greater than 2 are the same.

\item There are two more wallpaper groups---$pmg$ and $pgg$---that satisfy the condition of Theorem~\ref{thm:periodic}. They also have a periodic resolution and the mod-2 cohomology groups in degrees equal to or greater than 2 are the same as well.

\end{enumerate}

We will see that these patterns reappear in the study of 3D space groups.

\section{Mod-2 Cohomology of 3D space groups}\label{sec:cohomology}

In this section, we sketch how we perform the calculation for the cohomology of crystallographic groups in 2D and 3D, and highlight certain important features of the results. 

There are standard textbook methods for calculating group cohomology \cite{brown2012cohomology,adem2013cohomology}, many of which are already implemented in GAP \cite{ellis2019invitation}. The GAP routine of computing cohomology rings has so far been limited to $p$-groups. Nevertheless, we can use the elementary GAP routines as building blocks and calculate the mod-2 cohomology ring $H^*(G,\mathbb{Z}_2)$ of a space group $G$ algorithmically. Importantly, the algorithm requires us to calculate the \emph{contracting homotopy} of a resolution of space group $G$ to sufficiently high degrees. Following this strategy, we create a program tailored for computing the mod-2 cohomology ring of space groups \footnote{The implementation is in our GitHub repository \cite{github}; the main package files are \texttt{gap/data.gi} and \texttt{gap/functions.gi}.}. An important byproduct of our work is exactly the code to calculate the contracting homotopy of the resolution built from GAP function \texttt{ResolutionSpaceGroup}. We hope our algorithm will also be useful for calculating the cohomology rings of more general groups.

Even more explicitly, to connect these cohomology data to lattice data, we managed to write down the inhomogeneous functions representing the cohomology generators at degree-1, 2, and 3. We write additional code to obtain the mapping from these function-represented cocycles to the vector-represented cocycles, connecting the output data of these two separate methods. The explicit inhomogeneous functions allow us to evaluate topological invariants associated with IWPs, establishing the final link we aim for---a correspondence between 3-cocycles and IWPs.

\subsection{Methods of calculation}\label{subsec:methods}

The datum encoding the cohomology of any group $G$ is a \emph{free $\mathbb{Z}G$ resolution of $\mathbb{Z}$}: it is an acyclic chain complex consisting of free $\mathbb{Z}G$-modules $R^G_n$ and homomorphisms $\partial_n$
\begin{equation}\label{res}
R^G_*\colon \cdots \xrightarrow {\partial_{n+2}} R^G_{n+1}\xrightarrow{\partial_{n+1}} R^G_k\xrightarrow {\partial_n} R^G_{n-1}
\xrightarrow{\partial_{n-1}}\cdots \xrightarrow{\partial_2}
R^G_1\xrightarrow{\partial_1}
R^G_0 \xrightarrow{\partial_0} \Z,
\end{equation}
such that $\ker \partial_n = \im \partial_{n+1}$ for every $n$. The cohomology calculation we did relies exclusively on the construction and manipulation of the resolution. %

The resolution $R^G_*$ may be finitely or infinitely generated, depending on whether each $R^G_k$ is finitely or infinitely generated as a $\mathbb{Z}G$-module. For an infinite-order crystallographic group $G$, GAP always produces a finitely generated resolution, which is easy to work with but quite abstract. In contrast, the familiar bar resolution provides clearer meanings but is infinitely generated, making it more challenging to use. 
These two resolutions serve as the foundation for the two independent calculations presented in Secs.~\ref{method1} and \ref{method2}.

In Sec.~\ref{method1}, we outline the standard procedures implemented in GAP for computing resolutions for space groups $G$ and highlight the algorithm that we develop to calculate contracting homotopy. Then, in Sec.~\ref{method2}, we explain how the familiar bar resolution method can be applied to obtain the standard cocycle functions for crystallographic groups. The key to this calculation is a conjecture on the restriction of cohomology to a finite lattice space group that, if holds, allows to convert the computation of cocycle functions from an infinite problem to a finite problem. Finally, in Sec.~\ref{connect_method1_method2} we explain how to connect the outputs of the two methods using the contracting homotopy for the bar resolution.

\subsubsection{Resolution from GAP and contracting homotopy}\label{method1}

This computation leverages the full power of GAP \cite{GAP4}. 
The resolution \eqref{res} can be computed internally in GAP. Once the resolution is constructed, the cohomology $H^n(G,A)$ as an abelian group is then obtained in a straightforward manner by first taking the $\mathrm{Hom}$ functor with the coefficient module $A$ and then the homology of the cochain maps. The calculation of the ring structure in GAP relies crucially on the data of \emph{contracting homotopy}:

\begin{definition}
(Contracting homotopy. \cite{ellis2019invitation})
Let $R^G_*$ be a free $\mathbb{Z}G$-resolution of $\mathbb{Z}$. A contracting homotopy on $R^G_*$ consists of a sequence of abelian group homomorphisms $h_n\colon R^G_n \rightarrow  R^G_{n+1}, n \geq  0$, satisfying
\begin{equation}
h_{n-1}\partial_n + \partial_{n+1}h_n = 1,
\end{equation}
for $n \geq  0$ with $h_{-1} = 0$. The homomorphisms $h_k$ do not have to preserve the action of $G$.
\end{definition}

The detailed procedure to obtain the ring structure from the contracting homotopy is as follows. The contracting homotopy is used to construct a chain-level diagonal approximation
\begin{equation}
\Delta : R_\bullet \longrightarrow R_\bullet \otimes_{\mathbb Z} R_\bullet,
\end{equation}
which is a chain map lifting the ordinary diagonal map
\begin{equation}
\mathbb Z \longrightarrow \mathbb Z\otimes \mathbb Z,
\qquad
1\longmapsto 1\otimes 1.
\end{equation}
The diagonal approximation is constructed recursively. In degree \(0\), one sends the preferred generator \(e_0\in R_0\) to
\begin{equation}
\Delta(e_0)=e_0\otimes e_0.
\end{equation}
Suppose \(\Delta\) has already been defined in degrees \(<n\), and let \(x\in R_n\) be a basis element. The chain-map condition requires
\begin{equation}
\partial_{\otimes}\Delta(x)
=
\Delta(\partial x),
\end{equation}
where the tensor-product differential is
\begin{equation}
\partial_{\otimes}(a\otimes b)
=
\partial a\otimes b
+
(-1)^{|a|}a\otimes \partial b.
\end{equation}
The element \(\Delta(\partial x)\) is a cycle in \(R_\bullet\otimes R_\bullet\), because
\begin{equation}
\partial_{\otimes}\Delta(\partial x)
=
\Delta(\partial^2 x)
=
0.
\end{equation}
Therefore one needs to solve the equation
\begin{equation}
y~\text{ s.t. } ~~ \partial_{\otimes} y=\Delta(\partial x).
\end{equation}
This is exactly where the contracting homotopy is used. The contracting homotopy on \(R_\bullet\) induces a contracting homotopy \(S\) on the tensor product complex \(R_\bullet\otimes R_\bullet\), so one may define
\begin{equation}
\Delta(x)
=
S\bigl(\Delta(\partial x)\bigr).
\end{equation}
This formula is applied only to the chosen $\mathbb{Z}G$-basis and then extended by $\Delta(gx)=g\Delta(x)$, since $S$ need not be $G$-equivariant. Thus the contracting homotopy provides the chain-level fillings needed to construct the diagonal approximation degree by degree.

Once the diagonal approximation is known, the cup product of cochains is defined by composing with \(\Delta\). Explicitly, after constructing a free \(\mathbb ZG\)-resolution \(R_\bullet \to \mathbb Z\) together with a contracting homotopy, the mod \(2\) cohomology groups are computed from the cochain complex
\begin{equation}
C^n(G;\mathbb F_2)
=
\operatorname{Hom}_{\mathbb ZG}(R_n,\mathbb F_2).
\end{equation}
To obtain the ring structure on \(H^*(G;\mathbb F_2)\), one must compute cup products
\begin{equation}
H^p(G;\mathbb F_2)\times H^q(G;\mathbb F_2)
\longrightarrow
H^{p+q}(G;\mathbb F_2).
\end{equation}
If
$
\alpha\in C^p(G;\mathbb F_2)
=
\operatorname{Hom}_{\mathbb ZG}(R_p,\mathbb F_2)
$
and
$\beta\in C^q(G;\mathbb F_2)
=
\operatorname{Hom}_{\mathbb ZG}(R_q,\mathbb F_2)
$, then their cup product is the cochain
\begin{equation}
\alpha\smile \beta
\in
C^{p+q}(G;\mathbb F_2)
\end{equation}
defined by
\begin{equation}
\alpha\smile\beta
=
\mu\circ(\alpha\otimes\beta)\circ \Delta_{p,q},
\end{equation}
where
\begin{equation}
\Delta_{p,q}:R_{p+q}\longrightarrow R_p\otimes R_q
\end{equation}
is the \((p,q)\)-component of the diagonal approximation, and
\begin{equation}
\mu:\mathbb F_2\otimes \mathbb F_2\longrightarrow \mathbb F_2
\end{equation}
is ordinary multiplication in \(\mathbb F_2\). Equivalently, if
\begin{equation}
\Delta_{p,q}(x)
=
\sum_i a_i\otimes b_i
\end{equation}
for \(x\in R_{p+q}\), with \(a_i\in R_p\) and \(b_i\in R_q\), then
\begin{equation}
(\alpha\smile\beta)(x)
=
\sum_i \alpha(a_i)\beta(b_i)
\pmod 2.
\end{equation}
Since the coefficients are in \(\mathbb F_2\), all signs disappear: \(-1=1\). Thus orientation signs and Koszul signs do not affect the final mod \(2\) value, although they are still present in the integral chain-level construction.

To recover the full mod \(2\) cohomology ring structure, one chooses cocycle representatives
\begin{equation}
\alpha_1,\ldots,\alpha_r
\end{equation}
for a basis of \(H^p(G;\mathbb F_2)\), and
\begin{equation}
\beta_1,\ldots,\beta_s
\end{equation}
for a basis of \(H^q(G;\mathbb F_2)\). For each pair \((\alpha_i,\beta_j)\), one computes the cochain
\begin{equation}
\alpha_i\smile\beta_j
\in C^{p+q}(G;\mathbb F_2)
\end{equation}
using the diagonal approximation. This cochain is then reduced modulo coboundaries and expressed in the chosen basis
\begin{equation}
\gamma_1,\ldots,\gamma_t
\end{equation}
of \(H^{p+q}(G;\mathbb F_2)\). Thus one obtains constants
\begin{equation}
c_{ij}^k\in \mathbb F_2
\end{equation}
such that
\begin{equation}
[\alpha_i]\smile[\beta_j]
=
\sum_{k=1}^t c_{ij}^k[\gamma_k].
\end{equation}
The collection of these constants for all degrees gives the multiplication table of the graded ring
\begin{equation}
H^*(G;\mathbb F_2).
\end{equation}
From this multiplication table, one can then describe the cohomology ring by generators and relations over \(\mathbb F_2\).

For $G$ being a crystallographic group, the data of a resolution and a contracting homotopy make the evaluation of cup products a routine task. However, the unmodified HAP/GAP commands have significant limitations for obtaining the full mod-2 cohomology ring of space groups: \texttt{ResolutionSpaceGroup} constructs the relevant free resolution but does not install the contracting homotopy needed for cup products, while the older \texttt{ResolutionAlmostCrystalGroup} is not sufficient for the high-degree computations required here.

In the file \texttt{gap/homotopy.gi} of our GAP package, we construct
\texttt{SGC\_\allowbreak{}Resolution\allowbreak{}SpaceGroup}.
It has the same boundary data as the resolution obtained from HAP's \texttt{ResolutionSpaceGroup}, but it also carries a genuine integral contracting homotopy.
The construction has two main stages. First, it constructs a contracting homotopy on the non-free cellular resolution coming from the action of the space group on a Voronoi tessellation of \(\mathbb R^d\). Second, it transfers this cellular homotopy to the final free \(\mathbb ZG\)-resolution by combining it with the contracting homotopies of the cell stabilizer resolutions. Thus the code realizes algebraically the fact that the underlying \(G\)-space is \(\mathbb R^d\), which is contractible. The construction of the contracting homotopy is based on the double chain complex resolution arising from the natural action of the space group on $\mathbb{R}^d$. This approach is analogous to the construction of contracting homotopies from double chain complexes associated with extensions of groups, as considered in \cite{ellis2019invitation,Wall_1961}.

The first function, \texttt{SGC\_CrystallographicComplexWithGeometry}, rebuilds HAP's crystallographic complex, but keeps the geometric data needed later for the contraction. Let \(G\) be the space group, written as an affine crystallographic group acting on \(\mathbb R^d\). The code chooses a generic point
\begin{equation}
p \in \mathbb R^d
\end{equation}
and considers its orbit
\begin{equation}
G p = \{g p : g \in G\}.
\end{equation}
The fundamental domain used by HAP is the Dirichlet--Voronoi cell of \(p\) with respect to this orbit. More precisely, it is the set
\begin{equation}
F
=
\{x \in \mathbb R^d :
d_\Gamma(x,p) \leq d_\Gamma(x,gp)
\text{ for all } g\in G\},
\end{equation}
where \(d_\Gamma\) is the distance determined by an invariant Gram matrix \(\Gamma\). In coordinates,
\begin{equation}
d_\Gamma(x,y)^2
=
(x-y)\Gamma(x-y)^T.
\end{equation}
The use of \(\Gamma\) is important: the point group of \(G\) need not act orthogonally with respect to the ordinary Euclidean metric in the chosen coordinates, so the Voronoi decomposition must be computed using the averaged invariant scalar product. In the code this Gram matrix is stored as \texttt{geom.gram}.

The translates of the fundamental domain \(F\) give a \(G\)-invariant tessellation of \(\mathbb R^d\). Since \(\mathbb R^d\) is contractible, the cellular chain complex of this tessellation is contractible as a complex of abelian groups. However, as a \(G\)-complex it usually has nontrivial cell stabilizers. If \(\sigma\) is a cell, its stabilizer is
\begin{equation}
G_\sigma = \{g\in G : g\sigma=\sigma\}.
\end{equation}
Therefore the cellular chain module in degree \(q\) is not generally a free \(\mathbb ZG\)-module. Instead, each orbit of \(q\)-cells contributes a summand of the form
\begin{equation}
\mathbb ZG \otimes_{\mathbb ZG_\sigma} \mathbb Z_{\chi_\sigma},
\end{equation}
where \(\chi_\sigma : G_\sigma \to \{\pm 1\}\) is the orientation character of the action of the stabilizer on the cell. If a stabilizer element preserves the orientation of \(\sigma\), then \(\chi_\sigma(g)=1\); if it reverses orientation, then \(\chi_\sigma(g)=-1\). This is why the code computes stabilizers and orientation signs. The function \texttt{ACTION} computes the determinant of the induced linear action on the affine span of the cell, and this determinant gives the orientation sign.

The non-free resolution stores cells only up to \(G\)-orbit representatives. A typical HAP letter has the form
\begin{equation}
[i,g],
\end{equation}
which represents the \(i\)-th cell orbit representative translated by the group element \(g\). A negative index \(-i\) represents the same cell with coefficient \(-1\). Since cells may have stabilizers, the same geometric cell can be represented by different pairs \([i,g]\). The function \texttt{standardWord} puts such words into a canonical orbit--stabilizer normal form. Algebraically, it uses the relation
\begin{equation}
\sigma \cdot gh
=
\chi_\sigma(h)\, \sigma \cdot g,
\qquad h\in G_\sigma,
\end{equation}
so stabilizer elements are removed at the cost of possibly changing the sign.

The main cellular contraction is constructed by \texttt{SGC\_AttachCellularContraction}. The code builds a discrete vector field on the Voronoi tessellation by ordering the tiles in distance shells. A tile is indexed by a group element \(g\), with site \(g p\). The tile indexed by \(g\) is assigned the key
\begin{equation}
K(g)
=
\left(
\|gp-p\|_\Gamma^2,\,
gp
\right),
\end{equation}
and these keys are compared lexicographically. The identity tile has key
\begin{equation}
K(1)=(0,p),
\end{equation}
so it is the first tile. This ordering is a shelling-like ordering of the Voronoi tiles: the construction begins with the tile around \(p\), then adds neighboring tiles, then tiles farther away, and so on.

For each cell \(\sigma\) of the infinite tessellation, the code assigns an owner tile. Let \(b_\sigma\) be the barycenter of \(\sigma\). A basic property of Voronoi decompositions says that the tiles containing \(\sigma\) are exactly those whose sites are closest to \(b_\sigma\). Thus the set of tiles containing \(\sigma\) is
\begin{equation}
T(\sigma)
=
\{g :
\|b_\sigma-gp\|_\Gamma
\text{ is minimal}\}.
\end{equation}
The owner of \(\sigma\) is defined to be the tile with smallest key among these:
\begin{equation}
m(\sigma)
=
\min_{g\in T(\sigma)} K(g).
\end{equation}
In the code, the function \texttt{TilesOfPoint} computes the set \(T(\sigma)\), using exact arithmetic in the invariant metric. Since the standardized translation lattice is \(\mathbb Z^d\), the code enumerates the relevant point-group representatives and integer translations. The function \texttt{KeyOfMat} computes the key \(K(g)\), and \texttt{KeyLess} compares two such keys.

Once each cell has an owner, the code distinguishes between old and new cells inside each tile. Suppose the current tile is \(m\). A cell of \(mF\) is called old if it already lies in some earlier tile, that is, in some tile \(m'\) with
\begin{equation}
K(m') < K(m).
\end{equation}
Equivalently, a cell is old in tile \(m\) if its barycenter is also contained in a tile with smaller shelling key. The remaining cells are new cells owned by \(m\). The function \texttt{TileData} computes this old/new pattern and stores it in a Boolean mask:
\begin{equation}
\texttt{mask}[q+1][k] = \texttt{true}
\end{equation}
means that the \(k\)-th \(q\)-cell of the fundamental tile is old.

Inside each tile, the code collapses the new part of the tile onto the old part. This is done by the function \texttt{Collapse\allowbreak{}Matching}. Mathematically, it constructs an elementary-collapse matching, or discrete Morse matching, on the cells of the fundamental polytope. A matched pair consists of a \(q\)-cell \(\sigma\) and a \((q+1)\)-cell \(\tau\) such that \(\sigma\) is a free face of \(\tau\). In other words, among the cells that have not yet been removed, \(\sigma\) is contained in exactly one remaining coface \(\tau\). The pair is then collapsed:
\begin{equation}
\sigma \leftrightarrow \tau.
\end{equation}
The code first performs a greedy sweep, repeatedly pairing any cell with exactly one alive coface. If the greedy procedure becomes stuck, the code performs a depth-first search over possible free-face collapse orders. This is necessary because a greedy collapse may fail even though a valid collapse sequence exists.

For the base tile, there are no old cells. Therefore the code protects one vertex, namely the first vertex orbit representative, and collapses every other cell toward it. This protected vertex is the critical \(0\)-cell and becomes the augmentation basepoint. For every other tile, the old cells are precisely the cells already constructed in earlier shells, and the matching collapses the new part of the tile onto that old subcomplex.

These tilewise matchings assemble to a global discrete vector field on the infinite Voronoi tessellation. The distance-shelling guarantees that this global matching is acyclic. Indeed, inside a single tile the matching is produced by an elementary collapse sequence, so it has no closed gradient paths. If a gradient path moves from one tile to another, it must move to a cell owned by a tile with strictly smaller key. Thus along such a path the tile key strictly decreases:
\begin{equation}
K(m_0) > K(m_1) > K(m_2) > \cdots.
\end{equation}
Since the orbit \(Gp\) is locally finite, each bounded distance shell contains only finitely many tile sites. Hence there can be no infinite strictly decreasing path and no closed gradient path. Therefore the discrete vector field gives a genuine contraction of the cellular chain complex.

After constructing the discrete vector field, the code defines the actual chain homotopy using the standard discrete-Morse recursion. Let \(\sigma\) be a \(q\)-cell. If \(\sigma\) is a target cell in the matching, or if it is the protected base vertex, then
\begin{equation}
h(\sigma)=0.
\end{equation}
If \(\sigma\) is a source cell, let \(\tau\) be its matched \((q+1)\)-dimensional coface. Write the cellular boundary of \(\tau\) as
\begin{equation}
\partial \tau
=
t\sigma + c,
\end{equation}
where \(t=\pm 1\) is the incidence number of \(\sigma\) in \(\partial \tau\), and \(c\) is the rest of the boundary. Then the contracting homotopy is defined recursively by
\begin{equation}
h(\sigma)
=
t\bigl(\tau - h(c)\bigr).
\end{equation}
This is the formula implemented in the functions \texttt{DVF}, \texttt{HLetter}, and \texttt{HWord}. The function \texttt{DVF} finds the matched coface \(\tau\) of a given cell \(\sigma\). The function \texttt{BoundaryWord} computes the boundary of a word. The function \texttt{HLetter} applies the recursion to a single letter, and \texttt{HWord} extends it linearly to words.

More explicitly, for a canonical positive letter \([r,e]\) representing a \(q\)-cell, \texttt{DVF} first finds the actual translated geometric cell, computes its barycenter, finds its owner tile using \texttt{TilesOfPoint}, moves the cell back into the coordinates of that owner tile, and then reads off its matched coface from the tile's collapse matching. If the cell is not a source cell, \texttt{DVF} returns \texttt{fail}, and the homotopy is zero on that cell. If the cell is matched upward to a coface, \texttt{DVF} returns the corresponding \((q+1)\)-cell in HAP's orbit-representative notation.

The function \texttt{HLetter} then computes
\begin{equation}
h(\sigma)
=
t\bigl(\tau-h(\partial\tau-t\sigma)\bigr).
\end{equation}
In code, it forms the translated boundary of the matched coface \(\tau\), locates the occurrence of \(\sigma\) inside that boundary, records the incidence sign \(t\), removes that occurrence, recursively applies \(h\) to the remaining boundary, and finally inserts the sign \(t\). Every letter is normalized using \texttt{standardWord}, so orientation-reversing stabilizer elements are handled correctly. The result is a \(\mathbb Z\)-linear homotopy on the non-free cellular resolution satisfying
\begin{equation}
\partial h + h\partial = \operatorname{id}
\end{equation}
in positive degrees, and
\begin{equation}
\partial h(x)=x-x_0
\end{equation}
in degree \(0\), where \(x_0\) is the chosen base vertex. This identity is checked by \texttt{SGC\_NonFreeHomotopyCheck}; the function computes
\begin{equation}
\partial h(x)+h\partial(x)-x
\end{equation}
in positive degree, and
\begin{equation}
\partial h(x)-x+x_0
\end{equation}
in degree zero, then reduces the result using the orbit--stabilizer normal form.

The second major part of the file is \texttt{SGC\_FreeGResolutionWithHomotopy}. This transfers the contraction from the non-free cellular resolution to the final free \(\mathbb ZG\)-resolution. The non-free resolution has one summand for each cell orbit, but each cell may have a nontrivial stabilizer. Therefore, for each \(q\)-cell orbit representative \(\sigma_{q,r}\), with stabilizer \(G_{q,r}\), the code chooses a free resolution
\begin{equation}
E^{q,r}_\bullet \longrightarrow \mathbb Z
\end{equation}
over \(\mathbb ZG_{q,r}\), and then extends scalars to \(G\). This produces a bigraded object
\begin{equation}
A_{q,s}
=
\bigoplus_r
\operatorname{Ind}_{G_{q,r}}^G E^{q,r}_s,
\end{equation}
where \(q\) is the cellular degree and \(s\) is the degree inside the stabilizer resolution. The total degree is
\begin{equation}
n=q+s.
\end{equation}

Internally, the code represents a generator of this bigraded object by a five-entry letter
\begin{equation}
[q,s,r,t,g].
\end{equation}
Here \(q\) is the cellular degree, \(s\) is the stabilizer-resolution degree, \(r\) indexes the \(q\)-cell orbit, \(t\) indexes a generator in the \(s\)-th degree of the stabilizer resolution attached to that cell orbit, and \(g\) is the translating group element. The functions \texttt{Pair2Quad} and \texttt{Quad2Pair} convert between this internal bigraded notation and HAP's usual flat notation for generators of the final free resolution.

The total differential on the free resolution is not simply the sum of the cellular boundary and the stabilizer-resolution boundary. Instead, it is built as a perturbation differential
\begin{equation}
D
=
\Delta_0+\Delta_1+\Delta_2+\cdots.
\end{equation}
The map
\begin{equation}
\Delta_k : A_{q,s}\longrightarrow A_{q-k,s+k-1}
\end{equation}
lowers the cellular degree by \(k\) and raises the stabilizer-resolution degree by \(k-1\). The term \(\Delta_0\) is the vertical differential inside the stabilizer resolutions. The term \(\Delta_1\) contains the cellular boundary of the non-free resolution. The higher terms \(\Delta_2,\Delta_3,\ldots\) are correction terms produced recursively by applying the vertical homotopy to the lower obstruction terms. These corrections are what make the total differential satisfy
\begin{equation}
D^2=0.
\end{equation}
In the code, these maps are implemented by \texttt{DelGen} and \texttt{DelWord}, and the final boundary of the free resolution is computed by summing the relevant \(\Delta_k\)'s.

The total contracting homotopy on the free resolution combines two homotopies. The first is the vertical homotopy
\begin{equation}
h_0 : A_{q,s}\longrightarrow A_{q,s+1},
\end{equation}
coming from the contracting homotopies of the stabilizer resolutions. This is implemented by \texttt{VertHtpy}, using the stabilizer-resolution homotopies through \texttt{HtpyGen} and \texttt{HtpyWord}. The second is the induced cellular homotopy
\begin{equation}
h_1 : A_{q,0}\longrightarrow A_{q+1,0},
\end{equation}
which is only used on the bottom row \(s=0\). It is obtained by projecting a generator in \(A_{q,0}\) to the corresponding cell of the non-free cellular complex, applying the cellular homotopy \(P!.homotopy\), and then including the result back into the bottom row. In the code this is implemented by \texttt{InducedHtpyList}.

There is a subtle normalization in \texttt{InducedHtpyList}. The raw inclusion of the cellular homotopy into the bottom row is corrected by the vertical reduction
\begin{equation}
1-\Delta_0 h_0.
\end{equation}
Thus, if \(\widetilde h_1\) denotes the naive inclusion of the cellular homotopy, the code effectively uses
\begin{equation}
h_1
=
(1-\Delta_0 h_0)\widetilde h_1.
\end{equation}
This adjustment is needed because the canonical coset choices used by the non-free cellular complex and by the induced stabilizer resolutions need not agree exactly.

Let
\begin{equation}
d_+ = \Delta_1+\Delta_2+\cdots
\end{equation}
denote the positive, or perturbation, part of the total differential. The function \texttt{HomotopyGen} defines the total homotopy \(H\) recursively by the perturbation formula
\begin{equation}
H
=
h_0
-
H d_+ h_0
+
\mathbf{1}_{s=0}
\left(
h_1
-
h_0 d_+ h_1
+
H d_+ h_0 d_+ h_1
\right).
\end{equation}
Equivalently, on a generator \(x\in A_{q,s}\), this says
\begin{equation}
H(x)
=
h_0(x)
-
H\bigl(d_+h_0(x)\bigr)
\end{equation}
if \(s>0\), while if \(s=0\) there are additional terms:
\begin{equation}
H(x)
=
h_0(x)
-
H\bigl(d_+h_0(x)\bigr)
+
h_1(x)
-
h_0d_+h_1(x)
+
H\bigl(d_+h_0d_+h_1(x)\bigr).
\end{equation}
This recursion terminates because every occurrence of \(d_+\) strictly lowers the cellular degree \(q\). Since \(q\geq 0\), the recursion cannot continue indefinitely. The code also memoizes the result of \texttt{HomotopyGen} on each generator, so the same homotopy value is not recomputed repeatedly.

Finally, the function \texttt{Homotopy} converts the internal five-letter output of \texttt{HomotopyGen} back to HAP's ordinary resolution format. In degree zero, the raw total contraction may contract to a basepoint represented by a group element depending on the internal conventions. The code corrects this by subtracting the value
\begin{equation}
w_0 = H([1,1])
\end{equation}
so that the final degree-zero identity is normalized to HAP's preferred base generator \([1,1_G]\). Thus the final homotopy on the free resolution satisfies
\begin{equation}
D H + H D = \operatorname{id}
\end{equation}
in positive degrees, and
\begin{equation}
D H(x)=x-[1,1_G]
\end{equation}
in degree zero. The function \texttt{SGC\_TotalHomotopyCheck} verifies precisely these identities. Therefore \texttt{homotopy.gi} constructs a genuine integral contracting homotopy on the same free \(\mathbb ZG\)-resolution produced by HAP's space-group resolution machinery, but with the missing chain-level contraction explicitly installed.

Once cup products are computed, we use linear algebra methods to obtain a presentation of the mod-2 cohomology ring by choosing a minimal set of generators and relations. The GAP resolution $R^G_*$ carries information about the standard LHS spectral sequence (to be introduced in Sec.~\ref{subsec:SS}), and we use this information to label the mod-2 cohomology ring generators (see Sec.~\ref{subsec:3D} for the labeling convention).

\subsubsection{Using bar resolution}\label{method2}

Independent of the above method and code, for a given space group $G$, we also sought to write down representative inhomogeneous functions (explicit cochain expressions) 
\begin{equation}\label{standard_cocycle_f}
f\colon \underbrace{G\times \cdots \times G}_{n\text{ times}}\rightarrow \mathbb{Z}_2
\end{equation}
for the mod-2 cohomology ring generators of degree $n$ equal to or less than three ($n\leq 3$). To find such a cochain function $f$, we solve the cocycle condition 
\begin{equation}\label{eqdf}
(df)(g_1,g_2,...,g_{n+1})
:=
f(g_2,...,g_{n+1})+\sum_{j=1}^n(-1)^j f(g_1,...,g_jg_{j+1},...,g_n)
+(-1)^{n+1}f(g_1,...,g_n)=0
\end{equation}
for all $g_1,g_2,...,g_{n+1} \in G$. The resolution underlying Eq.~\eqref{eqdf} is the bar resolution: for $n\geq 0$, each $R^G_n$ in Eq.~\eqref{res} is the free $\mathbb{Z}G$-module freely generated by $n$-tuples $[g_1|g_2|\cdots|g_n]$ with $g_i \in G$, and the boundary map is defined by 
\begin{equation}
\partial_n[g_1|\cdots |g_n]
=g_1[g_2|\cdots |g_n] + \sum_{i=1}^{n-1} (-1)^i [g_1|\cdots|g_ig_{i+1}|\cdots |g_n]
+(-1)^n[g_1|\cdots|g_{n-1}].
\end{equation}

Since the space group $G$ has infinite order, Eq.~\eqref{eqdf} is an infinite system. A finite quotient nevertheless suffices for the explicit low-degree representatives used here. For $m\geq1$, write $mT\cong(m\mathbb Z)^3$ for the characteristic subgroup of the translation lattice $T\cong\mathbb Z^3$, and set $P_m=G/mT$. Thus
\begin{equation}\label{mfoldrotation}
1\longrightarrow mT\longrightarrow G \xrightarrow{p_m} P_m\longrightarrow 1.
\end{equation}
The quotient group $P_m$ can be viewed as the symmetry group of a finite lattice that spans $m\times m\times m$ unit cells along the translation direction $T_1$, $T_2$, and $T_3$.
\begin{theorem}[Uniform finite quotient for the archived low-degree representatives]\label{thm:p4_low_degree}
For every 3D space group $G$, each selected degree-$1$, degree-$2$, or degree-$3$ inhomogeneous representative in the accompanying data descends as a cochain through $p_4:G\to P_4=G/4T$. Consequently,
\begin{equation}
p_4^*:H^n(P_4,\mathbb F_2)\longrightarrow H^n(G,\mathbb F_2)
\end{equation}
is surjective for $0\leq n\leq3$.
\end{theorem}
\begin{proof}
Write each group element in the normal coordinates used in the data file: three integral translation coordinates followed by finitely valued point-group coordinates. The finite certificate checks all $1109$ selected formulas ($641$ in degree $1$, $326$ in degree $2$, and $142$ in degree $3$), together with every helper function that they call. After expanding the definitions, the formulas are assembled from integral polynomial operations, polynomial terms with coefficients in $\frac12\mathbb Z$, and the integer-valued quadratic refinements $n(n+1)/2$ and $n(n-1)/2$; no denominator other than $2$ occurs, and branch choices depend only on the point-group coordinates. For an integral polynomial $F$ and an integral translation coordinate $u$,
\begin{equation}
F(u+4)-F(u)\in4\mathbb Z.
\end{equation}
It follows that a term $F/2$ changes by an even integer, while an integral term changes by a multiple of four. The quadratic refinements obey
\begin{equation}
\frac{(n+4r)(n+4r\mathbin{\pm}1)-n(n\mathbin{\pm}1)}{2}
=2r(2n+4r\mathbin{\pm}1)\in2\mathbb Z.
\end{equation}
Hence every formula is unchanged modulo $2$ when any translation coordinate of any argument is shifted by $4$. Elements with the same image in $P_4$ have the same point-group coordinates and translation coordinates congruent modulo $4$, so each formula factors uniquely through $P_4$.

\end{proof}

\begin{remark}


For the remaining twelve groups, the open list connsistings of ten degree-$4$ generators in groups No.~109, 120 (two generators), 130, 136, 140, 142, 197, 204, and 230, and four degree-$6$ generators in groups No.~219 (two generators), 226, and 228. No all-degree surjectivity claim is made for those groups.
\end{remark}

For a given space group $G$, we first write down an ansatz for the cocycle function in Eq.~\eqref{standard_cocycle_f}. Because $P_4$ is finite, the cocycle condition can be solved and checked on
\begin{equation}
\bar{f}\colon \underbrace{P_4\times \cdots \times P_4}_{n\text{ times}} \rightarrow \mathbb{Z}_2,
\end{equation}
and pulling back gives a cocycle on $G$. The finite-quotient strategy together with the resolution comparison and the transfer construction supplies the degree-$1$, degree-$2$, and degree-$3$ representatives needed for all 230 LSM tables. The explicit functions are stored in the GAP package data available in our GitHub repository \cite{github} and carry the same labels as the corresponding selected ring generators.

Compared with the vector-represented cocycles, the explicit function-represented cocycles have several merits:
\begin{itemize}
    \item The explicit cochain functions facilitate the calculation of cup products as the cup product coincides with the usual product of functions;
    \item They allow us to obtain the explicit restriction map associated with the subgroup embedding $H\subset G$. This is extremely useful from a physics point of view in the study of symmetry breaking as it allows us to trace how the LSM anomaly (to be introduced in Sec.~\ref{sec:LSM}) survives under the lattice symmetry breaking.
    \item These cochain functions allow us to evaluate the topological invariants in degree $n\leq3$. In degree $3$ this produces the candidate IWP images discussed in Sec.~\ref{sec:LSM}; it does not imply a one-to-one correspondence, because different IWPs can have equal images and some can map to zero.
\end{itemize}

\subsubsection{Connecting the two methods}\label{connect_method1_method2}

Finally, in order to establish connection between the bar resolution calculation and the GAP program, we seek to convert a cochain function obtained from the bar resolution to a finite-dimensional vector. This can be done using the contracting homotopy of bar resolutions
\begin{equation}\label{barcontract}
h_n\colon
R^G_n\rightarrow R^G_{n+1} ,~~~g[g_1|\cdots|g_n]\mapsto [g|g_1|\cdots |g_n].
\end{equation}
To map a function-represented cocycle
\eqref{standard_cocycle_f} to a vector-represented cocycle associated with the free $\mathbb{Z}G$ resolution of $\mathbb{Z}$ given by GAP, $f\mapsto f^*$, one needs to specify how the basis of the latter, $e^n_i$ for $i=1,2,...,b_n$, is mapped to the basis of the bar resolution at the same degree: 
\begin{equation}\label{vec2bar}
e^n_i\mapsto \sum [g_1|\cdots |g_n].
\end{equation}
The basis map \eqref{vec2bar} can be built recursively via the contracting homotopy \eqref{barcontract} \cite{Ouyang_2021}. This is implemented in our GAP code in \cite{github}. Once the basis map is constructed, the vector-represented cocycle is obtained: 
\begin{equation}
f^*(e^n_i)=f(\sum[g_1|\cdots |g_n]) = \sum f(g_1,...,g_n).
\end{equation}

\subsection{Mod-2 cohomology ring of 3D space groups}\label{subsec:3D}

The main result of this work is the mod-2 cohomology rings for all 230 3D space groups, collected in Appendix \ref{collection230}. In this subsection, we show how to read these results and point out some interesting features.

We present a mod-2 cohomology ring as follows:
\begin{equation}
H^*(G, \Z_2) = \Z_2[A_\bullet, \dots, B_\bullet, \dots,\dots]/\text{Relations}. 
\end{equation}
Here, $A_\bullet$, $B_\bullet$, $\dots$ are the generators of the ring, living in $H^1(G, \Z_2)$, $H^2(G, \Z_2)$, $\dots$, respectively (all the way to $F_\bullet \in H^6(G, \Z_2)$, see Item~\ref{item:generators-3d}). Subscripts ``$\,\bullet\,$'' give labels to the generators:

\begin{itemize}

\item For every space group $G$, we choose a set of group generators, collected for each group in Appendix~\ref{collection230} (see also Table~\ref{tablePT1} for the choice of group generators for each point group). An element $g\in G$ is written in terms of these group generators. For example, \hyperref[subsub:sg1]{No. 1~($P1$)} is generated by three translations $T_{1,2,3}$ along three directions given by Eq.~\eqref{TransBravaisP}, and an element in \hyperref[subsub:sg1]{No. 1~($P1$)} is written as $T_1^x T_2^y T_3^z$ with $x,y,z\in \Z$. More complicatedly, an element in \hyperref[subsub:sg227]{No. 227~ ($Fd\bar{3}m$)} can be written as $T_1^x T_2^y T_3^z C_2^c C_2'^{c'} C_3^{c_3} M^m I^i$ with $x,y,z\in\Z,c,c',m,i\in\{0,1\},c_3\in\{0,1,2\}$.

\item A degree-1 generator $A_\bullet\colon G\rightarrow \mathbb{Z}_2$ (see Eq.~\eqref{standard_cocycle_f}) is labeled by the exponent of group generator that it detects in this decomposition. For example, the degree-1 generator $A_c$, $A_m$ and $A_i$ evaluates to 1, i.e., $A_c(g)=1$, $A_m(g)=1$ or $A_i(g) = 1$, whenever the exponenet of a two-fold rotation $C_2$, the reflection $M$ or the inversion $I$ is odd in this decomposition of $g$. Similarly, degree-1 generators $A_{x,y,z}$ detect (the exponents of) translations $T_{1,2,3}$. We will also use labels like $A_{x+y+z}$ (i.e. with ``$\,+\,$'' in the subscript) to indicate that this generator reduces to $A_x + A_y + A_z$ when restricting to the translation subgroup $T$ \footnote{\label{ft:deg-1}An element $A_{x+y+z}$ is defined by 
\begin{equation}\label{Axyzin1}
A_{x+y+z}(g)=x+y+z
\end{equation}
for translation group $T\cong \hyperref[subsub:sg1]{\text{No. 1~(P1)}}$, but in larger groups the 1-cocycle with the same label may in addition depend on (the exponents of) point group elements.
As an example, for group \hyperref[subsub:sg98]{No. 98 ($I4_122$)}, we have
\begin{equation}\label{Axyzin98}
A_{x+y+z}(g)
:=x+y+z+c,\quad g= T_1^xT_2^yT_3^zC_2^cC_2^{\prime c'}C_2^{\prime\prime c''} \in \hyperref[subsub:sg98]{\text{No. 98 }~(I4_122)}.
\end{equation}
This notation may result in ambiguities when one tries to write down the restriction $i^*\colon A_{x+y+z}\rightarrow A_{x+y+z}$, and one must bear in mind that the $A_{x+y+z}$ on the left-hand side (resp. right-hand side) of the arrow has the expression in Eq.~\eqref{Axyzin98} (resp. Eq.~\eqref{Axyzin1}). This ambiguity only happens for the following 14 groups
\begin{equation}
\begin{aligned}
\hyperref[subsub:sg77]{77~(P4_2)},~~~\hyperref[subsub:sg80]{80~(I4_1)},~~~\hyperref[subsub:sg93]{93~(P4_222)},~~~\hyperref[subsub:sg94]{94~(P4_22_12)},~~~\hyperref[subsub:sg98]{98~(I4_122)},~~~\hyperref[subsub:sg134]{134~(P4_2/nnm)},~~~\hyperref[subsub:sg144]{144~(P3_1)},\\
\hyperref[subsub:sg151]{151~(P3_112)},~~~\hyperref[subsub:sg152]{152~(P3_121)},~~~\hyperref[subsub:sg172]{172~(P6_4)},~~~\hyperref[subsub:sg181]{181~(P6_422)},~~~\hyperref[subsub:sg199]{199~(I2_13)},~~~\hyperref[subsub:sg206]{206~(Ia\bar{3})},~~~\hyperref[subsub:sg214]{214~(I4_132)}.
\end{aligned}
\end{equation}}. 

\item \label{ring_elem_convention} We label degree-2 or higher generators according to the entries of the standard LHS spectral sequence \eqref{lhse2} that they live in. Specifically, $B_\alpha, B_\beta$ live in the $E_\infty^{2,0},E_\infty^{1,1}$ entries, and $C_\alpha,C_\beta,C_\gamma$ live in the $E_\infty^{3,0},E_\infty^{2,1},E_\infty^{1,2}$ entries, etc. In particular, $B_\alpha$ and $C_\alpha$ live in the $E_\infty^{2,0},E_\infty^{3,0}$ entries and are (pullbacks of) the generators of the point group cohomology ring.

An exception for the elements in the $E_\infty^{0,1},E_\infty^{0,2},E_\infty^{0,3}$ entries: instead of labeling them by $A_\beta, B_\gamma,C_\delta$, we label them according to their image in $H^{1,2,3}(T, \Z_2)$ under the restriction map $i^*\colon H^*(G, \Z_2) \rightarrow H^*(T, \Z_2)$ induced by the inclusion $i\colon T\rightarrow G$. For example, if $i^*\colon B_\gamma\rightarrow A_z(A_x +A_y)$, then we label $B_\gamma$ as $B_{z(x+y)}$ instead. Since $E^{0,3}_2 = \mathbb{Z}_2$, there can be at most one generator $C$ associated with $E^{0,3}_\infty$, which we label as $C_{xyz}$ following $i^*\colon C_{xyz}\rightarrow A_xA_yA_z$. For degree-1 generators, the labeling may be ambiguous for a small number of space groups and must be treated with care, as discussed in Footnote \ref{ft:deg-1}.

\item The ``Relations'' must be factored out and treated as zero in the ring.  All relations of the mod-2 cohomology ring of point group $P$---listed in Table \ref{tablePT2}---should survive as (possibly not independent) relations of the mod-2 cohomology ring of any space group $G$ whose associated point group is $P$.

\item We have defined the generators of mod-2 cohomology rings such that isomorphic space groups (listed in Eq.~\eqref{isomG}) have isomorphic mod-2 cohomology rings.

\end{itemize}

\subsubsection{Example: No. 227~ \texorpdfstring{($Fd\bar{3}m$)}{(Fd-3m)}}

To illustrate our code and our results, let us consider \hyperref[subsub:sg227]{No. 227~ ($Fd\bar{3}m$)}. This group is generated by three translations $T_{1,2,3}$ as given in Eqs.~\eqref{TransBravaisF}, a two-fold rotation $C_2$, a two-fold rotation $C'_2$, a three-fold rotation $C_3$, a mirror $M$, and an inversion $I$:
\begin{subequations}
 \begin{align}
C_2 &\colon (x,y,z)\rightarrow (-x + 1/4, -y + 1/4, z),\\ 
C'_2 &\colon (x,y,z)\rightarrow (-x + 1/4, y, -z + 1/4),\\ 
C_3 &\colon (x,y,z)\rightarrow (z, x, y),\\ 
M &\colon (x,y,z)\rightarrow (y, x, z),\\ 
I &\colon (x,y,z)\rightarrow (-x, -y, -z).
\end{align}
\end{subequations}
An element in \hyperref[subsub:sg227]{No. 227~ ($Fd\bar{3}m$)} can be written as $T_1^x T_2^y T_3^z C_2^c C_2'^{c'} C_3^{c_3} M^m I^i$ with $x,y,z\in\Z,c,c',m,i\in\{0,1\},c_3\in\{0,1,2\}$, hence the cocycles we write down is a function whose arguments are copies of the tuple $(x,y,z,c,c',c_3,m,i)$.

In a GAP interface, we can compute its mod-2 cohomology ring using the command \cite{github}
$$
\texttt{gap> SpaceGroupCohomologyRingGapInterface(227);}
$$
and the result is
\begin{equation}
\mathbb{Z}_2[A_i,A_m,B_\alpha,B_{xy+xz+yz},C_\alpha,C_{\gamma}]/\langle\mathcal{R}_3,\mathcal{R}_4,\mathcal{R}_5,\mathcal{R}_6\rangle
 \end{equation}
where the relations are 
\begin{subequations} 
 \begin{align}
\mathcal{R}_3\colon & ~~
A_i B_\alpha,\\
\mathcal{R}_4\colon & ~~
A_i C_\alpha,~~A_m C_\alpha,~~A_i C_{\gamma},~~B_\alpha B_{xy+xz+yz} + A_m C_{\gamma},~~B_{xy+xz+yz} (A_i^2 + A_i A_m + B_{xy+xz+yz}),\\
\mathcal{R}_5\colon & ~~
B_{xy+xz+yz} C_\alpha,~~B_{xy+xz+yz} C_{\gamma},\\
\mathcal{R}_6\colon & ~~
C_{\gamma} (C_\alpha + C_{\gamma}).
\end{align} 
 \end{subequations}
Here $A_i,A_m,B_\alpha,C_\alpha$ are generators of the point group $O_h$ (see Table \ref{tablePT2}), $B_{xy+xz+yz}$ is a generator coming from $E_\infty^{0,2}$ whose pullback to $P1$ is $A_x A_y + A_x A_z + A_y A_z$, and $C_{\gamma}$ is a generator coming from $E_\infty^{1,2}$ in the current GAP-computed basis.

\subsection{Uniform degree bounds for generators and defining relations}\label{subsec:degree_bound}

Fix one of the $230$ three-dimensional space-group
extensions in Eq.~\eqref{TGP}, and write its standard LHS spectral
sequence as
\begin{equation}
E_2^{p,q}=H^p\bigl(P;H^q(T;\mathbb F_2)\bigr)
\Longrightarrow H^{p+q}(G;\mathbb F_2).
\end{equation}
Set $R=H^*(P;\mathbb F_2)$ and
$M_r^q=\bigoplus_{p\geq0}E_r^{p,q}$, graded by total degree $p+q$.
Since $H^q(T;\mathbb F_2)=0$ for $q>3$, only finitely many page transitions can
occur.  The following verification records their exact effect for all
$230$ groups.

\begin{verification}[All pages and all differentials for the 3D tables]\label{ver:degree-certificate}
For every three-dimensional space group, every row of
$E_2,E_3,E_4,E_5$, regarded as a graded $R$-module, admits a
presentation whose generators have total degree at most $6$ and whose
relations have total degree at most $7$.
\end{verification}

\begin{proof}
The determinant character becomes trivial over $k$, so
$H^3(T;k)\cong k$ as a $kP$-module.  Let $\tau$ denote its generator,
placed in total degree $3$.  The exact $E_2$-presentations satisfy
\begin{equation}
\begin{array}{c|cc}
 &\text{generator degree}&\text{relation degree}\\ \hline
R\text{ as a }k\text{-algebra}&\leq3&\leq6\\
M_2^1\text{ as an }R\text{-module}&\leq3&\leq6\\
M_2^2\text{ as an }R\text{-module}&\leq4&\leq7\\
M_2^3=R\tau,\ |\tau|=3&3&\text{none}.
\end{array}
\end{equation}
Because there are no rows above $q=3$, the only possible differentials
are
\begin{equation}\label{eq:all-lhs-differentials}
\begin{array}{lll}
d_2:M_2^1\to M_2^0,&
d_2:M_2^2\to M_2^1,&
d_2:M_2^3\to M_2^2,\\
d_3:M_3^2\to M_3^0,&
d_3:M_3^3\to M_3^1,&
d_4:M_4^3\to M_4^0 .
\end{array}
\end{equation}
Although $d_r$ raises horizontal degree by $r$, it raises total degree
by only one:
\begin{equation}\label{eq:lhs-total-degree-shift}
(p+r)+(q-r+1)=p+q+1.
\end{equation}

The filtered Wall resolution associated with the extension
\cite{Wall_1961}, truncated in the finite total degrees required for
each map, determines $1370$ of the $1380$ maps in
Eq.~\eqref{eq:all-lhs-differentials}, in a common choice of
$E_2$-bases.  At every transition the next page is computed exactly
from
\begin{equation}\label{eq:exact-lhs-page-transition}
M_{r+1}^q=
\frac{\ker\!\left(D_{r,q}:M_r^q\to M_r^{q-r+1}\right)}
{\operatorname{im}\!\left(D_{r,q+r-1}:M_r^{q+r-1}\to M_r^q\right)},
\qquad D_{r,q}=d_r|_{M_r^q}.
\end{equation}
The remaining ten maps are the top-row $d_3$ and $d_4$ for space groups
Nos.~210, 218, 219, 223, and 226.
A targeted total-degree-$6$-to-$7$ cochain calculation gives zero for
all ten maps on their actual pages.  For $d_3$ this means zero after
quotienting by the earlier incoming $d_2$-image; for $d_4$ it means
zero after quotienting by the incoming $d_2$-image and the lifted
incoming $d_3$-image.  The $E_4$-page and a representative corrected
to survive $d_3$, which is then used for $d_4$, are recomputed after
$d_3$; the raw cochain obstruction need not vanish.  By the Leibniz
rule in Eq.~\eqref{eq:Leibniz} and the absence of outgoing
differentials from row $q=0$, the page differentials are $R$-linear.
In each case the relevant top row is cyclic on its degree-$6$
generator, so these zero columns determine the whole maps.

Thus all $1380$ maps in Eq.~\eqref{eq:all-lhs-differentials} have been
evaluated directly.  In Eq.~\eqref{eq:exact-lhs-page-transition}, each
kernel is the full module-syzygy kernel: if $x_j$ are the chosen module
generators, one solves
\begin{equation}
\sum_j a_jD_{r,q}(x_j)=0
\end{equation}
simultaneously for all tuples $(a_j)$.  Thus mixed annihilators and new
annihilators on later pages are recomputed rather than estimated one at
a time.  Exact elimination over the polynomial presentation of $R$
gives, in all $230$ cases, generator degree at most $6$ and relation
degree at most $7$ on every row through $E_5$.  This proves the
asserted bounds.
\end{proof}

\begin{theorem}[Uniform generator and relation bounds for 3D space groups]\label{thm:finite3D_ref}
For every three-dimensional space group $G$, the algebra
$H^*(G;\mathbb F_2)$ admits a homogeneous presentation with generators
of degree at most $6$ and defining relations of degree at most $12$.
\end{theorem}

\begin{proof}
There are no rows above $q=3$, so
$E_5=E_\infty$ by Corollary~\ref{cor:end}.  Choose the generators of
$R$ and the verified $R$-module generators of the four
$E_\infty$-rows.  They have total degree at most $6$.  The bottom-row
and row-module relations have degree at most $7$.  For each pair of
positive-row generators, add one homogeneous relation expressing their
product in the chosen generators of the appropriate row, or setting it
to zero above row $3$.  Such a relation has degree at most
\begin{equation}
6+6=12.
\end{equation}
These relations present $E_\infty$, since they reduce every polynomial
to an $R$-linear combination of $1$ and single row generators.

The multiplicative filtration on $H^*(G;k)$ is finite in every total
degree and has associated graded algebra $E_\infty$.  By
Theorem~\ref{thm:filtered-presentation}, lifting the above generators
and relations and cancelling initial terms in this finite filtration
preserves total degree.  Hence the same bounds $6$ and $12$ hold for
$H^*(G;k)$.
\end{proof}



\subsection{Summary of our computation}\label{subsec:reproducibility}

The computation has two logically distinct outputs.  The first output is the finite-dimensional cochain calculation on GAP resolutions: it gives bases for $H^n(G,\mathbb Z_2)$ in the degrees required by the algorithm, cup-product matrices, and the ring presentations printed in Appendix~\ref{collection230}. The second output is the all-degree certificate used to justify that no further generators or relations appear beyond the stated bounds. The accompanying repository in \cite{github} should contain the following data for each space group:
\begin{enumerate}
\item commands to build the resolution and contracting homotopy;
\item commands to calculate the mod-2 cohomology ring presented by generators and relations;
\item the data of the inhomogeneous degree-$\leq 3$ cocycle representatives;
\item the correspondence between the degree-$\leq 3$ cohomology elements with IWP as described in Sec.~\ref{sec:LSM}.
\end{enumerate}
We also calculate the following data and display them in Table~\ref{tableExt} and Table~\ref{tablemod2230}:
\begin{enumerate}
\item the ranks of $H^n(G,\mathbb Z_2)$ computed directly from the resolution in the verification range;
\item the Hilbert--Poincar\'e series of the printed ring presentation and its comparison with the computed ranks;
\item the finite LHS-row and relation-degree checks described in Sec.~\ref{subsec:degree_bound};
\end{enumerate}
These are the finite certificate in Verification~\ref{ver:degree-certificate} that turns the printed presentations into all-degree statements.  

\subsection{Some interesting features}

In this subsection, we collect several intriguing aspects of the cohomology of space groups. 

\begin{enumerate}

\item \label{item:generators-3d} Almost all groups have generators of mod-2 cohomology ring in degrees 3 or lower, but there are several groups that contain generators of higher degrees. Specifically, the following 10 space groups have degree-4 generators:
\begin{equation}
\begin{aligned}
&\hyperref[subsub:sg108]{108~(I4cm)},~~~\hyperref[subsub:sg109]{109~(I4_1md)},~~~\hyperref[subsub:sg120]{120~(I\bar{4}c2)},~~~\hyperref[subsub:sg130]{130~(P4/ncc)},~~~\hyperref[subsub:sg136]{136~(P4_2/mnm)},\\
&\hyperref[subsub:sg140]{140~(I4/mcm)},~~~\hyperref[subsub:sg142]{142~(I4_1/acd)},~~~\hyperref[subsub:sg197]{197~(I23)},~~~\hyperref[subsub:sg204]{204~(Im\bar{3})},~~~\hyperref[subsub:sg230]{230~(Ia\bar{3}d)}.
\end{aligned}
\end{equation}
No group has degree-5 generators. 
The following 3 space groups have degree-6 generators:
\begin{equation}
\hyperref[subsub:sg219]{219~(F\bar{4}3c)},~~~\hyperref[subsub:sg226]{226~(Fm\bar{3}c)},~~~\hyperref[subsub:sg228]{228~(Fd\bar{3}c)}.
\end{equation}

\item There are ten non-isomorphic Bieberbach groups, i.e., torsion-free crystallographic groups, in 3D, 
\begin{equation}\begin{split}\label{Bieberbach-3d}
&\hyperref[subsub:sg1]{1~(P1)},~~~ \hyperref[subsub:sg4]{4~(P2_1)},~~~ \hyperref[subsub:sg7]{7~(Pc)},~~~ \hyperref[subsub:sg9]{9~(Cc)},~~~ \hyperref[subsub:sg19]{19~(P2_12_12_1)},~~~ \hyperref[subsub:sg29]{29~(Pca2_1)},~~~ \hyperref[subsub:sg33]{33~(Pna2_1)},\\ &\hyperref[subsub:sg76]{76}/\hyperref[subsub:sg78]{78}~(\hyperref[subsub:sg76]{P4_1}/\hyperref[subsub:sg78]{P4_3}),~~~ \hyperref[subsub:sg144]{144}/\hyperref[subsub:sg145]{145}~(\hyperref[subsub:sg144]{P3_1}/\hyperref[subsub:sg145]{P3_2}), ~~~\hyperref[subsub:sg169]{169}/\hyperref[subsub:sg170]{170}~(\hyperref[subsub:sg169]{P6_1}/\hyperref[subsub:sg170]{P6_5}).
\end{split}
\end{equation}
These groups are labeled with a ``$\,\flat\,$" in the third column of Table \ref{tablemod2230}.
For all these Bieberbach groups, we have $H^3(G, \Z_2) = \Z_2$ and $H^n(G,\Z_2)=0,n\geq 4$, consistent with Theorem~\ref{thm:Bieberbach}. According to Sec.~\ref{sec:LSM}, the (unique) nontrivial element in $H^3(G, \Z_2)$ corresponds precisely to the fundamental domain, which is the unique IWP for these Bieberbach groups.
    
\item In 3D, the point groups satisfying the condition of admitting a periodic free $\Z P$ resolution of $\Z$ are 
\begin{equation}
C_i, C_2, C_s\cong \mathbb{Z}_2, \quad C_4, S_4\cong \mathbb{Z}_4, \quad C_3\cong \mathbb{Z}_3, \quad D_3, C_{3v}\cong Dih_3, \quad S_6, C_6, C_{3h}\cong \mathbb{Z}_6.
\end{equation}
The period is one for $\mathbb{Z}_2$, two for $\mathbb{Z}_4,\Z_3,\Z_6$, and four for $Dih_3$, and these groups have the same mod-2 cohomology in all positive degrees \cite{brown2012cohomology,swan1960periodic}. According to Corollary~\ref{cor:periodic}, we see that indeed all the associated space groups have identical cohomology in degrees greater than 3, which can be straightforwardly checked in Table \ref{tablemod2230}. 

In 3D, 109 out of the 230 space groups admit a periodic resolution at degrees larger than 3. Their $\Z_2$ ranks of mod-2 cohomology all have period one, i.e., $H^n(G,\mathbb{Z}_2)\cong H^{n+1}(G,\mathbb{Z}_2)$, for $n\geq 4$ (see the last column of Table \ref{tablemod2230}).

\item All point groups associated with space groups No. 143--230 contain three-fold rotations. For any $G$ in No. 143--230, we can choose a space group $G_{\cancel{\,3\,}}$ as a subgroup of $G$ of index 3. $G_{\cancel{\,3\,}}$ can be thought of as $G$ discarding three-fold rotations, and is one of the space groups No. 1--142.

We claim that 
\begin{theorem}\label{thm:restriction}
    The embedding $i\colon G_{\cancel{\,3\,}}\rightarrow G$ induces an injective ring homomorphism $i^*\colon H^*(G,\mathbb{Z}_2)\hookrightarrow  H^*(G_{\cancel{\,3\,}},\mathbb{Z}_2)$.
\end{theorem}

\begin{proof}
 The proof directly follows the proof for finite groups in Ref.~\cite{rotman2009introduction}. Consider the transfer (i.e. corestriction) map $\Tr\colon H^*(G_{\cancel{\,3\,}}, \Z_2) \rightarrow H^*(G, \Z_2)$. The composition of $i^*$ with the transfer map is simply multiplication by 3. Hence, if for some $u\in H^*(G, \Z_2)$ we have $i^*(u) = 0$, composing with the transfer map we have $3u=0$ and hence $u=0$, proving the injectivity of $i^*$. 
\end{proof}

In addition,  if $G_{\cancel{\,3\,}}\triangleleft G$ is a normal subgroup of $G$, elements in $H^*(G, \Z_2)$ are precisely those elements in $H^*(G_{\cancel{\,3\,}}, \Z_2)$ which are invariant under the induced action of the three-fold rotation $C_3$. %

\end{enumerate}

\section{Deriving Lieb--Schultz--Mattis (LSM) constraints}\label{sec:LSM}

Equipped with knowledge of the mod-2 cohomology of crystallographic groups, we are now well-positioned to investigate various LSM constraints in 3D systems. As we will show below, for every IWP of the space group $G$, there is an element in the third group cohomology $H^3(G, \Z_2)$ that can be uniquely assigned to this IWP. This correspondence between IWPs and cohomology elements underlies the physics of LSM constraints when the coefficient, $\Z_2$, of the cohomology $H^3(G, \Z_2)$ is interpreted as a classification of \emph{anomalous textures} \footnote{We refrain from providing a precise definition of \emph{anomalous textures} and instead refer the reader to Ref.~\cite{Else2020} for a detailed exposition.} for the on-site degrees of freedom at this IWP: when the system has an additional internal symmetry, and the on-site degrees of freedom form a projective representation of this internal symmetry corresponding to the nontrivial element of the $\Z_2$ (hence an \emph{anomalous texture}), an \emph{LSM anomaly} is triggered, forbidding the physical system to have a unique, symmetric, gapped ground state.

Here, following the tradition of LSM theorems, we consider on-site degrees of freedom with a single $\mathbb{Z}_2$ classification under the action of internal symmetry. The most familiar example is spins with $\SO(3)$ rotation symmetry, where half-integer (resp. integer) spins correspond to a projective (resp. faithful) representation of $\SO(3)$ labeled by the nontrivial (resp. trivial) element of $H^2(\SO(3),U(1)) \cong \mathbb{Z}_2$.

Other internal symmetries---such as the $\Z_2\times \Z_2$ subgroup of $\SO(3)$ generated by two $\pi$ rotations along two perpendicular axes, or an anti-unitary time-reversal symmetry $\mathbb{Z}_2^{\mathcal{T}}$---can also lead to a $\mathbb{Z}_2$ classification for the on-site degree of freedom, the nontrivial class of which contains LSM constraints \cite{Ogata2019}. In this sense, the analysis of LSM constraints pertains to a broad class of models---including any exchange Hamiltonian written in terms of spin-1/2 operators in the absence of an external field and with or without spin--orbit coupling (XXZ, Dzyaloshinskii--Moriya, dipole-dipole, dipole-octopole etc.). 
For all these cases, the projective representation of the internal group $G_{\text{int}}$ triggers an LSM anomaly as a descendent of certain element of $H^5(G\times G_{\text{int}}, \U^{\text{or}})$, which classifies all the ('t Hooft) anomalies associated with the global symmetry $G\times G_{\text{int}}$.

For clarity, we will primarily assume the internal symmetry to be $\SO(3)$ in the rest of discussions; the results however apply to all the internal symmetries mentioned above.

\subsection{LSM(OH) theorem for translation symmetries}

The Lieb--Schultz--Mattis theorem in 1D, along with its later generalization to higher dimensions by Oshikawa and Hastings, provides important constraints on the ground state of a many-body lattice system's Hamiltonian based solely on basic symmetry properties, without relying on other specific details of the Hamiltonian. 
For simplicity, we will focus on the case where the system exhibits on-site $\SO(3)$ symmetry without spin--orbit coupling. A key feature of $\SO(3)$ symmetry is that its projective representations are classified by $\Z_2$, corresponding to half-integer or integer spins in physical terms. This discussion can be readily extended to cases where the projective representation of the internal symmetry is classified by $(\Z_2)^k$. We will briefly comment on situations where the projective representation is $\Z_3$-classified, with a more comprehensive treatment left for future work.

\begin{statement}\label{LSMOH}
(Lieb--Schultz--Mattis--Oshikawa--Hastings \cite{lieb1961two,PhysRevLett.84.1535,PhysRevB.69.104431})
In a crystal with translation symmetry and on-site $\SO(3)$ symmetry, if there are odd numbers of spin-1/2 degrees of freedom per unit cell, then there cannot be a unique, symmetric, gapped ground state.
\end{statement}

This constraint on the ground state wave function resembles that of the 't Hooft anomaly on the ground state wave function or the low-energy subspace of a quantum field theory (QFT). This connection was later explored in \cite{Cheng2015,Cho2017,Jian2017}, identifying LSMOH constraints as a mixed anomaly between (lattice) translation and internal $\SO(3)$ symmetries.

According to the Crystalline Equivalence Principle \cite{PhysRevX.8.011040,Else2020}, the 't Hooft anomaly is characterized by an element in 
\begin{equation}
    H^{k+2}(\Z^k\times \SO(3), \U) \cong \Z_2,
\end{equation}
where the nontrivial element can be written as
\begin{equation}\label{eq:coho_element}
    \exp(i \pi A_x A_y\cdots \cup w_2^{\SO(3)}).
\end{equation}
Here $\exp(i\pi \cdots)$ maps an element $\lambda \in H^{k+2}(G, \Z_2)$ to an element $\exp(i\pi \lambda) \in H^{k+2}(G, \U)$. $w_2^{\SO(3)}$ is the generator of $H^2(\SO(3), \Z_2)\cong \Z_2$ and $\exp(i\pi w_2^{\SO(3)}) \in H^2(\SO(3), \U)\cong \Z_2$ is the cohomology element corresponding to the projective representation of $\SO(3)$, i.e. spin-1/2. $A_x A_y \cdots$ is the generator of $H^k(\Z^k, \Z_2) \cong \Z_2$ and $A_x, A_y, \cdots$ are the generators of each individual $H^1(\Z, \Z_2)$, which can be roughly interpreted as the gauge fields corresponding to translations in each direction. 

\begin{figure}
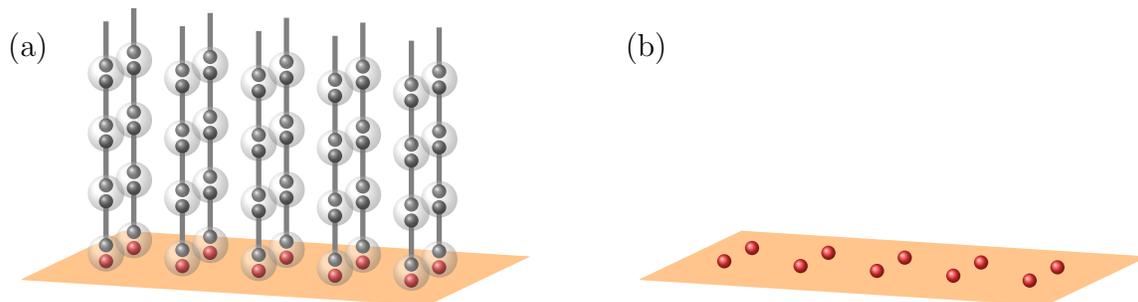


\caption{Bulk-edge correspondence for $k$-dimensional spin-1/2 lattice (the figure illustrates $k=2$). The on-site spin-1/2 lives on the boundary of a fictitious $(k+1)$-dimensional bulk. (a): the $(k+1)$-dimensional bulk. The bulk SPT consists of $S=1$ Haldane chains forming an $k$-dimensional array. The boundary of each Haldane chain has an $S=1/2$ edge mode (red dots). (b): the $k$-dimensional boundary on which the on-site $S=1/2$ degrees of freedom live.}\label{fig:bulk_edge_corresp}
\end{figure}

This cohomology element can be understood from the perspective of anomaly-inflow \cite{Jian2017}. From this perspective, consider placing the $k$-dimensional spin-1/2 system on the boundary of an $(k+1)$-dimensional crystalline SPT, constructed by stacking $\SO(3)$ Haldane chains, i.e., ($1+1$)-dimensional SPTs protected by $\SO(3)$ symmetry, while respecting the translation symmetries along the original $n$ directions, as illustrated in Figure \ref{fig:bulk_edge_corresp}. The crystalline SPT is clearly characterized by the cohomology element in Eq.~\eqref{eq:coho_element}, which can be thought of as the partition function of the crystalline SPT (or its TQFT), if we integrate it over the $(k+2)$-dimensional spacetime $\mathcal{M}_{k+2}$ and write it as $\exp(i \pi\int_{\mc{m}_{k+2}} A_x A_y\cdots \cup w_2^{\SO(3)})$.

Now we slice a boundary for this crystalline SPT perpendicular to the extra dimension. Because the boundary of each $\SO(3)$ Haldane chain has a spin-1/2 edge mode, the boundary is exactly characterized by a lattice of spin-1/2 degrees of freedom with one spin-1/2 per unit cell. Hence, from the perspective of anomaly-inflow, we see that the original system carries an anomaly characterized by Eq.~\eqref{eq:coho_element}.

We mention that the cohomology element Eq.~\eqref{eq:coho_element} can be derived from the \emph{anomalous texture} of the lattice system \cite{Else2020}, using the isomorphism between equivariant homology and group cohomology. In 1D, this cohomology element can also be understood through gauging \cite{Cheng2023,Sahand2024} or the \emph{anomaly index} of symmetry actions \cite{Kapustin2024}.

\subsection{LSM constraints for general crystallographic groups and IWP}

We now explore the general conditions for a generic crystallographic group to give rise to LSM constraints, in particular in 3D. 

Recall from Section \ref{subsec:IWP} that, performing lattice homotopy, which is a smooth deformation of the lattice structure that should not change the anomaly of the system, puts all physical degrees of freedom to the Irreducible Wyckoff Positions (IWP) of the crystallographic group $G$. It is then conjectured in Ref.~\cite{Po2017latticehomotopy} that 
\begin{conj} (General LSM constraints \cite{Po2017latticehomotopy})
    A symmetric short range entangled (sym-SRE) phase is possible only when the lattice is smoothly deformable to a trivial lattice.
\end{conj}

In a ``modern'' point of view, the LSM theorem is a statement about \emph{anomalous texture cancellation condition} \cite{Else2020}: the anomaly carried by the microscopic degree of freedom (the ``anomalous texture'') must be cancelled by the anomaly of the macroscopic phase of the matter (or more precisely, the ``defect network'') to make the system anomaly-free.

The traditional LSM theorem is the statement that:
The ground state is noninvertible (i.e. has symmetry enriched topological order) if and only if the equivariant pushforward map 
\begin{equation}
H^2(G,\Lambda=Z_0(\mathbb{R}^d,\U))\hookrightarrow \mc{H}^G_{-2}(X=\mathbb{R}^d,\U) \cong H^{d+2}(G,\U), [\omega]\mapsto [\mu]
\end{equation}
has nontrivial image, i.e. $[\mu]$ is a genuine cocycle and not a coboundary.
In words, $[\omega]\mapsto [0]$ means that the anomaly of the anomalous texture can be canceled by an invertible-substrate defect network, while $[\omega]\mapsto [\mu]\neq [0]$ means that this cancellation is impossible.  In the latter case a unique symmetric short-range-entangled ground state is forbidden; a symmetric gapped phase, if present, must be nontrivial, for example topologically ordered or otherwise anomalous.

From the form of the equivariant pushforward map, to generalize to crystalline symmetries beyond pure translations, we postulate that given any lattice symmetry $G$, we can associate the lattice structure by an element $\lambda \in H^n(G, \Z_2)$ corresponding to the IWPs where projective representations live. Then the  corresponding element in group cohomology is given by
\begin{equation}\label{eq:assemble}
\exp\left(i\pi \lambda \cup w_2^{\SO(3)}\right).
\end{equation}
For general internal symmetry groups $G_{\text{int}}$, we just need to change $w_2^{\SO(3)}$ to the element in $H^2(G_{\text{int}}, \U)$ that characterizes the projective representation of the on-site symmetry. The main part of latter discussion is to identify this $\lambda$ as an element in the cohomology ring.

\subsection{LSM constraints in 2D}

Specializing to 2D, the cohomology element in $H^2(G, \Z_2)$ corresponding to the IWPs is worked out in \cite{10.21468/SciPostPhys.13.3.066} in an explicit way. In this subsection, we review these results with an eye toward generalizing them to 3D scenarios.

Given a wallpaper group symmetry, its IWP can be decomposed into the following three classes based on its little group, i.e., subgroup of the wallpaper group that leaves individual points of the IWP invariant:

\begin{itemize}
\item A fundamental domain that tiles the 2D space under the actions of translation and glide symmetries, with the corresponding little group $C_1$ (trivial),
\item A translation unit cell along a mirror axis, with the corresponding little group $D_1$,
\item An $n$-fold rotation center, with the corresponding little group either $C_n$ or $D_n$.
\end{itemize}

Here $C_n$ and $D_n$ are groups of $n$-fold rotation, and $n$-fold rotation together with reflection. In particular, $D_1$ is reflection only.

We point out that we can simply view these three classes as three no-go theorems for symmetric short range entangled (SRE) states. Namely, from lattice homotopy, if there is a nontrivial $\Z_2$-classified projective representation in each class, then except the case where $n=3$ in the third item, they all give rise to nontrivial LSM constraints and a symmetric-SRE is forbidden \cite{Po2017latticehomotopy}. In particular, if there is a nontrivial $\Z_2$-classified projective representation in any IWP (possibly after the operation of lattice homotopy), a nontrivial LSM constraint is triggered and a symmetric-SRE is forbidden.

\begin{statement}
(LSM constraints in 2D \cite{Po2017latticehomotopy})
 A 2D LSM no go theorem is triggered when and only when there are odd numbers of spin-1/2's in any of the three categories of IWPs listed above.
\end{statement}

To identify an element in $H^2(G, \Z_2)$ corresponding to these IWPs, first we mention that the elements in $H^2(G, \Z_2)$ can be identified by its dual \emph{homology element}, or chain representative, in $H_2(G, \Z_2)$. The action of some cohomology element on a specific homology element will give a function of the cohomology element, which is called \emph{topological invariant} in \cite{10.21468/SciPostPhys.13.3.066}. 

Specializing to $G$ being a wallpaper group, we need to identify the \emph{minimal} set of group elements, on which the topological invariants completely determine the IWP. Given the three types of IWPs above, we have the following three possibilities:
\begin{enumerate}
\item Given two commuting elements $g_1,g_2\in G$ such that $g_1g_2 = g_2g_1$, we can define the following topological invariant,
\begin{equation}
\phi_2[g_1, g_2] := \lambda(g_1, g_2) + \lambda(g_2, g_1).
\end{equation}
\begin{itemize}
    \item When $g_1$ and $g_2$ are two translations $T_{1,2}$ in two directions, this topological invariant $\phi_2[T_1, T_2] = \lambda(T_1, T_2) + \lambda(T_2, T_1)$ corresponds to the first class of IWPs.
    \item When $g_2$ is a reflection $M$ and $g_1$ is a translation $T$ along the reflection axis, this topological invariant $\phi_2[T,M]=\lambda(T,M) + \lambda(M, T)$ corresponds to the second class of IWPs.
\end{itemize}

\item Given an order-2 element $g\in G$ such that $g^2=1$, we can define the following topological invariant
\begin{equation}
\phi_1[g] := \lambda(g, g) + \lambda(1, 1).
\end{equation}
\begin{itemize}
    \item When $g$ is a $C_2$ rotation, this topological invariant corresponds to the third class of IWPs
\end{itemize}

\end{enumerate}

In addition, the Bieberbach group $pg$ is special, whose unique IWP is the fundamental domain generated by a translation $T$ and a glide symmetry $G$. Its topological invariant, or simply the chain representative for the nontrivial element in $H_2(pg, \Z_2)$ has a slightly complicated form:
\begin{equation}
    \widetilde{\phi}_2[T, G] = \lambda(TG, TG^{-1}) + \lambda(T, G) + \lambda(T, G^{-1}) + \lambda(G, G^{-1}). 
\end{equation}

Finally, in order to identify an element in $H^2(G, \Z_2)$, we need a complete list of chain representatives or topological invariants of $H_2(G, \Z_2)$. It turns out that for 2D wallpaper groups, there is only one more possibility:
\begin{equation}
\phi_1[M] = \lambda(M, M) + \lambda(1, 1),
\end{equation}
where $M$ is the mirror reflection. We can immediately see that this is a topological invariant involving $M$ only, yet the reflection $M$ cannot specify an IWP. We will call this the non-LSM invariant. It is a highly nontrivial check that, for all wallpaper groups $G$, the three topological invariants corresponding to IWPs together with the one extra non-LSM invariant completely spans all elements in $H_2(G, \Z_2)$, hence completely specify elements in $H^2(G, \Z_2)$.

Hence, to associate every IWP of wallpaper groups with an element in $H^2(G, \Z_2)$, we need to (1) write down the complete list of topological invariants for $G$, including one for every IWP and all the non-LSM invariants, the number should be equal to the rank of $H^2(G, \Z_2)$; (2) find out the element in $H^2(G, \Z_2)$ whose value on the associated topological invariant is one, while zero on all the other topological invariants. This procedure is explicitly carried out in Ref.~\cite{10.21468/SciPostPhys.13.3.066} for all wallpaper groups. 

In 2D the identification of an IWP and an element in $H^2(G, \Z_2)$ can be made more rigorous by the argument of ``inserting a flux'' or either $\SO(3)$ rotation symmetry \cite{10.21468/SciPostPhys.13.3.066} or wallpaper group symmetry $G$ \cite{Po2017latticehomotopy}. Specifically, we can consider coupling the system to a probe gauge field of the $\SO(3)$ symmetry and examine the flux, or monopole, of this $\SO(3)$ gauge field. Because the wave function of the system acquires a $-1$ phase factor when an $\SO(3)$ monopole circles around a Haldane chain, if any of the 3 basic no-go theorems is triggered, the $\SO(3)$ monopole will carry a specific projective representation of $G$. Specifically, for the three types of IWPs, $T_1 T_2 T_1^{-1} T_2^{-1}$, $T M T^{-1} M^{-1}$ and $C_2^2$ acting on $\SO(3)$ will generate an extra $-1$ sign. Hence, we can view $\lambda \in H^2(G, \Z_2)$ as the projective representation, or the symmetry fractionalization pattern, of the $\SO(3)$ monopole, which completely encodes the LSM constraint. 

Finally, we have the following observation about the subgroup of $H^2(G, \Z_2)$ containing cohomology elements associated with LSM constraints,

\begin{statement} \label{LSM2DStatement}(LSM anomaly in 2D)
2D LSM constraints are associated with 
the subgroup $\ker{\beta_2}$ of $H^2(G, \Z_2)$, that is, the kernel of the Bockstein homomorphism:
\begin{equation}
    \beta_2\colon H^2(G, \Z_2) \rightarrow H^3(G, \Z^{\text{or}}).
\end{equation}

\end{statement}

For 2D wallpaper groups $G$, this condition is isomorphic to the condition that they vanish under the action of $\mathcal{SQ}^1:= Sq^1 + A_m\cup $, defined in Eq.~\eqref{eq:twisted_SQ1}. Ref.~\cite{10.21468/SciPostPhys.13.3.066} gives an argument by restricting to the $\U$ subgroup of $\SO(3)$.

\subsection{LSM constraints in 3D}

In this subsection, we go to 3D and outline the algorithm to identify the cohomology element in $H^3(G, \Z_2)$ corresponding to the IWPs of every 3D space groups. The full result for all 3D space groups is collected in the table in Appendix~\ref{collection230}.

Similar to the 2D scenarios, in 3D lattices the IWPs can be classified into the following five categories:
\begin{itemize}
    \item A fundamental domain that tiles the 3D space under the actions of translation, glide and screw symmetries, 
    \item A fundamental domain that tiles a 2D reflection plane under the actions of translation, glide and screw symmetries,
    \item A translation unit cell along a $C_2$ rotation axis,
    \item An intersection point of two perpendicular $C_2$ rotation axes,
    \item An inversion center.
\end{itemize}

We see that they roughly correspond to codimension $0$, $1$, $2$, and $3$ (for the last two items) regions in Euclidean space.  The following is the physical LSM input used in the rest of the paper.  It is supported by the cohomological anomaly construction and by known lower-dimensional and special-case LSM theorems, but a fully general many-body proof in 3D is beyond the present paper.

\begin{conj}[3D LSM constraints]\label{Thm:LSM3dphys}
 A 3D LSM no-go theorem is triggered precisely when there are odd numbers of spin-$1/2$ degrees of freedom on any of the five categories of IWPs listed above.
\end{conj}

Now, in 3D, our task is to associate an element in $H^3(G, \Z_2)$ to every IWP for all 3D space groups. To achieve this task, we define the following three types of topological invariants in parallel to our considerations in 2D:  
\begin{enumerate}
\item Given three mutually commuting elements $g_1, g_2, g_3\in G$, we define the following topological invariant
\begin{equation}\label{varphi3}
\varphi_3[g_1, g_2, g_3] := \lambda(g_1, g_2, g_3) + \lambda(g_1, g_3, g_2) + \lambda(g_2, g_1, g_3) + \lambda(g_2, g_3, g_1) + \lambda(g_3, g_1, g_2) + \lambda(g_3, g_2, g_1).
\end{equation}
\begin{itemize}
\item When $g_1$, $g_2$ and $g_3$ are translations in three directions, this topological invariant corresponds to the first class of IWPs. 
\item When $g_3$ is reflection and $g_1$ and $g_2$ are translations along two directions in the reflection plane, this topological invariant corresponds to the second class of IWPs.
\end{itemize}

\item Given two group commuting elements $g_1, g_2\in G$ such that $g_2^2=1$ and $g_1g_2 = g_2g_1$, we define the following topological invariant,
\begin{equation}\label{varphi2}
\varphi_2[g_1, g_2] := \lambda(g_1, g_2, g_2) + \lambda(g_2, g_1, g_2) + \lambda(g_2, g_2, g_1) + \lambda(g_1, 1, g_1).
\end{equation}
\begin{itemize}
    \item When $g_2$ is a $C_2$ rotation and $g_1$ is translation/glide/screw along the rotation axis, this topological invariant corresponds to the third class of IWPs.
    \item When $g_1$ and $g_2$ are two $C_2$ rotations along two perpendicular axes, this topological invariant corresponds to the fourth class of IWPs. 
\end{itemize}

\item Given an order-2 element $g\in G$ such that $g^2=1$, we define the following topological invariant
\begin{equation}\label{varphi1}
\varphi_1[g] := \lambda(g, g, g) + \lambda(g, 1, g).
\end{equation}
\begin{itemize}
    \item When $g$ is inversion, this topological invariant corresponds to the fifth class of IWPs. 
\end{itemize}

\end{enumerate}

In addition, we should consider Bieberbach groups separately. The topological invariant for each Bieberbach group is special, and it is listed in Appendix~\ref{collection230} under the table for each Bieberbach group separately.

For topological invariants not associated with IWP, i.e., the non-LSM invariants, we can restrict to the following four possibilities:
\begin{itemize}
\item $\varphi_1[C_2]$, where $C_2$ is some two-fold rotation,
\item $\varphi_1[M]$, where $M$ is some reflection,
\item $\varphi_2[X, M]$, where $X$ is any generator that commutes with reflection $M$,
\item $\varphi_2[C_4, C_4^2]$, where $C_4$ is some four-fold rotation.
\end{itemize}

We immediately see that for these non-LSM invariants, the group elements involved cannot specify an IWP. It turns out that the listed topological invariants combined, including LSM and non-LSM topological invariants, are enough to determine $H^3(G, \Z_2)$ completely, for every 3D space group $G$, which is a highly nontrivial consistency check.

Hence, to associate every IWP of 3D space groups $G$ with an element in $H^3(G, \Z_2)$, we have the algorithm (1) write down the
complete list of topological invariants for $G$, including one for every IWP and all the non-LSM invariant, the number
should be equal to the rank of $H^3(G, \Z_2)$; (2) find out the element in $H^3(G, \Z_2)$ whose value on the associated topological
invariant is one, while zero on all the other topological invariants. We implement this algorithm in our code and this indeed gives the cohomology element associated with every IWP, and the results are collected in each table in Appendix~\ref{collection230}. The tables contain basic information about all IWPs, including its name in ITC, its associated little group, and its coordinates. In the last two entries, we write down the topological invariant for each IWP, directly from the little group, as well as the cohomology class associated with the IWP.

As stated in the main text, given the internal symmetry to be e.g. $\SO(3)$, the value of the topological invariant for a given cohomology element in $H^3(G, \Z_2)$ can detect whether there is a spin-1/2 degree of freedom at the IWP. If we also insert a flux for $\SO(3)$ rotation symmetry in 3D, now the monopole should not be a point operator in 2D space, but a line in 3D. We can also analyze the symmetry fractionalization of this monopole, whose symmetry fractionalization pattern should be classified by an element in $H^3(G, \Z_2)$. This element is exactly the element associated with the IWP.

Finally, we record the following cohomological characterization of the IWP classes found in the tables.
\begin{statement}[Cohomological locus of the 3D IWP classes]\label{Thm:LSM3dmath}
The classes assigned in Appendix~\ref{collection230} to 3D IWPs lie in the subgroup
\begin{equation}
\ker{\beta_3^{\mathrm{or}}}\cap \ker\mathcal{SQ}^2\subset H^3(G, \Z_2),
\end{equation}
that is, in the intersection of kernels of the two maps
\begin{equation}
    \beta_3^{\mathrm{or}}\colon H^3(G, \Z_2) \rightarrow H^4(G, \mathbb{Z}^{\mathrm{or}}),
\end{equation}
and 
\begin{equation}
\mathcal{SQ}^2\colon H^3(G, \Z_2)\rightarrow H^5(G, \Z_2).
\end{equation}
Here, $\beta_3^{\mathrm{or}}$ is the Bockstein homomorphism associated with Eq.~\eqref{eq:twisted-short-exact} for the orientation local system.  The operation $\mc{SQ}^2$ is defined by
\begin{equation}
\mc{SQ}^2=Sq^2+w_1\cup Sq^1 + (w_2 + w_1^2)\cup,
\end{equation}
where $w_1\in H^1(G, \Z_2)$ and $w_2\in H^2(G, \Z_2)$ are the pullbacks of the Stiefel--Whitney classes of the natural 3-dimensional representation of $P$ acting on Euclidean space (see Table~\ref{tablePT2}). 
\end{statement}

Interestingly, $\mc{SQ}^1$ (defined in Eq.~\eqref{eq:twisted_SQ1}) and $\mc{SQ}^2$ are Steenrod operations on Thom space of $BG$ equipped with the 3-dimensional vector bundle explained in the statement, that is, the pullback of the natural 3-dimensional representation of $P$ acting on 3D Euclidean space \footnote{Given Eq.~\eqref{eq:twisted_SQ1} where $\beta_3$ appears, it might be visually appealing to change the criterion from $\ker{\beta_3^{\mathrm{or}}}\cap \ker\mathcal{SQ}^2$ to $\ker \mc{SQ}^1\cap \ker\mathcal{SQ}^2$. But the two sets are not exactly the same. This is different from the situation in 2D where, for all wallpaper groups, $\ker{\beta_2}$ is isomorphic to $\ker\mc{SQ}^1$. For example, in space group \hyperref[subsub:sg75]{No. 75~(P4)}, there exists an element $B_\alpha A_{\mathsf{q}}\in H^3(\hyperref[subsub:sg75]{P4}, \Z_2)$ that does not correspond to any LSM constraint, yet it is zero under the actions of both $\mc{SQ}^1$ and $\mc{SQ}^2$. Elements $\lambda\in H^3(G, \Z_2)$ which are zero under $\mc{SQ}^1$ and $\mc{SQ}^2$ but do not correspond to LSM constraints will yield a value of 1 for the fourth non-LSM topological invariant, i.e., $\varphi_2[C_4, C_4^2]$ for some $C_4$ rotation.}. Notably, $\mc{SQ}^2$ appears in the differential of the Atiyah-Hirzebruch spectral sequence, which classifies fermionic crystalline SPTs with spinless fermions, according to the fermionic Crystalline Equivalence Principle \cite{Debray2021}. It is intriguing to see how this fact may influence efforts to generalize LSM theorems to crystalline systems featuring on-site Majorana fermions \cite{Aksoy2021,Hsieh2016,Seiberg2024}.

\subsection{Examples}\label{subsec:examples}

Now, through a few examples,  we illustrate how to extract the data of LSM constraints from the tables we provide in Appendix \ref{collection230}. We also highlight certain interesting features of the groups considered in these examples.

\subsubsection{No. 16: \texorpdfstring{$P222$}{P222}}

\hyperref[subsub:sg16]{No. 16 ($P222$)} is generated by three translations and two two-fold rotations along two perpendicular axes. According to the IWP Table of \hyperref[subsub:sg16]{No. 16}, there are in total eight IWPs, corresponding to one site, three edge centers, three face centers, and one body center in a type-$P$ Bravais lattice, which are precisely eight intersection points of two two-fold rotations in a unit cell. 

From the Table, we can immediately write down the cohomology element in $H^3(G, \Z_2)$ for $2^8=256$ different kinds of LSM constraints for \hyperref[subsub:sg16]{No. 16}. Note that according to Ref.~\cite{Po2017latticehomotopy}, we can obtain 255 LSM constraints from considering lattice homotopy, yet there is one LSM constraint that cannot be derived this way, and it corresponds to putting spin-1/2's on all IWPs. We believe that the last one still corresponds to LSM constraints, and its corresponding cohomology element can be obtained by summing over the last column in the IWP Table of \hyperref[subsub:sg16]{No. 16}, which is $A_c A_{c'} (A_c + A_{c'})$.

\subsubsection{No. 19: \texorpdfstring{$P2_12_12_1$}{P212121}}

\hyperref[subsub:sg19]{No. 19 ($P2_12_12_1$)} is generated by three translations and two two-fold screws. This group is torsion-free, i.e., there is no group element other than identity such that some power of the group element is identity, hence it is a Bieberbach group in 3D. And we have observed that it satisfies $H^3(G, \Z_2) \cong \Z_2$ and $H^k(G, \Z_2)$ are all trivial for $k\geq 4$. 

For these Bieberbach groups, the topological invariant, or simply the chain representative for $H_3(G, \Z_2)\cong \Z_2$, is usually very complicated. This is a special feature of these Bieberbach groups. We provide one such candidate in Eq.~\eqref{TI_19}.

From the coordinate entry of the (unique) IWP in the IWP Table of \hyperref[subsub:sg19]{No. 19}, we see that every unit cell contains four regions related to each other by screw operations. Hence, every unit cell contains four fundamental domains. Let us restrict to the subgroup of \hyperref[subsub:sg19]{No. 19}, say just \hyperref[subsub:sg1]{No. 1~($P1$)}. Because each unit cell contains four fundamental domains, the IWP for \hyperref[subsub:sg19]{No. 19} no longer remains to be the IWP for \hyperref[subsub:sg1]{No. 1}, thus the restriction of the cohomology element $A_c B_{\beta 1}$ is also zero.

\subsubsection{No. 143: \texorpdfstring{$P3$}{P3}}

\hyperref[subsub:sg143]{No. 143 ($P3$)} is generated by three translations and a three-fold rotation. We see that there are three IWPs for this group, yet these three IWPs correspond to the same IWP once we remove the three-fold rotation. Hence, when considering mod-2 cohomology, or when the on-site degree of freedom is $\Z_2$ classified, we need to treat these three IWPs on equal footing. In particular, they correspond to the same cohomology element in $H^3(\hyperref[subsub:sg143]{P3}, \Z_2)$.

Still, when the on-site degrees of freedom are $\Z_3$-classified, we will see that the three IWPs correspond to three different cohomology elements, in fact three generators, in $H^3(\hyperref[subsub:sg143]{P3}, \Z_3)\cong (\Z_3)^6$. These three elements generate the subgroup $H^3(\hyperref[subsub:sg143]{P3}, \Z^{\text{or}})\otimes \Z_3 \cong (\Z_3)^3$.

\subsubsection{No. 147: \texorpdfstring{$P\bar{3}$}{P-3}}

\hyperref[subsub:sg147]{No. 147 ($P\bar{3}$)} is generated by three translations, a three-fold rotation and an inversion. There are five IWPs in total. If we forget about the three-fold rotation and restrict to the subgroup $\hyperref[subsub:sg2]{P\bar{1}}\subset \hyperref[subsub:sg147]{P\bar{3}}$, four of these IWPs still survive as nontrivial IWPs. However, there is one, with coordinate $(1/3,2/3,z),(2/3,1/3,-z)$, which does not survive as a nontrivial IWP, because it does not correspond to any inversion center of $\hyperref[subsub:sg2]{P\bar{1}}$. Hence, this IWP will not contribute to LSM constraints (for on-site $\Z_2$ degrees of freedom), and corresponds to no element in $H^3(\hyperref[subsub:sg147]{P\bar{3}}, \Z_2)$.

When the on-site degree of freedom is $\Z_3$-classified, we see that this IWP corresponds to a nontrivial cohomology element in $H^3(\hyperref[subsub:sg147]{P\bar{3}}, \Z_3)\cong (\Z_3)^3$. In particular, this element is one of the two generators for $H^3(\hyperref[subsub:sg226]{P\bar{3}}, \Z^{\text{or}})\otimes\Z_3\cong(\Z_3)^2$.

\section{Anomaly matching and relationship to PSG on the pyrochlore lattice}\label{sec:anomaly_matching}

In this section, to establish a connection with physics of lattice systems, we focus on space group \hyperref[subsub:sg227]{No. 227~($Fd\bar{3}m$)}, which contains the crystalline symmetry of both pyrochlore and diamond lattices. To illustrate the application of our results of the LSM anomaly, we present an example that shows how we can use the expression of LSM anomaly to give concrete prediction of the emergent infrared (IR) theory through anomaly matching. Specifically, motivated by Refs.~\cite{Liu2021pyrochlore}, we consider $\U$ quantum spin liquids (QSL) on the pyrochlore lattice. 

\subsection{Review of 3D Quantum Spin Liquid}

Let us begin with a brief overview of the physics underlying $\U$ quantum spin liquids (QSLs) in three dimensions, in particular the emergent IR field theory and its associated anomaly. Quantum spin liquids are zero-temperature phases of magnets in which the fluctuating quantum spins avoid magnetic long-range order and stay in liquid-like states \cite{Savary2017QSLReview}. These states are fundamentally characterized by intrinsic long-range entanglement: the non-local excitations interact with each other through an emergent gauge field, making gauge theories a natural framework for describing QSLs. Quite often, the elementary excitations in a QSL exhibit the phenomenon of \emph{symmetry fractionalization}---they carry fractional quantum numbers of the full symmetry $\mathcal{G}$ which label the projective representations of $\mathcal{G}$. 

Extensive research has been conducted on 2D QSLs, partly driven by the LSM anomaly and anomaly matching \cite{Wang2017,Song2018,Song2018a,10.21468/SciPostPhys.13.3.066,Metlitski2017,Zou2021,Cheng2015,Ye2024}. Notably, given the lattice structure we have been considering, the LSM anomaly should be equal to the anomaly of the emergent IR theory. From this simple idea, Refs.~\cite{10.21468/SciPostPhys.13.3.066,Ye2024} have initiated a comprehensive search for patterns that match the LSM anomaly (or the anomalies of UV systems more generally) with the anomalies of emergent QSL phases, regardless of whether these phases are gapped or gapless. Anomaly matching provides a valuable framework for understanding how crystalline symmetry acts on low-energy degrees of freedom in the infrared limit, ultimately allowing us to systematically classify all possible realizations of these QSL phases on a given lattice.

Focusing on three dimensions, one of the most extensively studied examples is the QSL on the pyrochlore lattice \cite{Gingras_2014}. This lattice, as illustrated in Fig.~\ref{fig:fd-3m}, is proposed to host a QSL phase since the very birth of the concept \cite{Anderson1956}. A significant theoretical breakthrough occurred in 2004, when Ref.~\cite{Hermele2004} established a mapping from the pyrochlore XXZ model with local Ising anisotropy to a $\U$ QSL phase—commonly referred to as the ``quantum spin ice'' or ``Coulomb phase'' \cite{PhysRevB.68.184512}—described by an emergent Maxwell's theory of electrodynamics. Since then, the properties of pyrochlore QSLs have been the focus of extensive study \cite{Savary2012,PhysRevX.9.011005,Yao2020,Yan2020}, with numerous experiments reporting liquid-like behaviors in rare-earth pyrochlore materials \cite{Sibille2015,Sibille2017,Gao2019,Gaudet2019,Leger2021}.

Now we move to the field-theoretic aspects of $(3+1)$-dimensional $\U$ Maxwell theory that describes the quantum spin ice phase. The theory has been studied in great detail in this context \cite{Wang2015QSL,Zou2017a,Zou2017,Ning2019,Hsin2020}. Here we provide the essential background for later consideration. Generally, a $(3+1)$-dimensional $\U$ gauge theory is described by the following Lagrangian at low energies:
\beq
\mathcal{L} = -\frac{1}{4g_0^2}f^{\mu\nu}f_{\mu\nu} + \frac{\theta}{32\pi^2}\epsilon_{\mu\nu\lambda\rho}f^{\mu\nu}f^{\lambda\rho}.
\eeq
Here, $g_0$ is the gauge coupling strength and $\theta$ is the axion angle, which is $2\pi$ or $4\pi$-periodic depending on whether the charges are bosonic or fermionic. At low energies, the theory simply describes propagating photons. Above
certain energy threshold, there are fractional excitations carrying
electric and magnetic charges, and we denote the excitations carrying one unit of electric (resp. magnetic) charge as $\mathcal{e}$ (resp.  $\mathcal{m}$). For simplicity from here on we restrict to the case where $\theta=0$. 

UV symmetries, including lattice symmetries and internal symmetries, can act on $\mathcal{e}$ and $\mathcal{m}$ in a nontrivial way. We call it the \emph{symmetry enrichment pattern} of $\mc{G}$. The symmetry enrichment pattern can exhibit a rich variety of phenomena \cite{Ning2019}: 
\begin{itemize}
\item Permutation of $\mc{e}$ and $\mc{m}$. Specifically, for a unitary symmetry, it either does not change the charges or acts as charge conjugation, i.e., $\mc{e}\rightarrow -\mc{e}, \mc{m}\rightarrow -\mc{m}$. For an anti-unitary symmetry, its action on $\mc{e}$ and $\mc{m}$ is either $\mc{e}\rightarrow -\mc{e}, \mc{m}\rightarrow \mc{m}$ or $\mc{e}\rightarrow \mc{e}, \mc{m}\rightarrow -\mc{m}$. In addition, for the reflection $M$ or the inversion $I$, from the Crystalline Equivalence Principle \cite{PhysRevX.8.011040} we should treat them as anti-unitary symmetries which also act on $\mc{e}$ and $\mc{m}$ as either $\mc{e}\rightarrow -\mc{e}, \mc{m}\rightarrow \mc{m}$ or $\mc{e}\rightarrow \mc{e}, \mc{m}\rightarrow -\mc{m}$. 

We mention that when both $\mc{e}$ and $\mc{m}$ are bosons, the permutation pattern can be highly nontrivial \cite{Ning2019}. Yet motivated by \cite{Liu2021pyrochlore}, we are primarily concerned with the case where $\mc{e}$ is fermionic while $\mc{m}$ is bosonic, and thus we do not need to consider them. Note that when both $\mc{e}$ and $\mc{m}$ are fermions, the theory has a nontrivial beyond-cohomology gravitational anomaly \cite{Wang2014Science,PhysRevD.92.085024,Wang2015QSL}, which is incompatible with the lattice structure we have.

\item Symmetry fractionalization of $\mc{e}$ and $\mc{m}$. We sometimes also say that $\mc{e}$ and $\mc{m}$ carry fractional quantum numbers of the symmetry, which goes beyond a mere sign flip of the charges. There are two pieces of data for symmetry fractionalization: $\omega_{\mc{e}}\in H^2_{\rho_{\mc{e}}}(\mathcal{G}, \U)$ and $\omega_{\mc{m}}\in H^2_{\rho_{\mc{m}}}(\mathcal{G}, \U)$, where the action of $\rho_{\mc{e}}$ and $\rho_{\mc{m}}$ of $\mc{G}$ on $\U$ module is determined by its action on the charges: a group element $g\in \mc{G}$ will act on $\U$ module for $\mc{e}$ by complex conjugation iff it flips the charge of $\mc{e}$, and similarly for $\mc{m}$. 

\end{itemize}

Now we connect these field-theoretic discussions with the parton construction for $\U$ QSLs. In the convention of Ref.~\cite{Liu2021pyrochlore}, a fermionic parton carries both physical spin-1/2 and the charge of the emergent $\U$ gauge field, and we identify it with the electric charge $\mc{e}$. The magnetic charge $\mc m$ becomes the $\U$ flux as seen by $\mc{e}$. Nevertheless, it is a nontrivial task to construct the operator corresponding to magnetic charge $\mc m$ \cite{zhanggaoliuchen2020} within this parton construction. 

When $\mc{e}$ and $\mc{m}$ are not swapped by the symmetry, one can consider the symmetry fractionalization patterns of individual $\mc{e}$ (or $\mc{m}$), and there is a completely general recipe to classify all the symmetry fractionalization patterns for crystalline symmetry, called the PSG classification \cite{wen_psg}. Yet it is difficult to obtain the symmetry fractionalization pattern of $\mc{e}$ and $\mc{m}$ simultaneously using PSG. Besides, in principle there can be symmetry fractionalization patterns beyond PSG calculation \cite{Ye2024}. Later, we will compare our results obtained from LSM anomaly on the 3D pyrochlore lattice with the results obtained from PSG.

With the information above, Ref.~\cite{Ning2019} proposes that the anomaly for the theory with both $\mc{e}$ and $\mc{m}$ bosons is given by
\beq\label{eq:anomaly_Maxwell_bosonic}
\exp\left(-2\pi i ~ \omega_{\mc{e}} \cup \beta_2(\omega_{\mc{m}})\right).
\eeq
Here $\beta_2$ is the Bockstein homomorphism associated with the short exact sequence $\Z \rightarrow \mathbb{R} \rightarrow \U$, which maps $\omega_{\mc{m}}\in H^2_{\rho_{\mc{m}}}(\mathcal{G}, \U)$ to $\beta_2(\omega_{\mc{m}})\in H^3_{\rho_{\mc{m}}}(\mathcal{G}, \Z)$. In the expression \eqref{eq:anomaly_Maxwell_bosonic}, $\omega_{\mc{e}}\in H^2_{\rho_{\mc{e}}}(\mathcal{G}, \U)$ takes values in $[0,1)$. From our requirement of permutation patterns, the total expression is an element in $H^5(\mc{G}, \U^{\text{or}})$, as expected. 

When $\mc{e}$ is a fermion and $\mc{m}$ a boson, according to Ref.~\cite{Hsin2020}, an extra $w_2$ term exists in \eqref{eq:anomaly_Maxwell_bosonic}, here $w_2$ is the second Stiefel--Whitney class for the tangent bundle of the spacetime manifold $\mc{M}_5$. From the Wu formula $Sq^2(\lambda) = (w_2+w_1^2)\lambda$, $\lambda\in H^3(\mc{M}_5, \Z_2)$, we propose that the anomaly for the theory with fermionic $\mc{e}$ and bosonic $\mc{m}$ is given by the following expression
\beq\label{eq:anomaly_Maxwell}
\exp\left(-2\pi i ~ \omega_{\mc{e}} \cup \beta_2(\omega_{\mc{m}})\right)\cdot \exp\left(-2\pi i ~ \left(Sq^2 + (\mc{SQ}^1(1))^2\right) \left(\beta_2(\omega_{\mc{m}})\right) \right),
\eeq
here the extra $Sq^2$ is the usual Steenrod operation on the group cohomology of $\mc{G}$, while $\mc{SQ}^1$ is the twisted Steenrod defined in Eq.~\eqref{eq:twisted_SQ1}. The total expression is still an element in $H^5(\mc{G}, \U^{\text{or}})$. This expression will serve as the IR anomaly $\Omega_{\text{IR}}$ of the system under consideration.

\subsection{Anomaly matching}

In this subsection, we attempt to see what kinds of symmetry fractionalization patterns can indeed match the LSM anomaly with the anomaly Eq.~\eqref{eq:anomaly_Maxwell} for the emergent $(3+1)$-dimensional $\U$ Maxwell theory (at $\theta=0$). We will focus on the case where the internal symmetry is $\SO(3)$ and the full symmetry is $\mc{G}=G\times \SO(3)$, where $G$ is space group \hyperref[subsub:sg227]{No. 227~($Fd\bar{3}m$)}. 

As discussed in Sec.~\ref{subsec:IWP}, the group \hyperref[subsub:sg227]{No. 227} contains four IWPs, $a$, $b$, $c$, and $d$ (see Fig.~\ref{fig:fd-3m}): $c$ and $d$ consist of inversion centers, whereas $a$ and $b$ consist of intersection points of two perpendicular axes. Theses IWPs determine two lattice types: placing physical degrees of freedom on either $c$ or $d$ generates a pyrochlore lattice, and placing them on either $a$ or $b$ generates a diamond lattice.

From the IWP Table of \hyperref[subsub:sg227]{No. 227}, we can explicitly write down the LSM anomaly in $H^5(G\times \SO(3), \U)$ for putting spin-1/2 degrees of freedom on IWP c (spin-1/2 pyrochlore lattice), 
\begin{equation}\label{eq:anomaly_pyrochlore}
\exp\left(i\pi A_i (A_i^2 + A_i A_m + B_{xy+xz+yz})w_2^{\SO(3)}\right).
\end{equation}
This is the UV anomaly $\Omega_{\text{UV}}$ for our system.

Now we start considering anomaly matching. To be specific, we ask the following question: 
\begin{flushleft}
~~~\emph{What are all the symmetry fractionalization patterns compatible with anomaly matching?}
\end{flushleft}
As discussed before, we check for a given symmetry action (symmetry fractionalization pattern), whether we can equate the LSM anomaly in Eq.~\eqref{eq:anomaly_pyrochlore} with Eq.~\eqref{eq:anomaly_Maxwell} or Eq.~\eqref{eq:anomaly_Maxwell_bosonic}. If the equation cannot hold, then the symmetry fractionalization pattern cannot appear on the given lattice, and the corresponding symmetry-enriched $\U$ QSL cannot emerge in this lattice. In contrast, if the equation can hold, we conjecture that the corresponding symmetry-enriched $\U$ QSL can emerge in this lattice, according to the hypothesis of emergibility \cite{Zou2021,10.21468/SciPostPhys.13.3.066}.

Because $H^2_{\rho_{\mc{e}}}(\mc{G}, \U) = H^2_{\rho_{\mc{e}}}(G, \U) \oplus H^2(\SO(3), \U)$ and similarly for $\mc{m}$, we can consider $G$ and $\SO(3)$ separately. For $\SO(3)$, we say that the nontrivial (trivial) element in $H^2(\SO(3), \U)$ suggests that $\mc{e}$ or $\mc{m}$ carries projective (integer) representation under $\SO(3)$. 
From the parton consideration of Refs.~\cite{Liu2021pyrochlore,Liu2023diamond}, we will assume that $\mc{e}$ is a fermion and carries projective representation under $\SO(3)$. This way the PSG classification in Refs.~\cite{Liu2021pyrochlore,Liu2023diamond} gives exactly the data about symmetry fractionalization on $\mc{e}$. In contrast, $\mc{m}$ is a boson and carries an integer representation under $\SO(3)$.

To consider the lattice symmetry $G$, we need to track how $G$ permutes these excitations. There are four possible actions of the crystalline symmetry on the excitations:
\begin{itemize}
\item $I\colon \mc{e}\rightarrow \mc{e},\mc{m}\rightarrow -\mc{m}$, $M\colon \mc{e}\rightarrow \mc{e},\mc{m}\rightarrow -\mc{m}$
\item $I\colon \mc{e}\rightarrow \mc{e},\mc{m}\rightarrow -\mc{m}$, $M\colon \mc{e}\rightarrow -\mc{e},\mc{m}\rightarrow \mc{m}$
\item $I\colon \mc{e}\rightarrow -\mc{e},\mc{m}\rightarrow \mc{m}$, $M\colon \mc{e}\rightarrow \mc{e},\mc{m}\rightarrow -\mc{m}$
\item $I\colon \mc{e}\rightarrow -\mc{e},\mc{m}\rightarrow \mc{m}$, $M\colon \mc{e}\rightarrow -\mc{e},\mc{m}\rightarrow \mc{m}$
\end{itemize}
We will use $(w_I, w_M)$ to label whether the generator $I$ or $M$ gives an extra $-1$ or not when acting on $\mc{e}$, and the four cases correspond to $(w_I, w_M) = (0,0), (0,1), (1,0), (1,1)$, respectively \footnote{In the notation of \cite{Liu2021pyrochlore}, $\overline{C}_6=C_3I$, $S=C_2MI$. Hence, $(w_{\overline{C}_6}, w_S)$ is equal to $(w_I, w_M+w_I \mod 2)$.}. We focus on $(w_I, w_M) = (0, 0)$, which can match the field theory for quantum spin ice with $\theta=0$.

Specializing to $(w_I, w_M) = (0, 0)$, we can immediately read all the possible symmetry fractionalization classes from Table~\ref{tableZZor230}. The symmetry fractionalization of the crystalline symmetry $G$ for $\mc{e}$ is classified by $H^2(G, \U)=H^3(G, \Z) \cong (\Z_2)^3$, where the action of $G$ on $\U$ or $\Z$ is trivial. Written in terms of mod-2 cohomology, it is generated by 
\beq\label{eq:SF_e_choice}
\exp(i \pi B_{xy+xz+yz}),~~~ \exp(i \pi B_\alpha),\text{~~~and~~~}\exp(i\pi A_i A_m).
\eeq
In the notation of Ref.~\cite{Liu2021pyrochlore}, the coefficient of the three generators are exactly $\chi_1$, $\chi_{\overline{C}_6 S}$ and $\chi_{ST_1}$ (divided by $\pi$). 

The symmetry fractionalization of the crystalline symmetry $G$ for $\mc{m}$ is classified by $H^2(G, \U^{\text{or}})=\text{Tor}\left(H^3(G, \Z^{\text{or}})\right) \cong (\Z_2)^4$, where the action of both $M$ and $I$ on $\U$ or $\Z$ is nontrivial. Written in terms of mod-2 cohomology, it is generated by 
\beq\label{eq:SF_m_choice}
\exp(i \pi B_{xy+xz+yz}), ~~~\exp(i \pi B_\alpha),~~~ \exp(i\pi A_i^2),\text{~~~and~~~} \exp(i\pi A_m^2).
\eeq
This information is not easily obtained from the PSG calculation.

Now we just need to enumerate, for each symmetry fractionalization pattern given by Eqs.~\eqref{eq:SF_e_choice} and \eqref{eq:SF_m_choice}, whether the obtained IR anomaly $\Omega_{\text{IR}}$ of the $(3+1)$-dimensional $\U$ gauge theory from Eq.~\eqref{eq:anomaly_Maxwell} is identical to the UV anomaly $\Omega_{\text{UV}}$ from Eq.~\eqref{eq:anomaly_pyrochlore}, i.e., 
\begin{equation}
    \Omega_{\text{UV}} = \Omega_{\text{IR}}.
\end{equation}
In the end, we have the following result. First, the symmetry fractionalization of $\mc{m}$ is fixed to be
\beq\label{eq:mag_SF}
\exp\left(i\pi (A_i^2 + B_{xy+xz+yz})\right).
\eeq
Physically, this means that, in a $\U$ QSL, the magnetic charge $\mc{m}$ must carry fractional charges of inversion and $T_i T_j T_i^{-1} T_j^{-1} = -1$ when acting on $\mc{m}$. The symmetry fractionalization of $\mc{e}$ can be written as 
\beq\label{eq:e_SF}
\exp\left(i\pi w_2^{\SO(3)} + i \chi_1 B_{xy+xz+yz} + i \chi_{\overline{C}_6 S} B_\alpha)\right),
\eeq
where $\chi_1$ and $\chi_{\overline{C}_6 S}$ can take both values $\{0,\pi\}$, and the first term involving $w_2^{\SO(3)}$ signals that $\mc{e}$ is a spin-1/2 under $\SO(3)$.\footnote{When we allow $\mc{m}$ to carry projective representation under $\SO(3)$, there is one more possibility that matches the anomaly of the $(3+1)$-dimensional $\U$ gauge theory. Here the symmetry fractionalization of $\mc{m}$ is fixed to be
\begin{equation*}\label{eq:mag_SF2}
\exp\left(i\pi (w_2^{\SO(3)} + A_m^2 + B_{xy+xz+yz})\right),
\end{equation*}
while the symmetry fractionalization of $\mc{e}$ is also fixed 
\begin{equation*}
\exp\left(i\pi w_2^{\SO(3)} \right),
\end{equation*}
corresponding to the situation where all three $\chi_1,\chi_{\overline{C}_6 S},\chi_{ST_1}$ are zero. This possibility seems to be incompatible with the lattice construction in e.g.~\cite{Hermele2004}. And we are curious if it can really be realized on the pyrochlore lattice.}

We conclude that, given the assumption that $\mc{e}$ is a half-integer-spin fermion that does not change sign under the action of $M$ and $I$, i.e., $(w_I,w_M)=(0,0)$, and $\mc{m}$ is an integer-spin boson, 
\begin{enumerate}
    \item The calculation from pure anomaly matching gives exactly the same result as the PSG calculation in Ref.~\cite{Liu2021pyrochlore}.
    \item It also gives the symmetry fractionalization of magnetic charges $\mc{m}$ to be Eq.~\eqref{eq:mag_SF}.
\end{enumerate} 

It is interesting to compare the PSG calculation for other options of $(w_I, w_M)$ with the calculation of anomaly matching shown above, and obtain further constraints on the emergent IR theory possibly with a nonzero $\theta$-term. We leave it to future work.

\section{Summary and discussion}

In this paper, we obtained the mod-2 cohomology rings for all 230 3D space groups and studied the connection between cohomology elements and real space lattice structures. For a given lattice symmetry $G$, we demonstrated a way to associate each irreducible Wyckoff position--high symmetry points on the lattice roughly speaking--with a unique element in the third group cohomology of $G$, i.e. a 3-cocycle $\lambda\in H^3(G, \Z_2)$, using the notion of topological invariants. We derived this lattice--cohomology (or, more precisely, IWP--3-cocycle) correspondence in the most explicit manner by obtaining the inhomogeneous functions or GAP-computed representative functions for the relevant 3-cocycles.

This lattice--cohomology (or IWP--3-cocycle) correspondence is the essential mathematical description for Lieb--Schultz--Mattis constraints---namely, the lattice constraints that forbid a magnetic crystal consisting of projective degrees of freedom (such as spin-1/2 moments) to have a unique, symmetric, gapped ground state. 
Using the 3-cocycles of the space group cohomology, $\lambda$, together with projective representations of the internal symmetry (assumed to be classified by powers of $\Z_2$), we assembled the cohomological anomaly data predicted by Conjecture~\ref{Thm:LSM3dphys}. We applied these explicit data of quantum anomaly to the study of $(3+1)$-dimensional $\U$ quantum spin liquids on the pyrochlore lattice and compared the results we obtained with projective symmetry group calculations. 

We conclude the paper with a few comments and suggestions for future directions.

Obtaining the full structure of group cohomology for 3D space groups is a difficult task. We hope that the techniques and code that we developed for mod-2 cohomology can be used to calculate other homological-algebraic aspects of 3D space groups, including (co)homology and (co)bordism for 3D space groups, and also the equivariant version of them, which are important in understanding the topological phases protected/enriched by these crystalline symmetries. For example, we believe that our result will be important in understanding 3D crystalline SPTs, and could ultimately lead to a comprehensive classification of bosonic and fermionic SPTs protected by 3D space groups, as discussed in Ref.~\cite{PhysRevX.8.011040} (also see Eq.~\eqref{eq:missing}) and Refs.~\cite{Debray2021,Zhang2022,zhang2022construction}. 

Another direction is to study the structure of group cohomology for higher dimensional crystallographic groups. Notable dimensions are $k=4$, where a full classification of crystallographic groups has been achieved, and $k=6$, which is of special interest to string theory and may inform the study of quasicrystals in 3D (or the fusion of these two subjects \cite{Baykara:2024vss}). While conceptually this is a straightforward generalization, development on the computational homotopy side awaits.

Turning to the story of LSM constraints, a general many-body proof of Conjecture~\ref{Thm:LSM3dphys} in full 3D generality is still lacking. We view the cohomological tables in this paper as precise anomaly data that should help formulate such a proof. Even more broadly, we seek to establish a clear connection between general symmetry actions and the elements in group cohomology that classify anomalies. We note that the connection has been rigorously established just recently for the 1D case \cite{Kapustin2024}.

Having obtained the cohomology data corresponding to various LSM constraints in 3D, we illustrate the application of our results through a specific example in Sec.~\ref{sec:anomaly_matching}. We expect that our findings will have significant implications for understanding how different low-energy phases can be realized in specific lattice systems, thereby expanding our knowledge of the complex nature of phase diagrams in real materials with diverse lattice structures. Below, we outline a few candidates.

\begin{itemize}
    \item We anticipate that our analysis of $(3+1)$-dimensional quantum spin liquids can be generalized to provide a comprehensive understanding of symmetry actions in various $\U$ or $\Z_2$ quantum spin liquids across different lattice systems, similar to the $(2+1)$-dimensional story in Ref.~\cite{Ye2024}. This approach complements the projected symmetry group (PSG) analysis for 3D lattices presented in prior works \cite{PhysRevB.95.054404,PhysRevB.97.195141,Liu2019,PhysRevB.102.125140,Liu2021pyrochlore,Chern2021,Desrochers2021,Desrochers2022,Chern2022,PhysRevB.105.125122,Liu2023diamond}. We believe that employing the method of anomaly matching will yield valuable extra information about how crystalline or UV symmetries act on the emergent degrees of freedom in these systems \cite{jian_unpublished}.

    \item A simple class of phases in these lattice systems is the ferromagnetic phase, or magnetic order in general, which falls under the category of general spontaneous symmetry breaking phases. Even though the ground state of ferromagnets on the lattice can be straightforward, understanding its low-energy dynamics, particularly elucidating the detailed matching of the symmetry actions in the lattice and in the continuum can be subtle, as emphasized in \cite{Seiberg2024translation}. Investigating spontaneous symmetry breaking phases through the lens of anomaly and anomaly matching has proven to provide valuable insights into these systems \cite{Hason2020,Debray2023,Debray2024,Else2021,Wang2024SSB}, and we expect that applying these methods to crystalline symmetries can yield equally fruitful results.
\end{itemize}

Finally, it is intriguing to explore how our results about LSM constraints can be generalized to other systems. Notable examples include systems with filling constraints \cite{parameswaran2013topological,Watanabe2015,LU2020168060,Cheng2023}, systems with magnetic space group symmetries \cite{Lu2017,xuyang2018,Jiang2019,Else2020}, systems with long-range interactions \cite{Liu2024,Ma2024,Zhou2024}, systems with conserved dipole moments \cite{PhysRevB.101.165145,PhysRevB.103.125133}, systems with disorders \cite{PhysRevX.8.031028}, systems with Majorana translations \cite{Aksoy2021,Cheng2019,Hsieh2016,Seiberg2024}, open systems \cite{Kawabata2024,Zhou2023}, and many more \cite{Kobayashi2018,xuyang2018,else_quasicrystal_lsm,Gioia2022,Yao2024,Pace:2024acq,musser2024fractionalization}.

\begin{acknowledgments}

We thank Leon Balents, Federico Becca, Andrea Cappelli, Hank Chen, Benoît Douçot, Graham Ellis, Dominic Else, Zheng-Cheng Gu, Zhaoyu Han, Yin-Chen He, Jiankang Huang, Yasir Iqbal, Ethan Lake, Siddhant Mal, Joel Moore, Adam Nahum, Shang-Qiang Ning, Frédéric Piéchon, Andrew C. Potter, Yang Qi, Nathan Seiberg, Ryan Thorngren, Cenke Xu, Xu Yang, Liujun Zou, and many others for helpful discussions. C.L. would like to express deep gratitude to Bill Jacob for the early discussions and guidance that inspired this project. This research was supported in part through computational resources and services provided by Quantum Advanced Research Computing (QuARC) at the Stewart Blusson Quantum Matter Institute. C.L. acknowledges the fellowship support from the Gordon and Betty Moore Foundation through the Emergent Phenomena in Quantum Systems (EPiQS) program.

\end{acknowledgments}

\begin{appendices}

\section{Information about point groups and their group cohomology in 3D}

In this appendix, we list necessary information about point groups and their group cohomology for 3D space groups. This information is organized in Table \ref{tablePT1} and Table \ref{tablePT2}.

In Table~\ref{tablePT1}, we list the basic information about the point groups, including their international (Hermann--Mauguin) notation, Sch\"{o}nflies notation, abstract group structure, order, and generators in ``polycyclic order" in permutation notation and notation we use in Appendix~\ref{collection230}. In the first column we provide the ITC numbering of the space groups associated with each point group, and in the second last column, the numbering of the generators in ITC \cite{aroyo2013international} corresponding to the polycyclic generators.

\addtolength{\tabcolsep}{0pt} 
\begin{longtable}{c|cc|c|c|c|c|c}
\caption{Information about the structure of the 32 point groups. $*$: the symmetry operation may differ from the ITC operation by a translation. $\dag$: the generators $(5)$ and $(8)$ needs to be switched for space groups No. 115--120. $\ddag$: $(10)$ instead of $(7)$ for space groups No. 187--188.}\label{tablePT1}\\
\hline\hline
\multirow{2}{*}{No. }& \multicolumn{7}{c}{Point Group Information} \\
\cline{2-8}
&Intl &Sch\"{o}nflies&Abstract&Order&Permutation Notation&ITC Numbering$^*$&Polycyclic Gens\\
\hline
1 &$1$                   &$C_1$   &$\mathbb{Z}_1$& 1 &Trivial &Trivial&Trivial\\
\cline{1-8}
2 & $\overline{1}$       &$C_i$   & \multirow{3}{*}{$\mathbb{Z}_2$} &\multirow{3}{*}{2} & \multirow{3}{*}{$\langle(12)\rangle$}& \multirow{3}{*}{$(2)$} & $\langle I\rangle$\\
3--5 &$2$                &$C_2$   &&&&& $\langle C_2\rangle$\\
6--9 &$m$                &$C_s$   &&&&& $\langle M\rangle$\\
\cline{1-8}
10--15 &$2/m$            &$C_{2h}$& \multirow{3}{*}{$(\mathbb{Z}_2)^2$} &\multirow{3}{*}{4}& \multirow{3}{*}{$\langle (12),(34)\rangle$} & \multirow{3}{*}{(2),(3)} &$\langle C_2,I\rangle$\\
\hline\hline
\end{longtable}
\addtolength{\tabcolsep}{0pt}

In Table \ref{tablePT2}, we present information about the group cohomology of these point groups, including their mod-2 cohomology rings, Stiefel--Whitney classes for the natural 3D representation, and topological invariants to detect the associated IWPs in $H^3(G, \Z_2)$.

Notably, point groups with the same abstract group structure---such as $C_i$, $C_2$ and $C_s$, all of which are isomorphic to $\Z_2$---have identical mod-2 cohomology rings. However, they differ by their action on the 3D Euclidean space. The 3D Euclidean space can be viewed as a natural 3D representation of the point group, and we list the first, second, and third Stiefel--Whitney classes for this representation. It is evident that these differing actions are reflected in the distinct Stiefel--Whitney classes associated with the 3D representations.

In the third column about the mod-2 cohomology ring of point groups, the degree-1 generator $A_\bullet$ always denotes (the mod-2 reduction of) the character associated with the group element indicated by the subscript. For example, $A_{c'}$ denotes the character associated with a $C'_2$ rotation.\footnote{We comment that, for degree-2 and degree-3 generators, although we use the same label (for example, $B_\alpha$) in many point groups, these elements should not be assumed the same. In fact, there is no way to compare them unless there is a subgroup relation. For two point groups $P$ and $P'$ with $P'\subset P$, even if their cohomology contains elements with the same label, we do not require that the generator of the cohomology of $P$ reduces to the one with the same label in the cohomology of $P'$. For example, although $D_{4h}\subset O_h$, and both of their mod-2 cohomology contain the element labeled by $B_\alpha$, $B_\alpha$ of $H^*(O_h,\mathbb{Z}_2)$ actually reduces to $B_\alpha+A^2_{c'}+A_iA_{c'}+A_iA_m+A_i^2$ (and not $B_\alpha$) of $H^*(D_{4h},\mathbb{Z}_2)$. Nevertheless, as far as the mod-2 cohomology ring of the \emph{space group} is concerned, we have made sure that, whenever a generator of the mod-2 cohomology of a space group is the pullback of a generator of its associated point group, they are labeled the same.}%

Finally, in the last column, we list the topological invariants used to identify the corresponding space group cohomology elements in $H^3(G, \Z_2)$, based on the little groups of the IWPs. The definition of $\varphi_1$, $\varphi_2$ and $\varphi_3$ are given in Eqs.~\eqref{varphi1}, \eqref{varphi2} and \eqref{varphi3}, respectively.

If the little group is neither $C_1$ nor $C_s$, the topological invariant associated with an IWP is fully determined by its corresponding little group. For these cases, we list the topological invariant we choose for the given little group. In particular, when $w_3$ is nonzero, the topological invariant just detects $w_3$ in $H^3(P, \Z_2)$, such that the action of $w_3$ on it gives 1. When $w_3$ is zero, the corresponding IWP is a translation unit cell along a $C_2$ rotation axis, and we choose the topological invariant to be $\varphi_2[X, C_2]$, where we use $X$ to label the translation, glide, or screw along the rotation axis.

If the little group is $C_1$ (i.e. trivial), the topological invariant must be that of one of the 10 3D Bieberbach groups, listed in Eq.~\eqref{Bieberbach-3d}, and we list the topological invariant for each case separately below their IWP tables in Appendix \ref{collection230}. When the little group is $C_s$, depending on whether the action of space group in the reflection plane is either isomorphic to $p1$ or $pg$, we choose the topological invariant to be either $\varphi_3[T_1, T_2, M]$ or $\widetilde{\varphi_3}[T_1, G_2, M]$, where 
\begin{equation}\label{wilde3}
\begin{aligned}
    \widetilde{\varphi_3}[T_1, G_2, M] = & \lambda(M, T_1G_2, T_1G_2^{-1}) + \lambda(M, T_1, G_2) + \lambda(M, T_1, G_2^{-1}) + \lambda(M, G_2, G_2^{-1})  \\
    & + \lambda(T_1G_2, M, T_1G_2^{-1}) + \lambda(T_1, M, G_2) + \lambda(T_1, M, G_2^{-1}) + \lambda(G_2, M, G_2^{-1}) \\ 
    & + \lambda(T_1G_2, T_1G_2^{-1}, M) + \lambda(T_1, G_2, M) + \lambda(T_1, G_2^{-1}, M) + \lambda(G_2, G_2^{-1}, M).
\end{aligned}
\end{equation}
In the first case, $T_1$ and $T_2$ are two translations along the reflection plane. In the second case, $T_1$, $G_2$ are translation and glide along the reflection plane, respectively, such that $G_2^{-1}T_1 G_2 = T_1^{-1}$.

For little groups that contain a $C_3$ rotation, the topological invariant remains the same as that of the little group after disregarding the $C_3$ rotation. But for the little groups labeled by $(s)$, multiple IWPs
may share the same cohomology element, or there may be no corresponding cohomology element at all.

\addtolength{\tabcolsep}{0pt} 
\begin{longtable}{c|c|l|l|l|l|c}
\caption[Information about the mod-2 cohomology of the 32 space groups]{Information about the mod-2 cohomology of the 32 space groups. We defined $\mathcal{R}:= B_\alpha^3+C^2_{\alpha1}+C_{\alpha1} C_{\alpha2}+C^2_{\alpha2}$ and $w_3^{\text{47--74}}:=(A_i\!+\!A_c)(A_i\!+\!A_{c'})(A_i\!+\!A_c\!+\!A_{c'})$. 
In the last column, $X$ represents a translation, glide, or screw along the $C_2$  rotation axis. We use $\star$ to indicate that when the little group for the IWP is $C_1$ or $C_s$, each case must be considered separately. Groups labeled with ``(s)" imply that when the little group for the IWP falls into this category, multiple IWPs may share the same cohomology element, or there may be no corresponding cohomology element at all.}\label{tablePT2}\\
\hline
\hline
 \multicolumn{2}{c|}{Point Group} & \multicolumn{4}{c|}{Mod-2 Cohomology Ring Information} &{Topo. Inv.}\\
\cline{1-6}
Sch\"{o}nflies&Abstract&Expression&$w_1$ & $w_2$ & $w_3$ & for IWP \\
\hline
$C_1$   &$\mathbb{Z}_1$&$\mathbb{Z}_2[]$ &0&0&0&$\star$\\
\hline
$C_i$   & \multirow{3}{*}{$\mathbb{Z}_2$}&$\mathbb{Z}_2[A_i]$&$A_i$& $A_i^2$ & $A_i^3$ & $\varphi_1[I]$\\
$C_2$   &&$\mathbb{Z}_2[A_c]$&0&$A_c^2$&0&$\varphi_1[X,C_2]$\\
$C_s$   &&$\mathbb{Z}_2[A_m]$&$A_m$ &0&0&$\star$\\
\hline
$C_{2h}$& \multirow{3}{*}{$(\mathbb{Z}_2)^2$} &$\mathbb{Z}_2[A_c,A_i]$&$A_i$ &$A_i^2\!+\!A_c^2$ & $A_i(A_i^2\!+\!A_c^2)$&$\varphi_1[I]$\\
\hline
\hline
\end{longtable}
\addtolength{\tabcolsep}{0pt}

\section{Information about space groups as group extensions}

In this appendix, we summarize how to view the space groups $G$ as its point group $P$ extended by the translation group $T\cong\Z^3$ (see Eq.~\eqref{TGP}). These properties are listed in Table~\ref{tableExt}. 

As introduced in the Sec.~\ref{subsec:crystal}, given a point group $P$, the conjugation action $\rho$ of $P$ on the translations $T$ is different depending on different Bravais lattice types. The group extension is characterized by the conjugation action $\rho$ of the point group $P$ on translations $T$ (Eq.~\eqref{intrep}) and, for a given $\rho$, the second cohomology group $H^2_\rho(P,T)$. For the 32 point groups, there are altogether 14 Bravais lattice types. The distinct actions $\rho$ across all the point groups give rise to 73 \emph{arithmetic crystal classes}. 

The 14 Bravais lattice types are $aP,mP,mS,oP,oS,oI,oF,tP,tI,hR,hP,cP,cI,cF$. The first letter is dependent on the point group, and the second letter labels different basis for the translation generators. We choose the generators of translation symmetry according to the following convention:
\begin{itemize}

\item $P$ (Primitive) lattices: the three translation generators are
\begin{subequations}\label{TransBravaisP}
\begin{align}
&T_1\colon (x,y,z)\rightarrow (x+1, y, z),\\
&T_2\colon (x,y,z)\rightarrow (x, y+1, z),\\
&T_3\colon (x,y,z)\rightarrow (x, y, z+1).
\end{align}
\end{subequations}

\item $S$ (Base-centered) lattices: it has three subtypes $A,B,C$. For $C$ lattices, the three translation generators are
\begin{subequations}\label{TransBravaisC}
\begin{align}
&T_1\colon (x,y,z)\rightarrow (x+1/2, y+1/2, z),\\
&T_2\colon (x,y,z)\rightarrow (x-1/2, y+1/2, z),\\
&T_3\colon (x,y,z)\rightarrow (x, y, z+1).
\end{align}
\end{subequations}
Similarly, for $A$ lattices, the three translation generators are
\begin{subequations}\label{TransBravaisA}
\begin{align}
&T_1\colon (x,y,z)\rightarrow (x+1, y, z),\\
&T_2\colon (x,y,z)\rightarrow (x, y+1/2, z+1/2),\\
&T_3\colon (x,y,z)\rightarrow (x, y-1/2, z+1/2).
\end{align}
\end{subequations}

\item $I$ (Body-centered) lattices: the three translation generators are
\begin{subequations}\label{TransBravaisI}
\begin{align}
&T_1\colon (x,y,z)\rightarrow (x-1/2, y+1/2, z+1/2),\\
&T_2\colon (x,y,z)\rightarrow (x+1/2, y-1/2, z+1/2),\\
&T_3\colon (x,y,z)\rightarrow (x+1/2, y+1/2, z-1/2).
\end{align}
\end{subequations}

\item $F$ (Face-centered) lattices: the three translation generators are
\begin{subequations}\label{TransBravaisF}
\begin{align}
&T_1\colon (x,y,z)\rightarrow (x, y+1/2, z+1/2),\\
&T_2\colon (x,y,z)\rightarrow (x+1/2, y, z+1/2),\\
&T_3\colon (x,y,z)\rightarrow (x+1/2, y+1/2, z).
\end{align}
\end{subequations}

\item $R$ (Rhombohedral) lattices: the three translation generators are
\begin{subequations}\label{TransBravaisR}
\begin{align}
&T_1\colon (x,y,z)\rightarrow (x+1, y, z),\\
&T_2\colon (x,y,z)\rightarrow (x, y+1, z),\\
&T_3\colon (x,y,z)\rightarrow (x+2/3, y+1/3, z+1/3).
\end{align}
\end{subequations}

\end{itemize}

The space groups belonging to each arithmetic crystal class are given in the fourth column of Table \ref{tableExt} \footnote{Note that the first group in each row of the column ``Arithmetic Crystal Class $\rho$" is not necessarily the symmorphic (i.e. split) one.}. In the fifth column we give the second cohomology group $H^2_\rho(P,T)$, whose elements determine the space groups in the column next to it on the left. In fact, there is the so-called \emph{main theorem of mathematical crystallography}. 

\begin{theorem} (Main theorem of mathematical crystallography \cite{Schwarzenberger_1974,HillerCrystallography})
For $k$-dimensional crystallography groups with arithmetic crystal class determined by point group $P$ and an integral representation $\rho$ on $T\cong\Z^k$, consider the action of $GL(k, \Z)$ on $H^2_\rho(P, T)$, such that given a function $\omega$ representing $[\omega]\in H^2_\rho(P, T)$, we have
\begin{equation}
    (g.\omega)(p_1, p_2) = g.(\omega(g^{-1}p_1 g, g^{-1}p_2 g)),\quad g\in GL(k),~p_{1,2}\in P.
\end{equation}
There exists a one-to-one correspondence between space groups in the arithmetic class $(P, \rho)$ and the orbits of the action in $H^2_\rho(P, T)$.    
\end{theorem}
We tabulate the results for the 3D space groups in Table~\ref{tableExt} as we have not found them elsewhere. We note that the results for the 2D wallpaper groups are nicely summarized by Morandi \cite{morandi2003classification}.

Furthermore, in the last three columns of Table \ref{tableExt} we give the $\Z_2$ rank of cohomology $ H^p(P,H^q(T,\mathbb{Z}_2))$ associated with the arithmetic crystal classes. They appear in the $(p,q)$ entry of the $E_2$ page of the standard LHS spectral sequence \eqref{lhse2}.

\begin{longtable}{c|c|c|c|c|c|c|c}
\caption{Information on the group extension of the point group $P$ by translations $T=\Z^3$. $\rho$ denotes the action of the point group $P$ on the translation $T$, and $\rho^*$ denotes its induced action of $P$ on the cohomology $H^{1,2}(\mathbb{Z}^3,\mathbb{Z}_2)$. $\dag$: this lattice type corresponds to two arithmetic crystal classes. $a_n =(n-1)-2\lfloor (n-1)/3\rfloor$, $s_n =\lfloor 2n/3 \rfloor+1$,
where the floor function $\lfloor n \rfloor$ gives the integer equal to or less than $n$. }\label{tableExt}\\
\hline
\hline
\multicolumn{2}{c|}{Point Group} & {Lattice}&\multicolumn{2}{c|}{Arithmetic Crystal Class $\rho$}& \multicolumn{3}{c}{ Cohomology Dimension in $\mathbb{Z}_2$}\\
\cline{1-2}
\cline{4-8}
Abstract & Sch\"{o}nflies &Type&Space Group No.&$H^2_\rho(P,T)$&$H^n(P,\mathbb{Z}_2)$&$H^n_{\rho^*}(P,H^1(T,\mathbb{Z}_2))$&$H^n_{\rho^*}(P,H^2(T,\mathbb{Z}_2))$\\
\hline
Trivial& $C_1$ &$P$ &1&Trivial&$\delta_{n,0}$&$3\delta_{n,0}$&$3\delta_{n,0}$\\
\hline
\multirow{5}{*}{$\mathbb{Z}_2$}&$C_i$ &$P$ & 2&Trivial&\multirow{5}{*}{$1$}&3&3\\
\cline{2-5}\cline{7-8}
& \multirow{2}{*}{$C_2$}& $P$ &3--4&$\mathbb{Z}_2$&&3&3\\

&&$C$ &5&Trivial&&$\delta_{n,0}+1$&$\delta_{n=0}+1$\\
\cline{2-5}\cline{7-8}
&\multirow{2}{*}{$C_s$}& $P$  &6--7&$(\mathbb{Z}_2)^2$&&$3$&$3$\\
\hline
\hline
\end{longtable}

\section{Information about the cohomology of space groups}

In Table \ref{tablemod2230}, we list the basic information about the mod-2 cohomology of all 230 space groups. In Table \ref{tableZZor230}, we list the integral cohomology of all 230 space groups.

Specifically, for the mod-2 cohomology, the $\mathbb{Z}_2$ rank in each degree can be summarized by a \emph{Poincar\'e series}. By Taylor expanding the series around $x=0$, the $\mathbb{Z}_2$ rank for each degree $n$ can be extracted from the coefficient in front of $x^n$. In the fourth column we list the Poincar\'e series for each space group. 

From this we can immediately assess how fast the $\Z_2$ rank grows with respect to $n$: write the Poincar\'e series as $P(x)/Q(x)$ where $P(x)$ and $Q(x)$ are integer polynomials of $x$, the $\Z_2$ rank grows as fast as $n^{d-1}$ with respect to $n$ where $d$ is the degree of the integer polynomial $Q(x)$. This $d$ is exactly the \emph{Krull dimension} of the graded ring $H^*(G, \Z_2)$ \cite{adem2013cohomology}. Specially, when $Q(x)=1$ and the Poincar\'e series itself is an integer polynomial, the $\mathbb{Z}_2$ rank will be zero for sufficiently large $n$, and this happens for Bieberbach groups. When $\text{deg}~Q(x)=1$, the $\mathbb{Z}_2$ rank will be a constant for sufficiently large $n$. We discuss the two phenomena in Sec.~\ref{subsec:3D}.

In the fifth column, we explicitly list the dimension in the first six degrees, and use a superscript to label the number of the mod-2 cohomology ring generators at this degree (if nonzero).

Moreover, in the third column, we list three features of the group: a ``$\,\rtimes\,$'' means the group is symmorphic (i.e. the group splits as point group extended by translation), a ``\,\customyinyang\,'' means that the standard LHS spectral sequence \eqref{lhse2} of this group collapses at the $E_2$ page, and a ``$\,\flat\,$'' means the group is a Bieberbach group.

In Table \ref{tableZZor230}, we analyze both trivial and orientation-reversing action of space groups $G$ on $\Z$. In the orientation-reversing action, reflection $M$ and inversion $I$ act on $\Z$ by multiplication of $-1$, and we use a superscript $^\text{or}$ to indicate this nontrivial action. The fifth twisted cohomology, $H^5(G,\mathbb{Z}^{\text{or}})$ listed in the $n=5$ column of orientation-reversing case agrees with the so-called classification of in-cohomology bosonic SPT protected by crystalline symmetries given in Ref.~\cite{PhysRevX.8.011040} for all 227 groups that they obtained, following the fact that $H^5(G,\mathbb{Z}^{\text{or}})\cong H^4(G,\U^{\text{or}})$. Notably, the results for the remaining three groups---No. \hyperref[subsub:sg227]{227}, \hyperref[subsub:sg228]{228}, and \hyperref[subsub:sg230]{230}---are missing in Ref.~\cite{PhysRevX.8.011040}, where we give their classification results to be 
\begin{equation}\label{eq:missing}
H^5(\hyperref[subsub:sg227]{\text{No. 227}},\mathbb{Z}^{\text{or}})=
\mathbb{Z}_2^7,\quad
H^5(\hyperref[subsub:sg228]{\text{No. 228}},\mathbb{Z}^{\text{or}})=
\mathbb{Z}_2^3,\quad
\text{and }
H^5(\hyperref[subsub:sg230]{\text{No. 230}},\mathbb{Z}^{\text{or}})=
\mathbb{Z}_2^4.
\end{equation}

\addtolength{\tabcolsep}{-1pt} 
\begin{longtable}{c|c|c|c|l||c|c|c|c|l} 
\caption{Additional information about space groups and their mod-2 cohomology. The ``Feat.'' column lists the features of the group: ``$\,\rtimes\,$'', ``\,\protect\customyinyang\,'',
and ``$\,\flat\,$'' mean the group ``is symmorphic'', ``has a standard LHS spectral sequence that collapses at the $E_2$ page'', and ``is a Bieberbach group'', respectively. $\mathcal{P}(x)$: generating function of the Poincaré series for the $\mathbb{Z}_2$ rank of the mod-2 cohomology $H^n(G,\mathbb{Z}_2)$. The ranks for the first degrees ($n=1\text{--}6$) are listed in the last column, with the number of the mod-2 cohomology ring generators at this degree (if there are any) written as a superscript.\label{tablemod2230}}\\
\hline
\hline
No. & Name & Feat.
& $\mathcal{P}(x)$ & Mod-2 Coh. Dim.${}^{\# \text{Gens}}$ & No. & Name & Feat.
& $\mathcal{P}(x)$ & Mod-2 Coh. Dim.${}^{\# \text{Gens}}$\\\hline
\hyperref[subsub:sg1]{1} & \hyperref[subsub:sg1]{$P1$} & $\rtimes$ \customyinyang\, $\flat$ & $(x\!+\!1)^3$ & $3^{\mathbf{3}},3,1,0,0,0,...$ & \hyperref[subsub:sg116]{116} & \hyperref[subsub:sg116]{$P\bar{4}c2$} &    & $\frac{-x^3+x^2+x+1}{(x-1)^2}$ & $3^{\mathbf{3}},6^{\mathbf{3}},8,10,12,14,...$ \\
\hyperref[subsub:sg2]{2} & \hyperref[subsub:sg2]{$P\bar{1}$} & $\rtimes$ \customyinyang\,  & $-\frac{(x+1)^3}{x-1}$ & $4^{\mathbf{4}},7,8,8,8,8,...$ & \hyperref[subsub:sg117]{117} & \hyperref[subsub:sg117]{$P\bar{4}b2$} &    & $\frac{x+1}{(x-1)^2}$ & $3^{\mathbf{3}},5^{\mathbf{2}},7^{\mathbf{1}},9,11,13,...$ \\
\hyperref[subsub:sg3]{3} & \hyperref[subsub:sg3]{$P2$} & $\rtimes$ \customyinyang\,  & $-\frac{(x+1)^3}{x-1}$ & $4^{\mathbf{4}},7,8,8,8,8,...$ & \hyperref[subsub:sg118]{118} & \hyperref[subsub:sg118]{$P\bar{4}n2$} &    & $\frac{x+1}{(x-1)^2}$ & $3^{\mathbf{3}},5^{\mathbf{1}},7^{\mathbf{1}},9,11,13,...$ \\
\hline\hline
\end{longtable} 
 \addtolength{\tabcolsep}{1pt}

\addtolength{\tabcolsep}{-1pt} 
 \begin{longtable}{c|l|l|l|l|l|l||l|l|l|l|l|l} 
\caption{Integral cohomology groups for the 230 space groups $G$ with trivial action ($H^n(G,\mathbb{Z})$) or the orientation-reversing action ($H^n(G,\mathbb{Z}^{\text{or}})$) up to $n=6$.\label{tableZZor230}}\\
\hline
\hline
\multirow{2}{*}{No.} &  \multicolumn{6}{c||}{$H^n(G,\mathbb{Z})$} & \multicolumn{6}{c}{$H^n(G,\mathbb{Z}^{\text{or}})$} \\
\cline{2-13}
& $n=1$ & $n=2$ & $n=3$ & $n=4$ & $n=5$ & $n=6$ & $n=1$ & $n=2$ & $n=3$ & $n=4$ & $n=5$ & $n=6$  \\
\hline
\hyperref[subsub:sg1]{1} & $\mathbb{Z}^{3}$ & $\mathbb{Z}^{3}$ & $\mathbb{Z}$& $\mathbb{Z}_1$& $\mathbb{Z}_1$& $\mathbb{Z}_1$ & $\mathbb{Z}^{3}$ & $\mathbb{Z}^{3}$ & $\mathbb{Z}$& $\mathbb{Z}_1$& $\mathbb{Z}_1$& $\mathbb{Z}_1$\\
\hyperref[subsub:sg2]{2}& $\mathbb{Z}_1$ & $\mathbb{Z}^{3}\!\oplus\!\mathbb{Z}_2^{4}$& $\mathbb{Z}_1$ & $\mathbb{Z}_2^{8}$& $\mathbb{Z}_1$ & $\mathbb{Z}_2^{8}$ & $\mathbb{Z}^{3}\!\oplus\!\mathbb{Z}_2$& $\mathbb{Z}_1$ & $\mathbb{Z}\!\oplus\!\mathbb{Z}_2^{7}$& $\mathbb{Z}_1$ & $\mathbb{Z}_2^{8}$& $\mathbb{Z}_1$\\
\hyperref[subsub:sg3]{3} & $\mathbb{Z}$ & $\mathbb{Z}\!\oplus\!\mathbb{Z}_2^{3}$ & $\mathbb{Z}\!\oplus\!\mathbb{Z}_2^{3}$ & $\mathbb{Z}_2^{4}$ & $\mathbb{Z}_2^{4}$ & $\mathbb{Z}_2^{4}$ & $\mathbb{Z}$ & $\mathbb{Z}\!\oplus\!\mathbb{Z}_2^{3}$ & $\mathbb{Z}\!\oplus\!\mathbb{Z}_2^{3}$ & $\mathbb{Z}_2^{4}$ & $\mathbb{Z}_2^{4}$ & $\mathbb{Z}_2^{4}$\\
\hyperref[subsub:sg4]{4} & $\mathbb{Z}$ & $\mathbb{Z}\!\oplus\!\mathbb{Z}_2^{2}$ & $\mathbb{Z}$& $\mathbb{Z}_1$& $\mathbb{Z}_1$& $\mathbb{Z}_1$ & $\mathbb{Z}$ & $\mathbb{Z}\!\oplus\!\mathbb{Z}_2^{2}$ & $\mathbb{Z}$& $\mathbb{Z}_1$& $\mathbb{Z}_1$& $\mathbb{Z}_1$\\
\hyperref[subsub:sg5]{5} & $\mathbb{Z}$ & $\mathbb{Z}\!\oplus\!\mathbb{Z}_2^{2}$ & $\mathbb{Z}\!\oplus\!\mathbb{Z}_2$ & $\mathbb{Z}_2^{2}$ & $\mathbb{Z}_2^{2}$ & $\mathbb{Z}_2^{2}$ & $\mathbb{Z}$ & $\mathbb{Z}\!\oplus\!\mathbb{Z}_2^{2}$ & $\mathbb{Z}\!\oplus\!\mathbb{Z}_2$ & $\mathbb{Z}_2^{2}$ & $\mathbb{Z}_2^{2}$ & $\mathbb{Z}_2^{2}$\\
\hline\hline
\end{longtable} 
 \addtolength{\tabcolsep}{1pt}

\section{Nontrivial differential \texorpdfstring{$d_2^{1,2}$}{d2(1,2)} in the standard LHS spectral sequence for No. 42 \texorpdfstring{($Fmm2$)}{(Fmm2)}}\label{app:calculation_d2}

In this appendix, we perform the calculation
\begin{equation}\label{42d12}
    d^{1,2}_2\colon H^1(P,H^2(T,\mathbb{Z}_2)) \rightarrow H^3(P, H^1(T, \Z_2))
\end{equation}
for group $G=\hyperref[subsub:sg42]{\text{No. 42}~(Fmm2)}$, which gives 
\begin{equation}
d_2(\omega_{12}) = (A_c^3 + A_c A_m^2) \omega_{01}.
\end{equation}
This group is a symmorphic group whose corresponding extension class $H^2_\rho(P, T)$ is trivial (see Table~\ref{tableExt}). Yet, following \cite{Wang2021,lyndon1948cohomology}, we prove that $d^{1,2}_2$ is still nontrivial because of the nontrivial action of $P$ on $T$.

First let us write down an explicit cochain expression $\omega_{12} \in C^1(P, C^2(T, \Z_2))$ which represents the nontrivial element of $H^1(P, H^2(T, \Z_2))\cong \Z_2$. Note that $H^2(T, \Z_2)\cong (\Z_2)^3$, and is generated by three elements whose cochain representatives can be chosen to be
\begin{equation}
b_1(t_1,t_2) = y_1z_2, \quad b_2(t_1,t_2)=x_1z_2, \quad b_3(t_1,t_2)=x_1y_2,
\end{equation}
for group elements $t_i=T_1^{x_i}T_2^{y_i}T_3^{z_i} \in T,i=1,2$. 
To consider the action of $P$ on $b_{1,2,3}$, first we write down its action on the generators $T_{1,2,3}$ of $T$,
\begin{subequations}
\begin{align}
C_2&\colon ~T_1\rightarrow T_2T_3^{-1},\quad T_2\rightarrow T_1T_3^{-1},\quad T_3\rightarrow T_3^{-1},\\
M&\colon ~T_1\rightarrow T_2 T_3^{-1},\quad T_2\rightarrow T_2,\quad T_3\rightarrow T_1^{-1} T_2,\\
C_2M&\colon ~T_1\rightarrow T_1,\quad T_2\rightarrow T_1T_3^{-1},\quad T_3\rightarrow T_1 T_2^{-1}.
\end{align}
\end{subequations}
Then its action on $b_{1,2,3}$ is dictated by
\begin{subequations}
\begin{align}
(C_2.b)(t_1,t_2)&=b(C_2^{-1}t_1C_2,C_2^{-1}t_2C_2)=b(T_1^{y_1}T_2^{x_1}T_3^{-x_1-y_1-z_1}
,T_1^{y_2}T_2^{x_2}T_3^{-x_2-y_2-z_2}),\\
(M.b)(t_1,t_2)&=b(M^{-1}t_1M,M^{-1}t_2M)=b(T_1^{-z_1}T_2^{x_1+y_1+z_1}T_3^{-x_1},T_1^{-z_2}T_2^{x_2+y_2+z_2}T_3^{-x_2}),\\
((C_2M).b)(t_1,t_2)&=b(M^{-1}C_2^{-1}t_1C_2M,M^{-1}C_2^{-1}t_2C_2M)=b(T_1^{x_1+y_1+z_1}T_2^{-z_1}T_3^{-y_1},T_1^{x_2+y_2+z_2}T_2^{-z_2}T_3^{-y_2}).
\end{align}
\end{subequations}
Thus, the elements of $P$ map $[b_{1,2,3}]\in H^2(T, \Z_2)$ in the following way,
\begin{subequations}
\begin{align}
C_2&\colon [b_1]\rightarrow [b_2+b_3],\quad
[b_2]\rightarrow [b_1+b_3],\quad [b_3]\rightarrow [b_3],\\
M&\colon [b_1]\rightarrow [b_2+b_3],\quad
[b_2]\rightarrow [b_2],\quad [b_3]\rightarrow [b_1+b_2],\\
C_2M&\colon [b_1]\rightarrow [b_1],\quad
[b_2]\rightarrow [b_1+b_3],\quad [b_3]\rightarrow [b_1+b_2],
\end{align}
\end{subequations}
 For $\omega_{12}\colon P\rightarrow H^2(T,\mathbb{Z}_2)$, define \begin{equation}\label{dpomega12}
 (\delta_P\omega_{12})(p_1, p_2):=p_1.(\omega_{12}(p_2))+\omega_{12}(p_1)-\omega_{12}(p_1p_2),\quad p_1,p_2\in P.
 \end{equation}
 we should have $[(\delta_P\omega_{12})(p_1,p_2)]$ equals 0 in $H^2(T, \Z_2)$.
Accordingly, we can choose the cochain representative $\omega_{12}$ to be
\begin{equation}
\omega_{12}\colon ~C_2\mapsto b_3,~~~M\mapsto b_2,~~~ C_2M\mapsto b_1.
\end{equation}
The nontriviality of this cochain representative can be seen by restricting to the subgroup generated by, e.g., $C_2$, which remains nontrivial in the subgroup.

To obtain the image of the differential of LHS spectral sequence, we attempt to promote $\omega_{12}$ into a cocycle in $H^3(G, \Z_2)$, following the ansatz outlined in \cite{Wang2021}. In particular, for the given $\omega_{12}\in C^1(P, C^2(T, \Z_2))$, we attempt to solve the following sets of equations for some $f_{21}\in C^2(P, C^1(T, \Z_2))$ \footnote{These are equations (B83), (B84) and (B85) in \cite{Wang2021}, and we simplify it for the special group structure and module that we are considering.},
\begin{subequations}
\begin{align}
\label{eq:LHS_eq1}\delta_T \omega_{12} &= 0,\\
\label{eq:LHS_eq2}\delta_P \omega_{12} + \delta_T f_{21} &=0,\\
\label{eq:LHS_eq3}\delta_P f_{21} &=0.
\end{align}
\end{subequations}

Eq.~\eqref{eq:LHS_eq1} is automatically satisfied for our choice of $\omega_{12}$. However, $ (\delta_P\omega_{12})(p_1, p_2)$ may not be a zero cochain in $C^2(T, \Z_2)$, and we have to find a nontrivial $f_{21}$ that satisfies Eq.~\eqref{eq:LHS_eq2}. This fact will contribute to $d_2^{1,2}$. Now we explicitly perform the calculation of the cochains $ (\delta_P\omega_{12})(p_1, p_2)$ using Eq.~\eqref{dpomega12} to demonstrate this fact:
\begin{equation}
\begin{aligned}
((\delta_P \omega_{12})(C_2,C_2))(t_1,t_2)
&=y_1x_2+x_1y_2,\\
((\delta_P \omega_{12})(M,M))(t_1,t_2)
&=z_1x_2+x_1z_2,\\
((\delta_P \omega_{12})(C_2M,C_2M))(t_1,t_2)
&=z_1y_2+y_1z_2,\\
((\delta_P \omega_{12})(C_2,M))(t_1,t_2)
&=y_1(x_2+y_2+z_2)+x_1y_2+y_1z_2,\\
((\delta_P \omega_{12})(M,C_2))(t_1,t_2)
&=z_1(x_2+y_2+z_2)+x_1z_2+y_1z_2,\\
((\delta_P \omega_{12})(C_2,C_2M))(t_1,t_2)
&=x_1(x_2+y_2+z_2)+x_1y_2+x_1z_2,\\
((\delta_P \omega_{12})(C_2M,C_2))(t_1,t_2)
&=(x_1+y_1+z_1)z_2+y_1z_2+x_1z_2,\\
((\delta_P \omega_{12})(M,C_2M))(t_1,t_2)
&=(x_1+y_1+z_1)x_2+x_1z_2+x_1y_2,\\
((\delta_P \omega_{12})(C_2M,M))(t_1,t_2)
&
=(x_1+y_1+z_1)y_2+y_1z_2+x_1y_2,
\end{aligned}
\end{equation}
we can check that 
\begin{equation}
(\delta_P\omega_{12})(p_1,p_2)(t_1,t_2)
=
(\delta_T (f_{21}(p_1,p_2)))(t_1,t_2),\quad p_1,p_2 \in P,~~t_1,t_2\in T,
\end{equation}
where we defined $f_{21}(id,p)=f_{21}(p,id)=0$ for $p \in P$ and
\begin{equation}
\begin{aligned}
&f_{21}(C_2,C_2)(t) = xy, && f_{21}(M,M)(t) = xz, && f_{21}(C_2M,C_2M)(t) = yz,\\
&f_{21}(C_2,M)(t) = \frac{y(y+1)}{2}+xy,&&
f_{21}(M,C_2)(t) = \frac{z(z+1)}{2}+xz+yz,&&
f_{21}(C_2,C_2M)(t) = \frac{x(x+1)}{2},\\
&f_{21}(C_2M,M)(t) = \frac{y(y+1)}{2}+yz,&&
f_{21}(M,C_2M)(t) = \frac{x(x+1)}{2}+xy+xz,&&f_{21}(C_2M,C_2)(t) = \frac{z(z+1)}{2}.
\end{aligned}
\end{equation}
However, the obtained $f_{21}\in C^2(P, C^1(T, \Z_2))$ does not satisfy Eq.~\eqref{eq:LHS_eq3}. We continue to define
$$
f_{31}(p_1,p_2,p_3)(t) := (\delta_P f_{21})(p_1,p_2,p_3)(t) = [p_1.f_{21}(p_2,p_3)+f_{21}(p_1p_2,p_3)+f_{21}(p_1,p_2p_3)+f_{21}(p_1,p_2)](t).$$
$f_{31}$ is a cochain in $C^3(P, C^1(T, \Z_2))$. Moreover, one can explicitly check using elementary methods that $[f_{31}]\colon P\times P\times P\rightarrow H^1(T,\mathbb{Z}_2)$ is a representative cocycle for a nontrivial element in $E_2^{3,1} = H^3(P,H^1(T,\mathbb{Z}_2))$. Here we list all the nonzero maps of $f_{31}$:
\begin{equation}\label{eq:f31}
\begin{aligned}
&f_{31}(M,C_2,M)=A_z,&&f_{31}(M,C_2M,M)=A_x,&&f_{31}(M,C_2,C_2)=A_z,\\
&f_{31}(M,C_2,C_2M)=A_z,&&f_{31}(M,C_2M,C_2)=A_x,&&f_{31}(M,C_2M,C_2M)=A_x,\\
&f_{31}(C_2,M,M)=A_y,&&f_{31}(C_2M,M,M)=A_y,&&f_{31}(C_2,M,C_2)=A_z,\\
&f_{31}(C_2,M,C_2M)=A_y,&&f_{31}(C_2M,M,C_2M)=A_{y+z},&&f_{31}(C_2,C_2M,M)=A_x,\\
&f_{31}(C_2M,C_2M,M)=A_z,&&f_{31}(C_2,C_2M,C_2)=A_{x+y+z},&&f_{31}(C_2,C_2M,C_2M)=A_x,\\
&f_{31}(C_2M,C_2,C_2)=A_z,&&f_{31}(C_2M,C_2M,C_2)=A_y.
\end{aligned}
\end{equation}

$f_{31}$ is exactly the obstruction of promoting $\omega_{12}$ to a cocycle in $H^3(G, \Z_2)$, and from the explicit cochain representative in Eq.~\eqref{eq:f31}, we can check that it indeed represents $(A_c^3 + A_c A_m^2)\omega_{01}$. To summarize, we have the nonzero differential 
\begin{equation}
d_2^{1,2}\colon \omega_{12} \mapsto (A_c^3 + A_c A_m^2)\omega_{01}.
\end{equation}
This nonzero differential serves as an example for the statement of Corollary 1 of Ref.~\cite{totaro1996cohomology}.

\section{Collection of results for space groups No.~1--230}\label{collection230}
\addtocontents{toc}{\protect\SGCHideSubsectionsInTOC}

This appendix contains the main results of this paper: a complete list of the mod-2 cohomology rings for space groups No. 1--230; and for each space group $G$, its IWPs, their associated cohomology element in $H^3(G, \Z_2)$, and their associated topological invariants.

For completeness, we also give the group generators as coordinate transformations, where the coordinate setup agrees with the ``Standard/Default Setting" on Bilbao Crystallographic Server \cite{Bilbao} or International Tables for Crystallography (ITC) \cite{aroyo2013international}.

\subsection*{No. 1: $P1$}\label{subsub:sg1}

This group is generated by three translations $T_{1,2,3}$ as given in Eqs.~\eqref{TransBravaisP}.
The $\mathbb{Z}_2$ cohomology ring is given by

\begin{equation}
\mathbb{Z}_2[A_x,A_y,A_z]/\langle\mathcal{R}_2\rangle
 \end{equation}
where the relations are 
\begin{subequations} 
 \begin{align}
\mathcal{R}_2\colon & ~~
A_x^2,~~A_y^2,~~A_z^2.
\end{align} 
 \end{subequations}
We have the following table regarding IWPs and group cohomology at degree 3.
\begin{center}
\begin{tabular}{c|cc|c|c|c}\hline\hline {Wyckoff}&\multicolumn{2}{c|}{Little group}& \multirow{2}{*}{Coordinates}&\multirow{2}{*}{LSM anomaly class}&\multirow{2}{*}{Topo. inv.} \\ \cline{2-3} position & Intl. & Sch\"{o}nflies & & & \\ \hline
1a&$1$&$C_1$& $(x,y,z)$ & $A_x A_y A_z$ & $\varphi_3[T_1, T_2, T_3]$\\ 
\hline
\hline 
 \end{tabular} 
 \end{center}

\subsection*{No. 2: $P\overline1$}\label{subsub:sg2}

This group is generated by three translations $T_{1,2,3}$ as given in Eqs.~\eqref{TransBravaisP}, and an inversion $I$:
\begin{subequations}
 \begin{align}
I &\colon (x,y,z)\rightarrow (-x, -y, -z).
\end{align}
\end{subequations}

The $\mathbb{Z}_2$ cohomology ring is given by

\begin{equation}
\mathbb{Z}_2[A_i,A_x,A_y,A_z]/\langle\mathcal{R}_2\rangle
 \end{equation}
where the relations are 
\begin{subequations} 
 \begin{align}
\mathcal{R}_2\colon & ~~
A_x (A_i + A_x),~~A_y (A_i + A_y),~~A_z (A_i + A_z).
\end{align} 
 \end{subequations}
We have the following table regarding IWPs and group cohomology at degree 3.
\begin{center}
\begin{tabular}{c|cc|c|c|c}\hline\hline {Wyckoff}&\multicolumn{2}{c|}{Little group}& \multirow{2}{*}{Coordinates}&\multirow{2}{*}{LSM anomaly class}&\multirow{2}{*}{Topo. inv.} \\ \cline{2-3} position & Intl. & Sch\"{o}nflies & & & \\ \hline
1a&$\overline{1}$&$C_i$& $(0,0,0)$ & $(A_i + A_x) (A_i + A_y) (A_i + A_z)$ & $\varphi_1[I]$\\ 
1b&$\overline{1}$&$C_i$& $(0,0,1/2)$ & $(A_i + A_x) (A_i + A_y) A_z$ & $\varphi_1[T_3I]$\\ 
1c&$\overline{1}$&$C_i$& $(0,1/2,0)$ & $(A_i + A_x) A_y (A_i + A_z)$ & $\varphi_1[T_2I]$\\ 
1d&$\overline{1}$&$C_i$& $(1/2,0,0)$ & $A_x (A_i + A_y) (A_i + A_z)$ & $\varphi_1[T_1I]$\\ 
1e&$\overline{1}$&$C_i$& $(1/2,1/2,0)$ & $A_x A_y (A_i + A_z)$ & $\varphi_1[T_1T_2I]$\\ 
1f&$\overline{1}$&$C_i$& $(1/2,0,1/2)$ & $A_x (A_i + A_y) A_z$ & $\varphi_1[T_1T_3I]$\\ 
1g&$\overline{1}$&$C_i$& $(0,1/2,1/2)$ & $(A_i + A_x) A_y A_z$ & $\varphi_1[T_2T_3I]$\\ 
1h&$\overline{1}$&$C_i$& $(1/2,1/2,1/2)$ & $A_x A_y A_z$ & $\varphi_1[T_1T_2T_3I]$\\ 
\hline
\hline 
 \end{tabular} 
 \end{center}

\subsection*{No. 3: $P2$}\label{subsub:sg3}

This group is generated by three translations $T_{1,2,3}$ as given in Eqs.~\eqref{TransBravaisP}, and a two-fold rotation $C_2$:
\begin{subequations}
 \begin{align}
C_2 &\colon (x,y,z)\rightarrow (-x, y, -z).
\end{align}
\end{subequations}

The $\mathbb{Z}_2$ cohomology ring is given by

\begin{equation}
\mathbb{Z}_2[A_c,A_x,A_y,A_z]/\langle\mathcal{R}_2\rangle
 \end{equation}
where the relations are 
\begin{subequations} 
 \begin{align}
\mathcal{R}_2\colon & ~~
A_x (A_c + A_x),~~A_y^2,~~A_z (A_c + A_z).
\end{align} 
 \end{subequations}
We have the following table regarding IWPs and group cohomology at degree 3.
\begin{center}
\begin{tabular}{c|cc|c|c|c}\hline\hline {Wyckoff}&\multicolumn{2}{c|}{Little group}& \multirow{2}{*}{Coordinates}&\multirow{2}{*}{LSM anomaly class}&\multirow{2}{*}{Topo. inv.} \\ \cline{2-3} position & Intl. & Sch\"{o}nflies & & & \\ \hline
1a&$2$&$C_2$& $(0,y,0)$ & $(A_c + A_x) A_y (A_c + A_z)$ & $\varphi_2[T_2, C_2]$\\ 
1b&$2$&$C_2$& $(0,y,1/2)$ & $(A_c + A_x) A_y A_z$ & $\varphi_2[T_2, T_3C_2]$\\ 
1c&$2$&$C_2$& $(1/2,y,0)$ & $A_x A_y (A_c + A_z)$ & $\varphi_2[T_2, T_1C_2]$\\ 
1d&$2$&$C_2$& $(1/2,y,1/2)$ & $A_x A_y A_z$ & $\varphi_2[T_2, T_1T_3C_2]$\\ 
\hline
\hline 
 \end{tabular} 
 \end{center}

\subsection*{No. 4: $P2_1$}\label{subsub:sg4}

This group is generated by three translations $T_{1,2,3}$ as given in Eqs.~\eqref{TransBravaisP}, and a two-fold screw $S_2$:
\begin{subequations}
 \begin{align}
S_2 &\colon (x,y,z)\rightarrow (-x, y + 1/2, -z).
\end{align}
\end{subequations}

The $\mathbb{Z}_2$ cohomology ring is given by

\begin{equation}
\mathbb{Z}_2[A_c,A_x,A_z]/\langle\mathcal{R}_2\rangle
 \end{equation}
where the relations are 
\begin{subequations} 
 \begin{align}
\mathcal{R}_2\colon & ~~
A_c^2,~~A_x (A_c + A_x),~~A_z (A_c + A_z).
\end{align} 
 \end{subequations}
We have the following table regarding IWPs and group cohomology at degree 3.
\begin{center}
\begin{tabular}{c|cc|c|c|c}\hline\hline {Wyckoff}&\multicolumn{2}{c|}{Little group}& \multirow{2}{*}{Coordinates}&\multirow{2}{*}{LSM anomaly class}&\multirow{2}{*}{Topo. inv.} \\ \cline{2-3} position & Intl. & Sch\"{o}nflies & & & \\ \hline
2a&$1$&$C_1$& $(x,y,z)$, $(-x,y+1/2,-z)$ & $A_c A_x A_z$ & $\widehat{\varphi_3}[T_1, T_3, S_2]$\\ 
\hline
\hline 
 \end{tabular} 
 \end{center}

Here the topological invariant can be chosen to be
\begin{equation}\label{TI_4}
\begin{aligned}
\widehat{\varphi_3}[T_1, T_3, S_2] = &\lambda(T_1,T_3,T_1^{-1}T_3^{-1}S_2)+\lambda(T_3,T_1,T_1^{-1}T_3^{-1}S_2)
+\lambda(T_1,T_1^{-1}S_2,T_3)
+\lambda(T_3,T_3^{-1}S_2,T_1)\\
&+\lambda(S_2,T_1,T_3)
+\lambda(S_2,T_3,T_1).
\end{aligned}
\end{equation}

\subsection*{No. 5: $C2$}\label{subsub:sg5}

This group is generated by three translations $T_{1,2,3}$ as given in Eqs.~\eqref{TransBravaisC}, and a two-fold rotation $C_2$:
\begin{subequations}
 \begin{align}
C_2 &\colon (x,y,z)\rightarrow (-x, y, -z).
\end{align}
\end{subequations}

The $\mathbb{Z}_2$ cohomology ring is given by

\begin{equation}
\mathbb{Z}_2[A_c,A_{x+y},A_z,B_{xy}]/\langle\mathcal{R}_2,\mathcal{R}_3,\mathcal{R}_4\rangle
 \end{equation}
where the relations are 
\begin{subequations} 
 \begin{align}
\mathcal{R}_2\colon & ~~
A_c A_{x+y},~~A_{x+y}^2,~~A_z (A_c + A_z),\\
\mathcal{R}_3\colon & ~~
A_{x+y} B_{xy},\\
\mathcal{R}_4\colon & ~~
B_{xy}^2.
\end{align} 
 \end{subequations}
We have the following table regarding IWPs and group cohomology at degree 3.
\begin{center}
\begin{tabular}{c|cc|c|c|c}\hline\hline {Wyckoff}&\multicolumn{2}{c|}{Little group}& {Coordinates}&\multirow{2}{*}{LSM anomaly class}&\multirow{2}{*}{Topo. inv.} \\ \cline{2-4} position & Intl. & Sch\"{o}nflies & $ (0,0,0) + (1/2,1/2,0) + $ & &\\ \hline
2a&$2$&$C_2$& $(0,y,0)$ & $(A_c + A_z) B_{xy}$ & $\varphi_2[T_1T_2, C_2]$\\ 
2b&$2$&$C_2$& $(0,y,1/2)$ & $A_z B_{xy}$ & $\varphi_2[T_1T_2, T_3C_2]$\\ 
\hline
\hline 
 \end{tabular} 
 \end{center}

\subsection*{No. 6: $Pm$}\label{subsub:sg6}

This group is generated by three translations $T_{1,2,3}$ as given in Eqs.~\eqref{TransBravaisP}, and a mirror $M$:
\begin{subequations}
 \begin{align}
M &\colon (x,y,z)\rightarrow (x, -y, z).
\end{align}
\end{subequations}

The $\mathbb{Z}_2$ cohomology ring is given by

\begin{equation}
\mathbb{Z}_2[A_m,A_x,A_y,A_z]/\langle\mathcal{R}_2\rangle
 \end{equation}
where the relations are 
\begin{subequations} 
 \begin{align}
\mathcal{R}_2\colon & ~~
A_x^2,~~A_y (A_m + A_y),~~A_z^2.
\end{align} 
 \end{subequations}
We have the following table regarding IWPs and group cohomology at degree 3.
\begin{center}
\begin{tabular}{c|cc|c|c|c}\hline\hline {Wyckoff}&\multicolumn{2}{c|}{Little group}& \multirow{2}{*}{Coordinates}&\multirow{2}{*}{LSM anomaly class}&\multirow{2}{*}{Topo. inv.} \\ \cline{2-3} position & Intl. & Sch\"{o}nflies & & & \\ \hline
1a&$m$&$C_s$& $(x,0,z)$ & $A_x (A_m + A_y) A_z$ & $\varphi_3[T_1, T_3, M]$\\ 
1b&$m$&$C_s$& $(x,1/2,z)$ & $A_x A_y A_z$ & $\varphi_3[T_1, T_3, T_2M]$\\ 
\hline
\hline 
 \end{tabular} 
 \end{center}

\subsection*{No. 7: $Pc$}\label{subsub:sg7}

This group is generated by three translations $T_{1,2,3}$ as given in Eqs.~\eqref{TransBravaisP}, and a glide $G$:
\begin{subequations}
 \begin{align}
G &\colon (x,y,z)\rightarrow (x, -y, z + 1/2).
\end{align}
\end{subequations}

The $\mathbb{Z}_2$ cohomology ring is given by

\begin{equation}
\mathbb{Z}_2[A_m,A_x,A_y]/\langle\mathcal{R}_2\rangle
 \end{equation}
where the relations are 
\begin{subequations} 
 \begin{align}
\mathcal{R}_2\colon & ~~
A_m^2,~~A_x^2,~~A_y (A_m + A_y).
\end{align} 
 \end{subequations}
We have the following table regarding IWPs and group cohomology at degree 3.
\begin{center}
\begin{tabular}{c|cc|c|c|c}\hline\hline {Wyckoff}&\multicolumn{2}{c|}{Little group}& \multirow{2}{*}{Coordinates}&\multirow{2}{*}{LSM anomaly class}&\multirow{2}{*}{Topo. inv.} \\ \cline{2-3} position & Intl. & Sch\"{o}nflies & & & \\ \hline
2a&$1$&$C_1$& $(x,y,z)$, $(x,-y,z+1/2)$ & $A_m A_x A_y$ & $\widetilde{\varphi_3}[T_2, G, T_1]$\\ 
\hline
\hline 
 \end{tabular} 
 \end{center}
The expression of $\widetilde{\varphi_3}$ is given in Eq.~\eqref{wilde3}.

\subsection*{No. 8: $Cm$}\label{subsub:sg8}

This group is generated by three translations $T_{1,2,3}$ as given in Eqs.~\eqref{TransBravaisC}, and a mirror $M$:
\begin{subequations}
 \begin{align}
M &\colon (x,y,z)\rightarrow (x, -y, z).
\end{align}
\end{subequations}

The $\mathbb{Z}_2$ cohomology ring is given by

\begin{equation}
\mathbb{Z}_2[A_m,A_{x+y},A_z,B_{xy}]/\langle\mathcal{R}_2,\mathcal{R}_3,\mathcal{R}_4\rangle
 \end{equation}
where the relations are 
\begin{subequations} 
 \begin{align}
\mathcal{R}_2\colon & ~~
A_m A_{x+y},~~A_{x+y}^2,~~A_z^2,\\
\mathcal{R}_3\colon & ~~
A_{x+y} B_{xy},\\
\mathcal{R}_4\colon & ~~
B_{xy}^2.
\end{align} 
 \end{subequations}
We have the following table regarding IWPs and group cohomology at degree 3.
\begin{center}
\begin{tabular}{c|cc|c|c|c}\hline\hline {Wyckoff}&\multicolumn{2}{c|}{Little group}& {Coordinates}&\multirow{2}{*}{LSM anomaly class}&\multirow{2}{*}{Topo. inv.} \\ \cline{2-4} position & Intl. & Sch\"{o}nflies & $ (0,0,0) + (1/2,1/2,0) + $ & &\\ \hline
2a&$m$&$C_s$& $(x,0,z)$ & $A_z B_{xy}$ & $\varphi_3[T_1T_2^{-1}, T_3, M]$\\ 
\hline
\hline 
 \end{tabular} 
 \end{center}

\subsection*{No. 9: $Cc$}\label{subsub:sg9}

This group is generated by three translations $T_{1,2,3}$ as given in Eqs.~\eqref{TransBravaisC}, and a glide $G$:
\begin{subequations}
 \begin{align}
G &\colon (x,y,z)\rightarrow (x, -y, z + 1/2).
\end{align}
\end{subequations}

The $\mathbb{Z}_2$ cohomology ring is given by

\begin{equation}
\mathbb{Z}_2[A_m,A_{x+y},B_{xy},B_{z(x+y)}]/\langle\mathcal{R}_2,\mathcal{R}_3,\mathcal{R}_4\rangle
 \end{equation}
where the relations are 
\begin{subequations} 
 \begin{align}
\mathcal{R}_2\colon & ~~
A_m A_{x+y},~~A_m^2,~~A_{x+y}^2,\\
\mathcal{R}_3\colon & ~~
A_{x+y} B_{xy},~~A_m B_{z(x+y)},~~A_m B_{xy} + A_{x+y} B_{z(x+y)},\\
\mathcal{R}_4\colon & ~~
B_{xy}^2,~~B_{xy} B_{z(x+y)},~~B_{z(x+y)}^2.
\end{align} 
 \end{subequations}
We have the following table regarding IWPs and group cohomology at degree 3.
\begin{center}
\begin{tabular}{c|cc|c|c|c}\hline\hline {Wyckoff}&\multicolumn{2}{c|}{Little group}& {Coordinates}&\multirow{2}{*}{LSM anomaly class}&\multirow{2}{*}{Topo. inv.} \\ \cline{2-4} position & Intl. & Sch\"{o}nflies & $ (0,0,0) + (1/2,1/2,0) + $ & &\\ \hline
4a&$1$&$C_1$& $(x,y,z)$, $(x,-y,z+1/2)$ & $A_m B_{xy}$ & $\widehat{\varphi_3}[T_1, T_2, G]$\\ 
\hline
\hline 
 \end{tabular} 
 \end{center}
 Here the topological invariant can be chosen to be
\begin{equation}\label{TI_9}
\begin{aligned}
\widehat{\varphi_3}[T_1,T_2,G]=&
\lambda(T_1,T_2,T_1^{-1}T_2^{-1}G)+
\lambda(T_2,T_1,T_1^{-1}T_2^{-1}G)+
\lambda(T_1,T_1^{-1}G,T_1)+
\lambda(T_2,T_2^{-1}G,T_2)\\
&+
\lambda(G,T_1,T_2)+
\lambda(G,T_2,T_1).
\end{aligned}
\end{equation}

\subsection*{No. 10: $P2/m$}\label{subsub:sg10}

This group is generated by three translations $T_{1,2,3}$ as given in Eqs.~\eqref{TransBravaisP}, a two-fold rotation $C_2$, and an inversion $I$:
\begin{subequations}
 \begin{align}
C_2 &\colon (x,y,z)\rightarrow (-x, y, -z),\\ 
I &\colon (x,y,z)\rightarrow (-x, -y, -z).
\end{align}
\end{subequations}

The $\mathbb{Z}_2$ cohomology ring is given by

\begin{equation}
\mathbb{Z}_2[A_c,A_i,A_x,A_y,A_z]/\langle\mathcal{R}_2\rangle
 \end{equation}
where the relations are 
\begin{subequations} 
 \begin{align}
\mathcal{R}_2\colon & ~~
A_x (A_c + A_i + A_x),~~A_y (A_i + A_y),~~A_z (A_c + A_i + A_z).
\end{align} 
 \end{subequations}
We have the following table regarding IWPs and group cohomology at degree 3.
\begin{center}
\begin{tabular}{c|cc|c|c|c}\hline\hline {Wyckoff}&\multicolumn{2}{c|}{Little group}& \multirow{2}{*}{Coordinates}&\multirow{2}{*}{LSM anomaly class}&\multirow{2}{*}{Topo. inv.} \\ \cline{2-3} position & Intl. & Sch\"{o}nflies & & & \\ \hline
1a&$2/m$&$C_{2h}$& $(0,0,0)$ & $(A_c + A_i + A_x) (A_i + A_y) (A_c + A_i + A_z)$ & $\varphi_1[I]$\\ 
1b&$2/m$&$C_{2h}$& $(0,1/2,0)$ & $(A_c + A_i + A_x) A_y (A_c + A_i + A_z)$ & $\varphi_1[T_2I]$\\ 
1c&$2/m$&$C_{2h}$& $(0,0,1/2)$ & $(A_c + A_i + A_x) (A_i + A_y) A_z$ & $\varphi_1[T_3I]$\\ 
1d&$2/m$&$C_{2h}$& $(1/2,0,0)$ & $A_x (A_i + A_y) (A_c + A_i + A_z)$ & $\varphi_1[T_1I]$\\ 
1e&$2/m$&$C_{2h}$& $(1/2,1/2,0)$ & $A_x A_y (A_c + A_i + A_z)$ & $\varphi_1[T_1T_2I]$\\ 
1f&$2/m$&$C_{2h}$& $(0,1/2,1/2)$ & $(A_c + A_i + A_x) A_y A_z$ & $\varphi_1[T_2T_3I]$\\ 
1g&$2/m$&$C_{2h}$& $(1/2,0,1/2)$ & $A_x (A_i + A_y) A_z$ & $\varphi_1[T_1T_3I]$\\ 
1h&$2/m$&$C_{2h}$& $(1/2,1/2,1/2)$ & $A_x A_y A_z$ & $\varphi_1[T_1T_2T_3I]$\\ 
\hline
\hline 
 \end{tabular} 
 \end{center}

\subsection*{No. 11: $P2_1/m$}\label{subsub:sg11}

This group is generated by three translations $T_{1,2,3}$ as given in Eqs.~\eqref{TransBravaisP}, a two-fold screw $S_2$, and an inversion $I$:
\begin{subequations}
 \begin{align}
S_2 &\colon (x,y,z)\rightarrow (-x, y + 1/2, -z),\\ 
I &\colon (x,y,z)\rightarrow (-x, -y, -z).
\end{align}
\end{subequations}

The $\mathbb{Z}_2$ cohomology ring is given by

\begin{equation}
\mathbb{Z}_2[A_c,A_i,A_x,A_z]/\langle\mathcal{R}_2\rangle
 \end{equation}
where the relations are 
\begin{subequations} 
 \begin{align}
\mathcal{R}_2\colon & ~~
A_c (A_c + A_i),~~A_x (A_c + A_i + A_x),~~A_z (A_c + A_i + A_z).
\end{align} 
 \end{subequations}
We have the following table regarding IWPs and group cohomology at degree 3.
\begin{center}
\begin{tabular}{c|cc|c|c|c}\hline\hline {Wyckoff}&\multicolumn{2}{c|}{Little group}& \multirow{2}{*}{Coordinates}&\multirow{2}{*}{LSM anomaly class}&\multirow{2}{*}{Topo. inv.} \\ \cline{2-3} position & Intl. & Sch\"{o}nflies & & & \\ \hline
2a&$\overline{1}$&$C_i$& $(0,0,0)$, $(0,1/2,0)$ & 
$(A_c + A_i) (A_i + A_x)(A_i + A_z)$ & $\varphi_1[I]$\\ 
2b&$\overline{1}$&$C_i$& $(1/2,0,0)$, $(1/2,1/2,0)$ & $(A_c + A_i) A_x (A_i + A_z)$ & $\varphi_1[T_1I]$\\ 
2c&$\overline{1}$&$C_i$& $(0,0,1/2)$, $(0,1/2,1/2)$ & $(A_c + A_i) (A_i + A_x) A_z$ & $\varphi_1[T_3I]$\\ 
2d&$\overline{1}$&$C_i$& $(1/2,0,1/2)$, $(1/2,1/2,1/2)$ & $(A_c + A_i) A_x A_z$ & $\varphi_1[T_1T_3I]$\\ 
2e&$m$&$C_s$& $(x,1/4,z)$, $(-x,3/4,-z)$ & $A_c A_x A_z$ & $\varphi_3[T_1, T_3, S_2I]$\\ 
\hline
\hline 
 \end{tabular} 
 \end{center}

\subsection*{No. 12: $C2/m$}\label{subsub:sg12}

This group is generated by three translations $T_{1,2,3}$ as given in Eqs.~\eqref{TransBravaisC}, a two-fold rotation $C_2$, and an inversion $I$:
\begin{subequations}
 \begin{align}
C_2 &\colon (x,y,z)\rightarrow (-x, y, -z),\\ 
I &\colon (x,y,z)\rightarrow (-x, -y, -z).
\end{align}
\end{subequations}

The $\mathbb{Z}_2$ cohomology ring is given by

\begin{equation}
\mathbb{Z}_2[A_c,A_i,A_{x+y},A_z,B_{xy}]/\langle\mathcal{R}_2,\mathcal{R}_3,\mathcal{R}_4\rangle
 \end{equation}
where the relations are 
\begin{subequations} 
 \begin{align}
\mathcal{R}_2\colon & ~~
A_c A_{x+y},~~A_{x+y} (A_i + A_{x+y}),~~A_z (A_c + A_i + A_z),\\
\mathcal{R}_3\colon & ~~
A_{x+y} B_{xy},\\
\mathcal{R}_4\colon & ~~
B_{xy} (A_c A_i + A_i^2 + B_{xy}).
\end{align} 
 \end{subequations}
We have the following table regarding IWPs and group cohomology at degree 3.
\begin{center}
\resizebox{\columnwidth}{!}{
\begin{tabular}{c|cc|c|c|c}\hline\hline {Wyckoff}&\multicolumn{2}{c|}{Little group}& {Coordinates}&\multirow{2}{*}{LSM anomaly class}&\multirow{2}{*}{Topo. inv.} \\ \cline{2-4} position & Intl. & Sch\"{o}nflies & $ (0,0,0) + (1/2,1/2,0) + $ & &\\ \hline
2a&$2/m$&$C_{2h}$& $(0,0,0)$ & 
$(A_c+A_i+A_z)(A_cA_i+A_i^2+A_i A_{x+y} + B_{xy})$ & $\varphi_1[I]$\\ 
2b&$2/m$&$C_{2h}$& $(0,1/2,0)$ & $(A_c + A_i + A_z) B_{xy}$ & $\varphi_1[T_1T_2I]$\\ 
2c&$2/m$&$C_{2h}$& $(0,0,1/2)$ & $A_z (A_c A_i + A_i^2 + A_i A_{x+y} + B_{xy})$ & $\varphi_1[T_3I]$\\ 
2d&$2/m$&$C_{2h}$& $(0,1/2,1/2)$ & $A_z B_{xy}$ & $\varphi_1[T_1T_2T_3I]$\\ 
4e&$\overline{1}$&$C_i$& $(1/4,1/4,0)$, $(3/4,1/4,0)$ & $A_i A_{x+y} (A_i + A_z)$ & $\varphi_1[T_1I]$\\ 
4f&$\overline{1}$&$C_i$& $(1/4,1/4,1/2)$, $(3/4,1/4,1/2)$ & $A_i A_{x+y} A_z$ & $\varphi_1[T_1T_3I]$\\ 
\hline
\hline 
 \end{tabular} 
 }
 \end{center}

\subsection*{No. 13: $P2/c$}\label{subsub:sg13}

This group is generated by three translations $T_{1,2,3}$ as given in Eqs.~\eqref{TransBravaisP}, a two-fold rotation $C_2$, and an inversion $I$:
\begin{subequations}
 \begin{align}
C_2 &\colon (x,y,z)\rightarrow (-x, y, -z + 1/2),\\ 
I &\colon (x,y,z)\rightarrow (-x, -y, -z).
\end{align}
\end{subequations}

The $\mathbb{Z}_2$ cohomology ring is given by

\begin{equation}
\mathbb{Z}_2[A_c,A_i,A_x,A_y]/\langle\mathcal{R}_2\rangle
 \end{equation}
where the relations are 
\begin{subequations} 
 \begin{align}
\mathcal{R}_2\colon & ~~
A_c A_i,~~A_x (A_c + A_i + A_x),~~A_y (A_i + A_y).
\end{align} 
 \end{subequations}
We have the following table regarding IWPs and group cohomology at degree 3.
\begin{center}
\begin{tabular}{c|cc|c|c|c}\hline\hline {Wyckoff}&\multicolumn{2}{c|}{Little group}& \multirow{2}{*}{Coordinates}&\multirow{2}{*}{LSM anomaly class}&\multirow{2}{*}{Topo. inv.} \\ \cline{2-3} position & Intl. & Sch\"{o}nflies & & & \\ \hline
2a&$\overline{1}$&$C_i$& $(0,0,0)$, $(0,0,1/2)$ & $A_i (A_i + A_x) (A_i + A_y)$ & $\varphi_1[I]$\\ 
2b&$\overline{1}$&$C_i$& $(1/2,1/2,0)$, $(1/2,1/2,1/2)$ & $A_i A_x A_y$ & $\varphi_1[T_1T_2I]$\\ 
2c&$\overline{1}$&$C_i$& $(0,1/2,0)$, $(0,1/2,1/2)$ & $A_i (A_i + A_x) A_y$ & $\varphi_1[T_2I]$\\ 
2d&$\overline{1}$&$C_i$& $(1/2,0,0)$, $(1/2,0,1/2)$ & $A_i A_x (A_i + A_y)$ & $\varphi_1[T_1I]$\\ 
2e&$2$&$C_2$& $(0,y,1/4)$, $(0,-y,3/4)$ & $A_c (A_c + A_x) A_y$ & $\varphi_2[T_2, C_2]$\\ 
2f&$2$&$C_2$& $(1/2,y,1/4)$, $(1/2,-y,3/4)$ & $A_c A_x A_y$ & $\varphi_2[T_2, T_1C_2]$\\ 
\hline
\hline 
 \end{tabular} 
 \end{center}

\subsection*{No. 14: $P2_1/c$}\label{subsub:sg14}

This group is generated by three translations $T_{1,2,3}$ as given in Eqs.~\eqref{TransBravaisP}, a two-fold screw $S_2$, and an inversion $I$:
\begin{subequations}
 \begin{align}
S_2 &\colon (x,y,z)\rightarrow (-x, y + 1/2, -z + 1/2),\\ 
I &\colon (x,y,z)\rightarrow (-x, -y, -z).
\end{align}
\end{subequations}

The $\mathbb{Z}_2$ cohomology ring is given by

\begin{equation}
\mathbb{Z}_2[A_c,A_i,A_x,B_\beta]/\langle\mathcal{R}_2,\mathcal{R}_3,\mathcal{R}_4\rangle
 \end{equation}
where the relations are 
\begin{subequations} 
 \begin{align}
\mathcal{R}_2\colon & ~~
A_c A_i,~~A_c^2,~~A_x (A_c + A_i + A_x),\\
\mathcal{R}_3\colon & ~~
A_c B_\beta,\\
\mathcal{R}_4\colon & ~~
B_\beta (A_i^2 + B_\beta).
\end{align} 
 \end{subequations}
We have the following table regarding IWPs and group cohomology at degree 3.
\begin{center}
\begin{tabular}{c|cc|c|c|c}\hline\hline {Wyckoff}&\multicolumn{2}{c|}{Little group}& \multirow{2}{*}{Coordinates}&\multirow{2}{*}{LSM anomaly class}&\multirow{2}{*}{Topo. inv.} \\ \cline{2-3} position & Intl. & Sch\"{o}nflies & & & \\ \hline
2a&$\overline{1}$&$C_i$& $(0,0,0)$, $(0,1/2,1/2)$ & $(A_i + A_x) B_\beta$ & $\varphi_1[I]$\\ 
2b&$\overline{1}$&$C_i$& $(1/2,0,0)$, $(1/2,1/2,1/2)$ & $A_x B_\beta$ & $\varphi_1[T_1I]$\\ 
2c&$\overline{1}$&$C_i$& $(0,0,1/2)$, $(0,1/2,0)$ & $(A_i + A_x) (A_i^2 + B_\beta)$ & $\varphi_1[T_2I]$\\ 
2d&$\overline{1}$&$C_i$& $(1/2,0,1/2)$, $(1/2,1/2,0)$ & $A_x (A_i^2 + B_\beta)$ & $\varphi_1[T_1T_2I]$\\ 
\hline
\hline 
 \end{tabular} 
 \end{center}

\subsection*{No. 15: $C2/c$}\label{subsub:sg15}

This group is generated by three translations $T_{1,2,3}$ as given in Eqs.~\eqref{TransBravaisC}, a two-fold rotation $C_2$, and an inversion $I$:
\begin{subequations}
 \begin{align}
C_2 &\colon (x,y,z)\rightarrow (-x, y, -z + 1/2),\\ 
I &\colon (x,y,z)\rightarrow (-x, -y, -z).
\end{align}
\end{subequations}

The $\mathbb{Z}_2$ cohomology ring is given by

\begin{equation}
\mathbb{Z}_2[A_c,A_i,A_{x+y},B_{xy},B_{z(x+y)}]/\langle\mathcal{R}_2,\mathcal{R}_3,\mathcal{R}_4\rangle
 \end{equation}
where the relations are 
\begin{subequations} 
 \begin{align}
\mathcal{R}_2\colon & ~~
A_c A_i,~~A_c A_{x+y},~~A_{x+y} (A_i + A_{x+y}),\\
\mathcal{R}_3\colon & ~~
A_{x+y} B_{xy},~~A_c B_{z(x+y)},~~A_i B_{xy} + A_i B_{z(x+y)} + A_{x+y} B_{z(x+y)},\\
\mathcal{R}_4\colon & ~~
B_{xy} (A_i^2 + B_{xy}),~~B_{xy} (A_i^2 + B_{z(x+y)}),~~B_{z(x+y)} (A_i^2 + B_{z(x+y)}).
\end{align} 
 \end{subequations}
We have the following table regarding IWPs and group cohomology at degree 3.
\begin{center}
\begin{tabular}{c|cc|c|c|c}\hline\hline {Wyckoff}&\multicolumn{2}{c|}{Little group}& {Coordinates}&\multirow{2}{*}{LSM anomaly class}&\multirow{2}{*}{Topo. inv.} \\ \cline{2-4} position & Intl. & Sch\"{o}nflies & $ (0,0,0) + (1/2,1/2,0) + $ & &\\ \hline
4a&$\overline{1}$&$C_i$& $(0,0,0)$, $(0,0,1/2)$ & $A_i (A_i^2 + A_i A_{x+y} + B_{xy})$ & $\varphi_1[I]$\\ 
4b&$\overline{1}$&$C_i$& $(0,1/2,0)$, $(0,1/2,1/2)$ & $A_i B_{xy}$ & $\varphi_1[T_1T_2I]$\\ 
4c&$\overline{1}$&$C_i$& $(1/4,1/4,0)$, $(3/4,1/4,1/2)$ & $A_i (B_{xy} + B_{z(x+y)})$ & $\varphi_1[T_1I]$\\ 
4d&$\overline{1}$&$C_i$& $(1/4,1/4,1/2)$, $(3/4,1/4,0)$ & $A_i (A_i A_{x+y} + B_{xy} + B_{z(x+y)})$ & $\varphi_1[T_1T_3I]$\\ 
4e&$2$&$C_2$& $(0,y,1/4)$, $(0,-y,3/4)$ & $A_c B_{xy}$ & $\varphi_2[T_1T_2, C_2]$\\ 
\hline
\hline 
 \end{tabular} 
 \end{center}

\subsection*{No. 16: $P222$}\label{subsub:sg16}

This group is generated by three translations $T_{1,2,3}$ as given in Eqs.~\eqref{TransBravaisP}, a two-fold rotation $C_2$, and a two-fold rotation $C'_2$:
\begin{subequations}
 \begin{align}
C_2 &\colon (x,y,z)\rightarrow (-x, -y, z),\\ 
C'_2 &\colon (x,y,z)\rightarrow (-x, y, -z).
\end{align}
\end{subequations}

The $\mathbb{Z}_2$ cohomology ring is given by

\begin{equation}
\mathbb{Z}_2[A_c,A_{c'},A_x,A_y,A_z]/\langle\mathcal{R}_2\rangle
 \end{equation}
where the relations are 
\begin{subequations} 
 \begin{align}
\mathcal{R}_2\colon & ~~
A_x (A_c + A_{c'} + A_x),~~A_y (A_c + A_y),~~A_z (A_{c'} + A_z).
\end{align} 
 \end{subequations}
We have the following table regarding IWPs and group cohomology at degree 3.
\begin{center}
\begin{tabular}{c|cc|c|c|c}\hline\hline {Wyckoff}&\multicolumn{2}{c|}{Little group}& \multirow{2}{*}{Coordinates}&\multirow{2}{*}{LSM anomaly class}&\multirow{2}{*}{Topo. inv.} \\ \cline{2-3} position & Intl. & Sch\"{o}nflies & & & \\ \hline
1a&$222$&$D_2$& $(0,0,0)$ & $(A_c + A_{c'} + A_x) (A_c + A_y) (A_{c'} + A_z)$ & $\varphi_2[C_2, C'_2]$\\ 
1b&$222$&$D_2$& $(1/2,0,0)$ & $A_x (A_c + A_y) (A_{c'} + A_z)$ & $\varphi_2[T_1C_2, T_1C'_2]$\\ 
1c&$222$&$D_2$& $(0,1/2,0)$ & $(A_c + A_{c'} + A_x) A_y (A_{c'} + A_z)$ & $\varphi_2[T_2C_2, C'_2]$\\ 
1d&$222$&$D_2$& $(0,0,1/2)$ & $(A_c + A_{c'} + A_x) (A_c + A_y) A_z$ & $\varphi_2[C_2, T_3C'_2]$\\ 
1e&$222$&$D_2$& $(1/2,1/2,0)$ & $A_x A_y (A_{c'} + A_z)$ & $\varphi_2[T_1T_2C_2, T_1C'_2]$\\ 
1f&$222$&$D_2$& $(1/2,0,1/2)$ & $A_x (A_c + A_y) A_z$ & $\varphi_2[T_1C_2, T_1T_3C'_2]$\\ 
1g&$222$&$D_2$& $(0,1/2,1/2)$ & $(A_c + A_{c'} + A_x) A_y A_z$ & $\varphi_2[T_2C_2, T_3C'_2]$\\ 
1h&$222$&$D_2$& $(1/2,1/2,1/2)$ & $A_x A_y A_z$ & $\varphi_2[T_1T_2C_2, T_1T_3C'_2]$\\ 
\hline
\hline 
 \end{tabular} 
 \end{center}

\subsection*{No. 17: $P222_1$}\label{subsub:sg17}

This group is generated by three translations $T_{1,2,3}$ as given in Eqs.~\eqref{TransBravaisP}, a two-fold screw $S_2$, and a two-fold rotation $C'_2$:
\begin{subequations}
 \begin{align}
S_2 &\colon (x,y,z)\rightarrow (-x, -y, z + 1/2),\\ 
C'_2 &\colon (x,y,z)\rightarrow (-x, y, -z + 1/2).
\end{align}
\end{subequations}

The $\mathbb{Z}_2$ cohomology ring is given by

\begin{equation}
\mathbb{Z}_2[A_c,A_{c'},A_x,A_y]/\langle\mathcal{R}_2\rangle
 \end{equation}
where the relations are 
\begin{subequations} 
 \begin{align}
\mathcal{R}_2\colon & ~~
A_c (A_c + A_{c'}),~~A_x (A_c + A_{c'} + A_x),~~A_y (A_c + A_y).
\end{align} 
 \end{subequations}
We have the following table regarding IWPs and group cohomology at degree 3.
\begin{center}
\begin{tabular}{c|cc|c|c|c}\hline\hline {Wyckoff}&\multicolumn{2}{c|}{Little group}& \multirow{2}{*}{Coordinates}&\multirow{2}{*}{LSM anomaly class}&\multirow{2}{*}{Topo. inv.} \\ \cline{2-3} position & Intl. & Sch\"{o}nflies & & & \\ \hline
2a&$2$&$C_2$& $(x,0,0)$, $(-x,0,1/2)$ & $A_c A_x (A_{c'} + A_y)$ & $\varphi_2[T_1, S_2C'_2]$\\ 
2b&$2$&$C_2$& $(x,1/2,0)$, $(-x,1/2,1/2)$ & $A_c A_x A_y$ & $\varphi_2[T_1, T_2S_2C'_2]$\\ 
2c&$2$&$C_2$& $(0,y,1/4)$, $(0,-y,3/4)$ & $(A_c + A_{c'}) (A_{c'} + A_x) A_y$ & $\varphi_2[T_2, C'_2]$\\ 
2d&$2$&$C_2$& $(1/2,y,1/4)$, $(1/2,-y,3/4)$ & $(A_c + A_{c'}) A_x A_y$ & $\varphi_2[T_2, T_1C'_2]$\\ 
\hline
\hline 
 \end{tabular} 
 \end{center}

\subsection*{No. 18: $P2_12_12$}\label{subsub:sg18}

This group is generated by three translations $T_{1,2,3}$ as given in Eqs.~\eqref{TransBravaisP}, a two-fold rotation $C_2$, and a two-fold screw $S'_2$:
\begin{subequations}
 \begin{align}
C_2 &\colon (x,y,z)\rightarrow (-x, -y, z),\\ 
S'_2 &\colon (x,y,z)\rightarrow (-x + 1/2, y + 1/2, -z).
\end{align}
\end{subequations}

The $\mathbb{Z}_2$ cohomology ring is given by

\begin{equation}
\mathbb{Z}_2[A_c,A_{c'},A_z,B_\beta]/\langle\mathcal{R}_2,\mathcal{R}_3,\mathcal{R}_4\rangle
 \end{equation}
where the relations are 
\begin{subequations} 
 \begin{align}
\mathcal{R}_2\colon & ~~
A_c A_{c'},~~A_{c'}^2,~~A_z (A_{c'} + A_z),\\
\mathcal{R}_3\colon & ~~
A_{c'} B_\beta,\\
\mathcal{R}_4\colon & ~~
B_\beta (A_c^2 + B_\beta).
\end{align} 
 \end{subequations}
We have the following table regarding IWPs and group cohomology at degree 3.
\begin{center}
\begin{tabular}{c|cc|c|c|c}\hline\hline {Wyckoff}&\multicolumn{2}{c|}{Little group}& \multirow{2}{*}{Coordinates}&\multirow{2}{*}{LSM anomaly class}&\multirow{2}{*}{Topo. inv.} \\ \cline{2-3} position & Intl. & Sch\"{o}nflies & & & \\ \hline
2a&$2$&$C_2$& $(0,0,z)$, $(1/2,1/2,-z)$ & $A_z (A_c^2 + B_\beta)$ & $\varphi_2[T_3, C_2]$\\ 
2b&$2$&$C_2$& $(0,1/2,z)$, $(1/2,0,-z)$ & $A_z B_\beta$ & $\varphi_2[T_3, T_2C_2]$\\ 
\hline
\hline 
 \end{tabular} 
 \end{center}

\subsection*{No. 19: $P2_12_12_1$}\label{subsub:sg19}

This group is generated by three translations $T_{1,2,3}$ as given in Eqs.~\eqref{TransBravaisP}, a two-fold screw $S_2$, and a two-fold screw $S'_2$:
\begin{subequations}
 \begin{align}
S_2 &\colon (x,y,z)\rightarrow (-x + 1/2, -y, z + 1/2),\\ 
S'_2 &\colon (x,y,z)\rightarrow (-x, y + 1/2, -z + 1/2).
\end{align}
\end{subequations}

The $\mathbb{Z}_2$ cohomology ring is given by

\begin{equation}
\mathbb{Z}_2[A_c,A_{c'},B_{\beta 1},B_{\beta 2}]/\langle\mathcal{R}_2,\mathcal{R}_3,\mathcal{R}_4\rangle
 \end{equation}
where the relations are 
\begin{subequations} 
 \begin{align}
\mathcal{R}_2\colon & ~~
A_c A_{c'},~~A_c^2,~~A_{c'}^2,\\
\mathcal{R}_3\colon & ~~
(A_c + A_{c'}) B_{\beta 1},~~A_c (B_{\beta 1} + B_{\beta 2}),~~A_{c'} B_{\beta 2},\\
\mathcal{R}_4\colon & ~~
B_{\beta 1}^2,~~B_{\beta 1} B_{\beta 2},~~B_{\beta 2}^2.
\end{align} 
 \end{subequations}
We have the following table regarding IWPs and group cohomology at degree 3.
\begin{center}
\resizebox{\columnwidth}{!}{
\begin{tabular}{c|cc|c|c|c}\hline\hline {Wyckoff}&\multicolumn{2}{c|}{Little group}& \multirow{2}{*}{Coordinates}&\multirow{2}{*}{LSM anomaly class}&\multirow{2}{*}{Topo. inv.} \\ \cline{2-3} position & Intl. & Sch\"{o}nflies & & & \\ \hline
\multirow{2}{*}{4a} & \multirow{2}{*}{$1$} & \multirow{2}{*}{$C_1$} & $(x,y,z)$, $(-x+1/2,-y,z+1/2)$, & \multirow{2}{*}{$A_c B_{\beta 1}$} & \multirow{2}{*}{$\widehat{\varphi_4}[T_2, T_3, S_2, S'_2]$}\\
& & & $(-x,y+1/2,-z+1/2)$, $(x+1/2,-y+1/2,-z)$ & & \\ \hline
\hline 
 \end{tabular} 
 }
 \end{center}
 
Here the topological invariant can be chosen to be
\begin{equation}\label{TI_19}
\begin{aligned}
\widehat{\varphi_4}[T_2, T_3, S_2, S_2'] =&
\lambda(T_3, T_2, T_2^{-1})+ 
\lambda(T_2, T_3, T_2^{-1})+ 
\lambda(T_2, T_2^{-1}, T_3)+ 
\lambda(T_2, T_3, T_3^{-1})+\lambda(T_3, T_2, T_3^{-1})+ 
\lambda(T_3, T_3^{-1}, T_2)
\\
& 
+\lambda(T_2, T_2^{-1}T_3^{-1}S_2', T_2^{-1}T_3^{-1}S_2')+\lambda(T_3^{-1}S_2', T_2, T_2^{-1}T_3^{-1}S_2')+
\lambda(T_3^{-1}S_2', T_3^{-1}S_2', T_2)\\
&+ \lambda(T_3, T_2^{-1}T_3^{-1}S_{2}, T_2^{-1}T_3^{-1}S_{2})+\lambda(T_2^{-1}S_{2}, T_3, T_2^{-1}T_3^{-1}S_{2})+
\lambda(T_2^{-1}S_{2}, T_2^{-1}S_{2}, T_3)\\&+\lambda(T_3,1,T_3)+\lambda(T_2,1,T_2).
\end{aligned}
\end{equation}

\subsection*{No. 20: $C222_1$}\label{subsub:sg20}

This group is generated by three translations $T_{1,2,3}$ as given in Eqs.~\eqref{TransBravaisC}, a two-fold screw $S_2$, and a two-fold rotation $C'_2$:
\begin{subequations}
 \begin{align}
S_2 &\colon (x,y,z)\rightarrow (-x, -y, z + 1/2),\\ 
C'_2 &\colon (x,y,z)\rightarrow (-x, y, -z + 1/2).
\end{align}
\end{subequations}

The $\mathbb{Z}_2$ cohomology ring is given by

\begin{equation}
\mathbb{Z}_2[A_c,A_{c'},A_{x+y},B_{xy}]/\langle\mathcal{R}_2,\mathcal{R}_3,\mathcal{R}_4\rangle
 \end{equation}
where the relations are 
\begin{subequations} 
 \begin{align}
\mathcal{R}_2\colon & ~~
A_{c'} A_{x+y},~~A_c (A_c + A_{c'}),~~A_{x+y} (A_c + A_{x+y}),\\
\mathcal{R}_3\colon & ~~
A_{x+y} B_{xy},\\
\mathcal{R}_4\colon & ~~
B_{xy}^2.
\end{align} 
 \end{subequations}
 We have the following table regarding IWPs and group cohomology at degree 3.
\begin{center}
\begin{tabular}{c|cc|c|c|c}\hline\hline {Wyckoff}&\multicolumn{2}{c|}{Little group}& {Coordinates}&\multirow{2}{*}{LSM anomaly class}&\multirow{2}{*}{Topo. inv.} \\ \cline{2-4} position & Intl. & Sch\"{o}nflies & $ (0,0,0) + (1/2,1/2,0) + $ & &\\ \hline
4a&$2$&$C_2$& $(x,0,0)$, $(-x,0,1/2)$ & $A_c B_{xy}$ & $\varphi_2[T_1T_2^{-1}, S_2C'_2]$\\ 
4b&$2$&$C_2$& $(0,y,1/4)$, $(0,-y,3/4)$ & $(A_c + A_{c'}) B_{xy}$ & $\varphi_2[T_1T_2, C'_2]$\\ 
\hline
\hline 
 \end{tabular} 
 \end{center}
 
\subsection*{No. 21: $C222$}\label{subsub:sg21}

This group is generated by three translations $T_{1,2,3}$ as given in Eqs.~\eqref{TransBravaisC}, a two-fold rotation $C_2$, and a two-fold rotation $C'_2$:
\begin{subequations}
 \begin{align}
C_2 &\colon (x,y,z)\rightarrow (-x, -y, z),\\ 
C'_2 &\colon (x,y,z)\rightarrow (-x, y, -z).
\end{align}
\end{subequations}

The $\mathbb{Z}_2$ cohomology ring is given by

\begin{equation}
\mathbb{Z}_2[A_c,A_{c'},A_{x+y},A_z,B_{xy}]/\langle\mathcal{R}_2,\mathcal{R}_3,\mathcal{R}_4\rangle
 \end{equation}
where the relations are 
\begin{subequations} 
 \begin{align}
\mathcal{R}_2\colon & ~~
A_{c'} A_{x+y},~~A_{x+y} (A_c + A_{x+y}),~~A_z (A_{c'} + A_z),\\
\mathcal{R}_3\colon & ~~
A_{x+y} B_{xy},\\
\mathcal{R}_4\colon & ~~
B_{xy} (A_c^2 + A_c A_{c'} + B_{xy}).
\end{align} 
 \end{subequations}
We have the following table regarding IWPs and group cohomology at degree 3.
\begin{center}
\resizebox{\columnwidth}{!}{
\begin{tabular}{c|cc|c|c|c}\hline\hline {Wyckoff}&\multicolumn{2}{c|}{Little group}& {Coordinates}&\multirow{2}{*}{LSM anomaly class}&\multirow{2}{*}{Topo. inv.} \\ \cline{2-4} position & Intl. & Sch\"{o}nflies & $ (0,0,0) + (1/2,1/2,0) + $ & &\\ \hline
2a&$222$&$D_2$& $(0,0,0)$ & 
$(A_{c'} + A_z)(A_c^2 + A_c A_{c'} + A_c A_{x+y} + B_{xy})$ & $\varphi_2[C_2, C'_2]$\\ 
2b&$222$&$D_2$& $(0,1/2,0)$ & $(A_{c'} + A_z) B_{xy}$ & $\varphi_2[T_1T_2C_2, C'_2]$\\ 
2c&$222$&$D_2$& $(1/2,0,1/2)$ & $A_z B_{xy}$ & $\varphi_2[T_1T_2C_2, T_3C'_2]$\\ 
2d&$222$&$D_2$& $(0,0,1/2)$ & $A_z (A_c^2 + A_c A_{c'} + A_c A_{x+y} + B_{xy})$ & $\varphi_2[C_2, T_3C'_2]$\\ 
4k&$2$&$C_2$& $(1/4,1/4,z)$, $(3/4,1/4,-z)$ & $A_c A_{x+y} A_z$ & $\varphi_2[T_3, T_1C_2]$\\ 
\hline
\hline 
 \end{tabular} }
 \end{center}

\subsection*{No. 22: $F222$}\label{subsub:sg22}

This group is generated by three translations $T_{1,2,3}$ as given in Eqs.~\eqref{TransBravaisF}, a two-fold rotation $C_2$, and a two-fold rotation $C'_2$:
\begin{subequations}
 \begin{align}
C_2 &\colon (x,y,z)\rightarrow (-x, -y, z),\\ 
C'_2 &\colon (x,y,z)\rightarrow (-x, y, -z).
\end{align}
\end{subequations}

The $\mathbb{Z}_2$ cohomology ring is given by

\begin{equation}
\mathbb{Z}_2[A_c,A_{c'},A_{x+y},A_{x+z},C_\gamma,C_{xyz}]/\langle\mathcal{R}_2,\mathcal{R}_4,\mathcal{R}_6\rangle
 \end{equation}
where the relations are 
\begin{subequations} 
 \begin{align}
\mathcal{R}_2\colon & ~~
A_c A_{x+y} + A_{c'} A_{x+z},~~A_{x+y} (A_{c'} + A_{x+y}),~~A_{x+z} (A_c + A_{x+z}),\\
\mathcal{R}_4\colon & ~~
A_c C_\gamma + A_{x+z} C_\gamma + A_c C_{xyz},~~A_{c'} C_\gamma + A_{x+y} C_\gamma + A_{c'} C_{xyz},~~A_{x+y} C_{xyz},~~A_{x+z} C_{xyz},\\
\mathcal{R}_6\colon & ~~
C_\gamma (A_c^2 A_{c'} + A_c A_{c'}^2 + C_\gamma),~~C_{xyz} (A_c^2 A_{c'} + A_c A_{c'}^2 + C_\gamma),~~C_{xyz} (A_c^2 A_{c'}  + A_c A_{c'}^2  + C_{xyz}).
\end{align} 
 \end{subequations}
We have the following table regarding IWPs and group cohomology at degree 3.
\begin{center}
\begin{tabular}{c|cc|c|c|c}\hline\hline \multirow{3}{*}{\shortstack[l]{Wyckoff\\position}}&\multicolumn{2}{c|}{Little group}& {Coordinates}&\multirow{3}{*}{LSM anomaly class}&\multirow{3}{*}{Topo. inv.} \\ \cline{2-4} & \multirow{2}{*}{Intl.} & \multirow{2}{*}{Sch\"{o}nflies} & $(0,0,0) + ~(0,1/2,1/2) + $ & & \\ & & & $ (1/2,0,1/2) + ~(1/2,1/2,0) +$ & &\\ \hline
4a&$222$&$D_2$& $(0,0,0)$ & 
$A_c(A_c + A_{c'})(A_{c'} + A_{x+y}) + C_{xyz}$ & $\varphi_2[C_2, C'_2]$\\ 
4b&$222$&$D_2$& $(0,0,1/2)$ & $C_{xyz}$ & $\varphi_2[C_2, T_1T_2T_3^{-1}C'_2]$\\ 
4c&$222$&$D_2$& $(1/4,1/4,1/4)$ & $C_\gamma + C_{xyz}$ & $\varphi_2[T_3C_2, T_2C'_2]$\\ 
4d&$222$&$D_2$& $(1/4,1/4,3/4)$ & 
$A_c A_{c'} (A_{x+y} + A_{x+z}) + C_\gamma + C_{xyz}$ & $\varphi_2[T_1^{-1}T_2C_2, T_2C'_2]$\\ 
\hline
\hline 
 \end{tabular} 
 \end{center}

\subsection*{No. 23: $I222$}\label{subsub:sg23}

This group is generated by three translations $T_{1,2,3}$ as given in Eqs.~\eqref{TransBravaisI}, a two-fold rotation $C_2$, and a two-fold rotation $C'_2$:
\begin{subequations}
 \begin{align}
C_2 &\colon (x,y,z)\rightarrow (-x, -y, z),\\ 
C'_2 &\colon (x,y,z)\rightarrow (-x, y, -z).
\end{align}
\end{subequations}

The $\mathbb{Z}_2$ cohomology ring is given by

\begin{equation}
\mathbb{Z}_2[A_c,A_{c'},A_{x+y+z},B_\beta,B_{x(y+z)},B_{y(x+z)},C_{xyz}]/\langle\mathcal{R}_2,\mathcal{R}_3,\mathcal{R}_4,\mathcal{R}_5,\mathcal{R}_6\rangle
 \end{equation}
where the relations are 
\begin{subequations} 
 \begin{align}
\mathcal{R}_2\colon & ~~
A_c A_{x+y+z},~~A_{c'} A_{x+y+z},~~A_{x+y+z}^2,\\
\mathcal{R}_3\colon & ~~
A_{x+y+z} B_\beta,~~A_{x+y+z} B_{x(y+z)},~~A_c B_\beta + A_c B_{x(y+z)} + A_c B_{y(x+z)} + A_{c'} B_{y(x+z)},~~A_{x+y+z} B_{y(x+z)},\\
\mathcal{R}_4\colon & ~~
A_{x+y+z} C_{xyz},~~A_c^2 B_\beta + A_{c'}^2 B_\beta + B_\beta^2 + A_c A_{c'} B_{x(y+z)},~~A_c A_{c'} B_\beta + B_\beta B_{x(y+z)} + A_c C_{xyz} + A_{c'} C_{xyz},\nonumber\\&~~A_c A_{c'} B_\beta + A_c^2 B_{x(y+z)} + A_c^2 B_{y(x+z)} + B_\beta B_{y(x+z)} + A_c C_{xyz},~~B_{x(y+z)} (A_c A_{c'} + B_{x(y+z)}),\nonumber\\&~~A_c^2 B_\beta + A_c^2 B_{x(y+z)} + A_c^2 B_{y(x+z)} + B_{x(y+z)} B_{y(x+z)} + A_c C_{xyz},~~A_c^2 B_\beta + B_{y(x+z)}^2,\\
\mathcal{R}_5\colon & ~~
(A_c^2 + A_{c'}^2 + B_\beta) C_{xyz},~~B_{x(y+z)} C_{xyz},~~(A_c^2 + A_c A_{c'} + B_{y(x+z)}) C_{xyz},\\
\mathcal{R}_6\colon & ~~
C_{xyz} (A_c^2 A_{c'} + A_c A_{c'}^2 + C_{xyz}).
\end{align} 
 \end{subequations}
We have the following table regarding IWPs and group cohomology at degree 3.
\begin{center}
\resizebox{\columnwidth}{!}{
\begin{tabular}{c|cc|c|c|c}\hline\hline {Wyckoff}&\multicolumn{2}{c|}{Little group}& {Coordinates}&\multirow{2}{*}{LSM anomaly class}&\multirow{2}{*}{Topo. inv.} \\ \cline{2-4} position & Intl. & Sch\"{o}nflies & $(0,0,0) + ~(1/2,1/2,1/2) + $ & &\\ \hline
2a&$222$&$D_2$& $(0,0,0)$ & 
$(A_c + A_{c'})(A_c A_{c'} + B_{x(y+z)}) + C_{xyz}$ & $\varphi_2[C_2, C'_2]$\\ 
2b&$222$&$D_2$& $(1/2,0,0)$ & $C_{xyz}$ & $\varphi_2[T_2T_3C_2, T_2T_3C'_2]$\\ 
2c&$222$&$D_2$& $(0,0,1/2)$ & $A_c B_\beta + A_c B_{y(x+z)} + C_{xyz}$ & $\varphi_2[C_2, T_1T_2C'_2]$\\ 
2d&$222$&$D_2$& $(0,1/2,0)$ & 
$A_{c'}(B_{x(y+z)} + B_{y(x+z)})+ C_{xyz}$ & $\varphi_2[T_1T_3C_2, C'_2]$\\ 
\hline
\hline 
 \end{tabular} }
 \end{center}

\subsection*{No. 24: $I2_12_12_1$}\label{subsub:sg24}

This group is generated by three translations $T_{1,2,3}$ as given in Eqs.~\eqref{TransBravaisI}, a two-fold rotation $C_2$, and a two-fold rotation $C'_2$:
\begin{subequations}
 \begin{align}
C_2 &\colon (x,y,z)\rightarrow (-x, -y + 1/2, z),\\ 
C'_2 &\colon (x,y,z)\rightarrow (-x + 1/2, y, -z).
\end{align}
\end{subequations}

The $\mathbb{Z}_2$ cohomology ring is given by

\begin{equation}
\mathbb{Z}_2[A_c,A_{c'},A_{x+y+z},B_{y(x+z)},B_{z(x+y)}]/\langle\mathcal{R}_2,\mathcal{R}_3,\mathcal{R}_4\rangle
 \end{equation}
where the relations are 
\begin{subequations} 
 \begin{align}
\mathcal{R}_2\colon & ~~
A_c (A_{c'} + A_{x+y+z}),~~A_{c'} (A_c + A_{x+y+z}),~~A_c A_{c'} + A_{x+y+z}^2,\\
\mathcal{R}_3\colon & ~~
(A_{c'} + A_{x+y+z}) B_{y(x+z)},~~A_{c'} B_{y(x+z)} + A_c B_{z(x+y)},~~A_{c'} B_{y(x+z)} + A_{x+y+z} B_{z(x+y)},\\
\mathcal{R}_4\colon & ~~
A_c^3 A_{c'} + B_{y(x+z)}^2,~~A_c^3 A_{c'} + B_{y(x+z)} B_{z(x+y)},~~A_c^3 A_{c'} + B_{z(x+y)}^2.
\end{align} 
 \end{subequations}
We have the following table regarding IWPs and group cohomology at degree 3.
\begin{center}
\begin{tabular}{c|cc|c|c|c}\hline\hline {Wyckoff}&\multicolumn{2}{c|}{Little group}& {Coordinates}&\multirow{2}{*}{LSM anomaly class}&\multirow{2}{*}{Topo. inv.} \\ \cline{2-4} position & Intl. & Sch\"{o}nflies & $(0,0,0) + ~(1/2,1/2,1/2) + $ & &\\ \hline
4a&$2$&$C_2$& $(x,0,1/4)$, $(-x+1/2,0,3/4)$ & $A_{c'} (A_c^2 + B_{y(x+z)})$ & $\varphi_2[T_2T_3, T_2C_2C'_2]$\\ 
4b&$2$&$C_2$& $(1/4,y,0)$, $(1/4,-y,1/2)$ & $A_{c'} (B_{y(x+z)} + B_{z(x+y)})$ & $\varphi_2[T_1T_3, C'_2]$\\ 
4c&$2$&$C_2$& $(0,1/4,z)$, $(0,3/4,-z+1/2)$ & $(A_c + A_{c'}) B_{y(x+z)}$ & $\varphi_2[T_1T_2, C_2]$\\ 
\hline
\hline 
 \end{tabular} 
 \end{center}

\subsection*{No. 25: $Pmm2$}\label{subsub:sg25}

This group is generated by three translations $T_{1,2,3}$ as given in Eqs.~\eqref{TransBravaisP}, a two-fold rotation $C_2$, and a mirror $M$:
\begin{subequations}
 \begin{align}
C_2 &\colon (x,y,z)\rightarrow (-x, -y, z),\\ 
M &\colon (x,y,z)\rightarrow (x, -y, z).
\end{align}
\end{subequations}

The $\mathbb{Z}_2$ cohomology ring is given by

\begin{equation}
\mathbb{Z}_2[A_c,A_m,A_x,A_y,A_z]/\langle\mathcal{R}_2\rangle
 \end{equation}
where the relations are 
\begin{subequations} 
 \begin{align}
\mathcal{R}_2\colon & ~~
A_x (A_c + A_x),~~A_y (A_c + A_m + A_y),~~A_z^2.
\end{align} 
 \end{subequations}
We have the following table regarding IWPs and group cohomology at degree 3.
\begin{center}
\begin{tabular}{c|cc|c|c|c}\hline\hline {Wyckoff}&\multicolumn{2}{c|}{Little group}& \multirow{2}{*}{Coordinates}&\multirow{2}{*}{LSM anomaly class}&\multirow{2}{*}{Topo. inv.} \\ \cline{2-3} position & Intl. & Sch\"{o}nflies & & & \\ \hline
1a&$mm2$&$C_{2v}$& $(0,0,z)$ & $(A_c + A_x) (A_c + A_m + A_y) A_z$ & $\varphi_2[T_3, C_2]$\\ 
1b&$mm2$&$C_{2v}$& $(0,1/2,z)$ & $(A_c + A_x) A_y A_z$ & $\varphi_2[T_3, T_2C_2]$\\ 
1c&$mm2$&$C_{2v}$& $(1/2,0,z)$ & $A_x (A_c + A_m + A_y) A_z$ & $\varphi_2[T_3, T_1C_2]$\\ 
1d&$mm2$&$C_{2v}$& $(1/2,1/2,z)$ & $A_x A_y A_z$ & $\varphi_2[T_3, T_1T_2C_2]$\\ 
\hline
\hline 
 \end{tabular} 
 \end{center}

\subsection*{No. 26: $Pmc2_1$}\label{subsub:sg26}

This group is generated by three translations $T_{1,2,3}$ as given in Eqs.~\eqref{TransBravaisP}, a two-fold screw $S_2$, and a glide $G$:
\begin{subequations}
 \begin{align}
S_2 &\colon (x,y,z)\rightarrow (-x, -y, z + 1/2),\\ 
G &\colon (x,y,z)\rightarrow (x, -y, z + 1/2).
\end{align}
\end{subequations}

The $\mathbb{Z}_2$ cohomology ring is given by

\begin{equation}
\mathbb{Z}_2[A_c,A_m,A_x,A_y]/\langle\mathcal{R}_2\rangle
 \end{equation}
where the relations are 
\begin{subequations} 
 \begin{align}
\mathcal{R}_2\colon & ~~
(A_c + A_m)^2,~~A_x (A_c + A_x),~~A_y (A_c + A_m + A_y).
\end{align} 
 \end{subequations}
We have the following table regarding IWPs and group cohomology at degree 3.
\begin{center}
\begin{tabular}{c|cc|c|c|c}\hline\hline {Wyckoff}&\multicolumn{2}{c|}{Little group}& \multirow{2}{*}{Coordinates}&\multirow{2}{*}{LSM anomaly class}&\multirow{2}{*}{Topo. inv.} \\ \cline{2-3} position & Intl. & Sch\"{o}nflies & & & \\ \hline
2a&$m$&$C_s$& $(0,y,z)$, $(0,-y,z+1/2)$ & $(A_c + A_m) (A_c + A_x) A_y$ & $\widetilde{\varphi_3}[T_2, G, T_3^{-1}S_2G]$\\ 
2b&$m$&$C_s$& $(1/2,y,z)$, $(1/2,-y,z+1/2)$ & $(A_c + A_m) A_x A_y$ & $\widetilde{\varphi_3}[T_2, G, T_1T_3^{-1}S_2G]$\\ 
\hline
\hline 
 \end{tabular} 
 \end{center}
The expression of $\widetilde{\varphi_3}$ is given in Eq.~\eqref{wilde3}.

\subsection*{No. 27: $Pcc2$}\label{subsub:sg27}

This group is generated by three translations $T_{1,2,3}$ as given in Eqs.~\eqref{TransBravaisP}, a two-fold rotation $C_2$, and a glide $G$:
\begin{subequations}
 \begin{align}
C_2 &\colon (x,y,z)\rightarrow (-x, -y, z),\\ 
G &\colon (x,y,z)\rightarrow (x, -y, z + 1/2).
\end{align}
\end{subequations}

The $\mathbb{Z}_2$ cohomology ring is given by

\begin{equation}
\mathbb{Z}_2[A_c,A_m,A_x,A_y]/\langle\mathcal{R}_2\rangle
 \end{equation}
where the relations are 
\begin{subequations} 
 \begin{align}
\mathcal{R}_2\colon & ~~
A_m^2,~~A_x (A_c + A_x),~~A_y (A_c + A_m + A_y).
\end{align} 
 \end{subequations}
We have the following table regarding IWPs and group cohomology at degree 3.
\begin{center}
\begin{tabular}{c|cc|c|c|c}\hline\hline {Wyckoff}&\multicolumn{2}{c|}{Little group}& \multirow{2}{*}{Coordinates}&\multirow{2}{*}{LSM anomaly class}&\multirow{2}{*}{Topo. inv.} \\ \cline{2-3} position & Intl. & Sch\"{o}nflies & & & \\ \hline
2a&$2$&$C_2$& $(0,0,z)$, $(0,0,z+1/2)$ & $A_m (A_c + A_x) (A_c + A_y)$ & $\varphi_2[G, C_2]$\\ 
2b&$2$&$C_2$& $(0,1/2,z)$, $(0,1/2,z+1/2)$ & $A_m (A_c + A_x) A_y$ & $\varphi_2[C_2G, T_2C_2]$\\ 
2c&$2$&$C_2$& $(1/2,0,z)$, $(1/2,0,z+1/2)$ & $A_m A_x (A_c + A_y)$ & $\varphi_2[G, T_1C_2]$\\ 
2d&$2$&$C_2$& $(1/2,1/2,z)$, $(1/2,1/2,z+1/2)$ & $A_m A_x A_y$ & $\varphi_2[T_2G, T_1T_2C_2]$\\ 
\hline
\hline 
 \end{tabular} 
 \end{center}

\subsection*{No. 28: $Pma2$}\label{subsub:sg28}

This group is generated by three translations $T_{1,2,3}$ as given in Eqs.~\eqref{TransBravaisP}, a two-fold rotation $C_2$, and a glide $G$:
\begin{subequations}
 \begin{align}
C_2 &\colon (x,y,z)\rightarrow (-x, -y, z),\\ 
G &\colon (x,y,z)\rightarrow (x + 1/2, -y, z).
\end{align}
\end{subequations}

The $\mathbb{Z}_2$ cohomology ring is given by

\begin{equation}
\mathbb{Z}_2[A_c,A_m,A_y,A_z]/\langle\mathcal{R}_2\rangle
 \end{equation}
where the relations are 
\begin{subequations} 
 \begin{align}
\mathcal{R}_2\colon & ~~
A_m (A_c + A_m),~~A_y (A_c + A_m + A_y),~~A_z^2.
\end{align} 
 \end{subequations}
We have the following table regarding IWPs and group cohomology at degree 3.
\begin{center}
\begin{tabular}{c|cc|c|c|c}\hline\hline {Wyckoff}&\multicolumn{2}{c|}{Little group}& \multirow{2}{*}{Coordinates}&\multirow{2}{*}{LSM anomaly class}&\multirow{2}{*}{Topo. inv.} \\ \cline{2-3} position & Intl. & Sch\"{o}nflies & & & \\ \hline
2a&$2$&$C_2$& $(0,0,z)$, $(1/2,0,z)$ & $(A_c + A_m) (A_c + A_y) A_z$ & $\varphi_2[T_3, C_2]$\\ 
2b&$2$&$C_2$& $(0,1/2,z)$, $(1/2,1/2,z)$ & $(A_c + A_m) A_y A_z$ & $\varphi_2[T_3, T_2C_2]$\\ 
2c&$m$&$C_s$& $(1/4,y,z)$, $(3/4,-y,z)$ & $A_m A_y A_z$ & $\varphi_3[T_2, T_3, C_2G]$\\ 
\hline
\hline 
 \end{tabular} 
 \end{center}

\subsection*{No. 29: $Pca2_1$}\label{subsub:sg29}

This group is generated by three translations $T_{1,2,3}$ as given in Eqs.~\eqref{TransBravaisP}, a two-fold screw $S_2$, and a glide $G$:
\begin{subequations}
 \begin{align}
S_2 &\colon (x,y,z)\rightarrow (-x, -y, z + 1/2),\\ 
G &\colon (x,y,z)\rightarrow (x + 1/2, -y, z).
\end{align}
\end{subequations}

The $\mathbb{Z}_2$ cohomology ring is given by

\begin{equation}
\mathbb{Z}_2[A_c,A_m,A_y]/\langle\mathcal{R}_2\rangle
 \end{equation}
where the relations are 
\begin{subequations} 
 \begin{align}
\mathcal{R}_2\colon & ~~
A_c^2,~~A_m (A_c + A_m),~~A_y (A_c + A_m + A_y).
\end{align} 
 \end{subequations}
We have the following table regarding IWPs and group cohomology at degree 3.
\begin{center}
\begin{tabular}{c|cc|c|c|c}\hline\hline {Wyckoff}&\multicolumn{2}{c|}{Little group}& \multirow{2}{*}{Coordinates}&\multirow{2}{*}{LSM anomaly class}&\multirow{2}{*}{Topo. inv.} \\ \cline{2-3} position & Intl. & Sch\"{o}nflies & & & \\ \hline
\multirow{2}{*}{4a} & \multirow{2}{*}{$1$} & \multirow{2}{*}{$C_1$} & $(x,y,z)$, $(-x,-y,z+1/2)$, & \multirow{2}{*}{$A_c A_m A_y$} & \multirow{2}{*}{$\widehat{\varphi_4}[T_1, T_2, T_3^{-1}S_2, G]$}\\
& & & $(x+1/2,-y,z)$, $(-x+1/2,y,z+1/2)$ & & \\ \hline
\hline 
 \end{tabular} 
 \end{center}
 Here the topological invariant can be chosen to be
\begin{equation}
\begin{aligned}
\widehat{\varphi_4}[T_1,T_2,T_3^{-1}S_2,&G]=
\lambda((T_2,T_1,T_1^{-1})+
\lambda(T_1,T_2,T_1^{-1})+\lambda(T_1,T_1^{-1},T_2)+\lambda(T_2,1,T_2)\\
&+\lambda(T_2,T_1^{-1}T_2^{-1}G,T_1^{-1}T_2^{-1}G)+\lambda(T_1^{-1}G,T_2,T_1^{-1}T_2^{-1}G)+
\lambda(T_1^{-1}G,T_1^{-1}G,T_2)\\
&+
\lambda(T_1,T_2,T_3^{-1}S_2G)+\lambda(T_2,T_1,T_3^{-1}S_2G)+\lambda(T_1,T_3^{-1}S_2G,T_2)\\
&+\lambda(T_3^{-1}S_2,T_1^{-1}G,T_2)+\lambda(T_1^{-1}G,T_3^{-1}S_2,T_2)+
\lambda(T_3^{-1}S_2,T_2,T_1^{-1}T_2^{-1}G))
+\lambda(T_1^{-1}G,T_2,T_2^{-1}T_3^{-1}S_2)\\
&+\lambda(T_2,T_2^{-1}T_3^{-1}S_2,T_1^{-1}T_2^{-1}G)+\lambda(T_2,T_1^{-1}T_2^{-1}G,T_2^{-1}T_3^{-1}S_2).
\end{aligned}
\end{equation}

\subsection*{No. 30: $Pnc2$}\label{subsub:sg30}

This group is generated by three translations $T_{1,2,3}$ as given in Eqs.~\eqref{TransBravaisP}, a two-fold rotation $C_2$, and a glide $G$:
\begin{subequations}
 \begin{align}
C_2 &\colon (x,y,z)\rightarrow (-x, -y, z),\\ 
G &\colon (x,y,z)\rightarrow (x, -y + 1/2, z + 1/2).
\end{align}
\end{subequations}

The $\mathbb{Z}_2$ cohomology ring is given by

\begin{equation}
\mathbb{Z}_2[A_c,A_m,A_x,B_\beta]/\langle\mathcal{R}_2,\mathcal{R}_3,\mathcal{R}_4\rangle
 \end{equation}
where the relations are 
\begin{subequations} 
 \begin{align}
\mathcal{R}_2\colon & ~~
A_c A_m,~~A_m^2,~~A_x (A_c + A_x),\\
\mathcal{R}_3\colon & ~~
A_m B_\beta,\\
\mathcal{R}_4\colon & ~~
B_\beta^2.
\end{align} 
 \end{subequations}
We have the following table regarding IWPs and group cohomology at degree 3.
\begin{center}
\begin{tabular}{c|cc|c|c|c}\hline\hline {Wyckoff}&\multicolumn{2}{c|}{Little group}& \multirow{2}{*}{Coordinates}&\multirow{2}{*}{LSM anomaly class}&\multirow{2}{*}{Topo. inv.} \\ \cline{2-3} position & Intl. & Sch\"{o}nflies & & & \\ \hline
2a&$2$&$C_2$& $(0,0,z)$, $(0,1/2,z+1/2)$ & $(A_c + A_x) B_\beta$ & $\varphi_2[T_3, C_2]$\\ 
2b&$2$&$C_2$& $(1/2,0,z)$, $(1/2,1/2,z+1/2)$ & $A_x B_\beta$ & $\varphi_2[T_3, T_1C_2]$\\ 
\hline
\hline 
 \end{tabular} 
 \end{center}

\subsection*{No. 31: $Pmn2_1$}\label{subsub:sg31}

This group is generated by three translations $T_{1,2,3}$ as given in Eqs.~\eqref{TransBravaisP}, a two-fold screw $S_2$, and a glide $G$:
\begin{subequations}
 \begin{align}
S_2 &\colon (x,y,z)\rightarrow (-x + 1/2, -y, z + 1/2),\\ 
G &\colon (x,y,z)\rightarrow (x + 1/2, -y, z + 1/2).
\end{align}
\end{subequations}

The $\mathbb{Z}_2$ cohomology ring is given by

\begin{equation}
\mathbb{Z}_2[A_c,A_m,A_y,B_\beta]/\langle\mathcal{R}_2,\mathcal{R}_3,\mathcal{R}_4\rangle
 \end{equation}
where the relations are 
\begin{subequations} 
 \begin{align}
\mathcal{R}_2\colon & ~~
A_c (A_c + A_m),~~A_m (A_c + A_m),~~A_y (A_c + A_m + A_y),\\
\mathcal{R}_3\colon & ~~
(A_c + A_m) B_\beta,\\
\mathcal{R}_4\colon & ~~
B_\beta^2.
\end{align} 
 \end{subequations}
We have the following table regarding IWPs and group cohomology at degree 3.
\begin{center}
\begin{tabular}{c|cc|c|c|c}\hline\hline {Wyckoff}&\multicolumn{2}{c|}{Little group}& \multirow{2}{*}{Coordinates}&\multirow{2}{*}{LSM anomaly class}&\multirow{2}{*}{Topo. inv.} \\ \cline{2-3} position & Intl. & Sch\"{o}nflies & & & \\ \hline
2a&$m$&$C_s$& $(0,y,z)$, $(1/2,-y,z+1/2)$ & $A_y B_\beta$ & $\varphi_3[T_2, T_3, T_3^{-1}S_2G]$\\ 
\hline
\hline 
 \end{tabular} 
 \end{center}

\subsection*{No. 32: $Pba2$}\label{subsub:sg32}

This group is generated by three translations $T_{1,2,3}$ as given in Eqs.~\eqref{TransBravaisP}, a two-fold rotation $C_2$, and a glide $G$:
\begin{subequations}
 \begin{align}
C_2 &\colon (x,y,z)\rightarrow (-x, -y, z),\\ 
G &\colon (x,y,z)\rightarrow (x + 1/2, -y + 1/2, z).
\end{align}
\end{subequations}

The $\mathbb{Z}_2$ cohomology ring is given by

\begin{equation}
\mathbb{Z}_2[A_c,A_m,A_z,B_\beta]/\langle\mathcal{R}_2,\mathcal{R}_3,\mathcal{R}_4\rangle
 \end{equation}
where the relations are 
\begin{subequations} 
 \begin{align}
\mathcal{R}_2\colon & ~~
A_c A_m,~~A_m^2,~~A_z^2,\\
\mathcal{R}_3\colon & ~~
A_m B_\beta,\\
\mathcal{R}_4\colon & ~~
B_\beta (A_c^2 + B_\beta).
\end{align} 
 \end{subequations}
We have the following table regarding IWPs and group cohomology at degree 3.
\begin{center}
\begin{tabular}{c|cc|c|c|c}\hline\hline {Wyckoff}&\multicolumn{2}{c|}{Little group}& \multirow{2}{*}{Coordinates}&\multirow{2}{*}{LSM anomaly class}&\multirow{2}{*}{Topo. inv.} \\ \cline{2-3} position & Intl. & Sch\"{o}nflies & & & \\ \hline
2a&$2$&$C_2$& $(0,0,z)$, $(1/2,1/2,z)$ & $A_z (A_c^2 + B_\beta)$ & $\varphi_2[T_3, C_2]$\\ 
2b&$2$&$C_2$& $(0,1/2,z)$, $(1/2,0,z)$ & $A_z B_\beta$ & $\varphi_2[T_3, T_1C_2]$\\ 
\hline
\hline 
 \end{tabular} 
 \end{center}

\subsection*{No. 33: $Pna2_1$}\label{subsub:sg33}

This group is generated by three translations $T_{1,2,3}$ as given in Eqs.~\eqref{TransBravaisP}, a two-fold screw $S_2$, and a glide $G$:
\begin{subequations}
 \begin{align}
S_2 &\colon (x,y,z)\rightarrow (-x, -y, z + 1/2),\\ 
G &\colon (x,y,z)\rightarrow (x + 1/2, -y + 1/2, z).
\end{align}
\end{subequations}

The $\mathbb{Z}_2$ cohomology ring is given by

\begin{equation}
\mathbb{Z}_2[A_c,A_m,B_{\beta 1},B_{\beta 2}]/\langle\mathcal{R}_2,\mathcal{R}_3,\mathcal{R}_4\rangle
 \end{equation}
where the relations are 
\begin{subequations} 
 \begin{align}
\mathcal{R}_2\colon & ~~
A_c A_m,~~A_c^2,~~A_m^2,\\
\mathcal{R}_3\colon & ~~
A_c B_{\beta 1},~~A_m B_{\beta 1} + A_c B_{\beta 2},~~A_m (B_{\beta 1} + B_{\beta 2}),\\
\mathcal{R}_4\colon & ~~
B_{\beta 1}^2,~~B_{\beta 1} B_{\beta 2},~~B_{\beta 2}^2.
\end{align} 
 \end{subequations}
We have the following table regarding IWPs and group cohomology at degree 3.
\begin{center}
\resizebox{\columnwidth}{!}{
\begin{tabular}{c|cc|c|c|c}\hline\hline {Wyckoff}&\multicolumn{2}{c|}{Little group}& \multirow{2}{*}{Coordinates}&\multirow{2}{*}{LSM anomaly class}&\multirow{2}{*}{Topo. inv.} \\ \cline{2-3} position & Intl. & Sch\"{o}nflies & & & \\ \hline
\multirow{2}{*}{4a} & \multirow{2}{*}{$1$} & \multirow{2}{*}{$C_1$} & $(x,y,z)$, $(-x,-y,z+1/2)$, & \multirow{2}{*}{$A_m B_{\beta 1}$} & \multirow{2}{*}{$\widehat{\varphi_4}[T_1, T_2, T_3^{-1}S_2, G]$}\\
& & & $(x+1/2,-y+1/2,z)$, $(-x+1/2,y+1/2,z+1/2)$ & & \\ \hline
\hline 
 \end{tabular} }
 \end{center}
Here the topological invariant can be chosen to be
\begin{equation}
\begin{aligned}
\widehat{\varphi_4}[T_1, T_2, T_3^{-1}S_2, &G]=
\lambda((T_2,T_1,T_1^{-1})+
\lambda(T_1,T_2,T_1^{-1})+\lambda(T_1,T_1^{-1},T_2)+\lambda(T_2,1,T_2)\\
&+\lambda(T_2,T_1^{-1}T_2^{-1}G_1,T_1^{-1}T_2^{-1}G_1)+\lambda(T_1^{-1}G_1,T_2,T_1^{-1}T_2^{-1}G_1)+
\lambda(T_1^{-1}G_1,T_1^{-1}G_1,T_2)\\
&+
\lambda(T_1,T_2,T_3^{-1}S_3G_1)+\lambda(T_2,T_1,T_3^{-1}S_3G_1)+\lambda(T_1,T_3^{-1}S_3G_1,T_2)+\lambda(T_2,T_3^{-1}S_2G_1,T_2)\\
&+\lambda(T_3^{-1}S_3,T_1^{-1}G_1,T_2)+\lambda(T_1^{-1}G_1,T_3^{-1}S_3,T_2)+
\lambda(T_3^{-1}S_3,T_2,T_1^{-1}T_2^{-1}G_1))\\
&+\lambda(T_1^{-1}G_1,T_2,T_2^{-1}T_3^{-1}S_3)+
\lambda(T_2,T_2^{-1}T_3^{-1}S_3,T_1^{-1}T_2^{-1}G_1)+\lambda(T_2,T_1^{-1}T_2^{-1}G_1,T_2^{-1}T_3^{-1}S_3).
\end{aligned}
\end{equation}

\subsection*{No. 34: $Pnn2$}\label{subsub:sg34}

This group is generated by three translations $T_{1,2,3}$ as given in Eqs.~\eqref{TransBravaisP}, a two-fold rotation $C_2$, and a glide $G$:
\begin{subequations}
 \begin{align}
C_2 &\colon (x,y,z)\rightarrow (-x, -y, z),\\ 
G &\colon (x,y,z)\rightarrow (x + 1/2, -y + 1/2, z + 1/2).
\end{align}
\end{subequations}

The $\mathbb{Z}_2$ cohomology ring is given by

\begin{equation}
\mathbb{Z}_2[A_c,A_m,A_{x+y+z}]/\langle\mathcal{R}_2,\mathcal{R}_3,\mathcal{R}_4\rangle
 \end{equation}
where the relations are 
\begin{subequations} 
 \begin{align}
\mathcal{R}_2\colon & ~~
A_c A_m,~~A_m^2,\\
\mathcal{R}_3\colon & ~~
A_m A_{x+y+z}^2,\\
\mathcal{R}_4\colon & ~~
A_{x+y+z}^2 (A_c + A_{x+y+z})^2.
\end{align} 
 \end{subequations}
We have the following table regarding IWPs and group cohomology at degree 3.
\begin{center}
\begin{tabular}{c|cc|c|c|c}\hline\hline {Wyckoff}&\multicolumn{2}{c|}{Little group}& \multirow{2}{*}{Coordinates}&\multirow{2}{*}{LSM anomaly class}&\multirow{2}{*}{Topo. inv.} \\ \cline{2-3} position & Intl. & Sch\"{o}nflies & & & \\ \hline
2a&$2$&$C_2$& $(0,0,z)$, $(1/2,1/2,z+1/2)$ & $A_{x+y+z} (A_c + A_{x+y+z})^2$ & $\varphi_2[T_3, C_2]$\\ 
2b&$2$&$C_2$& $(0,1/2,z)$, $(1/2,0,z+1/2)$ & $A_{x+y+z}^2 (A_c + A_{x+y+z})$ & $\varphi_2[T_3, T_1C_2]$\\ 
\hline
\hline 
 \end{tabular} 
 \end{center}

\subsection*{No. 35: $Cmm2$}\label{subsub:sg35}

This group is generated by three translations $T_{1,2,3}$ as given in Eqs.~\eqref{TransBravaisC}, a two-fold rotation $C_2$, and a mirror $M$:
\begin{subequations}
 \begin{align}
C_2 &\colon (x,y,z)\rightarrow (-x, -y, z),\\ 
M &\colon (x,y,z)\rightarrow (x, -y, z).
\end{align}
\end{subequations}

The $\mathbb{Z}_2$ cohomology ring is given by

\begin{equation}
\mathbb{Z}_2[A_c,A_m,A_{x+y},A_z,B_{xy}]/\langle\mathcal{R}_2,\mathcal{R}_3,\mathcal{R}_4\rangle
 \end{equation}
where the relations are 
\begin{subequations} 
 \begin{align}
\mathcal{R}_2\colon & ~~
A_m A_{x+y},~~A_{x+y} (A_c + A_{x+y}),~~A_z^2,\\
\mathcal{R}_3\colon & ~~
A_{x+y} B_{xy},\\
\mathcal{R}_4\colon & ~~
B_{xy} (A_c^2 + A_c A_m + B_{xy}).
\end{align} 
 \end{subequations}
We have the following table regarding IWPs and group cohomology at degree 3.
\begin{center}
\begin{tabular}{c|cc|c|c|c}\hline\hline {Wyckoff}&\multicolumn{2}{c|}{Little group}& {Coordinates}&\multirow{2}{*}{LSM anomaly class}&\multirow{2}{*}{Topo. inv.} \\ \cline{2-4} position & Intl. & Sch\"{o}nflies & $ (0,0,0) + (1/2,1/2,0) + $ & &\\ \hline
2a&$mm2$&$C_{2v}$& $(0,0,z)$ & $A_z (A_c^2 + A_c A_m + A_c A_{x+y} + B_{xy})$ & $\varphi_2[T_3, C_2]$\\ 
2b&$mm2$&$C_{2v}$& $(0,1/2,z)$ & $A_z B_{xy}$ & $\varphi_2[T_3, T_1T_2C_2]$\\ 
4c&$2$&$C_2$& $(1/4,1/4,z)$, $(1/4,3/4,z)$ & $A_c A_{x+y} A_z$ & $\varphi_2[T_3, T_1C_2]$\\ 
\hline
\hline 
 \end{tabular} 
 \end{center}

\subsection*{No. 36: $Cmc2_1$}\label{subsub:sg36}

This group is generated by three translations $T_{1,2,3}$ as given in Eqs.~\eqref{TransBravaisC}, a two-fold screw $S_2$, and a glide $G$:
\begin{subequations}
 \begin{align}
S_2 &\colon (x,y,z)\rightarrow (-x, -y, z + 1/2),\\ 
G &\colon (x,y,z)\rightarrow (x, -y, z + 1/2).
\end{align}
\end{subequations}

The $\mathbb{Z}_2$ cohomology ring is given by

\begin{equation}
\mathbb{Z}_2[A_c,A_m,A_{x+y},B_{xy}]/\langle\mathcal{R}_2,\mathcal{R}_3,\mathcal{R}_4\rangle
 \end{equation}
where the relations are 
\begin{subequations} 
 \begin{align}
\mathcal{R}_2\colon & ~~
A_m A_{x+y},~~(A_c + A_m)^2,~~A_{x+y} (A_c + A_{x+y}),\\
\mathcal{R}_3\colon & ~~
A_{x+y} B_{xy},\\
\mathcal{R}_4\colon & ~~
B_{xy} (A_c^2 + A_c A_m + B_{xy}).
\end{align} 
 \end{subequations}
We have the following table regarding IWPs and group cohomology at degree 3.
\begin{center}
\begin{tabular}{c|cc|c|c|c}\hline\hline {Wyckoff}&\multicolumn{2}{c|}{Little group}& {Coordinates}&\multirow{2}{*}{LSM anomaly class}&\multirow{2}{*}{Topo. inv.} \\ \cline{2-4} position & Intl. & Sch\"{o}nflies & $ (0,0,0) + (1/2,1/2,0) + $ & &\\ \hline
4a&$m$&$C_s$& $(0,y,z)$, $(0,-y,z+1/2)$ & $(A_c + A_m) B_{xy}$ & $\widetilde{\varphi_3}[T_1T_2, G, T_3^{-1}S_2G]$\\ 
\hline
\hline 
 \end{tabular} 
 \end{center}
 The expression of $\widetilde{\varphi_3}$ is given in Eq.~\eqref{wilde3}.

\subsection*{No. 37: $Ccc2$}\label{subsub:sg37}

This group is generated by three translations $T_{1,2,3}$ as given in Eqs.~\eqref{TransBravaisC}, a two-fold rotation $C_2$, and a glide $G$:
\begin{subequations}
 \begin{align}
C_2 &\colon (x,y,z)\rightarrow (-x, -y, z),\\ 
G &\colon (x,y,z)\rightarrow (x, -y, z + 1/2).
\end{align}
\end{subequations}

The $\mathbb{Z}_2$ cohomology ring is given by

\begin{equation}
\mathbb{Z}_2[A_c,A_m,A_{x+y},B_{xy},B_{z(x+y)}]/\langle\mathcal{R}_2,\mathcal{R}_3,\mathcal{R}_4\rangle
 \end{equation}
where the relations are 
\begin{subequations} 
 \begin{align}
\mathcal{R}_2\colon & ~~
A_m A_{x+y},~~A_m^2,~~A_{x+y} (A_c + A_{x+y}),\\
\mathcal{R}_3\colon & ~~
A_{x+y} B_{xy},~~A_m B_{z(x+y)},~~A_m B_{xy} + A_c B_{z(x+y)} + A_{x+y} B_{z(x+y)},\\
\mathcal{R}_4\colon & ~~
B_{xy} (A_c^2 + A_c A_m + B_{xy}),~~B_{xy} (A_c A_m + B_{z(x+y)}),~~B_{z(x+y)}^2.
\end{align} 
 \end{subequations}
We have the following table regarding IWPs and group cohomology at degree 3.
\begin{center}
\begin{tabular}{c|cc|c|c|c}\hline\hline {Wyckoff}&\multicolumn{2}{c|}{Little group}& {Coordinates}&\multirow{2}{*}{LSM anomaly class}&\multirow{2}{*}{Topo. inv.} \\ \cline{2-4} position & Intl. & Sch\"{o}nflies & $ (0,0,0) + (1/2,1/2,0) + $ & &\\ \hline
4a&$2$&$C_2$& $(0,0,z)$, $(0,0,z+1/2)$ & $A_m (A_c^2 + B_{xy})$ & $\varphi_2[G, C_2]$\\ 
4b&$2$&$C_2$& $(0,1/2,z)$, $(0,1/2,z+1/2)$ & $A_m B_{xy}$ & $\varphi_2[G, T_1T_2^{-1}C_2]$\\ 
4c&$2$&$C_2$& $(1/4,1/4,z)$, $(1/4,3/4,z+1/2)$ & $A_m B_{xy} + A_c B_{z(x+y)}$ & $\varphi_2[T_3, T_1C_2]$\\ 
\hline
\hline 
 \end{tabular} 
 \end{center}

\subsection*{No. 38: $Amm2$}\label{subsub:sg38}

This group is generated by three translations $T_{1,2,3}$ as given in Eqs.~\eqref{TransBravaisA}, a two-fold rotation $C_2$, and a mirror $M$:
\begin{subequations}
 \begin{align}
C_2 &\colon (x,y,z)\rightarrow (-x, -y, z),\\ 
M &\colon (x,y,z)\rightarrow (x, -y, z).
\end{align}
\end{subequations}

The $\mathbb{Z}_2$ cohomology ring is given by

\begin{equation}
\mathbb{Z}_2[A_c,A_m,A_x,A_{y+z},B_{yz}]/\langle\mathcal{R}_2,\mathcal{R}_3,\mathcal{R}_4\rangle
 \end{equation}
where the relations are 
\begin{subequations} 
 \begin{align}
\mathcal{R}_2\colon & ~~
(A_c + A_m) A_{y+z},~~A_x (A_c + A_x),~~A_{y+z}^2,\\
\mathcal{R}_3\colon & ~~
A_{y+z} B_{yz},\\
\mathcal{R}_4\colon & ~~
B_{yz}^2.
\end{align} 
 \end{subequations}
We have the following table regarding IWPs and group cohomology at degree 3.
\begin{center}
\begin{tabular}{c|cc|c|c|c}\hline\hline {Wyckoff}&\multicolumn{2}{c|}{Little group}& {Coordinates}&\multirow{2}{*}{LSM anomaly class}&\multirow{2}{*}{Topo. inv.} \\ \cline{2-4} position & Intl. & Sch\"{o}nflies & $(0,0,0) + ~(0,1/2,1/2) + $ & &\\ \hline
2a&$mm2$&$C_{2v}$& $(0,0,z)$ & $(A_c + A_x) B_{yz}$ & $\varphi_2[T_2T_3, C_2]$\\ 
2b&$mm2$&$C_{2v}$& $(1/2,0,z)$ & $A_x B_{yz}$ & $\varphi_2[T_2T_3, T_1C_2]$\\ 
\hline
\hline 
 \end{tabular} 
 \end{center}

\subsection*{No. 39: $Aem2$}\label{subsub:sg39}

This group is generated by three translations $T_{1,2,3}$ as given in Eqs.~\eqref{TransBravaisA}, a two-fold rotation $C_2$, and a mirror $M$:
\begin{subequations}
 \begin{align}
C_2 &\colon (x,y,z)\rightarrow (-x, -y, z),\\ 
M &\colon (x,y,z)\rightarrow (x, -y + 1/2, z).
\end{align}
\end{subequations}

The $\mathbb{Z}_2$ cohomology ring is given by

\begin{equation}
\mathbb{Z}_2[A_c,A_m,A_x,A_{y+z}]/\langle\mathcal{R}_2\rangle
 \end{equation}
where the relations are 
\begin{subequations} 
 \begin{align}
\mathcal{R}_2\colon & ~~
A_c A_m + A_c A_{y+z} + A_m A_{y+z},~~A_x (A_c + A_x),~~A_{y+z}^2.
\end{align} 
 \end{subequations}
We have the following table regarding IWPs and group cohomology at degree 3.
\begin{center}
\begin{tabular}{c|cc|c|c|c}\hline\hline {Wyckoff}&\multicolumn{2}{c|}{Little group}& {Coordinates}&\multirow{2}{*}{LSM anomaly class}&\multirow{2}{*}{Topo. inv.} \\ \cline{2-4} position & Intl. & Sch\"{o}nflies & $(0,0,0) + ~(0,1/2,1/2) + $ & &\\ \hline
4a&$2$&$C_2$& $(0,0,z)$, $(0,1/2,z)$ & $A_c (A_c A_m + A_x A_{y+z})$ & $\varphi_2[T_3M, C_2]$\\ 
4b&$2$&$C_2$& $(1/2,0,z)$, $(1/2,1/2,z)$ & $A_c A_x A_{y+z}$ & $\varphi_2[T_3M, T_1C_2]$\\ 
4c&$m$&$C_s$& $(x,1/4,z)$, $(-x,3/4,z)$ & $A_c A_x (A_m + A_{y+z})$ & $\widetilde{\varphi_3}[T_1, T_2C_2M, M]$\\ 
\hline
\hline 
 \end{tabular} 
 \end{center}
 The expression of $\widetilde{\varphi_3}$ is given in Eq.~\eqref{wilde3}.

\subsection*{No. 40: $Ama2$}\label{subsub:sg40}

This group is generated by three translations $T_{1,2,3}$ as given in Eqs.~\eqref{TransBravaisA}, a two-fold rotation $C_2$, and a glide $G$:
\begin{subequations}
 \begin{align}
C_2 &\colon (x,y,z)\rightarrow (-x, -y, z),\\ 
G &\colon (x,y,z)\rightarrow (x + 1/2, -y, z).
\end{align}
\end{subequations}

The $\mathbb{Z}_2$ cohomology ring is given by

\begin{equation}
\mathbb{Z}_2[A_c,A_m,A_{y+z},B_{yz},B_{x(y+z)}]/\langle\mathcal{R}_2,\mathcal{R}_3,\mathcal{R}_4\rangle
 \end{equation}
where the relations are 
\begin{subequations} 
 \begin{align}
\mathcal{R}_2\colon & ~~
(A_c + A_m) A_{y+z},~~A_m (A_c + A_m),~~A_{y+z}^2,\\
\mathcal{R}_3\colon & ~~
A_{y+z} B_{yz},~~(A_c + A_m) B_{x(y+z)},~~A_{y+z} B_{x(y+z)} + A_m B_{yz},\\
\mathcal{R}_4\colon & ~~
B_{yz}^2,~~B_{x(y+z)} B_{yz},~~B_{x(y+z)}^2.
\end{align} 
 \end{subequations}
We have the following table regarding IWPs and group cohomology at degree 3.
\begin{center}
\begin{tabular}{c|cc|c|c|c}\hline\hline {Wyckoff}&\multicolumn{2}{c|}{Little group}& {Coordinates}&\multirow{2}{*}{LSM anomaly class}&\multirow{2}{*}{Topo. inv.} \\ \cline{2-4} position & Intl. & Sch\"{o}nflies & $(0,0,0) + ~(0,1/2,1/2) + $ & &\\ \hline
4a&$2$&$C_2$& $(0,0,z)$, $(1/2,0,z)$ & $(A_c + A_m) B_{yz}$ & $\varphi_2[T_2T_3, C_2]$\\ 
4b&$m$&$C_s$& $(1/4,y,z)$, $(3/4,-y,z)$ & $A_m B_{yz}$ & $\varphi_3[T_2, T_3, C_2G]$\\ 
\hline
\hline 
 \end{tabular} 
 \end{center}

\subsection*{No. 41: $Aea2$}\label{subsub:sg41}

This group is generated by three translations $T_{1,2,3}$ as given in Eqs.~\eqref{TransBravaisA}, a two-fold rotation $C_2$, and a glide $G$:
\begin{subequations}
 \begin{align}
C_2 &\colon (x,y,z)\rightarrow (-x, -y, z),\\ 
G &\colon (x,y,z)\rightarrow (x + 1/2, -y + 1/2, z).
\end{align}
\end{subequations}

The $\mathbb{Z}_2$ cohomology ring is given by

\begin{equation}
\mathbb{Z}_2[A_c,A_m,A_{y+z},C_\gamma]/\langle\mathcal{R}_2,\mathcal{R}_4,\mathcal{R}_6\rangle
 \end{equation}
where the relations are 
\begin{subequations} 
 \begin{align}
\mathcal{R}_2\colon & ~~
A_c A_m + A_c A_{y+z} + A_m A_{y+z},~~A_m (A_c + A_m),~~A_{y+z}^2,\\
\mathcal{R}_4\colon & ~~
A_m C_\gamma,~~A_{y+z} C_\gamma,\\
\mathcal{R}_6\colon & ~~
C_\gamma^2.
\end{align} 
 \end{subequations}
We have the following table regarding IWPs and group cohomology at degree 3.
\begin{center}
\begin{tabular}{c|cc|c|c|c}\hline\hline {Wyckoff}&\multicolumn{2}{c|}{Little group}& {Coordinates}&\multirow{2}{*}{LSM anomaly class}&\multirow{2}{*}{Topo. inv.} \\ \cline{2-4} position & Intl. & Sch\"{o}nflies & $(0,0,0) + ~(0,1/2,1/2) + $ & &\\ \hline
4a&$2$&$C_2$& $(0,0,z)$, $(1/2,1/2,z)$ & $C_\gamma$ & $\varphi_2[T_2T_3, C_2]$\\ 
\hline
\hline 
 \end{tabular} 
 \end{center}

\subsection*{No. 42: $Fmm2$}\label{subsub:sg42}

This group is generated by three translations $T_{1,2,3}$ as given in Eqs.~\eqref{TransBravaisF}, a two-fold rotation $C_2$, and a mirror $M$:
\begin{subequations}
 \begin{align}
C_2 &\colon (x,y,z)\rightarrow (-x, -y, z),\\ 
M &\colon (x,y,z)\rightarrow (x, -y, z).
\end{align}
\end{subequations}

The $\mathbb{Z}_2$ cohomology ring is given by

\begin{equation}
\mathbb{Z}_2[A_c,A_m,A_{x+z},A_{y+z},C_{xyz}]/\langle\mathcal{R}_2,\mathcal{R}_4,\mathcal{R}_6\rangle
 \end{equation}
where the relations are 
\begin{subequations} 
 \begin{align}
\mathcal{R}_2\colon & ~~
A_c A_{x+z} + A_m A_{x+z} + A_c A_{y+z},~~A_{x+z}^2 + A_c A_{y+z},~~A_{y+z} (A_c + A_{y+z}),\\
\mathcal{R}_4\colon & ~~
A_{x+z} C_{xyz},~~A_{y+z} C_{xyz},\\
\mathcal{R}_6\colon & ~~
C_{xyz}^2.
\end{align} 
 \end{subequations}
We have the following table regarding IWPs and group cohomology at degree 3.
\begin{center}
\begin{tabular}{c|cc|c|c|c}\hline\hline \multirow{3}{*}{\shortstack[l]{Wyckoff\\position}}&\multicolumn{2}{c|}{Little group}& {Coordinates}&\multirow{3}{*}{LSM anomaly class}&\multirow{3}{*}{Topo. inv.} \\ \cline{2-4} & \multirow{2}{*}{Intl.} & \multirow{2}{*}{Sch\"{o}nflies} & $(0,0,0) + ~(0,1/2,1/2) + $ & & \\ & & & $ (1/2,0,1/2) + ~(1/2,1/2,0) +$ & &\\ \hline
4a&$mm2$&$C_{2v}$& $(0,0,z)$ & $C_{xyz}$ & $\varphi_2[T_1T_2T_3^{-1}, C_2]$\\ 
8b&$2$&$C_2$& $(1/4,1/4,z)$, $(1/4,3/4,z)$ & $A_c A_m A_{y+z}$ & $\varphi_2[T_1M, T_3C_2]$\\ 
\hline
\hline 
 \end{tabular} 
 \end{center}

\subsection*{No. 43: $Fdd2$}\label{subsub:sg43}

This group is generated by three translations $T_{1,2,3}$ as given in Eqs.~\eqref{TransBravaisF}, a two-fold rotation $C_2$, and a glide $G$:
\begin{subequations}
 \begin{align}
C_2 &\colon (x,y,z)\rightarrow (-x, -y, z),\\ 
G &\colon (x,y,z)\rightarrow (x + 1/4, -y + 1/4, z + 1/4).
\end{align}
\end{subequations}

The $\mathbb{Z}_2$ cohomology ring is given by

\begin{equation}
\mathbb{Z}_2[A_c,A_m,B_{xy+xz+yz},C_\gamma]/\langle\mathcal{R}_2,\mathcal{R}_3,\mathcal{R}_4,\mathcal{R}_5,\mathcal{R}_6\rangle
 \end{equation}
where the relations are 
\begin{subequations} 
 \begin{align}
\mathcal{R}_2\colon & ~~
A_c A_m,~~A_m^2,\\
\mathcal{R}_3\colon & ~~
A_c B_{xy+xz+yz},~~A_m B_{xy+xz+yz},\\
\mathcal{R}_4\colon & ~~
A_m C_\gamma,~~B_{xy+xz+yz}^2,\\
\mathcal{R}_5\colon & ~~
B_{xy+xz+yz} C_\gamma,\\
\mathcal{R}_6\colon & ~~
C_\gamma^2.
\end{align} 
 \end{subequations}
We have the following table regarding IWPs and group cohomology at degree 3.
\begin{center}
\begin{tabular}{c|cc|c|c|c}\hline\hline \multirow{3}{*}{\shortstack[l]{Wyckoff\\position}}&\multicolumn{2}{c|}{Little group}& {Coordinates}&\multirow{3}{*}{LSM anomaly class}&\multirow{3}{*}{Topo. inv.} \\ \cline{2-4} & \multirow{2}{*}{Intl.} & \multirow{2}{*}{Sch\"{o}nflies} & $(0,0,0) + ~(0,1/2,1/2) + $ & & \\ & & & $ (1/2,0,1/2) + ~(1/2,1/2,0) +$ & &\\ \hline
8a&$2$&$C_2$& $(0,0,z)$, $(1/4,1/4,z+1/4)$ & $C_\gamma$ & $\varphi_2[T_1T_2T_3^{-1}, C_2]$\\ 
\hline
\hline 
 \end{tabular} 
 \end{center}

\subsection*{No. 44: $Imm2$}\label{subsub:sg44}

This group is generated by three translations $T_{1,2,3}$ as given in Eqs.~\eqref{TransBravaisI}, a two-fold rotation $C_2$, and a mirror $M$:
\begin{subequations}
 \begin{align}
C_2 &\colon (x,y,z)\rightarrow (-x, -y, z),\\ 
M &\colon (x,y,z)\rightarrow (x, -y, z).
\end{align}
\end{subequations}

The $\mathbb{Z}_2$ cohomology ring is given by

\begin{equation}
\mathbb{Z}_2[A_c,A_m,A_{x+y+z},B_\beta,B_{x(y+z)},B_{y(x+z)},C_{xyz}]/\langle\mathcal{R}_2,\mathcal{R}_3,\mathcal{R}_4,\mathcal{R}_5,\mathcal{R}_6\rangle
 \end{equation}
where the relations are 
\begin{subequations} 
 \begin{align}
\mathcal{R}_2\colon & ~~
A_c A_{x+y+z},~~A_m A_{x+y+z},~~A_{x+y+z}^2,\\
\mathcal{R}_3\colon & ~~
A_{x+y+z} B_\beta,~~A_{x+y+z} B_{x(y+z)},~~A_c B_\beta + A_m B_\beta + A_c B_{x(y+z)} + A_c B_{y(x+z)} + A_m B_{y(x+z)},~~A_{x+y+z} B_{y(x+z)},\\
\mathcal{R}_4\colon & ~~
A_{x+y+z} C_{xyz},~~A_c^2 B_\beta + A_c A_m B_\beta + B_\beta^2 + A_c A_m B_{x(y+z)},~~B_\beta B_{x(y+z)} + A_c C_{xyz} + A_m C_{xyz},\nonumber\\
&~~A_c^2 B_\beta + A_c A_m B_\beta + A_c A_m B_{x(y+z)} + B_\beta B_{y(x+z)} + A_c C_{xyz},~~B_{x(y+z)}^2,~~B_{x(y+z)} B_{y(x+z)} + A_c C_{xyz} + A_m C_{xyz},\nonumber\\&
~~A_c^2 B_\beta + A_c A_m B_\beta + A_c A_m B_{x(y+z)} + B_{y(x+z)}^2,\\
\mathcal{R}_5\colon & ~~
(A_c^2 + A_c A_m + B_\beta) C_{xyz},~~B_{x(y+z)} C_{xyz},~~(A_c^2 + A_c A_m + B_{y(x+z)}) C_{xyz},\\
\mathcal{R}_6\colon & ~~
C_{xyz}^2.
\end{align} 
 \end{subequations}
We have the following table regarding IWPs and group cohomology at degree 3.
\begin{center}
\begin{tabular}{c|cc|c|c|c}\hline\hline {Wyckoff}&\multicolumn{2}{c|}{Little group}& {Coordinates}&\multirow{2}{*}{LSM anomaly class}&\multirow{2}{*}{Topo. inv.} \\ \cline{2-4} position & Intl. & Sch\"{o}nflies & $(0,0,0) + ~(1/2,1/2,1/2) + $ & &\\ \hline
2a&$mm2$&$C_{2v}$& $(0,0,z)$ & $A_c B_{x(y+z)} + C_{xyz}$ & $\varphi_2[T_1T_2, C_2]$\\ 
2b&$mm2$&$C_{2v}$& $(0,1/2,z)$ & $C_{xyz}$ & $\varphi_2[T_1T_2, T_1T_3C_2]$\\ 
\hline
\hline 
 \end{tabular} 
 \end{center}

\subsection*{No. 45: $Iba2$}\label{subsub:sg45}

This group is generated by three translations $T_{1,2,3}$ as given in Eqs.~\eqref{TransBravaisI}, a two-fold rotation $C_2$, and a glide $G$:
\begin{subequations}
 \begin{align}
C_2 &\colon (x,y,z)\rightarrow (-x, -y, z),\\ 
G &\colon (x,y,z)\rightarrow (x, -y, z + 1/2).
\end{align}
\end{subequations}

The $\mathbb{Z}_2$ cohomology ring is given by

\begin{equation}
\mathbb{Z}_2[A_c,A_m,A_{x+y+z},B_{z(x+y)}]/\langle\mathcal{R}_2,\mathcal{R}_3,\mathcal{R}_4\rangle
 \end{equation}
where the relations are 
\begin{subequations} 
 \begin{align}
\mathcal{R}_2\colon & ~~
A_m A_{x+y+z},~~A_m^2 + A_c A_{x+y+z},~~A_{x+y+z} (A_c + A_{x+y+z}),\\
\mathcal{R}_3\colon & ~~
A_{x+y+z} B_{z(x+y)},\\
\mathcal{R}_4\colon & ~~
B_{z(x+y)} (A_c^2 + A_c A_m + B_{z(x+y)}).
\end{align} 
 \end{subequations}
We have the following table regarding IWPs and group cohomology at degree 3.
\begin{center}
\begin{tabular}{c|cc|c|c|c}\hline\hline {Wyckoff}&\multicolumn{2}{c|}{Little group}& {Coordinates}&\multirow{2}{*}{LSM anomaly class}&\multirow{2}{*}{Topo. inv.} \\ \cline{2-4} position & Intl. & Sch\"{o}nflies & $(0,0,0) + ~(1/2,1/2,1/2) + $ & &\\ \hline
4a&$2$&$C_2$& $(0,0,z)$, $(1/2,1/2,z)$ & $A_m (A_c^2 + B_{z(x+y)})$ & $\varphi_2[G, C_2]$\\ 
4b&$2$&$C_2$& $(0,1/2,z)$, $(1/2,0,z)$ & $A_m B_{z(x+y)}$ & $\varphi_2[G, T_2T_3C_2]$\\ 
\hline
\hline 
 \end{tabular} 
 \end{center}

\subsection*{No. 46: $Ima2$}\label{subsub:sg46}

This group is generated by three translations $T_{1,2,3}$ as given in Eqs.~\eqref{TransBravaisI}, a two-fold rotation $C_2$, and a glide $G$:
\begin{subequations}
 \begin{align}
C_2 &\colon (x,y,z)\rightarrow (-x, -y, z),\\ 
G &\colon (x,y,z)\rightarrow (x + 1/2, -y, z).
\end{align}
\end{subequations}

The $\mathbb{Z}_2$ cohomology ring is given by

\begin{equation}
\mathbb{Z}_2[A_c,A_m,A_{x+y+z},B_{x(y+z)},B_{y(x+z)}]/\langle\mathcal{R}_2,\mathcal{R}_3,\mathcal{R}_4\rangle
 \end{equation}
where the relations are 
\begin{subequations} 
 \begin{align}
\mathcal{R}_2\colon & ~~
(A_c + A_m) A_{x+y+z},~~A_c A_m + A_m^2 + A_c A_{x+y+z},~~A_{x+y+z}^2,\\
\mathcal{R}_3\colon & ~~
A_{x+y+z} B_{x(y+z)},~~A_c^3 + A_c A_m^2 + A_c B_{x(y+z)} + A_c B_{y(x+z)} + A_m B_{y(x+z)},~~A_m B_{x(y+z)} + A_{x+y+z} B_{y(x+z)},\\
\mathcal{R}_4\colon & ~~
B_{x(y+z)}^2,~~B_{x(y+z)} (A_c^2 + A_c A_m + B_{y(x+z)}),~~A_c^4 + A_c^2 A_m^2 + A_c A_m B_{x(y+z)} + B_{y(x+z)}^2.
\end{align} 
 \end{subequations}
We have the following table regarding IWPs and group cohomology at degree 3.
\begin{center}
\begin{tabular}{c|cc|c|c|c}\hline\hline {Wyckoff}&\multicolumn{2}{c|}{Little group}& {Coordinates}&\multirow{2}{*}{LSM anomaly class}&\multirow{2}{*}{Topo. inv.} \\ \cline{2-4} position & Intl. & Sch\"{o}nflies & $(0,0,0) + ~(1/2,1/2,1/2) + $ & &\\ \hline
4a&$2$&$C_2$& $(0,0,z)$, $(1/2,0,z)$ & $(A_c + A_m) B_{x(y+z)}$ & $\varphi_2[T_1T_2, C_2]$\\ 
4b&$m$&$C_s$& $(1/4,y,z)$, $(3/4,-y,z)$ & $A_m B_{x(y+z)}$ & $\widetilde{\varphi_3}[T_1T_3, T_1G, C_2G]$\\ 
\hline
\hline 
 \end{tabular} 
 \end{center}
 The expression of $\widetilde{\varphi_3}$ is given in Eq.~\eqref{wilde3}.

\subsection*{No. 47: $Pmmm$}\label{subsub:sg47}

This group is generated by three translations $T_{1,2,3}$ as given in Eqs.~\eqref{TransBravaisP}, a two-fold rotation $C_2$, a two-fold rotation $C'_2$, and an inversion $I$:
\begin{subequations}
 \begin{align}
C_2 &\colon (x,y,z)\rightarrow (-x, -y, z),\\ 
C'_2 &\colon (x,y,z)\rightarrow (-x, y, -z),\\ 
I &\colon (x,y,z)\rightarrow (-x, -y, -z).
\end{align}
\end{subequations}

The $\mathbb{Z}_2$ cohomology ring is given by

\begin{equation}
\mathbb{Z}_2[A_i,A_{c'},A_c,A_x,A_y,A_z]/\langle\mathcal{R}_2\rangle
 \end{equation}
where the relations are 
\begin{subequations} 
 \begin{align}
\mathcal{R}_2\colon & ~~
A_x (A_c + A_{c'} + A_i + A_x),~~A_y (A_c + A_i + A_y),~~A_z (A_{c'} + A_i + A_z).
\end{align} 
 \end{subequations}
We have the following table regarding IWPs and group cohomology at degree 3.
\begin{center}
\resizebox{\columnwidth}{!}{
\begin{tabular}{c|cc|c|c|c}\hline\hline {Wyckoff}&\multicolumn{2}{c|}{Little group}& \multirow{2}{*}{Coordinates}&\multirow{2}{*}{LSM anomaly class}&\multirow{2}{*}{Topo. inv.} \\ \cline{2-3} position & Intl. & Sch\"{o}nflies & & & \\ \hline
1a&$mmm$&$D_{2h}$& $(0,0,0)$ & $(A_c + A_{c'} + A_i + A_x) (A_c + A_i + A_y) (A_{c'} + A_i + A_z)$ & $\varphi_2[C_2, C'_2]$\\ 
1b&$mmm$&$D_{2h}$& $(1/2,0,0)$ & $A_x (A_c + A_i + A_y) (A_{c'} + A_i + A_z)$ & $\varphi_2[T_1C_2, T_1C'_2]$\\ 
1c&$mmm$&$D_{2h}$& $(0,0,1/2)$ & $(A_c + A_{c'} + A_i + A_x) (A_c + A_i + A_y) A_z$ & $\varphi_2[C_2, T_3C'_2]$\\ 
1d&$mmm$&$D_{2h}$& $(1/2,0,1/2)$ & $A_x (A_c + A_i + A_y) A_z$ & $\varphi_2[T_1C_2, T_1T_3C'_2]$\\ 
1e&$mmm$&$D_{2h}$& $(0,1/2,0)$ & $(A_c + A_{c'} + A_i + A_x) A_y (A_{c'} + A_i + A_z)$ & $\varphi_2[T_2C_2, C'_2]$\\ 
1f&$mmm$&$D_{2h}$& $(1/2,1/2,0)$ & $A_x A_y (A_{c'} + A_i + A_z)$ & $\varphi_2[T_1T_2C_2, T_1C'_2]$\\ 
1g&$mmm$&$D_{2h}$& $(0,1/2,1/2)$ & $(A_c + A_{c'} + A_i + A_x) A_y A_z$ & $\varphi_2[T_2C_2, T_3C'_2]$\\ 
1h&$mmm$&$D_{2h}$& $(1/2,1/2,1/2)$ & $A_x A_y A_z$ & $\varphi_2[T_1T_2C_2, T_1T_3C'_2]$\\ 
\hline
\hline 
 \end{tabular} }
 \end{center}

\subsection*{No. 48: $Pnnn$}\label{subsub:sg48}

This group is generated by three translations $T_{1,2,3}$ as given in Eqs.~\eqref{TransBravaisP}, a two-fold rotation $C_2$, a two-fold rotation $C'_2$, and an inversion $I$:
\begin{subequations}
 \begin{align}
C_2 &\colon (x,y,z)\rightarrow (-x + 1/2, -y + 1/2, z),\\ 
C'_2 &\colon (x,y,z)\rightarrow (-x + 1/2, y, -z + 1/2),\\ 
I &\colon (x,y,z)\rightarrow (-x, -y, -z).
\end{align}
\end{subequations}

The $\mathbb{Z}_2$ cohomology ring is given by

\begin{equation}
\mathbb{Z}_2[A_i,A_{c'},A_c,A_{x+y+z}]/\langle\mathcal{R}_2,\mathcal{R}_3,\mathcal{R}_4\rangle
 \end{equation}
where the relations are 
\begin{subequations} 
 \begin{align}
\mathcal{R}_2\colon & ~~
A_{c'} A_i,~~A_c A_i,\\
\mathcal{R}_3\colon & ~~
A_i A_{x+y+z} (A_i + A_{x+y+z}),\\
\mathcal{R}_4\colon & ~~
A_{x+y+z} (A_c^2 A_{c'} + A_c A_{c'}^2 + A_i^3 + A_c^2 A_{x+y+z} + A_c A_{c'} A_{x+y+z} + A_{c'}^2 A_{x+y+z} + A_{x+y+z}^3).
\end{align} 
 \end{subequations}
We have the following table regarding IWPs and group cohomology at degree 3.
\begin{center}
\resizebox{\columnwidth}{!}{
\begin{tabular}{c|cc|c|c|c}\hline\hline {Wyckoff}&\multicolumn{2}{c|}{Little group}& \multirow{2}{*}{Coordinates}&\multirow{2}{*}{LSM anomaly class}&\multirow{2}{*}{Topo. inv.} \\ \cline{2-3} position & Intl. & Sch\"{o}nflies & & & \\ \hline
2a&$222$&$D_2$& $(1/4,1/4,1/4)$, $(3/4,3/4,3/4)$ & $A_c^2 A_{c'} + A_c A_{c'}^2 + A_c^2 A_{x+y+z} + A_c A_{c'} A_{x+y+z} + A_{c'}^2 A_{x+y+z} + A_i^2 A_{x+y+z} + A_{x+y+z}^3$ & $\varphi_2[C_2, C'_2]$\\ 
2b&$222$&$D_2$& $(3/4,1/4,1/4)$, $(1/4,3/4,3/4)$ & $A_{x+y+z} (A_c A_{c'} + A_i^2 + A_c A_{x+y+z} + A_{c'} A_{x+y+z} + A_{x+y+z}^2)$ & $\varphi_2[T_1C_2, T_1C'_2]$\\ 
2c&$222$&$D_2$& $(1/4,1/4,3/4)$, $(3/4,3/4,1/4)$ & $A_{x+y+z} (A_c^2 + A_c A_{c'} + A_i^2 + A_{c'} A_{x+y+z} + A_{x+y+z}^2)$ & $\varphi_2[C_2, T_3C'_2]$\\ 
2d&$222$&$D_2$& $(1/4,3/4,1/4)$, $(3/4,1/4,3/4)$ & $A_{x+y+z} (A_c A_{c'} + A_{c'}^2 + A_i^2 + A_c A_{x+y+z} + A_{x+y+z}^2)$ & $\varphi_2[T_1C_2, T_1T_3C'_2]$\\ 
\hline
\multirow{2}{*}{4e} & \multirow{2}{*}{$\overline{1}$} & \multirow{2}{*}{$C_i$} & $(1/2,1/2,1/2)$, $(0,0,1/2)$, & \multirow{2}{*}{$A_i^2 A_{x+y+z}$} & \multirow{2}{*}{$\varphi_1[T_1T_2T_3I]$}\\
& & & $(0,1/2,0)$, $(1/2,0,0)$ & & \\ \hline
\multirow{2}{*}{4f} & \multirow{2}{*}{$\overline{1}$} & \multirow{2}{*}{$C_i$} & $(0,0,0)$, $(1/2,1/2,0)$, & \multirow{2}{*}{$A_i^2 (A_i + A_{x+y+z})$} & \multirow{2}{*}{$\varphi_1[I]$}\\
& & & $(1/2,0,1/2)$, $(0,1/2,1/2)$ & & \\ \hline
\hline 
 \end{tabular} }
 \end{center}

\subsection*{No. 49: $Pccm$}\label{subsub:sg49}

This group is generated by three translations $T_{1,2,3}$ as given in Eqs.~\eqref{TransBravaisP}, a two-fold rotation $C_2$, a two-fold rotation $C'_2$, and an inversion $I$:
\begin{subequations}
 \begin{align}
C_2 &\colon (x,y,z)\rightarrow (-x, -y, z),\\ 
C'_2 &\colon (x,y,z)\rightarrow (-x, y, -z + 1/2),\\ 
I &\colon (x,y,z)\rightarrow (-x, -y, -z).
\end{align}
\end{subequations}

The $\mathbb{Z}_2$ cohomology ring is given by

\begin{equation}
\mathbb{Z}_2[A_i,A_{c'},A_c,A_x,A_y]/\langle\mathcal{R}_2\rangle
 \end{equation}
where the relations are 
\begin{subequations} 
 \begin{align}
\mathcal{R}_2\colon & ~~
A_{c'} A_i,~~A_x (A_c + A_{c'} + A_i + A_x),~~A_y (A_c + A_i + A_y).
\end{align} 
 \end{subequations}
We have the following table regarding IWPs and group cohomology at degree 3.
\begin{center}
\begin{tabular}{c|cc|c|c|c}\hline\hline {Wyckoff}&\multicolumn{2}{c|}{Little group}& \multirow{2}{*}{Coordinates}&\multirow{2}{*}{LSM anomaly class}&\multirow{2}{*}{Topo. inv.} \\ \cline{2-3} position & Intl. & Sch\"{o}nflies & & & \\ \hline
2a&$2/m$&$C_{2h}$& $(0,0,0)$, $(0,0,1/2)$ & $A_i (A_c + A_i + A_x) (A_c + A_i + A_y)$ & $\varphi_1[I]$\\ 
2b&$2/m$&$C_{2h}$& $(1/2,1/2,0)$, $(1/2,1/2,1/2)$ & $A_i A_x A_y$ & $\varphi_1[T_1T_2I]$\\ 
2c&$2/m$&$C_{2h}$& $(0,1/2,0)$, $(0,1/2,1/2)$ & $A_i (A_c + A_i + A_x) A_y$ & $\varphi_1[T_2I]$\\ 
2d&$2/m$&$C_{2h}$& $(1/2,0,0)$, $(1/2,0,1/2)$ & $A_i A_x (A_c + A_i + A_y)$ & $\varphi_1[T_1I]$\\ 
2e&$222$&$D_2$& $(0,0,1/4)$, $(0,0,3/4)$ & $A_{c'} (A_c + A_{c'} + A_x) (A_c + A_y)$ & $\varphi_2[C_2, C'_2]$\\ 
2f&$222$&$D_2$& $(1/2,0,1/4)$, $(1/2,0,3/4)$ & $A_{c'} A_x (A_c + A_y)$ & $\varphi_2[T_1C_2, T_1C'_2]$\\ 
2g&$222$&$D_2$& $(0,1/2,1/4)$, $(0,1/2,3/4)$ & $A_{c'} (A_c + A_{c'} + A_x) A_y$ & $\varphi_2[T_2C_2, C'_2]$\\ 
2h&$222$&$D_2$& $(1/2,1/2,1/4)$, $(1/2,1/2,3/4)$ & $A_{c'} A_x A_y$ & $\varphi_2[T_1T_2C_2, T_1C'_2]$\\ 
\hline
\hline 
 \end{tabular} 
 \end{center}

\subsection*{No. 50: $Pban$}\label{subsub:sg50}

This group is generated by three translations $T_{1,2,3}$ as given in Eqs.~\eqref{TransBravaisP}, a two-fold rotation $C_2$, a two-fold rotation $C'_2$, and an inversion $I$:
\begin{subequations}
 \begin{align}
C_2 &\colon (x,y,z)\rightarrow (-x + 1/2, -y + 1/2, z),\\ 
C'_2 &\colon (x,y,z)\rightarrow (-x + 1/2, y, -z),\\ 
I &\colon (x,y,z)\rightarrow (-x, -y, -z).
\end{align}
\end{subequations}

The $\mathbb{Z}_2$ cohomology ring is given by

\begin{equation}
\mathbb{Z}_2[A_i,A_{c'},A_c,A_z,B_\beta]/\langle\mathcal{R}_2,\mathcal{R}_3,\mathcal{R}_4\rangle
 \end{equation}
where the relations are 
\begin{subequations} 
 \begin{align}
\mathcal{R}_2\colon & ~~
A_{c'} A_i,~~A_c A_i,~~A_z (A_{c'} + A_i + A_z),\\
\mathcal{R}_3\colon & ~~
A_i B_\beta,\\
\mathcal{R}_4\colon & ~~
B_\beta (A_c^2 + A_c A_{c'} + B_\beta).
\end{align} 
 \end{subequations}
We have the following table regarding IWPs and group cohomology at degree 3.
\begin{center}
\begin{tabular}{c|cc|c|c|c}\hline\hline {Wyckoff}&\multicolumn{2}{c|}{Little group}& \multirow{2}{*}{Coordinates}&\multirow{2}{*}{LSM anomaly class}&\multirow{2}{*}{Topo. inv.} \\ \cline{2-3} position & Intl. & Sch\"{o}nflies & & & \\ \hline
2a&$222$&$D_2$& $(1/4,1/4,0)$, $(3/4,3/4,0)$ & $(A_{c'} + A_z) (A_c^2 + A_c A_{c'} + B_\beta)$ & $\varphi_2[C_2, C'_2]$\\ 
2b&$222$&$D_2$& $(3/4,1/4,0)$, $(1/4,3/4,0)$ & $(A_{c'} + A_z) B_\beta$ & $\varphi_2[T_2C_2, C'_2]$\\ 
2c&$222$&$D_2$& $(3/4,1/4,1/2)$, $(1/4,3/4,1/2)$ & $A_z B_\beta$ & $\varphi_2[T_2C_2, T_3C'_2]$\\ 
2d&$222$&$D_2$& $(1/4,1/4,1/2)$, $(3/4,3/4,1/2)$ & $A_z (A_c^2 + A_c A_{c'} + B_\beta)$ & $\varphi_2[C_2, T_3C'_2]$\\ 
\hline
\multirow{2}{*}{4e} & \multirow{2}{*}{$\overline{1}$} & \multirow{2}{*}{$C_i$} & $(0,0,0)$, $(1/2,1/2,0)$, & \multirow{2}{*}{$A_i^2 (A_i + A_z)$} & \multirow{2}{*}{$\varphi_1[I]$}\\
& & & $(1/2,0,0)$, $(0,1/2,0)$ & & \\ \hline
\multirow{2}{*}{4f} & \multirow{2}{*}{$\overline{1}$} & \multirow{2}{*}{$C_i$} & $(0,0,1/2)$, $(1/2,1/2,1/2)$, & \multirow{2}{*}{$A_i^2 A_z$} & \multirow{2}{*}{$\varphi_1[T_3I]$}\\
& & & $(1/2,0,1/2)$, $(0,1/2,1/2)$ & & \\ \hline
\hline 
 \end{tabular} 
 \end{center}

\subsection*{No. 51: $Pmma$}\label{subsub:sg51}

This group is generated by three translations $T_{1,2,3}$ as given in Eqs.~\eqref{TransBravaisP}, a two-fold rotation $C_2$, a two-fold rotation $C'_2$, and an inversion $I$:
\begin{subequations}
 \begin{align}
C_2 &\colon (x,y,z)\rightarrow (-x + 1/2, -y, z),\\ 
C'_2 &\colon (x,y,z)\rightarrow (-x, y, -z),\\ 
I &\colon (x,y,z)\rightarrow (-x, -y, -z).
\end{align}
\end{subequations}

The $\mathbb{Z}_2$ cohomology ring is given by

\begin{equation}
\mathbb{Z}_2[A_i,A_{c'},A_c,A_y,A_z]/\langle\mathcal{R}_2\rangle
 \end{equation}
where the relations are 
\begin{subequations} 
 \begin{align}
\mathcal{R}_2\colon & ~~
A_c (A_{c'} + A_i),~~A_y (A_c + A_i + A_y),~~A_z (A_{c'} + A_i + A_z).
\end{align} 
 \end{subequations}
We have the following table regarding IWPs and group cohomology at degree 3.
\begin{center}
\begin{tabular}{c|cc|c|c|c}\hline\hline {Wyckoff}&\multicolumn{2}{c|}{Little group}& \multirow{2}{*}{Coordinates}&\multirow{2}{*}{LSM anomaly class}&\multirow{2}{*}{Topo. inv.} \\ \cline{2-3} position & Intl. & Sch\"{o}nflies & & & \\ \hline
2a&$2m$&$C_{2h}$& $(0,0,0)$, $(1/2,0,0)$ & $(A_{c'} + A_i) (A_i + A_y) (A_{c'} + A_i + A_z)$ & $\varphi_1[I]$\\ 
2b&$2m$&$C_{2h}$& $(0,1/2,0)$, $(1/2,1/2,0)$ & $(A_{c'} + A_i) A_y (A_{c'} + A_i + A_z)$ & $\varphi_1[T_2I]$\\ 
2c&$2m$&$C_{2h}$& $(0,0,1/2)$, $(1/2,0,1/2)$ & $(A_{c'} + A_i) (A_i + A_y) A_z$ & $\varphi_1[T_3I]$\\ 
2d&$2m$&$C_{2h}$& $(0,1/2,1/2)$, $(1/2,1/2,1/2)$ & $(A_{c'} + A_i) A_y A_z$ & $\varphi_1[T_2T_3I]$\\ 
2e&$mm2$&$C_{2v}$& $(1/4,0,z)$, $(3/4,0,-z)$ & $A_c (A_c + A_i + A_y) A_z$ & $\varphi_2[T_3, C_2]$\\ 
2f&$mm2$&$C_{2v}$& $(1/4,1/2,z)$, $(3/4,1/2,-z)$ & $A_c A_y A_z$ & $\varphi_2[T_3, T_2C_2]$\\ 
\hline
\hline 
 \end{tabular} 
 \end{center}

\subsection*{No. 52: $Pnna$}\label{subsub:sg52}

This group is generated by three translations $T_{1,2,3}$ as given in Eqs.~\eqref{TransBravaisP}, a two-fold rotation $C_2$, a two-fold screw $S'_2$, and an inversion $I$:
\begin{subequations}
 \begin{align}
C_2 &\colon (x,y,z)\rightarrow (-x + 1/2, -y, z),\\ 
S'_2 &\colon (x,y,z)\rightarrow (-x + 1/2, y + 1/2, -z + 1/2),\\ 
I &\colon (x,y,z)\rightarrow (-x, -y, -z).
\end{align}
\end{subequations}

The $\mathbb{Z}_2$ cohomology ring is given by

\begin{equation}
\mathbb{Z}_2[A_i,A_{c'},A_c,B_{\beta 1},B_{\beta 2}]/\langle\mathcal{R}_2,\mathcal{R}_3,\mathcal{R}_4\rangle
 \end{equation}
where the relations are 
\begin{subequations} 
 \begin{align}
\mathcal{R}_2\colon & ~~
A_{c'} A_i,~~A_c A_i,~~A_{c'} (A_c + A_{c'}),\\
\mathcal{R}_3\colon & ~~
A_{c'} B_{\beta 1},~~A_i (B_{\beta 1} + B_{\beta 2}),~~(A_c + A_{c'}) B_{\beta 2},\\
\mathcal{R}_4\colon & ~~
B_{\beta 1} (A_i^2 + B_{\beta 1}),~~B_{\beta 1} (A_i^2 + B_{\beta 2}),~~A_i^2 B_{\beta 1} + B_{\beta 2}^2.
\end{align} 
 \end{subequations}
We have the following table regarding IWPs and group cohomology at degree 3.
\begin{center}
\begin{tabular}{c|cc|c|c|c}\hline\hline {Wyckoff}&\multicolumn{2}{c|}{Little group}& \multirow{2}{*}{Coordinates}&\multirow{2}{*}{LSM anomaly class}&\multirow{2}{*}{Topo. inv.} \\ \cline{2-3} position & Intl. & Sch\"{o}nflies & & & \\ \hline
\multirow{2}{*}{4a} & \multirow{2}{*}{$\overline{1}$} & \multirow{2}{*}{$C_i$} & $(0,0,0)$, $(1/2,0,0)$, & \multirow{2}{*}{$A_i B_{\beta 1}$} & \multirow{2}{*}{$\varphi_1[I]$}\\
& & & $(1/2,1/2,1/2)$, $(0,1/2,1/2)$ & & \\ \hline
\multirow{2}{*}{4b} & \multirow{2}{*}{$\overline{1}$} & \multirow{2}{*}{$C_i$} & $(0,0,1/2)$, $(1/2,0,1/2)$, & \multirow{2}{*}{$A_i (A_i^2 + B_{\beta 1})$} & \multirow{2}{*}{$\varphi_1[T_2I]$}\\
& & & $(1/2,1/2,0)$, $(0,1/2,0)$ & & \\ \hline
\multirow{2}{*}{4c} & \multirow{2}{*}{$2$} & \multirow{2}{*}{$C_2$} & $(1/4,0,z)$, $(1/4,1/2,-z+1/2)$, & \multirow{2}{*}{$A_c B_{\beta 1}$} & \multirow{2}{*}{$\varphi_2[T_3, C_2]$}\\
& & & $(3/4,0,-z)$, $(3/4,1/2,z+1/2)$ & & \\ \hline
\multirow{2}{*}{4d} & \multirow{2}{*}{$2$} & \multirow{2}{*}{$C_2$} & $(x,1/4,1/4)$, $(-x+1/2,3/4,1/4)$, & \multirow{2}{*}{$A_{c'} B_{\beta 2}$} & \multirow{2}{*}{$\varphi_2[T_1, C_2S'_2]$}\\
& & & $(-x,3/4,3/4)$, $(x+1/2,1/4,3/4)$ & & \\ \hline
\hline 
 \end{tabular} 
 \end{center}

\subsection*{No. 53: $Pmna$}\label{subsub:sg53}

This group is generated by three translations $T_{1,2,3}$ as given in Eqs.~\eqref{TransBravaisP}, a two-fold screw $S_2$, a two-fold rotation $C'_2$, and an inversion $I$:
\begin{subequations}
 \begin{align}
S_2 &\colon (x,y,z)\rightarrow (-x + 1/2, -y, z + 1/2),\\ 
C'_2 &\colon (x,y,z)\rightarrow (-x + 1/2, y, -z + 1/2),\\ 
I &\colon (x,y,z)\rightarrow (-x, -y, -z).
\end{align}
\end{subequations}

The $\mathbb{Z}_2$ cohomology ring is given by

\begin{equation}
\mathbb{Z}_2[A_i,A_{c'},A_c,A_y,B_\beta]/\langle\mathcal{R}_2,\mathcal{R}_3,\mathcal{R}_4\rangle
 \end{equation}
where the relations are 
\begin{subequations} 
 \begin{align}
\mathcal{R}_2\colon & ~~
(A_c + A_{c'}) A_i,~~A_c (A_c + A_{c'}),~~A_y (A_c + A_i + A_y),\\
\mathcal{R}_3\colon & ~~
(A_c + A_{c'}) B_\beta,\\
\mathcal{R}_4\colon & ~~
B_\beta (A_{c'} A_i + A_i^2 + B_\beta).
\end{align} 
 \end{subequations}
We have the following table regarding IWPs and group cohomology at degree 3.
\begin{center}
\begin{tabular}{c|cc|c|c|c}\hline\hline {Wyckoff}&\multicolumn{2}{c|}{Little group}& \multirow{2}{*}{Coordinates}&\multirow{2}{*}{LSM anomaly class}&\multirow{2}{*}{Topo. inv.} \\ \cline{2-3} position & Intl. & Sch\"{o}nflies & & & \\ \hline
2a&$2/m$&$C_{2h}$& $(0,0,0)$, $(1/2,0,1/2)$ & $(A_{c'} + A_i + A_y) B_\beta$ & $\varphi_1[I]$\\ 
2b&$2/m$&$C_{2h}$& $(1/2,0,0)$, $(0,0,1/2)$ & $(A_{c'} + A_i + A_y) (A_{c'} A_i + A_i^2 + B_\beta)$ & $\varphi_1[T_1I]$\\ 
2c&$2/m$&$C_{2h}$& $(1/2,1/2,0)$, $(0,1/2,1/2)$ & $A_y (A_{c'} A_i + A_i^2 + B_\beta)$ & $\varphi_1[T_1T_2I]$\\ 
2d&$2/m$&$C_{2h}$& $(0,1/2,0)$, $(1/2,1/2,1/2)$ & $A_y B_\beta$ & $\varphi_1[T_2I]$\\ 
\hline
\multirow{2}{*}{4g} & \multirow{2}{*}{$2$} & \multirow{2}{*}{$C_2$} & $(1/4,y,1/4)$, $(1/4,-y,3/4)$, & \multirow{2}{*}{$A_{c'} (A_c + A_{c'}) A_y$} & \multirow{2}{*}{$\varphi_2[T_2, C'_2]$}\\
& & & $(3/4,-y,3/4)$, $(3/4,y,1/4)$ & & \\ \hline
\hline 
 \end{tabular} 
 \end{center}

\subsection*{No. 54: $Pcca$}\label{subsub:sg54}

This group is generated by three translations $T_{1,2,3}$ as given in Eqs.~\eqref{TransBravaisP}, a two-fold rotation $C_2$, a two-fold rotation $C'_2$, and an inversion $I$:
\begin{subequations}
 \begin{align}
C_2 &\colon (x,y,z)\rightarrow (-x + 1/2, -y, z),\\ 
C'_2 &\colon (x,y,z)\rightarrow (-x, y, -z + 1/2),\\ 
I &\colon (x,y,z)\rightarrow (-x, -y, -z).
\end{align}
\end{subequations}

The $\mathbb{Z}_2$ cohomology ring is given by

\begin{equation}
\mathbb{Z}_2[A_i,A_{c'},A_c,A_y]/\langle\mathcal{R}_2\rangle
 \end{equation}
where the relations are 
\begin{subequations} 
 \begin{align}
\mathcal{R}_2\colon & ~~
A_{c'} A_i,~~A_c (A_{c'} + A_i),~~A_y (A_c + A_i + A_y).
\end{align} 
 \end{subequations}
We have the following table regarding IWPs and group cohomology at degree 3.
\begin{center}
\begin{tabular}{c|cc|c|c|c}\hline\hline {Wyckoff}&\multicolumn{2}{c|}{Little group}& \multirow{2}{*}{Coordinates}&\multirow{2}{*}{LSM anomaly class}&\multirow{2}{*}{Topo. inv.} \\ \cline{2-3} position & Intl. & Sch\"{o}nflies & & & \\ \hline
\multirow{2}{*}{4a} & \multirow{2}{*}{$\overline{1}$} & \multirow{2}{*}{$C_i$} & $(0,0,0)$, $(1/2,0,0)$, & \multirow{2}{*}{$A_i^2 (A_i + A_y)$} & \multirow{2}{*}{$\varphi_1[I]$}\\
& & & $(0,0,1/2)$, $(1/2,0,1/2)$ & & \\ \hline
\multirow{2}{*}{4b} & \multirow{2}{*}{$\overline{1}$} & \multirow{2}{*}{$C_i$} & $(0,1/2,0)$, $(1/2,1/2,0)$, & \multirow{2}{*}{$A_i^2 A_y$} & \multirow{2}{*}{$\varphi_1[T_2I]$}\\
& & & $(0,1/2,1/2)$, $(1/2,1/2,1/2)$ & & \\ \hline
\multirow{2}{*}{4c} & \multirow{2}{*}{$2$} & \multirow{2}{*}{$C_2$} & $(0,y,1/4)$, $(1/2,-y,1/4)$, & \multirow{2}{*}{$A_{c'}^2 A_y$} & \multirow{2}{*}{$\varphi_2[T_2, C'_2]$}\\
& & & $(0,-y,3/4)$, $(1/2,y,3/4)$ & & \\ \hline
\multirow{2}{*}{4d} & \multirow{2}{*}{$2$} & \multirow{2}{*}{$C_2$} & $(1/4,0,z)$, $(3/4,0,-z+1/2)$, & \multirow{2}{*}{$A_c A_i (A_c + A_y)$} & \multirow{2}{*}{$\varphi_2[C'_2I, C_2]$}\\
& & & $(3/4,0,-z)$, $(1/4,0,z+1/2)$ & & \\ \hline
\multirow{2}{*}{4e} & \multirow{2}{*}{$2$} & \multirow{2}{*}{$C_2$} & $(1/4,1/2,z)$, $(3/4,1/2,-z+1/2)$, & \multirow{2}{*}{$A_c A_i A_y$} & \multirow{2}{*}{$\varphi_2[T_2C'_2I, T_2C_2]$}\\
& & & $(3/4,1/2,-z)$, $(1/4,1/2,z+1/2)$ & & \\ \hline
\hline 
 \end{tabular} 
 \end{center}

\subsection*{No. 55: $Pbam$}\label{subsub:sg55}

This group is generated by three translations $T_{1,2,3}$ as given in Eqs.~\eqref{TransBravaisP}, a two-fold rotation $C_2$, a two-fold screw $S'_2$, and an inversion $I$:
\begin{subequations}
 \begin{align}
C_2 &\colon (x,y,z)\rightarrow (-x, -y, z),\\ 
S'_2 &\colon (x,y,z)\rightarrow (-x + 1/2, y + 1/2, -z),\\ 
I &\colon (x,y,z)\rightarrow (-x, -y, -z).
\end{align}
\end{subequations}

The $\mathbb{Z}_2$ cohomology ring is given by

\begin{equation}
\mathbb{Z}_2[A_i,A_{c'},A_c,A_z,B_\beta]/\langle\mathcal{R}_2,\mathcal{R}_3,\mathcal{R}_4\rangle
 \end{equation}
where the relations are 
\begin{subequations} 
 \begin{align}
\mathcal{R}_2\colon & ~~
A_{c'} (A_c + A_i),~~A_{c'}^2,~~A_z (A_{c'} + A_i + A_z),\\
\mathcal{R}_3\colon & ~~
A_{c'} B_\beta,\\
\mathcal{R}_4\colon & ~~
B_\beta (A_c^2 + A_i^2 + B_\beta).
\end{align} 
 \end{subequations}
We have the following table regarding IWPs and group cohomology at degree 3.
\begin{center}
\begin{tabular}{c|cc|c|c|c}\hline\hline {Wyckoff}&\multicolumn{2}{c|}{Little group}& \multirow{2}{*}{Coordinates}&\multirow{2}{*}{LSM anomaly class}&\multirow{2}{*}{Topo. inv.} \\ \cline{2-3} position & Intl. & Sch\"{o}nflies & & & \\ \hline
2a&$2/m$&$C_{2h}$& $(0,0,0)$, $(1/2,1/2,0)$ & $(A_i + A_z) (A_c^2 + A_i^2 + B_\beta)$ & $\varphi_1[I]$\\ 
2b&$2/m$&$C_{2h}$& $(0,0,1/2)$, $(1/2,1/2,1/2)$ & $A_z (A_c^2 + A_i^2 + B_\beta)$ & $\varphi_1[T_3I]$\\ 
2c&$2/m$&$C_{2h}$& $(0,1/2,0)$, $(1/2,0,0)$ & $(A_i + A_z) B_\beta$ & $\varphi_1[T_1I]$\\ 
2d&$2/m$&$C_{2h}$& $(0,1/2,1/2)$, $(1/2,0,1/2)$ & $A_z B_\beta$ & $\varphi_1[T_1T_3I]$\\ 
\hline
\hline 
 \end{tabular} 
 \end{center}

\subsection*{No. 56: $Pccn$}\label{subsub:sg56}

This group is generated by three translations $T_{1,2,3}$ as given in Eqs.~\eqref{TransBravaisP}, a two-fold rotation $C_2$, a two-fold screw $S'_2$, and an inversion $I$:
\begin{subequations}
 \begin{align}
C_2 &\colon (x,y,z)\rightarrow (-x + 1/2, -y + 1/2, z),\\ 
S'_2 &\colon (x,y,z)\rightarrow (-x, y + 1/2, -z + 1/2),\\ 
I &\colon (x,y,z)\rightarrow (-x, -y, -z).
\end{align}
\end{subequations}

The $\mathbb{Z}_2$ cohomology ring is given by

\begin{equation}
\mathbb{Z}_2[A_i,A_{c'},A_c,B_{\beta 1},B_{\beta 2}]/\langle\mathcal{R}_2,\mathcal{R}_3,\mathcal{R}_4\rangle
 \end{equation}
where the relations are 
\begin{subequations} 
 \begin{align}
\mathcal{R}_2\colon & ~~
A_{c'} A_i,~~A_c (A_{c'} + A_i),~~A_{c'}^2,\\
\mathcal{R}_3\colon & ~~
(A_{c'} + A_i) B_{\beta 1},~~A_{c'} B_{\beta 2},~~A_i B_{\beta 1} + A_c B_{\beta 2},\\
\mathcal{R}_4\colon & ~~
(A_c^2 + B_{\beta 1}) (A_c A_i + B_{\beta 1}),~~B_{\beta 1} (A_c A_i + B_{\beta 2}),~~B_{\beta 2} (A_i^2 + B_{\beta 2}).
\end{align} 
 \end{subequations}
We have the following table regarding IWPs and group cohomology at degree 3.
\begin{center}
\begin{tabular}{c|cc|c|c|c}\hline\hline {Wyckoff}&\multicolumn{2}{c|}{Little group}& \multirow{2}{*}{Coordinates}&\multirow{2}{*}{LSM anomaly class}&\multirow{2}{*}{Topo. inv.} \\ \cline{2-3} position & Intl. & Sch\"{o}nflies & & & \\ \hline
\multirow{2}{*}{4a} & \multirow{2}{*}{$\overline{1}$} & \multirow{2}{*}{$C_i$} & $(0,0,0)$, $(1/2,1/2,0)$, & \multirow{2}{*}{$A_i (A_i^2 + B_{\beta 2})$} & \multirow{2}{*}{$\varphi_1[I]$}\\
& & & $(0,1/2,1/2)$, $(1/2,0,1/2)$ & & \\ \hline
\multirow{2}{*}{4b} & \multirow{2}{*}{$\overline{1}$} & \multirow{2}{*}{$C_i$} & $(0,0,1/2)$, $(1/2,1/2,1/2)$, & \multirow{2}{*}{$A_i B_{\beta 2}$} & \multirow{2}{*}{$\varphi_1[T_1I]$}\\
& & & $(0,1/2,0)$, $(1/2,0,0)$ & & \\ \hline
\multirow{2}{*}{4c} & \multirow{2}{*}{$2$} & \multirow{2}{*}{$C_2$} & $(1/4,1/4,z)$, $(3/4,3/4,-z+1/2)$, & \multirow{2}{*}{$A_i (A_c^2 + B_{\beta 1})$} & \multirow{2}{*}{$\varphi_2[S'_2I, C_2]$}\\
& & & $(3/4,3/4,-z)$, $(1/4,1/4,z+1/2)$ & & \\ \hline
\multirow{2}{*}{4d} & \multirow{2}{*}{$2$} & \multirow{2}{*}{$C_2$} & $(1/4,3/4,z)$, $(3/4,1/4,-z+1/2)$, & \multirow{2}{*}{$A_i B_{\beta 1}$} & \multirow{2}{*}{$\varphi_2[S'_2I, T_1C_2]$}\\
& & & $(3/4,1/4,-z)$, $(1/4,3/4,z+1/2)$ & & \\ \hline
\hline 
 \end{tabular} 
 \end{center}

\subsection*{No. 57: $Pbcm$}\label{subsub:sg57}

This group is generated by three translations $T_{1,2,3}$ as given in Eqs.~\eqref{TransBravaisP}, a two-fold screw $S_2$, a two-fold screw $S'_2$, and an inversion $I$:
\begin{subequations}
 \begin{align}
S_2 &\colon (x,y,z)\rightarrow (-x, -y, z + 1/2),\\ 
S'_2 &\colon (x,y,z)\rightarrow (-x, y + 1/2, -z + 1/2),\\ 
I &\colon (x,y,z)\rightarrow (-x, -y, -z).
\end{align}
\end{subequations}

The $\mathbb{Z}_2$ cohomology ring is given by

\begin{equation}
\mathbb{Z}_2[A_i,A_{c'},A_c,A_x]/\langle\mathcal{R}_2\rangle
 \end{equation}
where the relations are 
\begin{subequations} 
 \begin{align}
\mathcal{R}_2\colon & ~~
A_{c'} (A_c + A_{c'} + A_i),~~(A_c + A_{c'}) (A_c + A_i),~~A_x (A_c + A_{c'} + A_i + A_x).
\end{align} 
 \end{subequations}
We have the following table regarding IWPs and group cohomology at degree 3.
\begin{center}
\begin{tabular}{c|cc|c|c|c}\hline\hline {Wyckoff}&\multicolumn{2}{c|}{Little group}& \multirow{2}{*}{Coordinates}&\multirow{2}{*}{LSM anomaly class}&\multirow{2}{*}{Topo. inv.} \\ \cline{2-3} position & Intl. & Sch\"{o}nflies & & & \\ \hline
\multirow{2}{*}{4a} & \multirow{2}{*}{$\overline{1}$} & \multirow{2}{*}{$C_i$} & $(0,0,0)$, $(0,0,1/2)$, & \multirow{2}{*}{$A_i (A_c + A_i) (A_i + A_x)$} & \multirow{2}{*}{$\varphi_1[I]$}\\
& & & $(0,1/2,1/2)$, $(0,1/2,0)$ & & \\ \hline
\multirow{2}{*}{4b} & \multirow{2}{*}{$\overline{1}$} & \multirow{2}{*}{$C_i$} & $(1/2,0,0)$, $(1/2,0,1/2)$, & \multirow{2}{*}{$A_i (A_c + A_i) A_x$} & \multirow{2}{*}{$\varphi_1[T_1I]$}\\
& & & $(1/2,1/2,1/2)$, $(1/2,1/2,0)$ & & \\ \hline
\multirow{2}{*}{4c} & \multirow{2}{*}{$2$} & \multirow{2}{*}{$C_2$} & $(x,1/4,0)$, $(-x,3/4,1/2)$, & \multirow{2}{*}{$A_{c'} (A_c + A_i) A_x$} & \multirow{2}{*}{$\varphi_2[T_1, S_2S'_2]$}\\
& & & $(-x,3/4,0)$, $(x,1/4,1/2)$ & & \\ \hline
\multirow{2}{*}{4d} & \multirow{2}{*}{$m$} & \multirow{2}{*}{$C_s$} & $(x,y,1/4)$, $(-x,-y,3/4)$, & \multirow{2}{*}{$A_{c'} A_i A_x$} & \multirow{2}{*}{$\widetilde{\varphi_3}[T_1, T_3^{-1}S_2S'_2I, S_2I]$}\\
& & & $(-x,y+1/2,1/4)$, $(x,-y+1/2,3/4)$ & & \\ \hline
\hline 
 \end{tabular} 
 \end{center}
 The expression of $\widetilde{\varphi_3}$ is given in Eq.~\eqref{wilde3}.

\subsection*{No. 58: $Pnnm$}\label{subsub:sg58}

This group is generated by three translations $T_{1,2,3}$ as given in Eqs.~\eqref{TransBravaisP}, a two-fold rotation $C_2$, a two-fold screw $S'_2$, and an inversion $I$:
\begin{subequations}
 \begin{align}
C_2 &\colon (x,y,z)\rightarrow (-x, -y, z),\\ 
S'_2 &\colon (x,y,z)\rightarrow (-x + 1/2, y + 1/2, -z + 1/2),\\ 
I &\colon (x,y,z)\rightarrow (-x, -y, -z).
\end{align}
\end{subequations}

The $\mathbb{Z}_2$ cohomology ring is given by

\begin{equation}
\mathbb{Z}_2[A_i,A_{c'},A_c,B_{\beta 1},B_{\beta 2},B_{\beta 3},C_\gamma]/\langle\mathcal{R}_2,\mathcal{R}_3,\mathcal{R}_4,\mathcal{R}_5,\mathcal{R}_6\rangle
 \end{equation}
where the relations are 
\begin{subequations} 
 \begin{align}
\mathcal{R}_2\colon & ~~
A_{c'} A_i,~~A_c A_{c'},~~A_{c'}^2,\\
\mathcal{R}_3\colon & ~~
A_{c'} B_{\beta 1},~~A_{c'} B_{\beta 2},~~A_c B_{\beta 1} + A_i B_{\beta 1} + A_c B_{\beta 2},~~A_{c'} B_{\beta 3},\\
\mathcal{R}_4\colon & ~~
A_{c'} C_\gamma,~~B_{\beta 1} (A_c^2 + A_c A_i + B_{\beta 1}),~~B_{\beta 1} (A_c^2 + A_i^2 + B_{\beta 2}),~~A_c^2 B_{\beta 3} + A_c A_i B_{\beta 3} + B_{\beta 1} B_{\beta 3} + A_c C_\gamma,\nonumber\\
&~~A_c^2 B_{\beta 1} + A_c A_i B_{\beta 1} + A_i^2 B_{\beta 2} + B_{\beta 2}^2,~~A_c^2 B_{\beta 3} + A_i^2 B_{\beta 3} + B_{\beta 2} B_{\beta 3} + A_c C_\gamma + A_i C_\gamma,~~B_{\beta 3} (A_c A_i + A_i^2 + B_{\beta 3}),\\
\mathcal{R}_5\colon & ~~
B_{\beta 1} C_\gamma,~~B_{\beta 2} C_\gamma,~~(A_c A_i + A_i^2 + B_{\beta 3}) C_\gamma,\\
\mathcal{R}_6\colon & ~~
C_\gamma (A_c^2 A_i + A_i^3 + C_\gamma).
\end{align} 
 \end{subequations}
We have the following table regarding IWPs and group cohomology at degree 3.
\begin{center}
\begin{tabular}{c|cc|c|c|c}\hline\hline {Wyckoff}&\multicolumn{2}{c|}{Little group}& \multirow{2}{*}{Coordinates}&\multirow{2}{*}{LSM anomaly class}&\multirow{2}{*}{Topo. inv.} \\ \cline{2-3} position & Intl. & Sch\"{o}nflies & & & \\ \hline
2a&$2/m$&$C_{2h}$& $(0,0,0)$, $(1/2,1/2,1/2)$ & $A_c^2 A_i + A_i^3 + A_i B_{\beta 2} + C_\gamma$ & $\varphi_1[I]$\\ 
2b&$2/m$&$C_{2h}$& $(0,0,1/2)$, $(1/2,1/2,0)$ & $C_\gamma$ & $\varphi_1[T_3I]$\\ 
2c&$2/m$&$C_{2h}$& $(0,1/2,0)$, $(1/2,0,1/2)$ & $A_i B_{\beta 2} + A_c B_{\beta 3} + A_i B_{\beta 3} + C_\gamma$ & $\varphi_1[T_2I]$\\ 
2d&$2/m$&$C_{2h}$& $(0,1/2,1/2)$, $(1/2,0,0)$ & $A_c B_{\beta 3} + A_i B_{\beta 3} + C_\gamma$ & $\varphi_1[T_1I]$\\ 
\hline
\hline 
 \end{tabular} 
 \end{center}

\subsection*{No. 59: $Pmmn$}\label{subsub:sg59}

This group is generated by three translations $T_{1,2,3}$ as given in Eqs.~\eqref{TransBravaisP}, a two-fold rotation $C_2$, a two-fold screw $S'_2$, and an inversion $I$:
\begin{subequations}
 \begin{align}
C_2 &\colon (x,y,z)\rightarrow (-x + 1/2, -y + 1/2, z),\\ 
S'_2 &\colon (x,y,z)\rightarrow (-x, y + 1/2, -z),\\ 
I &\colon (x,y,z)\rightarrow (-x, -y, -z).
\end{align}
\end{subequations}

The $\mathbb{Z}_2$ cohomology ring is given by

\begin{equation}
\mathbb{Z}_2[A_i,A_{c'},A_c,A_z,B_\beta]/\langle\mathcal{R}_2,\mathcal{R}_3,\mathcal{R}_4\rangle
 \end{equation}
where the relations are 
\begin{subequations} 
 \begin{align}
\mathcal{R}_2\colon & ~~
A_c (A_{c'} + A_i),~~A_{c'} (A_{c'} + A_i),~~A_z (A_{c'} + A_i + A_z),\\
\mathcal{R}_3\colon & ~~
(A_{c'} + A_i) B_\beta,\\
\mathcal{R}_4\colon & ~~
B_\beta (A_c^2 + A_c A_i + B_\beta).
\end{align} 
 \end{subequations}
We have the following table regarding IWPs and group cohomology at degree 3.
\begin{center}
\begin{tabular}{c|cc|c|c|c}\hline\hline {Wyckoff}&\multicolumn{2}{c|}{Little group}& \multirow{2}{*}{Coordinates}&\multirow{2}{*}{LSM anomaly class}&\multirow{2}{*}{Topo. inv.} \\ \cline{2-3} position & Intl. & Sch\"{o}nflies & & & \\ \hline
2a&$mm2$&$C_{2v}$& $(1/4,1/4,z)$, $(3/4,3/4,-z)$ & $A_z B_\beta$ & $\varphi_2[T_3, C_2]$\\ 
2b&$mm2$&$C_{2v}$& $(1/4,3/4,z)$, $(3/4,1/4,-z)$ & $A_z (A_c^2 + A_c A_i + B_\beta)$ & $\varphi_2[T_3, T_1C_2]$\\ 
\hline
\multirow{2}{*}{4c} & \multirow{2}{*}{$\overline{1}$} & \multirow{2}{*}{$C_i$} & $(0,0,0)$, $(1/2,1/2,0)$, & \multirow{2}{*}{$A_i (A_{c'} + A_i) (A_i + A_z)$} & \multirow{2}{*}{$\varphi_1[I]$}\\
& & & $(0,1/2,0)$, $(1/2,0,0)$ & & \\ \hline
\multirow{2}{*}{4d} & \multirow{2}{*}{$\overline{1}$} & \multirow{2}{*}{$C_i$} & $(0,0,1/2)$, $(1/2,1/2,1/2)$, & \multirow{2}{*}{$A_i (A_{c'} + A_i) A_z$} & \multirow{2}{*}{$\varphi_1[T_3I]$}\\
& & & $(0,1/2,1/2)$, $(1/2,0,1/2)$ & & \\ \hline
\hline 
 \end{tabular} 
 \end{center}

\subsection*{No. 60: $Pbcn$}\label{subsub:sg60}

This group is generated by three translations $T_{1,2,3}$ as given in Eqs.~\eqref{TransBravaisP}, a two-fold screw $S_2$, a two-fold rotation $C'_2$, and an inversion $I$:
\begin{subequations}
 \begin{align}
S_2 &\colon (x,y,z)\rightarrow (-x + 1/2, -y + 1/2, z + 1/2),\\ 
C'_2 &\colon (x,y,z)\rightarrow (-x, y, -z + 1/2),\\ 
I &\colon (x,y,z)\rightarrow (-x, -y, -z).
\end{align}
\end{subequations}

The $\mathbb{Z}_2$ cohomology ring is given by

\begin{equation}
\mathbb{Z}_2[A_i,A_{c'},A_c,B_\beta]/\langle\mathcal{R}_2,\mathcal{R}_3,\mathcal{R}_4\rangle
 \end{equation}
where the relations are 
\begin{subequations} 
 \begin{align}
\mathcal{R}_2\colon & ~~
A_c A_i,~~A_c A_{c'},~~A_c^2 + A_{c'} A_i,\\
\mathcal{R}_3\colon & ~~
A_c B_\beta,\\
\mathcal{R}_4\colon & ~~
B_\beta (A_i^2 + B_\beta).
\end{align} 
 \end{subequations}
We have the following table regarding IWPs and group cohomology at degree 3.
\begin{center}
\begin{tabular}{c|cc|c|c|c}\hline\hline {Wyckoff}&\multicolumn{2}{c|}{Little group}& \multirow{2}{*}{Coordinates}&\multirow{2}{*}{LSM anomaly class}&\multirow{2}{*}{Topo. inv.} \\ \cline{2-3} position & Intl. & Sch\"{o}nflies & & & \\ \hline
\multirow{2}{*}{4a} & \multirow{2}{*}{$\overline{1}$} & \multirow{2}{*}{$C_i$} & $(0,0,0)$, $(1/2,1/2,1/2)$, & \multirow{2}{*}{$A_i B_\beta$} & \multirow{2}{*}{$\varphi_1[I]$}\\
& & & $(0,0,1/2)$, $(1/2,1/2,0)$ & & \\ \hline
\multirow{2}{*}{4b} & \multirow{2}{*}{$\overline{1}$} & \multirow{2}{*}{$C_i$} & $(0,1/2,0)$, $(1/2,0,1/2)$, & \multirow{2}{*}{$A_i (A_i^2 + B_\beta)$} & \multirow{2}{*}{$\varphi_1[T_1I]$}\\
& & & $(0,1/2,1/2)$, $(1/2,0,0)$ & & \\ \hline
\multirow{2}{*}{4c} & \multirow{2}{*}{$2$} & \multirow{2}{*}{$C_2$} & $(0,y,1/4)$, $(1/2,-y+1/2,3/4)$, & \multirow{2}{*}{$A_{c'} B_\beta$} & \multirow{2}{*}{$\varphi_2[T_2, C'_2]$}\\
& & & $(0,-y,3/4)$, $(1/2,y+1/2,1/4)$ & & \\ \hline
\hline 
 \end{tabular} 
 \end{center}

\subsection*{No. 61: $Pbca$}\label{subsub:sg61}

This group is generated by three translations $T_{1,2,3}$ as given in Eqs.~\eqref{TransBravaisP}, a two-fold screw $S_2$, a two-fold screw $S'_2$, and an inversion $I$:
\begin{subequations}
 \begin{align}
S_2 &\colon (x,y,z)\rightarrow (-x + 1/2, -y, z + 1/2),\\ 
S'_2 &\colon (x,y,z)\rightarrow (-x, y + 1/2, -z + 1/2),\\ 
I &\colon (x,y,z)\rightarrow (-x, -y, -z).
\end{align}
\end{subequations}

The $\mathbb{Z}_2$ cohomology ring is given by

\begin{equation}
\mathbb{Z}_2[A_i,A_{c'},A_c,C_\beta]/\langle\mathcal{R}_2,\mathcal{R}_4,\mathcal{R}_6\rangle
 \end{equation}
where the relations are 
\begin{subequations} 
 \begin{align}
\mathcal{R}_2\colon & ~~
A_c (A_{c'} + A_i),~~A_{c'}^2 + A_c A_i + A_{c'} A_i,~~A_c^2 + A_{c'} A_i,\\
\mathcal{R}_4\colon & ~~
A_{c'} C_\beta,~~A_c C_\beta,\\
\mathcal{R}_6\colon & ~~
C_\beta (A_i^3 + C_\beta).
\end{align} 
 \end{subequations}
We have the following table regarding IWPs and group cohomology at degree 3.
\begin{center}
\begin{tabular}{c|cc|c|c|c}\hline\hline {Wyckoff}&\multicolumn{2}{c|}{Little group}& \multirow{2}{*}{Coordinates}&\multirow{2}{*}{LSM anomaly class}&\multirow{2}{*}{Topo. inv.} \\ \cline{2-3} position & Intl. & Sch\"{o}nflies & & & \\ \hline
\multirow{2}{*}{4a} & \multirow{2}{*}{$\overline{1}$} & \multirow{2}{*}{$C_i$} & $(0,0,0)$, $(1/2,0,1/2)$, & \multirow{2}{*}{$A_i^3 + C_\beta$} & \multirow{2}{*}{$\varphi_1[I]$}\\
& & & $(0,1/2,1/2)$, $(1/2,1/2,0)$ & & \\ \hline
\multirow{2}{*}{4b} & \multirow{2}{*}{$\overline{1}$} & \multirow{2}{*}{$C_i$} & $(0,0,1/2)$, $(1/2,0,0)$, & \multirow{2}{*}{$C_\beta$} & \multirow{2}{*}{$\varphi_1[T_1I]$}\\
& & & $(0,1/2,0)$, $(1/2,1/2,1/2)$ & & \\ \hline
\hline 
 \end{tabular} 
 \end{center}

\subsection*{No. 62: $Pnma$}\label{subsub:sg62}

This group is generated by three translations $T_{1,2,3}$ as given in Eqs.~\eqref{TransBravaisP}, a two-fold screw $S_2$, a two-fold screw $S'_2$, and an inversion $I$:
\begin{subequations}
 \begin{align}
S_2 &\colon (x,y,z)\rightarrow (-x + 1/2, -y, z + 1/2),\\ 
S'_2 &\colon (x,y,z)\rightarrow (-x, y + 1/2, -z),\\ 
I &\colon (x,y,z)\rightarrow (-x, -y, -z).
\end{align}
\end{subequations}

The $\mathbb{Z}_2$ cohomology ring is given by

\begin{equation}
\mathbb{Z}_2[A_i,A_{c'},A_c,B_{\beta 1},B_{\beta 2}]/\langle\mathcal{R}_2,\mathcal{R}_3,\mathcal{R}_4\rangle
 \end{equation}
where the relations are 
\begin{subequations} 
 \begin{align}
\mathcal{R}_2\colon & ~~
A_c (A_{c'} + A_i),~~A_{c'}^2 + A_c A_i + A_{c'} A_i,~~A_c^2,\\
\mathcal{R}_3\colon & ~~
(A_c + A_{c'} + A_i) B_{\beta 1},~~A_i B_{\beta 1} + A_{c'} B_{\beta 2},~~A_{c'} B_{\beta 1} + A_i B_{\beta 1} + A_c B_{\beta 2},\\
\mathcal{R}_4\colon & ~~
B_{\beta 1} (A_{c'} A_i + A_i^2 + B_{\beta 1}),~~B_{\beta 1} (A_{c'} A_i + A_i^2 + B_{\beta 2}),~~A_i^2 B_{\beta 1} + A_i^2 B_{\beta 2} + B_{\beta 2}^2.
\end{align} 
 \end{subequations}
We have the following table regarding IWPs and group cohomology at degree 3.
\begin{center}
\begin{tabular}{c|cc|c|c|c}\hline\hline {Wyckoff}&\multicolumn{2}{c|}{Little group}& \multirow{2}{*}{Coordinates}&\multirow{2}{*}{LSM anomaly class}&\multirow{2}{*}{Topo. inv.} \\ \cline{2-3} position & Intl. & Sch\"{o}nflies & & & \\ \hline
\multirow{2}{*}{4a} & \multirow{2}{*}{$\overline{1}$} & \multirow{2}{*}{$C_i$} & $(0,0,0)$, $(1/2,0,1/2)$, & \multirow{2}{*}{$A_{c'}^2 A_i + A_i^3 + A_{c'} B_{\beta 1} + A_i B_{\beta 2}$} & \multirow{2}{*}{$\varphi_1[I]$}\\
& & & $(0,1/2,0)$, $(1/2,1/2,1/2)$ & & \\ \hline
\multirow{2}{*}{4b} & \multirow{2}{*}{$\overline{1}$} & \multirow{2}{*}{$C_i$} & $(0,0,1/2)$, $(1/2,0,0)$, & \multirow{2}{*}{$A_{c'} B_{\beta 1} + A_i B_{\beta 2}$} & \multirow{2}{*}{$\varphi_1[T_1I]$}\\
& & & $(0,1/2,1/2)$, $(1/2,1/2,0)$ & & \\ \hline
\multirow{2}{*}{4c} & \multirow{2}{*}{$m$} & \multirow{2}{*}{$C_s$} & $(x,1/4,z)$, $(-x+1/2,3/4,z+1/2)$, & \multirow{2}{*}{$(A_{c'} + A_i) B_{\beta 1}$} & \multirow{2}{*}{$\widetilde{\varphi_3}[T_3, S_2I, S'_2I]$}\\
& & & $(-x,3/4,-z)$, $(x+1/2,1/4,-z+1/2)$ & & \\ \hline
\hline 
 \end{tabular} 
 \end{center}
The expression of $\widetilde{\varphi_3}$ is given in Eq.~\eqref{wilde3}.

\subsection*{No. 63: $Cmcm$}\label{subsub:sg63}

This group is generated by three translations $T_{1,2,3}$ as given in Eqs.~\eqref{TransBravaisC}, a two-fold screw $S_2$, a two-fold rotation $C'_2$, and an inversion $I$:
\begin{subequations}
 \begin{align}
S_2 &\colon (x,y,z)\rightarrow (-x, -y, z + 1/2),\\ 
C'_2 &\colon (x,y,z)\rightarrow (-x, y, -z + 1/2),\\ 
I &\colon (x,y,z)\rightarrow (-x, -y, -z).
\end{align}
\end{subequations}

The $\mathbb{Z}_2$ cohomology ring is given by

\begin{equation}
\mathbb{Z}_2[A_i,A_{c'},A_c,A_{x+y},B_{xy}]/\langle\mathcal{R}_2,\mathcal{R}_3,\mathcal{R}_4\rangle
 \end{equation}
where the relations are 
\begin{subequations} 
 \begin{align}
\mathcal{R}_2\colon & ~~
A_{c'} A_{x+y},~~(A_c + A_{c'}) (A_c + A_i),~~A_{x+y} (A_c + A_i + A_{x+y}),\\
\mathcal{R}_3\colon & ~~
A_{x+y} B_{xy},\\
\mathcal{R}_4\colon & ~~
B_{xy} (A_c A_i + A_i^2 + B_{xy}).
\end{align} 
 \end{subequations}
We have the following table regarding IWPs and group cohomology at degree 3.
\begin{center}
\begin{tabular}{c|cc|c|c|c}\hline\hline {Wyckoff}&\multicolumn{2}{c|}{Little group}& {Coordinates}&\multirow{2}{*}{LSM anomaly class}&\multirow{2}{*}{Topo. inv.} \\ \cline{2-4} position & Intl. & Sch\"{o}nflies & $ (0,0,0) + (1/2,1/2,0) + $ & &\\ \hline
4a&$2/m$&$C_{2h}$& $(0,0,0)$, $(0,0,1/2)$ & $(A_c + A_i) (A_{c'} A_i + A_i^2 + A_i A_{x+y} + B_{xy})$ & $\varphi_1[I]$\\ 
4b&$2/m$&$C_{2h}$& $(0,1/2,0)$, $(0,1/2,1/2)$ & $(A_c + A_i) B_{xy}$ & $\varphi_1[T_1T_2I]$\\ 
4c&$mm2$&$C_{2v}$& $(0,y,1/4)$, $(0,-y,3/4)$ & $(A_c + A_{c'}) B_{xy}$ & $\varphi_2[T_1T_2, C'_2]$\\ 
\hline
\multirow{2}{*}{8d} & \multirow{2}{*}{$\overline{1}$} & \multirow{2}{*}{$C_i$} & $(1/4,1/4,0)$, $(3/4,3/4,1/2)$, & \multirow{2}{*}{$A_i (A_c + A_i) A_{x+y}$} & \multirow{2}{*}{$\varphi_1[T_1I]$}\\
& & & $(3/4,1/4,1/2)$, $(1/4,3/4,0)$ & & \\ \hline
\hline 
 \end{tabular} 
 \end{center}

\subsection*{No. 64: $Cmce$}\label{subsub:sg64}

This group is generated by three translations $T_{1,2,3}$ as given in Eqs.~\eqref{TransBravaisC}, a two-fold screw $S_2$, a two-fold rotation $C'_2$, and an inversion $I$:
\begin{subequations}
 \begin{align}
S_2 &\colon (x,y,z)\rightarrow (-x, -y + 1/2, z + 1/2),\\ 
C'_2 &\colon (x,y,z)\rightarrow (-x + 1/2, y, -z + 1/2),\\ 
I &\colon (x,y,z)\rightarrow (-x, -y, -z).
\end{align}
\end{subequations}

The $\mathbb{Z}_2$ cohomology ring is given by

\begin{equation}
\mathbb{Z}_2[A_i,A_{c'},A_c,A_{x+y},C_\gamma]/\langle\mathcal{R}_2,\mathcal{R}_4,\mathcal{R}_6\rangle
 \end{equation}
where the relations are 
\begin{subequations} 
 \begin{align}
\mathcal{R}_2\colon & ~~
A_c A_{c'} + A_c A_i + A_{c'} A_i + A_{c'} A_{x+y},~~(A_c + A_{c'}) (A_c + A_i),~~(A_c + A_{x+y}) (A_i + A_{x+y}),\\
\mathcal{R}_4\colon & ~~
(A_c + A_{c'}) C_\gamma,~~(A_{c'} + A_{x+y}) C_\gamma,\\
\mathcal{R}_6\colon & ~~
C_\gamma (A_{c'}^2 A_i + A_i^3 + C_\gamma).
\end{align} 
 \end{subequations}
We have the following table regarding IWPs and group cohomology at degree 3.
\begin{center}
\begin{tabular}{c|cc|c|c|c}\hline\hline {Wyckoff}&\multicolumn{2}{c|}{Little group}& {Coordinates}&\multirow{2}{*}{LSM anomaly class}&\multirow{2}{*}{Topo. inv.} \\ \cline{2-4} position & Intl. & Sch\"{o}nflies & $ (0,0,0) + (1/2,1/2,0) + $ & &\\ \hline
4a&$2/m$&$C_{2h}$& $(0,0,0)$, $(0,1/2,1/2)$ & 
$A_i (A_{c'} + A_i) (A_i + A_{x+y}) + C_\gamma$ & $\varphi_1[I]$\\ 
4b&$2/m$&$C_{2h}$& $(1/2,0,0)$, $(1/2,1/2,1/2)$ & $C_\gamma$ & $\varphi_1[T_1T_2I]$\\ 
\hline
\multirow{2}{*}{8c} & \multirow{2}{*}{$\overline{1}$} & \multirow{2}{*}{$C_i$} & $(1/4,1/4,0)$, $(3/4,1/4,1/2)$, & \multirow{2}{*}{$A_i^2 (A_{c'} + A_{x+y})$} & \multirow{2}{*}{$\varphi_1[T_1I]$}\\
& & & $(3/4,3/4,1/2)$, $(1/4,3/4,0)$ & & \\ \hline
\multirow{2}{*}{8e} & \multirow{2}{*}{$2$} & \multirow{2}{*}{$C_2$} & $(1/4,y,1/4)$, $(3/4,-y+1/2,3/4)$, & \multirow{2}{*}{$A_{c'} A_i (A_{c'} + A_{x+y})$} & \multirow{2}{*}{$\varphi_2[S_2I, C'_2]$}\\
& & & $(3/4,-y,3/4)$, $(1/4,y+1/2,1/4)$ & & \\ \hline
\hline 
 \end{tabular} 
 \end{center}

\subsection*{No. 65: $Cmmm$}\label{subsub:sg65}

This group is generated by three translations $T_{1,2,3}$ as given in Eqs.~\eqref{TransBravaisC}, a two-fold rotation $C_2$, a two-fold rotation $C'_2$, and an inversion $I$:
\begin{subequations}
 \begin{align}
C_2 &\colon (x,y,z)\rightarrow (-x, -y, z),\\ 
C'_2 &\colon (x,y,z)\rightarrow (-x, y, -z),\\ 
I &\colon (x,y,z)\rightarrow (-x, -y, -z).
\end{align}
\end{subequations}

The $\mathbb{Z}_2$ cohomology ring is given by

\begin{equation}
\mathbb{Z}_2[A_i,A_{c'},A_c,A_{x+y},A_z,B_{xy}]/\langle\mathcal{R}_2,\mathcal{R}_3,\mathcal{R}_4\rangle
 \end{equation}
where the relations are 
\begin{subequations} 
 \begin{align}
\mathcal{R}_2\colon & ~~
A_{c'} A_{x+y},~~A_{x+y} (A_c + A_i + A_{x+y}),~~A_z (A_{c'} + A_i + A_z),\\
\mathcal{R}_3\colon & ~~
A_{x+y} B_{xy},\\
\mathcal{R}_4\colon & ~~
B_{xy} (A_c^2 + A_c A_{c'} + A_{c'} A_i + A_i^2 + B_{xy}).
\end{align} 
 \end{subequations}
We have the following table regarding IWPs and group cohomology at degree 3.
\begin{center}
\resizebox{\columnwidth}{!}{
\begin{tabular}{c|cc|c|c|c}\hline\hline {Wyckoff}&\multicolumn{2}{c|}{Little group}& {Coordinates}&\multirow{2}{*}{LSM anomaly class}&\multirow{2}{*}{Topo. inv.} \\ \cline{2-4} position & Intl. & Sch\"{o}nflies & $ (0,0,0) + (1/2,1/2,0) + $ & &\\ \hline
2a&$mmm$&$D_{2h}$& $(0,0,0)$ & 
$(A_{c'} + A_i + A_z) ((A_c+A_i)(A_c+A_{c'}+A_i+A_{x+y}) + B_{xy})$ & $\varphi_2[C_2, C'_2]$\\ 
2b&$mmm$&$D_{2h}$& $(1/2,0,0)$ & $(A_{c'} + A_i + A_z) B_{xy}$ & $\varphi_2[T_1T_2C_2, C'_2]$\\ 
2c&$mmm$&$D_{2h}$& $(1/2,0,1/2)$ & $A_z B_{xy}$ & $\varphi_2[T_1T_2C_2, T_3C'_2]$\\ 
2d&$mmm$&$D_{2h}$& $(0,0,1/2)$ & 
$A_z((A_c+A_i)(A_c+A_{c'}+A_i+A_{x+y}) + B_{xy})$
& $\varphi_2[C_2, T_3C'_2]$\\ 
4e&$2/m$&$C_{2h}$& $(1/4,1/4,0)$, $(3/4,1/4,0)$ & $(A_c + A_i) A_{x+y} (A_i + A_z)$ & $\varphi_1[T_1I]$\\ 
4f&$2/m$&$C_{2h}$& $(1/4,1/4,1/2)$, $(3/4,1/4,1/2)$ & $(A_c + A_i) A_{x+y} A_z$ & $\varphi_1[T_1T_3I]$\\ 
\hline
\hline 
 \end{tabular} }
 \end{center}

\subsection*{No. 66: $Cccm$}\label{subsub:sg66}

This group is generated by three translations $T_{1,2,3}$ as given in Eqs.~\eqref{TransBravaisC}, a two-fold rotation $C_2$, a two-fold rotation $C'_2$, and an inversion $I$:
\begin{subequations}
 \begin{align}
C_2 &\colon (x,y,z)\rightarrow (-x, -y, z),\\ 
C'_2 &\colon (x,y,z)\rightarrow (-x, y, -z + 1/2),\\ 
I &\colon (x,y,z)\rightarrow (-x, -y, -z).
\end{align}
\end{subequations}

The $\mathbb{Z}_2$ cohomology ring is given by

\begin{equation}
\mathbb{Z}_2[A_i,A_{c'},A_c,A_{x+y},B_{xy},B_{z(x+y)}]/\langle\mathcal{R}_2,\mathcal{R}_3,\mathcal{R}_4\rangle
 \end{equation}
where the relations are 
\begin{subequations} 
 \begin{align}
\mathcal{R}_2\colon & ~~
A_{c'} A_i,~~A_{c'} A_{x+y},~~A_{x+y} (A_c + A_i + A_{x+y}),\\
\mathcal{R}_3\colon & ~~
A_{x+y} B_{xy},~~A_{c'} B_{z(x+y)},~~A_i B_{xy} + A_c B_{z(x+y)} + A_i B_{z(x+y)} + A_{x+y} B_{z(x+y)},\\
\mathcal{R}_4\colon & ~~
B_{xy} (A_c^2 + A_c A_{c'} + A_i^2 + B_{xy}),~~B_{xy} (A_c A_i + A_i^2 + B_{z(x+y)}),~~B_{z(x+y)} (A_c A_i + A_i^2 + B_{z(x+y)}).
\end{align} 
 \end{subequations}
We have the following table regarding IWPs and group cohomology at degree 3.
\begin{center}
\resizebox{\columnwidth}{!}{
\begin{tabular}{c|cc|c|c|c}\hline\hline {Wyckoff}&\multicolumn{2}{c|}{Little group}& {Coordinates}&\multirow{2}{*}{LSM anomaly class}&\multirow{2}{*}{Topo. inv.} \\ \cline{2-4} position & Intl. & Sch\"{o}nflies & $ (0,0,0) + (1/2,1/2,0) + $ & &\\ \hline
4a&$222$&$D_2$& $(0,0,1/4)$, $(0,0,3/4)$ & $A_{c'} (A_c^2 + A_c A_{c'} + B_{xy})$ & $\varphi_2[C_2, C'_2]$\\ 
4b&$222$&$D_2$& $(0,1/2,1/4)$, $(0,1/2,3/4)$ & $A_{c'} B_{xy}$ & $\varphi_2[T_1T_2C_2, C'_2]$\\ 
4c&$2/m$&$C_{2h}$& $(0,0,0)$, $(0,0,1/2)$ & $A_i (A_c^2 + A_i^2 + A_c A_{x+y} + A_i A_{x+y} + B_{xy})$ & $\varphi_1[I]$\\ 
4d&$2/m$&$C_{2h}$& $(0,1/2,0)$, $(0,1/2,1/2)$ & $A_i B_{xy}$ & $\varphi_1[T_1T_2I]$\\ 
4e&$2/m$&$C_{2h}$& $(1/4,1/4,0)$, $(3/4,1/4,1/2)$ & $A_i B_{xy} + A_c B_{z(x+y)} + A_i B_{z(x+y)}$ & $\varphi_1[T_1I]$\\ 
4f&$2/m$&$C_{2h}$& $(1/4,3/4,0)$, $(3/4,3/4,1/2)$ & 
$A_i(A_c  A_{x+y} + A_i A_{x+y} + B_{xy}) + (A_c + A_i) B_{z(x+y)}$ & $\varphi_1[T_2I]$\\ 
\hline
\hline 
 \end{tabular} }
 \end{center}

\subsection*{No. 67: $Cmme$}\label{subsub:sg67}

This group is generated by three translations $T_{1,2,3}$ as given in Eqs.~\eqref{TransBravaisC}, a two-fold rotation $C_2$, a two-fold rotation $C'_2$, and an inversion $I$:
\begin{subequations}
 \begin{align}
C_2 &\colon (x,y,z)\rightarrow (-x + 1/2, -y, z),\\ 
C'_2 &\colon (x,y,z)\rightarrow (-x + 1/2, y, -z),\\ 
I &\colon (x,y,z)\rightarrow (-x, -y, -z).
\end{align}
\end{subequations}

The $\mathbb{Z}_2$ cohomology ring is given by

\begin{equation}
\mathbb{Z}_2[A_i,A_{c'},A_c,A_{x+y},A_z]/\langle\mathcal{R}_2\rangle
 \end{equation}
where the relations are 
\begin{subequations} 
 \begin{align}
\mathcal{R}_2\colon & ~~
A_c A_i + A_{c'} A_i + A_{c'} A_{x+y},~~A_{x+y} (A_c + A_i + A_{x+y}),~~A_z (A_{c'} + A_i + A_z).
\end{align} 
 \end{subequations}
We have the following table regarding IWPs and group cohomology at degree 3.
\begin{center}
\resizebox{\columnwidth}{!}{
\begin{tabular}{c|cc|c|c|c}\hline\hline {Wyckoff}&\multicolumn{2}{c|}{Little group}& {Coordinates}&\multirow{2}{*}{LSM anomaly class}&\multirow{2}{*}{Topo. inv.} \\ \cline{2-4} position & Intl. & Sch\"{o}nflies & $ (0,0,0) + (1/2,1/2,0) + $ & &\\ \hline
4a&$222$&$D_2$& $(1/4,0,0)$, $(3/4,0,0)$ & $A_c (A_c A_{c'} + A_{c'}^2 + A_c A_i + A_{c'} A_i + A_c A_z + A_{c'} A_z + A_{x+y} A_z)$ & $\varphi_2[C_2, C'_2]$\\ 
4b&$222$&$D_2$& $(1/4,0,1/2)$, $(3/4,0,1/2)$ & $A_c (A_c + A_{c'} + A_{x+y}) A_z$ & $\varphi_2[C_2, T_3C'_2]$\\ 
4c&$2/m$&$C_{2h}$& $(0,0,0)$, $(0,1/2,0)$ & $A_i (A_c A_{c'} + A_i^2 + A_i A_{x+y} + A_c A_z + A_i A_z + A_{x+y} A_z)$ & $\varphi_1[I]$\\ 
4d&$2/m$&$C_{2h}$& $(0,0,1/2)$, $(0,1/2,1/2)$ & $A_i (A_c + A_i + A_{x+y}) A_z$ & $\varphi_1[T_3I]$\\ 
4e&$2m$&$C_{2h}$& $(1/4,1/4,0)$, $(3/4,1/4,0)$ & $A_i (A_c A_{c'} + A_{c'}^2 + A_i A_{x+y} + A_c A_z + A_{c'} A_z + A_{x+y} A_z)$ & $\varphi_1[T_1I]$\\ 
4f&$2m$&$C_{2h}$& $(1/4,1/4,1/2)$, $(3/4,1/4,1/2)$ & $A_i (A_c + A_{c'} + A_{x+y}) A_z$ & $\varphi_1[T_1T_3I]$\\ 
4g&$mm2$&$C_{2v}$& $(0,1/4,z)$, $(0,3/4,-z)$ & $A_c A_{x+y} A_z$ & $\varphi_2[T_3, T_2C_2]$\\ 
\hline
\hline 
 \end{tabular} }
 \end{center}

\subsection*{No. 68: $Ccce$}\label{subsub:sg68}

This group is generated by three translations $T_{1,2,3}$ as given in Eqs.~\eqref{TransBravaisC}, a two-fold rotation $C_2$, a two-fold rotation $C'_2$, and an inversion $I$:
\begin{subequations}
 \begin{align}
C_2 &\colon (x,y,z)\rightarrow (-x, -y + 1/2, z),\\ 
C'_2 &\colon (x,y,z)\rightarrow (-x, y, -z + 1/2),\\ 
I &\colon (x,y,z)\rightarrow (-x, -y, -z).
\end{align}
\end{subequations}

The $\mathbb{Z}_2$ cohomology ring is given by

\begin{equation}
\mathbb{Z}_2[A_i,A_{c'},A_c,A_{x+y},C_\gamma]/\langle\mathcal{R}_2,\mathcal{R}_4,\mathcal{R}_6\rangle
 \end{equation}
where the relations are 
\begin{subequations} 
 \begin{align}
\mathcal{R}_2\colon & ~~
A_{c'} A_i,~~A_c A_i + A_{c'} A_{x+y},~~(A_c + A_{x+y}) (A_i + A_{x+y}),\\
\mathcal{R}_4\colon & ~~
A_i C_\gamma,~~A_{x+y} C_\gamma,\\
\mathcal{R}_6\colon & ~~
C_\gamma (A_c^2 A_{c'} + A_c A_{c'}^2 + C_\gamma).
\end{align} 
 \end{subequations}
We have the following table regarding IWPs and group cohomology at degree 3.
\begin{center}
\begin{tabular}{c|cc|c|c|c}\hline\hline {Wyckoff}&\multicolumn{2}{c|}{Little group}& {Coordinates}&\multirow{2}{*}{LSM anomaly class}&\multirow{2}{*}{Topo. inv.} \\ \cline{2-4} position & Intl. & Sch\"{o}nflies & $ (0,0,0) + (1/2,1/2,0) + $ & &\\ \hline
4a&$222$&$D_2$& $(0,1/4,1/4)$, $(0,3/4,3/4)$ & $A_c^2 A_{c'} + A_c A_{c'}^2 + A_c^2 A_i + C_\gamma$ & $\varphi_2[C_2, C'_2]$\\ 
4b&$222$&$D_2$& $(0,1/4,3/4)$, $(0,3/4,1/4)$ & $C_\gamma$ & $\varphi_2[T_1T_2C_2, C'_2]$\\ 
\hline
\multirow{2}{*}{8c} & \multirow{2}{*}{$\overline{1}$} & \multirow{2}{*}{$C_i$} & $(1/4,3/4,0)$, $(1/4,1/4,0)$, & \multirow{2}{*}{$A_i^2 A_{x+y}$} & \multirow{2}{*}{$\varphi_1[T_1I]$}\\
& & & $(3/4,3/4,1/2)$, $(3/4,1/4,1/2)$ & & \\ \hline
\multirow{2}{*}{8d} & \multirow{2}{*}{$\overline{1}$} & \multirow{2}{*}{$C_i$} & $(0,0,0)$, $(1/2,0,0)$, & \multirow{2}{*}{$A_i^2 (A_i + A_{x+y})$} & \multirow{2}{*}{$\varphi_1[I]$}\\
& & & $(0,0,1/2)$, $(1/2,0,1/2)$ & & \\ \hline
\multirow{2}{*}{8h} & \multirow{2}{*}{$2$} & \multirow{2}{*}{$C_2$} & $(1/4,0,z)$, $(3/4,0,-z+1/2)$, & \multirow{2}{*}{$A_c^2 A_i$} & \multirow{2}{*}{$\varphi_2[T_2^{-1}C_2C'_2I, T_2^{-1}C_2]$}\\
& & & $(3/4,0,-z)$, $(1/4,0,z+1/2)$ & & \\ \hline
\hline 
 \end{tabular} 
 \end{center}

\subsection*{No. 69: $Fmmm$}\label{subsub:sg69}

This group is generated by three translations $T_{1,2,3}$ as given in Eqs.~\eqref{TransBravaisF}, a two-fold rotation $C_2$, a two-fold rotation $C'_2$, and an inversion $I$:
\begin{subequations}
 \begin{align}
C_2 &\colon (x,y,z)\rightarrow (-x, -y, z),\\ 
C'_2 &\colon (x,y,z)\rightarrow (-x, y, -z),\\ 
I &\colon (x,y,z)\rightarrow (-x, -y, -z).
\end{align}
\end{subequations}

The $\mathbb{Z}_2$ cohomology ring is given by

\begin{equation}
\mathbb{Z}_2[A_i,A_{c'},A_c,A_{x+z},A_{y+z},C_{xyz}]/\langle\mathcal{R}_2,\mathcal{R}_4,\mathcal{R}_6\rangle
 \end{equation}
where the relations are 
\begin{subequations} 
 \begin{align}
\mathcal{R}_2\colon & ~~
A_c A_{x+z} + A_{c'} A_{x+z} + A_c A_{y+z},~~A_{x+z} (A_c + A_i + A_{x+z}),~~A_c A_{x+z} + A_{c'} A_{x+z} + A_{c'} A_{y+z} + A_i A_{y+z} + A_{y+z}^2,\\
\mathcal{R}_4\colon & ~~
A_{x+z} C_{xyz},~~A_{y+z} C_{xyz},\\
\mathcal{R}_6\colon & ~~
C_{xyz} (A_c^2 A_{c'} + A_c A_{c'}^2 + A_c^2 A_i + A_c A_{c'} A_i + A_{c'}^2 A_i + A_i^3 + C_{xyz}).
\end{align} 
 \end{subequations}
We have the following table regarding IWPs and group cohomology at degree 3.
\begin{center}
\resizebox{\columnwidth}{!}{
\begin{tabular}{c|cc|c|c|c}\hline\hline \multirow{3}{*}{\shortstack[l]{Wyckoff\\position}}&\multicolumn{2}{c|}{Little group}& {Coordinates}&\multirow{3}{*}{LSM anomaly class}&\multirow{3}{*}{Topo. inv.} \\ \cline{2-4} & \multirow{2}{*}{Intl.} & \multirow{2}{*}{Sch\"{o}nflies} & $(0,0,0) + ~(0,1/2,1/2) + $ & & \\ & & & $ (1/2,0,1/2) + ~(1/2,1/2,0) +$ & &\\ \hline
4a&$mmm$&$D_{2h}$& $(0,0,0)$ & \begin{tabular}{c}$A_c^2 A_{c'} + A_c A_{c'}^2 + A_c^2 A_i + A_c A_{c'} A_i + A_{c'}^2 A_i + A_i^3 + A_{c'} A_i A_{x+z} + A_i^2 A_{x+z} $\\
$+ A_c A_{c'} A_{y+z} + A_{c'} A_i A_{y+z} + A_i^2 A_{y+z} + A_i A_{x+z} A_{y+z} + C_{xyz}$\end{tabular} & $\varphi_2[C_2, C'_2]$\\ 
4b&$mmm$&$D_{2h}$& $(0,0,1/2)$ & $C_{xyz}$ & $\varphi_2[C_2, T_1T_2T_3^{-1}C'_2]$\\ 
8c&$2/m$&$C_{2h}$& $(0,1/4,1/4)$, $(0,3/4,1/4)$ & $A_i A_{x+z} (A_c + A_i + A_{y+z})$ & $\varphi_1[T_1I]$\\ 
8d&$2m$&$C_{2h}$& $(1/4,0,1/4)$, $(3/4,0,1/4)$ & $A_i (A_c A_{x+z} + A_{c'} A_{x+z} + A_{c'} A_{y+z} + A_i A_{y+z} + A_{x+z} A_{y+z})$ & $\varphi_1[T_2I]$\\ 
8e&$2/m$&$C_{2h}$& $(1/4,1/4,0)$, $(3/4,1/4,0)$ & $A_i A_{x+z} A_{y+z}$ & $\varphi_1[T_3I]$\\ 
8f&$222$&$D_2$& $(1/4,1/4,1/4)$, $(3/4,3/4,3/4)$ & $A_c A_{c'} A_{y+z}$ & $\varphi_2[T_3C_2, T_2C'_2]$\\ 
\hline
\hline 
 \end{tabular} }
 \end{center}

\subsection*{No. 70: $Fddd$}\label{subsub:sg70}

This group is generated by three translations $T_{1,2,3}$ as given in Eqs.~\eqref{TransBravaisF}, a two-fold rotation $C_2$, a two-fold rotation $C'_2$, and an inversion $I$:
\begin{subequations}
 \begin{align}
C_2 &\colon (x,y,z)\rightarrow (-x + 1/4, -y + 1/4, z),\\ 
C'_2 &\colon (x,y,z)\rightarrow (-x + 1/4, y, -z + 1/4),\\ 
I &\colon (x,y,z)\rightarrow (-x, -y, -z).
\end{align}
\end{subequations}

The $\mathbb{Z}_2$ cohomology ring is given by

\begin{equation}
\mathbb{Z}_2[A_i,A_{c'},A_c,B_{xy+xz+yz},C_\gamma]/\langle\mathcal{R}_2,\mathcal{R}_3,\mathcal{R}_4,\mathcal{R}_5,\mathcal{R}_6\rangle
 \end{equation}
where the relations are 
\begin{subequations} 
 \begin{align}
\mathcal{R}_2\colon & ~~
A_{c'} A_i,~~A_c A_i,\\
\mathcal{R}_3\colon & ~~
A_{c'} (A_c A_{c'} + B_{xy+xz+yz}),~~A_c (A_c A_{c'} + B_{xy+xz+yz}),\\
\mathcal{R}_4\colon & ~~
A_i C_\gamma,~~A_c^2 A_{c'}^2 + A_i^2 B_{xy+xz+yz} + B_{xy+xz+yz}^2,\\
\mathcal{R}_5\colon & ~~
(A_c A_{c'} + B_{xy+xz+yz}) C_\gamma,\\
\mathcal{R}_6\colon & ~~
C_\gamma (A_c^2 A_{c'} + A_c A_{c'}^2 + C_\gamma).
\end{align} 
 \end{subequations}
We have the following table regarding IWPs and group cohomology at degree 3.
\begin{center}
\begin{tabular}{c|cc|c|c|c}\hline\hline \multirow{3}{*}{\shortstack[l]{Wyckoff\\position}}&\multicolumn{2}{c|}{Little group}& {Coordinates}&\multirow{3}{*}{LSM anomaly class}&\multirow{3}{*}{Topo. inv.} \\ \cline{2-4} & \multirow{2}{*}{Intl.} & \multirow{2}{*}{Sch\"{o}nflies} & $(0,0,0) + ~(0,1/2,1/2) + $ & & \\ & & & $ (1/2,0,1/2) + ~(1/2,1/2,0) +$ & &\\ \hline
8a&$222$&$D_2$& $(1/8,1/8,1/8)$, $(7/8,7/8,7/8)$ & $A_c^2 A_{c'} + A_c A_{c'}^2 + C_\gamma$ & $\varphi_2[C_2, C'_2]$\\ 
8b&$222$&$D_2$& $(1/8,1/8,5/8)$, $(7/8,7/8,3/8)$ & $C_\gamma$ & $\varphi_2[C_2, T_1T_2T_3^{-1}C'_2]$\\ 
\hline
\multirow{2}{*}{16c} & \multirow{2}{*}{$\overline{1}$} & \multirow{2}{*}{$C_i$} & $(0,0,0)$, $(3/4,3/4,0)$, & \multirow{2}{*}{$A_i (A_i^2 + B_{xy+xz+yz})$} & \multirow{2}{*}{$\varphi_1[I]$}\\
& & & $(3/4,0,3/4)$, $(0,3/4,3/4)$ & & \\ \hline
\multirow{2}{*}{16d} & \multirow{2}{*}{$\overline{1}$} & \multirow{2}{*}{$C_i$} & $(1/2,1/2,1/2)$, $(1/4,1/4,1/2)$, & \multirow{2}{*}{$A_i B_{xy+xz+yz}$} & \multirow{2}{*}{$\varphi_1[T_1T_2T_3I]$}\\
& & & $(1/4,1/2,1/4)$, $(1/2,1/4,1/4)$ & & \\ \hline
\hline 
 \end{tabular} 
 \end{center}

\subsection*{No. 71: $Immm$}\label{subsub:sg71}

This group is generated by three translations $T_{1,2,3}$ as given in Eqs.~\eqref{TransBravaisI}, a two-fold rotation $C_2$, a two-fold rotation $C'_2$, and an inversion $I$:
\begin{subequations}
 \begin{align}
C_2 &\colon (x,y,z)\rightarrow (-x, -y, z),\\ 
C'_2 &\colon (x,y,z)\rightarrow (-x, y, -z),\\ 
I &\colon (x,y,z)\rightarrow (-x, -y, -z).
\end{align}
\end{subequations}

The $\mathbb{Z}_2$ cohomology ring is given by

\begin{equation}
\mathbb{Z}_2[A_i,A_{c'},A_c,A_{x+y+z},B_\beta,B_{x(y+z)},B_{y(x+z)},C_{xyz}]/\langle\mathcal{R}_2,\mathcal{R}_3,\mathcal{R}_4,\mathcal{R}_5,\mathcal{R}_6\rangle
 \end{equation}
where the relations are 
\begin{subequations} 
 \begin{align}
\mathcal{R}_2\colon & ~~
A_{c'} A_{x+y+z},~~A_c A_{x+y+z},~~A_{x+y+z} (A_i + A_{x+y+z}),\\
\mathcal{R}_3\colon & ~~
A_{x+y+z} B_\beta,~~A_{x+y+z} B_{x(y+z)},~~A_c B_\beta + A_i B_\beta + A_c B_{x(y+z)} + A_c B_{y(x+z)} + A_{c'} B_{y(x+z)},~~A_{x+y+z} B_{y(x+z)},\\
\mathcal{R}_4\colon & ~~
A_{x+y+z} C_{xyz},~~A_c^2 B_\beta + A_{c'}^2 B_\beta + A_c A_i B_\beta + A_{c'} A_i B_\beta + B_\beta^2 + A_c A_{c'} B_{x(y+z)},\nonumber\\&~~A_c A_{c'} B_\beta + A_c A_i B_\beta + A_{c'} A_i B_\beta + A_i^2 B_\beta + B_\beta B_{x(y+z)} + A_c C_{xyz} + A_{c'} C_{xyz},\nonumber\\&~~A_c^2 B_\beta + A_c A_i B_\beta + A_c A_{c'} B_{x(y+z)} + A_{c'}^2 B_{y(x+z)} + A_{c'} A_i B_{y(x+z)} + B_\beta B_{y(x+z)} + A_c C_{xyz},\nonumber\\&~~B_{x(y+z)} (A_c A_{c'} + A_c A_i + A_{c'} A_i + A_i^2 + B_{x(y+z)}),\nonumber\\&~~
(A_c A_{c'} + A_c A_i +A_{c'} A_i + A_i^2) B_\beta + A_c(A_{c'} + A_i)B_{x(y+z)} + (A_{c'}^2 + A_i^2) B_{y(x+z)}  + B_{x(y+z)} B_{y(x+z)}  + (A_c + A_i) C_{xyz},
\nonumber\\&~~A_c^2 B_\beta + A_c A_i B_\beta + A_c A_i B_{x(y+z)} + A_{c'} A_i B_{y(x+z)} + A_i^2 B_{y(x+z)} + B_{y(x+z)}^2,\\
\mathcal{R}_5\colon & ~~
(A_c^2 + A_{c'}^2 + A_c A_i + A_{c'} A_i + B_\beta) C_{xyz},~~B_{x(y+z)} C_{xyz},~~(A_c^2 + A_c A_{c'} + A_{c'} A_i + A_i^2 + B_{y(x+z)}) C_{xyz},\\
\mathcal{R}_6\colon & ~~
C_{xyz} (A_c^2 A_{c'} + A_c A_{c'}^2 + A_c^2 A_i + A_c A_{c'} A_i + A_{c'}^2 A_i + A_i^3 + C_{xyz}).
\end{align} 
 \end{subequations}
We have the following table regarding IWPs and group cohomology at degree 3.
\begin{center}
\resizebox{\columnwidth}{!}{
\begin{tabular}{c|cc|c|c|c}\hline\hline {Wyckoff}&\multicolumn{2}{c|}{Little group}& {Coordinates}&\multirow{2}{*}{LSM anomaly class}&\multirow{2}{*}{Topo. inv.} \\ \cline{2-4} position & Intl. & Sch\"{o}nflies & $(0,0,0) + ~(1/2,1/2,1/2) + $ & &\\ \hline
2a&$mmm$&$D_{2h}$& $(0,0,0)$ & \begin{tabular}{c}$A_c^2 A_{c'} + A_c A_{c'}^2 + A_c^2 A_i + A_c A_{c'} A_i + A_{c'}^2 A_i + A_i^3 $\\
$ + A_i^2 A_{x+y+z} + (A_c  + A_{c'}  + A_i) B_{x(y+z)} + C_{xyz}$ \end{tabular} & $\varphi_2[C_2, C'_2]$\\\hline 
2b&$mmm$&$D_{2h}$& $(0,1/2,1/2)$ & $C_{xyz}$ & $\varphi_2[T_2T_3C_2, T_2T_3C'_2]$\\ 
2c&$mmm$&$D_{2h}$& $(1/2,1/2,0)$ & $A_c B_{x(y+z)} + A_{c'} B_{y(x+z)} + A_i B_{y(x+z)} + C_{xyz}$ & $\varphi_2[C_2, T_1T_2C'_2]$\\ 
2d&$mmm$&$D_{2h}$& $(1/2,0,1/2)$ & $(A_{c'}  + A_i)(B_{x(y+z)} + B_{y(x+z)}) + C_{xyz}$ & $\varphi_2[T_1T_3C_2, C'_2]$\\ 
\hline
\multirow{2}{*}{8k} & \multirow{2}{*}{$\overline{1}$} & \multirow{2}{*}{$C_i$} & $(1/4,1/4,1/4)$, $(3/4,3/4,1/4)$, & \multirow{2}{*}{$A_i^2 A_{x+y+z}$} & \multirow{2}{*}{$\varphi_1[T_1T_2T_3I]$}\\
& & & $(3/4,1/4,3/4)$, $(1/4,3/4,3/4)$ & & \\ \hline
\hline 
 \end{tabular} }
 \end{center}

\subsection*{No. 72: $Ibam$}\label{subsub:sg72}

This group is generated by three translations $T_{1,2,3}$ as given in Eqs.~\eqref{TransBravaisI}, a two-fold rotation $C_2$, a two-fold rotation $C'_2$, and an inversion $I$:
\begin{subequations}
 \begin{align}
C_2 &\colon (x,y,z)\rightarrow (-x, -y, z),\\ 
C'_2 &\colon (x,y,z)\rightarrow (-x, y, -z + 1/2),\\ 
I &\colon (x,y,z)\rightarrow (-x, -y, -z).
\end{align}
\end{subequations}

The $\mathbb{Z}_2$ cohomology ring is given by

\begin{equation}
\mathbb{Z}_2[A_i,A_{c'},A_c,A_{x+y+z},B_{z(x+y)}]/\langle\mathcal{R}_2,\mathcal{R}_3,\mathcal{R}_4\rangle
 \end{equation}
where the relations are 
\begin{subequations} 
 \begin{align}
\mathcal{R}_2\colon & ~~
A_{c'} A_{x+y+z},~~A_{c'} A_i + A_c A_{x+y+z},~~A_{c'} A_i + A_i A_{x+y+z} + A_{x+y+z}^2,\\
\mathcal{R}_3\colon & ~~
A_{x+y+z} B_{z(x+y)},\\
\mathcal{R}_4\colon & ~~
B_{z(x+y)} (A_c^2 + A_c A_{c'} + A_i^2 + B_{z(x+y)}).
\end{align} 
 \end{subequations}
We have the following table regarding IWPs and group cohomology at degree 3.
\begin{center}
\resizebox{\columnwidth}{!}{
\begin{tabular}{c|cc|c|c|c}\hline\hline {Wyckoff}&\multicolumn{2}{c|}{Little group}& {Coordinates}&\multirow{2}{*}{LSM anomaly class}&\multirow{2}{*}{Topo. inv.} \\ \cline{2-4} position & Intl. & Sch\"{o}nflies & $(0,0,0) + ~(1/2,1/2,1/2) + $ & &\\ \hline
4a&$222$&$D_2$& $(0,0,1/4)$, $(0,0,3/4)$ & $A_{c'} (A_c^2 + A_c A_{c'} + A_i^2 + B_{z(x+y)})$ & $\varphi_2[C_2, C'_2]$\\ 
4b&$222$&$D_2$& $(1/2,0,1/4)$, $(1/2,0,3/4)$ & $A_{c'} B_{z(x+y)}$ & $\varphi_2[T_2T_3C_2, T_2T_3C'_2]$\\ 
4c&$2/m$&$C_{2h}$& $(0,0,0)$, $(1/2,1/2,0)$ & $A_i (A_c^2 + A_{c'} A_i + A_i^2 + A_i A_{x+y+z} + B_{z(x+y)})$ & $\varphi_1[I]$\\ 
4d&$2/m$&$C_{2h}$& $(1/2,0,0)$, $(0,1/2,0)$ & $A_i B_{z(x+y)}$ & $\varphi_1[T_2T_3I]$\\ 
\hline
\multirow{2}{*}{8e} & \multirow{2}{*}{$\overline{1}$} & \multirow{2}{*}{$C_i$} & $(1/4,1/4,1/4)$, $(3/4,3/4,1/4)$, & \multirow{2}{*}{$A_i^2 (A_{c'} + A_{x+y+z})$} & \multirow{2}{*}{$\varphi_1[T_1T_2T_3I]$}\\
& & & $(1/4,3/4,3/4)$, $(3/4,1/4,3/4)$ & & \\ \hline
\hline 
 \end{tabular} }
 \end{center}

\subsection*{No. 73: $Ibca$}\label{subsub:sg73}

This group is generated by three translations $T_{1,2,3}$ as given in Eqs.~\eqref{TransBravaisI}, a two-fold rotation $C_2$, a two-fold rotation $C'_2$, and an inversion $I$:
\begin{subequations}
 \begin{align}
C_2 &\colon (x,y,z)\rightarrow (-x, -y + 1/2, z),\\ 
C'_2 &\colon (x,y,z)\rightarrow (-x + 1/2, y, -z),\\ 
I &\colon (x,y,z)\rightarrow (-x, -y, -z).
\end{align}
\end{subequations}

The $\mathbb{Z}_2$ cohomology ring is given by

\begin{equation}
\mathbb{Z}_2[A_i,A_{c'},A_c,A_{x+y+z}]/\langle\mathcal{R}_2\rangle
 \end{equation}
where the relations are 
\begin{subequations} 
 \begin{align}
\mathcal{R}_2\colon & ~~
A_c A_{c'} + A_c A_i + A_{c'} A_i + A_{c'} A_{x+y+z},~A_c A_{c'} + A_{c'} A_i + A_c A_{x+y+z},~A_c A_{c'} + A_c A_i + A_{c'} A_i + A_i A_{x+y+z} + A_{x+y+z}^2.
\end{align} 
 \end{subequations}
We have the following table regarding IWPs and group cohomology at degree 3.
\begin{center}
\begin{tabular}{c|cc|c|c|c}\hline\hline {Wyckoff}&\multicolumn{2}{c|}{Little group}& {Coordinates}&\multirow{2}{*}{LSM anomaly class}&\multirow{2}{*}{Topo. inv.} \\ \cline{2-4} position & Intl. & Sch\"{o}nflies & $(0,0,0) + ~(1/2,1/2,1/2) + $ & &\\ \hline
\multirow{2}{*}{8a} & \multirow{2}{*}{$\overline{1}$} & \multirow{2}{*}{$C_i$} & $(0,0,0)$, $(1/2,0,1/2)$, & \multirow{2}{*}{$A_i^2 (A_i + A_{x+y+z})$} & \multirow{2}{*}{$\varphi_1[I]$}\\
& & & $(0,1/2,1/2)$, $(1/2,1/2,0)$ & & \\ \hline
\multirow{2}{*}{8b} & \multirow{2}{*}{$\overline{1}$} & \multirow{2}{*}{$C_i$} & $(1/4,1/4,1/4)$, $(1/4,3/4,3/4)$, & \multirow{2}{*}{$A_i^2 A_{x+y+z}$} & \multirow{2}{*}{$\varphi_1[T_1T_2T_3I]$}\\
& & & $(3/4,3/4,1/4)$, $(3/4,1/4,3/4)$ & & \\ \hline
\multirow{2}{*}{8c} & \multirow{2}{*}{$2$} & \multirow{2}{*}{$C_2$} & $(x,0,1/4)$, $(-x+1/2,0,3/4)$, & \multirow{2}{*}{$A_c A_{c'} A_i$} & \multirow{2}{*}{$\varphi_2[C'_2I, T_2C_2C'_2]$}\\
& & & $(-x,0,3/4)$, $(x+1/2,0,1/4)$ & & \\ \hline
\multirow{2}{*}{8d} & \multirow{2}{*}{$2$} & \multirow{2}{*}{$C_2$} & $(1/4,y,0)$, $(1/4,-y,1/2)$, & \multirow{2}{*}{$A_{c'} (A_c + A_{c'}) A_i$} & \multirow{2}{*}{$\varphi_2[C_2I, C'_2]$}\\
& & & $(3/4,-y,0)$, $(3/4,y,1/2)$ & & \\ \hline
\multirow{2}{*}{8e} & \multirow{2}{*}{$2$} & \multirow{2}{*}{$C_2$} & $(0,1/4,z)$, $(0,3/4,-z+1/2)$, & \multirow{2}{*}{$A_c (A_c + A_{c'}) A_i$} & \multirow{2}{*}{$\varphi_2[T_2C_2C'_2I, C_2]$}\\
& & & $(0,3/4,-z)$, $(0,1/4,z+1/2)$ & & \\ \hline
\hline 
 \end{tabular} 
 \end{center}

\subsection*{No. 74: $Imma$}\label{subsub:sg74}

This group is generated by three translations $T_{1,2,3}$ as given in Eqs.~\eqref{TransBravaisI}, a two-fold rotation $C_2$, a two-fold rotation $C'_2$, and an inversion $I$:
\begin{subequations}
 \begin{align}
C_2 &\colon (x,y,z)\rightarrow (-x, -y + 1/2, z),\\ 
C'_2 &\colon (x,y,z)\rightarrow (-x + 1/2, y, -z + 1/2),\\ 
I &\colon (x,y,z)\rightarrow (-x, -y, -z).
\end{align}
\end{subequations}

The $\mathbb{Z}_2$ cohomology ring is given by

\begin{equation}
\mathbb{Z}_2[A_i,A_{c'},A_c,A_{x+y+z},B_{y(x+z)},B_{z(x+y)}]/\langle\mathcal{R}_2,\mathcal{R}_3,\mathcal{R}_4\rangle
 \end{equation}
where the relations are 
\begin{subequations} 
 \begin{align}
\mathcal{R}_2\colon & ~~
A_c A_{c'} + A_c A_i + A_{c'} A_i + A_{c'} A_{x+y+z},~~A_c (A_{c'} + A_{x+y+z}),~~A_c A_{c'} + A_c A_i + A_i A_{x+y+z} + A_{x+y+z}^2,\\
\mathcal{R}_3\colon & ~~
(A_{c'} + A_{x+y+z}) B_{y(x+z)},~~A_{c'} B_{y(x+z)} + A_c B_{z(x+y)},~~A_{c'} B_{y(x+z)} + A_i B_{y(x+z)} + A_i B_{z(x+y)} + A_{x+y+z} B_{z(x+y)},\\
\mathcal{R}_4\colon & ~~
A_c A_{c'}^3 + A_c A_{c'}^2 A_i + A_{c'} A_i^3 + A_{c'} A_i^2 A_{x+y+z} + A_{c'} A_i B_{y(x+z)} + A_i^2 B_{y(x+z)} + B_{y(x+z)}^2,\nonumber\\&~~A_c A_{c'}^3 + A_c A_{c'}^2 A_i + A_{c'} A_i^3 + A_{c'} A_i^2 A_{x+y+z} + A_{c'} A_i B_{y(x+z)} + A_i^2 B_{y(x+z)} + B_{y(x+z)} B_{z(x+y)},\nonumber\\&~~A_c A_{c'}^3 + A_c A_{c'}^2 A_i + A_{c'} A_i^3 + A_{c'} A_i^2 A_{x+y+z} + A_{c'} A_i B_{z(x+y)} + A_i^2 B_{z(x+y)} + B_{z(x+y)}^2.
\end{align} 
 \end{subequations}
We have the following table regarding IWPs and group cohomology at degree 3.
\begin{center}
\resizebox{\columnwidth}{!}{
\begin{tabular}{c|cc|c|c|c}\hline\hline {Wyckoff}&\multicolumn{2}{c|}{Little group}& {Coordinates}&\multirow{2}{*}{LSM anomaly class}&\multirow{2}{*}{Topo. inv.} \\ \cline{2-4} position & Intl. & Sch\"{o}nflies & $(0,0,0) + ~(1/2,1/2,1/2) + $ & &\\ \hline
4a&$2/m$&$C_{2h}$& $(0,0,0)$, $(0,1/2,0)$ & $A_c A_{c'}^2 + A_c A_{c'} A_i + A_{c'} A_i^2 + A_{c'} A_i A_{x+y+z} + A_{c'} B_{y(x+z)} + A_i B_{y(x+z)}$ & $\varphi_1[I]$\\ 
4b&$2/m$&$C_{2h}$& $(0,0,1/2)$, $(0,1/2,1/2)$ & $A_c A_{c'}^2 + A_c A_{c'} A_i + A_i^3 + A_i^2 A_{x+y+z} + A_{c'} B_{y(x+z)} + A_i B_{y(x+z)}$ & $\varphi_1[T_1T_2I]$\\ 
4c&$2m$&$C_{2h}$& $(1/4,1/4,1/4)$, $(3/4,1/4,1/4)$ & $(A_{c'} + A_i) (A_{c'} A_i + A_i A_{x+y+z} + B_{y(x+z)} + B_{z(x+y)})$ & $\varphi_1[T_1T_2T_3I]$\\ 
4d&$2m$&$C_{2h}$& $(1/4,1/4,3/4)$, $(3/4,1/4,3/4)$ & $(A_{c'} + A_i) (B_{y(x+z)} + B_{z(x+y)})$ & $\varphi_1[T_3I]$\\ 
4e&$mm2$&$C_{2v}$& $(0,1/4,z)$, $(0,3/4,-z)$ & $(A_c + A_{c'}) B_{y(x+z)}$ & $\varphi_2[T_1T_2, C_2]$\\ 
\hline
\hline 
 \end{tabular} }
 \end{center}

\subsection*{No. 75: $P4$}\label{subsub:sg75}

This group is generated by three translations $T_{1,2,3}$ as given in Eqs.~\eqref{TransBravaisP}, a two-fold rotation $C_2$, and a four-fold rotation $C_4$:
\begin{subequations}
 \begin{align}
C_2 &\colon (x,y,z)\rightarrow (-x, -y, z),\\ 
C_4 &\colon (x,y,z)\rightarrow (-y, x, z).
\end{align}
\end{subequations}

The $\mathbb{Z}_2$ cohomology ring is given by

\begin{equation}
\mathbb{Z}_2[A_{\mathsf{q}},A_{x+y},A_z,B_\alpha,B_{xy}]/\langle\mathcal{R}_2,\mathcal{R}_3,\mathcal{R}_4\rangle
 \end{equation}
where the relations are 
\begin{subequations} 
 \begin{align}
\mathcal{R}_2\colon & ~~
A_{\mathsf{q}} A_{x+y},~~A_{\mathsf{q}}^2,~~A_z^2,\\
\mathcal{R}_3\colon & ~~
A_{x+y}^3 + A_{x+y} B_\alpha + A_{\mathsf{q}} B_{xy},~~A_{x+y} (A_{x+y}^2 + B_\alpha + B_{xy}),\\
\mathcal{R}_4\colon & ~~
B_{xy} (B_\alpha + B_{xy}).
\end{align} 
 \end{subequations}
We have the following table regarding IWPs and group cohomology at degree 3.
\begin{center}
\begin{tabular}{c|cc|c|c|c}\hline\hline {Wyckoff}&\multicolumn{2}{c|}{Little group}& \multirow{2}{*}{Coordinates}&\multirow{2}{*}{LSM anomaly class}&\multirow{2}{*}{Topo. inv.} \\ \cline{2-3} position & Intl. & Sch\"{o}nflies & & & \\ \hline
1a&$4$&$C_4$& $(0,0,z)$ & $A_z (A_{x+y}^2 + B_\alpha + B_{xy})$ & $\varphi_2[T_3, C_2]$\\ 
1b&$4$&$C_4$& $(1/2,1/2,z)$ & $A_z B_{xy}$ & $\varphi_2[T_3, T_1T_2C_2]$\\ 
2c&$2$&$C_2$& $(0,1/2,z)$, $(1/2,0,z)$ & $A_{x+y}^2 A_z$ & $\varphi_2[T_3, T_1C_2]$\\ 
\hline
\hline 
 \end{tabular} 
 \end{center}

\subsection*{No. 76: $P4_1$}\label{subsub:sg76}

This group is generated by three translations $T_{1,2,3}$ as given in Eqs.~\eqref{TransBravaisP}, a two-fold screw $S_2$, and a four-fold screw $S_4$:
\begin{subequations}
 \begin{align}
S_2 &\colon (x,y,z)\rightarrow (-x, -y, z + 1/2),\\ 
S_4 &\colon (x,y,z)\rightarrow (-y, x, z + 1/4).
\end{align}
\end{subequations}

The $\mathbb{Z}_2$ cohomology ring is given by

\begin{equation}
\mathbb{Z}_2[A_{\mathsf{q}},A_{x+y},B_{xy}]/\langle\mathcal{R}_2,\mathcal{R}_3,\mathcal{R}_4\rangle
 \end{equation}
where the relations are 
\begin{subequations} 
 \begin{align}
\mathcal{R}_2\colon & ~~
A_{\mathsf{q}} A_{x+y},~~A_{\mathsf{q}}^2,\\
\mathcal{R}_3\colon & ~~
A_{x+y}^3 + A_{\mathsf{q}} B_{xy},~~A_{x+y} (A_{x+y}^2 + B_{xy}),\\
\mathcal{R}_4\colon & ~~
B_{xy}^2.
\end{align} 
 \end{subequations}
We have the following table regarding IWPs and group cohomology at degree 3.
\begin{center}
\begin{tabular}{c|cc|c|c|c}\hline\hline {Wyckoff}&\multicolumn{2}{c|}{Little group}& \multirow{2}{*}{Coordinates}&\multirow{2}{*}{LSM anomaly class}&\multirow{2}{*}{Topo. inv.} \\ \cline{2-3} position & Intl. & Sch\"{o}nflies & & & \\ \hline
\multirow{2}{*}{4a} & \multirow{2}{*}{$1$} & \multirow{2}{*}{$C_1$} & $(x,y,z)$, $(-x,-y,z+1/2)$, & \multirow{2}{*}{$A_{x+y}^3$} & \multirow{2}{*}{$\widehat{\varphi_3}[T_1, T_2, S_4]$}\\
& & & $(-y,x,z+1/4)$, $(y,-x,z+3/4)$ & & \\ \hline
\hline 
 \end{tabular} 
 \end{center}
 Here the topological invariant can be chosen to be
\begin{equation}\label{TI_76}
\begin{aligned}
\widehat{\varphi_3}[T_1, T_2, S_4] =& 
\lambda(T_1,T_2,T_1^{-1}S_4)+\lambda(T_2,T_1,T_1^{-1}S_4)+
\lambda(T_1,T_1^{-1}S_4,T_1)+
\lambda(T_2,S_4,T_2)\\
&+\lambda(S_4,T_1,T_2)+
\lambda(S_4,T_2,T_1).
\end{aligned}
\end{equation}

\subsection*{No. 77: $P4_2$}\label{subsub:sg77}

This group is generated by three translations $T_{1,2,3}$ as given in Eqs.~\eqref{TransBravaisP}, a two-fold rotation $C_2$, and a four-fold screw $S_4$:
\begin{subequations}
 \begin{align}
C_2 &\colon (x,y,z)\rightarrow (-x, -y, z),\\ 
S_4 &\colon (x,y,z)\rightarrow (-y, x, z + 1/2).
\end{align}
\end{subequations}

The $\mathbb{Z}_2$ cohomology ring is given by

\begin{equation}
\mathbb{Z}_2[A_{\mathsf{q}},A_{x+y},A_z,B_{xy}]/\langle\mathcal{R}_2,\mathcal{R}_3,\mathcal{R}_4\rangle
 \end{equation}
where the relations are 
\begin{subequations} 
 \begin{align}
\mathcal{R}_2\colon & ~~
A_{\mathsf{q}} A_{x+y},~~A_{\mathsf{q}}^2,\\
\mathcal{R}_3\colon & ~~
A_{x+y}^3 + A_{x+y} A_z^2 + A_{\mathsf{q}} B_{xy},~~A_{x+y} (A_{x+y}^2 + A_z^2 + B_{xy}),\\
\mathcal{R}_4\colon & ~~
B_{xy} (A_z^2 + B_{xy}).
\end{align} 
 \end{subequations}
We have the following table regarding IWPs and group cohomology at degree 3.
\begin{center}
\begin{tabular}{c|cc|c|c|c}\hline\hline {Wyckoff}&\multicolumn{2}{c|}{Little group}& \multirow{2}{*}{Coordinates}&\multirow{2}{*}{LSM anomaly class}&\multirow{2}{*}{Topo. inv.} \\ \cline{2-3} position & Intl. & Sch\"{o}nflies & & & \\ \hline
2a&$2$&$C_2$& $(0,0,z)$, $(0,0,z+1/2)$ & $A_{x+y}^3 + A_{\mathsf{q}} A_z^2 + A_{x+y} A_z^2$ & $\varphi_2[S_4, C_2]$\\ 
2b&$2$&$C_2$& $(1/2,1/2,z)$, $(1/2,1/2,z+1/2)$ & $A_{x+y} (A_{x+y}^2 + A_z^2)$ & $\varphi_2[T_1S_4, T_1T_2C_2]$\\ 
2c&$2$&$C_2$& $(0,1/2,z)$, $(1/2,0,z+1/2)$ & $A_{x+y}^2 (A_{x+y} + A_z)$ & $\varphi_2[T_3, T_1C_2]$\\ 
\hline
\hline 
 \end{tabular} 
 \end{center}

\subsection*{No. 78: $P4_3$}\label{subsub:sg78}

This group is generated by three translations $T_{1,2,3}$ as given in Eqs.~\eqref{TransBravaisP}, a two-fold screw $S_2$, and a four-fold screw $S_4$:
\begin{subequations}
 \begin{align}
S_2 &\colon (x,y,z)\rightarrow (-x, -y, z + 1/2),\\ 
S_4 &\colon (x,y,z)\rightarrow (-y, x, z + 3/4).
\end{align}
\end{subequations}

The $\mathbb{Z}_2$ cohomology ring is given by

\begin{equation}
\mathbb{Z}_2[A_{\mathsf{q}},A_{x+y},B_{xy}]/\langle\mathcal{R}_2,\mathcal{R}_3,\mathcal{R}_4\rangle
 \end{equation}
where the relations are 
\begin{subequations} 
 \begin{align}
\mathcal{R}_2\colon & ~~
A_{\mathsf{q}} A_{x+y},~~A_{\mathsf{q}}^2,\\
\mathcal{R}_3\colon & ~~
A_{x+y}^3 + A_{\mathsf{q}} B_{xy},~~A_{x+y} (A_{x+y}^2 + B_{xy}),\\
\mathcal{R}_4\colon & ~~
B_{xy}^2.
\end{align} 
 \end{subequations}
We have the following table regarding IWPs and group cohomology at degree 3.
\begin{center}
\begin{tabular}{c|cc|c|c|c}\hline\hline {Wyckoff}&\multicolumn{2}{c|}{Little group}& \multirow{2}{*}{Coordinates}&\multirow{2}{*}{LSM anomaly class}&\multirow{2}{*}{Topo. inv.} \\ \cline{2-3} position & Intl. & Sch\"{o}nflies & & & \\ \hline
\multirow{2}{*}{4a} & \multirow{2}{*}{$1$} & \multirow{2}{*}{$C_1$} & $(x,y,z)$, $(-x,-y,z+1/2)$, & \multirow{2}{*}{$A_{x+y}^3$} & \multirow{2}{*}{$\widehat{\varphi_3}[T_1, T_2, S_4]$}\\
& & & $(-y,x,z+3/4)$, $(y,-x,z+1/4)$ & & \\ \hline
\hline 
 \end{tabular} 
 \end{center}
  Here the topological invariant $\widehat{\varphi_3}$ can be chosen to be the same as that of group No.76 $P4_1$, given by Eq.~\eqref{TI_76}.

\subsection*{No. 79: $I4$}\label{subsub:sg79}

This group is generated by three translations $T_{1,2,3}$ as given in Eqs.~\eqref{TransBravaisI}, a two-fold rotation $C_2$, and a four-fold rotation $C_4$:
\begin{subequations}
 \begin{align}
C_2 &\colon (x,y,z)\rightarrow (-x, -y, z),\\ 
C_4 &\colon (x,y,z)\rightarrow (-y, x, z).
\end{align}
\end{subequations}

The $\mathbb{Z}_2$ cohomology ring is given by

\begin{equation}
\mathbb{Z}_2[A_{\mathsf{q}},A_{x+y+z},B_\alpha,B_\beta,B_{z(x+y)},C_\gamma,C_{xyz}]/\langle\mathcal{R}_2,\mathcal{R}_3,\mathcal{R}_4,\mathcal{R}_5,\mathcal{R}_6\rangle
 \end{equation}
where the relations are 
\begin{subequations} 
 \begin{align}
\mathcal{R}_2\colon & ~~
A_{\mathsf{q}} A_{x+y+z},~~A_{\mathsf{q}}^2,~~A_{x+y+z}^2,\\
\mathcal{R}_3\colon & ~~
A_{x+y+z} B_\alpha + A_{\mathsf{q}} B_\beta,~~A_{x+y+z} (B_\alpha + B_\beta),~~A_{x+y+z} B_\alpha + A_{\mathsf{q}} B_{z(x+y)},~~A_{x+y+z} (B_\alpha + B_{z(x+y)}),\\
\mathcal{R}_4\colon & ~~
A_{\mathsf{q}} C_\gamma + A_{x+y+z} C_\gamma + A_{\mathsf{q}} C_{xyz},~~A_{x+y+z} C_{xyz},~~B_\alpha B_\beta + B_\alpha B_{z(x+y)} + A_{\mathsf{q}} C_\gamma,~~B_\alpha B_\beta + B_\beta^2 + A_{\mathsf{q}} C_\gamma,\nonumber\\&~~B_\alpha B_\beta + B_\beta B_{z(x+y)} + A_{\mathsf{q}} C_\gamma + A_{x+y+z} C_\gamma,~~B_\alpha B_\beta + B_{z(x+y)}^2 + A_{\mathsf{q}} C_\gamma,\\
\mathcal{R}_5\colon & ~~
A_{x+y+z} B_\alpha^2 + B_\beta C_\gamma + B_{z(x+y)} C_\gamma,~~A_{x+y+z} B_\alpha^2 + B_\alpha C_\gamma + B_\beta C_\gamma + B_\alpha C_{xyz},~~B_\beta C_{xyz},~~B_{z(x+y)} C_{xyz},\\
\mathcal{R}_6\colon & ~~
B_\alpha^2 B_\beta + A_{\mathsf{q}} B_\alpha C_\gamma + C_\gamma^2,~~C_\gamma C_{xyz},~~C_{xyz}^2.
\end{align} 
 \end{subequations}
We have the following table regarding IWPs and group cohomology at degree 3.
\begin{center}
\begin{tabular}{c|cc|c|c|c}\hline\hline {Wyckoff}&\multicolumn{2}{c|}{Little group}& {Coordinates}&\multirow{2}{*}{LSM anomaly class}&\multirow{2}{*}{Topo. inv.} \\ \cline{2-4} position & Intl. & Sch\"{o}nflies & $(0,0,0) + ~(1/2,1/2,1/2) + $ & &\\ \hline
2a&$4$&$C_4$& $(0,0,z)$ & $C_{xyz}$ & $\varphi_2[T_1T_2, C_2]$\\ 
4b&$2$&$C_2$& $(0,1/2,z)$, $(1/2,0,z)$ & $A_{x+y+z} B_\alpha$ & $\varphi_2[T_2C_4, T_2T_3C_2]$\\ 
\hline
\hline 
 \end{tabular} 
 \end{center}

\subsection*{No. 80: $I4_1$}\label{subsub:sg80}

This group is generated by three translations $T_{1,2,3}$ as given in Eqs.~\eqref{TransBravaisI}, a two-fold rotation $C_2$, and a four-fold screw $S_4$:
\begin{subequations}
 \begin{align}
C_2 &\colon (x,y,z)\rightarrow (-x, -y, z),\\ 
S_4 &\colon (x,y,z)\rightarrow (-y -1/2, x, z + 3/4).
\end{align}
\end{subequations}

The $\mathbb{Z}_2$ cohomology ring is given by

\begin{equation}
\mathbb{Z}_2[A_{\mathsf{q}},A_{x+y+z},B_{z(x+y)},C_\gamma]/\langle\mathcal{R}_2,\mathcal{R}_3,\mathcal{R}_4,\mathcal{R}_5,\mathcal{R}_6\rangle
 \end{equation}
where the relations are 
\begin{subequations} 
 \begin{align}
\mathcal{R}_2\colon & ~~
A_{\mathsf{q}} A_{x+y+z},~~A_{\mathsf{q}}^2,\\
\mathcal{R}_3\colon & ~~
A_{\mathsf{q}} B_{z(x+y)},~~A_{x+y+z} B_{z(x+y)},\\
\mathcal{R}_4\colon & ~~
A_{\mathsf{q}} C_\gamma,~~B_{z(x+y)}^2,\\
\mathcal{R}_5\colon & ~~
B_{z(x+y)} C_\gamma,\\
\mathcal{R}_6\colon & ~~
C_\gamma^2.
\end{align} 
 \end{subequations}
We have the following table regarding IWPs and group cohomology at degree 3.
\begin{center}
\begin{tabular}{c|cc|c|c|c}\hline\hline {Wyckoff}&\multicolumn{2}{c|}{Little group}& {Coordinates}&\multirow{2}{*}{LSM anomaly class}&\multirow{2}{*}{Topo. inv.} \\ \cline{2-4} position & Intl. & Sch\"{o}nflies & $(0,0,0) + ~(1/2,1/2,1/2) + $ & &\\ \hline
4a&$2$&$C_2$& $(0,0,z)$, $(0,1/2,z+1/4)$ & $C_\gamma$ & $\varphi_2[T_1T_2, C_2]$\\ 
\hline
\hline 
 \end{tabular} 
 \end{center}

\subsection*{No. 81: $P\overline4$}\label{subsub:sg81}

This group is generated by three translations $T_{1,2,3}$ as given in Eqs.~\eqref{TransBravaisP}, a two-fold rotation $C_2$, and a four-fold roto-inversion $\overline{C}_4$:
\begin{subequations}
 \begin{align}
C_2 &\colon (x,y,z)\rightarrow (-x, -y, z),\\ 
\overline{C}_4 &\colon (x,y,z)\rightarrow (y, -x, -z).
\end{align}
\end{subequations}

The $\mathbb{Z}_2$ cohomology ring is given by

\begin{equation}
\mathbb{Z}_2[A_{\mathsf{q}},A_{x+y},A_z,B_\alpha,B_{xy}]/\langle\mathcal{R}_2,\mathcal{R}_3,\mathcal{R}_4\rangle
 \end{equation}
where the relations are 
\begin{subequations} 
 \begin{align}
\mathcal{R}_2\colon & ~~
A_{\mathsf{q}} A_{x+y},~~A_{\mathsf{q}}^2,~~A_z (A_{\mathsf{q}} + A_z),\\
\mathcal{R}_3\colon & ~~
A_{x+y}^3 + A_{x+y} B_\alpha + A_{\mathsf{q}} B_{xy},~~A_{x+y} (A_{x+y}^2 + B_\alpha + B_{xy}),\\
\mathcal{R}_4\colon & ~~
B_{xy} (B_\alpha + B_{xy}).
\end{align} 
 \end{subequations}
We have the following table regarding IWPs and group cohomology at degree 3.
\begin{center}
\begin{tabular}{c|cc|c|c|c}\hline\hline {Wyckoff}&\multicolumn{2}{c|}{Little group}& \multirow{2}{*}{Coordinates}&\multirow{2}{*}{LSM anomaly class}&\multirow{2}{*}{Topo. inv.} \\ \cline{2-3} position & Intl. & Sch\"{o}nflies & & & \\ \hline
1a&$\overline{4}$&$S_4$& $(0,0,0)$ & %
$(A_{\mathsf{q}} + A_z)(A_{x+y}^2 + B_\alpha + B_{xy})$
& $\varphi_2[\overline{C}_4, C_2]$\\ 
1b&$\overline{4}$&$S_4$& $(0,0,1/2)$ & $A_z (A_{x+y}^2 + B_\alpha + B_{xy})$ & $\varphi_2[T_3\overline{C}_4, C_2]$\\ 
1c&$\overline{4}$&$S_4$& $(1/2,1/2,0)$ & $A_{x+y}^3 + A_{x+y} B_\alpha + A_z B_{xy}$ & $\varphi_2[T_2\overline{C}_4, T_1T_2C_2]$\\ 
1d&$\overline{4}$&$S_4$& $(1/2,1/2,1/2)$ & $A_z B_{xy}$ & $\varphi_2[T_2T_3\overline{C}_4, T_1T_2C_2]$\\ 
2g&$2$&$C_2$& $(0,1/2,z)$, $(1/2,0,-z)$ & $A_{x+y}^2 A_z$ & $\varphi_2[T_3, T_1C_2]$\\ 
\hline
\hline 
 \end{tabular} 
 \end{center}

\subsection*{No. 82: $I\overline4$}\label{subsub:sg82}

This group is generated by three translations $T_{1,2,3}$ as given in Eqs.~\eqref{TransBravaisI}, a two-fold rotation $C_2$, and a four-fold roto-inversion $\overline{C}_4$:
\begin{subequations}
 \begin{align}
C_2 &\colon (x,y,z)\rightarrow (-x, -y, z),\\ 
\overline{C}_4 &\colon (x,y,z)\rightarrow (y, -x, -z).
\end{align}
\end{subequations}

The $\mathbb{Z}_2$ cohomology ring is given by

\begin{equation}
\mathbb{Z}_2[A_{\mathsf{q}},A_{x+y+z},B_\alpha,B_\beta,B_{z(x+y)},C_\gamma,C_{xyz}]/\langle\mathcal{R}_2,\mathcal{R}_3,\mathcal{R}_4,\mathcal{R}_5,\mathcal{R}_6\rangle
 \end{equation}
where the relations are 
\begin{subequations} 
 \begin{align}
\mathcal{R}_2\colon & ~~
A_{\mathsf{q}} A_{x+y+z},~~A_{\mathsf{q}}^2,~~A_{x+y+z}^2,\\
\mathcal{R}_3\colon & ~~
A_{x+y+z} B_\alpha + A_{\mathsf{q}} B_\beta,~~A_{x+y+z} (B_\alpha + B_\beta),~~A_{x+y+z} B_\alpha + A_{\mathsf{q}} B_{z(x+y)},~~A_{x+y+z} (B_\alpha + B_{z(x+y)}),\\
\mathcal{R}_4\colon & ~~
A_{\mathsf{q}} C_\gamma + A_{x+y+z} C_\gamma + A_{\mathsf{q}} C_{xyz},~~A_{x+y+z} C_{xyz},~~B_\alpha B_\beta + B_\alpha B_{z(x+y)} + A_{\mathsf{q}} C_\gamma,\nonumber\\&~~B_\alpha B_\beta + B_\beta^2 + A_{\mathsf{q}} C_\gamma,~~B_\alpha B_\beta + B_\beta B_{z(x+y)} + A_{\mathsf{q}} C_\gamma + A_{x+y+z} C_\gamma,~~B_\alpha B_\beta + B_{z(x+y)}^2 + A_{\mathsf{q}} C_\gamma,\\
\mathcal{R}_5\colon & ~~
(B_\beta + B_{z(x+y)}) C_\gamma,~~B_\alpha C_\gamma + B_\beta C_\gamma + B_\alpha C_{xyz},~~B_\beta C_{xyz},~~B_{z(x+y)} C_{xyz},\\
\mathcal{R}_6\colon & ~~
C_\gamma (A_{\mathsf{q}} B_\alpha + C_\gamma),~~C_\gamma (A_{\mathsf{q}} B_\alpha + A_{x+y+z} B_\alpha + C_{xyz}),~~A_{\mathsf{q}} B_\alpha C_\gamma + A_{x+y+z} B_\alpha C_\gamma + C_{xyz}^2.
\end{align} 
 \end{subequations}
We have the following table regarding IWPs and group cohomology at degree 3.
\begin{center}
\begin{tabular}{c|cc|c|c|c}\hline\hline {Wyckoff}&\multicolumn{2}{c|}{Little group}& {Coordinates}&\multirow{2}{*}{LSM anomaly class}&\multirow{2}{*}{Topo. inv.} \\ \cline{2-4} position & Intl. & Sch\"{o}nflies & $(0,0,0) + ~(1/2,1/2,1/2) + $ & &\\ \hline
2a&$\overline{4}$&$S_4$& $(0,0,0)$ & $A_{\mathsf{q}} B_\alpha + A_{x+y+z} B_\alpha + C_{xyz}$ & $\varphi_2[\overline{C}_4, C_2]$\\ 
2b&$\overline{4}$&$S_4$& $(0,0,1/2)$ & $C_{xyz}$ & $\varphi_2[T_1T_2\overline{C}_4, C_2]$\\ 
2c&$\overline{4}$&$S_4$& $(0,1/2,1/4)$ & $A_{x+y+z} B_\alpha + C_\gamma + C_{xyz}$ & $\varphi_2[T_1\overline{C}_4, T_1T_3C_2]$\\ 
2d&$\overline{4}$&$S_4$& $(0,1/2,3/4)$ & $C_\gamma + C_{xyz}$ & $\varphi_2[T_2^{-1}\overline{C}_4, T_1T_3C_2]$\\ 
\hline
\hline 
 \end{tabular} 
 \end{center}

\subsection*{No. 83: $P4/m$}\label{subsub:sg83}

This group is generated by three translations $T_{1,2,3}$ as given in Eqs.~\eqref{TransBravaisP}, a two-fold rotation $C_2$, a four-fold rotation $C_4$, and an inversion $I$:
\begin{subequations}
 \begin{align}
C_2 &\colon (x,y,z)\rightarrow (-x, -y, z),\\ 
C_4 &\colon (x,y,z)\rightarrow (-y, x, z),\\ 
I &\colon (x,y,z)\rightarrow (-x, -y, -z).
\end{align}
\end{subequations}

The $\mathbb{Z}_2$ cohomology ring is given by

\begin{equation}
\mathbb{Z}_2[A_i,A_{\mathsf{q}},A_{x+y},A_z,B_\alpha,B_{xy}]/\langle\mathcal{R}_2,\mathcal{R}_3,\mathcal{R}_4\rangle
 \end{equation}
where the relations are 
\begin{subequations} 
 \begin{align}
\mathcal{R}_2\colon & ~~
A_{\mathsf{q}} A_{x+y},~~A_{\mathsf{q}}^2,~~A_z (A_i + A_z),\\
\mathcal{R}_3\colon & ~~
A_{x+y}^3 + A_{x+y} B_\alpha + A_{\mathsf{q}} B_{xy},~~A_{x+y} (A_{x+y}^2 + B_\alpha + B_{xy}),\\
\mathcal{R}_4\colon & ~~
B_{xy} (B_\alpha + B_{xy}).
\end{align} 
 \end{subequations}
We have the following table regarding IWPs and group cohomology at degree 3.
\begin{center}
\begin{tabular}{c|cc|c|c|c}\hline\hline {Wyckoff}&\multicolumn{2}{c|}{Little group}& \multirow{2}{*}{Coordinates}&\multirow{2}{*}{LSM anomaly class}&\multirow{2}{*}{Topo. inv.} \\ \cline{2-3} position & Intl. & Sch\"{o}nflies & & & \\ \hline
1a&$4/m$&$C_{4h}$& $(0,0,0)$ & $(A_i + A_z) (A_{x+y}^2 + B_\alpha + B_{xy})$ & $\varphi_1[I]$\\ 
1b&$4/m$&$C_{4h}$& $(0,0,1/2)$ & $A_z (A_{x+y}^2 + B_\alpha + B_{xy})$ & $\varphi_1[T_3I]$\\ 
1c&$4/m$&$C_{4h}$& $(1/2,1/2,0)$ & $(A_i + A_z) B_{xy}$ & $\varphi_1[T_1T_2I]$\\ 
1d&$4/m$&$C_{4h}$& $(1/2,1/2,1/2)$ & $A_z B_{xy}$ & $\varphi_1[T_1T_2T_3I]$\\ 
2e&$2/m$&$C_{2h}$& $(0,1/2,0)$, $(1/2,0,0)$ & $A_{x+y}^2 (A_i + A_z)$ & $\varphi_1[T_1I]$\\ 
2f&$2/m$&$C_{2h}$& $(0,1/2,1/2)$, $(1/2,0,1/2)$ & $A_{x+y}^2 A_z$ & $\varphi_1[T_1T_3I]$\\ 
\hline
\hline 
 \end{tabular} 
 \end{center}

\subsection*{No. 84: $P4_2/m$}\label{subsub:sg84}

This group is generated by three translations $T_{1,2,3}$ as given in Eqs.~\eqref{TransBravaisP}, a two-fold rotation $C_2$, a four-fold screw $S_4$, and an inversion $I$:
\begin{subequations}
 \begin{align}
C_2 &\colon (x,y,z)\rightarrow (-x, -y, z),\\ 
S_4 &\colon (x,y,z)\rightarrow (-y, x, z + 1/2),\\ 
I &\colon (x,y,z)\rightarrow (-x, -y, -z).
\end{align}
\end{subequations}

The $\mathbb{Z}_2$ cohomology ring is given by

\begin{equation}
\mathbb{Z}_2[A_i,A_{\mathsf{q}},A_{x+y},B_\alpha,B_\beta,B_{xy},B_{z(x+y)}]/\langle\mathcal{R}_2,\mathcal{R}_3,\mathcal{R}_4\rangle
 \end{equation}
where the relations are 
\begin{subequations} 
 \begin{align}
\mathcal{R}_2\colon & ~~
A_i A_{\mathsf{q}},~~A_{\mathsf{q}} A_{x+y},~~A_{\mathsf{q}}^2,\\
\mathcal{R}_3\colon & ~~
A_{\mathsf{q}} B_\beta,~~A_{x+y} (A_i^2 + A_i A_{x+y} + B_\beta),~~A_{x+y}^3 + A_{x+y} B_\alpha + A_{\mathsf{q}} B_{xy},~~A_{x+y} (A_{x+y}^2 + B_\alpha + B_{xy}),~~A_{\mathsf{q}} B_{z(x+y)},\\
\mathcal{R}_4\colon & ~~
A_i^4 + A_i^2 B_\alpha + B_\beta^2,~~A_i^2 B_{xy} + B_\beta B_{xy} + A_{x+y}^2 B_{z(x+y)} + B_\alpha B_{z(x+y)},\nonumber\\&~~A_i^2 B_{xy} + A_i^2 B_{z(x+y)} + A_i A_{x+y} B_{z(x+y)} + B_\beta B_{z(x+y)},~~B_{xy} (B_\alpha + B_{xy}),\nonumber\\&~~(A_{x+y}^2 + B_\alpha + B_{xy}) B_{z(x+y)},~~A_i A_{x+y}^3 + A_{x+y}^4 + A_i^2 B_{xy} + A_i A_{x+y} B_{z(x+y)} + B_{z(x+y)}^2.
\end{align} 
 \end{subequations}
We have the following table regarding IWPs and group cohomology at degree 3.
\begin{center}
\resizebox{\columnwidth}{!}{
\begin{tabular}{c|cc|c|c|c}\hline\hline {Wyckoff}&\multicolumn{2}{c|}{Little group}& \multirow{2}{*}{Coordinates}&\multirow{2}{*}{LSM anomaly class}&\multirow{2}{*}{Topo. inv.} \\ \cline{2-3} position & Intl. & Sch\"{o}nflies & & & \\ \hline
2a&$2/m$&$C_{2h}$& $(0,0,0)$, $(0,0,1/2)$ & 
$(A_i+A_{\mathsf{q}})(A_{x+y}^2+B_\alpha + B_{xy})$
& $\varphi_1[I]$\\ 
2b&$2/m$&$C_{2h}$& $(1/2,1/2,0)$, $(1/2,1/2,1/2)$ & $A_{x+y}^3 + A_{x+y} B_\alpha + A_i B_{xy}$ & $\varphi_1[T_1T_2I]$\\ 
2c&$2/m$&$C_{2h}$& $(0,1/2,0)$, $(1/2,0,1/2)$ & $A_{x+y} (A_i A_{x+y} + A_{x+y}^2 + B_{z(x+y)})$ & $\varphi_1[T_2I]$\\ 
2d&$2/m$&$C_{2h}$& $(0,1/2,1/2)$, $(1/2,0,0)$ & $A_{x+y} (A_{x+y}^2 + B_{z(x+y)})$ & $\varphi_1[T_1I]$\\ 
2e&$\overline{4}$&$S_4$& $(0,0,1/4)$, $(0,0,3/4)$ & $A_{x+y}^3 + A_{\mathsf{q}} B_\alpha + A_{x+y} B_\alpha$ & $\varphi_2[S_4I, C_2]$\\ 
2f&$\overline{4}$&$S_4$& $(1/2,1/2,1/4)$, $(1/2,1/2,3/4)$ & $A_{x+y} (A_{x+y}^2 + B_\alpha)$ & $\varphi_2[T_2S_4I, T_1T_2C_2]$\\ 
\hline
\hline 
 \end{tabular} }
 \end{center}

\subsection*{No. 85: $P4/n$}\label{subsub:sg85}

This group is generated by three translations $T_{1,2,3}$ as given in Eqs.~\eqref{TransBravaisP}, a two-fold rotation $C_2$, a four-fold rotation $C_4$, and an inversion $I$:
\begin{subequations}
 \begin{align}
C_2 &\colon (x,y,z)\rightarrow (-x + 1/2, -y + 1/2, z),\\ 
C_4 &\colon (x,y,z)\rightarrow (-y + 1/2, x, z),\\ 
I &\colon (x,y,z)\rightarrow (-x, -y, -z).
\end{align}
\end{subequations}

The $\mathbb{Z}_2$ cohomology ring is given by

\begin{equation}
\mathbb{Z}_2[A_i,A_{\mathsf{q}},A_z,B_\alpha,B_\beta]/\langle\mathcal{R}_2,\mathcal{R}_3,\mathcal{R}_4\rangle
 \end{equation}
where the relations are 
\begin{subequations} 
 \begin{align}
\mathcal{R}_2\colon & ~~
A_i A_{\mathsf{q}},~~A_{\mathsf{q}}^2,~~A_z (A_i + A_z),\\
\mathcal{R}_3\colon & ~~
A_i (B_\alpha + B_\beta),~~A_i^3 + A_i B_\alpha + A_{\mathsf{q}} B_\beta,\\
\mathcal{R}_4\colon & ~~
B_\beta (B_\alpha + B_\beta).
\end{align} 
 \end{subequations}
We have the following table regarding IWPs and group cohomology at degree 3.
\begin{center}
\begin{tabular}{c|cc|c|c|c}\hline\hline {Wyckoff}&\multicolumn{2}{c|}{Little group}& \multirow{2}{*}{Coordinates}&\multirow{2}{*}{LSM anomaly class}&\multirow{2}{*}{Topo. inv.} \\ \cline{2-3} position & Intl. & Sch\"{o}nflies & & & \\ \hline
2a&$\overline{4}$&$S_4$& $(1/4,3/4,0)$, $(3/4,1/4,0)$ & $A_i^3 + A_i^2 A_z + A_i B_\alpha + A_z B_\beta$ & $\varphi_2[C_4I, T_2^{-1}C_2]$\\ 
2b&$\overline{4}$&$S_4$& $(1/4,3/4,1/2)$, $(3/4,1/4,1/2)$ & $A_z (A_i^2 + B_\beta)$ & $\varphi_2[T_3C_4I, T_2^{-1}C_2]$\\ 
2c&$4$&$C_4$& $(1/4,1/4,z)$, $(3/4,3/4,-z)$ & $A_z (B_\alpha + B_\beta)$ & $\varphi_2[T_3, C_2]$\\ 
\hline
\multirow{2}{*}{4d} & \multirow{2}{*}{$\overline{1}$} & \multirow{2}{*}{$C_i$} & $(0,0,0)$, $(1/2,1/2,0)$, & \multirow{2}{*}{$A_i^2 (A_i + A_z)$} & \multirow{2}{*}{$\varphi_1[I]$}\\
& & & $(1/2,0,0)$, $(0,1/2,0)$ & & \\ \hline
\multirow{2}{*}{4e} & \multirow{2}{*}{$\overline{1}$} & \multirow{2}{*}{$C_i$} & $(0,0,1/2)$, $(1/2,1/2,1/2)$, & \multirow{2}{*}{$A_i^2 A_z$} & \multirow{2}{*}{$\varphi_1[T_3I]$}\\
& & & $(1/2,0,1/2)$, $(0,1/2,1/2)$ & & \\ \hline
\hline 
 \end{tabular} 
 \end{center}

\subsection*{No. 86: $P4_2/n$}\label{subsub:sg86}

This group is generated by three translations $T_{1,2,3}$ as given in Eqs.~\eqref{TransBravaisP}, a two-fold screw $S_2$, a four-fold screw $S_4$, and an inversion $I$:
\begin{subequations}
 \begin{align}
S_2 &\colon (x,y,z)\rightarrow (-x -1/2, -y + 1/2, z + 1),\\ 
S_4 &\colon (x,y,z)\rightarrow (-y, x + 1/2, z + 1/2),\\ 
I &\colon (x,y,z)\rightarrow (-x, -y, -z).
\end{align}
\end{subequations}

The $\mathbb{Z}_2$ cohomology ring is given by

\begin{equation}
\mathbb{Z}_2[A_i,A_{\mathsf{q}},A_{x+y+z},B_\alpha]/\langle\mathcal{R}_2,\mathcal{R}_3,\mathcal{R}_4\rangle
 \end{equation}
where the relations are 
\begin{subequations} 
 \begin{align}
\mathcal{R}_2\colon & ~~
A_i A_{\mathsf{q}},~~A_{\mathsf{q}}^2,\\
\mathcal{R}_3\colon & ~~
A_i A_{x+y+z} (A_i + A_{x+y+z}),~~A_i^3 + A_{\mathsf{q}} A_{x+y+z}^2 + A_i B_\alpha + A_{\mathsf{q}} B_\alpha,\\
\mathcal{R}_4\colon & ~~
A_{x+y+z} (A_i + A_{x+y+z}) (A_{x+y+z}^2 + B_\alpha).
\end{align} 
 \end{subequations}
We have the following table regarding IWPs and group cohomology at degree 3.
\begin{center}
\resizebox{\columnwidth}{!}{
\begin{tabular}{c|cc|c|c|c}\hline\hline {Wyckoff}&\multicolumn{2}{c|}{Little group}& \multirow{2}{*}{Coordinates}&\multirow{2}{*}{LSM anomaly class}&\multirow{2}{*}{Topo. inv.} \\ \cline{2-3} position & Intl. & Sch\"{o}nflies & & & \\ \hline
2a&$\overline{4}$&$S_4$& $(1/4,1/4,1/4)$, $(3/4,3/4,3/4)$ & $(A_i + A_{x+y+z}) (A_i^2 + A_i A_{x+y+z} + A_{x+y+z}^2 + B_\alpha)$ & $\varphi_2[S_4I, T_1T_3^{-1}S_2]$\\ 
2b&$\overline{4}$&$S_4$& $(1/4,1/4,3/4)$, $(3/4,3/4,1/4)$ & $A_{x+y+z} (A_{x+y+z}^2 + B_\alpha)$ & $\varphi_2[T_3S_4I, T_1T_3^{-1}S_2]$\\ 
\hline
\multirow{2}{*}{4c} & \multirow{2}{*}{$\overline{1}$} & \multirow{2}{*}{$C_i$} & $(0,0,0)$, $(1/2,1/2,0)$, & \multirow{2}{*}{$A_i^2 (A_i + A_{x+y+z})$} & \multirow{2}{*}{$\varphi_1[I]$}\\
& & & $(0,1/2,1/2)$, $(1/2,0,1/2)$ & & \\ \hline
\multirow{2}{*}{4d} & \multirow{2}{*}{$\overline{1}$} & \multirow{2}{*}{$C_i$} & $(0,0,1/2)$, $(1/2,1/2,1/2)$, & \multirow{2}{*}{$A_i^2 A_{x+y+z}$} & \multirow{2}{*}{$\varphi_1[T_3I]$}\\
& & & $(0,1/2,0)$, $(1/2,0,0)$ & & \\ \hline
\multirow{2}{*}{4e} & \multirow{2}{*}{$2$} & \multirow{2}{*}{$C_2$} & $(3/4,1/4,z)$, $(3/4,1/4,z+1/2)$, & \multirow{2}{*}{$A_{\mathsf{q}} A_{x+y+z}^2$} & \multirow{2}{*}{$\varphi_2[S_4, T_3^{-1}S_2]$}\\
& & & $(1/4,3/4,-z)$, $(1/4,3/4,-z+1/2)$ & & \\ \hline
\hline 
 \end{tabular} }
 \end{center}

\subsection*{No. 87: $I4/m$}\label{subsub:sg87}

This group is generated by three translations $T_{1,2,3}$ as given in Eqs.~\eqref{TransBravaisI}, a two-fold rotation $C_2$, a four-fold rotation $C_4$, and an inversion $I$:
\begin{subequations}
 \begin{align}
C_2 &\colon (x,y,z)\rightarrow (-x, -y, z),\\ 
C_4 &\colon (x,y,z)\rightarrow (-y, x, z),\\ 
I &\colon (x,y,z)\rightarrow (-x, -y, -z).
\end{align}
\end{subequations}

The $\mathbb{Z}_2$ cohomology ring is given by

\begin{equation}
\mathbb{Z}_2[A_i,A_{\mathsf{q}},A_{x+y+z},B_\alpha,B_\beta,B_{z(x+y)},C_{xyz}]/\langle\mathcal{R}_2,\mathcal{R}_3,\mathcal{R}_4,\mathcal{R}_5,\mathcal{R}_6\rangle
 \end{equation}
where the relations are 
\begin{subequations} 
 \begin{align}
\mathcal{R}_2\colon & ~~
A_{\mathsf{q}} A_{x+y+z},~~A_{\mathsf{q}}^2,~~A_{x+y+z} (A_i + A_{x+y+z}),\\
\mathcal{R}_3\colon & ~~
A_i^2 A_{x+y+z} + A_{x+y+z} B_\alpha + A_{\mathsf{q}} B_\beta,~~A_{x+y+z} (A_i^2 + B_\alpha + B_\beta),~~A_i^2 A_{x+y+z} + A_{x+y+z} B_\alpha + A_{\mathsf{q}} B_{z(x+y)},\nonumber\\&~~A_{x+y+z} (A_i^2 + B_\alpha + B_{z(x+y)}),\\
\mathcal{R}_4\colon & ~~
A_{x+y+z} C_{xyz},~~B_\beta^2 + A_i^2 B_{z(x+y)} + B_\alpha B_{z(x+y)},~~B_\alpha B_\beta + B_\beta B_{z(x+y)} + A_{\mathsf{q}} C_{xyz},~~B_{z(x+y)} (B_\alpha + B_{z(x+y)}),\\
\mathcal{R}_5\colon & ~~
(A_i A_{\mathsf{q}} + B_\beta) C_{xyz},~~B_{z(x+y)} C_{xyz},\\
\mathcal{R}_6\colon & ~~
C_{xyz} (A_i B_\alpha + C_{xyz}).
\end{align} 
 \end{subequations}
We have the following table regarding IWPs and group cohomology at degree 3.
\begin{center}
\resizebox{\columnwidth}{!}{
\begin{tabular}{c|cc|c|c|c}\hline\hline {Wyckoff}&\multicolumn{2}{c|}{Little group}& {Coordinates}&\multirow{2}{*}{LSM anomaly class}&\multirow{2}{*}{Topo. inv.} \\ \cline{2-4} position & Intl. & Sch\"{o}nflies & $(0,0,0) + ~(1/2,1/2,1/2) + $ & &\\ \hline
2a&$4/m$&$C_{4h}$& $(0,0,0)$ & $A_i^2 A_{x+y+z} + A_i B_\alpha + A_i B_{z(x+y)} + C_{xyz}$ & $\varphi_1[I]$\\ 
2b&$4/m$&$C_{4h}$& $(0,0,1/2)$ & $C_{xyz}$ & $\varphi_1[T_1T_2I]$\\ 
4c&$2/m$&$C_{2h}$& $(0,1/2,0)$, $(1/2,0,0)$ & $A_i^2 A_{x+y+z} + A_{x+y+z} B_\alpha + A_i B_{z(x+y)}$ & $\varphi_1[T_2T_3I]$\\ 
4d&$\overline{4}$&$S_4$& $(0,1/2,1/4)$, $(1/2,0,1/4)$ & $A_{x+y+z} (A_i^2 + B_\alpha)$ & $\varphi_2[T_1C_4I, T_1T_3C_2]$\\ 
\hline
\multirow{2}{*}{8f} & \multirow{2}{*}{$\overline{1}$} & \multirow{2}{*}{$C_i$} & $(1/4,1/4,1/4)$, $(3/4,3/4,1/4)$, & \multirow{2}{*}{$A_i^2 A_{x+y+z}$} & \multirow{2}{*}{$\varphi_1[T_1T_2T_3I]$}\\
& & & $(3/4,1/4,1/4)$, $(1/4,3/4,1/4)$ & & \\ \hline
\hline 
 \end{tabular} }
 \end{center}

\subsection*{No. 88: $I4_1/a$}\label{subsub:sg88}

This group is generated by three translations $T_{1,2,3}$ as given in Eqs.~\eqref{TransBravaisI}, a two-fold rotation $C_2$, a four-fold screw $S_4$, and an inversion $I$:
\begin{subequations}
 \begin{align}
C_2 &\colon (x,y,z)\rightarrow (-x, -y + 1/2, z),\\ 
S_4 &\colon (x,y,z)\rightarrow (-y + 3/4, x + 1/4, z + 1/4),\\ 
I &\colon (x,y,z)\rightarrow (-x, -y, -z).
\end{align}
\end{subequations}

The $\mathbb{Z}_2$ cohomology ring is given by

\begin{equation}
\mathbb{Z}_2[A_i,A_{\mathsf{q}},B_\alpha,B_{z(x+y)},C_\gamma]/\langle\mathcal{R}_2,\mathcal{R}_3,\mathcal{R}_4,\mathcal{R}_5,\mathcal{R}_6\rangle
 \end{equation}
where the relations are 
\begin{subequations} 
 \begin{align}
\mathcal{R}_2\colon & ~~
A_i A_{\mathsf{q}},~~A_{\mathsf{q}}^2,\\
\mathcal{R}_3\colon & ~~
A_i^3 + A_i B_\alpha + A_{\mathsf{q}} B_\alpha,~~A_i^3 + A_i B_\alpha + A_{\mathsf{q}} B_{z(x+y)},\\
\mathcal{R}_4\colon & ~~
(A_i + A_{\mathsf{q}}) C_\gamma,~~A_i^4 + B_\alpha^2 + A_i^2 B_{z(x+y)} + B_\alpha B_{z(x+y)} + A_i C_\gamma,~~A_i^4 + B_\alpha^2 + A_i^2 B_{z(x+y)} + B_{z(x+y)}^2,\\
\mathcal{R}_5\colon & ~~
(B_\alpha + B_{z(x+y)}) C_\gamma,\\
\mathcal{R}_6\colon & ~~
C_\gamma (A_i B_\alpha + C_\gamma).
\end{align} 
 \end{subequations}
We have the following table regarding IWPs and group cohomology at degree 3.
\begin{center}
\begin{tabular}{c|cc|c|c|c}\hline\hline {Wyckoff}&\multicolumn{2}{c|}{Little group}& {Coordinates}&\multirow{2}{*}{LSM anomaly class}&\multirow{2}{*}{Topo. inv.} \\ \cline{2-4} position & Intl. & Sch\"{o}nflies & $(0,0,0) + ~(1/2,1/2,1/2) + $ & &\\ \hline
4a&$\overline{4}$&$S_4$& $(0,1/4,1/8)$, $(1/2,1/4,3/8)$ & $A_i^3 + A_i B_\alpha + C_\gamma$ & $\varphi_2[S_4I, T_1^{-1}T_2C_2]$\\ 
4b&$\overline{4}$&$S_4$& $(0,1/4,5/8)$, $(1/2,1/4,7/8)$ & $C_\gamma$ & $\varphi_2[T_1T_2S_4I, T_1^{-1}T_2C_2]$\\ 
\hline
\multirow{2}{*}{8c} & \multirow{2}{*}{$\overline{1}$} & \multirow{2}{*}{$C_i$} & $(0,0,0)$, $(1/2,0,1/2)$, & \multirow{2}{*}{$A_i (A_i^2 + B_\alpha + B_{z(x+y)})$} & \multirow{2}{*}{$\varphi_1
[I]$}\\
& & & $(3/4,1/4,1/4)$, $(3/4,3/4,3/4)$ & & \\ \hline
\multirow{2}{*}{8d} & \multirow{2}{*}{$\overline{1}$} & \multirow{2}{*}{$C_i$} & $(0,0,1/2)$, $(1/2,0,0)$, & \multirow{2}{*}{$A_i (B_\alpha + B_{z(x+y)})$} & \multirow{2}{*}{$\varphi_1[T_1T_2I]$}\\
& & & $(3/4,1/4,3/4)$, $(3/4,3/4,1/4)$ & & \\ \hline
\hline 
 \end{tabular} 
 \end{center}

\subsection*{No. 89: $P422$}\label{subsub:sg89}

This group is generated by three translations $T_{1,2,3}$ as given in Eqs.~\eqref{TransBravaisP}, a two-fold rotation $C_2$, a two-fold rotation $C'_2$, and a two-fold rotation $C''_2$:
\begin{subequations}
 \begin{align}
C_2 &\colon (x,y,z)\rightarrow (-x, -y, z),\\ 
C'_2 &\colon (x,y,z)\rightarrow (-x, y, -z),\\ 
C''_2 &\colon (x,y,z)\rightarrow (-y, -x, -z).
\end{align}
\end{subequations}

The $\mathbb{Z}_2$ cohomology ring is given by

\begin{equation}
\mathbb{Z}_2[A_{c'},A_{c''},A_{x+y},A_z,B_\alpha,B_{xy}]/\langle\mathcal{R}_2,\mathcal{R}_3,\mathcal{R}_4\rangle
 \end{equation}
where the relations are 
\begin{subequations} 
 \begin{align}
\mathcal{R}_2\colon & ~~
A_{c'} A_{c''},~~A_{c''} A_{x+y},~~A_z (A_{c'} + A_{c''} + A_z),\\
\mathcal{R}_3\colon & ~~
A_{c'} A_{x+y}^2 + A_{x+y}^3 + A_{x+y} B_\alpha + A_{c'} B_{xy},~~A_{x+y} (A_{c'} A_{x+y} + A_{x+y}^2 + B_\alpha + B_{xy}),\\
\mathcal{R}_4\colon & ~~
B_{xy} (B_\alpha + B_{xy}).
\end{align} 
 \end{subequations}
We have the following table regarding IWPs and group cohomology at degree 3.
\begin{center}
\resizebox{\columnwidth}{!}{
\begin{tabular}{c|cc|c|c|c}\hline\hline {Wyckoff}&\multicolumn{2}{c|}{Little group}& \multirow{2}{*}{Coordinates}&\multirow{2}{*}{LSM anomaly class}&\multirow{2}{*}{Topo. inv.} \\ \cline{2-3} position & Intl. & Sch\"{o}nflies & & & \\ \hline
1a&$422$&$D_4$& $(0,0,0)$ & 
$(A_{c'} + A_{c''} + A_z)(A_{c'} A_{x+y} + A_{x+y}^2 + B_\alpha + B_{xy})$
& $\varphi_2[C_2, C'_2]$\\ 
1b&$422$&$D_4$& $(0,0,1/2)$ & $A_z (A_{c'} A_{x+y} + A_{x+y}^2 + B_\alpha + B_{xy})$ & $\varphi_2[C_2, T_3C'_2]$\\ 
1c&$422$&$D_4$& $(1/2,1/2,0)$ & 
$(A_{c'} + A_{c''} + A_z) B_{xy}$ & $\varphi_2[T_1T_2C_2, T_1C'_2]$\\ 
1d&$422$&$D_4$& $(1/2,1/2,1/2)$ & $A_z B_{xy}$ & $\varphi_2[T_1T_2C_2, T_1T_3C'_2]$\\ 
2e&$222$&$D_2$& $(1/2,0,0)$, $(0,1/2,0)$ & $A_{x+y} (A_{c'} + A_{x+y}) (A_{c'} + A_z)$ & $\varphi_2[T_1C_2, T_1C'_2]$\\ 
2f&$222$&$D_2$& $(1/2,0,1/2)$, $(0,1/2,1/2)$ & $A_{x+y} (A_{c'} + A_{x+y}) A_z$ & $\varphi_2[T_1C_2, T_1T_3C'_2]$\\ 
\hline
\hline 
 \end{tabular} }
 \end{center}

\subsection*{No. 90: $P42_12$}\label{subsub:sg90}

This group is generated by three translations $T_{1,2,3}$ as given in Eqs.~\eqref{TransBravaisP}, a two-fold rotation $C_2$, a two-fold screw $S'_2$, and a two-fold rotation $C''_2$:
\begin{subequations}
 \begin{align}
C_2 &\colon (x,y,z)\rightarrow (-x, -y, z),\\ 
S'_2 &\colon (x,y,z)\rightarrow (-x + 1/2, y + 1/2, -z),\\ 
C''_2 &\colon (x,y,z)\rightarrow (-y, -x, -z).
\end{align}
\end{subequations}

The $\mathbb{Z}_2$ cohomology ring is given by

\begin{equation}
\mathbb{Z}_2[A_{c'},A_{c''},A_z,B_\alpha,B_\beta,C_\gamma]/\langle\mathcal{R}_2,\mathcal{R}_3,\mathcal{R}_4,\mathcal{R}_5,\mathcal{R}_6\rangle
 \end{equation}
where the relations are 
\begin{subequations} 
 \begin{align}
\mathcal{R}_2\colon & ~~
A_{c'} A_{c''},~~A_{c'}^2,~~A_z (A_{c'} + A_{c''} + A_z),\\
\mathcal{R}_3\colon & ~~
A_{c'} (B_\alpha + B_\beta),~~A_{c'} B_\alpha + A_{c''} B_\beta,\\
\mathcal{R}_4\colon & ~~
A_{c'} C_\gamma,~~B_\beta (B_\alpha + B_\beta),\\
\mathcal{R}_5\colon & ~~
B_\beta C_\gamma,\\
\mathcal{R}_6\colon & ~~
B_\alpha^3 + B_\alpha^2 B_\beta + A_{c''} B_\alpha C_\gamma + C_\gamma^2.
\end{align} 
 \end{subequations}
We have the following table regarding IWPs and group cohomology at degree 3.
\begin{center}
\begin{tabular}{c|cc|c|c|c}\hline\hline {Wyckoff}&\multicolumn{2}{c|}{Little group}& \multirow{2}{*}{Coordinates}&\multirow{2}{*}{LSM anomaly class}&\multirow{2}{*}{Topo. inv.} \\ \cline{2-3} position & Intl. & Sch\"{o}nflies & & & \\ \hline
2a&$222$&$D_2$& $(0,0,0)$, $(1/2,1/2,0)$ & 
$(A_{c'} + A_{c''} + A_z) B_\alpha + A_z B_\beta$ & $\varphi_2[C_2, C''_2]$\\ 
2b&$222$&$D_2$& $(0,0,1/2)$, $(1/2,1/2,1/2)$ & $A_z (B_\alpha + B_\beta)$ & $\varphi_2[C_2, T_3C''_2]$\\ 
2c&$4$&$C_4$& $(0,1/2,z)$, $(1/2,0,-z)$ & $A_z B_\beta$ & $\varphi_2[T_3, T_1C_2]$\\ 
\hline
\hline 
 \end{tabular} 
 \end{center}

\subsection*{No. 91: $P4_122$}\label{subsub:sg91}

This group is generated by three translations $T_{1,2,3}$ as given in Eqs.~\eqref{TransBravaisP}, a two-fold screw $S_2$, a two-fold rotation $C'_2$, and a two-fold rotation $C''_2$:
\begin{subequations}
 \begin{align}
S_2 &\colon (x,y,z)\rightarrow (-x, -y, z + 1/2),\\ 
C'_2 &\colon (x,y,z)\rightarrow (-x, y, -z),\\ 
C''_2 &\colon (x,y,z)\rightarrow (-y, -x, -z + 1/4).
\end{align}
\end{subequations}

The $\mathbb{Z}_2$ cohomology ring is given by

\begin{equation}
\mathbb{Z}_2[A_{c'},A_{c''},A_{x+y},B_{xy}]/\langle\mathcal{R}_2,\mathcal{R}_3,\mathcal{R}_4\rangle
 \end{equation}
where the relations are 
\begin{subequations} 
 \begin{align}
\mathcal{R}_2\colon & ~~
A_{c'} A_{c''},~~A_{c''} A_{x+y},\\
\mathcal{R}_3\colon & ~~
A_{c'} A_{x+y}^2 + A_{x+y}^3 + A_{c'} B_{xy},~~A_{x+y} (A_{c'} A_{x+y} + A_{x+y}^2 + B_{xy}),\\
\mathcal{R}_4\colon & ~~
B_{xy}^2.
\end{align} 
 \end{subequations}
We have the following table regarding IWPs and group cohomology at degree 3.
\begin{center}
\begin{tabular}{c|cc|c|c|c}\hline\hline {Wyckoff}&\multicolumn{2}{c|}{Little group}& \multirow{2}{*}{Coordinates}&\multirow{2}{*}{LSM anomaly class}&\multirow{2}{*}{Topo. inv.} \\ \cline{2-3} position & Intl. & Sch\"{o}nflies & & & \\ \hline
\multirow{2}{*}{4a} & \multirow{2}{*}{$2$} & \multirow{2}{*}{$C_2$} & $(0,y,0)$, $(0,-y,1/2)$, & \multirow{2}{*}{$A_{x+y} (A_{c'}^2 + A_{x+y}^2)$} & \multirow{2}{*}{$\varphi_2[T_2, C'_2]$}\\
& & & $(-y,0,1/4)$, $(y,0,3/4)$ & & \\ \hline
\multirow{2}{*}{4b} & \multirow{2}{*}{$2$} & \multirow{2}{*}{$C_2$} & $(1/2,y,0)$, $(1/2,-y,1/2)$, & \multirow{2}{*}{$A_{x+y}^2 (A_{c'} + A_{x+y})$} & \multirow{2}{*}{$\varphi_2[T_2, T_1C'_2]$}\\
& & & $(-y,1/2,1/4)$, $(y,1/2,3/4)$ & & \\ \hline
\multirow{2}{*}{4c} & \multirow{2}{*}{$2$} & \multirow{2}{*}{$C_2$} & $(x,x,3/8)$, $(-x,-x,7/8)$, & \multirow{2}{*}{$A_{c''} B_{xy}$} & \multirow{2}{*}{$\varphi_2[T_1T_2^{-1}, C''_2]$}\\
& & & $(-x,x,5/8)$, $(x,-x,1/8)$ & & \\ \hline
\hline 
 \end{tabular} 
 \end{center}

\subsection*{No. 92: $P4_12_12$}\label{subsub:sg92}

This group is generated by three translations $T_{1,2,3}$ as given in Eqs.~\eqref{TransBravaisP}, a two-fold screw $S_2$, a two-fold screw $S'_2$, and a two-fold rotation $C''_2$:
\begin{subequations}
 \begin{align}
S_2 &\colon (x,y,z)\rightarrow (-x, -y, z + 1/2),\\ 
S'_2 &\colon (x,y,z)\rightarrow (-x + 1/2, y + 1/2, -z + 1/4),\\ 
C''_2 &\colon (x,y,z)\rightarrow (-y, -x, -z + 1/2).
\end{align}
\end{subequations}

The $\mathbb{Z}_2$ cohomology ring is given by

\begin{equation}
\mathbb{Z}_2[A_{c'},A_{c''},B_\beta,C_\gamma]/\langle\mathcal{R}_2,\mathcal{R}_3,\mathcal{R}_4,\mathcal{R}_5,\mathcal{R}_6\rangle
 \end{equation}
where the relations are 
\begin{subequations} 
 \begin{align}
\mathcal{R}_2\colon & ~~
A_{c'} A_{c''},~~A_{c'}^2,\\
\mathcal{R}_3\colon & ~~
A_{c'} B_\beta,~~A_{c''} B_\beta,\\
\mathcal{R}_4\colon & ~~
A_{c'} C_\gamma,~~B_\beta^2,\\
\mathcal{R}_5\colon & ~~
B_\beta C_\gamma,\\
\mathcal{R}_6\colon & ~~
C_\gamma^2.
\end{align} 
 \end{subequations}
We have the following table regarding IWPs and group cohomology at degree 3.
\begin{center}
\begin{tabular}{c|cc|c|c|c}\hline\hline {Wyckoff}&\multicolumn{2}{c|}{Little group}& \multirow{2}{*}{Coordinates}&\multirow{2}{*}{LSM anomaly class}&\multirow{2}{*}{Topo. inv.} \\ \cline{2-3} position & Intl. & Sch\"{o}nflies & & & \\ \hline
\multirow{2}{*}{4a} & \multirow{2}{*}{$2$} & \multirow{2}{*}{$C_2$} & $(x,x,0)$, $(-x,-x,1/2)$, & \multirow{2}{*}{$C_\gamma$} & \multirow{2}{*}{$\varphi_2[T_1T_2^{-1}, C''_2]$}\\
& & & $(-x+1/2,x+1/2,1/4)$, $(x+1/2,-x+1/2,3/4)$ & & \\ \hline
\hline 
 \end{tabular} 
 \end{center}

\subsection*{No. 93: $P4_222$}\label{subsub:sg93}

This group is generated by three translations $T_{1,2,3}$ as given in Eqs.~\eqref{TransBravaisP}, a two-fold rotation $C_2$, a two-fold rotation $C'_2$, and a two-fold rotation $C''_2$:
\begin{subequations}
 \begin{align}
C_2 &\colon (x,y,z)\rightarrow (-x, -y, z),\\ 
C'_2 &\colon (x,y,z)\rightarrow (-x, y, -z),\\ 
C''_2 &\colon (x,y,z)\rightarrow (-y, -x, -z + 1/2).
\end{align}
\end{subequations}

The $\mathbb{Z}_2$ cohomology ring is given by

\begin{equation}
\mathbb{Z}_2[A_{c'},A_{c''},A_{x+y},A_z,B_{xy}]/\langle\mathcal{R}_2,\mathcal{R}_3,\mathcal{R}_4\rangle
 \end{equation}
where the relations are 
\begin{subequations} 
 \begin{align}
\mathcal{R}_2\colon & ~~
A_{c'} A_{c''},~~A_{c''} A_{x+y},\\
\mathcal{R}_3\colon & ~~
A_{c'} A_{x+y}^2 + A_{x+y}^3 + A_{c'} A_{x+y} A_z + A_{x+y} A_z^2 + A_{c'} B_{xy},~~A_{x+y} (A_{c'} A_{x+y} + A_{x+y}^2 + A_{c'} A_z + A_z^2 + B_{xy}),\\
\mathcal{R}_4\colon & ~~
A_{c'} A_{x+y}^2 A_z + A_{x+y}^3 A_z + A_{c'} A_{x+y} A_z^2 + A_{x+y} A_z^3 + A_{c''} A_z B_{xy} + A_z^2 B_{xy} + B_{xy}^2.
\end{align} 
 \end{subequations}
We have the following table regarding IWPs and group cohomology at degree 3.
\begin{center}
\resizebox{\columnwidth}{!}{
\begin{tabular}{c|cc|c|c|c}\hline\hline {Wyckoff}&\multicolumn{2}{c|}{Little group}& \multirow{2}{*}{Coordinates}&\multirow{2}{*}{LSM anomaly class}&\multirow{2}{*}{Topo. inv.} \\ \cline{2-3} position & Intl. & Sch\"{o}nflies & & & \\ \hline
2a&$222$&$D_2$& $(0,0,0)$, $(0,0,1/2)$ & $(A_{c'} + A_{x+y}) (A_{x+y} + A_z) (A_{c'} + A_{x+y} + A_z)$ & $\varphi_2[C_2, C'_2]$\\ 
2b&$222$&$D_2$& $(1/2,1/2,0)$, $(1/2,1/2,1/2)$ & $A_{x+y} (A_{x+y} + A_z) (A_{c'} + A_{x+y} + A_z)$ & $\varphi_2[T_1T_2C_2, T_1C'_2]$\\ 
2c&$222$&$D_2$& $(0,1/2,0)$, $(1/2,0,1/2)$ & $A_{x+y} (A_{c'} + A_{x+y}) (A_{c'} + A_{x+y} + A_z)$ & $\varphi_2[T_2C_2, C'_2]$\\ 
2d&$222$&$D_2$& $(0,1/2,1/2)$, $(1/2,0,0)$ & $A_{x+y} (A_{c'} + A_{x+y}) (A_{x+y} + A_z)$ & $\varphi_2[T_2C_2, T_3C'_2]$\\ 
2e&$222$&$D_2$& $(0,0,1/4)$, $(0,0,3/4)$ & $A_{c''} (A_{c''} A_z + A_z^2 + B_{xy})$ & $\varphi_2[C_2, C''_2]$\\ 
2f&$222$&$D_2$& $(1/2,1/2,1/4)$, $(1/2,1/2,3/4)$ & $A_{c''} B_{xy}$ & $\varphi_2[T_1T_2C_2, T_1T_2C''_2]$\\ 
\hline
\hline 
 \end{tabular} }
 \end{center}

\subsection*{No. 94: $P4_22_12$}\label{subsub:sg94}

This group is generated by three translations $T_{1,2,3}$ as given in Eqs.~\eqref{TransBravaisP}, a two-fold rotation $C_2$, a two-fold screw $S'_2$, and a two-fold rotation $C''_2$:
\begin{subequations}
 \begin{align}
C_2 &\colon (x,y,z)\rightarrow (-x, -y, z),\\ 
S'_2 &\colon (x,y,z)\rightarrow (-x + 1/2, y + 1/2, -z + 1/2),\\ 
C''_2 &\colon (x,y,z)\rightarrow (-y, -x, -z).
\end{align}
\end{subequations}

The $\mathbb{Z}_2$ cohomology ring is given by

\begin{equation}
\mathbb{Z}_2[A_{c'},A_{c''},A_z,B_\beta,C_\gamma]/\langle\mathcal{R}_2,\mathcal{R}_3,\mathcal{R}_4,\mathcal{R}_5,\mathcal{R}_6\rangle
 \end{equation}
where the relations are 
\begin{subequations} 
 \begin{align}
\mathcal{R}_2\colon & ~~
A_{c'} A_{c''},~~A_{c'}^2,\\
\mathcal{R}_3\colon & ~~
A_{c'} (A_z^2 + B_\beta),~~A_{c'} A_z^2 + A_{c''} B_\beta,\\
\mathcal{R}_4\colon & ~~
A_{c'} C_\gamma,~~B_\beta (A_z^2 + B_\beta),\\
\mathcal{R}_5\colon & ~~
B_\beta C_\gamma,\\
\mathcal{R}_6\colon & ~~
C_\gamma (A_{c''}^2 A_z + A_{c''} A_z^2 + C_\gamma).
\end{align} 
 \end{subequations}
We have the following table regarding IWPs and group cohomology at degree 3.
\begin{center}
\begin{tabular}{c|cc|c|c|c}\hline\hline {Wyckoff}&\multicolumn{2}{c|}{Little group}& \multirow{2}{*}{Coordinates}&\multirow{2}{*}{LSM anomaly class}&\multirow{2}{*}{Topo. inv.} \\ \cline{2-3} position & Intl. & Sch\"{o}nflies & & & \\ \hline
2a&$222$&$D_2$& $(0,0,0)$, $(1/2,1/2,1/2)$ & $A_{c''}^2 A_z + A_{c'} A_z^2 + A_{c''} A_z^2 + C_\gamma$ & $\varphi_2[C_2, C''_2]$\\ 
2b&$222$&$D_2$& $(0,0,1/2)$, $(1/2,1/2,0)$ & $C_\gamma$ & $\varphi_2[C_2, T_3C''_2]$\\ 
\hline
\multirow{2}{*}{4d} & \multirow{2}{*}{$2$} & \multirow{2}{*}{$C_2$} & $(0,1/2,z)$, $(0,1/2,z+1/2)$, & \multirow{2}{*}{$A_{c'} A_z^2$} & \multirow{2}{*}{$\varphi_2[S'_2C''_2, T_1C_2]$}\\
& & & $(1/2,0,-z+1/2)$, $(1/2,0,-z)$ & & \\ \hline
\hline 
 \end{tabular} 
 \end{center}

\subsection*{No. 95: $P4_322$}\label{subsub:sg95}

This group is generated by three translations $T_{1,2,3}$ as given in Eqs.~\eqref{TransBravaisP}, a two-fold screw $S_2$, a two-fold rotation $C'_2$, and a two-fold rotation $C''_2$:
\begin{subequations}
 \begin{align}
S_2 &\colon (x,y,z)\rightarrow (-x, -y, z + 1/2),\\ 
C'_2 &\colon (x,y,z)\rightarrow (-x, y, -z),\\ 
C''_2 &\colon (x,y,z)\rightarrow (-y, -x, -z + 3/4).
\end{align}
\end{subequations}

The $\mathbb{Z}_2$ cohomology ring is given by

\begin{equation}
\mathbb{Z}_2[A_{c'},A_{c''},A_{x+y},B_{xy}]/\langle\mathcal{R}_2,\mathcal{R}_3,\mathcal{R}_4\rangle
 \end{equation}
where the relations are 
\begin{subequations} 
 \begin{align}
\mathcal{R}_2\colon & ~~
A_{c'} A_{c''},~~A_{c''} A_{x+y},\\
\mathcal{R}_3\colon & ~~
A_{c'} A_{x+y}^2 + A_{x+y}^3 + A_{c'} B_{xy},~~A_{x+y} (A_{c'} A_{x+y} + A_{x+y}^2 + B_{xy}),\\
\mathcal{R}_4\colon & ~~
B_{xy}^2.
\end{align} 
 \end{subequations}
We have the following table regarding IWPs and group cohomology at degree 3.
\begin{center}
\begin{tabular}{c|cc|c|c|c}\hline\hline {Wyckoff}&\multicolumn{2}{c|}{Little group}& \multirow{2}{*}{Coordinates}&\multirow{2}{*}{LSM anomaly class}&\multirow{2}{*}{Topo. inv.} \\ \cline{2-3} position & Intl. & Sch\"{o}nflies & & & \\ \hline
\multirow{2}{*}{4a} & \multirow{2}{*}{$2$} & \multirow{2}{*}{$C_2$} & $(0,y,0)$, $(0,-y,1/2)$, & \multirow{2}{*}{$A_{x+y} (A_{c'}^2 + A_{x+y}^2)$} & \multirow{2}{*}{$\varphi_2[T_2, C'_2]$}\\
& & & $(-y,0,3/4)$, $(y,0,1/4)$ & & \\ \hline
\multirow{2}{*}{4b} & \multirow{2}{*}{$2$} & \multirow{2}{*}{$C_2$} & $(1/2,y,0)$, $(1/2,-y,1/2)$, & \multirow{2}{*}{$A_{x+y}^2 (A_{c'} + A_{x+y})$} & \multirow{2}{*}{$\varphi_2[T_2, T_1C'_2]$}\\
& & & $(-y,1/2,3/4)$, $(y,1/2,1/4)$ & & \\ \hline
\multirow{2}{*}{4c} & \multirow{2}{*}{$2$} & \multirow{2}{*}{$C_2$} & $(x,x,5/8)$, $(-x,-x,1/8)$, & \multirow{2}{*}{$A_{c''} B_{xy}$} & \multirow{2}{*}{$\varphi_2[T_1T_2^{-1}, C''_2]$}\\
& & & $(-x,x,3/8)$, $(x,-x,7/8)$ & & \\ \hline
\hline 
 \end{tabular} 
 \end{center}

\subsection*{No. 96: $P4_32_12$}\label{subsub:sg96}

This group is generated by three translations $T_{1,2,3}$ as given in Eqs.~\eqref{TransBravaisP}, a two-fold screw $S_2$, a two-fold screw $S'_2$, and a two-fold rotation $C''_2$:
\begin{subequations}
 \begin{align}
S_2 &\colon (x,y,z)\rightarrow (-x, -y, z + 1/2),\\ 
S'_2 &\colon (x,y,z)\rightarrow (-x + 1/2, y + 1/2, -z + 3/4),\\ 
C''_2 &\colon (x,y,z)\rightarrow (-y, -x, -z + 1/2).
\end{align}
\end{subequations}

The $\mathbb{Z}_2$ cohomology ring is given by

\begin{equation}
\mathbb{Z}_2[A_{c'},A_{c''},B_\beta,C_\gamma]/\langle\mathcal{R}_2,\mathcal{R}_3,\mathcal{R}_4,\mathcal{R}_5,\mathcal{R}_6\rangle
 \end{equation}
where the relations are 
\begin{subequations} 
 \begin{align}
\mathcal{R}_2\colon & ~~
A_{c'} A_{c''},~~A_{c'}^2,\\
\mathcal{R}_3\colon & ~~
A_{c'} B_\beta,~~A_{c''} B_\beta,\\
\mathcal{R}_4\colon & ~~
A_{c'} C_\gamma,~~B_\beta^2,\\
\mathcal{R}_5\colon & ~~
B_\beta C_\gamma,\\
\mathcal{R}_6\colon & ~~
C_\gamma^2.
\end{align} 
 \end{subequations}
We have the following table regarding IWPs and group cohomology at degree 3.
\begin{center}
\resizebox{\columnwidth}{!}{
\begin{tabular}{c|cc|c|c|c}\hline\hline {Wyckoff}&\multicolumn{2}{c|}{Little group}& \multirow{2}{*}{Coordinates}&\multirow{2}{*}{LSM anomaly class}&\multirow{2}{*}{Topo. inv.} \\ \cline{2-3} position & Intl. & Sch\"{o}nflies & & & \\ \hline
\multirow{2}{*}{4a} & \multirow{2}{*}{$2$} & \multirow{2}{*}{$C_2$} & $(x,x,0)$, $(-x,-x,1/2)$, & \multirow{2}{*}{$C_\gamma$} & \multirow{2}{*}{$\varphi_2[T_1T_2, T_3^{-1}S_2C''_2]$}\\
& & & $(-x+1/2,x+1/2,3/4)$, $(x+1/2,-x+1/2,1/4)$ & & \\ \hline
\hline 
 \end{tabular} }
 \end{center}

\subsection*{No. 97: $I422$}\label{subsub:sg97}

This group is generated by three translations $T_{1,2,3}$ as given in Eqs.~\eqref{TransBravaisI}, a two-fold rotation $C_2$, a two-fold rotation $C'_2$, and a two-fold rotation $C''_2$:
\begin{subequations}
 \begin{align}
C_2 &\colon (x,y,z)\rightarrow (-x, -y, z),\\ 
C'_2 &\colon (x,y,z)\rightarrow (-x, y, -z),\\ 
C''_2 &\colon (x,y,z)\rightarrow (-y, -x, -z).
\end{align}
\end{subequations}

The $\mathbb{Z}_2$ cohomology ring is given by

\begin{equation}
\mathbb{Z}_2[A_{c'},A_{c''},A_{x+y+z},B_\alpha,B_\beta,B_{z(x+y)},C_\gamma,C_{xyz}]/\langle\mathcal{R}_2,\mathcal{R}_3,\mathcal{R}_4,\mathcal{R}_5,\mathcal{R}_6\rangle
 \end{equation}
where the relations are 
\begin{subequations} 
 \begin{align}
\mathcal{R}_2\colon & ~~
A_{c'} A_{c''},~~A_{c'} A_{x+y+z},~~A_{x+y+z} (A_{c''} + A_{x+y+z}),\\
\mathcal{R}_3\colon & ~~
A_{x+y+z} B_\alpha + A_{c''} B_\beta,~~A_{x+y+z} (B_\alpha + B_\beta),~~A_{x+y+z} B_\alpha + A_{c''} B_{z(x+y)},~~A_{x+y+z} (B_\alpha + B_{z(x+y)}),\\
\mathcal{R}_4\colon & ~~
A_{c''} C_\gamma + A_{x+y+z} C_\gamma + A_{c''} C_{xyz},~~A_{x+y+z} C_{xyz},~~A_{c'}^2 B_\beta + B_\alpha B_\beta + A_{c'}^2 B_{z(x+y)} + B_\alpha B_{z(x+y)} + A_{c'} C_\gamma,\nonumber\\&~~B_\alpha B_\beta + B_\beta^2 + A_{c'}^2 B_{z(x+y)} + A_{c'} C_\gamma,~~B_\alpha B_\beta + B_\beta B_{z(x+y)} + A_{c'} C_{xyz},\nonumber\\&~~A_{c'}^2 B_\beta + B_\alpha B_\beta + A_{c'}^2 B_{z(x+y)} + B_{z(x+y)}^2 + A_{c'} C_\gamma,\\
\mathcal{R}_5\colon & ~~
A_{c'}^3 B_\beta + A_{c'} B_\alpha B_\beta + B_\beta C_\gamma + B_{z(x+y)} C_\gamma,\nonumber\\&~~A_{c'}^3 B_\beta + A_{c'} B_\alpha B_\beta + B_\alpha C_\gamma + B_\beta C_\gamma + A_{c'}^2 C_{xyz} + B_\alpha C_{xyz},~~(A_{c'}^2 + B_\beta) C_{xyz},~~B_{z(x+y)} C_{xyz},\\
\mathcal{R}_6\colon & ~~
A_{c'}^4 B_\beta + B_\alpha^2 B_\beta + A_{c''} B_\alpha C_\gamma + C_\gamma^2,\nonumber\\&~~A_{c'}^4 B_\beta + A_{c'}^2 B_\alpha B_\beta + A_{c'} B_\alpha C_\gamma + A_{c''} B_\alpha C_\gamma + A_{x+y+z} B_\alpha C_\gamma + A_{c'} B_\beta C_\gamma + C_\gamma C_{xyz},\nonumber\\&~~A_{c'}^4 B_\beta + A_{c'}^2 B_\alpha B_\beta + A_{c'} B_\alpha C_\gamma + A_{c''} B_\alpha C_\gamma + A_{x+y+z} B_\alpha C_\gamma + A_{c'} B_\beta C_\gamma + A_{c'}^3 C_{xyz} + C_{xyz}^2.
\end{align} 
 \end{subequations}
We have the following table regarding IWPs and group cohomology at degree 3.
\begin{center}
\resizebox{\columnwidth}{!}{
\begin{tabular}{c|cc|c|c|c}\hline\hline {Wyckoff}&\multicolumn{2}{c|}{Little group}& {Coordinates}&\multirow{2}{*}{LSM anomaly class}&\multirow{2}{*}{Topo. inv.} \\ \cline{2-4} position & Intl. & Sch\"{o}nflies & $(0,0,0) + ~(1/2,1/2,1/2) + $ & &\\ \hline
2a&$422$&$D_4$& $(0,0,0)$ & $(A_{c'}  + A_{c''} + A_{x+y+z}) B_\alpha + A_{c'} B_{z(x+y)} + C_{xyz}$ & $\varphi_2[C_2, C'_2]$\\ 
2b&$422$&$D_4$& $(0,0,1/2)$ & $C_{xyz}$ & $\varphi_2[C_2, T_1T_2C'_2]$\\ 
4c&$222$&$D_2$& $(0,1/2,0)$, $(1/2,0,0)$ & $A_{c'} B_{z(x+y)}$ & $\varphi_2[T_2T_3C_2, T_2T_3C'_2]$\\ 
4d&$222$&$D_2$& $(0,1/2,1/4)$, $(1/2,0,1/4)$ & $A_{x+y+z} B_\alpha$ & $\varphi_2[T_2T_3C_2, T_1T_2T_3C''_2]$\\ 
\hline
\hline 
 \end{tabular} }
 \end{center}

\subsection*{No. 98: $I4_122$}\label{subsub:sg98}

This group is generated by three translations $T_{1,2,3}$ as given in Eqs.~\eqref{TransBravaisI}, a two-fold rotation $C_2$, a two-fold rotation $C'_2$, and a two-fold rotation $C''_2$:
\begin{subequations}
 \begin{align}
C_2 &\colon (x,y,z)\rightarrow (-x, -y, z),\\ 
C'_2 &\colon (x,y,z)\rightarrow (-x + 1/2, y, -z + 3/4),\\ 
C''_2 &\colon (x,y,z)\rightarrow (-y, -x, -z).
\end{align}
\end{subequations}

The $\mathbb{Z}_2$ cohomology ring is given by

\begin{equation}
\mathbb{Z}_2[A_{c'},A_{c''},A_{x+y+z},B_{z(x+y)},C_\gamma]/\langle\mathcal{R}_2,\mathcal{R}_3,\mathcal{R}_4,\mathcal{R}_5,\mathcal{R}_6\rangle
 \end{equation}
where the relations are 
\begin{subequations} 
 \begin{align}
\mathcal{R}_2\colon & ~~
A_{c'} A_{c''},~~A_{c'} A_{x+y+z},\\
\mathcal{R}_3\colon & ~~
A_{c''} B_{z(x+y)},~~A_{x+y+z} B_{z(x+y)},\\
\mathcal{R}_4\colon & ~~
A_{c'} C_\gamma,~~B_{z(x+y)}^2,\\
\mathcal{R}_5\colon & ~~
B_{z(x+y)} C_\gamma,\\
\mathcal{R}_6\colon & ~~
C_\gamma (A_{c''}^2 A_{x+y+z} + A_{c''} A_{x+y+z}^2 + C_\gamma).
\end{align} 
 \end{subequations}
We have the following table regarding IWPs and group cohomology at degree 3.
\begin{center}
\begin{tabular}{c|cc|c|c|c}\hline\hline {Wyckoff}&\multicolumn{2}{c|}{Little group}& {Coordinates}&\multirow{2}{*}{LSM anomaly class}&\multirow{2}{*}{Topo. inv.} \\ \cline{2-4} position & Intl. & Sch\"{o}nflies & $(0,0,0) + ~(1/2,1/2,1/2) + $ & &\\ \hline
4a&$222$&$D_2$& $(0,0,0)$, $(0,1/2,1/4)$ & $A_{c''}^2 A_{x+y+z} + A_{c''} A_{x+y+z}^2 + C_\gamma$ & $\varphi_2[C_2, C''_2]$\\ 
4b&$222$&$D_2$& $(0,0,1/2)$, $(0,1/2,3/4)$ & $C_\gamma$ & $\varphi_2[C_2, T_1T_2C''_2]$\\ 
\hline
\multirow{2}{*}{8f} & \multirow{2}{*}{$2$} & \multirow{2}{*}{$C_2$} & $(x,1/4,1/8)$, $(-x+1/2,1/4,5/8)$, & \multirow{2}{*}{$A_{c'} B_{z(x+y)}$} & \multirow{2}{*}{$\varphi_2[T_1T_3, C'_2]$}\\
& & & $(3/4,x+1/2,3/8)$, $(3/4,-x,7/8)$ & & \\ \hline
\hline 
 \end{tabular} 
 \end{center}

\subsection*{No. 99: $P4mm$}\label{subsub:sg99}

This group is generated by three translations $T_{1,2,3}$ as given in Eqs.~\eqref{TransBravaisP}, a two-fold rotation $C_2$, a mirror $M'$, and a mirror $M$:
\begin{subequations}
 \begin{align}
C_2 &\colon (x,y,z)\rightarrow (-x, -y, z),\\ 
M' &\colon (x,y,z)\rightarrow (x, -y, z),\\ 
M &\colon (x,y,z)\rightarrow (y, x, z).
\end{align}
\end{subequations}

The $\mathbb{Z}_2$ cohomology ring is given by

\begin{equation}
\mathbb{Z}_2[A_{m'},A_m,A_{x+y},A_z,B_\alpha,B_{xy}]/\langle\mathcal{R}_2,\mathcal{R}_3,\mathcal{R}_4\rangle
 \end{equation}
where the relations are 
\begin{subequations} 
 \begin{align}
\mathcal{R}_2\colon & ~~
A_m A_{m'},~~A_m A_{x+y},~~A_z^2,\\
\mathcal{R}_3\colon & ~~
A_{m'} A_{x+y}^2 + A_{x+y}^3 + A_{x+y} B_\alpha + A_{m'} B_{xy},~~A_{x+y} (A_{m'} A_{x+y} + A_{x+y}^2 + B_\alpha + B_{xy}),\\
\mathcal{R}_4\colon & ~~
B_{xy} (B_\alpha + B_{xy}).
\end{align} 
 \end{subequations}
We have the following table regarding IWPs and group cohomology at degree 3.
\begin{center}
\begin{tabular}{c|cc|c|c|c}\hline\hline {Wyckoff}&\multicolumn{2}{c|}{Little group}& \multirow{2}{*}{Coordinates}&\multirow{2}{*}{LSM anomaly class}&\multirow{2}{*}{Topo. inv.} \\ \cline{2-3} position & Intl. & Sch\"{o}nflies & & & \\ \hline
1a&$4mm$&$C_{4v}$& $(0,0,z)$ & $A_z (A_{m'} A_{x+y} + A_{x+y}^2 + B_\alpha + B_{xy})$ & $\varphi_2[T_3, C_2]$\\ 
1b&$4mm$&$C_{4v}$& $(1/2,1/2,z)$ & $A_z B_{xy}$ & $\varphi_2[T_3, T_1T_2C_2]$\\ 
2c&$mm2$&$C_{2v}$& $(1/2,0,z)$, $(0,1/2,z)$ & $A_{x+y} (A_{m'} + A_{x+y}) A_z$ & $\varphi_2[T_3, T_1C_2]$\\ 
\hline
\hline 
 \end{tabular} 
 \end{center}

\subsection*{No. 100: $P4bm$}\label{subsub:sg100}

This group is generated by three translations $T_{1,2,3}$ as given in Eqs.~\eqref{TransBravaisP}, a two-fold rotation $C_2$, a glide $G'$, and a glide $G$:
\begin{subequations}
 \begin{align}
C_2 &\colon (x,y,z)\rightarrow (-x, -y, z),\\ 
G' &\colon (x,y,z)\rightarrow (x + 1/2, -y + 1/2, z),\\ 
G &\colon (x,y,z)\rightarrow (y + 1/2, x + 1/2, z).
\end{align}
\end{subequations}

The $\mathbb{Z}_2$ cohomology ring is given by

\begin{equation}
\mathbb{Z}_2[A_{m'},A_m,A_z,B_\alpha,B_\beta,C_\gamma]/\langle\mathcal{R}_2,\mathcal{R}_3,\mathcal{R}_4,\mathcal{R}_5,\mathcal{R}_6\rangle
 \end{equation}
where the relations are 
\begin{subequations} 
 \begin{align}
\mathcal{R}_2\colon & ~~
A_m A_{m'},~~A_{m'}^2,~~A_z^2,\\
\mathcal{R}_3\colon & ~~
A_{m'} (B_\alpha + B_\beta),~~A_{m'} B_\alpha + A_m B_\beta,\\
\mathcal{R}_4\colon & ~~
A_{m'} C_\gamma,~~B_\beta (B_\alpha + B_\beta),\\
\mathcal{R}_5\colon & ~~
B_\beta C_\gamma,\\
\mathcal{R}_6\colon & ~~
A_m^6 + A_m^4 B_\alpha + B_\alpha^3 + B_\alpha^2 B_\beta + A_m B_\alpha C_\gamma + C_\gamma^2.
\end{align} 
 \end{subequations}
We have the following table regarding IWPs and group cohomology at degree 3.
\begin{center}
\begin{tabular}{c|cc|c|c|c}\hline\hline {Wyckoff}&\multicolumn{2}{c|}{Little group}& \multirow{2}{*}{Coordinates}&\multirow{2}{*}{LSM anomaly class}&\multirow{2}{*}{Topo. inv.} \\ \cline{2-3} position & Intl. & Sch\"{o}nflies & & & \\ \hline
2a&$4$&$C_4$& $(0,0,z)$, $(1/2,1/2,z)$ & $A_z B_\beta$ & $\varphi_2[T_3, C_2]$\\ 
2b&$mm2$&$C_{2v}$& $(1/2,0,z)$, $(0,1/2,z)$ & $A_z (B_\alpha + B_\beta)$ & $\varphi_2[T_3, T_1C_2]$\\ 
\hline
\hline 
 \end{tabular} 
 \end{center}

\subsection*{No. 101: $P4_2cm$}\label{subsub:sg101}

This group is generated by three translations $T_{1,2,3}$ as given in Eqs.~\eqref{TransBravaisP}, a two-fold rotation $C_2$, a glide $G'$, and a mirror $M$:
\begin{subequations}
 \begin{align}
C_2 &\colon (x,y,z)\rightarrow (-x, -y, z),\\ 
G' &\colon (x,y,z)\rightarrow (x, -y, z + 1/2),\\ 
M &\colon (x,y,z)\rightarrow (y, x, z).
\end{align}
\end{subequations}

The $\mathbb{Z}_2$ cohomology ring is given by

\begin{equation}
\mathbb{Z}_2[A_{m'},A_m,A_{x+y},B_\alpha,B_\beta,B_{xy}]/\langle\mathcal{R}_2,\mathcal{R}_3,\mathcal{R}_4\rangle
 \end{equation}
where the relations are 
\begin{subequations} 
 \begin{align}
\mathcal{R}_2\colon & ~~
A_m A_{m'},~~A_m A_{x+y},~~A_{m'}^2,\\
\mathcal{R}_3\colon & ~~
A_{m'} B_\beta,~~A_{x+y} (A_{m'} A_{x+y} + B_\beta),~~A_{m'} A_{x+y}^2 + A_{x+y}^3 + A_{x+y} B_\alpha + A_{m'} B_{xy},~~A_{x+y} (A_{m'} A_{x+y} + A_{x+y}^2 + B_\alpha + B_{xy}),\\
\mathcal{R}_4\colon & ~~
B_\beta^2,~~B_{xy} (B_\alpha + B_{xy}).
\end{align} 
 \end{subequations}
We have the following table regarding IWPs and group cohomology at degree 3.
\begin{center}
\begin{tabular}{c|cc|c|c|c}\hline\hline {Wyckoff}&\multicolumn{2}{c|}{Little group}& \multirow{2}{*}{Coordinates}&\multirow{2}{*}{LSM anomaly class}&\multirow{2}{*}{Topo. inv.} \\ \cline{2-3} position & Intl. & Sch\"{o}nflies & & & \\ \hline
2a&$mm2$&$C_{2v}$& $(0,0,z)$, $(0,0,z+1/2)$ & $A_{x+y}^3 + A_{m'} B_\alpha + A_{x+y} B_\alpha$ & $\varphi_2[G', C_2]$\\ 
2b&$mm2$&$C_{2v}$& $(1/2,1/2,z)$, $(1/2,1/2,z+1/2)$ & $A_{x+y} (A_{m'} A_{x+y} + A_{x+y}^2 + B_\alpha)$ & $\varphi_2[T_2G', T_1T_2C_2]$\\ 
\hline
\multirow{2}{*}{4c} & \multirow{2}{*}{$2$} & \multirow{2}{*}{$C_2$} & $(0,1/2,z)$, $(1/2,0,z+1/2)$, & \multirow{2}{*}{$A_{m'} A_{x+y}^2$} & \multirow{2}{*}{$\varphi_2[T_2G', T_2C_2]$}\\
& & & $(0,1/2,z+1/2)$, $(1/2,0,z)$ & & \\ \hline
\hline 
 \end{tabular} 
 \end{center}

\subsection*{No. 102: $P4_2nm$}\label{subsub:sg102}

This group is generated by three translations $T_{1,2,3}$ as given in Eqs.~\eqref{TransBravaisP}, a two-fold rotation $C_2$, a glide $G'$, and a mirror $M$:
\begin{subequations}
 \begin{align}
C_2 &\colon (x,y,z)\rightarrow (-x, -y, z),\\ 
G' &\colon (x,y,z)\rightarrow (x + 1/2, -y + 1/2, z + 1/2),\\ 
M &\colon (x,y,z)\rightarrow (y, x, z).
\end{align}
\end{subequations}

The $\mathbb{Z}_2$ cohomology ring is given by

\begin{equation}
\mathbb{Z}_2[A_{m'},A_m,A_{x+y+z},B_\alpha,C_\gamma]/\langle\mathcal{R}_2,\mathcal{R}_3,\mathcal{R}_4,\mathcal{R}_5,\mathcal{R}_6\rangle
 \end{equation}
where the relations are 
\begin{subequations} 
 \begin{align}
\mathcal{R}_2\colon & ~~
A_m A_{m'},~~A_{m'}^2,\\
\mathcal{R}_3\colon & ~~
(A_m + A_{m'}) A_{x+y+z}^2,~~A_{m'} (A_{x+y+z}^2 + B_\alpha),\\
\mathcal{R}_4\colon & ~~
A_{x+y+z}^2 (A_{x+y+z}^2 + B_\alpha),~~A_{m'} C_\gamma,\\
\mathcal{R}_5\colon & ~~
A_{x+y+z}^2 C_\gamma,\\
\mathcal{R}_6\colon & ~~
A_m^6 + A_{x+y+z}^6 + A_m^4 B_\alpha + B_\alpha^3 + A_m B_\alpha C_\gamma + C_\gamma^2.
\end{align} 
 \end{subequations}
We have the following table regarding IWPs and group cohomology at degree 3.
\begin{center}
\begin{tabular}{c|cc|c|c|c}\hline\hline {Wyckoff}&\multicolumn{2}{c|}{Little group}& \multirow{2}{*}{Coordinates}&\multirow{2}{*}{LSM anomaly class}&\multirow{2}{*}{Topo. inv.} \\ \cline{2-3} position & Intl. & Sch\"{o}nflies & & & \\ \hline
2a&$mm2$&$C_{2v}$& $(0,0,z)$, $(1/2,1/2,z+1/2)$ & $A_{x+y+z} (A_{x+y+z}^2 + B_\alpha)$ & $\varphi_2[T_3, C_2]$\\ 
\hline
\multirow{2}{*}{4b} & \multirow{2}{*}{$2$} & \multirow{2}{*}{$C_2$} & $(0,1/2,z)$, $(0,1/2,z+1/2)$, & \multirow{2}{*}{$A_{m'} A_{x+y+z}^2$} & \multirow{2}{*}{$\varphi_2[G'M, T_1C_2]$}\\
& & & $(1/2,0,z+1/2)$, $(1/2,0,z)$ & & \\ \hline
\hline 
 \end{tabular} 
 \end{center}

\subsection*{No. 103: $P4cc$}\label{subsub:sg103}

This group is generated by three translations $T_{1,2,3}$ as given in Eqs.~\eqref{TransBravaisP}, a two-fold rotation $C_2$, a glide $G'$, and a glide $G$:
\begin{subequations}
 \begin{align}
C_2 &\colon (x,y,z)\rightarrow (-x, -y, z),\\ 
G' &\colon (x,y,z)\rightarrow (x, -y, z + 1/2),\\ 
G &\colon (x,y,z)\rightarrow (y, x, z + 1/2).
\end{align}
\end{subequations}

The $\mathbb{Z}_2$ cohomology ring is given by

\begin{equation}
\mathbb{Z}_2[A_{m'},A_m,A_{x+y},B_\alpha,B_{xy}]/\langle\mathcal{R}_2,\mathcal{R}_3,\mathcal{R}_4\rangle
 \end{equation}
where the relations are 
\begin{subequations} 
 \begin{align}
\mathcal{R}_2\colon & ~~
A_m A_{m'},~~A_m A_{x+y},~~(A_m + A_{m'})^2,\\
\mathcal{R}_3\colon & ~~
A_{m'} A_{x+y}^2 + A_{x+y}^3 + A_{x+y} B_\alpha + A_{m'} B_{xy},~~A_{x+y} (A_{m'} A_{x+y} + A_{x+y}^2 + B_\alpha + B_{xy}),\\
\mathcal{R}_4\colon & ~~
B_{xy} (B_\alpha + B_{xy}).
\end{align} 
 \end{subequations}
We have the following table regarding IWPs and group cohomology at degree 3.
\begin{center}
\resizebox{\columnwidth}{!}{
\begin{tabular}{c|cc|c|c|c}\hline\hline {Wyckoff}&\multicolumn{2}{c|}{Little group}& \multirow{2}{*}{Coordinates}&\multirow{2}{*}{LSM anomaly class}&\multirow{2}{*}{Topo. inv.} \\ \cline{2-3} position & Intl. & Sch\"{o}nflies & & & \\ \hline
2a&$4$&$C_4$& $(0,0,z)$, $(0,0,z+1/2)$ & $A_{x+y}^3 + (A_m + A_{m'} + A_{x+y}) B_\alpha + A_m B_{xy}$ & $\varphi_2[G', C_2]$\\ 
2b&$4$&$C_4$& $(1/2,1/2,z)$, $(1/2,1/2,z+1/2)$ & $A_{m'} A_{x+y}^2 + A_{x+y}^3 + A_{x+y} B_\alpha + A_m B_{xy}$ & $\varphi_2[T_2G', T_1T_2C_2]$\\ 
\hline
\multirow{2}{*}{4c} & \multirow{2}{*}{$2$} & \multirow{2}{*}{$C_2$} & $(0,1/2,z)$, $(1/2,0,z)$, & \multirow{2}{*}{$A_{m'} A_{x+y}^2$} & \multirow{2}{*}{$\varphi_2[G', T_1C_2]$}\\
& & & $(0,1/2,z+1/2)$, $(1/2,0,z+1/2)$ & & \\ \hline
\hline 
 \end{tabular} }
 \end{center}

\subsection*{No. 104: $P4nc$}\label{subsub:sg104}

This group is generated by three translations $T_{1,2,3}$ as given in Eqs.~\eqref{TransBravaisP}, a two-fold rotation $C_2$, a glide $G'$, and a glide $G$:
\begin{subequations}
 \begin{align}
C_2 &\colon (x,y,z)\rightarrow (-x, -y, z),\\ 
G' &\colon (x,y,z)\rightarrow (x + 1/2, -y + 1/2, z + 1/2),\\ 
G &\colon (x,y,z)\rightarrow (y + 1/2, x + 1/2, z + 1/2).
\end{align}
\end{subequations}

The $\mathbb{Z}_2$ cohomology ring is given by

\begin{equation}
\mathbb{Z}_2[A_{m'},A_m,B_\alpha,B_{\beta 1},B_{\beta 2},C_{\gamma1},C_{\gamma2}]/\langle\mathcal{R}_2,\mathcal{R}_3,\mathcal{R}_4,\mathcal{R}_5,\mathcal{R}_6\rangle
 \end{equation}
where the relations are 
\begin{subequations} 
 \begin{align}
\mathcal{R}_2\colon & ~~
A_m A_{m'},~~A_{m'}^2,~~A_m^2,\\
\mathcal{R}_3\colon & ~~
A_{m'} (B_\alpha + B_{\beta 1}),~~A_{m'} B_\alpha + A_m B_{\beta 1},~~A_{m'} B_{\beta 2},~~A_m B_\alpha + A_{m'} B_\alpha + A_m B_{\beta 2},\\
\mathcal{R}_4\colon & ~~
(A_m + A_{m'}) C_{\gamma1},~~A_{m'} (C_{\gamma1} + C_{\gamma2}),~~B_\alpha^2 + B_\alpha B_{\beta 1} + B_\alpha B_{\beta 2} + A_m C_{\gamma2},~~B_{\beta 1} (B_\alpha + B_{\beta 1}),~~B_{\beta 1} B_{\beta 2} + A_{m'} C_{\gamma1},\nonumber\\&~~B_\alpha^2 + B_\alpha B_{\beta 1} + B_{\beta 2}^2,\\
\mathcal{R}_5\colon & ~~
(B_\alpha + B_{\beta 1}) C_{\gamma1},~~B_{\beta 2} C_{\gamma1},~~A_{m'} B_\alpha^2 + B_\alpha C_{\gamma1} + B_{\beta 1} C_{\gamma2},~~A_m B_\alpha^2 + B_\alpha C_{\gamma1} + B_\alpha C_{\gamma2} + B_{\beta 2} C_{\gamma2},\\
\mathcal{R}_6\colon & ~~
C_{\gamma1}^2,~~C_{\gamma1} (A_{m'} B_\alpha + C_{\gamma2}),~~B_\alpha^3 + B_\alpha^2 B_{\beta 1} + A_{m'} B_\alpha C_{\gamma1} + A_m B_\alpha C_{\gamma2} + C_{\gamma2}^2.
\end{align} 
 \end{subequations}
We have the following table regarding IWPs and group cohomology at degree 3.
\begin{center}
\begin{tabular}{c|cc|c|c|c}\hline\hline {Wyckoff}&\multicolumn{2}{c|}{Little group}& \multirow{2}{*}{Coordinates}&\multirow{2}{*}{LSM anomaly class}&\multirow{2}{*}{Topo. inv.} \\ \cline{2-3} position & Intl. & Sch\"{o}nflies & & & \\ \hline
2a&$4$&$C_4$& $(0,0,z)$, $(1/2,1/2,z+1/2)$ & $C_{\gamma1}$ & $\varphi_2[T_3, C_2]$\\ 
\hline
\multirow{2}{*}{4b} & \multirow{2}{*}{$2$} & \multirow{2}{*}{$C_2$} & $(0,1/2,z)$, $(1/2,0,z)$, & \multirow{2}{*}{$(A_m + A_{m'}) B_\alpha$} & \multirow{2}{*}{$\varphi_2[T_1^{-1}G, T_2C_2]$}\\
& & & $(1/2,0,z+1/2)$, $(0,1/2,z+1/2)$ & & \\ \hline
\hline 
 \end{tabular} 
 \end{center}

\subsection*{No. 105: $P4_2mc$}\label{subsub:sg105}

This group is generated by three translations $T_{1,2,3}$ as given in Eqs.~\eqref{TransBravaisP}, a two-fold rotation $C_2$, a mirror $M'$, and a glide $G$:
\begin{subequations}
 \begin{align}
C_2 &\colon (x,y,z)\rightarrow (-x, -y, z),\\ 
M' &\colon (x,y,z)\rightarrow (x, -y, z),\\ 
G &\colon (x,y,z)\rightarrow (y, x, z + 1/2).
\end{align}
\end{subequations}

The $\mathbb{Z}_2$ cohomology ring is given by

\begin{equation}
\mathbb{Z}_2[A_{m'},A_m,A_{x+y},B_\alpha,B_\beta,B_{xy},B_{z(x+y)}]/\langle\mathcal{R}_2,\mathcal{R}_3,\mathcal{R}_4\rangle
 \end{equation}
where the relations are 
\begin{subequations} 
 \begin{align}
\mathcal{R}_2\colon & ~~
A_m A_{m'},~~A_m A_{x+y},~~A_m^2,\\
\mathcal{R}_3\colon & ~~
A_m B_\beta,~~A_{m'} A_{x+y}^2 + A_{x+y}^3 + A_{x+y} B_\alpha + A_{m'} B_{xy},~~A_{x+y} (A_{m'} A_{x+y} + A_{x+y}^2 + B_\alpha + B_{xy}),\nonumber\\&~~A_{m'} A_{x+y}^2 + A_{x+y} B_\beta + A_{m'} B_{z(x+y)},~~A_m B_{z(x+y)},\\
\mathcal{R}_4\colon & ~~
B_\beta^2,~~A_{m'}^2 A_{x+y}^2 + A_{m'} A_{x+y} B_\alpha + A_{x+y}^2 B_\beta + B_\beta B_{xy} + A_{x+y}^2 B_{z(x+y)} + B_\alpha B_{z(x+y)},~~B_\beta (A_{x+y}^2 + B_{z(x+y)}),\nonumber\\&~~B_{xy} (B_\alpha + B_{xy}),~~A_{m'} A_{x+y}^3 + A_{x+y}^2 B_\beta + A_{x+y}^2 B_{z(x+y)} + B_\alpha B_{z(x+y)} + B_{xy} B_{z(x+y)},~~(A_{x+y}^2 + B_{z(x+y)})^2.
\end{align} 
 \end{subequations}
We have the following table regarding IWPs and group cohomology at degree 3.
\begin{center}
\begin{tabular}{c|cc|c|c|c}\hline\hline {Wyckoff}&\multicolumn{2}{c|}{Little group}& \multirow{2}{*}{Coordinates}&\multirow{2}{*}{LSM anomaly class}&\multirow{2}{*}{Topo. inv.} \\ \cline{2-3} position & Intl. & Sch\"{o}nflies & & & \\ \hline
2a&$mm2$&$C_{2v}$& $(0,0,z)$, $(0,0,z+1/2)$ & $A_m (B_\alpha + B_{xy})$ & $\varphi_2[G, C_2]$\\ 
2b&$mm2$&$C_{2v}$& $(1/2,1/2,z)$, $(1/2,1/2,z+1/2)$ & $A_m B_{xy}$ & $\varphi_2[G, T_1T_2C_2]$\\ 
2c&$mm2$&$C_{2v}$& $(0,1/2,z)$, $(1/2,0,z+1/2)$ & $A_{x+y} (A_{x+y}^2 + B_\beta + B_{z(x+y)})$ & $\varphi_2[T_3, T_1C_2]$\\ 
\hline
\hline 
 \end{tabular} 
 \end{center}

\subsection*{No. 106: $P4_2bc$}\label{subsub:sg106}

This group is generated by three translations $T_{1,2,3}$ as given in Eqs.~\eqref{TransBravaisP}, a two-fold rotation $C_2$, a glide $G'$, and a glide $G$:
\begin{subequations}
 \begin{align}
C_2 &\colon (x,y,z)\rightarrow (-x, -y, z),\\ 
G' &\colon (x,y,z)\rightarrow (x + 1/2, -y + 1/2, z),\\ 
G &\colon (x,y,z)\rightarrow (y + 1/2, x + 1/2, z + 1/2).
\end{align}
\end{subequations}

The $\mathbb{Z}_2$ cohomology ring is given by

\begin{equation}
\mathbb{Z}_2[A_{m'},A_m,B_\alpha,B_{\beta 1},B_{\beta 2},C_{\gamma1},C_{\gamma2}]/\langle\mathcal{R}_2,\mathcal{R}_3,\mathcal{R}_4,\mathcal{R}_5,\mathcal{R}_6\rangle
 \end{equation}
where the relations are 
\begin{subequations} 
 \begin{align}
\mathcal{R}_2\colon & ~~
A_m A_{m'},~~A_{m'}^2,~~A_m^2,\\
\mathcal{R}_3\colon & ~~
A_{m'} (B_\alpha + B_{\beta 1}),~~A_{m'} B_\alpha + A_m B_{\beta 1},~~A_{m'} B_{\beta 2},~~A_m B_{\beta 2},\\
\mathcal{R}_4\colon & ~~
A_{m'} C_{\gamma1},~~A_m C_{\gamma1} + A_m C_{\gamma2} + A_{m'} C_{\gamma2},~~B_\alpha B_{\beta 2} + A_m C_{\gamma1} + A_{m'} C_{\gamma2},~~B_{\beta 1} (B_\alpha + B_{\beta 1}),~~B_{\beta 1} B_{\beta 2} + A_{m'} C_{\gamma2},~~B_{\beta 2}^2,\\
\mathcal{R}_5\colon & ~~
A_{m'} B_\alpha^2 + B_{\beta 1} C_{\gamma1},~~A_m B_\alpha^2 + A_{m'} B_\alpha^2 + B_{\beta 2} C_{\gamma1},~~A_{m'} B_\alpha^2 + B_\alpha C_{\gamma1} + B_\alpha C_{\gamma2} + B_{\beta 1} C_{\gamma2},~~A_m B_\alpha^2 + B_{\beta 2} C_{\gamma2},\\
\mathcal{R}_6\colon & ~~
B_\alpha^3 + B_\alpha^2 B_{\beta 1} + A_m B_\alpha C_{\gamma1} + C_{\gamma1}^2,~~B_\alpha^3 + B_\alpha^2 B_{\beta 1} + A_m B_\alpha C_{\gamma1} + A_{m'} B_\alpha C_{\gamma2} + C_{\gamma1} C_{\gamma2},~~B_\alpha^3 + A_m B_\alpha C_{\gamma1} + C_{\gamma2}^2.
\end{align} 
 \end{subequations}
We have the following table regarding IWPs and group cohomology at degree 3.
\begin{center}
\begin{tabular}{c|cc|c|c|c}\hline\hline {Wyckoff}&\multicolumn{2}{c|}{Little group}& \multirow{2}{*}{Coordinates}&\multirow{2}{*}{LSM anomaly class}&\multirow{2}{*}{Topo. inv.} \\ \cline{2-3} position & Intl. & Sch\"{o}nflies & & & \\ \hline
\multirow{2}{*}{4a} & \multirow{2}{*}{$2$} & \multirow{2}{*}{$C_2$} & $(0,0,z)$, $(0,0,z+1/2)$, & \multirow{2}{*}{$A_{m'} B_\alpha$} & \multirow{2}{*}{$\varphi_2[T_1^{-1}G'G, C_2]$}\\
& & & $(1/2,1/2,z)$, $(1/2,1/2,z+1/2)$ & & \\ \hline
\multirow{2}{*}{4b} & \multirow{2}{*}{$2$} & \multirow{2}{*}{$C_2$} & $(0,1/2,z)$, $(1/2,0,z+1/2)$, & \multirow{2}{*}{$(A_m + A_{m'}) B_\alpha$} & \multirow{2}{*}{$\varphi_2[T_1T_2C_2G, T_1C_2]$}\\
& & & $(1/2,0,z)$, $(0,1/2,z+1/2)$ & & \\ \hline
\hline 
 \end{tabular} 
 \end{center}

\subsection*{No. 107: $I4mm$}\label{subsub:sg107}

This group is generated by three translations $T_{1,2,3}$ as given in Eqs.~\eqref{TransBravaisI}, a two-fold rotation $C_2$, a mirror $M'$, and a mirror $M$:
\begin{subequations}
 \begin{align}
C_2 &\colon (x,y,z)\rightarrow (-x, -y, z),\\ 
M' &\colon (x,y,z)\rightarrow (x, -y, z),\\ 
M &\colon (x,y,z)\rightarrow (y, x, z).
\end{align}
\end{subequations}

The $\mathbb{Z}_2$ cohomology ring is given by

\begin{equation}
\mathbb{Z}_2[A_{m'},A_m,A_{x+y+z},B_\alpha,B_\beta,B_{z(x+y)},C_{xyz}]/\langle\mathcal{R}_2,\mathcal{R}_3,\mathcal{R}_4,\mathcal{R}_5,\mathcal{R}_6\rangle
 \end{equation}
where the relations are 
\begin{subequations} 
 \begin{align}
\mathcal{R}_2\colon & ~~
A_m A_{m'},~~A_{m'} A_{x+y+z},~~A_{x+y+z}^2,\\
\mathcal{R}_3\colon & ~~
A_{x+y+z} B_\alpha + A_m B_\beta,~~A_{x+y+z} (B_\alpha + B_\beta),~~A_{x+y+z} B_\alpha + A_m B_{z(x+y)},~~A_{x+y+z} (B_\alpha + B_{z(x+y)}),\\
\mathcal{R}_4\colon & ~~
A_{x+y+z} C_{xyz},~~B_\beta^2 + B_\alpha B_{z(x+y)},~~B_\alpha B_\beta + B_\beta B_{z(x+y)} + A_{m'} C_{xyz},~~B_{z(x+y)} (B_\alpha + B_{z(x+y)}),\\
\mathcal{R}_5\colon & ~~
B_\beta C_{xyz},~~B_{z(x+y)} C_{xyz},\\
\mathcal{R}_6\colon & ~~
C_{xyz}^2.
\end{align} 
 \end{subequations}
We have the following table regarding IWPs and group cohomology at degree 3.
\begin{center}
\begin{tabular}{c|cc|c|c|c}\hline\hline {Wyckoff}&\multicolumn{2}{c|}{Little group}& {Coordinates}&\multirow{2}{*}{LSM anomaly class}&\multirow{2}{*}{Topo. inv.} \\ \cline{2-4} position & Intl. & Sch\"{o}nflies & $(0,0,0) + ~(1/2,1/2,1/2) + $ & &\\ \hline
2a&$4mm$&$C_{4v}$& $(0,0,z)$ & $C_{xyz}$ & $\varphi_2[T_1T_2, C_2]$\\ 
4b&$mm2$&$C_{2v}$& $(0,1/2,z)$, $(1/2,0,z)$ & $A_{x+y+z} B_\alpha$ & $\varphi_2[T_2M, T_2T_3C_2]$\\ 
\hline
\hline 
 \end{tabular} 
 \end{center}

\subsection*{No. 108: $I4cm$}\label{subsub:sg108}

This group is generated by three translations $T_{1,2,3}$ as given in Eqs.~\eqref{TransBravaisI}, a two-fold rotation $C_2$, a glide $G'$, and a glide $G$:
\begin{subequations}
 \begin{align}
C_2 &\colon (x,y,z)\rightarrow (-x, -y, z),\\ 
G' &\colon (x,y,z)\rightarrow (x, -y, z + 1/2),\\ 
G &\colon (x,y,z)\rightarrow (y, x, z + 1/2).
\end{align}
\end{subequations}

The $\mathbb{Z}_2$ cohomology ring is given by

\begin{equation}
\mathbb{Z}_2[A_{m'},A_m,A_{x+y+z},B_\alpha,B_{z(x+y)},D_\delta]/\langle\mathcal{R}_2,\mathcal{R}_3,\mathcal{R}_4,\mathcal{R}_5,\mathcal{R}_6,\mathcal{R}_8\rangle
 \end{equation}
where the relations are 
\begin{subequations} 
 \begin{align}
\mathcal{R}_2\colon & ~~
A_m A_{m'},~~A_{m'} A_{x+y+z},~~A_m^2 + A_{m'}^2 + A_{x+y+z}^2,\\
\mathcal{R}_3\colon & ~~
A_m^2 A_{x+y+z} + A_m A_{x+y+z}^2 + A_{x+y+z} B_\alpha + A_m B_{z(x+y)},~~A_{x+y+z} (A_m^2 + A_m A_{x+y+z} + B_\alpha + B_{z(x+y)}),\\
\mathcal{R}_4\colon & ~~
B_{z(x+y)} (B_\alpha + B_{z(x+y)}),\\
\mathcal{R}_5\colon & ~~
A_{m'} D_\delta,~~(A_m + A_{x+y+z}) D_\delta,\\
\mathcal{R}_6\colon & ~~
(B_\alpha + B_{z(x+y)}) D_\delta,\\
\mathcal{R}_8\colon & ~~
A_m^4 B_\alpha^2 + D_\delta^2.
\end{align} 
 \end{subequations}
We have the following table regarding IWPs and group cohomology at degree 3.
\begin{center}
\begin{tabular}{c|cc|c|c|c}\hline\hline {Wyckoff}&\multicolumn{2}{c|}{Little group}& {Coordinates}&\multirow{2}{*}{LSM anomaly class}&\multirow{2}{*}{Topo. inv.} \\ \cline{2-4} position & Intl. & Sch\"{o}nflies & $(0,0,0) + ~(1/2,1/2,1/2) + $ & &\\ \hline
4a&$4$&$C_4$& $(0,0,z)$, $(0,0,z+1/2)$ & $(A_m + A_{m'} + A_{x+y+z}) B_\alpha + A_{m'} B_{z(x+y)}$ & $\varphi_2[G', C_2]$\\ 
4b&$mm2$&$C_{2v}$& $(1/2,0,z)$, $(0,1/2,z)$ & $A_{m'} B_{z(x+y)}$ & $\varphi_2[G', T_2T_3C_2]$\\ 
\hline
\hline 
 \end{tabular} 
 \end{center}

\subsection*{No. 109: $I4_1md$}\label{subsub:sg109}

This group is generated by three translations $T_{1,2,3}$ as given in Eqs.~\eqref{TransBravaisI}, a two-fold rotation $C_2$, a mirror $M'$, and a glide $G$:
\begin{subequations}
 \begin{align}
C_2 &\colon (x,y,z)\rightarrow (-x, -y, z),\\ 
M' &\colon (x,y,z)\rightarrow (x, -y, z),\\ 
G &\colon (x,y,z)\rightarrow (y + 1/2, x, z + 3/4).
\end{align}
\end{subequations}

The $\mathbb{Z}_2$ cohomology ring is given by

\begin{equation}
\mathbb{Z}_2[A_{m'},A_m,B_\alpha,B_{z(x+y)},C_\beta,C_\gamma,D_\delta]/\langle\mathcal{R}_2,\mathcal{R}_3,\mathcal{R}_4,\mathcal{R}_5,\mathcal{R}_6,\mathcal{R}_7,\mathcal{R}_8\rangle
 \end{equation}
where the relations are 
\begin{subequations} 
 \begin{align}
\mathcal{R}_2\colon & ~~
A_m A_{m'},~~A_m^2,\\
\mathcal{R}_3\colon & ~~
A_m B_\alpha,~~A_m B_{z(x+y)},\\
\mathcal{R}_4\colon & ~~
A_m C_\beta,~~A_m C_\gamma,~~B_\alpha B_{z(x+y)} + A_{m'} C_\gamma,~~B_{z(x+y)}^2,\\
\mathcal{R}_5\colon & ~~
A_m D_\delta,~~A_{m'}^2 C_\beta + B_{z(x+y)} C_\beta + A_{m'}^2 C_\gamma + A_{m'} D_\delta,~~B_{z(x+y)} C_\gamma,\\
\mathcal{R}_6\colon & ~~
A_{m'}^3 C_\beta + A_{m'}^3 C_\gamma + A_{m'}^2 D_\delta + B_{z(x+y)} D_\delta,~~B_\alpha^3 + A_{m'} B_\alpha C_\beta + C_\beta^2 + A_{m'}^3 C_\gamma,\nonumber\\&~~A_{m'} B_\alpha C_\beta + A_{m'} B_\alpha C_\gamma + C_\beta C_\gamma + B_\alpha D_\delta,~~C_\gamma^2,\\
\mathcal{R}_7\colon & ~~
A_{m'} B_\alpha^3 + A_{m'}^2 B_\alpha C_\beta + A_{m'}^4 C_\gamma + B_\alpha^2 C_\gamma + C_\beta D_\delta,~~A_{m'}^2 B_\alpha C_\beta + A_{m'}^2 B_\alpha C_\gamma + A_{m'} B_\alpha D_\delta + C_\gamma D_\delta,\\
\mathcal{R}_8\colon & ~~
A_{m'}^2 B_\alpha^3 + A_{m'}^3 B_\alpha C_\beta + A_{m'}^5 C_\gamma + D_\delta^2.
\end{align} 
 \end{subequations}
We have the following table regarding IWPs and group cohomology at degree 3.
\begin{center}
\begin{tabular}{c|cc|c|c|c}\hline\hline {Wyckoff}&\multicolumn{2}{c|}{Little group}& {Coordinates}&\multirow{2}{*}{LSM anomaly class}&\multirow{2}{*}{Topo. inv.} \\ \cline{2-4} position & Intl. & Sch\"{o}nflies & $(0,0,0) + ~(1/2,1/2,1/2) + $ & &\\ \hline
4a&$mm2$&$C_{2v}$& $(0,0,z)$, $(0,1/2,z+1/4)$ & $C_\gamma$ & $\varphi_2[T_1T_2, C_2]$\\ 
\hline
\hline 
 \end{tabular} 
 \end{center}

\subsection*{No. 110: $I4_1cd$}\label{subsub:sg110}

This group is generated by three translations $T_{1,2,3}$ as given in Eqs.~\eqref{TransBravaisI}, a two-fold rotation $C_2$, a glide $G'$, and a glide $G$:
\begin{subequations}
 \begin{align}
C_2 &\colon (x,y,z)\rightarrow (-x, -y, z),\\ 
G' &\colon (x,y,z)\rightarrow (x, -y, z + 1/2),\\ 
G &\colon (x,y,z)\rightarrow (y + 1/2, x, z + 1/4).
\end{align}
\end{subequations}

The $\mathbb{Z}_2$ cohomology ring is given by

\begin{equation}
\mathbb{Z}_2[A_{m'},A_m,B_\alpha,C_\gamma]/\langle\mathcal{R}_2,\mathcal{R}_3,\mathcal{R}_4,\mathcal{R}_5,\mathcal{R}_6\rangle
 \end{equation}
where the relations are 
\begin{subequations} 
 \begin{align}
\mathcal{R}_2\colon & ~~
A_m A_{m'},~~A_m^2,\\
\mathcal{R}_3\colon & ~~
A_{m'}^3,~~A_m B_\alpha,\\
\mathcal{R}_4\colon & ~~
A_{m'}^2 B_\alpha,~~A_m C_\gamma,\\
\mathcal{R}_5\colon & ~~
A_{m'}^2 C_\gamma,\\
\mathcal{R}_6\colon & ~~
B_\alpha^3 + A_{m'} B_\alpha C_\gamma + C_\gamma^2.
\end{align} 
 \end{subequations}
We have the following table regarding IWPs and group cohomology at degree 3.
\begin{center}
\begin{tabular}{c|cc|c|c|c}\hline\hline {Wyckoff}&\multicolumn{2}{c|}{Little group}& {Coordinates}&\multirow{2}{*}{LSM anomaly class}&\multirow{2}{*}{Topo. inv.} \\ \cline{2-4} position & Intl. & Sch\"{o}nflies & $(0,0,0) + ~(1/2,1/2,1/2) + $ & &\\ \hline
\multirow{2}{*}{8a} & \multirow{2}{*}{$2$} & \multirow{2}{*}{$C_2$} & $(0,0,z)$, $(0,1/2,z+1/4)$, & \multirow{2}{*}{$A_{m'} B_\alpha$} & \multirow{2}{*}{$\varphi_2[G', C_2]$}\\
& & & $(0,0,z+1/2)$, $(0,1/2,z+3/4)$ & & \\ \hline
\hline 
 \end{tabular} 
 \end{center}

\subsection*{No. 111: $P\overline42m$}\label{subsub:sg111}

This group is generated by three translations $T_{1,2,3}$ as given in Eqs.~\eqref{TransBravaisP}, a two-fold rotation $C_2$, a two-fold rotation $C'_2$, and a mirror $M$:
\begin{subequations}
 \begin{align}
C_2 &\colon (x,y,z)\rightarrow (-x, -y, z),\\ 
C'_2 &\colon (x,y,z)\rightarrow (-x, y, -z),\\ 
M &\colon (x,y,z)\rightarrow (y, x, z).
\end{align}
\end{subequations}

The $\mathbb{Z}_2$ cohomology ring is given by

\begin{equation}
\mathbb{Z}_2[A_{c'},A_m,A_{x+y},A_z,B_\alpha,B_{xy}]/\langle\mathcal{R}_2,\mathcal{R}_3,\mathcal{R}_4\rangle
 \end{equation}
where the relations are 
\begin{subequations} 
 \begin{align}
\mathcal{R}_2\colon & ~~
A_{c'} A_m,~~A_m A_{x+y},~~A_z (A_{c'} + A_z),\\
\mathcal{R}_3\colon & ~~
A_{c'} A_{x+y}^2 + A_{x+y}^3 + A_{x+y} B_\alpha + A_{c'} B_{xy},~~A_{x+y} (A_{c'} A_{x+y} + A_{x+y}^2 + B_\alpha + B_{xy}),\\
\mathcal{R}_4\colon & ~~
B_{xy} (B_\alpha + B_{xy}).
\end{align} 
 \end{subequations}
We have the following table regarding IWPs and group cohomology at degree 3.
\begin{center}
\resizebox{\columnwidth}{!}{
\begin{tabular}{c|cc|c|c|c}\hline\hline {Wyckoff}&\multicolumn{2}{c|}{Little group}& \multirow{2}{*}{Coordinates}&\multirow{2}{*}{LSM anomaly class}&\multirow{2}{*}{Topo. inv.} \\ \cline{2-3} position & Intl. & Sch\"{o}nflies & & & \\ \hline
1a&$\overline{4}2m$&$D_{2d}$& $(0,0,0)$ & 
$(A_{c'} + A_z)(A_{c'} A_{x+y}  + A_{x+y}^2 + B_\alpha + B_{xy})$ 
& $\varphi_2[C_2, C'_2]$\\ 
1b&$\overline{4}2m$&$D_{2d}$& $(1/2,1/2,1/2)$ & $A_z B_{xy}$ & $\varphi_2[T_1T_2C_2, T_1T_3C'_2]$\\ 
1c&$\overline{4}2m$&$D_{2d}$& $(0,0,1/2)$ & $A_z (A_{c'} A_{x+y} + A_{x+y}^2 + B_\alpha + B_{xy})$ & $\varphi_2[C_2, T_3C'_2]$\\ 
1d&$\overline{4}2m$&$D_{2d}$& $(1/2,1/2,0)$ & 
$(A_{c'} + A_z) B_{xy}$ & $\varphi_2[T_1T_2C_2, T_1C'_2]$\\ 
2e&$222$&$D_2$& $(1/2,0,0)$, $(0,1/2,0)$ & $A_{x+y} (A_{c'} + A_{x+y}) (A_{c'} + A_z)$ & $\varphi_2[T_1C_2, T_1C'_2]$\\ 
2f&$222$&$D_2$& $(1/2,0,1/2)$, $(0,1/2,1/2)$ & $A_{x+y} (A_{c'} + A_{x+y}) A_z$ & $\varphi_2[T_1C_2, T_1T_3C'_2]$\\ 
\hline
\hline 
 \end{tabular} }
 \end{center}

\subsection*{No. 112: $P\overline42c$}\label{subsub:sg112}

This group is generated by three translations $T_{1,2,3}$ as given in Eqs.~\eqref{TransBravaisP}, a two-fold rotation $C_2$, a two-fold rotation $C'_2$, and a glide $G$:
\begin{subequations}
 \begin{align}
C_2 &\colon (x,y,z)\rightarrow (-x, -y, z),\\ 
C'_2 &\colon (x,y,z)\rightarrow (-x, y, -z),\\ 
G &\colon (x,y,z)\rightarrow (y, x, z + 1/2).
\end{align}
\end{subequations}

The $\mathbb{Z}_2$ cohomology ring is given by

\begin{equation}
\mathbb{Z}_2[A_{c'},A_m,A_{x+y},B_\alpha,B_{xy},B_\beta,B_{z(x+y)}]/\langle\mathcal{R}_2,\mathcal{R}_3,\mathcal{R}_4\rangle
 \end{equation}
where the relations are 
\begin{subequations} 
 \begin{align}
\mathcal{R}_2\colon & ~~
A_{c'} A_m,~~A_m A_{x+y},~~A_m^2,\\
\mathcal{R}_3\colon & ~~
A_{c'} A_{x+y}^2 + A_{x+y}^3 + A_{x+y} B_\alpha + A_{c'} B_{xy},~~A_{x+y} (A_{c'} A_{x+y} + A_{x+y}^2 + B_\alpha + B_{xy}),\nonumber\\&~~A_m B_\beta,~~A_{x+y} B_\beta + A_{c'} B_{z(x+y)},~~A_m B_{z(x+y)},\\
\mathcal{R}_4\colon & ~~
B_{xy} (B_\alpha + B_{xy}),~~A_{x+y}^2 B_\beta + B_\beta B_{xy} + A_{x+y}^2 B_{z(x+y)} + B_\alpha B_{z(x+y)},\nonumber\\&~~A_{x+y}^2 B_\beta + A_{x+y}^2 B_{z(x+y)} + B_\alpha B_{z(x+y)} + B_{xy} B_{z(x+y)},~~A_{c'}^2 B_\alpha + A_{c'}^2 B_\beta + B_\beta^2,\nonumber\\&~~A_{c'} A_{x+y} B_\alpha + A_{c'} A_{x+y} B_\beta + B_\beta B_{z(x+y)},~~A_{c'}^2 A_{x+y}^2 + A_{x+y}^4 + A_{c'} A_{x+y} B_\alpha + A_{x+y}^2 B_\beta + B_{z(x+y)}^2.
\end{align} 
 \end{subequations}
We have the following table regarding IWPs and group cohomology at degree 3.
\begin{center}
\resizebox{\columnwidth}{!}{
\begin{tabular}{c|cc|c|c|c}\hline\hline {Wyckoff}&\multicolumn{2}{c|}{Little group}& \multirow{2}{*}{Coordinates}&\multirow{2}{*}{LSM anomaly class}&\multirow{2}{*}{Topo. inv.} \\ \cline{2-3} position & Intl. & Sch\"{o}nflies & & & \\ \hline
2a&$222$&$D_2$& $(0,0,1/4)$, $(0,0,3/4)$ & 
$(A_{c'} + A_m)(A_{c'} A_{x+y} + A_{x+y}^2 + B_\alpha + B_{xy})$ & $\varphi_2[C_2, C'_2]$\\ 
2b&$222$&$D_2$& $(1/2,0,1/4)$, $(0,1/2,3/4)$ & $(A_{c'} + A_{x+y}) (A_{x+y}^2 + B_{z(x+y)})$ & $\varphi_2[T_1C_2, T_1C'_2]$\\ 
2c&$222$&$D_2$& $(1/2,1/2,1/4)$, $(1/2,1/2,3/4)$ & 
$(A_{c'} + A_m) B_{xy}$ & $\varphi_2[T_1T_2C_2, T_1C'_2]$\\ 
2d&$222$&$D_2$& $(0,1/2,1/4)$, $(1/2,0,3/4)$ & 
$(A_{c'} + A_{x+y}) (A_{c'}A_{x+y} + A_{x+y}^2 + B_{z(x+y)})$ & $\varphi_2[T_2C_2, C'_2]$\\ 
2e&$\overline{4}$&$S_4$& $(0,0,0)$, $(0,0,1/2)$ & $A_m (B_\alpha + B_{xy})$ & $\varphi_2[C'_2G, C_2]$\\ 
2f&$\overline{4}$&$S_4$& $(1/2,1/2,0)$, $(1/2,1/2,1/2)$ & $A_m B_{xy}$ & $\varphi_2[T_1C'_2G, T_1T_2C_2]$\\ 
\hline
\hline 
 \end{tabular} }
 \end{center}

\subsection*{No. 113: $P\overline42_1m$}\label{subsub:sg113}

This group is generated by three translations $T_{1,2,3}$ as given in Eqs.~\eqref{TransBravaisP}, a two-fold rotation $C_2$, a two-fold screw $S'_2$, and a glide $G$:
\begin{subequations}
 \begin{align}
C_2 &\colon (x,y,z)\rightarrow (-x, -y, z),\\ 
S'_2 &\colon (x,y,z)\rightarrow (-x + 1/2, y + 1/2, -z),\\ 
G &\colon (x,y,z)\rightarrow (y + 1/2, x + 1/2, z).
\end{align}
\end{subequations}

The $\mathbb{Z}_2$ cohomology ring is given by

\begin{equation}
\mathbb{Z}_2[A_{c'},A_m,A_z,B_\alpha,B_\beta,C_\gamma]/\langle\mathcal{R}_2,\mathcal{R}_3,\mathcal{R}_4,\mathcal{R}_5,\mathcal{R}_6\rangle
 \end{equation}
where the relations are 
\begin{subequations} 
 \begin{align}
\mathcal{R}_2\colon & ~~
A_{c'} A_m,~~A_{c'}^2,~~A_z (A_{c'} + A_z),\\
\mathcal{R}_3\colon & ~~
A_{c'} (B_\alpha + B_\beta),~~A_{c'} B_\alpha + A_m B_\beta,\\
\mathcal{R}_4\colon & ~~
A_{c'} C_\gamma,~~B_\beta (B_\alpha + B_\beta),\\
\mathcal{R}_5\colon & ~~
A_{c'} B_\alpha^2 + B_\beta C_\gamma,\\
\mathcal{R}_6\colon & ~~
B_\alpha^3 + B_\alpha^2 B_\beta + A_m B_\alpha C_\gamma + C_\gamma^2.
\end{align} 
 \end{subequations}
We have the following table regarding IWPs and group cohomology at degree 3.
\begin{center}
\begin{tabular}{c|cc|c|c|c}\hline\hline {Wyckoff}&\multicolumn{2}{c|}{Little group}& \multirow{2}{*}{Coordinates}&\multirow{2}{*}{LSM anomaly class}&\multirow{2}{*}{Topo. inv.} \\ \cline{2-3} position & Intl. & Sch\"{o}nflies & & & \\ \hline
2a&$\overline{4}$&$S_4$& $(0,0,0)$, $(1/2,1/2,0)$ & $A_{c'} B_\alpha + A_z B_\beta$ & $\varphi_2[T_2^{-1}S'_2G, C_2]$\\ 
2b&$\overline{4}$&$S_4$& $(0,0,1/2)$, $(1/2,1/2,1/2)$ & $A_z B_\beta$ & $\varphi_2[T_2^{-1}T_3S'_2G, C_2]$\\ 
2c&$mm2$&$C_{2v}$& $(0,1/2,z)$, $(1/2,0,-z)$ & $A_z (B_\alpha + B_\beta)$ & $\varphi_2[T_3, T_1C_2]$\\ 
\hline
\hline 
 \end{tabular} 
 \end{center}

\subsection*{No. 114: $P\overline42_1c$}\label{subsub:sg114}

This group is generated by three translations $T_{1,2,3}$ as given in Eqs.~\eqref{TransBravaisP}, a two-fold rotation $C_2$, a two-fold screw $S'_2$, and a glide $G$:
\begin{subequations}
 \begin{align}
C_2 &\colon (x,y,z)\rightarrow (-x, -y, z),\\ 
S'_2 &\colon (x,y,z)\rightarrow (-x + 1/2, y + 1/2, -z + 1/2),\\ 
G &\colon (x,y,z)\rightarrow (y + 1/2, x + 1/2, z + 1/2).
\end{align}
\end{subequations}

The $\mathbb{Z}_2$ cohomology ring is given by

\begin{equation}
\mathbb{Z}_2[A_{c'},A_m,B_\alpha,B_{\beta 1},B_{\beta 2},C_{\gamma1},C_{\gamma2}]/\langle\mathcal{R}_2,\mathcal{R}_3,\mathcal{R}_4,\mathcal{R}_5,\mathcal{R}_6\rangle
 \end{equation}
where the relations are 
\begin{subequations} 
 \begin{align}
\mathcal{R}_2\colon & ~~
A_{c'} A_m,~~A_{c'}^2,~~A_m^2,\\
\mathcal{R}_3\colon & ~~
A_{c'} (B_\alpha + B_{\beta 1}),~~A_{c'} B_\alpha + A_m B_{\beta 1},~~A_{c'} B_{\beta 2},~~A_{c'} B_\alpha + A_m B_\alpha + A_m B_{\beta 2},\\
\mathcal{R}_4\colon & ~~
(A_{c'} + A_m) C_{\gamma1},~~A_{c'} (C_{\gamma1} + C_{\gamma2}),~~B_\alpha^2 + B_\alpha B_{\beta 1} + B_\alpha B_{\beta 2} + A_m C_{\gamma2},~~B_{\beta 1} (B_\alpha + B_{\beta 1}),~~B_{\beta 1} B_{\beta 2} + A_{c'} C_{\gamma1},\nonumber\\&~~B_\alpha^2 + B_\alpha B_{\beta 1} + B_{\beta 2}^2,\\
\mathcal{R}_5\colon & ~~
(B_\alpha + B_{\beta 1}) C_{\gamma1},~~B_{\beta 2} C_{\gamma1},~~B_\alpha C_{\gamma1} + B_{\beta 1} C_{\gamma2},~~A_{c'} B_\alpha^2 + A_m B_\alpha^2 + B_\alpha C_{\gamma1} + B_\alpha C_{\gamma2} + B_{\beta 2} C_{\gamma2},\\
\mathcal{R}_6\colon & ~~
C_{\gamma1} (A_{c'} B_\alpha + C_{\gamma1}),~~C_{\gamma1} (A_{c'} B_\alpha + C_{\gamma2}),~~B_\alpha^3 + B_\alpha^2 B_{\beta 1} + A_m B_\alpha C_{\gamma2} + C_{\gamma2}^2.
\end{align} 
 \end{subequations}
We have the following table regarding IWPs and group cohomology at degree 3.
\begin{center}
\begin{tabular}{c|cc|c|c|c}\hline\hline {Wyckoff}&\multicolumn{2}{c|}{Little group}& \multirow{2}{*}{Coordinates}&\multirow{2}{*}{LSM anomaly class}&\multirow{2}{*}{Topo. inv.} \\ \cline{2-3} position & Intl. & Sch\"{o}nflies & & & \\ \hline
2a&$\overline{4}$&$S_4$& $(0,0,0)$, $(1/2,1/2,1/2)$ & $A_{c'} B_\alpha + C_{\gamma1}$ & $\varphi_2[T_2^{-1}S'_2G, C_2]$\\ 
2b&$\overline{4}$&$S_4$& $(0,0,1/2)$, $(1/2,1/2,0)$ & $C_{\gamma1}$ & $\varphi_2[T_2^{-1}T_3S'_2G, C_2]$\\ 
\hline
\multirow{2}{*}{4d} & \multirow{2}{*}{$2$} & \multirow{2}{*}{$C_2$} & $(0,1/2,z)$, $(1/2,0,-z)$, & \multirow{2}{*}{$(A_{c'} + A_m) B_\alpha$} & \multirow{2}{*}{$\varphi_2[T_2^{-1}G, T_1C_2]$}\\
& & & $(1/2,0,-z+1/2)$, $(0,1/2,z+1/2)$ & & \\ \hline
\hline 
 \end{tabular} 
 \end{center}

\subsection*{No. 115: $P\overline4m2$}\label{subsub:sg115}

This group is generated by three translations $T_{1,2,3}$ as given in Eqs.~\eqref{TransBravaisP}, a two-fold rotation $C_2$, a two-fold rotation $C'_2$, and a mirror $M$:
\begin{subequations}
 \begin{align}
C_2 &\colon (x,y,z)\rightarrow (-x, -y, z),\\ 
C'_2 &\colon (x,y,z)\rightarrow (-y, -x, -z),\\ 
M &\colon (x,y,z)\rightarrow (x, -y, z).
\end{align}
\end{subequations}

The $\mathbb{Z}_2$ cohomology ring is given by

\begin{equation}
\mathbb{Z}_2[A_{c'},A_m,A_{x+y},A_z,B_\alpha,B_{xy}]/\langle\mathcal{R}_2,\mathcal{R}_3,\mathcal{R}_4\rangle
 \end{equation}
where the relations are 
\begin{subequations} 
 \begin{align}
\mathcal{R}_2\colon & ~~
A_{c'} A_m,~~A_{c'} A_{x+y},~~A_z (A_{c'} + A_z),\\
\mathcal{R}_3\colon & ~~
A_m A_{x+y}^2 + A_{x+y}^3 + A_{x+y} B_\alpha + A_m B_{xy},~~A_{x+y} (A_m A_{x+y} + A_{x+y}^2 + B_\alpha + B_{xy}),\\
\mathcal{R}_4\colon & ~~
B_{xy} (B_\alpha + B_{xy}).
\end{align} 
 \end{subequations}
We have the following table regarding IWPs and group cohomology at degree 3.
\begin{center}
\resizebox{\columnwidth}{!}{
\begin{tabular}{c|cc|c|c|c}\hline\hline {Wyckoff}&\multicolumn{2}{c|}{Little group}& \multirow{2}{*}{Coordinates}&\multirow{2}{*}{LSM anomaly class}&\multirow{2}{*}{Topo. inv.} \\ \cline{2-3} position & Intl. & Sch\"{o}nflies & & & \\ \hline
1a&$\overline{4}m2$&$D_{2d}$& $(0,0,0)$ & $A_m A_{x+y} A_z + A_{x+y}^2 A_z + (A_{c'} + A_z) (B_\alpha + B_{xy})$ & $\varphi_2[C_2, C'_2]$\\ 
1b&$\overline{4}m2$&$D_{2d}$& $(1/2,1/2,0)$ & $(A_{c'} + A_z) B_{xy}$ & $\varphi_2[T_1T_2C_2, T_1T_2C'_2]$\\ 
1c&$\overline{4}m2$&$D_{2d}$& $(1/2,1/2,1/2)$ & $A_z B_{xy}$ & $\varphi_2[T_1T_2C_2, T_1T_2T_3C'_2]$\\ 
1d&$\overline{4}m2$&$D_{2d}$& $(0,0,1/2)$ & $A_z (A_m A_{x+y} + A_{x+y}^2 + B_\alpha + B_{xy})$ & $\varphi_2[C_2, T_3C'_2]$\\ 
2g&$mm2$&$C_{2v}$& $(0,1/2,z)$, $(1/2,0,-z)$ & $A_{x+y} (A_m + A_{x+y}) A_z$ & $\varphi_2[T_3, T_1C_2]$\\ 
\hline
\hline 
 \end{tabular} }
 \end{center}

\subsection*{No. 116: $P\overline4c2$}\label{subsub:sg116}

This group is generated by three translations $T_{1,2,3}$ as given in Eqs.~\eqref{TransBravaisP}, a two-fold rotation $C_2$, a two-fold rotation $C'_2$, and a glide $G$:
\begin{subequations}
 \begin{align}
C_2 &\colon (x,y,z)\rightarrow (-x, -y, z),\\ 
C'_2 &\colon (x,y,z)\rightarrow (-y, -x, -z + 1/2),\\ 
G &\colon (x,y,z)\rightarrow (x, -y, z + 1/2).
\end{align}
\end{subequations}

The $\mathbb{Z}_2$ cohomology ring is given by

\begin{equation}
\mathbb{Z}_2[A_{c'},A_m,A_{x+y},B_\alpha,B_{xy},B_\beta]/\langle\mathcal{R}_2,\mathcal{R}_3,\mathcal{R}_4\rangle
 \end{equation}
where the relations are 
\begin{subequations} 
 \begin{align}
\mathcal{R}_2\colon & ~~
A_{c'} A_m,~~A_{c'} A_{x+y},~~A_m^2,\\
\mathcal{R}_3\colon & ~~
A_m A_{x+y}^2 + A_{x+y}^3 + A_{x+y} B_\alpha + A_m B_{xy},~~A_{x+y} (A_m A_{x+y} + A_{x+y}^2 + B_\alpha + B_{xy}),~~A_m B_\beta,~~A_{x+y} (A_m A_{x+y} + B_\beta),\\
\mathcal{R}_4\colon & ~~
B_{xy} (B_\alpha + B_{xy}),~~A_{c'}^2 B_\alpha + A_{c'}^2 B_\beta + B_\beta^2.
\end{align} 
 \end{subequations}
We have the following table regarding IWPs and group cohomology at degree 3.
\begin{center}
\resizebox{\columnwidth}{!}{
\begin{tabular}{c|cc|c|c|c}\hline\hline {Wyckoff}&\multicolumn{2}{c|}{Little group}& \multirow{2}{*}{Coordinates}&\multirow{2}{*}{LSM anomaly class}&\multirow{2}{*}{Topo. inv.} \\ \cline{2-3} position & Intl. & Sch\"{o}nflies & & & \\ \hline
2a&$222$&$D_2$& $(0,0,1/4)$, $(0,0,3/4)$ & 
$(A_{c'} + A_m)(A_{x+y}^2+ B_\alpha + B_{xy})$ & $\varphi_2[C_2, C'_2]$\\ 
2b&$222$&$D_2$& $(1/2,1/2,1/4)$, $(1/2,1/2,3/4)$ & 
$( A_{c'} + A_m) B_{xy}$ & $\varphi_2[T_1T_2C_2, T_1T_2C'_2]$\\ 
2c&$\overline{4}$&$S_4$& $(0,0,0)$, $(0,0,1/2)$ & 
$A_m(A_{x+y}^2+ B_\alpha + B_{xy})$ & $\varphi_2[C'_2G, C_2]$\\ 
2d&$\overline{4}$&$S_4$& $(1/2,1/2,0)$, $(1/2,1/2,1/2)$ & 
$A_m B_{xy}$ & $\varphi_2[T_2C'_2G, T_1T_2C_2]$\\ 
\hline
\multirow{2}{*}{4i} & \multirow{2}{*}{$2$} & \multirow{2}{*}{$C_2$} & $(0,1/2,z)$, $(1/2,0,-z)$, & \multirow{2}{*}{$A_m A_{x+y}^2$} & \multirow{2}{*}{$\varphi_2[G, T_1C_2]$}\\
& & & $(0,1/2,z+1/2)$, $(1/2,0,-z+1/2)$ & & \\ \hline
\hline 
 \end{tabular} }
 \end{center}

\subsection*{No. 117: $P\overline4b2$}\label{subsub:sg117}

This group is generated by three translations $T_{1,2,3}$ as given in Eqs.~\eqref{TransBravaisP}, a two-fold rotation $C_2$, a two-fold rotation $C'_2$, and a glide $G$:
\begin{subequations}
 \begin{align}
C_2 &\colon (x,y,z)\rightarrow (-x, -y, z),\\ 
C'_2 &\colon (x,y,z)\rightarrow (-y + 1/2, -x + 1/2, -z),\\ 
G &\colon (x,y,z)\rightarrow (x + 1/2, -y + 1/2, z).
\end{align}
\end{subequations}

The $\mathbb{Z}_2$ cohomology ring is given by

\begin{equation}
\mathbb{Z}_2[A_{c'},A_m,A_z,B_\alpha,B_\beta,C_\gamma]/\langle\mathcal{R}_2,\mathcal{R}_3,\mathcal{R}_4,\mathcal{R}_5,\mathcal{R}_6\rangle
 \end{equation}
where the relations are 
\begin{subequations} 
 \begin{align}
\mathcal{R}_2\colon & ~~
A_{c'} A_m,~~A_m^2,~~A_z (A_{c'} + A_z),\\
\mathcal{R}_3\colon & ~~
A_m B_\alpha + A_{c'} B_\beta,~~A_m (B_\alpha + B_\beta),\\
\mathcal{R}_4\colon & ~~
A_m C_\gamma,~~B_\beta (B_\alpha + B_\beta),\\
\mathcal{R}_5\colon & ~~
B_\beta C_\gamma,\\
\mathcal{R}_6\colon & ~~
A_{c'}^6 + A_{c'}^4 B_\alpha + B_\alpha^3 + B_\alpha^2 B_\beta + A_{c'} B_\alpha C_\gamma + C_\gamma^2.
\end{align} 
 \end{subequations}
We have the following table regarding IWPs and group cohomology at degree 3.
\begin{center}
\begin{tabular}{c|cc|c|c|c}\hline\hline {Wyckoff}&\multicolumn{2}{c|}{Little group}& \multirow{2}{*}{Coordinates}&\multirow{2}{*}{LSM anomaly class}&\multirow{2}{*}{Topo. inv.} \\ \cline{2-3} position & Intl. & Sch\"{o}nflies & & & \\ \hline
2a&$\overline{4}$&$S_4$& $(0,0,0)$, $(1/2,1/2,0)$ & $A_m B_\alpha + A_z B_\beta$ & $\varphi_2[C'_2G, C_2]$\\ 
2b&$\overline{4}$&$S_4$& $(0,0,1/2)$, $(1/2,1/2,1/2)$ & $A_z B_\beta$ & $\varphi_2[T_3C'_2G, C_2]$\\ 
2c&$222$&$D_2$& $(0,1/2,0)$, $(1/2,0,0)$ & $A_{c'} B_\alpha + A_m B_\alpha + A_z B_\alpha + A_z B_\beta$ & $\varphi_2[T_1C_2, C'_2]$\\ 
2d&$222$&$D_2$& $(0,1/2,1/2)$, $(1/2,0,1/2)$ & $A_z (B_\alpha + B_\beta)$ & $\varphi_2[T_1C_2, T_3C'_2]$\\ 
\hline
\hline 
 \end{tabular} 
 \end{center}

\subsection*{No. 118: $P\overline4n2$}\label{subsub:sg118}

This group is generated by three translations $T_{1,2,3}$ as given in Eqs.~\eqref{TransBravaisP}, a two-fold rotation $C_2$, a two-fold rotation $C'_2$, and a glide $G$:
\begin{subequations}
 \begin{align}
C_2 &\colon (x,y,z)\rightarrow (-x, -y, z),\\ 
C'_2 &\colon (x,y,z)\rightarrow (-y + 1/2, -x + 1/2, -z + 1/2),\\ 
G &\colon (x,y,z)\rightarrow (x + 1/2, -y + 1/2, z + 1/2).
\end{align}
\end{subequations}

The $\mathbb{Z}_2$ cohomology ring is given by

\begin{equation}
\mathbb{Z}_2[A_{c'},A_m,A_{x+y+z},B_\alpha,C_\gamma]/\langle\mathcal{R}_2,\mathcal{R}_3,\mathcal{R}_4,\mathcal{R}_5,\mathcal{R}_6\rangle
 \end{equation}
where the relations are 
\begin{subequations} 
 \begin{align}
\mathcal{R}_2\colon & ~~
A_{c'} A_m,~~A_m^2,\\
\mathcal{R}_3\colon & ~~
A_m A_{x+y+z}^2,~~A_{c'}^2 A_{x+y+z} + A_{c'} A_{x+y+z}^2 + A_{c'} B_\alpha + A_m B_\alpha,\\
\mathcal{R}_4\colon & ~~
A_{x+y+z} (A_{c'} + A_{x+y+z}) (A_{c'} A_{x+y+z} + A_{x+y+z}^2 + B_\alpha),~~A_m C_\gamma,\\
\mathcal{R}_5\colon & ~~
(A_{c'} A_{x+y+z} + A_{x+y+z}^2 + B_\alpha) C_\gamma,\\
\mathcal{R}_6\colon & ~~
C_\gamma (A_{c'}^2 A_{x+y+z} + A_{c'} A_{x+y+z}^2 + C_\gamma).
\end{align} 
 \end{subequations}
We have the following table regarding IWPs and group cohomology at degree 3.
\begin{center}
\resizebox{\columnwidth}{!}{
\begin{tabular}{c|cc|c|c|c}\hline\hline {Wyckoff}&\multicolumn{2}{c|}{Little group}& \multirow{2}{*}{Coordinates}&\multirow{2}{*}{LSM anomaly class}&\multirow{2}{*}{Topo. inv.} \\ \cline{2-3} position & Intl. & Sch\"{o}nflies & & & \\ \hline
2a&$\overline{4}$&$S_4$& $(0,0,0)$, $(1/2,1/2,1/2)$ & $(A_{c'} + A_{x+y+z}) (A_{c'} A_{x+y+z} + A_{x+y+z}^2 + B_\alpha)$ & $\varphi_2[C'_2G, C_2]$\\ 
2b&$\overline{4}$&$S_4$& $(0,0,1/2)$, $(1/2,1/2,0)$ & $A_{x+y+z} (A_{c'} A_{x+y+z} + A_{x+y+z}^2 + B_\alpha)$ & $\varphi_2[T_3C'_2G, C_2]$\\ 
2c&$222$&$D_2$& $(0,1/2,1/4)$, $(1/2,0,3/4)$ & $A_{c'}^2 A_{x+y+z} + A_{c'} A_{x+y+z}^2 + C_\gamma$ & $\varphi_2[T_1C_2, T_3C'_2]$\\ 
2d&$222$&$D_2$& $(0,1/2,3/4)$, $(1/2,0,1/4)$ & $C_\gamma$ & $\varphi_2[T_1C_2, C'_2]$\\ 
\hline
\hline 
 \end{tabular} }
 \end{center}

\subsection*{No. 119: $I\overline4m2$}\label{subsub:sg119}

This group is generated by three translations $T_{1,2,3}$ as given in Eqs.~\eqref{TransBravaisI}, a two-fold rotation $C_2$, a two-fold rotation $C'_2$, and a mirror $M$:
\begin{subequations}
 \begin{align}
C_2 &\colon (x,y,z)\rightarrow (-x, -y, z),\\ 
C'_2 &\colon (x,y,z)\rightarrow (-y, -x, -z),\\ 
M &\colon (x,y,z)\rightarrow (x, -y, z).
\end{align}
\end{subequations}

The $\mathbb{Z}_2$ cohomology ring is given by

\begin{equation}
\mathbb{Z}_2[A_{c'},A_m,A_{x+y+z},B_\alpha,B_\beta,B_{z(x+y)},C_\gamma,C_{xyz}]/\langle\mathcal{R}_2,\mathcal{R}_3,\mathcal{R}_4,\mathcal{R}_5,\mathcal{R}_6\rangle
 \end{equation}
where the relations are 
\begin{subequations} 
 \begin{align}
\mathcal{R}_2\colon & ~~
A_{c'} A_m,~~A_m A_{x+y+z},~~A_{x+y+z} (A_{c'} + A_{x+y+z}),\\
\mathcal{R}_3\colon & ~~
A_{x+y+z} B_\alpha + A_{c'} B_\beta,~~A_{x+y+z} (B_\alpha + B_\beta),~~A_{x+y+z} B_\alpha + A_{c'} B_{z(x+y)},~~A_{x+y+z} (B_\alpha + B_{z(x+y)}),\\
\mathcal{R}_4\colon & ~~
A_{c'} C_\gamma + A_{x+y+z} C_\gamma + A_{c'} C_{xyz},~~A_{x+y+z} C_{xyz},~~B_\alpha B_\beta + B_\alpha B_{z(x+y)} + A_m C_\gamma,~~B_\alpha B_\beta + B_\beta^2 + A_m C_\gamma,\nonumber\\&~~B_\alpha B_\beta + B_\beta B_{z(x+y)} + A_m C_{xyz},~~B_\alpha B_\beta + B_{z(x+y)}^2 + A_m C_\gamma,\\
\mathcal{R}_5\colon & ~~
(B_\beta + B_{z(x+y)}) C_\gamma,~~B_\alpha C_\gamma + B_\beta C_\gamma + B_\alpha C_{xyz},~~B_\beta C_{xyz},~~B_{z(x+y)} C_{xyz},\\
\mathcal{R}_6\colon & ~~
C_\gamma (A_{c'} B_\alpha + C_\gamma),~~C_\gamma (A_{c'} B_\alpha + A_{x+y+z} B_\alpha + C_{xyz}),~~A_{c'} B_\alpha C_\gamma + A_{x+y+z} B_\alpha C_\gamma + C_{xyz}^2.
\end{align} 
 \end{subequations}
We have the following table regarding IWPs and group cohomology at degree 3.
\begin{center}
\begin{tabular}{c|cc|c|c|c}\hline\hline {Wyckoff}&\multicolumn{2}{c|}{Little group}& {Coordinates}&\multirow{2}{*}{LSM anomaly class}&\multirow{2}{*}{Topo. inv.} \\ \cline{2-4} position & Intl. & Sch\"{o}nflies & $(0,0,0) + ~(1/2,1/2,1/2) + $ & &\\ \hline
2a&$\overline{4}m2$&$D_{2d}$& $(0,0,0)$ & $A_{c'} B_\alpha + A_{x+y+z} B_\alpha + C_{xyz}$ & $\varphi_2[C_2, C'_2]$\\ 
2b&$\overline{4}m2$&$D_{2d}$& $(0,0,1/2)$ & $C_{xyz}$ & $\varphi_2[C_2, T_1T_2C'_2]$\\ 
2c&$\overline{4}m2$&$D_{2d}$& $(0,1/2,1/4)$ & $A_{x+y+z} B_\alpha + C_\gamma + C_{xyz}$ & $\varphi_2[T_1T_3C_2, T_1T_2T_3C'_2]$\\ 
2d&$\overline{4}m2$&$D_{2d}$& $(0,1/2,3/4)$ & $C_\gamma + C_{xyz}$ & $\varphi_2[T_1T_3C_2, T_3C'_2]$\\ 
\hline
\hline 
 \end{tabular} 
 \end{center}

\subsection*{No. 120: $I\overline4c2$}\label{subsub:sg120}

This group is generated by three translations $T_{1,2,3}$ as given in Eqs.~\eqref{TransBravaisI}, a two-fold rotation $C_2$, a two-fold rotation $C'_2$, and a glide $G$:
\begin{subequations}
 \begin{align}
C_2 &\colon (x,y,z)\rightarrow (-x, -y, z),\\ 
C'_2 &\colon (x,y,z)\rightarrow (-y, -x, -z + 1/2),\\ 
G &\colon (x,y,z)\rightarrow (x, -y, z + 1/2).
\end{align}
\end{subequations}

The $\mathbb{Z}_2$ cohomology ring is given by

\begin{equation}
\mathbb{Z}_2[A_{c'},A_m,A_{x+y+z},B_\alpha,B_{z(x+y)},D_\gamma,D_\delta]/\langle\mathcal{R}_2,\mathcal{R}_3,\mathcal{R}_4,\mathcal{R}_5,\mathcal{R}_6,\mathcal{R}_8\rangle
 \end{equation}
where the relations are 
\begin{subequations} 
 \begin{align}
\mathcal{R}_2\colon & ~~
A_{c'} A_m,~~A_m A_{x+y+z},~~A_m^2 + A_{c'} A_{x+y+z} + A_{x+y+z}^2,\\
\mathcal{R}_3\colon & ~~
A_{x+y+z} B_\alpha + A_{c'} B_{z(x+y)},~~A_{x+y+z} (B_\alpha + B_{z(x+y)}),\\
\mathcal{R}_4\colon & ~~
B_{z(x+y)} (B_\alpha + B_{z(x+y)}),\\
\mathcal{R}_5\colon & ~~
A_m (B_\alpha^2 + B_\alpha B_{z(x+y)} + D_\gamma),~~A_{x+y+z} D_\gamma + A_{c'} D_\delta,~~A_m D_\delta,~~A_{x+y+z} (D_\gamma + D_\delta),\\
\mathcal{R}_6\colon & ~~
B_{z(x+y)} D_\gamma + B_\alpha D_\delta,~~B_{z(x+y)} (D_\gamma + D_\delta),\\
\mathcal{R}_8\colon & ~~
A_{c'} A_{x+y+z} B_\alpha^3 + B_\alpha^4 + B_\alpha^3 B_{z(x+y)} + A_{c'}^2 B_\alpha D_\gamma + D_\gamma^2,~~A_{c'} A_{x+y+z} B_\alpha^3 + A_{c'} A_{x+y+z} B_\alpha D_\gamma + D_\gamma D_\delta,\nonumber\\&~~A_{c'} A_{x+y+z} B_\alpha^3 + A_{c'} A_{x+y+z} B_\alpha D_\gamma + D_\delta^2.
\end{align} 
 \end{subequations}
We have the following table regarding IWPs and group cohomology at degree 3.
\begin{center}
\resizebox{\columnwidth}{!}{
\begin{tabular}{c|cc|c|c|c}\hline\hline {Wyckoff}&\multicolumn{2}{c|}{Little group}& {Coordinates}&\multirow{2}{*}{LSM anomaly class}&\multirow{2}{*}{Topo. inv.} \\ \cline{2-4} position & Intl. & Sch\"{o}nflies & $(0,0,0) + ~(1/2,1/2,1/2) + $ & &\\ \hline
4a&$222$&$D_2$& $(0,0,1/4)$, $(0,0,3/4)$ & $(A_{c'}  + A_m  + A_{x+y+z}) B_\alpha + A_m B_{z(x+y)}$ & $\varphi_2[C_2, C'_2]$\\ 
4b&$\overline{4}$&$S_4$& $(0,0,0)$, $(0,0,1/2)$ & $A_m (B_\alpha + B_{z(x+y)})$ & $\varphi_2[C'_2G, C_2]$\\ 
4c&$\overline{4}$&$S_4$& $(0,1/2,1/4)$, $(0,1/2,3/4)$ & $A_m B_{z(x+y)}$ & $\varphi_2[T_1C'_2G, T_1T_3C_2]$\\ 
4d&$222$&$D_2$& $(0,1/2,0)$, $(1/2,0,0)$ & $A_{x+y+z} B_\alpha + A_m B_{z(x+y)}$ & $\varphi_2[T_1T_3C_2, T_3C'_2]$\\ 
\hline
\hline 
 \end{tabular} }
 \end{center}

\subsection*{No. 121: $I\overline42m$}\label{subsub:sg121}

This group is generated by three translations $T_{1,2,3}$ as given in Eqs.~\eqref{TransBravaisI}, a two-fold rotation $C_2$, a two-fold rotation $C'_2$, and a mirror $M$:
\begin{subequations}
 \begin{align}
C_2 &\colon (x,y,z)\rightarrow (-x, -y, z),\\ 
C'_2 &\colon (x,y,z)\rightarrow (-x, y, -z),\\ 
M &\colon (x,y,z)\rightarrow (y, x, z).
\end{align}
\end{subequations}

The $\mathbb{Z}_2$ cohomology ring is given by

\begin{equation}
\mathbb{Z}_2[A_{c'},A_m,A_{x+y+z},B_\alpha,B_\beta,B_{z(x+y)},C_{xyz}]/\langle\mathcal{R}_2,\mathcal{R}_3,\mathcal{R}_4,\mathcal{R}_5,\mathcal{R}_6\rangle
 \end{equation}
where the relations are 
\begin{subequations} 
 \begin{align}
\mathcal{R}_2\colon & ~~
A_{c'} A_m,~~A_{c'} A_{x+y+z},~~A_{x+y+z}^2,\\
\mathcal{R}_3\colon & ~~
A_{x+y+z} B_\alpha + A_m B_\beta,~~A_{x+y+z} (B_\alpha + B_\beta),~~A_{x+y+z} B_\alpha + A_m B_{z(x+y)},~~A_{x+y+z} (B_\alpha + B_{z(x+y)}),\\
\mathcal{R}_4\colon & ~~
A_{x+y+z} C_{xyz},~~A_{c'}^2 B_\beta + B_\beta^2 + B_\alpha B_{z(x+y)},~~B_\alpha B_\beta + B_\beta B_{z(x+y)} + A_{c'} C_{xyz},~~B_{z(x+y)} (B_\alpha + B_{z(x+y)}),\\
\mathcal{R}_5\colon & ~~
(A_{c'}^2 + B_\beta) C_{xyz},~~B_{z(x+y)} C_{xyz},\\
\mathcal{R}_6\colon & ~~
C_{xyz} (A_{c'} B_\alpha + C_{xyz}).
\end{align} 
 \end{subequations}
We have the following table regarding IWPs and group cohomology at degree 3.
\begin{center}
\begin{tabular}{c|cc|c|c|c}\hline\hline {Wyckoff}&\multicolumn{2}{c|}{Little group}& {Coordinates}&\multirow{2}{*}{LSM anomaly class}&\multirow{2}{*}{Topo. inv.} \\ \cline{2-4} position & Intl. & Sch\"{o}nflies & $(0,0,0) + ~(1/2,1/2,1/2) + $ & &\\ \hline
2a&$\overline{4}2m$&$D_{2d}$& $(0,0,0)$ & $A_{c'} B_\alpha + A_{c'} B_{z(x+y)} + C_{xyz}$ & $\varphi_2[C_2, C'_2]$\\ 
2b&$\overline{4}2m$&$D_{2d}$& $(0,0,1/2)$ & $C_{xyz}$ & $\varphi_2[C_2, T_1T_2C'_2]$\\ 
4c&$222$&$D_2$& $(0,1/2,0)$, $(1/2,0,0)$ & $A_{x+y+z} B_\alpha + A_{c'} B_{z(x+y)}$ & $\varphi_2[T_1T_3C_2, C'_2]$\\ 
4d&$\overline{4}$&$S_4$& $(0,1/2,1/4)$, $(0,1/2,3/4)$ & $A_{x+y+z} B_\alpha$ & $\varphi_2[T_1T_2T_3C'_2M, T_1T_3C_2]$\\ 
\hline
\hline 
 \end{tabular} 
 \end{center}

\subsection*{No. 122: $I\overline42d$}\label{subsub:sg122}

This group is generated by three translations $T_{1,2,3}$ as given in Eqs.~\eqref{TransBravaisI}, a two-fold rotation $C_2$, a two-fold rotation $C'_2$, and a glide $G$:
\begin{subequations}
 \begin{align}
C_2 &\colon (x,y,z)\rightarrow (-x, -y, z),\\ 
C'_2 &\colon (x,y,z)\rightarrow (-x + 1/2, y, -z + 3/4),\\ 
G &\colon (x,y,z)\rightarrow (y, x + 1/2, z + 1/4).
\end{align}
\end{subequations}

The $\mathbb{Z}_2$ cohomology ring is given by

\begin{equation}
\mathbb{Z}_2[A_{c'},A_m,B_\alpha,B_{z(x+y)},C_\gamma]/\langle\mathcal{R}_2,\mathcal{R}_3,\mathcal{R}_4,\mathcal{R}_5,\mathcal{R}_6\rangle
 \end{equation}
where the relations are 
\begin{subequations} 
 \begin{align}
\mathcal{R}_2\colon & ~~
A_{c'} A_m,~~A_m^2,\\
\mathcal{R}_3\colon & ~~
(A_{c'} + A_m) B_\alpha,~~A_{c'} B_\alpha + A_m B_{z(x+y)},\\
\mathcal{R}_4\colon & ~~
(A_{c'} + A_m) C_\gamma,~~B_\alpha^2 + B_\alpha B_{z(x+y)} + A_{c'} C_\gamma,~~(B_\alpha + B_{z(x+y)})^2,\\
\mathcal{R}_5\colon & ~~
(B_\alpha + B_{z(x+y)}) C_\gamma,\\
\mathcal{R}_6\colon & ~~
C_\gamma (A_{c'} B_\alpha + C_\gamma).
\end{align} 
 \end{subequations}
We have the following table regarding IWPs and group cohomology at degree 3.
\begin{center}
\begin{tabular}{c|cc|c|c|c}\hline\hline {Wyckoff}&\multicolumn{2}{c|}{Little group}& {Coordinates}&\multirow{2}{*}{LSM anomaly class}&\multirow{2}{*}{Topo. inv.} \\ \cline{2-4} position & Intl. & Sch\"{o}nflies & $(0,0,0) + ~(1/2,1/2,1/2) + $ & &\\ \hline
4a&$\overline{4}$&$S_4$& $(0,0,0)$, $(1/2,0,3/4)$ & $C_\gamma$ & $\varphi_2[C'_2G, T_1T_3C_2]$\\ 
4b&$\overline{4}$&$S_4$& $(0,0,1/2)$, $(1/2,0,1/4)$ & $A_{c'} B_\alpha + C_\gamma$ & $\varphi_2[T_3^{-1}C'_2G, C_2]$\\ 
\hline
\multirow{2}{*}{8c} & \multirow{2}{*}{$2$} & \multirow{2}{*}{$C_2$} & $(0,0,z)$, $(0,0,-z)$, & \multirow{2}{*}{$A_{c'} (B_\alpha + B_{z(x+y)})$} & \multirow{2}{*}{$\varphi_2[T_2T_3, T_3C_2C'_2]$}\\
& & & $(1/2,0,-z+3/4)$, $(1/2,0,z+3/4)$ & & \\ \hline
\hline 
 \end{tabular} 
 \end{center}

\subsection*{No. 123: $P4/mmm$}\label{subsub:sg123}

This group is generated by three translations $T_{1,2,3}$ as given in Eqs.~\eqref{TransBravaisP}, a two-fold rotation $C_2$, a two-fold rotation $C'_2$, a mirror $M$, and an inversion $I$:
\begin{subequations}
 \begin{align}
C_2 &\colon (x,y,z)\rightarrow (-x, -y, z),\\ 
C'_2 &\colon (x,y,z)\rightarrow (-x, y, -z),\\ 
M &\colon (x,y,z)\rightarrow (y, x, z),\\ 
I &\colon (x,y,z)\rightarrow (-x, -y, -z).
\end{align}
\end{subequations}

The $\mathbb{Z}_2$ cohomology ring is given by

\begin{equation}
\mathbb{Z}_2[A_i,A_m,A_{c'},A_{x+y},A_z,B_\alpha,B_{xy}]/\langle\mathcal{R}_2,\mathcal{R}_3,\mathcal{R}_4\rangle
 \end{equation}
where the relations are 
\begin{subequations} 
 \begin{align}
\mathcal{R}_2\colon & ~~
A_{c'} A_m,~~A_m A_{x+y},~~A_z (A_{c'} + A_i + A_z),\\
\mathcal{R}_3\colon & ~~
A_{c'} A_{x+y}^2 + A_{x+y}^3 + A_{x+y} B_\alpha + A_{c'} B_{xy},~~A_{x+y} (A_{c'} A_{x+y} + A_{x+y}^2 + B_\alpha + B_{xy}),\\
\mathcal{R}_4\colon & ~~
B_{xy} (B_\alpha + B_{xy}).
\end{align} 
 \end{subequations}
We have the following table regarding IWPs and group cohomology at degree 3.
\begin{center}
\resizebox{\columnwidth}{!}{
\begin{tabular}{c|cc|c|c|c}\hline\hline {Wyckoff}&\multicolumn{2}{c|}{Little group}& \multirow{2}{*}{Coordinates}&\multirow{2}{*}{LSM anomaly class}&\multirow{2}{*}{Topo. inv.} \\ \cline{2-3} position & Intl. & Sch\"{o}nflies & & & \\ \hline
1a&$4/mmm$&$D_{4h}$& $(0,0,0)$ & 
$(A_{c'} + A_i + A_z)(A_{c'} A_{x+y} + A_{x+y}^2 + B_\alpha + B_{xy})$ & $\varphi_2[C_2, C'_2]$\\ 
1b&$4/mmm$&$D_{4h}$& $(0,0,1/2)$ & $A_z (A_{c'} A_{x+y} + A_{x+y}^2 + B_\alpha + B_{xy})$ & $\varphi_2[C_2, T_3C'_2]$\\ 
1c&$4/mmm$&$D_{4h}$& $(1/2,1/2,0)$ & 
$(A_{c'} + A_i + A_z) B_{xy}$ & $\varphi_2[T_1T_2C_2, T_1C'_2]$\\ 
1d&$4/mmm$&$D_{4h}$& $(1/2,1/2,1/2)$ & $A_z B_{xy}$ & $\varphi_2[T_1T_2C_2, T_1T_3C'_2]$\\ 
2e&$mmm$&$D_{2h}$& $(0,1/2,1/2)$, $(1/2,0,1/2)$ & $A_{x+y} (A_{c'} + A_{x+y}) A_z$ & $\varphi_2[T_1C_2, T_1T_3C'_2]$\\ 
2f&$mmm$&$D_{2h}$& $(0,1/2,0)$, $(1/2,0,0)$ & $A_{x+y} (A_{c'} + A_{x+y}) (A_{c'} + A_i + A_z)$ & $\varphi_2[T_1C_2, T_1C'_2]$\\ 
\hline
\hline 
 \end{tabular} }
 \end{center}

\subsection*{No. 124: $P4/mcc$}\label{subsub:sg124}

This group is generated by three translations $T_{1,2,3}$ as given in Eqs.~\eqref{TransBravaisP}, a two-fold rotation $C_2$, a two-fold rotation $C'_2$, a glide $G$, and an inversion $I$:
\begin{subequations}
 \begin{align}
C_2 &\colon (x,y,z)\rightarrow (-x, -y, z),\\ 
C'_2 &\colon (x,y,z)\rightarrow (-x, y, -z + 1/2),\\ 
G &\colon (x,y,z)\rightarrow (y, x, z + 1/2),\\ 
I &\colon (x,y,z)\rightarrow (-x, -y, -z).
\end{align}
\end{subequations}

The $\mathbb{Z}_2$ cohomology ring is given by

\begin{equation}
\mathbb{Z}_2[A_i,A_m,A_{c'},A_{x+y},B_\alpha,B_{xy}]/\langle\mathcal{R}_2,\mathcal{R}_3,\mathcal{R}_4\rangle
 \end{equation}
where the relations are 
\begin{subequations} 
 \begin{align}
\mathcal{R}_2\colon & ~~
A_{c'} A_m,~~A_m A_{x+y},~~A_{c'} A_i + A_i A_m + A_m^2,\\
\mathcal{R}_3\colon & ~~
A_{c'} A_{x+y}^2 + A_{x+y}^3 + A_{x+y} B_\alpha + A_{c'} B_{xy},~~A_{x+y} (A_{c'} A_{x+y} + A_{x+y}^2 + B_\alpha + B_{xy}),\\
\mathcal{R}_4\colon & ~~
B_{xy} (B_\alpha + B_{xy}).
\end{align} 
 \end{subequations}
We have the following table regarding IWPs and group cohomology at degree 3.
\begin{center}
\resizebox{\columnwidth}{!}{
\begin{tabular}{c|cc|c|c|c}\hline\hline {Wyckoff}&\multicolumn{2}{c|}{Little group}& \multirow{2}{*}{Coordinates}&\multirow{2}{*}{LSM anomaly class}&\multirow{2}{*}{Topo. inv.} \\ \cline{2-3} position & Intl. & Sch\"{o}nflies & & & \\ \hline
2a&$422$&$D_4$& $(0,0,1/4)$, $(0,0,3/4)$ & 
$(A_{c'} + A_m)(A_{c'} A_{x+y}  + A_{x+y}^2 + B_\alpha + B_{xy})$ & $\varphi_2[C_2, GI]$\\ 
2b&$4/m$&$C_{4h}$& $(0,0,0)$, $(0,0,1/2)$ & 
$A_i A_{x+y}^2 + (A_{i} + A_m) (B_\alpha + B_{xy})$ & $\varphi_1[I]$\\ 
2c&$422$&$D_4$& $(1/2,1/2,1/4)$, $(1/2,1/2,3/4)$ & 
$(A_{c'} + A_m) B_{xy}$ & $\varphi_2[T_1T_2C_2, T_1T_2GI]$\\ 
2d&$4/m$&$C_{4h}$& $(1/2,1/2,0)$, $(1/2,1/2,1/2)$ & $(A_i + A_m) B_{xy}$ & $\varphi_1[T_1T_2I]$\\ 
\hline
\multirow{2}{*}{4e} & \multirow{2}{*}{$2/m$} & \multirow{2}{*}{$C_{2h}$} & $(0,1/2,0)$, $(1/2,0,0)$, & \multirow{2}{*}{$A_i A_{x+y}^2$} & \multirow{2}{*}{$\varphi_1[T_1I]$}\\
& & & $(0,1/2,1/2)$, $(1/2,0,1/2)$ & & \\ \hline
\multirow{2}{*}{4f} & \multirow{2}{*}{$222$} & \multirow{2}{*}{$D_2$} & $(0,1/2,1/4)$, $(1/2,0,1/4)$, & \multirow{2}{*}{$A_{c'} A_{x+y} (A_{c'} + A_{x+y})$} & \multirow{2}{*}{$\varphi_2[T_1C_2, T_1C'_2]$}\\
& & & $(0,1/2,3/4)$, $(1/2,0,3/4)$ & & \\ \hline
\hline 
 \end{tabular} }
 \end{center}

\subsection*{No. 125: $P4/nbm$}\label{subsub:sg125}

This group is generated by three translations $T_{1,2,3}$ as given in Eqs.~\eqref{TransBravaisP}, a two-fold rotation $C_2$, a two-fold rotation $C'_2$, a glide $G$, and an inversion $I$:
\begin{subequations}
 \begin{align}
C_2 &\colon (x,y,z)\rightarrow (-x + 1/2, -y + 1/2, z),\\ 
C'_2 &\colon (x,y,z)\rightarrow (-x + 1/2, y, -z),\\ 
G &\colon (x,y,z)\rightarrow (y + 1/2, x + 1/2, z),\\ 
I &\colon (x,y,z)\rightarrow (-x, -y, -z).
\end{align}
\end{subequations}

The $\mathbb{Z}_2$ cohomology ring is given by

\begin{equation}
\mathbb{Z}_2[A_i,A_m,A_{c'},A_z,B_\alpha,B_\beta]/\langle\mathcal{R}_2,\mathcal{R}_3,\mathcal{R}_4\rangle
 \end{equation}
where the relations are 
\begin{subequations} 
 \begin{align}
\mathcal{R}_2\colon & ~~
A_{c'} A_i,~~A_{c'} A_m,~~A_z (A_{c'} + A_i + A_z),\\
\mathcal{R}_3\colon & ~~
A_i B_\beta,~~A_i^3 + A_i A_m^2 + A_i B_\alpha + A_m B_\alpha + A_m B_\beta,\\
\mathcal{R}_4\colon & ~~
B_\beta (B_\alpha + B_\beta).
\end{align} 
 \end{subequations}
We have the following table regarding IWPs and group cohomology at degree 3.
\begin{center}
\resizebox{\columnwidth}{!}{
\begin{tabular}{c|cc|c|c|c}\hline\hline {Wyckoff}&\multicolumn{2}{c|}{Little group}& \multirow{2}{*}{Coordinates}&\multirow{2}{*}{LSM anomaly class}&\multirow{2}{*}{Topo. inv.} \\ \cline{2-3} position & Intl. & Sch\"{o}nflies & & & \\ \hline
2a&$422$&$D_4$& $(1/4,1/4,0)$, $(3/4,3/4,0)$ & 
$(A_{c'} + A_i + A_z)(A_i^2 + A_i A_m + B_\alpha + B_\beta)$ & $\varphi_2[C_2, C'_2]$\\ 
2b&$422$&$D_4$& $(1/4,1/4,1/2)$, $(3/4,3/4,1/2)$ & $A_z (A_i^2 + A_i A_m + B_\alpha + B_\beta)$ & $\varphi_2[C_2, T_3C'_2]$\\ 
2c&$\overline{4}2m$&$D_{2d}$& $(3/4,1/4,0)$, $(1/4,3/4,0)$ & $(A_{c'} + A_z) B_\beta$ & $\varphi_2[T_2C_2, C'_2]$\\ 
2d&$\overline{4}2m$&$D_{2d}$& $(3/4,1/4,1/2)$, $(1/4,3/4,1/2)$ & $A_z B_\beta$ & $\varphi_2[T_2C_2, T_3C'_2]$\\ 
\hline
\multirow{2}{*}{4e} & \multirow{2}{*}{$2/m$} & \multirow{2}{*}{$C_{2h}$} & $(0,0,0)$, $(1/2,1/2,0)$, & \multirow{2}{*}{$A_i (A_i + A_m) (A_i + A_z)$} & \multirow{2}{*}{$\varphi_1[I]$}\\
& & & $(1/2,0,0)$, $(0,1/2,0)$ & & \\ \hline
\multirow{2}{*}{4f} & \multirow{2}{*}{$2/m$} & \multirow{2}{*}{$C_{2h}$} & $(0,0,1/2)$, $(1/2,1/2,1/2)$, & \multirow{2}{*}{$A_i (A_i + A_m) A_z$} & \multirow{2}{*}{$\varphi_1[T_3I]$}\\
& & & $(1/2,0,1/2)$, $(0,1/2,1/2)$ & & \\ \hline
\hline 
 \end{tabular} }
 \end{center}

\subsection*{No. 126: $P4/nnc$}\label{subsub:sg126}

This group is generated by three translations $T_{1,2,3}$ as given in Eqs.~\eqref{TransBravaisP}, a two-fold rotation $C_2$, a two-fold rotation $C'_2$, a glide $G$, and an inversion $I$:
\begin{subequations}
 \begin{align}
C_2 &\colon (x,y,z)\rightarrow (-x + 1/2, -y + 1/2, z),\\ 
C'_2 &\colon (x,y,z)\rightarrow (-x + 1/2, y, -z + 1/2),\\ 
G &\colon (x,y,z)\rightarrow (y + 1/2, x + 1/2, z + 1/2),\\ 
I &\colon (x,y,z)\rightarrow (-x, -y, -z).
\end{align}
\end{subequations}

The $\mathbb{Z}_2$ cohomology ring is given by

\begin{equation}
\mathbb{Z}_2[A_i,A_{c'},A_m,B_\alpha,B_{\beta 1},B_{\beta 2},C_\gamma]/\langle\mathcal{R}_2,\mathcal{R}_3,\mathcal{R}_4,\mathcal{R}_5,\mathcal{R}_6\rangle
 \end{equation}
where the relations are 
\begin{subequations} 
 \begin{align}
\mathcal{R}_2\colon & ~~
A_{c'} A_i,~~A_{c'} A_m,~~A_m (A_i + A_m),\\
\mathcal{R}_3\colon & ~~
A_i B_{\beta 1},~~A_i B_\alpha + A_m B_\alpha + A_m B_{\beta 1},~~A_i B_{\beta 2},~~A_m B_{\beta 2},\\
\mathcal{R}_4\colon & ~~
(A_i + A_m) C_\gamma,~~B_{\beta 1} (A_{c'}^2 + B_\alpha + B_{\beta 1}),~~A_{c'}^2 B_{\beta 2} + B_\alpha B_{\beta 2} + B_{\beta 1} B_{\beta 2} + A_{c'} C_\gamma,~~A_{c'}^2 B_{\beta 1} + A_{c'}^2 B_{\beta 2} + B_{\beta 2}^2,\\
\mathcal{R}_5\colon & ~~
B_{\beta 1} C_\gamma,~~(A_{c'}^2 + B_{\beta 2}) C_\gamma,\\
\mathcal{R}_6\colon & ~~
C_\gamma (A_{c'}^3 + A_{c'} B_\alpha + A_i B_\alpha + C_\gamma).
\end{align} 
 \end{subequations}
We have the following table regarding IWPs and group cohomology at degree 3.
\begin{center}
\resizebox{\columnwidth}{!}{
\begin{tabular}{c|cc|c|c|c}\hline\hline {Wyckoff}&\multicolumn{2}{c|}{Little group}& \multirow{2}{*}{Coordinates}&\multirow{2}{*}{LSM anomaly class}&\multirow{2}{*}{Topo. inv.} \\ \cline{2-3} position & Intl. & Sch\"{o}nflies & & & \\ \hline
2a&$422$&$D_4$& $(1/4,1/4,1/4)$, $(3/4,3/4,3/4)$ & $A_{c'}^3 + A_{c'} B_\alpha + A_i B_\alpha + A_{c'} B_{\beta 1} + C_\gamma$ & $\varphi_2[C_2, C'_2]$\\ 
2b&$422$&$D_4$& $(1/4,1/4,3/4)$, $(3/4,3/4,1/4)$ & $C_\gamma$ & $\varphi_2[C_2, T_3C'_2]$\\ 
\hline
\multirow{2}{*}{4c} & \multirow{2}{*}{$222$} & \multirow{2}{*}{$D_2$} & $(1/4,3/4,3/4)$, $(3/4,1/4,3/4)$, & \multirow{2}{*}{$A_i B_\alpha + A_m B_\alpha + A_{c'} B_{\beta 1}$} & \multirow{2}{*}{$\varphi_2[T_1C_2, T_1T_3C'_2]$}\\
& & & $(3/4,1/4,1/4)$, $(1/4,3/4,1/4)$ & & \\ \hline
\multirow{2}{*}{4d} & \multirow{2}{*}{$\overline{4}$} & \multirow{2}{*}{$S_4$} & $(1/4,3/4,0)$, $(3/4,1/4,0)$, & \multirow{2}{*}{$(A_i + A_m) B_\alpha$} & \multirow{2}{*}{$\varphi_2[C'_2G, T_1^{-1}C_2]$}\\
& & & $(1/4,3/4,1/2)$, $(3/4,1/4,1/2)$ & & \\ \hline
\multirow{4}{*}{8f} & \multirow{4}{*}{$\overline{1}$} & \multirow{4}{*}{$C_i$} & $(0,0,0)$, $(1/2,1/2,0)$, & \multirow{4}{*}{$A_i^2 (A_i + A_m)$} & \multirow{4}{*}{$\varphi_1[I]$}\\
& & & $(1/2,0,0)$, $(0,1/2,0)$, & & \\
& & & $(1/2,0,1/2)$, $(0,1/2,1/2)$, & & \\
& & & $(0,0,1/2)$, $(1/2,1/2,1/2)$ & & \\ \hline
\hline 
 \end{tabular} }
 \end{center}

\subsection*{No. 127: $P4/mbm$}\label{subsub:sg127}

This group is generated by three translations $T_{1,2,3}$ as given in Eqs.~\eqref{TransBravaisP}, a two-fold rotation $C_2$, a two-fold screw $S'_2$, a glide $G$, and an inversion $I$:
\begin{subequations}
 \begin{align}
C_2 &\colon (x,y,z)\rightarrow (-x, -y, z),\\ 
S'_2 &\colon (x,y,z)\rightarrow (-x + 1/2, y + 1/2, -z),\\ 
G &\colon (x,y,z)\rightarrow (y + 1/2, x + 1/2, z),\\ 
I &\colon (x,y,z)\rightarrow (-x, -y, -z).
\end{align}
\end{subequations}

The $\mathbb{Z}_2$ cohomology ring is given by

\begin{equation}
\mathbb{Z}_2[A_i,A_m,A_{c'},A_z,B_\alpha,B_\beta,C_\gamma]/\langle\mathcal{R}_2,\mathcal{R}_3,\mathcal{R}_4,\mathcal{R}_5,\mathcal{R}_6\rangle
 \end{equation}
where the relations are 
\begin{subequations} 
 \begin{align}
\mathcal{R}_2\colon & ~~
A_{c'} A_m,~~A_{c'}^2,~~A_z (A_{c'} + A_i + A_z),\\
\mathcal{R}_3\colon & ~~
A_{c'} B_\alpha + A_m B_\beta,~~A_{c'} (B_\alpha + B_\beta),\\
\mathcal{R}_4\colon & ~~
A_{c'} (A_i B_\alpha + C_\gamma),~~B_\beta (B_\alpha + B_\beta),\\
\mathcal{R}_5\colon & ~~
A_{c'} B_\alpha^2 + A_i B_\alpha B_\beta + B_\beta C_\gamma,\\
\mathcal{R}_6\colon & ~~
A_i^4 A_m^2 + A_m^6 + A_i^2 A_m^2 B_\alpha + A_m^4 B_\alpha + A_{c'} A_i B_\alpha^2 + B_\alpha^3 + A_i^2 B_\alpha B_\beta + B_\alpha^2 B_\beta + A_m B_\alpha C_\gamma + C_\gamma^2.
\end{align} 
 \end{subequations}
We have the following table regarding IWPs and group cohomology at degree 3.
\begin{center}
\begin{tabular}{c|cc|c|c|c}\hline\hline {Wyckoff}&\multicolumn{2}{c|}{Little group}& \multirow{2}{*}{Coordinates}&\multirow{2}{*}{LSM anomaly class}&\multirow{2}{*}{Topo. inv.} \\ \cline{2-3} position & Intl. & Sch\"{o}nflies & & & \\ \hline
2a&$4/m$&$C_{4h}$& $(0,0,0)$, $(1/2,1/2,0)$ & $A_{c'} B_\alpha + A_i B_\beta + A_z B_\beta$ & $\varphi_1[I]$\\ 
2b&$4/m$&$C_{4h}$& $(0,0,1/2)$, $(1/2,1/2,1/2)$ & $A_z B_\beta$ & $\varphi_1[T_3I]$\\ 
2c&$mmm$&$D_{2h}$& $(0,1/2,1/2)$, $(1/2,0,1/2)$ & $A_z (B_\alpha + B_\beta)$ & $\varphi_2[T_2C_2, T_3GI]$\\ 
2d&$mmm$&$D_{2h}$& $(0,1/2,0)$, $(1/2,0,0)$ & $(A_i + A_z) (B_\alpha + B_\beta)$ & $\varphi_2[T_2C_2, GI]$\\ 
\hline
\hline 
 \end{tabular} 
 \end{center}

\subsection*{No. 128: $P4/mnc$}\label{subsub:sg128}

This group is generated by three translations $T_{1,2,3}$ as given in Eqs.~\eqref{TransBravaisP}, a two-fold rotation $C_2$, a two-fold screw $S'_2$, a glide $G$, and an inversion $I$:
\begin{subequations}
 \begin{align}
C_2 &\colon (x,y,z)\rightarrow (-x, -y, z),\\ 
S'_2 &\colon (x,y,z)\rightarrow (-x + 1/2, y + 1/2, -z + 1/2),\\ 
G &\colon (x,y,z)\rightarrow (y + 1/2, x + 1/2, z + 1/2),\\ 
I &\colon (x,y,z)\rightarrow (-x, -y, -z).
\end{align}
\end{subequations}

The $\mathbb{Z}_2$ cohomology ring is given by

\begin{equation}
\mathbb{Z}_2[A_i,A_m,A_{c'},B_\alpha,B_{\beta 1},B_{\beta 2},C_{\gamma1},C_{\gamma2}]/\langle\mathcal{R}_2,\mathcal{R}_3,\mathcal{R}_4,\mathcal{R}_5,\mathcal{R}_6\rangle
 \end{equation}
where the relations are 
\begin{subequations} 
 \begin{align}
\mathcal{R}_2\colon & ~~
A_{c'} A_m,~~A_{c'} A_i + A_i A_m + A_m^2,~~A_{c'}^2,\\
\mathcal{R}_3\colon & ~~
A_{c'} B_\alpha + A_m B_{\beta 1},~~A_{c'} (B_\alpha + B_{\beta 1}),~~A_{c'} B_\alpha + A_m B_\alpha + A_m B_{\beta 2},~~A_{c'} B_{\beta 2},\\
\mathcal{R}_4\colon & ~~
(A_{c'} + A_m) C_{\gamma1},~~A_m C_{\gamma1} + A_{c'} C_{\gamma2},\nonumber\\&~~A_i^3 A_m + A_i^2 A_m^2 + A_i A_m B_\alpha + B_\alpha^2 + B_\alpha B_{\beta 1} + A_i^2 B_{\beta 2} + B_\alpha B_{\beta 2} + A_i C_{\gamma1} + A_i C_{\gamma2} + A_m C_{\gamma2},~~B_{\beta 1} (B_\alpha + B_{\beta 1}),\nonumber\\&~~B_{\beta 1} B_{\beta 2} + A_m C_{\gamma1},~~A_{c'} A_i B_\alpha + A_i^2 B_\alpha + A_i A_m B_\alpha + B_\alpha^2 + A_i^2 B_{\beta 1} + B_\alpha B_{\beta 1} + B_{\beta 2}^2,\\
\mathcal{R}_5\colon & ~~
(B_\alpha + B_{\beta 1}) C_{\gamma1},~~(A_i A_m + B_{\beta 2}) C_{\gamma1},~~A_{c'} A_i^2 B_\alpha + A_{c'} B_\alpha^2 + A_i A_m C_{\gamma1} + B_\alpha C_{\gamma1} + B_{\beta 1} C_{\gamma2},\nonumber\\&~~A_i^4 A_m + A_i^3 A_m^2 + A_i^3 B_\alpha + A_i B_\alpha^2 + A_m B_\alpha^2 + A_i^3 B_{\beta 1} + A_i B_\alpha B_{\beta 1} + A_i^3 B_{\beta 2} + A_i^2 C_{\gamma1} + A_i A_m C_{\gamma1} \nonumber\\
&\quad + B_\alpha C_{\gamma1} + A_i^2 C_{\gamma2} + A_i A_m C_{\gamma2} + B_\alpha C_{\gamma2} + B_{\beta 2} C_{\gamma2},\\
\mathcal{R}_6\colon & ~~
C_{\gamma1} (A_i B_\alpha + A_m B_\alpha + C_{\gamma1}),~~C_{\gamma1} (A_i B_\alpha + C_{\gamma2}),\nonumber\\&~~A_i^4 A_m^2 + A_i^4 B_\alpha + A_{c'} A_i B_\alpha^2 + A_i^2 B_\alpha^2 + A_i A_m B_\alpha^2 + B_\alpha^3 + A_i^4 B_{\beta 1} + A_i^2 B_\alpha B_{\beta 1} + B_\alpha^2 B_{\beta 1} + A_i B_\alpha C_{\gamma1} \nonumber\\
&~~+ A_m B_\alpha C_{\gamma2} + C_{\gamma2}^2.
\end{align} 
 \end{subequations}
We have the following table regarding IWPs and group cohomology at degree 3.
\begin{center}
\begin{tabular}{c|cc|c|c|c}\hline\hline {Wyckoff}&\multicolumn{2}{c|}{Little group}& \multirow{2}{*}{Coordinates}&\multirow{2}{*}{LSM anomaly class}&\multirow{2}{*}{Topo. inv.} \\ \cline{2-3} position & Intl. & Sch\"{o}nflies & & & \\ \hline
2a&$4/m$&$C_{4h}$& $(0,0,0)$, $(1/2,1/2,1/2)$ & $A_{c'} B_\alpha + A_i B_{\beta 1} + C_{\gamma1}$ & $\varphi_1[I]$\\ 
2b&$4/m$&$C_{4h}$& $(0,0,1/2)$, $(1/2,1/2,0)$ & $C_{\gamma1}$ & $\varphi_1[T_3I]$\\ 
\hline
\multirow{2}{*}{4c} & \multirow{2}{*}{$2/m$} & \multirow{2}{*}{$C_{2h}$} & $(0,1/2,0)$, $(1/2,0,0)$, & \multirow{2}{*}{$A_{c'} B_\alpha + A_i B_\alpha + A_m B_\alpha + A_i B_{\beta 1}$} & \multirow{2}{*}{$\varphi_1[T_1I]$}\\
& & & $(1/2,0,1/2)$, $(0,1/2,1/2)$ & & \\ \hline
\multirow{2}{*}{4d} & \multirow{2}{*}{$222$} & \multirow{2}{*}{$D_2$} & $(0,1/2,1/4)$, $(1/2,0,1/4)$, & \multirow{2}{*}{$(A_{c'} + A_m) B_\alpha$} & \multirow{2}{*}{$\varphi_2[T_1C_2, GI]$}\\
& & & $(0,1/2,3/4)$, $(1/2,0,3/4)$ & & \\ \hline
\hline 
 \end{tabular} 
 \end{center}

\subsection*{No. 129: $P4/nmm$}\label{subsub:sg129}

This group is generated by three translations $T_{1,2,3}$ as given in Eqs.~\eqref{TransBravaisP}, a two-fold rotation $C_2$, a two-fold screw $S'_2$, a mirror $M$, and an inversion $I$:
\begin{subequations}
 \begin{align}
C_2 &\colon (x,y,z)\rightarrow (-x + 1/2, -y + 1/2, z),\\ 
S'_2 &\colon (x,y,z)\rightarrow (-x, y + 1/2, -z),\\ 
M &\colon (x,y,z)\rightarrow (y, x, z),\\ 
I &\colon (x,y,z)\rightarrow (-x, -y, -z).
\end{align}
\end{subequations}

The $\mathbb{Z}_2$ cohomology ring is given by

\begin{equation}
\mathbb{Z}_2[A_i,A_m,A_{c'},A_z,B_\alpha,B_\beta]/\langle\mathcal{R}_2,\mathcal{R}_3,\mathcal{R}_4\rangle
 \end{equation}
where the relations are 
\begin{subequations} 
 \begin{align}
\mathcal{R}_2\colon & ~~
A_{c'} A_m,~~A_{c'} (A_{c'} + A_i),~~A_z (A_{c'} + A_i + A_z),\\
\mathcal{R}_3\colon & ~~
A_{c'} A_i^2 + A_i^3 + A_i^2 A_m + A_{c'} B_\alpha + A_i B_\alpha + A_m B_\beta,~~A_{c'} A_i^2 + A_i^3 + A_i^2 A_m + A_{c'} B_\alpha + A_i B_\alpha + A_{c'} B_\beta + A_i B_\beta,\\
\mathcal{R}_4\colon & ~~
B_\beta (B_\alpha + B_\beta).
\end{align} 
 \end{subequations}
We have the following table regarding IWPs and group cohomology at degree 3.
\begin{center}
\resizebox{\columnwidth}{!}{
\begin{tabular}{c|cc|c|c|c}\hline\hline {Wyckoff}&\multicolumn{2}{c|}{Little group}& \multirow{2}{*}{Coordinates}&\multirow{2}{*}{LSM anomaly class}&\multirow{2}{*}{Topo. inv.} \\ \cline{2-3} position & Intl. & Sch\"{o}nflies & & & \\ \hline
2a&$\overline{4}m2$&$D_{2d}$& $(3/4,1/4,0)$, $(1/4,3/4,0)$ & 
$ (A_{c'} + A_i + A_z) B_\beta$ & $\varphi_2[T_2C_2, T_1T_2MI]$\\ 
2b&$\overline{4}m2$&$D_{2d}$& $(3/4,1/4,1/2)$, $(1/4,3/4,1/2)$ & $A_z B_\beta$ & $\varphi_2[T_2C_2, T_1T_2T_3MI]$\\ 
2c&$4mm$&$C_{4v}$& $(1/4,1/4,z)$, $(3/4,3/4,-z)$ & $A_z (A_{c'} A_i + A_i^2 + A_i A_m + B_\alpha + B_\beta)$ & $\varphi_2[T_3, C_2]$\\ 
\hline
\multirow{2}{*}{4d} & \multirow{2}{*}{$2/m$} & \multirow{2}{*}{$C_{2h}$} & $(0,0,0)$, $(1/2,1/2,0)$, & \multirow{2}{*}{$A_i (A_{c'} + A_i + A_m) (A_i + A_z)$} & \multirow{2}{*}{$\varphi_1[I]$}\\
& & & $(1/2,0,0)$, $(0,1/2,0)$ & & \\ \hline
\multirow{2}{*}{4e} & \multirow{2}{*}{$2/m$} & \multirow{2}{*}{$C_{2h}$} & $(0,0,1/2)$, $(1/2,1/2,1/2)$, & \multirow{2}{*}{$A_i (A_{c'} + A_i + A_m) A_z$} & \multirow{2}{*}{$\varphi_1[T_3I]$}\\
& & & $(1/2,0,1/2)$, $(0,1/2,1/2)$ & & \\ \hline
\hline 
 \end{tabular} }
 \end{center}

\subsection*{No. 130: $P4/ncc$}\label{subsub:sg130}

This group is generated by three translations $T_{1,2,3}$ as given in Eqs.~\eqref{TransBravaisP}, a two-fold rotation $C_2$, a two-fold screw $S'_2$, a glide $G$, and an inversion $I$:
\begin{subequations}
 \begin{align}
C_2 &\colon (x,y,z)\rightarrow (-x + 1/2, -y + 1/2, z),\\ 
S'_2 &\colon (x,y,z)\rightarrow (-x, y + 1/2, -z + 1/2),\\ 
G &\colon (x,y,z)\rightarrow (y, x, z + 1/2),\\ 
I &\colon (x,y,z)\rightarrow (-x, -y, -z).
\end{align}
\end{subequations}

The $\mathbb{Z}_2$ cohomology ring is given by

\begin{equation}
\mathbb{Z}_2[A_i,A_m,A_{c'},B_\alpha,B_\beta,D_\gamma]/\langle\mathcal{R}_2,\mathcal{R}_3,\mathcal{R}_4,\mathcal{R}_5,\mathcal{R}_6,\mathcal{R}_8\rangle
 \end{equation}
where the relations are 
\begin{subequations} 
 \begin{align}
\mathcal{R}_2\colon & ~~
A_{c'} A_m,~~A_{c'} A_i + A_i A_m + A_m^2,~~A_{c'} (A_{c'} + A_i),\\
\mathcal{R}_3\colon & ~~
A_i^3 + A_i^2 A_m + A_{c'} B_\alpha + A_i B_\alpha + A_m B_\beta,~~A_i^3 + A_i^2 A_m + A_{c'} B_\alpha + A_i B_\alpha + A_{c'} B_\beta + A_i B_\beta,\\
\mathcal{R}_4\colon & ~~
B_\beta (B_\alpha + B_\beta),\\
\mathcal{R}_5\colon & ~~
(A_i + A_m) D_\gamma,~~A_{c'} D_\gamma,\\
\mathcal{R}_6\colon & ~~
(B_\alpha + B_\beta) D_\gamma,\\
\mathcal{R}_8\colon & ~~
A_i^8 + A_i^6 B_\alpha + A_i A_m B_\alpha^3 + A_i^2 B_\alpha D_\gamma + D_\gamma^2.
\end{align} 
 \end{subequations}
We have the following table regarding IWPs and group cohomology at degree 3.
\begin{center}
\resizebox{\columnwidth}{!}{
\begin{tabular}{c|cc|c|c|c}\hline\hline {Wyckoff}&\multicolumn{2}{c|}{Little group}& \multirow{2}{*}{Coordinates}&\multirow{2}{*}{LSM anomaly class}&\multirow{2}{*}{Topo. inv.} \\ \cline{2-3} position & Intl. & Sch\"{o}nflies & & & \\ \hline
\multirow{2}{*}{4a} & \multirow{2}{*}{$222$} & \multirow{2}{*}{$D_2$} & $(3/4,1/4,1/4)$, $(1/4,3/4,1/4)$, & \multirow{2}{*}{$A_i B_\beta$} & \multirow{2}{*}{$\varphi_2[T_2C_2, T_1T_2GI]$}\\
& & & $(1/4,3/4,3/4)$, $(3/4,1/4,3/4)$ & & \\ \hline
\multirow{2}{*}{4b} & \multirow{2}{*}{$\overline{4}$} & \multirow{2}{*}{$S_4$} & $(3/4,1/4,0)$, $(1/4,3/4,0)$, & \multirow{2}{*}{$A_i^3 + A_i^2 A_m + A_{c'} B_\alpha + A_i B_\alpha + A_i B_\beta$} & \multirow{2}{*}{$\varphi_2[S'_2G, T_1^{-1}C_2]$}\\
& & & $(1/4,3/4,1/2)$, $(3/4,1/4,1/2)$ & & \\ \hline
\multirow{2}{*}{4c} & \multirow{2}{*}{$4$} & \multirow{2}{*}{$C_4$} & $(1/4,1/4,z)$, $(3/4,3/4,-z+1/2)$, & \multirow{2}{*}{$A_{c'} B_\alpha + A_m B_\alpha + A_i B_\beta$} & \multirow{2}{*}{$\varphi_2[G, C_2]$}\\
& & & $(3/4,3/4,-z)$, $(1/4,1/4,z+1/2)$ & & \\ \hline
\multirow{4}{*}{8d} & \multirow{4}{*}{$\overline{1}$} & \multirow{4}{*}{$C_i$} & $(0,0,0)$, $(1/2,1/2,0)$, & \multirow{4}{*}{$A_i^2 (A_i + A_m)$} & \multirow{4}{*}{$\varphi_1[I]$}\\
& & & $(1/2,0,0)$, $(0,1/2,0)$, & & \\
& & & $(0,1/2,1/2)$, $(1/2,0,1/2)$, & & \\
& & & $(1/2,1/2,1/2)$, $(0,0,1/2)$ & & \\ \hline
\hline 
 \end{tabular} }
 \end{center}

\subsection*{No. 131: $P4_2/mmc$}\label{subsub:sg131}

This group is generated by three translations $T_{1,2,3}$ as given in Eqs.~\eqref{TransBravaisP}, a two-fold rotation $C_2$, a two-fold rotation $C'_2$, a glide $G$, and an inversion $I$:
\begin{subequations}
 \begin{align}
C_2 &\colon (x,y,z)\rightarrow (-x, -y, z),\\ 
C'_2 &\colon (x,y,z)\rightarrow (-x, y, -z),\\ 
G &\colon (x,y,z)\rightarrow (y, x, z + 1/2),\\ 
I &\colon (x,y,z)\rightarrow (-x, -y, -z).
\end{align}
\end{subequations}

The $\mathbb{Z}_2$ cohomology ring is given by

\begin{equation}
\mathbb{Z}_2[A_i,A_m,A_{c'},A_{x+y},B_\alpha,B_\beta,B_{xy},B_{z(x+y)}]/\langle\mathcal{R}_2,\mathcal{R}_3,\mathcal{R}_4\rangle
 \end{equation}
where the relations are 
\begin{subequations} 
 \begin{align}
\mathcal{R}_2\colon & ~~
A_{c'} A_m,~~A_m A_{x+y},~~A_m (A_i + A_m),\\
\mathcal{R}_3\colon & ~~
A_m B_\beta,~~A_{c'} A_{x+y}^2 + A_{x+y}^3 + A_{x+y} B_\alpha + A_{c'} B_{xy},~~A_{x+y} (A_{c'} A_{x+y} + A_{x+y}^2 + B_\alpha + B_{xy}),~~A_m B_{z(x+y)},\nonumber\\&~~A_{c'} A_i A_{x+y} + A_i^2 A_{x+y} + A_i A_{x+y}^2 + A_{x+y} B_\beta + A_{c'} B_{z(x+y)},\\
\mathcal{R}_4\colon & ~~B_{xy} (B_\alpha + B_{xy}),~~A_{c'} A_i A_{x+y}^2 + A_i^2 A_{x+y}^2 + A_i A_{x+y}^3 + A_{x+y}^2 B_\beta + A_{x+y}^2 B_{z(x+y)} + B_\alpha B_{z(x+y)} + B_{xy} B_{z(x+y)},\nonumber\\
&~~
A_{c'}^3 A_i + A_{c'}^2 A_i^2 + A_{c'} A_i^3 + A_i^4 + A_i^3 A_m + A_{c'}^2 B_\alpha + A_i^2 B_\alpha + A_i A_m B_\alpha + A_{c'}^2 B_\beta + A_{c'} A_i B_\beta + B_\beta^2,\nonumber\\&~~A_{c'} A_i A_{x+y}^2 + A_i^2 A_{x+y}^2 + A_i A_{x+y}^3 + A_{x+y}^2 B_\beta + A_i^2 B_{xy} + A_i A_m B_{xy} + B_\beta B_{xy} + A_{x+y}^2 B_{z(x+y)} + B_\alpha B_{z(x+y)},\nonumber\\&~~
A_{c'}^2 A_i A_{x+y} + A_{c'} A_i^2 A_{x+y} + A_{c'} A_{x+y} B_\alpha + A_{c'} A_{x+y} B_\beta + A_i^2 B_{xy} + A_i A_m B_{xy} + A_i^2 B_{z(x+y)} + A_i A_{x+y} B_{z(x+y)}\nonumber\\&~~
 + B_\beta B_{z(x+y)},\nonumber\\&~~
A_{c'}^2 A_{x+y}^2 + A_{c'} A_i A_{x+y}^2 + A_i^2 A_{x+y}^2 + A_{x+y}^4 + A_{c'} A_{x+y} B_\alpha + A_{x+y}^2 B_\beta + A_i^2 B_{xy} + A_i A_m B_{xy} + A_i A_{x+y} B_{z(x+y)} \nonumber\\&~~
 + B_{z(x+y)}^2.
\end{align} 
 \end{subequations}
We have the following table regarding IWPs and group cohomology at degree 3.
\begin{center}
\resizebox{\columnwidth}{!}{
\begin{tabular}{c|cc|c|c|c}\hline\hline {Wyckoff}&\multicolumn{2}{c|}{Little group}& \multirow{2}{*}{Coordinates}&\multirow{2}{*}{LSM anomaly class}&\multirow{2}{*}{Topo. inv.} \\ \cline{2-3} position & Intl. & Sch\"{o}nflies & & & \\ \hline
2a&$mmm$&$D_{2h}$& $(0,0,0)$, $(0,0,1/2)$ & 
$(A_{c'} + A_m + A_i)(A_{c'}A_{x+y} + A_{x+y}^2 + B_\alpha + B_{xy})$ & $\varphi_2[C_2, C'_2]$\\ 
2b&$mmm$&$D_{2h}$& $(1/2,1/2,0)$, $(1/2,1/2,1/2)$ & 
$(A_{c'} + A_m + A_i) B_{xy}$ & $\varphi_2[T_1T_2C_2, T_1C'_2]$\\ 
2c&$mmm$&$D_{2h}$& $(0,1/2,0)$, $(1/2,0,1/2)$ & $A_{x+y} (A_{c'}^2 + A_i^2 + A_{x+y}^2 + B_\beta + B_{z(x+y)})$ & $\varphi_2[T_2C_2, C'_2]$\\ 
2d&$mmm$&$D_{2h}$& $(0,1/2,1/2)$, $(1/2,0,0)$ & 
$(A_{c'}  + A_{x+y})(A_{x+y}^2 + B_{z(x+y)})$ & $\varphi_2[T_1C_2, T_1C'_2]$\\ 
2e&$\overline{4}m2$&$D_{2d}$& $(0,0,1/4)$, $(0,0,3/4)$ & $A_m (B_\alpha + B_{xy})$ & $\varphi_2[C_2, GI]$\\ 
2f&$\overline{4}m2$&$D_{2d}$& $(1/2,1/2,1/4)$, $(1/2,1/2,3/4)$ & $A_m B_{xy}$ & $\varphi_2[T_1T_2C_2, T_1T_2GI]$\\ 
\hline
\hline 
 \end{tabular} }
 \end{center}

\subsection*{No. 132: $P4_2/mcm$}\label{subsub:sg132}

This group is generated by three translations $T_{1,2,3}$ as given in Eqs.~\eqref{TransBravaisP}, a two-fold rotation $C_2$, a two-fold rotation $C'_2$, a mirror $M$, and an inversion $I$:
\begin{subequations}
 \begin{align}
C_2 &\colon (x,y,z)\rightarrow (-x, -y, z),\\ 
C'_2 &\colon (x,y,z)\rightarrow (-x, y, -z + 1/2),\\ 
M &\colon (x,y,z)\rightarrow (y, x, z),\\ 
I &\colon (x,y,z)\rightarrow (-x, -y, -z).
\end{align}
\end{subequations}

The $\mathbb{Z}_2$ cohomology ring is given by

\begin{equation}
\mathbb{Z}_2[A_i,A_m,A_{c'},A_{x+y},B_\alpha,B_\beta,B_{xy}]/\langle\mathcal{R}_2,\mathcal{R}_3,\mathcal{R}_4\rangle
 \end{equation}
where the relations are 
\begin{subequations} 
 \begin{align}
\mathcal{R}_2\colon & ~~
A_{c'} A_i,~~A_{c'} A_m,~~A_m A_{x+y},\\
\mathcal{R}_3\colon & ~~
A_{c'} B_\beta,~~A_{x+y} (A_i^2 + A_i A_{x+y} + B_\beta),~~A_{c'} A_{x+y}^2 + A_{x+y}^3 + A_{x+y} B_\alpha + A_{c'} B_{xy},~~A_{x+y} (A_{c'} A_{x+y} + A_{x+y}^2 + B_\alpha + B_{xy}),\\
\mathcal{R}_4\colon & ~~
A_i^4 + A_i^3 A_m + A_i^2 B_\alpha + A_i A_m B_\beta + B_\beta^2,~~B_{xy} (B_\alpha + B_{xy}).
\end{align} 
 \end{subequations}
We have the following table regarding IWPs and group cohomology at degree 3.
\begin{center}
\resizebox{\columnwidth}{!}{
\begin{tabular}{c|cc|c|c|c}\hline\hline {Wyckoff}&\multicolumn{2}{c|}{Little group}& \multirow{2}{*}{Coordinates}&\multirow{2}{*}{LSM anomaly class}&\multirow{2}{*}{Topo. inv.} \\ \cline{2-3} position & Intl. & Sch\"{o}nflies & & & \\ \hline
2a&$mmm$&$D_{2h}$& $(0,0,0)$, $(0,0,1/2)$ & $A_i (A_{x+y}^2 + B_\alpha + B_{xy})$ & $\varphi_2[C_2, MI]$\\ 
2b&$\overline{4}2m$&$D_{2d}$& $(0,0,1/4)$, $(0,0,3/4)$ & $(A_{c'} + A_{x+y}) (A_{c'} A_{x+y} + A_{x+y}^2 + B_\alpha)$ & $\varphi_2[C_2, C'_2]$\\ 
2c&$mmm$&$D_{2h}$& $(1/2,1/2,0)$, $(1/2,1/2,1/2)$ & $A_i B_{xy}$ & $\varphi_2[T_1T_2C_2, T_1T_2MI]$\\ 
2d&$\overline{4}2m$&$D_{2d}$& $(1/2,1/2,1/4)$, $(1/2,1/2,3/4)$ & 
$A_{c'} B_{xy}$ & $\varphi_2[T_1T_2C_2, T_1C'_2]$\\ 
\hline
\multirow{2}{*}{4e} & \multirow{2}{*}{$222$} & \multirow{2}{*}{$D_2$} & $(0,1/2,1/4)$, $(1/2,0,3/4)$, & \multirow{2}{*}{$A_{c'} A_{x+y} (A_{c'} + A_{x+y})$} & \multirow{2}{*}{$\varphi_2[T_1C_2, T_1C'_2]$}\\
& & & $(0,1/2,3/4)$, $(1/2,0,1/4)$ & & \\ \hline
\multirow{2}{*}{4f} & \multirow{2}{*}{$2/m$} & \multirow{2}{*}{$C_{2h}$} & $(0,1/2,0)$, $(1/2,0,1/2)$, & \multirow{2}{*}{$A_i A_{x+y}^2$} & \multirow{2}{*}{$\varphi_1[T_1I]$}\\
& & & $(0,1/2,1/2)$, $(1/2,0,0)$ & & \\ \hline
\hline 
 \end{tabular} }
 \end{center}

\subsection*{No. 133: $P4_2/nbc$}\label{subsub:sg133}

This group is generated by three translations $T_{1,2,3}$ as given in Eqs.~\eqref{TransBravaisP}, a two-fold rotation $C_2$, a two-fold rotation $C'_2$, a glide $G$, and an inversion $I$:
\begin{subequations}
 \begin{align}
C_2 &\colon (x,y,z)\rightarrow (-x + 1/2, -y + 1/2, z),\\ 
C'_2 &\colon (x,y,z)\rightarrow (-x + 1/2, y, -z),\\ 
G &\colon (x,y,z)\rightarrow (y + 1/2, x + 1/2, z + 1/2),\\ 
I &\colon (x,y,z)\rightarrow (-x, -y, -z).
\end{align}
\end{subequations}

The $\mathbb{Z}_2$ cohomology ring is given by

\begin{equation}
\mathbb{Z}_2[A_i,A_m,A_{c'},B_\alpha,B_{\beta 1},B_{\beta 2},C_\gamma]/\langle\mathcal{R}_2,\mathcal{R}_3,\mathcal{R}_4,\mathcal{R}_5,\mathcal{R}_6\rangle
 \end{equation}
where the relations are 
\begin{subequations} 
 \begin{align}
\mathcal{R}_2\colon & ~~
A_{c'} A_i,~~A_{c'} A_m,~~A_m (A_i + A_m),\\
\mathcal{R}_3\colon & ~~
A_i B_{\beta 1},~~A_i^3 + A_i^2 A_m + A_i B_\alpha + A_m B_\alpha + A_m B_{\beta 1},~~A_i B_{\beta 2},~~A_m B_{\beta 2},\\
\mathcal{R}_4\colon & ~~
(A_i + A_m) C_\gamma,~~B_{\beta 1} (B_\alpha + B_{\beta 1}),~~A_{c'}^2 B_\alpha + A_{c'}^2 B_{\beta 2} + B_\alpha B_{\beta 2} + B_{\beta 1} B_{\beta 2} + A_{c'} C_\gamma,~~A_{c'}^2 B_\alpha + A_{c'}^2 B_{\beta 2} + B_{\beta 2}^2,\\
\mathcal{R}_5\colon & ~~
A_i^5 + A_i^4 A_m + A_{c'}^3 B_\alpha + A_i B_\alpha^2 + A_m B_\alpha^2 + A_{c'} B_\alpha B_{\beta 1} + A_{c'}^3 B_{\beta 2} + A_{c'} B_\alpha B_{\beta 2} + A_{c'}^2 C_\gamma + B_{\beta 1} C_\gamma,\nonumber\\&~~A_{c'} B_\alpha^2 + A_{c'} B_\alpha B_{\beta 1} + A_{c'} B_\alpha B_{\beta 2} + A_{c'}^2 C_\gamma + B_{\beta 2} C_\gamma,\\
\mathcal{R}_6\colon & ~~
A_i^5 A_m + A_{c'}^4 B_\alpha + A_i^4 B_\alpha + B_\alpha^3 + B_\alpha^2 B_{\beta 1} + A_{c'}^4 B_{\beta 2} + A_{c'}^2 B_\alpha B_{\beta 2} + A_{c'} B_\alpha C_\gamma + A_i B_\alpha C_\gamma + C_\gamma^2.
\end{align} 
 \end{subequations}
We have the following table regarding IWPs and group cohomology at degree 3.
\begin{center}
\resizebox{\columnwidth}{!}{
\begin{tabular}{c|cc|c|c|c}\hline\hline {Wyckoff}&\multicolumn{2}{c|}{Little group}& \multirow{2}{*}{Coordinates}&\multirow{2}{*}{LSM anomaly class}&\multirow{2}{*}{Topo. inv.} \\ \cline{2-3} position & Intl. & Sch\"{o}nflies & & & \\ \hline
\multirow{2}{*}{4a} & \multirow{2}{*}{$222$} & \multirow{2}{*}{$D_2$} & $(1/4,1/4,0)$, $(1/4,1/4,1/2)$, & \multirow{2}{*}{$A_{c'} (B_\alpha + B_{\beta 1})$} & \multirow{2}{*}{$\varphi_2[C_2, C'_2]$}\\
& & & $(3/4,3/4,0)$, $(3/4,3/4,1/2)$ & & \\ \hline
\multirow{2}{*}{4b} & \multirow{2}{*}{$222$} & \multirow{2}{*}{$D_2$} & $(3/4,1/4,0)$, $(1/4,3/4,1/2)$, & \multirow{2}{*}{$A_i^3 + A_i^2 A_m + A_i B_\alpha + A_m B_\alpha + A_{c'} B_{\beta 1}$} & \multirow{2}{*}{$\varphi_2[T_1C_2, T_1C'_2]$}\\
& & & $(1/4,3/4,0)$, $(3/4,1/4,1/2)$ & & \\ \hline
\multirow{2}{*}{4c} & \multirow{2}{*}{$222$} & \multirow{2}{*}{$D_2$} & $(1/4,1/4,1/4)$, $(1/4,1/4,3/4)$, & \multirow{2}{*}{$A_i (A_i^2 + A_i A_m + B_\alpha)$} & \multirow{2}{*}{$\varphi_2[C_2, GI]$}\\
& & & $(3/4,3/4,3/4)$, $(3/4,3/4,1/4)$ & & \\ \hline
\multirow{2}{*}{4d} & \multirow{2}{*}{$\overline{4}$} & \multirow{2}{*}{$S_4$} & $(3/4,1/4,3/4)$, $(1/4,3/4,1/4)$, & \multirow{2}{*}{$(A_i + A_m) (A_i^2 + B_\alpha)$} & \multirow{2}{*}{$\varphi_2[C_2C'_2G, T_2^{-1}C_2]$}\\
& & & $(3/4,1/4,1/4)$, $(1/4,3/4,3/4)$ & & \\ \hline
\multirow{4}{*}{8e} & \multirow{4}{*}{$\overline{1}$} & \multirow{4}{*}{$C_i$} & $(0,0,0)$, $(1/2,1/2,0)$, & \multirow{4}{*}{$A_i^2 (A_i + A_m)$} & \multirow{4}{*}{$\varphi_1[I]$}\\
& & & $(1/2,0,1/2)$, $(0,1/2,1/2)$, & & \\
& & & $(1/2,0,0)$, $(0,1/2,0)$, & & \\
& & & $(0,0,1/2)$, $(1/2,1/2,1/2)$ & & \\ \hline
\hline 
 \end{tabular} }
 \end{center}

\subsection*{No. 134: $P4_2/nnm$}\label{subsub:sg134}

This group is generated by three translations $T_{1,2,3}$ as given in Eqs.~\eqref{TransBravaisP}, a two-fold rotation $C_2$, a two-fold rotation $C'_2$, a glide $G$, and an inversion $I$:
\begin{subequations}
 \begin{align}
C_2 &\colon (x,y,z)\rightarrow (-x + 1/2, -y + 1/2, z),\\ 
C'_2 &\colon (x,y,z)\rightarrow (-x + 1/2, y, -z + 1/2),\\ 
G &\colon (x,y,z)\rightarrow (y + 1/2, x + 1/2, z),\\ 
I &\colon (x,y,z)\rightarrow (-x, -y, -z).
\end{align}
\end{subequations}

The $\mathbb{Z}_2$ cohomology ring is given by

\begin{equation}
\mathbb{Z}_2[A_i,A_m,A_{c'},A_{x+y+z},B_\alpha]/\langle\mathcal{R}_2,\mathcal{R}_3,\mathcal{R}_4\rangle
 \end{equation}
where the relations are 
\begin{subequations} 
 \begin{align}
\mathcal{R}_2\colon & ~~
A_{c'} A_i,~~A_{c'} A_m,\\
\mathcal{R}_3\colon & ~~
(A_i + A_m) (A_m + A_{x+y+z}) (A_i + A_m + A_{x+y+z}),~~A_i (A_i^2 + A_m^2 + A_i A_{x+y+z} + A_{x+y+z}^2 + B_\alpha),\\
\mathcal{R}_4\colon & ~~
A_i^3 A_m + A_i A_m^3 + A_i^3 A_{x+y+z} + A_{c'}^2 A_{x+y+z}^2 + A_m^2 A_{x+y+z}^2 + A_i A_{x+y+z}^3 + A_{x+y+z}^4 + A_m^2 B_\alpha \nonumber\\
&~~+ A_{c'} A_{x+y+z} B_\alpha + A_{x+y+z}^2 B_\alpha.
\end{align} 
 \end{subequations}
We have the following table regarding IWPs and group cohomology at degree 3.
\begin{center}
\resizebox{\columnwidth}{!}{
\begin{tabular}{c|cc|c|c|c}\hline\hline {Wyckoff}&\multicolumn{2}{c|}{Little group}& \multirow{2}{*}{Coordinates}&\multirow{2}{*}{LSM anomaly class}&\multirow{2}{*}{Topo. inv.} \\ \cline{2-3} position & Intl. & Sch\"{o}nflies & & & \\ \hline
2a&$\overline{4}2m$&$D_{2d}$& $(1/4,3/4,1/4)$, $(3/4,1/4,3/4)$ & $(A_{c'} + A_m + A_{x+y+z}) (A_{c'} A_{x+y+z} + A_m A_{x+y+z} + A_{x+y+z}^2 + B_\alpha)$ & $\varphi_2[T_2C_2, C'_2]$\\ 
2b&$\overline{4}2m$&$D_{2d}$& $(3/4,1/4,1/4)$, $(1/4,3/4,3/4)$ & $A_m^2 A_{x+y+z} + A_{c'} A_{x+y+z}^2 + A_{x+y+z}^3 + A_m B_\alpha + A_{x+y+z} B_\alpha$ & $\varphi_2[T_1C_2, T_1C'_2]$\\ 
\hline
\multirow{2}{*}{4c} & \multirow{2}{*}{$222$} & \multirow{2}{*}{$D_2$} & $(1/4,1/4,1/4)$, $(1/4,1/4,3/4)$, & \multirow{2}{*}{$A_{c'} A_{x+y+z} (A_{c'} + A_{x+y+z})$} & \multirow{2}{*}{$\varphi_2[C_2, C'_2]$}\\
& & & $(3/4,3/4,3/4)$, $(3/4,3/4,1/4)$ & & \\ \hline
\multirow{2}{*}{4d} & \multirow{2}{*}{$222$} & \multirow{2}{*}{$D_2$} & $(1/4,1/4,0)$, $(1/4,1/4,1/2)$, & \multirow{2}{*}{$A_i (A_m + A_{x+y+z}) (A_i + A_m + A_{x+y+z})$} & \multirow{2}{*}{$\varphi_2[C_2, GI]$}\\
& & & $(3/4,3/4,0)$, $(3/4,3/4,1/2)$ & & \\ \hline
\multirow{2}{*}{4e} & \multirow{2}{*}{$2/m$} & \multirow{2}{*}{$C_{2h}$} & $(0,0,1/2)$, $(1/2,1/2,1/2)$, & \multirow{2}{*}{$A_i (A_i + A_m) (A_i + A_m + A_{x+y+z})$} & \multirow{2}{*}{$\varphi_1[T_3I]$}\\
& & & $(1/2,0,0)$, $(0,1/2,0)$ & & \\ \hline
\multirow{2}{*}{4f} & \multirow{2}{*}{$2/m$} & \multirow{2}{*}{$C_{2h}$} & $(0,0,0)$, $(1/2,1/2,0)$, & \multirow{2}{*}{$A_i (A_i + A_m) (A_m + A_{x+y+z})$} & \multirow{2}{*}{$\varphi_1[I]$}\\
& & & $(1/2,0,1/2)$, $(0,1/2,1/2)$ & & \\ \hline
\hline 
 \end{tabular} }
 \end{center}

\subsection*{No. 135: $P4_2/mbc$}\label{subsub:sg135}

This group is generated by three translations $T_{1,2,3}$ as given in Eqs.~\eqref{TransBravaisP}, a two-fold rotation $C_2$, a two-fold screw $S'_2$, a glide $G$, and an inversion $I$:
\begin{subequations}
 \begin{align}
C_2 &\colon (x,y,z)\rightarrow (-x, -y, z),\\ 
S'_2 &\colon (x,y,z)\rightarrow (-x + 1/2, y + 1/2, -z),\\ 
G &\colon (x,y,z)\rightarrow (y + 1/2, x + 1/2, z + 1/2),\\ 
I &\colon (x,y,z)\rightarrow (-x, -y, -z).
\end{align}
\end{subequations}

The $\mathbb{Z}_2$ cohomology ring is given by

\begin{equation}
\mathbb{Z}_2[A_i,A_m,A_{c'},B_\alpha,B_{\beta 1},B_{\beta 2},C_\gamma]/\langle\mathcal{R}_2,\mathcal{R}_3,\mathcal{R}_4,\mathcal{R}_5,\mathcal{R}_6\rangle
 \end{equation}
where the relations are 
\begin{subequations} 
 \begin{align}
\mathcal{R}_2\colon & ~~
A_{c'} A_m,~~A_m (A_i + A_m),~~A_{c'}^2,\\
\mathcal{R}_3\colon & ~~
A_{c'} B_\alpha + A_m B_{\beta 1},~~A_{c'} (A_i^2 + B_\alpha + B_{\beta 1}),~~A_m B_{\beta 2},~~A_{c'} (A_i^2 + B_{\beta 2}),\\
\mathcal{R}_4\colon & ~~
A_{c'} (A_i^3 + C_\gamma),~~A_i^4 + A_i^3 A_m + A_i^2 B_\alpha + A_i A_m B_\alpha + B_\alpha B_{\beta 1} + B_{\beta 1}^2,\nonumber\\&~~A_{c'} A_i^3 + A_i^4 + A_i^3 A_m + A_i^2 B_\alpha + A_i A_m B_\alpha + A_i^2 B_{\beta 2} + B_\alpha B_{\beta 2} + B_{\beta 1} B_{\beta 2} + A_i C_\gamma + A_m C_\gamma,\nonumber\\&~~A_i^4 + A_i^3 A_m + A_i^2 B_\alpha + A_i A_m B_\alpha + B_{\beta 2}^2,\\
\mathcal{R}_5\colon & ~~
A_i^5 + A_i^4 A_m + A_i^3 B_\alpha + A_i^2 A_m B_\alpha + A_{c'} B_\alpha^2 + A_i^3 B_{\beta 1} + A_i B_\alpha B_{\beta 1} + A_i^2 C_\gamma + A_i A_m C_\gamma + B_{\beta 1} C_\gamma,\nonumber\\&~~A_{c'} A_i^4 + A_i^5 + A_i^4 A_m + A_{c'} B_\alpha^2 + A_i B_\alpha^2 + A_m B_\alpha^2 + A_i^3 B_{\beta 1} + A_i B_\alpha B_{\beta 1} + A_i^3 B_{\beta 2} + A_i B_\alpha B_{\beta 2} + B_{\beta 2} C_\gamma,\\
\mathcal{R}_6\colon & ~~
A_i^6 + A_i^4 B_\alpha + A_i^2 B_\alpha^2 + A_i A_m B_\alpha^2 + B_\alpha^3 + A_i^2 B_\alpha B_{\beta 1} + B_\alpha^2 B_{\beta 1} + A_m B_\alpha C_\gamma + C_\gamma^2.
\end{align} 
 \end{subequations}
We have the following table regarding IWPs and group cohomology at degree 3.
\begin{center}
\resizebox{\columnwidth}{!}{
\begin{tabular}{c|cc|c|c|c}\hline\hline {Wyckoff}&\multicolumn{2}{c|}{Little group}& \multirow{2}{*}{Coordinates}&\multirow{2}{*}{LSM anomaly class}&\multirow{2}{*}{Topo. inv.} \\ \cline{2-3} position & Intl. & Sch\"{o}nflies & & & \\ \hline
\multirow{2}{*}{4a} & \multirow{2}{*}{$2/m$} & \multirow{2}{*}{$C_{2h}$} & $(0,0,0)$, $(0,0,1/2)$, & \multirow{2}{*}{$A_i (A_i^2 + A_i A_m + B_{\beta 1})$} & \multirow{2}{*}{$\varphi_1[I]$}\\
& & & $(1/2,1/2,0)$, $(1/2,1/2,1/2)$ & & \\ \hline
\multirow{2}{*}{4b} & \multirow{2}{*}{$\overline{4}$} & \multirow{2}{*}{$S_4$} & $(0,0,1/4)$, $(0,0,3/4)$, & \multirow{2}{*}{$A_{c'} B_\alpha$} & \multirow{2}{*}{$\varphi_2[T_2^{-1}S'_2G, C_2]$}\\
& & & $(1/2,1/2,3/4)$, $(1/2,1/2,1/4)$ & & \\ \hline
\multirow{2}{*}{4c} & \multirow{2}{*}{$2/m$} & \multirow{2}{*}{$C_{2h}$} & $(0,1/2,0)$, $(1/2,0,1/2)$, & \multirow{2}{*}{$A_i^3 + A_i^2 A_m + A_{c'} B_\alpha + A_i B_\alpha + A_m B_\alpha + A_i B_{\beta 1}$} & \multirow{2}{*}{$\varphi_1[T_1I]$}\\
& & & $(1/2,0,0)$, $(0,1/2,1/2)$ & & \\ \hline
\multirow{2}{*}{4d} & \multirow{2}{*}{$222$} & \multirow{2}{*}{$D_2$} & $(0,1/2,1/4)$, $(1/2,0,3/4)$, & \multirow{2}{*}{$(A_{c'} + A_m) B_\alpha$} & \multirow{2}{*}{$\varphi_2[T_1C_2, GI]$}\\
& & & $(0,1/2,3/4)$, $(1/2,0,1/4)$ & & \\ \hline
\hline 
 \end{tabular} }
 \end{center}

\subsection*{No. 136: $P4_2/mnm$}\label{subsub:sg136}

This group is generated by three translations $T_{1,2,3}$ as given in Eqs.~\eqref{TransBravaisP}, a two-fold rotation $C_2$, a two-fold screw $S'_2$, a mirror $M$, and an inversion $I$:
\begin{subequations}
 \begin{align}
C_2 &\colon (x,y,z)\rightarrow (-x, -y, z),\\ 
S'_2 &\colon (x,y,z)\rightarrow (-x + 1/2, y + 1/2, -z + 1/2),\\ 
M &\colon (x,y,z)\rightarrow (y, x, z),\\ 
I &\colon (x,y,z)\rightarrow (-x, -y, -z).
\end{align}
\end{subequations}

The $\mathbb{Z}_2$ cohomology ring is given by

\begin{equation}
\mathbb{Z}_2[A_i,A_m,A_{c'},B_\alpha,B_{\beta 1},B_{\beta 2},B_{\beta 3},C_{\gamma1},C_{\gamma2},D_\delta]/\langle\mathcal{R}_2,\mathcal{R}_3,\mathcal{R}_4,\mathcal{R}_5,\mathcal{R}_6,\mathcal{R}_7,\mathcal{R}_8\rangle
 \end{equation}
where the relations are 
\begin{subequations} 
 \begin{align}
\mathcal{R}_2\colon & ~~
A_{c'} A_i,~~A_{c'} A_m,~~A_{c'}^2,\\
\mathcal{R}_3\colon & ~~
A_{c'} B_\alpha + A_m B_{\beta 1},~~A_{c'} (B_\alpha + B_{\beta 1}),~~A_m B_{\beta 2},~~A_{c'} B_{\beta 2},~~A_{c'} B_{\beta 3},\\
\mathcal{R}_4\colon & ~~
A_{c'} C_{\gamma1},~~A_{c'} C_{\gamma2},~~A_i^2 A_m^2 + A_i^2 B_{\beta 1} + B_\alpha B_{\beta 2} + B_\alpha B_{\beta 3} + A_i C_{\gamma1} + A_m C_{\gamma2},~~B_{\beta 1} (B_\alpha + B_{\beta 1}),\nonumber\\
&~~(B_\alpha + B_{\beta 1}) B_{\beta 2},~~A_i^2 B_{\beta 1} + B_\alpha B_{\beta 2} + B_{\beta 1} B_{\beta 3},~~A_i^2 B_{\beta 1} + B_{\beta 2}^2,~~A_i^2 B_{\beta 1} + A_i^2 B_{\beta 2} + B_{\beta 2} B_{\beta 3},\nonumber\\
&~~A_i^4 + A_i^3 A_m + A_i^2 B_\alpha + A_i A_m B_{\beta 3} + B_{\beta 3}^2,\\
\mathcal{R}_5\colon & ~~
A_{c'} D_\delta,~~B_{\beta 1} C_{\gamma1},~~B_{\beta 2} C_{\gamma1},~~B_{\beta 1} C_{\gamma2},~~B_{\beta 2} C_{\gamma2},\nonumber\\
&~~A_i^4 A_m + A_i^3 A_m^2 + A_i A_m^4 + A_i^3 B_\alpha + A_i A_m^2 B_\alpha + A_i B_\alpha^2 + A_i^3 B_{\beta 1} + A_i B_\alpha B_{\beta 1} + B_{\beta 3} C_{\gamma1} + A_m D_\delta,\nonumber\\
&~~A_i^5 + A_i^4 A_m + A_i^3 A_m^2 + A_i^2 A_m^3 + A_i^3 B_\alpha + A_i^2 A_m B_\alpha + A_i^3 B_{\beta 1} + A_i^3 B_{\beta 2} + A_i^2 A_m B_{\beta 3} + A_i^2 C_{\gamma1} \nonumber\\
&~~+ A_i A_m C_{\gamma2} + B_{\beta 3} C_{\gamma2} + A_i D_\delta,\\
\mathcal{R}_6\colon & ~~
A_i^4 B_{\beta 1} + A_i^2 B_\alpha B_{\beta 2} + B_{\beta 1} D_\delta,~~A_i^4 B_{\beta 1} + A_i^4 B_{\beta 2} + B_{\beta 2} D_\delta,~~C_{\gamma2} (A_i B_\alpha + C_{\gamma2}),\nonumber\\
&~~A_i^4 A_m^2 + A_i^3 A_m^3 + A_i^2 A_m^4 + A_i^4 B_\alpha + A_i^3 A_m B_\alpha + A_i^2 A_m^2 B_\alpha + A_i^2 B_\alpha^2 + A_i^2 B_\alpha B_{\beta 1} + A_i^4 B_{\beta 2} + A_i^4 B_{\beta 3} + A_i^3 A_m B_{\beta 3} \nonumber\\
&~~ + A_i A_m^3 B_{\beta 3} + A_i^3 C_{\gamma1} + A_i^2 A_m C_{\gamma1} + A_i^3 C_{\gamma2} + A_i A_m^2 C_{\gamma2} + A_i B_\alpha C_{\gamma2} + A_i A_m D_\delta + B_{\beta 3} D_\delta,\nonumber\\
&~~A_i^2 A_m^4 + A_i A_m^3 B_\alpha + A_i^2 B_\alpha^2 + A_i A_m B_\alpha^2 + B_\alpha^3 + A_i^2 B_\alpha B_{\beta 1} + B_\alpha^2 B_{\beta 1} + A_m B_\alpha C_{\gamma1} + C_{\gamma1}^2,\nonumber\\
& ~~A_i^3 A_m^3 + A_i^4 B_\alpha + A_i^3 A_m B_\alpha + A_i^2 A_m^2 B_\alpha + A_i A_m^3 B_\alpha + A_i^2 B_\alpha^2 + A_i A_m B_\alpha^2 + A_i^2 B_\alpha B_{\beta 1} + A_i^2 B_\alpha B_{\beta 2} + A_i^2 A_m C_{\gamma1} \nonumber\\
&~~ + A_i B_\alpha C_{\gamma1} + C_{\gamma1} C_{\gamma2} + B_\alpha D_\delta,\\
\mathcal{R}_7\colon & ~~
A_i^3 A_m^4 + A_i A_m^4 B_\alpha + A_i^3 B_\alpha^2 + A_i^2 A_m B_\alpha^2 + A_i A_m^2 B_\alpha^2 + A_i B_\alpha^3 + A_i^3 B_\alpha B_{\beta 1} + A_i B_\alpha^2 B_{\beta 1} + A_i^2 A_m^3 B_{\beta 3} + A_i^4 C_{\gamma1}\nonumber\\
&~~+ A_i^3 A_m C_{\gamma1} + A_i^2 A_m^2 C_{\gamma1} + A_i A_m^3 C_{\gamma1} + A_i^2 B_\alpha C_{\gamma1} + A_i A_m B_\alpha C_{\gamma1} + A_i A_m^3 C_{\gamma2} + A_i^2 B_\alpha C_{\gamma2} + A_i A_m B_\alpha C_{\gamma2}\nonumber\\
&~~+ B_\alpha^2 C_{\gamma2} + A_m B_\alpha D_\delta + C_{\gamma1} D_\delta,\nonumber\\
&~~A_i^6 A_m + A_i^5 A_m^2 + A_i^4 A_m^3 + A_i^3 A_m^4 + A_i^4 A_m B_\alpha + A_i^3 A_m^2 B_\alpha + A_i^3 A_m^2 B_{\beta 3} + A_i^3 A_m C_{\gamma1} + A_i^4 C_{\gamma2} + A_i^3 A_m C_{\gamma2}\nonumber\\
&~~+ A_i^2 A_m^2 C_{\gamma2} + A_i A_m^3 C_{\gamma2} + A_i^2 B_\alpha C_{\gamma2} + A_i A_m B_\alpha C_{\gamma2} + A_i^2 A_m D_\delta + C_{\gamma2} D_\delta,\\
\mathcal{R}_8\colon & ~~
A_i^8 + A_i^5 A_m^3 + A_i^4 A_m^4 + A_i^2 A_m^6 + A_i^5 A_m B_\alpha + A_i^2 A_m^4 B_\alpha + A_i^2 B_\alpha^3 + A_i^6 B_{\beta 1} + A_i^2 B_\alpha^2 B_{\beta 1} + A_i^3 A_m^3 B_{\beta 3} \nonumber\\
&~~ + A_i^3 A_m^2 C_{\gamma1}+ A_i^2 A_m^3 C_{\gamma2} + A_i^3 B_\alpha C_{\gamma2} + A_i^2 A_m B_\alpha C_{\gamma2} + A_i B_\alpha^2 C_{\gamma2} + A_i A_m B_\alpha D_\delta + D_\delta^2.
\end{align} 
 \end{subequations}
We have the following table regarding IWPs and group cohomology at degree 3.
\begin{center}
\begin{tabular}{c|cc|c|c|c}\hline\hline {Wyckoff}&\multicolumn{2}{c|}{Little group}& \multirow{2}{*}{Coordinates}&\multirow{2}{*}{LSM anomaly class}&\multirow{2}{*}{Topo. inv.} \\ \cline{2-3} position & Intl. & Sch\"{o}nflies & & & \\ \hline
2a&$mmm$&$D_{2h}$& $(0,0,0)$, $(1/2,1/2,1/2)$ & $A_i B_\alpha + A_i B_{\beta 1} + C_{\gamma2}$ & $\varphi_2[C_2, MI]$\\ 
2b&$mmm$&$D_{2h}$& $(0,0,1/2)$, $(1/2,1/2,0)$ & $C_{\gamma2}$ & $\varphi_2[C_2, T_3MI]$\\ 
\hline
\multirow{2}{*}{4c} & \multirow{2}{*}{$2/m$} & \multirow{2}{*}{$C_{2h}$} & $(0,1/2,0)$, $(0,1/2,1/2)$, & \multirow{2}{*}{$A_i B_{\beta 1}$} & \multirow{2}{*}{$\varphi_1[T_1I]$}\\
& & & $(1/2,0,1/2)$, $(1/2,0,0)$ & & \\ \hline
\multirow{2}{*}{4d} & \multirow{2}{*}{$\overline{4}$} & \multirow{2}{*}{$S_4$} & $(0,1/2,1/4)$, $(0,1/2,3/4)$, & \multirow{2}{*}{$A_{c'} B_\alpha$} & \multirow{2}{*}{$\varphi_2[S'_2M, T_2C_2]$}\\
& & & $(1/2,0,1/4)$, $(1/2,0,3/4)$ & & \\ \hline
\hline 
 \end{tabular} 
 \end{center}

\subsection*{No. 137: $P4_2/nmc$}\label{subsub:sg137}

This group is generated by three translations $T_{1,2,3}$ as given in Eqs.~\eqref{TransBravaisP}, a two-fold rotation $C_2$, a two-fold screw $S'_2$, a glide $G$, and an inversion $I$:
\begin{subequations}
 \begin{align}
C_2 &\colon (x,y,z)\rightarrow (-x + 1/2, -y + 1/2, z),\\ 
S'_2 &\colon (x,y,z)\rightarrow (-x, y + 1/2, -z),\\ 
G &\colon (x,y,z)\rightarrow (y, x, z + 1/2),\\ 
I &\colon (x,y,z)\rightarrow (-x, -y, -z).
\end{align}
\end{subequations}

The $\mathbb{Z}_2$ cohomology ring is given by

\begin{equation}
\mathbb{Z}_2[A_i,A_{c'},A_m,B_\alpha,B_{\beta 1},B_{\beta 2},C_\gamma]/\langle\mathcal{R}_2,\mathcal{R}_3,\mathcal{R}_4,\mathcal{R}_5,\mathcal{R}_6\rangle
 \end{equation}
where the relations are 
\begin{subequations} 
 \begin{align}
\mathcal{R}_2\colon & ~~
A_{c'} A_m,~~A_{c'} (A_{c'} + A_i),~~A_m (A_i + A_m),\\
\mathcal{R}_3\colon & ~~
A_{c'} A_i^2 + A_i^3 + A_i^2 A_m + A_{c'} B_\alpha + A_i B_\alpha + A_{c'} B_{\beta 1} + A_i B_{\beta 1},~~A_{c'} A_i^2 + A_i^3 + A_i^2 A_m + A_{c'} B_\alpha + A_i B_\alpha + A_m B_{\beta 1},\nonumber\\&~~(A_{c'} + A_i) B_{\beta 2},~~A_m B_{\beta 2},\\
\mathcal{R}_4\colon & ~~
(A_{c'} + A_i + A_m) C_\gamma,~~B_{\beta 1} (B_\alpha + B_{\beta 1}),~~B_{\beta 1} B_{\beta 2} + A_{c'} C_\gamma,~~B_{\beta 2}^2,\\
\mathcal{R}_5\colon & ~~
(B_\alpha + B_{\beta 1}) C_\gamma,~~B_{\beta 2} C_\gamma,\\
\mathcal{R}_6\colon & ~~
C_\gamma (A_{c'} B_\alpha + A_i B_\alpha + C_\gamma).
\end{align} 
 \end{subequations}
We have the following table regarding IWPs and group cohomology at degree 3.
\begin{center}
\resizebox{\columnwidth}{!}{
\begin{tabular}{c|cc|c|c|c}\hline\hline {Wyckoff}&\multicolumn{2}{c|}{Little group}& \multirow{2}{*}{Coordinates}&\multirow{2}{*}{LSM anomaly class}&\multirow{2}{*}{Topo. inv.} \\ \cline{2-3} position & Intl. & Sch\"{o}nflies & & & \\ \hline
2a&$\overline{4}m2$&$D_{2d}$& $(3/4,1/4,3/4)$, $(1/4,3/4,1/4)$ & $A_{c'} A_i^2 + A_i^3 + A_i^2 A_m + A_{c'} B_\alpha + A_i B_\alpha + C_\gamma$ & $\varphi_2[T_1C_2, T_1T_2T_3GI]$\\ 
2b&$\overline{4}m2$&$D_{2d}$& $(3/4,1/4,1/4)$, $(1/4,3/4,3/4)$ & $C_\gamma$ & $\varphi_2[T_1C_2, T_1T_2GI]$\\ 
\hline
\multirow{2}{*}{4d} & \multirow{2}{*}{$mm2$} & \multirow{2}{*}{$C_{2v}$} & $(1/4,1/4,z)$, $(1/4,1/4,z+1/2)$, & \multirow{2}{*}{$(A_{c'} + A_i + A_m) (A_i^2 + B_\alpha)$} & \multirow{2}{*}{$\varphi_2[S'_2GI, C_2]$}\\
& & & $(3/4,3/4,-z)$, $(3/4,3/4,-z+1/2)$ & & \\ \hline
\multirow{4}{*}{8e} & \multirow{4}{*}{$\overline{1}$} & \multirow{4}{*}{$C_i$} & $(0,0,0)$, $(1/2,1/2,0)$, & \multirow{4}{*}{$A_i^2 (A_{c'} + A_i + A_m)$} & \multirow{4}{*}{$\varphi_1[I]$}\\
& & & $(1/2,0,1/2)$, $(0,1/2,1/2)$, & & \\
& & & $(0,1/2,0)$, $(1/2,0,0)$, & & \\
& & & $(1/2,1/2,1/2)$, $(0,0,1/2)$ & & \\ \hline
\hline 
 \end{tabular} }
 \end{center}

\subsection*{No. 138: $P4_2/ncm$}\label{subsub:sg138}

This group is generated by three translations $T_{1,2,3}$ as given in Eqs.~\eqref{TransBravaisP}, a two-fold rotation $C_2$, a two-fold screw $S'_2$, a mirror $M$, and an inversion $I$:
\begin{subequations}
 \begin{align}
C_2 &\colon (x,y,z)\rightarrow (-x + 1/2, -y + 1/2, z),\\ 
S'_2 &\colon (x,y,z)\rightarrow (-x, y + 1/2, -z + 1/2),\\ 
M &\colon (x,y,z)\rightarrow (y, x, z),\\ 
I &\colon (x,y,z)\rightarrow (-x, -y, -z).
\end{align}
\end{subequations}

The $\mathbb{Z}_2$ cohomology ring is given by

\begin{equation}
\mathbb{Z}_2[A_i,A_m,A_{c'},B_\alpha,B_{\beta 1},B_{\beta 2},B_{\beta 3}]/\langle\mathcal{R}_2,\mathcal{R}_3,\mathcal{R}_4\rangle
 \end{equation}
where the relations are 
\begin{subequations} 
 \begin{align}
\mathcal{R}_2\colon & ~~
A_{c'} A_i,~~A_{c'} A_m,~~A_{c'}^2,\\
\mathcal{R}_3\colon & ~~
A_{c'} B_{\beta 1},~~A_i^3 + A_i^2 A_m + A_{c'} B_\alpha + A_i B_\alpha + A_m B_{\beta 2},~~A_i^3 + A_i^2 A_m + A_{c'} B_\alpha + A_i B_\alpha + A_{c'} B_{\beta 2} + A_i B_{\beta 2},\nonumber\\&~~A_i B_{\beta 1} + A_m B_{\beta 3},~~A_{c'} B_{\beta 3},\\
\mathcal{R}_4\colon & ~~
A_i^4 + A_i^3 A_m + A_i^2 B_\alpha + A_i A_m B_{\beta 1} + B_{\beta 1}^2,~~A_i^2 B_{\beta 1} + B_{\beta 1} B_{\beta 2} + A_i^2 B_{\beta 3} + B_\alpha B_{\beta 3},\nonumber\\&~~A_i^4 + A_i^3 A_m + A_i^2 B_\alpha + A_i^2 B_{\beta 1} + B_{\beta 1} B_{\beta 3},~~B_{\beta 2} (B_\alpha + B_{\beta 2}),~~A_i^2 B_{\beta 1} + A_i^2 B_{\beta 3} + B_\alpha B_{\beta 3} + B_{\beta 2} B_{\beta 3},\nonumber\\&~~A_i^4 + A_i^3 A_m + A_i^2 B_\alpha + A_i^2 B_{\beta 3} + B_{\beta 3}^2.
\end{align} 
 \end{subequations}
We have the following table regarding IWPs and group cohomology at degree 3.
\begin{center}
\resizebox{\columnwidth}{!}{
\begin{tabular}{c|cc|c|c|c}\hline\hline {Wyckoff}&\multicolumn{2}{c|}{Little group}& \multirow{2}{*}{Coordinates}&\multirow{2}{*}{LSM anomaly class}&\multirow{2}{*}{Topo. inv.} \\ \cline{2-3} position & Intl. & Sch\"{o}nflies & & & \\ \hline
\multirow{2}{*}{4a} & \multirow{2}{*}{$222$} & \multirow{2}{*}{$D_2$} & $(3/4,1/4,0)$, $(1/4,3/4,1/2)$, & \multirow{2}{*}{$A_i B_{\beta 2}$} & \multirow{2}{*}{$\varphi_2[T_1^{-1}C_2, MI]$}\\
& & & $(1/4,3/4,0)$, $(3/4,1/4,1/2)$ & & \\ \hline
\multirow{2}{*}{4b} & \multirow{2}{*}{$\overline{4}$} & \multirow{2}{*}{$S_4$} & $(3/4,1/4,3/4)$, $(1/4,3/4,1/4)$, & \multirow{2}{*}{$A_i^3 + A_i^2 A_m + A_{c'} B_\alpha + A_i B_\alpha + A_i B_{\beta 2}$} & \multirow{2}{*}{$\varphi_2[S'_2M, T_1^{-1}C_2]$}\\
& & & $(1/4,3/4,3/4)$, $(3/4,1/4,1/4)$ & & \\ \hline
\multirow{2}{*}{4c} & \multirow{2}{*}{$2/m$} & \multirow{2}{*}{$C_{2h}$} & $(0,0,1/2)$, $(1/2,1/2,1/2)$, & \multirow{2}{*}{$A_i (B_{\beta 1} + B_{\beta 3})$} & \multirow{2}{*}{$\varphi_1[T_1I]$}\\
& & & $(1/2,0,0)$, $(0,1/2,0)$ & & \\ \hline
\multirow{2}{*}{4d} & \multirow{2}{*}{$2/m$} & \multirow{2}{*}{$C_{2h}$} & $(0,0,0)$, $(1/2,1/2,0)$, & \multirow{2}{*}{$A_i (A_i^2 + A_i A_m + B_{\beta 1} + B_{\beta 3})$} & \multirow{2}{*}{$\varphi_1[I]$}\\
& & & $(1/2,0,1/2)$, $(0,1/2,1/2)$ & & \\ \hline
\multirow{2}{*}{4e} & \multirow{2}{*}{$mm2$} & \multirow{2}{*}{$C_{2v}$} & $(1/4,1/4,z)$, $(1/4,1/4,z+1/2)$, & \multirow{2}{*}{$A_i (A_i^2 + A_i A_m + B_\alpha + B_{\beta 2})$} & \multirow{2}{*}{$\varphi_2[S'_2I, C_2]$}\\
& & & $(3/4,3/4,-z+1/2)$, $(3/4,3/4,-z)$ & & \\ \hline
\hline 
 \end{tabular} }
 \end{center}

\subsection*{No. 139: $I4/mmm$}\label{subsub:sg139}

This group is generated by three translations $T_{1,2,3}$ as given in Eqs.~\eqref{TransBravaisI}, a two-fold rotation $C_2$, a two-fold rotation $C'_2$, a mirror $M$, and an inversion $I$:
\begin{subequations}
 \begin{align}
C_2 &\colon (x,y,z)\rightarrow (-x, -y, z),\\ 
C'_2 &\colon (x,y,z)\rightarrow (-x, y, -z),\\ 
M &\colon (x,y,z)\rightarrow (y, x, z),\\ 
I &\colon (x,y,z)\rightarrow (-x, -y, -z).
\end{align}
\end{subequations}

The $\mathbb{Z}_2$ cohomology ring is given by

\begin{equation}
\mathbb{Z}_2[A_i,A_m,A_{c'},A_{x+y+z},B_\alpha,B_\beta,B_{z(x+y)},C_{xyz}]/\langle\mathcal{R}_2,\mathcal{R}_3,\mathcal{R}_4,\mathcal{R}_5,\mathcal{R}_6\rangle
 \end{equation}
where the relations are 
\begin{subequations} 
 \begin{align}
\mathcal{R}_2\colon & ~~
A_{c'} A_m,~~A_{c'} A_{x+y+z},~~A_{x+y+z} (A_i + A_{x+y+z}),\\
\mathcal{R}_3\colon & ~~
A_i^2 A_{x+y+z} + A_i A_m A_{x+y+z} + A_{x+y+z} B_\alpha + A_m B_\beta,~~A_{x+y+z} (A_i^2 + A_i A_m + B_\alpha + B_\beta),\nonumber\\&~~A_i^2 A_{x+y+z} + A_i A_m A_{x+y+z} + A_{x+y+z} B_\alpha + A_m B_{z(x+y)},~~A_{x+y+z} (A_i^2 + A_i A_m + B_\alpha + B_{z(x+y)}),\\
\mathcal{R}_4\colon & ~~
A_{x+y+z} C_{xyz},\nonumber\\&~~A_i^3 A_{x+y+z} + A_i^2 A_m A_{x+y+z} + A_i A_{x+y+z} B_\alpha + A_{c'}^2 B_\beta + A_{c'} A_i B_\beta + B_\beta^2 + A_{c'} A_i B_{z(x+y)} + A_i^2 B_{z(x+y)} + B_\alpha B_{z(x+y)},\nonumber\\&~~B_\alpha B_\beta + B_\beta B_{z(x+y)} + A_{c'} C_{xyz},~~B_{z(x+y)} (B_\alpha + B_{z(x+y)}),\\
\mathcal{R}_5\colon & ~~
(A_{c'}^2 + A_{c'} A_i + B_\beta) C_{xyz},~~B_{z(x+y)} C_{xyz},\\
\mathcal{R}_6\colon & ~~
C_{xyz} (A_{c'} B_\alpha + A_i B_\alpha + C_{xyz}).
\end{align} 
 \end{subequations}
We have the following table regarding IWPs and group cohomology at degree 3.
\begin{center}
\resizebox{\columnwidth}{!}{
\begin{tabular}{c|cc|c|c|c}\hline\hline {Wyckoff}&\multicolumn{2}{c|}{Little group}& {Coordinates}&\multirow{2}{*}{LSM anomaly class}&\multirow{2}{*}{Topo. inv.} \\ \cline{2-4} position & Intl. & Sch\"{o}nflies & $(0,0,0) + ~(1/2,1/2,1/2) + $ & &\\ \hline
2a&$4/mmm$&$D_{4h}$& $(0,0,0)$ & $A_i^2 A_{x+y+z} + A_i A_m A_{x+y+z} + A_{c'} B_\alpha + A_i B_\alpha + A_{c'} B_{z(x+y)} + A_i B_{z(x+y)} + C_{xyz}$ & $\varphi_2[C_2, C'_2]$\\ 
2b&$4/mmm$&$D_{4h}$& $(0,0,1/2)$ & $C_{xyz}$ & $\varphi_2[C_2, T_1T_2C'_2]$\\ 
4c&$mmm$&$D_{2h}$& $(0,1/2,0)$, $(1/2,0,0)$ & $A_i^2 A_{x+y+z} + A_i A_m A_{x+y+z} + A_{x+y+z} B_\alpha + A_{c'} B_{z(x+y)} + A_i B_{z(x+y)}$ & $\varphi_2[T_1T_3C_2, C'_2]$\\ 
4d&$\overline{4}m2$&$D_{2d}$& $(0,1/2,1/4)$, $(1/2,0,1/4)$ & $A_{x+y+z} (A_i^2 + A_i A_m + B_\alpha)$ & $\varphi_2[T_1T_3C_2, T_1T_2T_3MI]$\\ 
\hline
\multirow{2}{*}{8f} & \multirow{2}{*}{$2/m$} & \multirow{2}{*}{$C_{2h}$} & $(1/4,1/4,1/4)$, $(3/4,3/4,1/4)$, & \multirow{2}{*}{$A_i (A_i + A_m) A_{x+y+z}$} & \multirow{2}{*}{$\varphi_1[T_1T_2T_3I]$}\\
& & & $(3/4,1/4,1/4)$, $(1/4,3/4,1/4)$ & & \\ \hline
\hline 
 \end{tabular} }
 \end{center}

\subsection*{No. 140: $I4/mcm$}\label{subsub:sg140}

This group is generated by three translations $T_{1,2,3}$ as given in Eqs.~\eqref{TransBravaisI}, a two-fold rotation $C_2$, a two-fold rotation $C'_2$, a glide $G$, and an inversion $I$:
\begin{subequations}
 \begin{align}
C_2 &\colon (x,y,z)\rightarrow (-x, -y, z),\\ 
C'_2 &\colon (x,y,z)\rightarrow (-x, y, -z + 1/2),\\ 
G &\colon (x,y,z)\rightarrow (y, x, z + 1/2),\\ 
I &\colon (x,y,z)\rightarrow (-x, -y, -z).
\end{align}
\end{subequations}

The $\mathbb{Z}_2$ cohomology ring is given by

\begin{equation}
\mathbb{Z}_2[A_i,A_m,A_{c'},A_{x+y+z},B_\alpha,B_{z(x+y)},D_\delta]/\langle\mathcal{R}_2,\mathcal{R}_3,\mathcal{R}_4,\mathcal{R}_5,\mathcal{R}_6,\mathcal{R}_8\rangle
 \end{equation}
where the relations are 
\begin{subequations} 
 \begin{align}
\mathcal{R}_2\colon & ~~
A_{c'} A_m,~~A_{c'} A_{x+y+z},~~A_{c'} A_i + A_i A_m + A_m^2 + A_i A_{x+y+z} + A_{x+y+z}^2,\\
\mathcal{R}_3\colon & ~~
A_m A_{x+y+z}^2 + A_{x+y+z}^3 + A_{x+y+z} B_\alpha + A_m B_{z(x+y)},~~A_{x+y+z}(A_m A_{x+y+z} + A_{x+y+z}^2 + B_\alpha + B_{z(x+y)}),\\
\mathcal{R}_4\colon & ~~
B_{z(x+y)} (B_\alpha + B_{z(x+y)}),\\
\mathcal{R}_5\colon & ~~
A_{c'} (B_\alpha B_{z(x+y)} + D_\delta),~~(A_m + A_{x+y+z}) D_\delta,\\
\mathcal{R}_6\colon & ~~
(B_\alpha + B_{z(x+y)}) D_\delta,\\
\mathcal{R}_8\colon & ~~
A_i^4 A_m^4 + A_i^3 A_m^5 + A_i^6 A_m A_{x+y+z} + A_i^4 A_m^3 A_{x+y+z} + A_i^3 A_m^4 A_{x+y+z} + A_{c'} A_i^5 B_\alpha + A_i^5 A_m B_\alpha + A_i^4 A_m^2 B_\alpha \nonumber\\
& ~~ + A_i^5 A_{x+y+z} B_\alpha + A_i A_{x+y+z} B_\alpha^3 + A_i^2 B_\alpha^2 B_{z(x+y)} + B_\alpha^3 B_{z(x+y)} + A_i A_m B_\alpha D_\delta + D_\delta^2.
\end{align} 
 \end{subequations}
We have the following table regarding IWPs and group cohomology at degree 3.
\begin{center}
\resizebox{\columnwidth}{!}{
\begin{tabular}{c|cc|c|c|c}\hline\hline {Wyckoff}&\multicolumn{2}{c|}{Little group}& {Coordinates}&\multirow{2}{*}{LSM anomaly class}&\multirow{2}{*}{Topo. inv.} \\ \cline{2-4} position & Intl. & Sch\"{o}nflies & $(0,0,0) + ~(1/2,1/2,1/2) + $ & &\\ \hline
4a&$422$&$D_4$& $(0,0,1/4)$, $(0,0,3/4)$ & 
$(A_{c'} + A_m)(A_m A_{x+y+z} + A_{x+y+z}^2 +  B_\alpha + B_{z(x+y)})$
& $\varphi_2[C_2, C'_2]$\\ 
4b&$\overline{4}2m$&$D_{2d}$& $(0,1/2,1/4)$, $(1/2,0,1/4)$ & $A_{c'} B_{z(x+y)}$ & $\varphi_2[T_1T_3C_2, C'_2]$\\ 
4c&$4/m$&$C_{4h}$& $(0,0,0)$, $(0,0,1/2)$ & $A_i B_\alpha + A_m B_\alpha + A_{x+y+z} B_\alpha + A_i B_{z(x+y)}$ & $\varphi_1[I]$\\ 
4d&$mmm$&$D_{2h}$& $(0,1/2,0)$, $(1/2,0,0)$ & $A_i B_{z(x+y)}$ & $\varphi_2[T_1T_3C_2, T_3GI]$\\ 
\hline
\multirow{2}{*}{8e} & \multirow{2}{*}{$2/m$} & \multirow{2}{*}{$C_{2h}$} & $(1/4,1/4,1/4)$, $(3/4,3/4,1/4)$, & \multirow{2}{*}{$A_i ( A_m + A_{x+y+z})A_{x+y+z}$}
& \multirow{2}{*}{$\varphi_1[T_1T_2T_3I]$}\\
& & & $(3/4,1/4,1/4)$, $(1/4,3/4,1/4)$ & & \\ \hline
\hline 
 \end{tabular} }
 \end{center}

\subsection*{No. 141: $I4_1/amd$}\label{subsub:sg141}

This group is generated by three translations $T_{1,2,3}$ as given in Eqs.~\eqref{TransBravaisI}, a two-fold rotation $C_2$, a two-fold rotation $C'_2$, a glide $G$, and an inversion $I$:
\begin{subequations}
 \begin{align}
C_2 &\colon (x,y,z)\rightarrow (-x, -y + 1/2, z),\\ 
C'_2 &\colon (x,y,z)\rightarrow (-x + 1/2, y, -z + 1/2),\\ 
G &\colon (x,y,z)\rightarrow (y + 1/4, x + 1/4, z -1/4),\\ 
I &\colon (x,y,z)\rightarrow (-x, -y, -z).
\end{align}
\end{subequations}

The $\mathbb{Z}_2$ cohomology ring is given by

\begin{equation}
\mathbb{Z}_2[A_i,A_m,A_{c'},B_\alpha,B_{z(x+y)},C_\gamma]/\langle\mathcal{R}_2,\mathcal{R}_3,\mathcal{R}_4,\mathcal{R}_5,\mathcal{R}_6\rangle
 \end{equation}
where the relations are 
\begin{subequations} 
 \begin{align}
\mathcal{R}_2\colon & ~~
A_{c'} A_m,~~A_m (A_i + A_m),\\
\mathcal{R}_3\colon & ~~
A_{c'}^2 A_i + A_i^3 + A_i^2 A_m + A_{c'} B_\alpha + A_i B_\alpha + A_m B_\alpha,~~A_m (B_\alpha + B_{z(x+y)}),\\
\mathcal{R}_4\colon & ~~
(A_{c'} + A_i + A_m) C_\gamma,\nonumber\\&~~A_{c'}^2 A_i^2 + A_i^4 + A_i^3 A_m + A_i A_m B_\alpha + B_\alpha^2 + A_{c'} A_i B_{z(x+y)} + A_i^2 B_{z(x+y)} + B_\alpha B_{z(x+y)} + A_i C_\gamma + A_m C_\gamma,\nonumber\\&~~A_{c'}^2 A_i^2 + A_i^4 + A_i^3 A_m + A_i A_m B_\alpha + B_\alpha^2 + A_{c'} A_i B_{z(x+y)} + A_i^2 B_{z(x+y)} + B_{z(x+y)}^2,\\
\mathcal{R}_5\colon & ~~
(B_\alpha + B_{z(x+y)}) C_\gamma,\\
\mathcal{R}_6\colon & ~~
C_\gamma (A_m B_\alpha + C_\gamma).
\end{align} 
 \end{subequations}
We have the following table regarding IWPs and group cohomology at degree 3.
\begin{center}
\resizebox{\columnwidth}{!}{
\begin{tabular}{c|cc|c|c|c}\hline\hline {Wyckoff}&\multicolumn{2}{c|}{Little group}& {Coordinates}&\multirow{2}{*}{LSM anomaly class}&\multirow{2}{*}{Topo. inv.} \\ \cline{2-4} position & Intl. & Sch\"{o}nflies & $(0,0,0) + ~(1/2,1/2,1/2) + $ & &\\ \hline
4a&$\overline{4}m2$&$D_{2d}$& $(0,3/4,1/8)$, $(1/2,3/4,3/8)$ & $A_m B_\alpha + C_\gamma$ & $\varphi_2[C_2, GI]$\\ 
4b&$\overline{4}m2$&$D_{2d}$& $(0,1/4,3/8)$, $(0,3/4,5/8)$ & $C_\gamma$ & $\varphi_2[T_1T_3C_2, T_3GI]$\\ 
\hline
\multirow{2}{*}{8c} & \multirow{2}{*}{$2m$} & \multirow{2}{*}{$C_{2h}$} & $(0,0,0)$, $(1/2,0,1/2)$, & \multirow{2}{*}{$A_m B_\alpha + A_{c'} B_{z(x+y)} + A_i B_{z(x+y)}$} & \multirow{2}{*}{$\varphi_1[I]$}\\
& & & $(1/4,3/4,1/4)$, $(1/4,1/4,3/4)$ & & \\ \hline
\multirow{2}{*}{8d} & \multirow{2}{*}{$2m$} & \multirow{2}{*}{$C_{2h}$} & $(0,0,1/2)$, $(1/2,0,0)$, & \multirow{2}{*}{$A_{c'}^2 A_i + A_i^3 + A_i^2 A_m + A_m B_\alpha + A_{c'} B_{z(x+y)} + A_i B_{z(x+y)}$} & \multirow{2}{*}{$\varphi_1[T_1T_2I]$}\\
& & & $(1/4,3/4,3/4)$, $(1/4,1/4,1/4)$ & & \\ \hline
\hline 
 \end{tabular} }
 \end{center}

\subsection*{No. 142: $I4_1/acd$}\label{subsub:sg142}

This group is generated by three translations $T_{1,2,3}$ as given in Eqs.~\eqref{TransBravaisI}, a two-fold rotation $C_2$, a two-fold rotation $C'_2$, a glide $G$, and an inversion $I$:
\begin{subequations}
 \begin{align}
C_2 &\colon (x,y,z)\rightarrow (-x, -y + 1/2, z),\\ 
C'_2 &\colon (x,y,z)\rightarrow (-x + 1/2, y, -z),\\ 
G &\colon (x,y,z)\rightarrow (y + 1/4, x + 1/4, z + 1/4),\\ 
I &\colon (x,y,z)\rightarrow (-x, -y, -z).
\end{align}
\end{subequations}

The $\mathbb{Z}_2$ cohomology ring is given by

\begin{equation}
\mathbb{Z}_2[A_i,A_m,A_{c'},B_\alpha,D_\gamma]/\langle\mathcal{R}_2,\mathcal{R}_3,\mathcal{R}_4,\mathcal{R}_5,\mathcal{R}_8\rangle
 \end{equation}
where the relations are 
\begin{subequations} 
 \begin{align}
\mathcal{R}_2\colon & ~~
A_{c'} A_m,~~A_m (A_i + A_m),\\
\mathcal{R}_3\colon & ~~
A_{c'} A_i^2,~~A_i^3 + A_i^2 A_m + A_{c'} B_\alpha + A_i B_\alpha + A_m B_\alpha,\\
\mathcal{R}_4\colon & ~~
A_i (A_i + A_m) (A_i^2 + B_\alpha),\\
\mathcal{R}_5\colon & ~~
(A_i + A_m) D_\gamma,~~A_{c'} D_\gamma,\\
\mathcal{R}_8\colon & ~~
A_i^8+A_i^7 A_m+A_i^2 B_\alpha^3+A_i^2 B_\alpha D_\gamma+D_\gamma^2.
\end{align} 
 \end{subequations}
We have the following table regarding IWPs and group cohomology at degree 3.
\begin{center}
\begin{tabular}{c|cc|c|c|c}\hline\hline {Wyckoff}&\multicolumn{2}{c|}{Little group}& {Coordinates}&\multirow{2}{*}{LSM anomaly class}&\multirow{2}{*}{Topo. inv.} \\ \cline{2-4} position & Intl. & Sch\"{o}nflies & $(0,0,0) + ~(1/2,1/2,1/2) + $ & &\\ \hline
\multirow{2}{*}{8a} & \multirow{2}{*}{$\overline{4}$} & \multirow{2}{*}{$S_4$} & $(0,1/4,3/8)$, $(0,3/4,5/8)$, & \multirow{2}{*}{$(A_i + A_m) (A_i^2 + B_\alpha)$} & \multirow{2}{*}{$\varphi_2[C'_2G, C_2]$}\\
& & & $(1/2,1/4,5/8)$, $(1/2,3/4,3/8)$ & & \\ \hline
\multirow{2}{*}{8b} & \multirow{2}{*}{$222$} & \multirow{2}{*}{$D_2$} & $(0,1/4,1/8)$, $(0,3/4,3/8)$, & \multirow{2}{*}{$A_i (A_i^2 + A_i A_m + B_\alpha)$} & \multirow{2}{*}{$\varphi_2[C_2, GI]$}\\
& & & $(0,3/4,7/8)$, $(0,1/4,5/8)$ & & \\ \hline
\multirow{4}{*}{16c} & \multirow{4}{*}{$\overline{1}$} & \multirow{4}{*}{$C_i$} & $(0,0,0)$, $(1/2,0,1/2)$, & \multirow{4}{*}{$A_i^2 (A_i + A_m)$} & \multirow{4}{*}{$\varphi_1[I]$}\\
& & & $(1/4,3/4,1/4)$, $(1/4,1/4,3/4)$, & & \\
& & & $(1/2,0,0)$, $(0,0,1/2)$, & & \\
& & & $(1/4,3/4,3/4)$, $(1/4,1/4,1/4)$ & & \\ \hline
\multirow{4}{*}{16e} & \multirow{4}{*}{$2$} & \multirow{4}{*}{$C_2$} & $(x,0,1/4)$, $(-x+1/2,0,3/4)$, & \multirow{4}{*}{$A_{c'}^2 A_i$} & \multirow{4}{*}{$\varphi_2[C'_2I, T_2C_2C'_2]$}\\
& & & $(1/4,x+3/4,1/2)$, $(1/4,-x+1/4,0)$, & & \\
& & & $(-x,0,3/4)$, $(x+1/2,0,1/4)$, & & \\
& & & $(3/4,-x+1/4,1/2)$, $(3/4,x+3/4,0)$ & & \\ \hline
\hline 
 \end{tabular} 
 \end{center}

\subsection*{No. 143: $P3$}\label{subsub:sg143}

This group is generated by three translations $T_{1,2,3}$ as given in Eqs.~\eqref{TransBravaisP}, and a three-fold rotation $C_3$:
\begin{subequations}
 \begin{align}
C_3 &\colon (x,y,z)\rightarrow (-y, x - y, z).
\end{align}
\end{subequations}

The $\mathbb{Z}_2$ cohomology ring is given by

\begin{equation}
\mathbb{Z}_2[A_z,B_{xy}]/\langle\mathcal{R}_2,\mathcal{R}_4\rangle
 \end{equation}
where the relations are 
\begin{subequations} 
 \begin{align}
\mathcal{R}_2\colon & ~~
A_z^2,\\
\mathcal{R}_4\colon & ~~
B_{xy}^2.
\end{align} 
 \end{subequations}
We have the following table regarding IWPs and group cohomology at degree 3.
\begin{center}
\begin{tabular}{c|cc|c|c|c}\hline\hline {Wyckoff}&\multicolumn{2}{c|}{Little group}& \multirow{2}{*}{Coordinates}&\multirow{2}{*}{LSM anomaly class}&\multirow{2}{*}{Topo. inv.} \\ \cline{2-3} position & Intl. & Sch\"{o}nflies & & & \\ \hline
1a&$3$&$C_3$& $(0,0,z)$ & $A_z B_{xy}$ & $\varphi_3[T_1, T_2, T_3]$\\ 
1b&$3$&$C_3$& $(1/3,2/3,z)$ & Same as 1a  & Same as 1a\\ 
1c&$3$&$C_3$& $(2/3,1/3,z)$ & Same as 1a  & Same as 1a\\ 
\hline
\hline 
 \end{tabular} 
 \end{center}

\subsection*{No. 144: $P3_1$}\label{subsub:sg144}

This group is generated by three translations $T_{1,2,3}$ as given in Eqs.~\eqref{TransBravaisP}, and a three-fold screw $S_3$:
\begin{subequations}
 \begin{align}
S_3 &\colon (x,y,z)\rightarrow (-y, x - y, z + 1/3).
\end{align}
\end{subequations}

The $\mathbb{Z}_2$ cohomology ring is given by

\begin{equation}
\mathbb{Z}_2[A_z,B_{xy}]/\langle\mathcal{R}_2,\mathcal{R}_4\rangle
 \end{equation}
where the relations are 
\begin{subequations} 
 \begin{align}
\mathcal{R}_2\colon & ~~
A_z^2,\\
\mathcal{R}_4\colon & ~~
B_{xy}^2.
\end{align} 
 \end{subequations}
We have the following table regarding IWPs and group cohomology at degree 3.
\begin{center}
\begin{tabular}{c|cc|c|c|c}\hline\hline {Wyckoff}&\multicolumn{2}{c|}{Little group}& \multirow{2}{*}{Coordinates}&\multirow{2}{*}{LSM anomaly class}&\multirow{2}{*}{Topo. inv.} \\ \cline{2-3} position & Intl. & Sch\"{o}nflies & & & \\ \hline
\multirow{2}{*}{3a} & \multirow{2}{*}{$1$} & \multirow{2}{*}{$C_1$} & $(x,y,z)$, $(-y,x-y,z+1/3)$, & \multirow{2}{*}{$A_z B_{xy}$} & \multirow{2}{*}{$\varphi_3[T_1, T_2, T_3]$}\\
& & & $(-x+y,-x,z+2/3)$ & & \\ \hline
\hline 
 \end{tabular} 
 \end{center}

\subsection*{No. 145: $P3_2$}\label{subsub:sg145}

This group is generated by three translations $T_{1,2,3}$ as given in Eqs.~\eqref{TransBravaisP}, and a three-fold screw $S_3$:
\begin{subequations}
 \begin{align}
S_3 &\colon (x,y,z)\rightarrow (-y, x - y, z + 2/3).
\end{align}
\end{subequations}

The $\mathbb{Z}_2$ cohomology ring is given by

\begin{equation}
\mathbb{Z}_2[A_z,B_{xy}]/\langle\mathcal{R}_2,\mathcal{R}_4\rangle
 \end{equation}
where the relations are 
\begin{subequations} 
 \begin{align}
\mathcal{R}_2\colon & ~~
A_z^2,\\
\mathcal{R}_4\colon & ~~
B_{xy}^2.
\end{align} 
 \end{subequations}
We have the following table regarding IWPs and group cohomology at degree 3.
\begin{center}
\begin{tabular}{c|cc|c|c|c}\hline\hline {Wyckoff}&\multicolumn{2}{c|}{Little group}& \multirow{2}{*}{Coordinates}&\multirow{2}{*}{LSM anomaly class}&\multirow{2}{*}{Topo. inv.} \\ \cline{2-3} position & Intl. & Sch\"{o}nflies & & & \\ \hline
\multirow{2}{*}{3a} & \multirow{2}{*}{$1$} & \multirow{2}{*}{$C_1$} & $(x,y,z)$, $(-y,x-y,z+2/3)$, & \multirow{2}{*}{$A_z B_{xy}$} & \multirow{2}{*}{$\varphi_3[T_1, T_2, T_3]$}\\
& & & $(-x+y,-x,z+1/3)$ & & \\ \hline
\hline 
 \end{tabular} 
 \end{center}

\subsection*{No. 146: $R3$}\label{subsub:sg146}

This group is generated by three translations $T_{1,2,3}$ as given in Eqs.~\eqref{TransBravaisR}, and a three-fold rotation $C_3$:
\begin{subequations}
 \begin{align}
C_3 &\colon (x,y,z)\rightarrow (-y, x - y, z).
\end{align}
\end{subequations}

The $\mathbb{Z}_2$ cohomology ring is given by

\begin{equation}
\mathbb{Z}_2[A_z,B_{x(y+z)}]/\langle\mathcal{R}_2,\mathcal{R}_4\rangle
 \end{equation}
where the relations are 
\begin{subequations} 
 \begin{align}
\mathcal{R}_2\colon & ~~
A_z^2,\\
\mathcal{R}_4\colon & ~~
B_{x(y+z)}^2.
\end{align} 
 \end{subequations}
We have the following table regarding IWPs and group cohomology at degree 3.
\begin{center}
\begin{tabular}{c|cc|c|c|c}\hline\hline {Wyckoff}&\multicolumn{2}{c|}{Little group}& {Coordinates}&\multirow{2}{*}{LSM anomaly class}&\multirow{2}{*}{Topo. inv.} \\ \cline{2-4} position & Intl. & Sch\"{o}nflies & $(0,0,0) + ~(2/3,1/3,1/3) + ~(1/3,2/3,2/3) +$ & &\\ \hline
3a&$3$&$C_3$& $(0,0,z)$ & $A_z B_{x(y+z)}$ & $\varphi_3[T_1, T_2, T_3]$\\ 
\hline
\hline 
 \end{tabular} 
 \end{center}

\subsection*{No. 147: $P\overline3$}\label{subsub:sg147}

This group is generated by three translations $T_{1,2,3}$ as given in Eqs.~\eqref{TransBravaisP}, a three-fold rotation $C_3$, and an inversion $I$:
\begin{subequations}
 \begin{align}
C_3 &\colon (x,y,z)\rightarrow (-y, x - y, z),\\ 
I &\colon (x,y,z)\rightarrow (-x, -y, -z).
\end{align}
\end{subequations}

The $\mathbb{Z}_2$ cohomology ring is given by

\begin{equation}
\mathbb{Z}_2[A_i,A_z,B_{xy}]/\langle\mathcal{R}_2,\mathcal{R}_4\rangle
 \end{equation}
where the relations are 
\begin{subequations} 
 \begin{align}
\mathcal{R}_2\colon & ~~
A_z (A_i + A_z),\\
\mathcal{R}_4\colon & ~~
B_{xy} (A_i^2 + B_{xy}).
\end{align} 
 \end{subequations}
We have the following table regarding IWPs and group cohomology at degree 3.
\begin{center}
\begin{tabular}{c|cc|c|c|c}\hline\hline {Wyckoff}&\multicolumn{2}{c|}{Little group}& \multirow{2}{*}{Coordinates}&\multirow{2}{*}{LSM anomaly class}&\multirow{2}{*}{Topo. inv.} \\ \cline{2-3} position & Intl. & Sch\"{o}nflies & & & \\ \hline
1a&$\overline{3}$&$C_{3i}$& $(0,0,0)$ & $(A_i + A_z) (A_i^2 + B_{xy})$ & $\varphi_1[I]$\\ 
1b&$\overline{3}$&$C_{3i}$& $(0,0,1/2)$ & $A_z (A_i^2 + B_{xy})$ & $\varphi_1[T_3I]$\\ 
2d&$3$&$C_3$& $(1/3,2/3,z)$, $(2/3,1/3,-z)$ & N/A & N/A\\ 
\hline
\multirow{2}{*}{3e} & \multirow{2}{*}{$\overline{1}$} & \multirow{2}{*}{$C_i$} & $(1/2,0,0)$, $(0,1/2,0)$, & \multirow{2}{*}{$(A_i + A_z) B_{xy}$} & \multirow{2}{*}{$\varphi_1[T_1I]$}\\
& & & $(1/2,1/2,0)$ & & \\ \hline
\multirow{2}{*}{3f} & \multirow{2}{*}{$\overline{1}$} & \multirow{2}{*}{$C_i$} & $(1/2,0,1/2)$, $(0,1/2,1/2)$, & \multirow{2}{*}{$A_z B_{xy}$} & \multirow{2}{*}{$\varphi_1[T_1T_3I]$}\\
& & & $(1/2,1/2,1/2)$ & & \\ \hline
\hline 
 \end{tabular} 
 \end{center}

\subsection*{No. 148: $R\overline3$}\label{subsub:sg148}

This group is generated by three translations $T_{1,2,3}$ as given in Eqs.~\eqref{TransBravaisR}, a three-fold rotation $C_3$, and an inversion $I$:
\begin{subequations}
 \begin{align}
C_3 &\colon (x,y,z)\rightarrow (-y, x - y, z),\\ 
I &\colon (x,y,z)\rightarrow (-x, -y, -z).
\end{align}
\end{subequations}

The $\mathbb{Z}_2$ cohomology ring is given by

\begin{equation}
\mathbb{Z}_2[A_i,A_z,B_{x(y+z)}]/\langle\mathcal{R}_2,\mathcal{R}_4\rangle
 \end{equation}
where the relations are 
\begin{subequations} 
 \begin{align}
\mathcal{R}_2\colon & ~~
A_z (A_i + A_z),\\
\mathcal{R}_4\colon & ~~
B_{x(y+z)} (A_i^2 + B_{x(y+z)}).
\end{align} 
 \end{subequations}
We have the following table regarding IWPs and group cohomology at degree 3.
\begin{center}
\resizebox{\columnwidth}{!}{
\begin{tabular}{c|cc|c|c|c}\hline\hline {Wyckoff}&\multicolumn{2}{c|}{Little group}& {Coordinates}&\multirow{2}{*}{LSM anomaly class}&\multirow{2}{*}{Topo. inv.} \\ \cline{2-4} position & Intl. & Sch\"{o}nflies & $(0,0,0) + ~(2/3,1/3,1/3) + ~(1/3,2/3,2/3) +$ & &\\ \hline
3a&$\overline{3}$&$C_{3i}$& $(0,0,0)$ & $(A_i + A_z) (A_i^2 + B_{x(y+z)})$ & $\varphi_1[I]$\\ 
3b&$\overline{3}$&$C_{3i}$& $(0,0,1/2)$ & $A_z B_{x(y+z)}$ & $\varphi_1[T_1^{-2}T_2^{-1}T_3^{3}I]$\\ 
{9d} & {$\overline{1}$} & {$C_i$} & $(1/2,0,1/2)$, $(0,1/2,1/2)$, $(1/2,1/2,1/2)$ & {$A_z (A_i^2 + B_{x(y+z)})$} & {$\varphi_1[T_1^{-1}T_2^{-1}T_3^{3}I]$}\\
{9e} & {$\overline{1}$} & {$C_i$} & $(1/2,0,0)$, $(0,1/2,0)$, $(1/2,1/2,0)$ & {$(A_i + A_z) B_{x(y+z)}$} & {$\varphi_1[T_1I]$}\\
\hline
\hline 
 \end{tabular} }
 \end{center}

\subsection*{No. 149: $P312$}\label{subsub:sg149}

This group is generated by three translations $T_{1,2,3}$ as given in Eqs.~\eqref{TransBravaisP}, a three-fold rotation $C_3$, and a two-fold rotation $C'_2$:
\begin{subequations}
 \begin{align}
C_3 &\colon (x,y,z)\rightarrow (-y, x - y, z),\\ 
C'_2 &\colon (x,y,z)\rightarrow (-y, -x, -z).
\end{align}
\end{subequations}

The $\mathbb{Z}_2$ cohomology ring is given by

\begin{equation}
\mathbb{Z}_2[A_{c'},A_z,B_{xy}]/\langle\mathcal{R}_2,\mathcal{R}_4\rangle
 \end{equation}
where the relations are 
\begin{subequations} 
 \begin{align}
\mathcal{R}_2\colon & ~~
A_z (A_{c'} + A_z),\\
\mathcal{R}_4\colon & ~~
B_{xy}^2.
\end{align} 
 \end{subequations}
We have the following table regarding IWPs and group cohomology at degree 3.
\begin{center}
\begin{tabular}{c|cc|c|c|c}\hline\hline {Wyckoff}&\multicolumn{2}{c|}{Little group}& \multirow{2}{*}{Coordinates}&\multirow{2}{*}{LSM anomaly class}&\multirow{2}{*}{Topo. inv.} \\ \cline{2-3} position & Intl. & Sch\"{o}nflies & & & \\ \hline
1a&$32$&$D_3$& $(0,0,0)$ & $(A_{c'} + A_z) B_{xy}$ & $\varphi_2[T_1T_2^{-1}, C'_2]$\\ 
1b&$32$&$D_3$& $(0,0,1/2)$ & $A_z B_{xy}$ & $\varphi_2[T_1T_2^{-1}, T_3C'_2]$\\ 
1c&$32$&$D_3$& $(1/3,2/3,0)$ & Same as 1a  & Same as 1a\\ 
1d&$32$&$D_3$& $(1/3,2/3,1/2)$ & Same as 1b  & Same as 1b\\ 
1e&$32$&$D_3$& $(2/3,1/3,0)$ & Same as 1a  & Same as 1a\\ 
1f&$32$&$D_3$& $(2/3,1/3,1/2)$ & Same as 1b  & Same as 1b\\ 
\hline
\hline 
 \end{tabular} 
 \end{center}

\subsection*{No. 150: $P321$}\label{subsub:sg150}

This group is generated by three translations $T_{1,2,3}$ as given in Eqs.~\eqref{TransBravaisP}, a three-fold rotation $C_3$, and a two-fold rotation $C'_2$:
\begin{subequations}
 \begin{align}
C_3 &\colon (x,y,z)\rightarrow (-y, x - y, z),\\ 
C'_2 &\colon (x,y,z)\rightarrow (y, x, -z).
\end{align}
\end{subequations}

The $\mathbb{Z}_2$ cohomology ring is given by

\begin{equation}
\mathbb{Z}_2[A_{c'},A_z,B_{xy}]/\langle\mathcal{R}_2,\mathcal{R}_4\rangle
 \end{equation}
where the relations are 
\begin{subequations} 
 \begin{align}
\mathcal{R}_2\colon & ~~
A_z (A_{c'} + A_z),\\
\mathcal{R}_4\colon & ~~
B_{xy}^2.
\end{align} 
 \end{subequations}
We have the following table regarding IWPs and group cohomology at degree 3.
\begin{center}
\begin{tabular}{c|cc|c|c|c}\hline\hline {Wyckoff}&\multicolumn{2}{c|}{Little group}& \multirow{2}{*}{Coordinates}&\multirow{2}{*}{LSM anomaly class}&\multirow{2}{*}{Topo. inv.} \\ \cline{2-3} position & Intl. & Sch\"{o}nflies & & & \\ \hline
1a&$32$&$D_3$& $(0,0,0)$ & $(A_{c'} + A_z) B_{xy}$ & $\varphi_2[T_1T_2, C'_2]$\\ 
1b&$32$&$D_3$& $(0,0,1/2)$ & $A_z B_{xy}$ & $\varphi_2[T_1T_2, T_3C'_2]$\\ 
2d&$3$&$C_3$& $(1/3,2/3,z)$, $(2/3,1/3,-z)$ & N/A & N/A\\ 
\hline
\hline 
 \end{tabular} 
 \end{center}

\subsection*{No. 151: $P3_112$}\label{subsub:sg151}

This group is generated by three translations $T_{1,2,3}$ as given in Eqs.~\eqref{TransBravaisP}, a three-fold screw $S_3$, and a two-fold rotation $C'_2$:
\begin{subequations}
 \begin{align}
S_3 &\colon (x,y,z)\rightarrow (-y, x - y, z + 1/3),\\ 
C'_2 &\colon (x,y,z)\rightarrow (-y, -x, -z + 2/3).
\end{align}
\end{subequations}

The $\mathbb{Z}_2$ cohomology ring is given by

\begin{equation}
\mathbb{Z}_2[A_{c'},A_z,B_{xy}]/\langle\mathcal{R}_2,\mathcal{R}_4\rangle
 \end{equation}
where the relations are 
\begin{subequations} 
 \begin{align}
\mathcal{R}_2\colon & ~~
A_z (A_{c'} + A_z),\\
\mathcal{R}_4\colon & ~~
B_{xy}^2.
\end{align} 
 \end{subequations}
We have the following table regarding IWPs and group cohomology at degree 3.
\begin{center}
\begin{tabular}{c|cc|c|c|c}\hline\hline {Wyckoff}&\multicolumn{2}{c|}{Little group}& \multirow{2}{*}{Coordinates}&\multirow{2}{*}{LSM anomaly class}&\multirow{2}{*}{Topo. inv.} \\ \cline{2-3} position & Intl. & Sch\"{o}nflies & & & \\ \hline
{3a} & {$2$} & {$C_2$} & $(x,-x,1/3)$, $(x,2x,2/3)$, $(-2x,-x,0)$ & {$(A_{c'} + A_z) B_{xy}$} & {$\varphi_2[T_1T_2^{-1}, C'_2]$}\\
{3b} & {$2$} & {$C_2$} & $(x,-x,5/6)$, $(x,2x,1/6)$, $(-2x,-x,1/2)$ & {$A_z B_{xy}$} & {$\varphi_2[T_1T_2^{-1}, T_3C'_2]$}\\
\hline
\hline 
 \end{tabular} 
 \end{center}

\subsection*{No. 152: $P3_121$}\label{subsub:sg152}

This group is generated by three translations $T_{1,2,3}$ as given in Eqs.~\eqref{TransBravaisP}, a three-fold screw $S_3$, and a two-fold rotation $C'_2$:
\begin{subequations}
 \begin{align}
S_3 &\colon (x,y,z)\rightarrow (-y, x - y, z + 1/3),\\ 
C'_2 &\colon (x,y,z)\rightarrow (y, x, -z).
\end{align}
\end{subequations}

The $\mathbb{Z}_2$ cohomology ring is given by

\begin{equation}
\mathbb{Z}_2[A_{c'},A_z,B_{xy}]/\langle\mathcal{R}_2,\mathcal{R}_4\rangle
 \end{equation}
where the relations are 
\begin{subequations} 
 \begin{align}
\mathcal{R}_2\colon & ~~
A_z (A_{c'} + A_z),\\
\mathcal{R}_4\colon & ~~
B_{xy}^2.
\end{align} 
 \end{subequations}
We have the following table regarding IWPs and group cohomology at degree 3.
\begin{center}
\begin{tabular}{c|cc|c|c|c}\hline\hline {Wyckoff}&\multicolumn{2}{c|}{Little group}& \multirow{2}{*}{Coordinates}&\multirow{2}{*}{LSM anomaly class}&\multirow{2}{*}{Topo. inv.} \\ \cline{2-3} position & Intl. & Sch\"{o}nflies & & & \\ \hline
{3a} & {$2$} & {$C_2$} & $(x,0,1/3)$, $(0,x,2/3)$, $(-x,-x,0)$ & {$(A_{c'} + A_z) B_{xy}$} & {$\varphi_2[T_1T_2, C'_2]$}\\
{3b} & {$2$} & {$C_2$} & $(x,0,5/6)$, $(0,x,1/6)$, $(-x,-x,1/2)$ & {$A_z B_{xy}$} & {$\varphi_2[T_1T_2, T_3C'_2]$}\\
\hline
\hline 
 \end{tabular} 
 \end{center}

\subsection*{No. 153: $P3_212$}\label{subsub:sg153}

This group is generated by three translations $T_{1,2,3}$ as given in Eqs.~\eqref{TransBravaisP}, a three-fold screw $S_3$, and a two-fold rotation $C'_2$:
\begin{subequations}
 \begin{align}
S_3 &\colon (x,y,z)\rightarrow (-y, x - y, z + 2/3),\\ 
C'_2 &\colon (x,y,z)\rightarrow (-y, -x, -z + 1/3).
\end{align}
\end{subequations}

The $\mathbb{Z}_2$ cohomology ring is given by

\begin{equation}
\mathbb{Z}_2[A_{c'},A_z,B_{xy}]/\langle\mathcal{R}_2,\mathcal{R}_4\rangle
 \end{equation}
where the relations are 
\begin{subequations} 
 \begin{align}
\mathcal{R}_2\colon & ~~
A_z (A_{c'} + A_z),\\
\mathcal{R}_4\colon & ~~
B_{xy}^2.
\end{align} 
 \end{subequations}
We have the following table regarding IWPs and group cohomology at degree 3.
\begin{center}
\begin{tabular}{c|cc|c|c|c}\hline\hline {Wyckoff}&\multicolumn{2}{c|}{Little group}& \multirow{2}{*}{Coordinates}&\multirow{2}{*}{LSM anomaly class}&\multirow{2}{*}{Topo. inv.} \\ \cline{2-3} position & Intl. & Sch\"{o}nflies & & & \\ \hline
{3a} & {$2$} & {$C_2$} & $(x,-x,2/3)$, $(x,2x,1/3)$, $(-2x,-x,0)$ & {$A_z B_{xy}$} & {$\varphi_2[T_1T_2^{-1}, T_3C'_2]$}\\
{3b} & {$2$} & {$C_2$} & $(x,-x,1/6)$, $(x,2x,5/6)$, $(-2x,-x,1/2)$ & {$(A_{c'} + A_z) B_{xy}$} & {$\varphi_2[T_1T_2^{-1}, C'_2]$}\\
\hline
\hline 
 \end{tabular} 
 \end{center}

\subsection*{No. 154: $P3_221$}\label{subsub:sg154}

This group is generated by three translations $T_{1,2,3}$ as given in Eqs.~\eqref{TransBravaisP}, a three-fold screw $S_3$, and a two-fold rotation $C'_2$:
\begin{subequations}
 \begin{align}
S_3 &\colon (x,y,z)\rightarrow (-y, x - y, z + 2/3),\\ 
C'_2 &\colon (x,y,z)\rightarrow (y, x, -z).
\end{align}
\end{subequations}

The $\mathbb{Z}_2$ cohomology ring is given by

\begin{equation}
\mathbb{Z}_2[A_{c'},A_z,B_{xy}]/\langle\mathcal{R}_2,\mathcal{R}_4\rangle
 \end{equation}
where the relations are 
\begin{subequations} 
 \begin{align}
\mathcal{R}_2\colon & ~~
A_z (A_{c'} + A_z),\\
\mathcal{R}_4\colon & ~~
B_{xy}^2.
\end{align} 
 \end{subequations}
We have the following table regarding IWPs and group cohomology at degree 3.
\begin{center}
\begin{tabular}{c|cc|c|c|c}\hline\hline {Wyckoff}&\multicolumn{2}{c|}{Little group}& \multirow{2}{*}{Coordinates}&\multirow{2}{*}{LSM anomaly class}&\multirow{2}{*}{Topo. inv.} \\ \cline{2-3} position & Intl. & Sch\"{o}nflies & & & \\ \hline
{3a} & {$2$} & {$C_2$} & $(x,0,2/3)$, $(0,x,1/3)$, $(-x,-x,0)$ & {$(A_{c'} + A_z) B_{xy}$} & {$\varphi_2[T_1T_2, C'_2]$}\\
{3b} & {$2$} & {$C_2$} & $(x,0,1/6)$, $(0,x,5/6)$, $(-x,-x,1/2)$ & {$A_z B_{xy}$} & {$\varphi_2[T_1T_2, T_3C'_2]$}\\
\hline
\hline 
 \end{tabular} 
 \end{center}

\subsection*{No. 155: $R32$}\label{subsub:sg155}

This group is generated by three translations $T_{1,2,3}$ as given in Eqs.~\eqref{TransBravaisR}, a three-fold rotation $C_3$, and a two-fold rotation $C'_2$:
\begin{subequations}
 \begin{align}
C_3 &\colon (x,y,z)\rightarrow (-y, x - y, z),\\ 
C'_2 &\colon (x,y,z)\rightarrow (y, x, -z).
\end{align}
\end{subequations}

The $\mathbb{Z}_2$ cohomology ring is given by

\begin{equation}
\mathbb{Z}_2[A_{c'},A_z,B_{x(y+z)}]/\langle\mathcal{R}_2,\mathcal{R}_4\rangle
 \end{equation}
where the relations are 
\begin{subequations} 
 \begin{align}
\mathcal{R}_2\colon & ~~
A_z (A_{c'} + A_z),\\
\mathcal{R}_4\colon & ~~
B_{x(y+z)}^2.
\end{align} 
 \end{subequations}
We have the following table regarding IWPs and group cohomology at degree 3.
\begin{center}
\resizebox{\columnwidth}{!}{
\begin{tabular}{c|cc|c|c|c}\hline\hline {Wyckoff}&\multicolumn{2}{c|}{Little group}& {Coordinates}&\multirow{2}{*}{LSM anomaly class}&\multirow{2}{*}{Topo. inv.} \\ \cline{2-4} position & Intl. & Sch\"{o}nflies & $(0,0,0) + ~(2/3,1/3,1/3) + ~(1/3,2/3,2/3) +$ & &\\ \hline
3a&$32$&$D_3$& $(0,0,0)$ & $(A_{c'} + A_z) B_{x(y+z)}$ & $\varphi_2[T_1T_2, C'_2]$\\ 
3b&$32$&$D_3$& $(0,0,1/2)$ & $A_z B_{x(y+z)}$ & $\varphi_2[T_1T_2, T_1^{-2}T_2^{-1}T_3^{3}C'_2]$\\ 
\hline
\hline 
 \end{tabular} }
 \end{center}

\subsection*{No. 156: $P3m1$}\label{subsub:sg156}

This group is generated by three translations $T_{1,2,3}$ as given in Eqs.~\eqref{TransBravaisP}, a three-fold rotation $C_3$, and a mirror $M$:
\begin{subequations}
 \begin{align}
C_3 &\colon (x,y,z)\rightarrow (-y, x - y, z),\\ 
M &\colon (x,y,z)\rightarrow (-y, -x, z).
\end{align}
\end{subequations}

The $\mathbb{Z}_2$ cohomology ring is given by

\begin{equation}
\mathbb{Z}_2[A_m,A_z,B_{xy}]/\langle\mathcal{R}_2,\mathcal{R}_4\rangle
 \end{equation}
where the relations are 
\begin{subequations} 
 \begin{align}
\mathcal{R}_2\colon & ~~
A_z^2,\\
\mathcal{R}_4\colon & ~~
B_{xy}^2.
\end{align} 
 \end{subequations}
We have the following table regarding IWPs and group cohomology at degree 3.
\begin{center}
\begin{tabular}{c|cc|c|c|c}\hline\hline {Wyckoff}&\multicolumn{2}{c|}{Little group}& \multirow{2}{*}{Coordinates}&\multirow{2}{*}{LSM anomaly class}&\multirow{2}{*}{Topo. inv.} \\ \cline{2-3} position & Intl. & Sch\"{o}nflies & & & \\ \hline
1a&$3m$&$C_{3v}$& $(0,0,z)$ & $A_z B_{xy}$ & $\varphi_3[T_1T_2^{-1}, T_3, M]$\\ 
1b&$3m$&$C_{3v}$& $(1/3,2/3,z)$ & Same as 1a  & Same as 1a\\ 
1c&$3m$&$C_{3v}$& $(2/3,1/3,z)$ & Same as 1a  & Same as 1a\\ 
\hline
\hline 
 \end{tabular} 
 \end{center}

\subsection*{No. 157: $P31m$}\label{subsub:sg157}

This group is generated by three translations $T_{1,2,3}$ as given in Eqs.~\eqref{TransBravaisP}, a three-fold rotation $C_3$, and a mirror $M$:
\begin{subequations}
 \begin{align}
C_3 &\colon (x,y,z)\rightarrow (-y, x - y, z),\\ 
M &\colon (x,y,z)\rightarrow (y, x, z).
\end{align}
\end{subequations}

The $\mathbb{Z}_2$ cohomology ring is given by

\begin{equation}
\mathbb{Z}_2[A_m,A_z,B_{xy}]/\langle\mathcal{R}_2,\mathcal{R}_4\rangle
 \end{equation}
where the relations are 
\begin{subequations} 
 \begin{align}
\mathcal{R}_2\colon & ~~
A_z^2,\\
\mathcal{R}_4\colon & ~~
B_{xy}^2.
\end{align} 
 \end{subequations}
We have the following table regarding IWPs and group cohomology at degree 3.
\begin{center}
\begin{tabular}{c|cc|c|c|c}\hline\hline {Wyckoff}&\multicolumn{2}{c|}{Little group}& \multirow{2}{*}{Coordinates}&\multirow{2}{*}{LSM anomaly class}&\multirow{2}{*}{Topo. inv.} \\ \cline{2-3} position & Intl. & Sch\"{o}nflies & & & \\ \hline
1a&$3m$&$C_{3v}$& $(0,0,z)$ & $A_z B_{xy}$ & $\varphi_3[T_1T_2, T_3, M]$\\ 
2b&$3$&$C_3$& $(1/3,2/3,z)$, $(2/3,1/3,z)$ & N/A & N/A\\ 
\hline
\hline 
 \end{tabular} 
 \end{center}

\subsection*{No. 158: $P3c1$}\label{subsub:sg158}

This group is generated by three translations $T_{1,2,3}$ as given in Eqs.~\eqref{TransBravaisP}, a three-fold rotation $C_3$, and a glide $G$:
\begin{subequations}
 \begin{align}
C_3 &\colon (x,y,z)\rightarrow (-y, x - y, z),\\ 
G &\colon (x,y,z)\rightarrow (-y, -x, z + 1/2).
\end{align}
\end{subequations}

The $\mathbb{Z}_2$ cohomology ring is given by

\begin{equation}
\mathbb{Z}_2[A_m,B_{xy}]/\langle\mathcal{R}_2,\mathcal{R}_4\rangle
 \end{equation}
where the relations are 
\begin{subequations} 
 \begin{align}
\mathcal{R}_2\colon & ~~
A_m^2,\\
\mathcal{R}_4\colon & ~~
B_{xy}^2.
\end{align} 
 \end{subequations}
We have the following table regarding IWPs and group cohomology at degree 3.
\begin{center}
\begin{tabular}{c|cc|c|c|c}\hline\hline {Wyckoff}&\multicolumn{2}{c|}{Little group}& \multirow{2}{*}{Coordinates}&\multirow{2}{*}{LSM anomaly class}&\multirow{2}{*}{Topo. inv.} \\ \cline{2-3} position & Intl. & Sch\"{o}nflies & & & \\ \hline
2a&$3$&$C_3$& $(0,0,z)$, $(0,0,z+1/2)$ & $A_m B_{xy}$ & $\widehat{\varphi_3}[T_1, T_2, G]$\\ 
2b&$3$&$C_3$& $(1/3,2/3,z)$, $(1/3,2/3,z+1/2)$ & Same as 2a  & Same as 2a\\ 
2c&$3$&$C_3$& $(2/3,1/3,z)$, $(2/3,1/3,z+1/2)$ & Same as 2a  & Same as 2a\\ 
\hline
\hline 
 \end{tabular} 
 \end{center}
Here the topological invariant $\widehat{\varphi_3}[T_1, T_2, G]$ can be chosen to be the same as that of group No.9 $Cc$, given by Eq.~\eqref{TI_9}.

\subsection*{No. 159: $P31c$}\label{subsub:sg159}

This group is generated by three translations $T_{1,2,3}$ as given in Eqs.~\eqref{TransBravaisP}, a three-fold rotation $C_3$, and a glide $G$:
\begin{subequations}
 \begin{align}
C_3 &\colon (x,y,z)\rightarrow (-y, x - y, z),\\ 
G &\colon (x,y,z)\rightarrow (y, x, z + 1/2).
\end{align}
\end{subequations}

The $\mathbb{Z}_2$ cohomology ring is given by

\begin{equation}
\mathbb{Z}_2[A_m,B_{xy}]/\langle\mathcal{R}_2,\mathcal{R}_4\rangle
 \end{equation}
where the relations are 
\begin{subequations} 
 \begin{align}
\mathcal{R}_2\colon & ~~
A_m^2,\\
\mathcal{R}_4\colon & ~~
B_{xy}^2.
\end{align} 
 \end{subequations}
We have the following table regarding IWPs and group cohomology at degree 3.
\begin{center}
\begin{tabular}{c|cc|c|c|c}\hline\hline {Wyckoff}&\multicolumn{2}{c|}{Little group}& \multirow{2}{*}{Coordinates}&\multirow{2}{*}{LSM anomaly class}&\multirow{2}{*}{Topo. inv.} \\ \cline{2-3} position & Intl. & Sch\"{o}nflies & & & \\ \hline
2a&$3$&$C_3$& $(0,0,z)$, $(0,0,z+1/2)$ & $A_m B_{xy}$ & $\widehat{\varphi_3}[T_1, T_2^{-1}, G]$\\ 
2b&$3$&$C_3$& $(1/3,2/3,z)$, $(2/3,1/3,z+1/2)$ & Same as 2a  & Same as 2a\\ 
\hline
\hline 
 \end{tabular} 
 \end{center}
  Here the topological invariant $\widehat{\varphi_3}[T_1, T_2^{-1}, G]$ can be chosen to be the same as that of group No.9 $Cc$, given by Eq.~\eqref{TI_9}.
 
\subsection*{No. 160: $R3m$}\label{subsub:sg160}

This group is generated by three translations $T_{1,2,3}$ as given in Eqs.~\eqref{TransBravaisR}, a three-fold rotation $C_3$, and a mirror $M$:
\begin{subequations}
 \begin{align}
C_3 &\colon (x,y,z)\rightarrow (-y, x - y, z),\\ 
M &\colon (x,y,z)\rightarrow (-y, -x, z).
\end{align}
\end{subequations}

The $\mathbb{Z}_2$ cohomology ring is given by

\begin{equation}
\mathbb{Z}_2[A_m,A_z,B_{x(y+z)}]/\langle\mathcal{R}_2,\mathcal{R}_4\rangle
 \end{equation}
where the relations are 
\begin{subequations} 
 \begin{align}
\mathcal{R}_2\colon & ~~
A_z^2,\\
\mathcal{R}_4\colon & ~~
B_{x(y+z)}^2.
\end{align} 
 \end{subequations}
We have the following table regarding IWPs and group cohomology at degree 3.
\begin{center}
\resizebox{\columnwidth}{!}{
\begin{tabular}{c|cc|c|c|c}\hline\hline {Wyckoff}&\multicolumn{2}{c|}{Little group}& {Coordinates}&\multirow{2}{*}{LSM anomaly class}&\multirow{2}{*}{Topo. inv.} \\ \cline{2-4} position & Intl. & Sch\"{o}nflies & $(0,0,0) + ~(2/3,1/3,1/3) + ~(1/3,2/3,2/3) +$ & &\\ \hline
3a&$3m$&$C_{3v}$& $(0,0,z)$ & $A_z B_{x(y+z)}$ & $\varphi_3[T_1T_2^{-1}, T_1^{-2}T_2^{-1}T_3^{3}, M]$\\ 
\hline
\hline 
 \end{tabular} }
 \end{center}

\subsection*{No. 161: $R3c$}\label{subsub:sg161}

This group is generated by three translations $T_{1,2,3}$ as given in Eqs.~\eqref{TransBravaisR}, a three-fold rotation $C_3$, and a glide $G$:
\begin{subequations}
 \begin{align}
C_3 &\colon (x,y,z)\rightarrow (-y, x - y, z),\\ 
G &\colon (x,y,z)\rightarrow (-y + 1/3, -x + 2/3, z + 1/6).
\end{align}
\end{subequations}

The $\mathbb{Z}_2$ cohomology ring is given by

\begin{equation}
\mathbb{Z}_2[A_m,B_{x(y+z)}]/\langle\mathcal{R}_2,\mathcal{R}_4\rangle
 \end{equation}
where the relations are 
\begin{subequations} 
 \begin{align}
\mathcal{R}_2\colon & ~~
A_m^2,\\
\mathcal{R}_4\colon & ~~
B_{x(y+z)}^2.
\end{align} 
 \end{subequations}
We have the following table regarding IWPs and group cohomology at degree 3.
\begin{center}
\begin{tabular}{c|cc|c|c|c}\hline\hline {Wyckoff}&\multicolumn{2}{c|}{Little group}& {Coordinates}&\multirow{2}{*}{LSM anomaly class}&\multirow{2}{*}{Topo. inv.} \\ \cline{2-4} position & Intl. & Sch\"{o}nflies & $(0,0,0) + ~(2/3,1/3,1/3) + ~(1/3,2/3,2/3) +$ & &\\ \hline
6a&$3$&$C_3$& $(0,0,z)$, $(0,0,z+1/2)$ & $A_m B_{x(y+z)}$ & $\widehat{\varphi_3}[T_1, T_2, G]$\\ 
\hline
\hline 
 \end{tabular} 
 \end{center}
 Here the topological invariant $\widehat{\varphi_3}[T_1, T_2, G]$ can be chosen to be the same as that of group No.9 $Cc$, given by Eq.~\eqref{TI_9}.

\subsection*{No. 162: $P\overline31m$}\label{subsub:sg162}

This group is generated by three translations $T_{1,2,3}$ as given in Eqs.~\eqref{TransBravaisP}, a three-fold rotation $C_3$, a two-fold rotation $C'_2$, and an inversion $I$:
\begin{subequations}
 \begin{align}
C_3 &\colon (x,y,z)\rightarrow (-y, x - y, z),\\ 
C'_2 &\colon (x,y,z)\rightarrow (-y, -x, -z),\\ 
I &\colon (x,y,z)\rightarrow (-x, -y, -z).
\end{align}
\end{subequations}

The $\mathbb{Z}_2$ cohomology ring is given by

\begin{equation}
\mathbb{Z}_2[A_i,A_{c'},A_z,B_{xy}]/\langle\mathcal{R}_2,\mathcal{R}_4\rangle
 \end{equation}
where the relations are 
\begin{subequations} 
 \begin{align}
\mathcal{R}_2\colon & ~~
A_z (A_{c'} + A_i + A_z),\\
\mathcal{R}_4\colon & ~~
B_{xy} (A_{c'} A_i + A_i^2 + B_{xy}).
\end{align} 
 \end{subequations}
We have the following table regarding IWPs and group cohomology at degree 3.
\begin{center}
\begin{tabular}{c|cc|c|c|c}\hline\hline {Wyckoff}&\multicolumn{2}{c|}{Little group}& \multirow{2}{*}{Coordinates}&\multirow{2}{*}{LSM anomaly class}&\multirow{2}{*}{Topo. inv.} \\ \cline{2-3} position & Intl. & Sch\"{o}nflies & & & \\ \hline
1a&$\overline{3}m$&$D_{3d}$& $(0,0,0)$ & $(A_{c'} + A_i + A_z) (A_{c'} A_i + A_i^2 + B_{xy})$ & $\varphi_1[I]$\\ 
1b&$\overline{3}m$&$D_{3d}$& $(0,0,1/2)$ & $A_z (A_{c'} A_i + A_i^2 + B_{xy})$ & $\varphi_1[T_3I]$\\ 
2c&$32$&$D_3$& $(1/3,2/3,0)$, $(2/3,1/3,0)$ & N/A & N/A\\ 
2d&$32$&$D_3$& $(1/3,2/3,1/2)$, $(2/3,1/3,1/2)$ & N/A & N/A\\ 
\hline
\multirow{2}{*}{3f} & \multirow{2}{*}{$2/m$} & \multirow{2}{*}{$C_{2h}$} & $(1/2,0,0)$, $(0,1/2,0)$, & \multirow{2}{*}{$(A_{c'} + A_i + A_z) B_{xy}$} & \multirow{2}{*}{$\varphi_1[T_1I]$}\\
& & & $(1/2,1/2,0)$ & & \\ \hline
\multirow{2}{*}{3g} & \multirow{2}{*}{$2/m$} & \multirow{2}{*}{$C_{2h}$} & $(1/2,0,1/2)$, $(0,1/2,1/2)$, & \multirow{2}{*}{$A_z B_{xy}$} & \multirow{2}{*}{$\varphi_1[T_1T_3I]$}\\
& & & $(1/2,1/2,1/2)$ & & \\ \hline
\hline 
 \end{tabular} 
 \end{center}

\subsection*{No. 163: $P\overline31c$}\label{subsub:sg163}

This group is generated by three translations $T_{1,2,3}$ as given in Eqs.~\eqref{TransBravaisP}, a three-fold rotation $C_3$, a two-fold rotation $C'_2$, and an inversion $I$:
\begin{subequations}
 \begin{align}
C_3 &\colon (x,y,z)\rightarrow (-y, x - y, z),\\ 
C'_2 &\colon (x,y,z)\rightarrow (-y, -x, -z + 1/2),\\ 
I &\colon (x,y,z)\rightarrow (-x, -y, -z).
\end{align}
\end{subequations}

The $\mathbb{Z}_2$ cohomology ring is given by

\begin{equation}
\mathbb{Z}_2[A_i,A_{c'},B_{xy}]/\langle\mathcal{R}_2,\mathcal{R}_4\rangle
 \end{equation}
where the relations are 
\begin{subequations} 
 \begin{align}
\mathcal{R}_2\colon & ~~
A_{c'} A_i,\\
\mathcal{R}_4\colon & ~~
B_{xy} (A_i^2 + B_{xy}).
\end{align} 
 \end{subequations}
We have the following table regarding IWPs and group cohomology at degree 3.
\begin{center}
\begin{tabular}{c|cc|c|c|c}\hline\hline {Wyckoff}&\multicolumn{2}{c|}{Little group}& \multirow{2}{*}{Coordinates}&\multirow{2}{*}{LSM anomaly class}&\multirow{2}{*}{Topo. inv.} \\ \cline{2-3} position & Intl. & Sch\"{o}nflies & & & \\ \hline
2a&$32$&$D_3$& $(0,0,1/4)$, $(0,0,3/4)$ & $A_{c'} B_{xy}$ & $\varphi_2[T_1T_2^{-1}, C'_2]$\\ 
2b&$\overline{3}$&$C_{3i}$& $(0,0,0)$, $(0,0,1/2)$ & $A_i (A_i^2 + B_{xy})$ & $\varphi_1[I]$\\ 
2c&$32$&$D_3$& $(1/3,2/3,1/4)$, $(2/3,1/3,3/4)$ & Same as 2a  & Same as 2a\\ 
2d&$32$&$D_3$& $(2/3,1/3,1/4)$, $(1/3,2/3,3/4)$ & Same as 2b  & Same as 2b\\ 
\hline
\multirow{3}{*}{6g} & \multirow{3}{*}{$\overline{1}$} & \multirow{3}{*}{$C_i$} & $(1/2,0,0)$, $(0,1/2,0)$, & \multirow{3}{*}{$A_i B_{xy}$} & \multirow{3}{*}{$\varphi_1[T_1I]$}\\
& & & $(1/2,1/2,0)$, $(0,1/2,1/2)$, & & \\
& & & $(1/2,0,1/2)$, $(1/2,1/2,1/2)$ & & \\ \hline
\hline 
 \end{tabular} 
 \end{center}

\subsection*{No. 164: $P\overline3m1$}\label{subsub:sg164}

This group is generated by three translations $T_{1,2,3}$ as given in Eqs.~\eqref{TransBravaisP}, a three-fold rotation $C_3$, a two-fold rotation $C'_2$, and an inversion $I$:
\begin{subequations}
 \begin{align}
C_3 &\colon (x,y,z)\rightarrow (-y, x - y, z),\\ 
C'_2 &\colon (x,y,z)\rightarrow (y, x, -z),\\ 
I &\colon (x,y,z)\rightarrow (-x, -y, -z).
\end{align}
\end{subequations}

The $\mathbb{Z}_2$ cohomology ring is given by

\begin{equation}
\mathbb{Z}_2[A_i,A_{c'},A_z,B_{xy}]/\langle\mathcal{R}_2,\mathcal{R}_4\rangle
 \end{equation}
where the relations are 
\begin{subequations} 
 \begin{align}
\mathcal{R}_2\colon & ~~
A_z (A_{c'} + A_i + A_z),\\
\mathcal{R}_4\colon & ~~
B_{xy} (A_{c'} A_i + A_i^2 + B_{xy}).
\end{align} 
 \end{subequations}
We have the following table regarding IWPs and group cohomology at degree 3.
\begin{center}
\begin{tabular}{c|cc|c|c|c}\hline\hline {Wyckoff}&\multicolumn{2}{c|}{Little group}& \multirow{2}{*}{Coordinates}&\multirow{2}{*}{LSM anomaly class}&\multirow{2}{*}{Topo. inv.} \\ \cline{2-3} position & Intl. & Sch\"{o}nflies & & & \\ \hline
1a&$\overline{3}m$&$D_{3d}$& $(0,0,0)$ & $(A_{c'} + A_i + A_z) (A_{c'} A_i + A_i^2 + B_{xy})$ & $\varphi_1[I]$\\ 
1b&$\overline{3}m$&$D_{3d}$& $(0,0,1/2)$ & $A_z (A_{c'} A_i + A_i^2 + B_{xy})$ & $\varphi_1[T_3I]$\\ 
2d&$3m$&$C_{3v}$& $(1/3,2/3,z)$, $(2/3,1/3,-z)$ & N/A & N/A\\ 
\hline
\multirow{2}{*}{3e} & \multirow{2}{*}{$2m$} & \multirow{2}{*}{$C_{2h}$} & $(1/2,0,0)$, $(0,1/2,0)$, & \multirow{2}{*}{$(A_{c'} + A_i + A_z) B_{xy}$} & \multirow{2}{*}{$\varphi_1[T_1I]$}\\
& & & $(1/2,1/2,0)$ & & \\ \hline
\multirow{2}{*}{3f} & \multirow{2}{*}{$2m$} & \multirow{2}{*}{$C_{2h}$} & $(1/2,0,1/2)$, $(0,1/2,1/2)$, & \multirow{2}{*}{$A_z B_{xy}$} & \multirow{2}{*}{$\varphi_1[T_1T_3I]$}\\
& & & $(1/2,1/2,1/2)$ & & \\ \hline
\hline 
 \end{tabular} 
 \end{center}

\subsection*{No. 165: $P\overline3c1$}\label{subsub:sg165}

This group is generated by three translations $T_{1,2,3}$ as given in Eqs.~\eqref{TransBravaisP}, a three-fold rotation $C_3$, a two-fold rotation $C'_2$, and an inversion $I$:
\begin{subequations}
 \begin{align}
C_3 &\colon (x,y,z)\rightarrow (-y, x - y, z),\\ 
C'_2 &\colon (x,y,z)\rightarrow (y, x, -z + 1/2),\\ 
I &\colon (x,y,z)\rightarrow (-x, -y, -z).
\end{align}
\end{subequations}

The $\mathbb{Z}_2$ cohomology ring is given by

\begin{equation}
\mathbb{Z}_2[A_i,A_{c'},B_{xy}]/\langle\mathcal{R}_2,\mathcal{R}_4\rangle
 \end{equation}
where the relations are 
\begin{subequations} 
 \begin{align}
\mathcal{R}_2\colon & ~~
A_{c'} A_i,\\
\mathcal{R}_4\colon & ~~
B_{xy} (A_i^2 + B_{xy}).
\end{align} 
 \end{subequations}
We have the following table regarding IWPs and group cohomology at degree 3.
\begin{center}
\begin{tabular}{c|cc|c|c|c}\hline\hline {Wyckoff}&\multicolumn{2}{c|}{Little group}& \multirow{2}{*}{Coordinates}&\multirow{2}{*}{LSM anomaly class}&\multirow{2}{*}{Topo. inv.} \\ \cline{2-3} position & Intl. & Sch\"{o}nflies & & & \\ \hline
2a&$32$&$D_3$& $(0,0,1/4)$, $(0,0,3/4)$ & $A_{c'} B_{xy}$ & $\varphi_2[T_1T_2, C'_2]$\\ 
2b&$\overline{3}$&$C_{3i}$& $(0,0,0)$, $(0,0,1/2)$ & $A_i (A_i^2 + B_{xy})$ & $\varphi_1[I]$\\ 
\hline
\multirow{2}{*}{4d} & \multirow{2}{*}{$3$} & \multirow{2}{*}{$C_3$} & $(1/3,2/3,z)$, $(2/3,1/3,-z+1/2)$, & \multirow{2}{*}{N/A} & \multirow{2}{*}{N/A}\\
& & & $(2/3,1/3,-z)$, $(1/3,2/3,z+1/2)$ & & \\ \hline
\multirow{3}{*}{6e} & \multirow{3}{*}{$\overline{1}$} & \multirow{3}{*}{$C_i$} & $(1/2,0,0)$, $(0,1/2,0)$, & \multirow{3}{*}{$A_i B_{xy}$} & \multirow{3}{*}{$\varphi_1[T_1I]$}\\
& & & $(1/2,1/2,0)$, $(0,1/2,1/2)$, & & \\
& & & $(1/2,0,1/2)$, $(1/2,1/2,1/2)$ & & \\ \hline
\hline 
 \end{tabular} 
 \end{center}

\subsection*{No. 166: $R\overline3m$}\label{subsub:sg166}

This group is generated by three translations $T_{1,2,3}$ as given in Eqs.~\eqref{TransBravaisR}, a three-fold rotation $C_3$, a two-fold rotation $C'_2$, and an inversion $I$:
\begin{subequations}
 \begin{align}
C_3 &\colon (x,y,z)\rightarrow (-y, x - y, z),\\ 
C'_2 &\colon (x,y,z)\rightarrow (y, x, -z),\\ 
I &\colon (x,y,z)\rightarrow (-x, -y, -z).
\end{align}
\end{subequations}

The $\mathbb{Z}_2$ cohomology ring is given by

\begin{equation}
\mathbb{Z}_2[A_i,A_{c'},A_z,B_{x(y+z)}]/\langle\mathcal{R}_2,\mathcal{R}_4\rangle
 \end{equation}
where the relations are 
\begin{subequations} 
 \begin{align}
\mathcal{R}_2\colon & ~~
A_z (A_{c'} + A_i + A_z),\\
\mathcal{R}_4\colon & ~~
B_{x(y+z)} (A_{c'} A_i + A_i^2 + B_{x(y+z)}).
\end{align} 
 \end{subequations}
We have the following table regarding IWPs and group cohomology at degree 3.
\begin{center}
\resizebox{\columnwidth}{!}{
\begin{tabular}{c|cc|c|c|c}\hline\hline {Wyckoff}&\multicolumn{2}{c|}{Little group}& {Coordinates}&\multirow{2}{*}{LSM anomaly class}&\multirow{2}{*}{Topo. inv.} \\ \cline{2-4} position & Intl. & Sch\"{o}nflies & $(0,0,0) + ~(2/3,1/3,1/3) + ~(1/3,2/3,2/3) +$ & &\\ \hline
3a&$\overline{3}m$&$D_{3d}$& $(0,0,0)$ & $(A_{c'} + A_i + A_z) (A_{c'} A_i + A_i^2 + B_{x(y+z)})$ & $\varphi_1[I]$\\ 
3b&$\overline{3}m$&$D_{3d}$& $(0,0,1/2)$ & $A_z B_{x(y+z)}$ & $\varphi_1[T_1^{-2}T_2^{-1}T_3^{3}I]$\\ 
{9d} & {$2/m$} & {$C_{2h}$} & $(1/2,0,1/2)$, $(0,1/2,1/2)$, $(1/2,1/2,1/2)$ & {$A_z (A_{c'} A_i + A_i^2 + B_{x(y+z)})$} & {$\varphi_1[T_1^{-1}T_2^{-1}T_3^{3}I]$}\\
{9e} & {*}{$2/m$} & {$C_{2h}$} & $(1/2,0,0)$, $(0,1/2,0)$, $(1/2,1/2,0)$ & {$(A_{c'} + A_i + A_z) B_{x(y+z)}$} & {$\varphi_1[T_1I]$}\\
\hline
\hline 
 \end{tabular} }
 \end{center}

\subsection*{No. 167: $R\overline3c$}\label{subsub:sg167}

This group is generated by three translations $T_{1,2,3}$ as given in Eqs.~\eqref{TransBravaisR}, a three-fold rotation $C_3$, a two-fold screw $S'_2$, and an inversion $I$:
\begin{subequations}
 \begin{align}
C_3 &\colon (x,y,z)\rightarrow (-y, x - y, z),\\ 
S'_2 &\colon (x,y,z)\rightarrow (y + 1/3, x + 2/3, -z + 1/6),\\ 
I &\colon (x,y,z)\rightarrow (-x, -y, -z).
\end{align}
\end{subequations}

The $\mathbb{Z}_2$ cohomology ring is given by

\begin{equation}
\mathbb{Z}_2[A_i,A_{c'},B_{x(y+z)}]/\langle\mathcal{R}_2,\mathcal{R}_4\rangle
 \end{equation}
where the relations are 
\begin{subequations} 
 \begin{align}
\mathcal{R}_2\colon & ~~
A_{c'} A_i,\\
\mathcal{R}_4\colon & ~~
B_{x(y+z)} (A_i^2 + B_{x(y+z)}).
\end{align} 
 \end{subequations}
We have the following table regarding IWPs and group cohomology at degree 3.
\begin{center}
\resizebox{\columnwidth}{!}{
\begin{tabular}{c|cc|c|c|c}\hline\hline {Wyckoff}&\multicolumn{2}{c|}{Little group}& {Coordinates}&\multirow{2}{*}{LSM anomaly class}&\multirow{2}{*}{Topo. inv.} \\ \cline{2-4} position & Intl. & Sch\"{o}nflies & $(0,0,0) + ~(2/3,1/3,1/3) + ~(1/3,2/3,2/3) +$ & &\\ \hline
6a&$32$&$D_3$& $(0,0,1/4)$, $(0,0,3/4)$ & $A_{c'} B_{x(y+z)}$ & $\varphi_2[T_1T_2, T_1^{-1}T_2^{-1}T_3S'_2]$\\ 
6b&$\overline{3}$&$C_{3i}$& $(0,0,0)$, $(0,0,1/2)$ & $A_i (A_i^2 + B_{x(y+z)})$ & $\varphi_1[I]$\\ 
\hline
\multirow{3}{*}{18d} & \multirow{3}{*}{$\overline{1}$} & \multirow{3}{*}{$C_i$} & $(1/2,0,0)$, $(0,1/2,0)$, & \multirow{3}{*}{$A_i B_{x(y+z)}$} & \multirow{3}{*}{$\varphi_1[T_1I]$}\\
& & & $(1/2,1/2,0)$, $(0,1/2,1/2)$, & & \\
& & & $(1/2,0,1/2)$, $(1/2,1/2,1/2)$ & & \\ \hline
\hline 
 \end{tabular} }
 \end{center}

\subsection*{No. 168: $P6$}\label{subsub:sg168}

This group is generated by three translations $T_{1,2,3}$ as given in Eqs.~\eqref{TransBravaisP}, a three-fold rotation $C_3$, and a two-fold rotation $C_2$:
\begin{subequations}
 \begin{align}
C_3 &\colon (x,y,z)\rightarrow (-y, x - y, z),\\ 
C_2 &\colon (x,y,z)\rightarrow (-x, -y, z).
\end{align}
\end{subequations}

The $\mathbb{Z}_2$ cohomology ring is given by

\begin{equation}
\mathbb{Z}_2[A_c,A_z,B_{xy}]/\langle\mathcal{R}_2,\mathcal{R}_4\rangle
 \end{equation}
where the relations are 
\begin{subequations} 
 \begin{align}
\mathcal{R}_2\colon & ~~
A_z^2,\\
\mathcal{R}_4\colon & ~~
B_{xy} (A_c^2 + B_{xy}).
\end{align} 
 \end{subequations}
We have the following table regarding IWPs and group cohomology at degree 3.
\begin{center}
\begin{tabular}{c|cc|c|c|c}\hline\hline {Wyckoff}&\multicolumn{2}{c|}{Little group}& \multirow{2}{*}{Coordinates}&\multirow{2}{*}{LSM anomaly class}&\multirow{2}{*}{Topo. inv.} \\ \cline{2-3} position & Intl. & Sch\"{o}nflies & & & \\ \hline
1a&$6$&$C_6$& $(0,0,z)$ & $A_z (A_c^2 + B_{xy})$ & $\varphi_2[T_3, C_2]$\\ 
2b&$3$&$C_3$& $(1/3,2/3,z)$, $(2/3,1/3,z)$ & N/A & N/A\\ 
{3c} & {$2$} & {$C_2$} & $(1/2,0,z)$, $(0,1/2,z)$, $(1/2,1/2,z)$ & {$A_z B_{xy}$} & {$\varphi_2[T_3, T_1C_2]$}\\
\hline
\hline 
 \end{tabular} 
 \end{center}

\subsection*{No. 169: $P6_1$}\label{subsub:sg169}

This group is generated by three translations $T_{1,2,3}$ as given in Eqs.~\eqref{TransBravaisP}, a three-fold screw $S_3$, and a two-fold screw $S_2$:
\begin{subequations}
 \begin{align}
S_3 &\colon (x,y,z)\rightarrow (-y, x - y, z + 1/3),\\ 
S_2 &\colon (x,y,z)\rightarrow (-x, -y, z + 1/2).
\end{align}
\end{subequations}

The $\mathbb{Z}_2$ cohomology ring is given by

\begin{equation}
\mathbb{Z}_2[A_c,B_{xy}]/\langle\mathcal{R}_2,\mathcal{R}_4\rangle
 \end{equation}
where the relations are 
\begin{subequations} 
 \begin{align}
\mathcal{R}_2\colon & ~~
A_c^2,\\
\mathcal{R}_4\colon & ~~
B_{xy}^2.
\end{align} 
 \end{subequations}
We have the following table regarding IWPs and group cohomology at degree 3.
\begin{center}
\begin{tabular}{c|cc|c|c|c}\hline\hline {Wyckoff}&\multicolumn{2}{c|}{Little group}& \multirow{2}{*}{Coordinates}&\multirow{2}{*}{LSM anomaly class}&\multirow{2}{*}{Topo. inv.} \\ \cline{2-3} position & Intl. & Sch\"{o}nflies & & & \\ \hline
\multirow{3}{*}{6a} & \multirow{3}{*}{$1$} & \multirow{3}{*}{$C_1$} & $(x,y,z)$, $(-y,x-y,z+1/3)$, & \multirow{3}{*}{$A_c B_{xy}$} & \multirow{3}{*}{$\widehat{\varphi_3}[T_1, T_2, S_2]$}\\
& & & $(-x+y,-x,z+2/3)$, $(-x,-y,z+1/2)$, & & \\
& & & $(y,-x+y,z+5/6)$, $(x-y,x,z+1/6)$ & & \\ \hline
\hline 
 \end{tabular} 
 \end{center}
 Here the topological invariant $\widehat{\varphi_3}[T_1, T_2, S_2]$ can be chosen to be the same as that of group No.4 $P2_1$, given by Eq.~\eqref{TI_4}.

\subsection*{No. 170: $P6_5$}\label{subsub:sg170}

This group is generated by three translations $T_{1,2,3}$ as given in Eqs.~\eqref{TransBravaisP}, a three-fold screw $S_3$, and a two-fold screw $S_2$:
\begin{subequations}
 \begin{align}
S_3 &\colon (x,y,z)\rightarrow (-y, x - y, z + 2/3),\\ 
S_2 &\colon (x,y,z)\rightarrow (-x, -y, z + 1/2).
\end{align}
\end{subequations}

The $\mathbb{Z}_2$ cohomology ring is given by

\begin{equation}
\mathbb{Z}_2[A_c,B_{xy}]/\langle\mathcal{R}_2,\mathcal{R}_4\rangle
 \end{equation}
where the relations are 
\begin{subequations} 
 \begin{align}
\mathcal{R}_2\colon & ~~
A_c^2,\\
\mathcal{R}_4\colon & ~~
B_{xy}^2.
\end{align} 
 \end{subequations}
We have the following table regarding IWPs and group cohomology at degree 3.
\begin{center}
\begin{tabular}{c|cc|c|c|c}\hline\hline {Wyckoff}&\multicolumn{2}{c|}{Little group}& \multirow{2}{*}{Coordinates}&\multirow{2}{*}{LSM anomaly class}&\multirow{2}{*}{Topo. inv.} \\ \cline{2-3} position & Intl. & Sch\"{o}nflies & & & \\ \hline
\multirow{3}{*}{6a} & \multirow{3}{*}{$1$} & \multirow{3}{*}{$C_1$} & $(x,y,z)$, $(-y,x-y,z+2/3)$, & \multirow{3}{*}{$A_c B_{xy}$} & \multirow{3}{*}{$\widehat{\varphi_3}[T_1, T_2, S_2]$}\\
& & & $(-x+y,-x,z+1/3)$, $(-x,-y,z+1/2)$, & & \\
& & & $(y,-x+y,z+1/6)$, $(x-y,x,z+5/6)$ & & \\ \hline
\hline 
 \end{tabular} 
 \end{center}
  Here the topological invariant $\widehat{\varphi_3}[T_1, T_2, S_2]$ can be chosen to be the same as that of group No.4 $P2_1$, given by Eq.~\eqref{TI_4}.

\subsection*{No. 171: $P6_2$}\label{subsub:sg171}

This group is generated by three translations $T_{1,2,3}$ as given in Eqs.~\eqref{TransBravaisP}, a three-fold screw $S_3$, and a two-fold rotation $C_2$:
\begin{subequations}
 \begin{align}
S_3 &\colon (x,y,z)\rightarrow (-y, x - y, z + 2/3),\\ 
C_2 &\colon (x,y,z)\rightarrow (-x, -y, z).
\end{align}
\end{subequations}

The $\mathbb{Z}_2$ cohomology ring is given by

\begin{equation}
\mathbb{Z}_2[A_c,A_z,B_{xy}]/\langle\mathcal{R}_2,\mathcal{R}_4\rangle
 \end{equation}
where the relations are 
\begin{subequations} 
 \begin{align}
\mathcal{R}_2\colon & ~~
A_z^2,\\
\mathcal{R}_4\colon & ~~
B_{xy} (A_c^2 + B_{xy}).
\end{align} 
 \end{subequations}
We have the following table regarding IWPs and group cohomology at degree 3.
\begin{center}
\begin{tabular}{c|cc|c|c|c}\hline\hline {Wyckoff}&\multicolumn{2}{c|}{Little group}& \multirow{2}{*}{Coordinates}&\multirow{2}{*}{LSM anomaly class}&\multirow{2}{*}{Topo. inv.} \\ \cline{2-3} position & Intl. & Sch\"{o}nflies & & & \\ \hline
{3a} & {$2$} & {$C_2$} & $(0,0,z)$, $(0,0,z+2/3)$, $(0,0,z+1/3)$ & {$A_z (A_c^2 + B_{xy})$} & {$\varphi_2[T_3, C_2]$}\\
{3b} & {$2$} & {$C_2$} & $(1/2,1/2,z)$, $(1/2,0,z+2/3)$, $(0,1/2,z+1/3)$ & {$A_z B_{xy}$} & {$\varphi_2[T_3, T_1T_2C_2]$}\\
\hline
\hline 
 \end{tabular} 
 \end{center}

\subsection*{No. 172: $P6_4$}\label{subsub:sg172}

This group is generated by three translations $T_{1,2,3}$ as given in Eqs.~\eqref{TransBravaisP}, a three-fold screw $S_3$, and a two-fold rotation $C_2$:
\begin{subequations}
 \begin{align}
S_3 &\colon (x,y,z)\rightarrow (-y, x - y, z + 1/3),\\ 
C_2 &\colon (x,y,z)\rightarrow (-x, -y, z).
\end{align}
\end{subequations}

The $\mathbb{Z}_2$ cohomology ring is given by

\begin{equation}
\mathbb{Z}_2[A_c,A_z,B_{xy}]/\langle\mathcal{R}_2,\mathcal{R}_4\rangle
 \end{equation}
where the relations are 
\begin{subequations} 
 \begin{align}
\mathcal{R}_2\colon & ~~
A_z^2,\\
\mathcal{R}_4\colon & ~~
B_{xy} (A_c^2 + B_{xy}).
\end{align} 
 \end{subequations}
We have the following table regarding IWPs and group cohomology at degree 3.
\begin{center}
\begin{tabular}{c|cc|c|c|c}\hline\hline {Wyckoff}&\multicolumn{2}{c|}{Little group}& \multirow{2}{*}{Coordinates}&\multirow{2}{*}{LSM anomaly class}&\multirow{2}{*}{Topo. inv.} \\ \cline{2-3} position & Intl. & Sch\"{o}nflies & & & \\ \hline
{3a} & {$2$} & {$C_2$} & $(0,0,z)$, $(0,0,z+1/3)$, $(0,0,z+2/3)$ & {$A_z (A_c^2 + B_{xy})$} & {$\varphi_2[T_3, C_2]$}\\
{3b} & {$2$} & {$C_2$} & $(1/2,1/2,z)$, $(1/2,0,z+1/3)$, $(0,1/2,z+2/3)$ & {$A_z B_{xy}$} & {$\varphi_2[T_3, T_1T_2C_2]$}\\
\hline
\hline 
 \end{tabular} 
 \end{center}

\subsection*{No. 173: $P6_3$}\label{subsub:sg173}

This group is generated by three translations $T_{1,2,3}$ as given in Eqs.~\eqref{TransBravaisP}, a three-fold rotation $C_3$, and a two-fold screw $S_2$:
\begin{subequations}
 \begin{align}
C_3 &\colon (x,y,z)\rightarrow (-y, x - y, z),\\ 
S_2 &\colon (x,y,z)\rightarrow (-x, -y, z + 1/2).
\end{align}
\end{subequations}

The $\mathbb{Z}_2$ cohomology ring is given by

\begin{equation}
\mathbb{Z}_2[A_c,B_{xy}]/\langle\mathcal{R}_2,\mathcal{R}_4\rangle
 \end{equation}
where the relations are 
\begin{subequations} 
 \begin{align}
\mathcal{R}_2\colon & ~~
A_c^2,\\
\mathcal{R}_4\colon & ~~
B_{xy}^2.
\end{align} 
 \end{subequations}
We have the following table regarding IWPs and group cohomology at degree 3.
\begin{center}
\begin{tabular}{c|cc|c|c|c}\hline\hline {Wyckoff}&\multicolumn{2}{c|}{Little group}& \multirow{2}{*}{Coordinates}&\multirow{2}{*}{LSM anomaly class}&\multirow{2}{*}{Topo. inv.} \\ \cline{2-3} position & Intl. & Sch\"{o}nflies & & & \\ \hline
2a&$3$&$C_3$& $(0,0,z)$, $(0,0,z+1/2)$ & $A_c B_{xy}$ & $\widehat{\varphi_3}[T_1, T_2, S_2]$\\ 
2b&$3$&$C_3$& $(1/3,2/3,z)$, $(2/3,1/3,z+1/2)$ & Same as 2a  & Same as 2a\\ 
\hline
\hline 
 \end{tabular} 
 \end{center}
   Here the topological invariant $\widehat{\varphi_3}[T_1, T_2, S_2]$ can be chosen to be the same as that of group No.4 $P2_1$, given by Eq.~\eqref{TI_4}.

\subsection*{No. 174: $P\overline6$}\label{subsub:sg174}

This group is generated by three translations $T_{1,2,3}$ as given in Eqs.~\eqref{TransBravaisP}, a three-fold rotation $C_3$, and a mirror $M$:
\begin{subequations}
 \begin{align}
C_3 &\colon (x,y,z)\rightarrow (-y, x - y, z),\\ 
M &\colon (x,y,z)\rightarrow (x, y, -z).
\end{align}
\end{subequations}

The $\mathbb{Z}_2$ cohomology ring is given by

\begin{equation}
\mathbb{Z}_2[A_m,A_z,B_{xy}]/\langle\mathcal{R}_2,\mathcal{R}_4\rangle
 \end{equation}
where the relations are 
\begin{subequations} 
 \begin{align}
\mathcal{R}_2\colon & ~~
A_z (A_m + A_z),\\
\mathcal{R}_4\colon & ~~
B_{xy}^2.
\end{align} 
 \end{subequations}
We have the following table regarding IWPs and group cohomology at degree 3.
\begin{center}
\begin{tabular}{c|cc|c|c|c}\hline\hline {Wyckoff}&\multicolumn{2}{c|}{Little group}& \multirow{2}{*}{Coordinates}&\multirow{2}{*}{LSM anomaly class}&\multirow{2}{*}{Topo. inv.} \\ \cline{2-3} position & Intl. & Sch\"{o}nflies & & & \\ \hline
1a&$\overline{6}$&$C_{3h}$& $(0,0,0)$ & $(A_m + A_z) B_{xy}$ & $\varphi_3[T_1, T_2, M]$\\ 
1b&$\overline{6}$&$C_{3h}$& $(0,0,1/2)$ & $A_z B_{xy}$ & $\varphi_3[T_1, T_2, T_3M]$\\ 
1c&$\overline{6}$&$C_{3h}$& $(1/3,2/3,0)$ & Same as 1a  & Same as 1a\\ 
1d&$\overline{6}$&$C_{3h}$& $(1/3,2/3,1/2)$ & Same as 1b  & Same as 1b\\ 
1e&$\overline{6}$&$C_{3h}$& $(2/3,1/3,0)$ & Same as 1a  & Same as 1a\\ 
1f&$\overline{6}$&$C_{3h}$& $(2/3,1/3,1/2)$ & Same as 1b  & Same as 1b\\ 
\hline
\hline 
 \end{tabular} 
 \end{center}

\subsection*{No. 175: $P6/m$}\label{subsub:sg175}

This group is generated by three translations $T_{1,2,3}$ as given in Eqs.~\eqref{TransBravaisP}, a three-fold rotation $C_3$, a two-fold rotation $C_2$, and an inversion $I$:
\begin{subequations}
 \begin{align}
C_3 &\colon (x,y,z)\rightarrow (-y, x - y, z),\\ 
C_2 &\colon (x,y,z)\rightarrow (-x, -y, z),\\ 
I &\colon (x,y,z)\rightarrow (-x, -y, -z).
\end{align}
\end{subequations}

The $\mathbb{Z}_2$ cohomology ring is given by

\begin{equation}
\mathbb{Z}_2[A_i,A_c,A_z,B_{xy}]/\langle\mathcal{R}_2,\mathcal{R}_4\rangle
 \end{equation}
where the relations are 
\begin{subequations} 
 \begin{align}
\mathcal{R}_2\colon & ~~
A_z (A_i + A_z),\\
\mathcal{R}_4\colon & ~~
B_{xy} (A_c^2 + A_i^2 + B_{xy}).
\end{align} 
 \end{subequations}
We have the following table regarding IWPs and group cohomology at degree 3.
\begin{center}
\begin{tabular}{c|cc|c|c|c}\hline\hline {Wyckoff}&\multicolumn{2}{c|}{Little group}& \multirow{2}{*}{Coordinates}&\multirow{2}{*}{LSM anomaly class}&\multirow{2}{*}{Topo. inv.} \\ \cline{2-3} position & Intl. & Sch\"{o}nflies & & & \\ \hline
1a&$6/m$&$C_{6h}$& $(0,0,0)$ & $(A_i + A_z) (A_c^2 + A_i^2 + B_{xy})$ & $\varphi_1[I]$\\ 
1b&$6/m$&$C_{6h}$& $(0,0,1/2)$ & $A_z (A_c^2 + A_i^2 + B_{xy})$ & $\varphi_1[T_3I]$\\ 
2c&$\overline{6}$&$C_{3h}$& $(1/3,2/3,0)$, $(2/3,1/3,0)$ & N/A & N/A\\ 
2d&$\overline{6}$&$C_{3h}$& $(1/3,2/3,1/2)$, $(2/3,1/3,1/2)$ & N/A & N/A\\ 
{3f} & {$2/m$} & {$C_{2h}$} & $(1/2,0,0)$, $(0,1/2,0)$, $(1/2,1/2,0)$ & {$(A_i + A_z) B_{xy}$} & {$\varphi_1[T_1I]$}\\
{3g} & {$2/m$} & {$C_{2h}$} & $(1/2,0,1/2)$, $(0,1/2,1/2)$, $(1/2,1/2,1/2)$ & {$A_z B_{xy}$} & {$\varphi_1[T_1T_3I]$}\\
\hline
\hline 
 \end{tabular} 
 \end{center}

\subsection*{No. 176: $P6_3/m$}\label{subsub:sg176}

This group is generated by three translations $T_{1,2,3}$ as given in Eqs.~\eqref{TransBravaisP}, a three-fold rotation $C_3$, a two-fold screw $S_2$, and an inversion $I$:
\begin{subequations}
 \begin{align}
C_3 &\colon (x,y,z)\rightarrow (-y, x - y, z),\\ 
S_2 &\colon (x,y,z)\rightarrow (-x, -y, z + 1/2),\\ 
I &\colon (x,y,z)\rightarrow (-x, -y, -z).
\end{align}
\end{subequations}

The $\mathbb{Z}_2$ cohomology ring is given by

\begin{equation}
\mathbb{Z}_2[A_i,A_c,B_{xy}]/\langle\mathcal{R}_2,\mathcal{R}_4\rangle
 \end{equation}
where the relations are 
\begin{subequations} 
 \begin{align}
\mathcal{R}_2\colon & ~~
A_c (A_c + A_i),\\
\mathcal{R}_4\colon & ~~
B_{xy} (A_c A_i + A_i^2 + B_{xy}).
\end{align} 
 \end{subequations}
We have the following table regarding IWPs and group cohomology at degree 3.
\begin{center}
\begin{tabular}{c|cc|c|c|c}\hline\hline {Wyckoff}&\multicolumn{2}{c|}{Little group}& \multirow{2}{*}{Coordinates}&\multirow{2}{*}{LSM anomaly class}&\multirow{2}{*}{Topo. inv.} \\ \cline{2-3} position & Intl. & Sch\"{o}nflies & & & \\ \hline
2a&$\overline{6}$&$C_{3h}$& $(0,0,1/4)$, $(0,0,3/4)$ & $A_c B_{xy}$ & $\varphi_3[T_1, T_2, S_2I]$\\ 
2b&$\overline{3}$&$C_{3i}$& $(0,0,0)$, $(0,0,1/2)$ & $(A_c + A_i) (A_i^2 + B_{xy})$ & $\varphi_1[I]$\\ 
2c&$\overline{6}$&$C_{3h}$& $(1/3,2/3,1/4)$, $(2/3,1/3,3/4)$ & Same as 2a  & Same as 2a\\ 
2d&$\overline{6}$&$C_{3h}$& $(2/3,1/3,1/4)$, $(1/3,2/3,3/4)$ & Same as 2a  & Same as 2a\\ 
\hline
\multirow{3}{*}{6g} & \multirow{3}{*}{$\overline{1}$} & \multirow{3}{*}{$C_i$} & $(1/2,0,0)$, $(0,1/2,0)$, & \multirow{3}{*}{$(A_c + A_i) B_{xy}$} & \multirow{3}{*}{$\varphi_1[T_1I]$}\\
& & & $(1/2,1/2,0)$, $(1/2,0,1/2)$, & & \\
& & & $(0,1/2,1/2)$, $(1/2,1/2,1/2)$ & & \\ \hline
\hline 
 \end{tabular} 
 \end{center}

\subsection*{No. 177: $P622$}\label{subsub:sg177}

This group is generated by three translations $T_{1,2,3}$ as given in Eqs.~\eqref{TransBravaisP}, a three-fold rotation $C_3$, a two-fold rotation $C'_2$, and a two-fold rotation $C_2$:
\begin{subequations}
 \begin{align}
C_3 &\colon (x,y,z)\rightarrow (-y, x - y, z),\\ 
C'_2 &\colon (x,y,z)\rightarrow (y, x, -z),\\ 
C_2 &\colon (x,y,z)\rightarrow (-x, -y, z).
\end{align}
\end{subequations}

The $\mathbb{Z}_2$ cohomology ring is given by

\begin{equation}
\mathbb{Z}_2[A_c,A_{c'},A_z,B_{xy}]/\langle\mathcal{R}_2,\mathcal{R}_4\rangle
 \end{equation}
where the relations are 
\begin{subequations} 
 \begin{align}
\mathcal{R}_2\colon & ~~
A_z (A_{c'} + A_z),\\
\mathcal{R}_4\colon & ~~
B_{xy} (A_c^2 + A_c A_{c'} + B_{xy}).
\end{align} 
 \end{subequations}
We have the following table regarding IWPs and group cohomology at degree 3.
\begin{center}
\begin{tabular}{c|cc|c|c|c}\hline\hline {Wyckoff}&\multicolumn{2}{c|}{Little group}& \multirow{2}{*}{Coordinates}&\multirow{2}{*}{LSM anomaly class}&\multirow{2}{*}{Topo. inv.} \\ \cline{2-3} position & Intl. & Sch\"{o}nflies & & & \\ \hline
1a&$622$&$D_6$& $(0,0,0)$ & $(A_{c'} + A_z) (A_c^2 + A_c A_{c'} + B_{xy})$ & $\varphi_2[C'_2, C_2]$\\ 
1b&$622$&$D_6$& $(0,0,1/2)$ & $A_z (A_c^2 + A_c A_{c'} + B_{xy})$ & $\varphi_2[T_3C'_2, C_2]$\\ 
2c&$32$&$D_3$& $(1/3,2/3,0)$, $(2/3,1/3,0)$ & N/A & N/A\\ 
2d&$32$&$D_3$& $(1/3,2/3,1/2)$, $(2/3,1/3,1/2)$ & N/A & N/A\\ 
\hline
\multirow{2}{*}{3f} & \multirow{2}{*}{$222$} & \multirow{2}{*}{$D_2$} & $(1/2,0,0)$, $(0,1/2,0)$, & \multirow{2}{*}{$(A_{c'} + A_z) B_{xy}$} & \multirow{2}{*}{$\varphi_2[C'_2, T_1T_2C_2]$}\\
& & & $(1/2,1/2,0)$ & & \\ \hline
\multirow{2}{*}{3g} & \multirow{2}{*}{$222$} & \multirow{2}{*}{$D_2$} & $(1/2,0,1/2)$, $(0,1/2,1/2)$, & \multirow{2}{*}{$A_z B_{xy}$} & \multirow{2}{*}{$\varphi_2[T_3C'_2, T_1T_2C_2]$}\\
& & & $(1/2,1/2,1/2)$ & & \\ \hline
\hline 
 \end{tabular} 
 \end{center}

\subsection*{No. 178: $P6_122$}\label{subsub:sg178}

This group is generated by three translations $T_{1,2,3}$ as given in Eqs.~\eqref{TransBravaisP}, a three-fold screw $S_3$, a two-fold rotation $C'_2$, and a two-fold screw $S_2$:
\begin{subequations}
 \begin{align}
S_3 &\colon (x,y,z)\rightarrow (-y, x - y, z + 1/3),\\ 
C'_2 &\colon (x,y,z)\rightarrow (y, x, -z + 1/3),\\ 
S_2 &\colon (x,y,z)\rightarrow (-x, -y, z + 1/2).
\end{align}
\end{subequations}

The $\mathbb{Z}_2$ cohomology ring is given by

\begin{equation}
\mathbb{Z}_2[A_c,A_{c'},B_{xy}]/\langle\mathcal{R}_2,\mathcal{R}_4\rangle
 \end{equation}
where the relations are 
\begin{subequations} 
 \begin{align}
\mathcal{R}_2\colon & ~~
A_c (A_c + A_{c'}),\\
\mathcal{R}_4\colon & ~~
B_{xy}^2.
\end{align} 
 \end{subequations}
We have the following table regarding IWPs and group cohomology at degree 3.
\begin{center}
\begin{tabular}{c|cc|c|c|c}\hline\hline {Wyckoff}&\multicolumn{2}{c|}{Little group}& \multirow{2}{*}{Coordinates}&\multirow{2}{*}{LSM anomaly class}&\multirow{2}{*}{Topo. inv.} \\ \cline{2-3} position & Intl. & Sch\"{o}nflies & & & \\ \hline
\multirow{3}{*}{6a} & \multirow{3}{*}{$2$} & \multirow{3}{*}{$C_2$} & $(x,0,0)$, $(0,x,1/3)$, & \multirow{3}{*}{$(A_c + A_{c'}) B_{xy}$} & \multirow{3}{*}{$\varphi_2[T_1T_2, C'_2]$}\\
& & & $(-x,-x,2/3)$, $(-x,0,1/2)$, & & \\
& & & $(0,-x,5/6)$, $(x,x,1/6)$ & & \\ \hline
\multirow{3}{*}{6b} & \multirow{3}{*}{$2$} & \multirow{3}{*}{$C_2$} & $(x,2x,1/4)$, $(-2x,-x,7/12)$, & \multirow{3}{*}{$A_c B_{xy}$} & \multirow{3}{*}{$\varphi_2[T_1T_2^{-1}, C'_2S_2]$}\\
& & & $(x,-x,11/12)$, $(-x,-2x,3/4)$, & & \\
& & & $(2x,x,1/12)$, $(-x,x,5/12)$ & & \\ \hline
\hline 
 \end{tabular} 
 \end{center}

\subsection*{No. 179: $P6_522$}\label{subsub:sg179}

This group is generated by three translations $T_{1,2,3}$ as given in Eqs.~\eqref{TransBravaisP}, a three-fold screw $S_3$, a two-fold rotation $C'_2$, and a two-fold screw $S_2$:
\begin{subequations}
 \begin{align}
S_3 &\colon (x,y,z)\rightarrow (-y, x - y, z + 2/3),\\ 
C'_2 &\colon (x,y,z)\rightarrow (y, x, -z + 2/3),\\ 
S_2 &\colon (x,y,z)\rightarrow (-x, -y, z + 1/2).
\end{align}
\end{subequations}

The $\mathbb{Z}_2$ cohomology ring is given by

\begin{equation}
\mathbb{Z}_2[A_c,A_{c'},B_{xy}]/\langle\mathcal{R}_2,\mathcal{R}_4\rangle
 \end{equation}
where the relations are 
\begin{subequations} 
 \begin{align}
\mathcal{R}_2\colon & ~~
A_c (A_c + A_{c'}),\\
\mathcal{R}_4\colon & ~~
B_{xy}^2.
\end{align} 
 \end{subequations}
We have the following table regarding IWPs and group cohomology at degree 3.
\begin{center}
\begin{tabular}{c|cc|c|c|c}\hline\hline {Wyckoff}&\multicolumn{2}{c|}{Little group}& \multirow{2}{*}{Coordinates}&\multirow{2}{*}{LSM anomaly class}&\multirow{2}{*}{Topo. inv.} \\ \cline{2-3} position & Intl. & Sch\"{o}nflies & & & \\ \hline
\multirow{3}{*}{6a} & \multirow{3}{*}{$2$} & \multirow{3}{*}{$C_2$} & $(x,0,0)$, $(0,x,2/3)$, & \multirow{3}{*}{$(A_c + A_{c'}) B_{xy}$} & \multirow{3}{*}{$\varphi_2[T_1T_2, C'_2]$}\\
& & & $(-x,-x,1/3)$, $(-x,0,1/2)$, & & \\
& & & $(0,-x,1/6)$, $(x,x,5/6)$ & & \\ \hline
\multirow{3}{*}{6b} & \multirow{3}{*}{$2$} & \multirow{3}{*}{$C_2$} & $(x,2x,3/4)$, $(-2x,-x,5/12)$, & \multirow{3}{*}{$A_c B_{xy}$} & \multirow{3}{*}{$\varphi_2[T_1T_2^{-1}, C'_2S_2]$}\\
& & & $(x,-x,1/12)$, $(-x,-2x,1/4)$, & & \\
& & & $(2x,x,11/12)$, $(-x,x,7/12)$ & & \\ \hline
\hline 
 \end{tabular} 
 \end{center}

\subsection*{No. 180: $P6_222$}\label{subsub:sg180}

This group is generated by three translations $T_{1,2,3}$ as given in Eqs.~\eqref{TransBravaisP}, a three-fold screw $S_3$, a two-fold rotation $C'_2$, and a two-fold rotation $C_2$:
\begin{subequations}
 \begin{align}
S_3 &\colon (x,y,z)\rightarrow (-y, x - y, z + 2/3),\\ 
C'_2 &\colon (x,y,z)\rightarrow (y, x, -z + 2/3),\\ 
C_2 &\colon (x,y,z)\rightarrow (-x, -y, z).
\end{align}
\end{subequations}

The $\mathbb{Z}_2$ cohomology ring is given by

\begin{equation}
\mathbb{Z}_2[A_c,A_{c'},A_z,B_{xy}]/\langle\mathcal{R}_2,\mathcal{R}_4\rangle
 \end{equation}
where the relations are 
\begin{subequations} 
 \begin{align}
\mathcal{R}_2\colon & ~~
A_z (A_{c'} + A_z),\\
\mathcal{R}_4\colon & ~~
B_{xy} (A_c^2 + A_c A_{c'} + B_{xy}).
\end{align} 
 \end{subequations}
We have the following table regarding IWPs and group cohomology at degree 3.
\begin{center}
\resizebox{\columnwidth}{!}{
\begin{tabular}{c|cc|c|c|c}\hline\hline {Wyckoff}&\multicolumn{2}{c|}{Little group}& \multirow{2}{*}{Coordinates}&\multirow{2}{*}{LSM anomaly class}&\multirow{2}{*}{Topo. inv.} \\ \cline{2-3} position & Intl. & Sch\"{o}nflies & & & \\ \hline
{3a} & {$222$} & {$D_2$} & $(0,0,0)$, $(0,0,2/3)$, $(0,0,1/3)$ & {$(A_{c'} + A_z) (A_c^2 + A_c A_{c'} + B_{xy})$} & {$\varphi_2[C'_2, C_2]$}\\
{3b} & {$222$} & {$D_2$} & $(0,0,1/2)$, $(0,0,1/6)$, $(0,0,5/6)$ & {$A_z (A_c^2 + A_c A_{c'} + B_{xy})$} & {$\varphi_2[T_3C'_2, C_2]$}\\
{3c} & {$222$} & {$D_2$} & $(1/2,0,0)$, $(0,1/2,2/3)$, $(1/2,1/2,1/3)$ & {$(A_{c'} + A_z) B_{xy}$} & {$\varphi_2[C'_2, T_1T_2C_2]$}\\
{3d} & {$222$} & {$D_2$} & $(1/2,0,1/2)$, $(0,1/2,1/6)$, $(1/2,1/2,5/6)$ & {$A_z B_{xy}$} & {$\varphi_2[T_3C'_2, T_1T_2C_2]$}\\
\hline
\hline 
 \end{tabular} }
 \end{center}

\subsection*{No. 181: $P6_422$}\label{subsub:sg181}

This group is generated by three translations $T_{1,2,3}$ as given in Eqs.~\eqref{TransBravaisP}, a three-fold screw $S_3$, a two-fold rotation $C'_2$, and a two-fold rotation $C_2$:
\begin{subequations}
 \begin{align}
S_3 &\colon (x,y,z)\rightarrow (-y, x - y, z + 1/3),\\ 
C'_2 &\colon (x,y,z)\rightarrow (y, x, -z + 1/3),\\ 
C_2 &\colon (x,y,z)\rightarrow (-x, -y, z).
\end{align}
\end{subequations}

The $\mathbb{Z}_2$ cohomology ring is given by

\begin{equation}
\mathbb{Z}_2[A_c,A_{c'},A_z,B_{xy}]/\langle\mathcal{R}_2,\mathcal{R}_4\rangle
 \end{equation}
where the relations are 
\begin{subequations} 
 \begin{align}
\mathcal{R}_2\colon & ~~
A_z (A_{c'} + A_z),\\
\mathcal{R}_4\colon & ~~
B_{xy} (A_c^2 + A_c A_{c'} + B_{xy}).
\end{align} 
 \end{subequations}
We have the following table regarding IWPs and group cohomology at degree 3.
\begin{center}
\resizebox{\columnwidth}{!}{
\begin{tabular}{c|cc|c|c|c}\hline\hline {Wyckoff}&\multicolumn{2}{c|}{Little group}& \multirow{2}{*}{Coordinates}&\multirow{2}{*}{LSM anomaly class}&\multirow{2}{*}{Topo. inv.} \\ \cline{2-3} position & Intl. & Sch\"{o}nflies & & & \\ \hline
{3a} & {$222$} & {$D_2$} & $(0,0,0)$, $(0,0,1/3)$, $(0,0,2/3)$ & {$A_z (A_c^2 + A_c A_{c'} + B_{xy})$} & {$\varphi_2[T_3C'_2, C_2]$}\\
{3b} & {$222$} & {$D_2$} & $(0,0,1/2)$, $(0,0,5/6)$, $(0,0,1/6)$ & {$(A_{c'} + A_z) (A_c^2 + A_c A_{c'} + B_{xy})$} & {$\varphi_2[C'_2, C_2]$}\\
{3c} & {$222$} & {$D_2$} & $(1/2,0,0)$, $(0,1/2,1/3)$, $(1/2,1/2,2/3)$ & {$A_z B_{xy}$} & {$\varphi_2[T_3C'_2, T_1T_2C_2]$}\\
{3d} & {$222$} & {$D_2$} & $(1/2,0,1/2)$, $(0,1/2,5/6)$, $(1/2,1/2,1/6)$ & {$(A_{c'} + A_z) B_{xy}$} & {$\varphi_2[C'_2, T_1T_2C_2]$}\\
\hline
\hline 
 \end{tabular} }
 \end{center}

\subsection*{No. 182: $P6_322$}\label{subsub:sg182}

This group is generated by three translations $T_{1,2,3}$ as given in Eqs.~\eqref{TransBravaisP}, a three-fold rotation $C_3$, a two-fold rotation $C'_2$, and a two-fold screw $S_2$:
\begin{subequations}
 \begin{align}
C_3 &\colon (x,y,z)\rightarrow (-y, x - y, z),\\ 
C'_2 &\colon (x,y,z)\rightarrow (y, x, -z),\\ 
S_2 &\colon (x,y,z)\rightarrow (-x, -y, z + 1/2).
\end{align}
\end{subequations}

The $\mathbb{Z}_2$ cohomology ring is given by

\begin{equation}
\mathbb{Z}_2[A_c,A_{c'},B_{xy}]/\langle\mathcal{R}_2,\mathcal{R}_4\rangle
 \end{equation}
where the relations are 
\begin{subequations} 
 \begin{align}
\mathcal{R}_2\colon & ~~
A_c (A_c + A_{c'}),\\
\mathcal{R}_4\colon & ~~
B_{xy}^2.
\end{align} 
 \end{subequations}
We have the following table regarding IWPs and group cohomology at degree 3.
\begin{center}
\begin{tabular}{c|cc|c|c|c}\hline\hline {Wyckoff}&\multicolumn{2}{c|}{Little group}& \multirow{2}{*}{Coordinates}&\multirow{2}{*}{LSM anomaly class}&\multirow{2}{*}{Topo. inv.} \\ \cline{2-3} position & Intl. & Sch\"{o}nflies & & & \\ \hline
2a&$32$&$D_3$& $(0,0,0)$, $(0,0,1/2)$ & $(A_c + A_{c'}) B_{xy}$ & $\varphi_2[T_1T_2, C'_2]$\\ 
2b&$32$&$D_3$& $(0,0,1/4)$, $(0,0,3/4)$ & $A_c B_{xy}$ & $\varphi_2[T_1T_2^{-1}, C'_2S_2]$\\ 
2c&$32$&$D_3$& $(1/3,2/3,1/4)$, $(2/3,1/3,3/4)$ & Same as 2b  & Same as 2b\\ 
2d&$32$&$D_3$& $(1/3,2/3,3/4)$, $(2/3,1/3,1/4)$ & Same as 2b  & Same as 2b\\ 
\hline
\hline 
 \end{tabular} 
 \end{center}

\subsection*{No. 183: $P6mm$}\label{subsub:sg183}

This group is generated by three translations $T_{1,2,3}$ as given in Eqs.~\eqref{TransBravaisP}, a three-fold rotation $C_3$, a mirror $M$, and a two-fold rotation $C_2$:
\begin{subequations}
 \begin{align}
C_3 &\colon (x,y,z)\rightarrow (-y, x - y, z),\\ 
M &\colon (x,y,z)\rightarrow (-y, -x, z),\\ 
C_2 &\colon (x,y,z)\rightarrow (-x, -y, z).
\end{align}
\end{subequations}

The $\mathbb{Z}_2$ cohomology ring is given by

\begin{equation}
\mathbb{Z}_2[A_c,A_m,A_z,B_{xy}]/\langle\mathcal{R}_2,\mathcal{R}_4\rangle
 \end{equation}
where the relations are 
\begin{subequations} 
 \begin{align}
\mathcal{R}_2\colon & ~~
A_z^2,\\
\mathcal{R}_4\colon & ~~
B_{xy} (A_c^2 + A_c A_m + B_{xy}).
\end{align} 
 \end{subequations}
We have the following table regarding IWPs and group cohomology at degree 3.
\begin{center}
\begin{tabular}{c|cc|c|c|c}\hline\hline {Wyckoff}&\multicolumn{2}{c|}{Little group}& \multirow{2}{*}{Coordinates}&\multirow{2}{*}{LSM anomaly class}&\multirow{2}{*}{Topo. inv.} \\ \cline{2-3} position & Intl. & Sch\"{o}nflies & & & \\ \hline
1a&$6mm$&$C_{6v}$& $(0,0,z)$ & $A_z (A_c^2 + A_c A_m + B_{xy})$ & $\varphi_2[T_3, C_2]$\\ 
2b&$3m$&$C_{3v}$& $(1/3,2/3,z)$, $(2/3,1/3,z)$ & N/A & N/A\\ 
{3c} & {$2mm$} & {$C_{2v}$} & $(1/2,0,z)$, $(0,1/2,z)$, $(1/2,1/2,z)$ & {$A_z B_{xy}$} & {$\varphi_2[T_3, T_1C_2]$}\\
\hline
\hline 
 \end{tabular} 
 \end{center}

\subsection*{No. 184: $P6cc$}\label{subsub:sg184}

This group is generated by three translations $T_{1,2,3}$ as given in Eqs.~\eqref{TransBravaisP}, a three-fold rotation $C_3$, a glide $G$, and a two-fold rotation $C_2$:
\begin{subequations}
 \begin{align}
C_3 &\colon (x,y,z)\rightarrow (-y, x - y, z),\\ 
G &\colon (x,y,z)\rightarrow (-y, -x, z + 1/2),\\ 
C_2 &\colon (x,y,z)\rightarrow (-x, -y, z).
\end{align}
\end{subequations}

The $\mathbb{Z}_2$ cohomology ring is given by

\begin{equation}
\mathbb{Z}_2[A_c,A_m,B_{xy}]/\langle\mathcal{R}_2,\mathcal{R}_4\rangle
 \end{equation}
where the relations are 
\begin{subequations} 
 \begin{align}
\mathcal{R}_2\colon & ~~
A_m^2,\\
\mathcal{R}_4\colon & ~~
B_{xy} (A_c^2 + A_c A_m + B_{xy}).
\end{align} 
 \end{subequations}
We have the following table regarding IWPs and group cohomology at degree 3.
\begin{center}
\begin{tabular}{c|cc|c|c|c}\hline\hline {Wyckoff}&\multicolumn{2}{c|}{Little group}& \multirow{2}{*}{Coordinates}&\multirow{2}{*}{LSM anomaly class}&\multirow{2}{*}{Topo. inv.} \\ \cline{2-3} position & Intl. & Sch\"{o}nflies & & & \\ \hline
2a&$6$&$C_6$& $(0,0,z)$, $(0,0,z+1/2)$ & $A_m (A_c^2 + B_{xy})$ & $\varphi_2[G, C_2]$\\ 
\hline
\multirow{2}{*}{4b} & \multirow{2}{*}{$3$} & \multirow{2}{*}{$C_3$} & $(1/3,2/3,z)$, $(2/3,1/3,z)$, & \multirow{2}{*}{N/A} & \multirow{2}{*}{N/A}\\
& & & $(1/3,2/3,z+1/2)$, $(2/3,1/3,z+1/2)$ & & \\ \hline
\multirow{3}{*}{6c} & \multirow{3}{*}{$2$} & \multirow{3}{*}{$C_2$} & $(1/2,0,z)$, $(0,1/2,z)$, & \multirow{3}{*}{$A_m B_{xy}$} & \multirow{3}{*}{$\varphi_2[T_1T_2G, T_1T_2C_2]$}\\
& & & $(1/2,1/2,z)$, $(0,1/2,z+1/2)$, & & \\
& & & $(1/2,0,z+1/2)$, $(1/2,1/2,z+1/2)$ & & \\ \hline
\hline 
 \end{tabular} 
 \end{center}

\subsection*{No. 185: $P6_3cm$}\label{subsub:sg185}

This group is generated by three translations $T_{1,2,3}$ as given in Eqs.~\eqref{TransBravaisP}, a three-fold rotation $C_3$, a glide $G$, and a two-fold screw $S_2$:
\begin{subequations}
 \begin{align}
C_3 &\colon (x,y,z)\rightarrow (-y, x - y, z),\\ 
G &\colon (x,y,z)\rightarrow (-y, -x, z + 1/2),\\ 
S_2 &\colon (x,y,z)\rightarrow (-x, -y, z + 1/2).
\end{align}
\end{subequations}

The $\mathbb{Z}_2$ cohomology ring is given by

\begin{equation}
\mathbb{Z}_2[A_c,A_m,B_{xy}]/\langle\mathcal{R}_2,\mathcal{R}_4\rangle
 \end{equation}
where the relations are 
\begin{subequations} 
 \begin{align}
\mathcal{R}_2\colon & ~~
(A_c + A_m)^2,\\
\mathcal{R}_4\colon & ~~
B_{xy} (A_c^2 + A_c A_m + B_{xy}).
\end{align} 
 \end{subequations}
We have the following table regarding IWPs and group cohomology at degree 3.
\begin{center}
\begin{tabular}{c|cc|c|c|c}\hline\hline {Wyckoff}&\multicolumn{2}{c|}{Little group}& \multirow{2}{*}{Coordinates}&\multirow{2}{*}{LSM anomaly class}&\multirow{2}{*}{Topo. inv.} \\ \cline{2-3} position & Intl. & Sch\"{o}nflies & & & \\ \hline
2a&$3m$&$C_{3v}$& $(0,0,z)$, $(0,0,z+1/2)$ & $(A_c + A_m) B_{xy}$ & $\widehat{\varphi_3}[T_1, T_2, G]$\\ 
\hline
\multirow{2}{*}{4b} & \multirow{2}{*}{$3$} & \multirow{2}{*}{$C_3$} & $(1/3,2/3,z)$, $(2/3,1/3,z+1/2)$, & \multirow{2}{*}{N/A} & \multirow{2}{*}{N/A}\\
& & & $(1/3,2/3,z+1/2)$, $(2/3,1/3,z)$ & & \\ \hline
\hline 
 \end{tabular} 
 \end{center}
    Here the topological invariant $\widehat{\varphi_3}[T_1, T_2, G]$ can be chosen to be the same as that of group No.9 $Cc$, given by Eq.~\eqref{TI_9}.

\subsection*{No. 186: $P6_3mc$}\label{subsub:sg186}

This group is generated by three translations $T_{1,2,3}$ as given in Eqs.~\eqref{TransBravaisP}, a three-fold rotation $C_3$, a mirror $M$, and a two-fold screw $S_2$:
\begin{subequations}
 \begin{align}
C_3 &\colon (x,y,z)\rightarrow (-y, x - y, z),\\ 
M &\colon (x,y,z)\rightarrow (-y, -x, z),\\ 
S_2 &\colon (x,y,z)\rightarrow (-x, -y, z + 1/2).
\end{align}
\end{subequations}

The $\mathbb{Z}_2$ cohomology ring is given by

\begin{equation}
\mathbb{Z}_2[A_c,A_m,B_{xy}]/\langle\mathcal{R}_2,\mathcal{R}_4\rangle
 \end{equation}
where the relations are 
\begin{subequations} 
 \begin{align}
\mathcal{R}_2\colon & ~~
A_c^2,\\
\mathcal{R}_4\colon & ~~
B_{xy} (A_c A_m + B_{xy}).
\end{align} 
 \end{subequations}
We have the following table regarding IWPs and group cohomology at degree 3.
\begin{center}
\begin{tabular}{c|cc|c|c|c}\hline\hline {Wyckoff}&\multicolumn{2}{c|}{Little group}& \multirow{2}{*}{Coordinates}&\multirow{2}{*}{LSM anomaly class}&\multirow{2}{*}{Topo. inv.} \\ \cline{2-3} position & Intl. & Sch\"{o}nflies & & & \\ \hline
2a&$3m$&$C_{3v}$& $(0,0,z)$, $(0,0,z+1/2)$ & $A_c B_{xy}$ & $\widetilde{\varphi_3}[T_1T_2^{-1}, MS_2, M]$\\ 
2b&$3m$&$C_{3v}$& $(1/3,2/3,z)$, $(2/3,1/3,z+1/2)$ & Same as 2a  & Same as 2a\\ 
\hline
\hline 
 \end{tabular} 
 \end{center}
 The expression of $\widetilde{\varphi_3}$ is given in Eq.~\eqref{wilde3}.

\subsection*{No. 187: $P\overline6m2$}\label{subsub:sg187}

This group is generated by three translations $T_{1,2,3}$ as given in Eqs.~\eqref{TransBravaisP}, a three-fold rotation $C_3$, a two-fold rotation $C'_2$, and a mirror $M$:
\begin{subequations}
 \begin{align}
C_3 &\colon (x,y,z)\rightarrow (-y, x - y, z),\\ 
C'_2 &\colon (x,y,z)\rightarrow (-y, -x, -z),\\ 
M &\colon (x,y,z)\rightarrow (x, y, -z).
\end{align}
\end{subequations}

The $\mathbb{Z}_2$ cohomology ring is given by

\begin{equation}
\mathbb{Z}_2[A_{c'},A_m,A_z,B_{xy}]/\langle\mathcal{R}_2,\mathcal{R}_4\rangle
 \end{equation}
where the relations are 
\begin{subequations} 
 \begin{align}
\mathcal{R}_2\colon & ~~
A_z (A_{c'} + A_m + A_z),\\
\mathcal{R}_4\colon & ~~
B_{xy}^2.
\end{align} 
 \end{subequations}
We have the following table regarding IWPs and group cohomology at degree 3.
\begin{center}
\begin{tabular}{c|cc|c|c|c}\hline\hline {Wyckoff}&\multicolumn{2}{c|}{Little group}& \multirow{2}{*}{Coordinates}&\multirow{2}{*}{LSM anomaly class}&\multirow{2}{*}{Topo. inv.} \\ \cline{2-3} position & Intl. & Sch\"{o}nflies & & & \\ \hline
1a&$\overline{6}m2$&$D_{3h}$& $(0,0,0)$ & $(A_{c'} + A_m + A_z) B_{xy}$ & $\varphi_2[T_1T_2^{-1}, C'_2]$\\ 
1b&$\overline{6}m2$&$D_{3h}$& $(0,0,1/2)$ & $A_z B_{xy}$ & $\varphi_2[T_1T_2^{-1}, T_3C'_2]$\\ 
1c&$\overline{6}m2$&$D_{3h}$& $(1/3,2/3,0)$ & Same as 1a  & Same as 1a\\ 
1d&$\overline{6}m2$&$D_{3h}$& $(1/3,2/3,1/2)$ & Same as 1b  & Same as 1b\\ 
1e&$\overline{6}m2$&$D_{3h}$& $(2/3,1/3,0)$ & Same as 1a  & Same as 1a\\ 
1f&$\overline{6}m2$&$D_{3h}$& $(2/3,1/3,1/2)$ & Same as 1b  & Same as 1b\\ 
\hline
\hline 
 \end{tabular} 
 \end{center}

\subsection*{No. 188: $P\overline6c2$}\label{subsub:sg188}

This group is generated by three translations $T_{1,2,3}$ as given in Eqs.~\eqref{TransBravaisP}, a three-fold rotation $C_3$, a two-fold rotation $C'_2$, and a mirror $M$:
\begin{subequations}
 \begin{align}
C_3 &\colon (x,y,z)\rightarrow (-y, x - y, z),\\ 
C'_2 &\colon (x,y,z)\rightarrow (-y, -x, -z),\\ 
M &\colon (x,y,z)\rightarrow (x, y, -z + 1/2).
\end{align}
\end{subequations}

The $\mathbb{Z}_2$ cohomology ring is given by

\begin{equation}
\mathbb{Z}_2[A_{c'},A_m,B_{xy}]/\langle\mathcal{R}_2,\mathcal{R}_4\rangle
 \end{equation}
where the relations are 
\begin{subequations} 
 \begin{align}
\mathcal{R}_2\colon & ~~
A_{c'} A_m,\\
\mathcal{R}_4\colon & ~~
B_{xy}^2.
\end{align} 
 \end{subequations}
We have the following table regarding IWPs and group cohomology at degree 3.
\begin{center}
\begin{tabular}{c|cc|c|c|c}\hline\hline {Wyckoff}&\multicolumn{2}{c|}{Little group}& \multirow{2}{*}{Coordinates}&\multirow{2}{*}{LSM anomaly class}&\multirow{2}{*}{Topo. inv.} \\ \cline{2-3} position & Intl. & Sch\"{o}nflies & & & \\ \hline
2a&$32$&$D_3$& $(0,0,0)$, $(0,0,1/2)$ & $A_{c'} B_{xy}$ & $\varphi_2[T_1T_2^{-1}, C'_2]$\\ 
2b&$\overline{6}$&$C_{3h}$& $(0,0,1/4)$, $(0,0,3/4)$ & $A_m B_{xy}$ & $\varphi_3[T_1, T_2, M]$\\ 
2c&$32$&$D_3$& $(1/3,2/3,0)$, $(1/3,2/3,1/2)$ & Same as 2a  & Same as 2a\\ 
2d&$\overline{6}$&$C_{3h}$& $(1/3,2/3,1/4)$, $(1/3,2/3,3/4)$ & Same as 2b  & Same as 2b\\ 
2e&$32$&$D_3$& $(2/3,1/3,0)$, $(2/3,1/3,1/2)$ & Same as 2a  & Same as 2a\\ 
2f&$\overline{6}$&$C_{3h}$& $(2/3,1/3,1/4)$, $(2/3,1/3,3/4)$ & Same as 2b  & Same as 2b\\ 
\hline
\hline 
 \end{tabular} 
 \end{center}

\subsection*{No. 189: $P\overline62m$}\label{subsub:sg189}

This group is generated by three translations $T_{1,2,3}$ as given in Eqs.~\eqref{TransBravaisP}, a three-fold rotation $C_3$, a two-fold rotation $C'_2$, and a mirror $M$:
\begin{subequations}
 \begin{align}
C_3 &\colon (x,y,z)\rightarrow (-y, x - y, z),\\ 
C'_2 &\colon (x,y,z)\rightarrow (y, x, -z),\\ 
M &\colon (x,y,z)\rightarrow (x, y, -z).
\end{align}
\end{subequations}

The $\mathbb{Z}_2$ cohomology ring is given by

\begin{equation}
\mathbb{Z}_2[A_{c'},A_m,A_z,B_{xy}]/\langle\mathcal{R}_2,\mathcal{R}_4\rangle
 \end{equation}
where the relations are 
\begin{subequations} 
 \begin{align}
\mathcal{R}_2\colon & ~~
A_z (A_{c'} + A_m + A_z),\\
\mathcal{R}_4\colon & ~~
B_{xy}^2.
\end{align} 
 \end{subequations}
We have the following table regarding IWPs and group cohomology at degree 3.
\begin{center}
\begin{tabular}{c|cc|c|c|c}\hline\hline {Wyckoff}&\multicolumn{2}{c|}{Little group}& \multirow{2}{*}{Coordinates}&\multirow{2}{*}{LSM anomaly class}&\multirow{2}{*}{Topo. inv.} \\ \cline{2-3} position & Intl. & Sch\"{o}nflies & & & \\ \hline
1a&$\overline{6}2m$&$D_{3h}$& $(0,0,0)$ & $(A_{c'} + A_m + A_z) B_{xy}$ & $\varphi_2[T_1T_2, C'_2]$\\ 
1b&$\overline{6}2m$&$D_{3h}$& $(0,0,1/2)$ & $A_z B_{xy}$ & $\varphi_2[T_1T_2, T_3C'_2]$\\ 
2c&$\overline{6}$&$C_{3h}$& $(1/3,2/3,0)$, $(2/3,1/3,0)$ & N/A & N/A\\ 
2d&$\overline{6}$&$C_{3h}$& $(1/3,2/3,1/2)$, $(2/3,1/3,1/2)$ & N/A & N/A\\ 
\hline
\hline 
 \end{tabular} 
 \end{center}

\subsection*{No. 190: $P\overline62c$}\label{subsub:sg190}

This group is generated by three translations $T_{1,2,3}$ as given in Eqs.~\eqref{TransBravaisP}, a three-fold rotation $C_3$, a two-fold rotation $C'_2$, and a mirror $M$:
\begin{subequations}
 \begin{align}
C_3 &\colon (x,y,z)\rightarrow (-y, x - y, z),\\ 
C'_2 &\colon (x,y,z)\rightarrow (y, x, -z),\\ 
M &\colon (x,y,z)\rightarrow (x, y, -z + 1/2).
\end{align}
\end{subequations}

The $\mathbb{Z}_2$ cohomology ring is given by

\begin{equation}
\mathbb{Z}_2[A_{c'},A_m,B_{xy}]/\langle\mathcal{R}_2,\mathcal{R}_4\rangle
 \end{equation}
where the relations are 
\begin{subequations} 
 \begin{align}
\mathcal{R}_2\colon & ~~
A_{c'} A_m,\\
\mathcal{R}_4\colon & ~~
B_{xy}^2.
\end{align} 
 \end{subequations}
We have the following table regarding IWPs and group cohomology at degree 3.
\begin{center}
\begin{tabular}{c|cc|c|c|c}\hline\hline {Wyckoff}&\multicolumn{2}{c|}{Little group}& \multirow{2}{*}{Coordinates}&\multirow{2}{*}{LSM anomaly class}&\multirow{2}{*}{Topo. inv.} \\ \cline{2-3} position & Intl. & Sch\"{o}nflies & & & \\ \hline
2a&$32$&$D_3$& $(0,0,0)$, $(0,0,1/2)$ & $A_{c'} B_{xy}$ & $\varphi_2[T_1T_2, C'_2]$\\ 
2b&$\overline{6}$&$C_{3h}$& $(0,0,1/4)$, $(0,0,3/4)$ & $A_m B_{xy}$ & $\varphi_3[T_1, T_2, M]$\\ 
2c&$\overline{6}$&$C_{3h}$& $(1/3,2/3,1/4)$, $(2/3,1/3,3/4)$ & Same as 2a  & Same as 2a\\ 
2d&$\overline{6}$&$C_{3h}$& $(2/3,1/3,1/4)$, $(1/3,2/3,3/4)$ & Same as 2b  & Same as 2b\\ 
\hline
\hline 
 \end{tabular} 
 \end{center}

\subsection*{No. 191: $P6/mmm$}\label{subsub:sg191}

This group is generated by three translations $T_{1,2,3}$ as given in Eqs.~\eqref{TransBravaisP}, a three-fold rotation $C_3$, a two-fold rotation $C'_2$, a two-fold rotation $C_2$, and an inversion $I$:
\begin{subequations}
 \begin{align}
C_3 &\colon (x,y,z)\rightarrow (-y, x - y, z),\\ 
C'_2 &\colon (x,y,z)\rightarrow (y, x, -z),\\ 
C_2 &\colon (x,y,z)\rightarrow (-x, -y, z),\\ 
I &\colon (x,y,z)\rightarrow (-x, -y, -z).
\end{align}
\end{subequations}

The $\mathbb{Z}_2$ cohomology ring is given by

\begin{equation}
\mathbb{Z}_2[A_i,A_c,A_{c'},A_z,B_{xy}]/\langle\mathcal{R}_2,\mathcal{R}_4\rangle
 \end{equation}
where the relations are 
\begin{subequations} 
 \begin{align}
\mathcal{R}_2\colon & ~~
A_z (A_{c'} + A_i + A_z),\\
\mathcal{R}_4\colon & ~~
B_{xy} (A_c^2 + A_c A_{c'} + A_{c'} A_i + A_i^2 + B_{xy}).
\end{align} 
 \end{subequations}
We have the following table regarding IWPs and group cohomology at degree 3.
\begin{center}
\resizebox{\columnwidth}{!}{
\begin{tabular}{c|cc|c|c|c}\hline\hline {Wyckoff}&\multicolumn{2}{c|}{Little group}& \multirow{2}{*}{Coordinates}&\multirow{2}{*}{LSM anomaly class}&\multirow{2}{*}{Topo. inv.} \\ \cline{2-3} position & Intl. & Sch\"{o}nflies & & & \\ \hline
1a&$6/mmm$&$D_{6h}$& $(0,0,0)$ & $(A_{c'} + A_i + A_z) (A_c^2 + A_c A_{c'} + A_{c'} A_i + A_i^2 + B_{xy})$ & $\varphi_2[C'_2, C_2]$\\ 
1b&$6/mmm$&$D_{6h}$& $(0,0,1/2)$ & $A_z (A_c^2 + A_c A_{c'} + A_{c'} A_i + A_i^2 + B_{xy})$ & $\varphi_2[T_3C'_2, C_2]$\\ 
2c&$\overline{6}m2$&$D_{3h}$& $(1/3,2/3,0)$, $(2/3,1/3,0)$ & N/A & N/A\\ 
2d&$\overline{6}m2$&$D_{3h}$& $(1/3,2/3,1/2)$, $(2/3,1/3,1/2)$ & N/A & N/A\\ 
\hline
\multirow{2}{*}{3f} & \multirow{2}{*}{$mmm$} & \multirow{2}{*}{$D_{2h}$} & $(1/2,0,0)$, $(0,1/2,0)$, & \multirow{2}{*}{$(A_{c'} + A_i + A_z) B_{xy}$} & \multirow{2}{*}{$\varphi_2[C'_2, T_1T_2C_2]$}\\
& & & $(1/2,1/2,0)$ & & \\ \hline
\multirow{2}{*}{3g} & \multirow{2}{*}{$mmm$} & \multirow{2}{*}{$D_{2h}$} & $(1/2,0,1/2)$, $(0,1/2,1/2)$, & \multirow{2}{*}{$A_z B_{xy}$} & \multirow{2}{*}{$\varphi_2[T_3C'_2, T_1T_2C_2]$}\\
& & & $(1/2,1/2,1/2)$ & & \\ \hline
\hline 
 \end{tabular} }
 \end{center}

\subsection*{No. 192: $P6/mcc$}\label{subsub:sg192}

This group is generated by three translations $T_{1,2,3}$ as given in Eqs.~\eqref{TransBravaisP}, a three-fold rotation $C_3$, a two-fold rotation $C'_2$, a two-fold rotation $C_2$, and an inversion $I$:
\begin{subequations}
 \begin{align}
C_3 &\colon (x,y,z)\rightarrow (-y, x - y, z),\\ 
C'_2 &\colon (x,y,z)\rightarrow (y, x, -z + 1/2),\\ 
C_2 &\colon (x,y,z)\rightarrow (-x, -y, z),\\ 
I &\colon (x,y,z)\rightarrow (-x, -y, -z).
\end{align}
\end{subequations}

The $\mathbb{Z}_2$ cohomology ring is given by

\begin{equation}
\mathbb{Z}_2[A_i,A_c,A_{c'},B_{xy}]/\langle\mathcal{R}_2,\mathcal{R}_4\rangle
 \end{equation}
where the relations are 
\begin{subequations} 
 \begin{align}
\mathcal{R}_2\colon & ~~
A_{c'} A_i,\\
\mathcal{R}_4\colon & ~~
B_{xy} (A_c^2 + A_c A_{c'} + A_i^2 + B_{xy}).
\end{align} 
 \end{subequations}
We have the following table regarding IWPs and group cohomology at degree 3.
\begin{center}
\begin{tabular}{c|cc|c|c|c}\hline\hline {Wyckoff}&\multicolumn{2}{c|}{Little group}& \multirow{2}{*}{Coordinates}&\multirow{2}{*}{LSM anomaly class}&\multirow{2}{*}{Topo. inv.} \\ \cline{2-3} position & Intl. & Sch\"{o}nflies & & & \\ \hline
2a&$622$&$D_6$& $(0,0,1/4)$, $(0,0,3/4)$ & $A_{c'} (A_c^2 + A_c A_{c'} + B_{xy})$ & $\varphi_2[C'_2, C_2]$\\ 
2b&$6/m$&$C_{6h}$& $(0,0,0)$, $(0,0,1/2)$ & $A_i (A_c^2 + A_i^2 + B_{xy})$ & $\varphi_1[I]$\\ 
\hline
\multirow{2}{*}{4c} & \multirow{2}{*}{$32$} & \multirow{2}{*}{$D_3$} & $(1/3,2/3,1/4)$, $(2/3,1/3,1/4)$, & \multirow{2}{*}{N/A} & \multirow{2}{*}{N/A}\\
& & & $(2/3,1/3,3/4)$, $(1/3,2/3,3/4)$ & & \\ \hline
\multirow{2}{*}{4d} & \multirow{2}{*}{$\overline{6}$} & \multirow{2}{*}{$C_{3h}$} & $(1/3,2/3,0)$, $(2/3,1/3,0)$, & \multirow{2}{*}{N/A} & \multirow{2}{*}{N/A}\\
& & & $(2/3,1/3,1/2)$, $(1/3,2/3,1/2)$ & & \\ \hline
\multirow{3}{*}{6f} & \multirow{3}{*}{$222$} & \multirow{3}{*}{$D_2$} & $(1/2,0,1/4)$, $(0,1/2,1/4)$, & \multirow{3}{*}{$A_{c'} B_{xy}$} & \multirow{3}{*}{$\varphi_2[C'_2, T_1T_2C_2]$}\\
& & & $(1/2,1/2,1/4)$, $(1/2,0,3/4)$, & & \\
& & & $(0,1/2,3/4)$, $(1/2,1/2,3/4)$ & & \\ \hline
\multirow{3}{*}{6g} & \multirow{3}{*}{$2/m$} & \multirow{3}{*}{$C_{2h}$} & $(1/2,0,0)$, $(0,1/2,0)$, & \multirow{3}{*}{$A_i B_{xy}$} & \multirow{3}{*}{$\varphi_1[T_1I]$}\\
& & & $(1/2,1/2,0)$, $(0,1/2,1/2)$, & & \\
& & & $(1/2,0,1/2)$, $(1/2,1/2,1/2)$ & & \\ \hline
\hline 
 \end{tabular} 
 \end{center}

\subsection*{No. 193: $P6_3/mcm$}\label{subsub:sg193}

This group is generated by three translations $T_{1,2,3}$ as given in Eqs.~\eqref{TransBravaisP}, a three-fold rotation $C_3$, a two-fold rotation $C'_2$, a two-fold screw $S_2$, and an inversion $I$:
\begin{subequations}
 \begin{align}
C_3 &\colon (x,y,z)\rightarrow (-y, x - y, z),\\ 
C'_2 &\colon (x,y,z)\rightarrow (y, x, -z + 1/2),\\ 
S_2 &\colon (x,y,z)\rightarrow (-x, -y, z + 1/2),\\ 
I &\colon (x,y,z)\rightarrow (-x, -y, -z).
\end{align}
\end{subequations}

The $\mathbb{Z}_2$ cohomology ring is given by

\begin{equation}
\mathbb{Z}_2[A_i,A_c,A_{c'},B_{xy}]/\langle\mathcal{R}_2,\mathcal{R}_4\rangle
 \end{equation}
where the relations are 
\begin{subequations} 
 \begin{align}
\mathcal{R}_2\colon & ~~
(A_c + A_{c'}) (A_c + A_i),\\
\mathcal{R}_4\colon & ~~
B_{xy} (A_c A_i + A_i^2 + B_{xy}).
\end{align} 
 \end{subequations}
We have the following table regarding IWPs and group cohomology at degree 3.
\begin{center}
\begin{tabular}{c|cc|c|c|c}\hline\hline {Wyckoff}&\multicolumn{2}{c|}{Little group}& \multirow{2}{*}{Coordinates}&\multirow{2}{*}{LSM anomaly class}&\multirow{2}{*}{Topo. inv.} \\ \cline{2-3} position & Intl. & Sch\"{o}nflies & & & \\ \hline
2a&$\overline{6}2m$&$D_{3h}$& $(0,0,1/4)$, $(0,0,3/4)$ & $(A_c + A_{c'}) B_{xy}$ & $\varphi_2[T_1T_2, C'_2]$\\ 
2b&$\overline{3}m$&$D_{3d}$& $(0,0,0)$, $(0,0,1/2)$ & $(A_c + A_i) (A_c A_i + A_i^2 + B_{xy})$ & $\varphi_1[I]$\\ 
\hline
\multirow{2}{*}{4c} & \multirow{2}{*}{$\overline{6}$} & \multirow{2}{*}{$C_{3h}$} & $(1/3,2/3,1/4)$, $(2/3,1/3,3/4)$, & \multirow{2}{*}{N/A} & \multirow{2}{*}{N/A}\\
& & & $(2/3,1/3,1/4)$, $(1/3,2/3,3/4)$ & & \\ \hline
\multirow{2}{*}{4d} & \multirow{2}{*}{$32$} & \multirow{2}{*}{$D_3$} & $(1/3,2/3,0)$, $(2/3,1/3,1/2)$, & \multirow{2}{*}{N/A} & \multirow{2}{*}{N/A}\\
& & & $(2/3,1/3,0)$, $(1/3,2/3,1/2)$ & & \\ \hline
\multirow{3}{*}{6f} & \multirow{3}{*}{$2/m$} & \multirow{3}{*}{$C_{2h}$} & $(1/2,0,0)$, $(0,1/2,0)$, & \multirow{3}{*}{$(A_c + A_i) B_{xy}$} & \multirow{3}{*}{$\varphi_1[T_1I]$}\\
& & & $(1/2,1/2,0)$, $(1/2,0,1/2)$, & & \\
& & & $(0,1/2,1/2)$, $(1/2,1/2,1/2)$ & & \\ \hline
\hline 
 \end{tabular} 
 \end{center}

\subsection*{No. 194: $P6_3/mmc$}\label{subsub:sg194}

This group is generated by three translations $T_{1,2,3}$ as given in Eqs.~\eqref{TransBravaisP}, a three-fold rotation $C_3$, a two-fold rotation $C'_2$, a two-fold screw $S_2$, and an inversion $I$:
\begin{subequations}
 \begin{align}
C_3 &\colon (x,y,z)\rightarrow (-y, x - y, z),\\ 
C'_2 &\colon (x,y,z)\rightarrow (y, x, -z),\\ 
S_2 &\colon (x,y,z)\rightarrow (-x, -y, z + 1/2),\\ 
I &\colon (x,y,z)\rightarrow (-x, -y, -z).
\end{align}
\end{subequations}

The $\mathbb{Z}_2$ cohomology ring is given by

\begin{equation}
\mathbb{Z}_2[A_i,A_c,A_{c'},B_{xy}]/\langle\mathcal{R}_2,\mathcal{R}_4\rangle
 \end{equation}
where the relations are 
\begin{subequations} 
 \begin{align}
\mathcal{R}_2\colon & ~~
A_c (A_c + A_{c'} + A_i),\\
\mathcal{R}_4\colon & ~~
B_{xy} (A_c A_i + A_{c'} A_i + A_i^2 + B_{xy}).
\end{align} 
 \end{subequations}
We have the following table regarding IWPs and group cohomology at degree 3.
\begin{center}
\resizebox{\columnwidth}{!}{
\begin{tabular}{c|cc|c|c|c}\hline\hline {Wyckoff}&\multicolumn{2}{c|}{Little group}& \multirow{2}{*}{Coordinates}&\multirow{2}{*}{LSM anomaly class}&\multirow{2}{*}{Topo. inv.} \\ \cline{2-3} position & Intl. & Sch\"{o}nflies & & & \\ \hline
2a&$\overline{3}m$&$D_{3d}$& $(0,0,0)$, $(0,0,1/2)$ & $(A_c + A_{c'} + A_i) (A_c A_i + A_{c'} A_i + A_i^2 + B_{xy})$ & $\varphi_1[I]$\\ 
2b&$\overline{6}m2$&$D_{3h}$& $(0,0,1/4)$, $(0,0,3/4)$ & $A_c B_{xy}$ & $\varphi_3[T_1, T_2, S_2I]$\\ 
2c&$\overline{6}m2$&$D_{3h}$& $(1/3,2/3,1/4)$, $(2/3,1/3,3/4)$ & Same as 2b  & Same as 2b\\ 
2d&$\overline{6}m2$&$D_{3h}$& $(1/3,2/3,3/4)$, $(2/3,1/3,1/4)$ & Same as 2b  & Same as 2b\\ 
\hline
\multirow{3}{*}{6g} & \multirow{3}{*}{$2m$} & \multirow{3}{*}{$C_{2h}$} & $(1/2,0,0)$, $(0,1/2,0)$, & \multirow{3}{*}{$(A_c + A_{c'} + A_i) B_{xy}$} & \multirow{3}{*}{$\varphi_1[T_1I]$}\\
& & & $(1/2,1/2,0)$, $(1/2,0,1/2)$, & & \\
& & & $(0,1/2,1/2)$, $(1/2,1/2,1/2)$ & & \\ \hline
\hline 
 \end{tabular} }
 \end{center}

\subsection*{No. 195: $P23$}\label{subsub:sg195}

This group is generated by three translations $T_{1,2,3}$ as given in Eqs.~\eqref{TransBravaisP}, a two-fold rotation $C_2$, a two-fold rotation $C'_2$, and a three-fold rotation $C_3$:
\begin{subequations}
 \begin{align}
C_2 &\colon (x,y,z)\rightarrow (-x, -y, z),\\ 
C'_2 &\colon (x,y,z)\rightarrow (-x, y, -z),\\ 
C_3 &\colon (x,y,z)\rightarrow (z, x, y).
\end{align}
\end{subequations}

The $\mathbb{Z}_2$ cohomology ring is given by

\begin{equation}
\mathbb{Z}_2[A_{x+y+z},B_\alpha,B_\beta,B_{xy+xz+yz},C_{\alpha1},C_{\alpha2},C_\beta,C_{xyz}]/\langle\mathcal{R}_4,\mathcal{R}_5,\mathcal{R}_6\rangle
 \end{equation}
where the relations are 
\begin{subequations} 
 \begin{align}
\mathcal{R}_4\colon & ~~
A_{x+y+z} C_{\alpha2}+A_{x+y+z}^4+A_{x+y+z}^2B_\alpha+A_{x+y+z}C_{\alpha1},~~A_{x+y+z}C_{xyz},\nonumber\\
&~~B_\alpha B_{xy+xz+yz}+A_{x+y+z}^4+A_{x+y+z}^2B_\beta+A_{x+y+z}C_{\alpha1}+B_\alpha B_\beta,\nonumber\\
&~~B_\beta^2+A_{x+y+z}^2B_\alpha+A_{x+y+z}C_{\alpha1}+B_\alpha B_\beta,\nonumber\\
&~~B_\beta B_{xy+xz+yz}+A_{x+y+z}^4+A_{x+y+z}^2B_\alpha+A_{x+y+z}C_{\alpha1}+A_{x+y+z}C_\beta+B_\alpha B_\beta,\nonumber\\
&~~B_{xy+xz+yz}^2+A_{x+y+z}^4+A_{x+y+z}^2B_\beta+A_{x+y+z}^2B_{xy+xz+yz}+A_{x+y+z}C_{\alpha1}+B_\alpha B_\beta,\\
\mathcal{R}_5\colon & ~~
B_{xy+xz+yz}C_{\alpha1}+A_{x+y+z}^3B_\alpha+A_{x+y+z}^3B_{xy+xz+yz}+A_{x+y+z}^2C_{\alpha1}+A_{x+y+z}^2C_\beta+A_{x+y+z}B_\alpha^2+B_\beta C_{\alpha1},\nonumber\\
&~~B_{xy+xz+yz}C_{\alpha2}+A_{x+y+z}^5+A_{x+y+z}^3B_\alpha+A_{x+y+z}^2C_\beta+A_{x+y+z}B_\alpha^2+B_\beta C_{\alpha2},\nonumber\\
&~~B_\alpha C_\beta+A_{x+y+z}^2C_{\alpha1}+B_\beta C_{\alpha2},\nonumber\\
&~~B_\beta C_\beta+A_{x+y+z}^5+A_{x+y+z}^3B_{xy+xz+yz}+A_{x+y+z}^2C_{\alpha1}+A_{x+y+z}^2C_\beta+A_{x+y+z}B_\alpha^2+B_\beta C_{\alpha2},\nonumber\\
&~~B_{xy+xz+yz}C_\beta+A_{x+y+z}^2C_{\alpha1}+A_{x+y+z}B_\alpha^2+A_{x+y+z}B_\alpha B_\beta+B_\beta C_{\alpha2},\nonumber\\
&~~B_\alpha C_{xyz}+A_{x+y+z}^3B_\beta+A_{x+y+z}B_\alpha B_\beta+B_\beta C_{\alpha1}+B_\beta C_{\alpha2},\nonumber\\
&~~B_\beta C_{xyz}+A_{x+y+z}^3B_\beta+A_{x+y+z}B_\alpha B_\beta+B_\beta C_{\alpha1}+B_\beta C_{\alpha2},\nonumber\\
&~~B_{xy+xz+yz}C_{xyz}+A_{x+y+z}^3B_\beta+A_{x+y+z}B_\alpha B_\beta+B_\beta C_{\alpha1}+B_\beta C_{\alpha2},\\
\mathcal{R}_6\colon & ~~
C_{\alpha2}^2+B_\alpha^3+C_{\alpha1}^2+C_{\alpha1}C_{\alpha2},\nonumber\\
&~~C_{\alpha2}C_\beta+A_{x+y+z}^4B_\alpha+A_{x+y+z}^4B_{xy+xz+yz}+A_{x+y+z}^3C_\beta+A_{x+y+z}^2B_\alpha^2+A_{x+y+z}B_\beta C_{\alpha1}+B_\alpha^2B_\beta+C_{\alpha1}C_{xyz},\nonumber\\
&~~C_{\alpha2}C_{xyz}+A_{x+y+z}^4B_\alpha+A_{x+y+z}^4B_\beta+A_{x+y+z}^4B_{xy+xz+yz}+A_{x+y+z}^3C_{\alpha1}+A_{x+y+z}^2B_\alpha^2\nonumber\\
&~~+A_{x+y+z}^2B_\alpha B_\beta+B_\alpha^2B_\beta+C_{\alpha1}C_\beta+C_{\alpha1}C_{xyz},\nonumber\\
&~~C_\beta^2+A_{x+y+z}^4B_\alpha+A_{x+y+z}^4B_{xy+xz+yz}+A_{x+y+z}^3C_\beta+A_{x+y+z}B_\alpha C_{\alpha1}+A_{x+y+z}B_\beta C_{\alpha1}+B_\alpha^2B_\beta+C_{\alpha1}C_{xyz},\nonumber\\
&~~C_\beta C_{xyz}+A_{x+y+z}^4B_\alpha+A_{x+y+z}^4B_\beta+A_{x+y+z}^4B_{xy+xz+yz}+A_{x+y+z}^3C_{\alpha1}+A_{x+y+z}^2B_\alpha^2\nonumber\\
&~~+A_{x+y+z}^2B_\alpha B_\beta+B_\alpha^2B_\beta+C_{\alpha1}C_\beta+C_{\alpha1}C_{xyz},\nonumber\\
&~~C_{xyz}^2+A_{x+y+z}^4B_\alpha+A_{x+y+z}^4B_\beta+A_{x+y+z}^4B_{xy+xz+yz}+A_{x+y+z}^3C_{\alpha1}+A_{x+y+z}^2B_\alpha^2\nonumber\\
&~~+A_{x+y+z}^2B_\alpha B_\beta+B_\alpha^2B_\beta+C_{\alpha1}C_\beta.
\end{align} 
 \end{subequations}
We have the following table regarding IWPs and group cohomology at degree 3.
\begin{center}
\resizebox{\columnwidth}{!}{
\begin{tabular}{c|cc|c|c|c}\hline\hline {Wyckoff}&\multicolumn{2}{c|}{Little group}& \multirow{2}{*}{Coordinates}&\multirow{2}{*}{LSM anomaly class}&\multirow{2}{*}{Topo. inv.} \\ \cline{2-3} position & Intl. & Sch\"{o}nflies & & & \\ \hline
1a&$23$&$T$& $(0,0,0)$ & $A_{x+y+z}^3 + A_{x+y+z} B_\alpha + C_{\alpha1} + C_{\alpha2} + C_{xyz}$ & $\varphi_2[C_2, C'_2]$\\ 
1b&$23$&$T$& $(1/2,1/2,1/2)$ & $C_{xyz}$ & $\varphi_2[T_1T_2C_2, T_1T_3C'_2]$\\ 
\hline
\multirow{2}{*}{3c} & \multirow{2}{*}{$222$} & \multirow{2}{*}{$D_2$} & $(0,1/2,1/2)$, $(1/2,0,1/2)$, & \multirow{2}{*}{$A_{x+y+z} B_{xy+xz+yz}$} & \multirow{2}{*}{$\varphi_2[T_1T_2C_2, T_1C'_2]$}\\
& & & $(1/2,1/2,0)$ & & \\ \hline
\multirow{2}{*}{3d} & \multirow{2}{*}{$222$} & \multirow{2}{*}{$D_2$} & $(1/2,0,0)$, $(0,1/2,0)$, & \multirow{2}{*}{$A_{x+y+z}^3 + A_{x+y+z}B_\alpha + A_{x+y+z}B_{xy+xz+yz}$} & \multirow{2}{*}{$\varphi_2[C_2, T_3C'_2]$}\\
& & & $(0,0,1/2)$ & & \\ \hline
\hline 
 \end{tabular} }
 \end{center}

\subsection*{No. 196: $F23$}\label{subsub:sg196}

This group is generated by three translations $T_{1,2,3}$ as given in Eqs.~\eqref{TransBravaisF}, a two-fold rotation $C_2$, a two-fold rotation $C'_2$, and a three-fold rotation $C_3$:
\begin{subequations}
 \begin{align}
C_2 &\colon (x,y,z)\rightarrow (-x, -y, z),\\ 
C'_2 &\colon (x,y,z)\rightarrow (-x, y, -z),\\ 
C_3 &\colon (x,y,z)\rightarrow (z, x, y).
\end{align}
\end{subequations}

The $\mathbb{Z}_2$ cohomology ring is given by

\begin{equation}
\mathbb{Z}_2[B_\alpha,B_\beta,B_{xy+xz+yz},C_{\alpha1},C_{\alpha2},C_{\beta1},C_{\beta2},C_\gamma,C_{xyz}]/\langle\mathcal{R}_4,\mathcal{R}_5,\mathcal{R}_6\rangle
 \end{equation}
where the relations are 
\begin{subequations} 
 \begin{align}
\mathcal{R}_4\colon & ~~
B_\alpha B_{xy+xz+yz},~~B_\beta (B_\alpha + B_\beta),~~B_\beta B_{xy+xz+yz},~~B_{xy+xz+yz}^2,\\
\mathcal{R}_5\colon & ~~
B_{xy+xz+yz} C_{\alpha1},~~B_{xy+xz+yz} C_{\alpha2},~~B_\beta C_{\alpha1} + B_\beta C_{\alpha2} + B_\alpha C_{\beta1},~~B_\beta (C_{\alpha1} + C_{\alpha2} + C_{\beta1}),~~B_{xy+xz+yz} C_{\beta1},\nonumber\\&~~B_\beta C_{\alpha2} + B_\alpha C_{\beta2},~~B_\beta (C_{\alpha2} + C_{\beta2}),~~B_{xy+xz+yz} C_{\beta2},~~B_{xy+xz+yz} C_\gamma,~~B_\beta C_\gamma + B_\alpha C_{xyz},\nonumber\\&~~B_\beta (C_\gamma + C_{xyz}),~~B_{xy+xz+yz} C_{xyz},\\
\mathcal{R}_6\colon & ~~
B_\alpha^3 + C_{\alpha1}^2 + C_{\alpha1} C_{\alpha2} + C_{\alpha2}^2,~~B_\alpha^2 B_\beta + C_{\alpha1} C_{\beta1} + C_{\alpha2} C_{\beta1} + C_{\alpha1} C_{\beta2},~~B_\alpha^2 B_\beta + C_{\alpha1} C_{\beta1} + C_{\alpha2} C_{\beta2},\nonumber\\&~~B_\alpha^2 B_\beta + C_{\beta1}^2 + C_{\alpha1} C_{\beta2},~~B_\alpha^2 B_\beta + C_{\alpha1} C_{\beta1} + C_{\alpha1} C_{\beta2} + C_{\beta1} C_{\beta2},~~C_{\beta1} C_\gamma + C_{\alpha1} C_{xyz} + C_{\alpha2} C_{xyz},\nonumber\\&~~(C_{\alpha1} + C_{\alpha2} + C_{\beta1}) C_{xyz},~~B_\alpha^2 B_\beta + C_{\alpha1} C_{\beta1} + C_{\beta2}^2,~~C_{\beta2} C_\gamma + C_{\alpha2} C_{xyz},~~(C_{\alpha2} + C_{\beta2}) C_{xyz},~~C_\gamma (C_{\alpha1} + C_{\alpha2} + C_\gamma),\nonumber\\&~~(C_{\alpha1} + C_{\alpha2} + C_\gamma) C_{xyz},~~C_{xyz} (C_{\alpha1} + C_{\alpha2} + C_{xyz}).
\end{align} 
 \end{subequations}
We have the following table regarding IWPs and group cohomology at degree 3.
\begin{center}
\begin{tabular}{c|cc|c|c|c}\hline\hline \multirow{3}{*}{\shortstack[l]{Wyckoff\\position}}&\multicolumn{2}{c|}{Little group}& {Coordinates}&\multirow{3}{*}{LSM anomaly class}&\multirow{3}{*}{Topo. inv.} \\ \cline{2-4} & \multirow{2}{*}{Intl.} & \multirow{2}{*}{Sch\"{o}nflies} & $(0,0,0) + ~(0,1/2,1/2) + $ & & \\ & & & $ (1/2,0,1/2) + ~(1/2,1/2,0) +$ & &\\ \hline
4a&$23$&$T$& $(0,0,0)$ & $C_{\alpha1} + C_{\alpha2} + C_{\beta1} + C_\gamma + C_{xyz}$ & $\varphi_2[C_2, C'_2]$\\ 
4b&$23$&$T$& $(1/2,1/2,1/2)$ & $C_\gamma + C_{xyz}$ & $\varphi_2[C_2, T_1T_2T_3^{-1}C'_2]$\\ 
4c&$23$&$T$& $(1/4,1/4,1/4)$ & $C_{xyz}$ & $\varphi_2[T_3C_2, T_2C'_2]$\\ 
4d&$23$&$T$& $(3/4,3/4,3/4)$ & $C_{\beta1} + C_{xyz}$ & $\varphi_2[T_3^{-1}C_2, T_2^{-1}C'_2]$\\ 
\hline
\hline 
 \end{tabular} 
 \end{center}

\subsection*{No. 197: $I23$}\label{subsub:sg197}

This group is generated by three translations $T_{1,2,3}$ as given in Eqs.~\eqref{TransBravaisI}, a two-fold rotation $C_2$, a two-fold rotation $C'_2$, and a three-fold rotation $C_3$:
\begin{subequations}
 \begin{align}
C_2 &\colon (x,y,z)\rightarrow (-x, -y, z),\\ 
C'_2 &\colon (x,y,z)\rightarrow (-x, y, -z),\\ 
C_3 &\colon (x,y,z)\rightarrow (z, x, y).
\end{align}
\end{subequations}

The $\mathbb{Z}_2$ cohomology ring is given by

\begin{equation}
\mathbb{Z}_2[A_{x+y+z},B_\alpha,B_\beta,C_{\alpha1},C_{\alpha2},C_\gamma,C_{xyz},D_\gamma]/\langle\mathcal{R}_2,\mathcal{R}_3,\mathcal{R}_4,\mathcal{R}_5,\mathcal{R}_6,\mathcal{R}_7,\mathcal{R}_8\rangle
 \end{equation}
where the relations are 
\begin{subequations} 
 \begin{align}
\mathcal{R}_2\colon & ~~
A_{x+y+z}^2,\\
\mathcal{R}_3\colon & ~~
A_{x+y+z} B_\alpha,~~A_{x+y+z} B_\beta,\\
\mathcal{R}_4\colon & ~~
A_{x+y+z} C_{\alpha1},~~A_{x+y+z} C_{\alpha2},~~A_{x+y+z} C_\gamma,~~A_{x+y+z} C_{xyz},\\
\mathcal{R}_5\colon & ~~
A_{x+y+z} D_\gamma,~~B_\alpha C_\gamma + B_\beta C_\gamma + B_\alpha C_{xyz},~~B_\beta (C_{\alpha1} + C_{\alpha2} + C_{xyz}),\\
\mathcal{R}_6\colon & ~~
B_\alpha^2 B_\beta + B_\alpha B_\beta^2 + C_{\alpha1} C_\gamma + B_\alpha D_\gamma,~~B_\alpha^2 B_\beta + B_\beta^3 + C_{\alpha1} C_{xyz} + B_\alpha D_\gamma + B_\beta D_\gamma,~~B_\alpha^3 + C_{\alpha1}^2 + C_{\alpha1} C_{\alpha2} + C_{\alpha2}^2,\nonumber\\
&~~C_{\alpha2} C_\gamma + B_\alpha D_\gamma,~~C_{\alpha2} C_{xyz} + B_\alpha D_\gamma + B_\beta D_\gamma,~~B_\alpha^2 B_\beta + C_\gamma^2,~~B_\alpha^2 B_\beta + B_\alpha B_\beta^2 + C_\gamma C_{xyz},~~B_\alpha^2 B_\beta + B_\beta^3 + C_{xyz}^2,\\
\mathcal{R}_7\colon & ~~
B_\alpha B_\beta C_{\alpha1} + B_\beta^2 C_{\alpha1} + B_\alpha B_\beta C_{\alpha2} + B_\beta^2 C_{\alpha2} + B_\alpha^2 C_\gamma + C_{\alpha1} D_\gamma,~~B_\alpha B_\beta C_{\alpha1} + B_\beta^2 C_{\alpha1} + B_\alpha^2 C_\gamma + C_{\alpha2} D_\gamma,\nonumber\\
&~~B_\alpha B_\beta C_{\alpha2} + C_\gamma D_\gamma,~~B_\alpha B_\beta C_{\alpha2} + B_\beta^2 C_{\alpha2} + C_{xyz} D_\gamma,\\
\mathcal{R}_8\colon & ~~
B_\alpha^3 B_\beta + B_\alpha^2 B_\beta^2 + B_\beta^4 + B_\alpha B_\beta D_\gamma + B_\beta^2 D_\gamma + D_\gamma^2.
\end{align} 
 \end{subequations}
We have the following table regarding IWPs and group cohomology at degree 3.
\begin{center}
\begin{tabular}{c|cc|c|c|c}\hline\hline {Wyckoff}&\multicolumn{2}{c|}{Little group}& {Coordinates}&\multirow{2}{*}{LSM anomaly class}&\multirow{2}{*}{Topo. inv.} \\ \cline{2-4} position & Intl. & Sch\"{o}nflies & $(0,0,0) + ~(1/2,1/2,1/2) + $ & &\\ \hline
2a&$23$&$T$& $(0,0,0)$ & $C_{\alpha1} + C_{\alpha2} + C_{xyz}$ & $\varphi_2[C_2, C'_2]$\\ 
\hline
\multirow{2}{*}{6b} & \multirow{2}{*}{$222$} & \multirow{2}{*}{$D_2$} & $(0,1/2,1/2)$, $(1/2,0,1/2)$, & \multirow{2}{*}{$C_{xyz}$} & \multirow{2}{*}{$\varphi_2[C_2, T_1T_2C'_2]$}\\
& & & $(1/2,1/2,0)$ & & \\ \hline
\hline 
 \end{tabular} 
 \end{center}

\subsection*{No. 198: $P2_13$}\label{subsub:sg198}

This group is generated by three translations $T_{1,2,3}$ as given in Eqs.~\eqref{TransBravaisP}, a two-fold screw $S_2$, a two-fold screw $S'_2$, and a three-fold rotation $C_3$:
\begin{subequations}
 \begin{align}
S_2 &\colon (x,y,z)\rightarrow (-x + 1/2, -y, z + 1/2),\\ 
S'_2 &\colon (x,y,z)\rightarrow (-x, y + 1/2, -z + 1/2),\\ 
C_3 &\colon (x,y,z)\rightarrow (z, x, y).
\end{align}
\end{subequations}

The $\mathbb{Z}_2$ cohomology ring is given by

\begin{equation}
\mathbb{Z}_2[C_\beta]/\langle\mathcal{R}_6\rangle
 \end{equation}
where the relations are 
\begin{subequations} 
 \begin{align}
\mathcal{R}_6\colon & ~~
C_\beta^2.
\end{align} 
 \end{subequations}
We have the following table regarding IWPs and group cohomology at degree 3.
\begin{center}
\resizebox{\columnwidth}{!}{
\begin{tabular}{c|cc|c|c|c}\hline\hline {Wyckoff}&\multicolumn{2}{c|}{Little group}& \multirow{2}{*}{Coordinates}&\multirow{2}{*}{LSM anomaly class}&\multirow{2}{*}{Topo. inv.} \\ \cline{2-3} position & Intl. & Sch\"{o}nflies & & & \\ \hline
\multirow{2}{*}{4a} & \multirow{2}{*}{$3$} & \multirow{2}{*}{$C_3$} & $(x,x,x)$, $(-x+1/2,-x,x+1/2)$, & \multirow{2}{*}{$C_\beta$} & \multirow{2}{*}{$\widehat{\varphi_4}[T_2, T_3, S_2, S'_2]$}\\
& & & $(-x,x+1/2,-x+1/2)$, $(x+1/2,-x+1/2,-x)$ & & \\ \hline
\hline 
 \end{tabular} }
 \end{center}
 Here the topological invariant $\widehat{\varphi_4}[T_2, T_3, S_2, S'_2]$ can be chosen to be the same as that of group No.19 $P2_12_12_1$, given by Eq.~\eqref{TI_19}.

\subsection*{No. 199: $I2_13$}\label{subsub:sg199}

This group is generated by three translations $T_{1,2,3}$ as given in Eqs.~\eqref{TransBravaisI}, a two-fold rotation $C_2$, a two-fold rotation $C'_2$, and a three-fold rotation $C_3$:
\begin{subequations}
 \begin{align}
C_2 &\colon (x,y,z)\rightarrow (-x, -y + 1/2, z),\\ 
C'_2 &\colon (x,y,z)\rightarrow (-x + 1/2, y, -z),\\ 
C_3 &\colon (x,y,z)\rightarrow (z, x, y).
\end{align}
\end{subequations}

The $\mathbb{Z}_2$ cohomology ring is given by

\begin{equation}
\mathbb{Z}_2[A_{x+y+z},C_\gamma]/\langle\mathcal{R}_6\rangle
 \end{equation}
where the relations are 
\begin{subequations} 
 \begin{align}
\mathcal{R}_6\colon & ~~
C_\gamma^2.
\end{align} 
 \end{subequations}
We have the following table regarding IWPs and group cohomology at degree 3.
\begin{center}
\begin{tabular}{c|cc|c|c|c}\hline\hline {Wyckoff}&\multicolumn{2}{c|}{Little group}& {Coordinates}&\multirow{2}{*}{LSM anomaly class}&\multirow{2}{*}{Topo. inv.} \\ \cline{2-4} position & Intl. & Sch\"{o}nflies & $(0,0,0) + ~(1/2,1/2,1/2) + $ & &\\ \hline
\multirow{2}{*}{8a} & \multirow{2}{*}{$3$} & \multirow{2}{*}{$C_3$} & $(x,x,x)$, $(-x+1/2,-x,x+1/2)$, & \multirow{2}{*}{N/A} & \multirow{2}{*}{N/A}\\
& & & $(-x,x+1/2,-x+1/2)$, $(x+1/2,-x+1/2,-x)$ & & \\ \hline
\multirow{3}{*}{12b} & \multirow{3}{*}{$2$} & \multirow{3}{*}{$C_2$} & $(x,0,1/4)$, $(-x+1/2,0,3/4)$, & \multirow{3}{*}{$C_\gamma$} & \multirow{3}{*}{$\varphi_2[T_1T_2, C_2]$}\\
& & & $(1/4,x,0)$, $(3/4,-x+1/2,0)$, & & \\
& & & $(0,1/4,x)$, $(0,3/4,-x+1/2)$ & & \\ \hline
\hline 
 \end{tabular} 
 \end{center}

\subsection*{No. 200: $Pm\overline3$}\label{subsub:sg200}

This group is generated by three translations $T_{1,2,3}$ as given in Eqs.~\eqref{TransBravaisP}, a two-fold rotation $C_2$, a two-fold rotation $C'_2$, a three-fold rotation $C_3$, and an inversion $I$:
\begin{subequations}
 \begin{align}
C_2 &\colon (x,y,z)\rightarrow (-x, -y, z),\\ 
C'_2 &\colon (x,y,z)\rightarrow (-x, y, -z),\\ 
C_3 &\colon (x,y,z)\rightarrow (z, x, y),\\ 
I &\colon (x,y,z)\rightarrow (-x, -y, -z).
\end{align}
\end{subequations}

The $\mathbb{Z}_2$ cohomology ring is given by

\begin{equation}
\mathbb{Z}_2[A_i,A_{x+y+z},B_\alpha,B_\beta,B_{xy+xz+yz},C_{\alpha1},C_{\alpha2},C_\beta,C_{xyz}]/\langle\mathcal{R}_4,\mathcal{R}_5,\mathcal{R}_6\rangle
 \end{equation}
where the relations are 
\begin{subequations} 
 \begin{align}
\mathcal{R}_4\colon & ~~
A_i^3 A_{x+y+z} + A_i^2 A_{x+y+z}^2 + A_i A_{x+y+z}^3 + A_{x+y+z}^4 + A_i A_{x+y+z} B_\alpha + A_{x+y+z}^2 B_\alpha  \nonumber \\
&~~ + A_i A_{x+y+z} B_{xy+xz+yz} + A_{x+y+z} C_{\alpha1} + A_{x+y+z} C_{\alpha2} + A_i C_{xyz},\nonumber\\
&~~A_{x+y+z} (A_i^3 + A_i^2 A_{x+y+z} + A_i A_{x+y+z}^2 + A_{x+y+z}^3 + A_i B_\alpha + A_{x+y+z} B_\alpha + A_i B_{xy+xz+yz} + C_{\alpha1} + C_{\alpha2} + C_{xyz}),\nonumber\\
&~~A_i^2 A_{x+y+z}^2 + A_i A_{x+y+z}^3 + A_i A_{x+y+z} B_\alpha + A_{x+y+z}^2 B_\alpha + A_i A_{x+y+z} B_\beta + A_{x+y+z}^2 B_\beta \nonumber \\
&~~+ B_\alpha B_\beta 
+ B_\alpha B_{xy+xz+yz} + A_{x+y+z} C_{\alpha2} + A_i C_\beta,\nonumber\\
&~~A_{x+y+z}^4 + A_i A_{x+y+z} B_\alpha + B_\alpha B_\beta + B_\beta^2 + A_{x+y+z} C_{\alpha2} + A_i C_\beta,\nonumber\\
&~~A_i^3 A_{x+y+z} + A_i^2 A_{x+y+z}^2 + A_i^2 B_\beta + A_i A_{x+y+z} B_\beta + B_\alpha B_\beta + A_i A_{x+y+z} B_{xy+xz+yz} \nonumber \\
&~~ + B_\beta B_{xy+xz+yz} + A_{x+y+z} C_{\alpha2} + A_{x+y+z} C_\beta,\nonumber\\
&~~A_i^2 A_{x+y+z}^2 + A_i A_{x+y+z}^3 + A_i A_{x+y+z} B_\alpha + A_{x+y+z}^2 B_\alpha + A_i A_{x+y+z} B_\beta + A_{x+y+z}^2 B_\beta + B_\alpha B_\beta + A_i^2 B_{xy+xz+yz}\nonumber \\
&~~ + A_i A_{x+y+z} B_{xy+xz+yz} + A_{x+y+z}^2 B_{xy+xz+yz} + B_{xy+xz+yz}^2 + A_{x+y+z} C_{\alpha2} + A_i C_\beta,\\
\mathcal{R}_5\colon & ~~
A_i^4 A_{x+y+z} + A_i^3 A_{x+y+z}^2 + A_i^2 A_{x+y+z}^3 + A_i A_{x+y+z}^4 + A_i^2 A_{x+y+z} B_\alpha + A_i A_{x+y+z}^2 B_\alpha \nonumber\\
&~~ + A_{x+y+z}^3 B_\alpha + A_{x+y+z} B_\alpha^2 + A_i B_\alpha B_\beta + A_i^2 A_{x+y+z} B_{xy+xz+yz} + A_{x+y+z}^3 B_{xy+xz+yz}\nonumber\\
&~~+ A_i A_{x+y+z} C_{\alpha1} + A_{x+y+z}^2 C_{\alpha1} + B_\beta C_{\alpha1} + B_{xy+xz+yz} C_{\alpha1} + A_i A_{x+y+z} C_\beta + A_{x+y+z}^2 C_\beta,\nonumber\\
&~~A_i^4 A_{x+y+z} + A_{x+y+z}^5 + A_{x+y+z}^3 B_\alpha + A_{x+y+z} B_\alpha^2 + A_i B_\alpha B_\beta + A_i^2 A_{x+y+z} B_{xy+xz+yz} + A_i A_{x+y+z}^2 B_{xy+xz+yz}\nonumber\\
&~~+ B_\beta C_{\alpha2} + B_{xy+xz+yz} C_{\alpha2} + A_i A_{x+y+z} C_\beta + A_{x+y+z}^2 C_\beta,\nonumber\\
&~~A_i A_{x+y+z} C_{\alpha1} + A_{x+y+z}^2 C_{\alpha1} + A_i A_{x+y+z} C_{\alpha2} + B_\beta C_{\alpha2} + B_\alpha C_\beta,\nonumber\\
&~~A_i^3 A_{x+y+z}^2 + A_i^2 A_{x+y+z}^3 + A_i A_{x+y+z}^4 + A_{x+y+z}^5 + A_i^2 A_{x+y+z} B_\alpha + A_{x+y+z} B_\alpha^2 + A_i B_\alpha B_\beta + A_i A_{x+y+z}^2 B_{xy+xz+yz} \nonumber\\
&~~+ A_{x+y+z}^3 B_{xy+xz+yz} + A_i A_{x+y+z} C_{\alpha1} + A_{x+y+z}^2 C_{\alpha1} + A_i A_{x+y+z} C_{\alpha2} + B_\beta C_{\alpha2} + A_{x+y+z}^2 C_\beta + B_\beta C_\beta,\nonumber\\
&~~A_i A_{x+y+z}^2 B_\alpha + A_{x+y+z} B_\alpha^2 + A_{x+y+z} B_\alpha B_\beta + A_i A_{x+y+z} C_{\alpha1}\nonumber\\
&~~+ A_{x+y+z}^2 C_{\alpha1} + B_\beta C_{\alpha2} + A_i^2 C_\beta + A_i A_{x+y+z} C_\beta + B_{xy+xz+yz} C_\beta,\nonumber\\
&~~A_i^2 A_{x+y+z}^3 + A_i A_{x+y+z}^4 + A_i^2 A_{x+y+z} B_\alpha + A_i A_{x+y+z}^2 B_\alpha + A_i A_{x+y+z}^2 B_\beta + A_{x+y+z}^3 B_\beta + A_{x+y+z} B_\alpha B_\beta\nonumber\\
&~~+ A_i A_{x+y+z} C_{\alpha1} + B_\beta C_{\alpha1} + B_\beta C_{\alpha2} + A_i A_{x+y+z} C_\beta + B_\alpha C_{xyz},\nonumber\\
&~~(B_\beta + B_{xy+xz+yz}) C_{xyz},\nonumber\\
&~~A_i^4 A_{x+y+z} + A_i^3 A_{x+y+z}^2 + A_i A_{x+y+z}^2 B_\beta + A_{x+y+z}^3 B_\beta + A_{x+y+z} B_\alpha B_\beta + A_i^2 A_{x+y+z} B_{xy+xz+yz} + B_\beta C_{\alpha1}\nonumber\\
&~~+ A_i A_{x+y+z} C_{\alpha2} + B_\beta C_{\alpha2} + A_i A_{x+y+z} C_\beta + B_{xy+xz+yz} C_{xyz},\\
\mathcal{R}_6\colon & ~~
B_\alpha^3 + C_{\alpha1}^2 + C_{\alpha1} C_{\alpha2} + C_{\alpha2}^2,~~(C_{\alpha2} + C_\beta) C_{xyz},\nonumber\\
&~~A_i^4 A_{x+y+z}^2 + A_i^3 A_{x+y+z}^3 + A_i^2 A_{x+y+z}^4 + A_i A_{x+y+z}^5 + A_i^2 A_{x+y+z}^2 B_\alpha + A_i A_{x+y+z}^3 B_\alpha + A_{x+y+z}^4 B_\alpha + A_i A_{x+y+z} B_\alpha^2\nonumber\\
&~~+ A_{x+y+z}^2 B_\alpha^2 + A_i A_{x+y+z} B_\alpha B_\beta + B_\alpha^2 B_\beta + A_i^2 A_{x+y+z}^2 B_{xy+xz+yz} + A_{x+y+z}^4 B_{xy+xz+yz} + A_i^2 A_{x+y+z} C_{\alpha1}\nonumber\\
&~~+ A_i A_{x+y+z}^2 C_{\alpha1} + A_{x+y+z} B_\beta C_{\alpha1} + A_i A_{x+y+z}^2 C_\beta + A_{x+y+z}^3 C_\beta + C_{\alpha2} C_\beta + C_{\alpha1} C_{xyz},\nonumber\\
&~~A_i^5 A_{x+y+z} + A_i^4 A_{x+y+z}^2 + A_i^3 A_{x+y+z}^3 + A_i^2 A_{x+y+z}^4 + A_i^3 A_{x+y+z} B_\alpha + A_i^2 A_{x+y+z}^2 B_\alpha + A_{x+y+z}^4 B_\alpha \nonumber\\
&~~+ A_i A_{x+y+z} B_\alpha B_\beta + B_\alpha^2 B_\beta + A_i^3 A_{x+y+z} B_{xy+xz+yz} + A_{x+y+z}^4 B_{xy+xz+yz} + A_i^2 A_{x+y+z} C_{\alpha1}\nonumber\\
&~~+ A_{x+y+z} B_\alpha C_{\alpha1} + A_i B_\beta C_{\alpha1} + A_{x+y+z} B_\beta C_{\alpha1} + A_i A_{x+y+z}^2 C_\beta + A_{x+y+z}^3 C_\beta + C_\beta^2 + C_{\alpha1} C_{xyz},\nonumber\\&~~A_i^4 A_{x+y+z}^2 + A_i^2 A_{x+y+z}^4 + A_i^3 A_{x+y+z} B_\alpha + A_i^2 A_{x+y+z}^2 B_\alpha + A_i A_{x+y+z} B_\alpha^2 + A_i^2 A_{x+y+z}^2 B_\beta + A_{x+y+z}^4 B_\beta \nonumber\\
&~~+ A_i A_{x+y+z} B_\alpha B_\beta + A_{x+y+z}^2 B_\alpha B_\beta + B_\alpha^2 B_\beta + A_i^2 A_{x+y+z}^2 B_{xy+xz+yz} + A_{x+y+z}^4 B_{xy+xz+yz} + A_i^2 A_{x+y+z} C_{\alpha1}\nonumber\\
&~~+ A_i A_{x+y+z}^2 C_{\alpha1} + A_{x+y+z}^3 C_{\alpha1} + A_{x+y+z} B_\alpha C_{\alpha1} + A_i^2 A_{x+y+z} C_{\alpha2} + A_{x+y+z} B_\alpha C_{\alpha2} + A_i^2 A_{x+y+z} C_\beta\nonumber\\
&~~ + A_i A_{x+y+z}^2 C_\beta + C_{\alpha1} C_\beta + C_{\alpha1} C_{xyz} + C_\beta C_{xyz},\nonumber\\&~~A_i^2 A_{x+y+z}^4 + A_i A_{x+y+z}^5 + A_i A_{x+y+z}^3 B_\alpha + A_{x+y+z}^4 B_\alpha + A_{x+y+z}^2 B_\alpha^2 + A_i A_{x+y+z}^3 B_\beta + A_{x+y+z}^4 B_\beta + A_{x+y+z}^2 B_\alpha B_\beta \nonumber\\
&~~+ B_\alpha^2 B_\beta + A_{x+y+z}^4 B_{xy+xz+yz} + A_i^2 A_{x+y+z} C_{\alpha1} + A_{x+y+z}^3 C_{\alpha1} + A_i^2 A_{x+y+z} C_{\alpha2} + A_i A_{x+y+z}^2 C_\beta + C_{\alpha1} C_\beta + C_{xyz}^2.
\end{align} 
 \end{subequations}
We have the following table regarding IWPs and group cohomology at degree 3.
\begin{center}
\resizebox{\columnwidth}{!}{
\begin{tabular}{c|cc|c|c|c}\hline\hline {Wyckoff}&\multicolumn{2}{c|}{Little group}& \multirow{2}{*}{Coordinates}&\multirow{2}{*}{LSM anomaly class}&\multirow{2}{*}{Topo. inv.} \\ \cline{2-3} position & Intl. & Sch\"{o}nflies & & & \\ \hline
1a&$m\overline{3}$&$T_h$& $(0,0,0)$ & \begin{tabular}{c}
$(A_i + A_{x+y+z})^3 + (A_i + A_{x+y+z})B_\alpha+A_iB_{xy + xz + yz} + C_{\alpha1} + C_{\alpha2} + C_{xyz}$\end{tabular}
& $\varphi_1[I]$\\ 
1b&$m\overline{3}$&$T_h$& $(1/2,1/2,1/2)$ & $C_{xyz}$ & $\varphi_1[T_1T_2T_3I]$\\ 
\hline
\multirow{2}{*}{3c} & \multirow{2}{*}{$mmm$} & \multirow{2}{*}{$D_{2h}$} & $(0,1/2,1/2)$, $(1/2,0,1/2)$, & \multirow{2}{*}{$(A_i + A_{x+y+z}) B_{xy+xz+yz}$} & \multirow{2}{*}{$\varphi_2[T_1T_2C_2, T_1C'_2]$}\\
& & & $(1/2,1/2,0)$ & & \\ \hline
\multirow{2}{*}{3d} & \multirow{2}{*}{$mmm$} & \multirow{2}{*}{$D_{2h}$} & $(1/2,0,0)$, $(0,1/2,0)$, & \multirow{2}{*}{$A_{x+y+z} (A_i^2 + A_i A_{x+y+z} + A_{x+y+z}^2 + B_\alpha + B_{xy+xz+yz})$} & \multirow{2}{*}{$\varphi_2[T_1C_2, T_1C'_2]$}\\
& & & $(0,0,1/2)$ & & \\ \hline
\hline 
 \end{tabular} }
 \end{center}

\subsection*{No. 201: $Pn\overline3$}\label{subsub:sg201}

This group is generated by three translations $T_{1,2,3}$ as given in Eqs.~\eqref{TransBravaisP}, a two-fold rotation $C_2$, a two-fold rotation $C'_2$, a three-fold rotation $C_3$, and an inversion $I$:
\begin{subequations}
 \begin{align}
C_2 &\colon (x,y,z)\rightarrow (-x + 1/2, -y + 1/2, z),\\ 
C'_2 &\colon (x,y,z)\rightarrow (-x + 1/2, y, -z + 1/2),\\ 
C_3 &\colon (x,y,z)\rightarrow (z, x, y),\\ 
I &\colon (x,y,z)\rightarrow (-x, -y, -z).
\end{align}
\end{subequations}

The $\mathbb{Z}_2$ cohomology ring is given by

\begin{equation}
\mathbb{Z}_2[A_i,A_{x+y+z},B_\alpha,C_{\alpha1},C_{\alpha2}]/\langle\mathcal{R}_3,\mathcal{R}_4,\mathcal{R}_6\rangle
 \end{equation}
where the relations are 
\begin{subequations} 
 \begin{align}
\mathcal{R}_3\colon & ~~
A_i A_{x+y+z} (A_i + A_{x+y+z}),~~A_i B_\alpha,\\
\mathcal{R}_4\colon & ~~
A_i C_{\alpha1},~~A_i C_{\alpha2},~~A_{x+y+z} (A_i^3 + A_{x+y+z}^3 + A_{x+y+z} B_\alpha + C_{\alpha1} + C_{\alpha2}),\\
\mathcal{R}_6\colon & ~~
B_\alpha^3 + C_{\alpha1}^2 + C_{\alpha1} C_{\alpha2} + C_{\alpha2}^2.
\end{align} 
 \end{subequations}
We have the following table regarding IWPs and group cohomology at degree 3.
\begin{center}
\resizebox{\columnwidth}{!}{
\begin{tabular}{c|cc|c|c|c}\hline\hline {Wyckoff}&\multicolumn{2}{c|}{Little group}& \multirow{2}{*}{Coordinates}&\multirow{2}{*}{LSM anomaly class}&\multirow{2}{*}{Topo. inv.} \\ \cline{2-3} position & Intl. & Sch\"{o}nflies & & & \\ \hline
2a&$23$&$T$& $(1/4,1/4,1/4)$, $(3/4,3/4,3/4)$ & $A_i^2 A_{x+y+z} + A_{x+y+z}^3 + A_{x+y+z} B_\alpha + C_{\alpha1} + C_{\alpha2}$ & $\varphi_2[C_2, C'_2]$\\ 
\hline
\multirow{2}{*}{4b} & \multirow{2}{*}{$\overline{3}$} & \multirow{2}{*}{$C_{3i}$} & $(0,0,0)$, $(1/2,1/2,0)$, & \multirow{2}{*}{$A_i^2 (A_i + A_{x+y+z})$} & \multirow{2}{*}{$\varphi_1[I]$}\\
& & & $(1/2,0,1/2)$, $(0,1/2,1/2)$ & & \\ \hline
\multirow{2}{*}{4c} & \multirow{2}{*}{$\overline{3}$} & \multirow{2}{*}{$C_{3i}$} & $(1/2,1/2,1/2)$, $(0,0,1/2)$, & \multirow{2}{*}{$A_i^2 A_{x+y+z}$} & \multirow{2}{*}{$\varphi_1[T_1I]$}\\
& & & $(0,1/2,0)$, $(1/2,0,0)$ & & \\ \hline
\multirow{3}{*}{6d} & \multirow{3}{*}{$222$} & \multirow{3}{*}{$D_2$} & $(1/4,3/4,3/4)$, $(3/4,1/4,3/4)$, & \multirow{3}{*}{$A_{x+y+z} (A_i^2 + A_{x+y+z}^2 + B_\alpha)$} & \multirow{3}{*}{$\varphi_2[C_2, T_3C'_2]$}\\
& & & $(3/4,3/4,1/4)$, $(3/4,1/4,1/4)$, & & \\
& & & $(1/4,3/4,1/4)$, $(1/4,1/4,3/4)$ & & \\ \hline
\hline 
 \end{tabular} }
 \end{center}

\subsection*{No. 202: $Fm\overline3$}\label{subsub:sg202}

This group is generated by three translations $T_{1,2,3}$ as given in Eqs.~\eqref{TransBravaisF}, a two-fold rotation $C_2$, a two-fold rotation $C'_2$, a three-fold rotation $C_3$, and an inversion $I$:
\begin{subequations}
 \begin{align}
C_2 &\colon (x,y,z)\rightarrow (-x, -y, z),\\ 
C'_2 &\colon (x,y,z)\rightarrow (-x, y, -z),\\ 
C_3 &\colon (x,y,z)\rightarrow (z, x, y),\\ 
I &\colon (x,y,z)\rightarrow (-x, -y, -z).
\end{align}
\end{subequations}

The $\mathbb{Z}_2$ cohomology ring is given by

\begin{equation}
\mathbb{Z}_2[A_i,B_\alpha,B_\beta,B_{xy+xz+yz},C_{\alpha1},C_{\alpha2},C_{\beta1},C_{\beta2},C_{xyz}]/\langle\mathcal{R}_4,\mathcal{R}_5,\mathcal{R}_6\rangle
 \end{equation}
where the relations are 
\begin{subequations} 
 \begin{align}
\mathcal{R}_4\colon & ~~
A_i C_{\beta2},~~A_i^2 B_\beta + B_\alpha B_{xy+xz+yz} + A_i C_{\beta1},\nonumber\\&~~A_i^2 B_\beta + B_\alpha B_\beta + B_\beta^2 + A_i C_{\beta1},~~B_\beta B_{xy+xz+yz} + A_i C_{\beta1},~~A_i^2 B_\beta + A_i^2 B_{xy+xz+yz} + B_{xy+xz+yz}^2,\\
\mathcal{R}_5\colon & ~~
A_i B_\alpha B_\beta + B_{xy+xz+yz} C_{\alpha1},~~A_i B_\alpha B_\beta + B_{xy+xz+yz} C_{\alpha2},~~A_i B_\alpha B_\beta + B_\beta C_{\alpha2} + B_\alpha C_{\beta1},\nonumber\\&~~A_i^3 B_\beta + B_\beta C_{\alpha2} + A_i^2 C_{\beta1} + B_\beta C_{\beta1},~~A_i^3 B_\beta + A_i B_\alpha B_\beta + B_{xy+xz+yz} C_{\beta1},~~B_\beta C_{\alpha1} + B_\beta C_{\alpha2} + B_\alpha C_{\beta2},\nonumber\\&~~B_\beta (C_{\alpha1} + C_{\alpha2} + C_{\beta2}),~~B_{xy+xz+yz} C_{\beta2},~~B_\beta (A_i^3 + A_i B_\alpha + C_{xyz}),\nonumber\\&~~A_i^3 B_\beta + A_i^3 B_{xy+xz+yz} + A_i^2 C_{\beta1} + B_{xy+xz+yz} C_{xyz},\\
\mathcal{R}_6\colon & ~~
B_\alpha^3 + C_{\alpha1}^2 + C_{\alpha1} C_{\alpha2} + C_{\alpha2}^2,~~B_\alpha^2 B_\beta + A_i B_\beta C_{\alpha1} + C_{\alpha2} C_{\beta1} + C_{\alpha1} C_{\beta2},\nonumber\\&~~B_\alpha^2 B_\beta + A_i B_\beta C_{\alpha1} + C_{\alpha1} C_{\beta1} + C_{\alpha1} C_{\beta2} + C_{\alpha2} C_{\beta2},\nonumber\\&~~A_i^4 B_\beta + A_i^2 B_\alpha B_\beta + B_\alpha^2 B_\beta + A_i B_\beta C_{\alpha1} + A_i^3 C_{\beta1} + C_{\beta1}^2 + C_{\alpha1} C_{\beta2},\nonumber\\&~~B_\alpha^2 B_\beta + A_i B_\beta C_{\alpha1} + C_{\alpha1} C_{\beta1} + C_{\alpha1} C_{\beta2} + C_{\beta1} C_{\beta2},~~A_i^2 B_\alpha B_\beta + A_i B_\beta C_{\alpha1} + A_i^3 C_{\beta1} + C_{\beta1} C_{xyz},\nonumber\\&~~B_\alpha^2 B_\beta + A_i B_\beta C_{\alpha1} + C_{\alpha1} C_{\beta1} + C_{\beta2}^2,~~C_{\beta2} C_{xyz},~~C_{xyz} (A_i^3 + A_i B_\alpha + C_{\alpha1} + C_{\alpha2} + C_{xyz}).
\end{align} 
 \end{subequations}
We have the following table regarding IWPs and group cohomology at degree 3.
\begin{center}
\resizebox{\columnwidth}{!}{
\begin{tabular}{c|cc|c|c|c}\hline\hline \multirow{3}{*}{\shortstack[l]{Wyckoff\\position}}&\multicolumn{2}{c|}{Little group}& {Coordinates}&\multirow{3}{*}{LSM anomaly class}&\multirow{3}{*}{Topo. inv.} \\ \cline{2-4} & \multirow{2}{*}{Intl.} & \multirow{2}{*}{Sch\"{o}nflies} & $(0,0,0) + ~(0,1/2,1/2) + $ & & \\ & & & $ (1/2,0,1/2) + ~(1/2,1/2,0) +$ & &\\ \hline
4a&$m\overline{3}$&$T_h$& $(0,0,0)$ & $A_i^3 + A_i B_\alpha + A_i B_\beta + A_i B_{xy+xz+yz} + C_{\alpha1} + C_{\alpha2} + C_{\beta2} + C_{xyz}$ & $\varphi_1[I]$\\ 
4b&$m\overline{3}$&$T_h$& $(1/2,1/2,1/2)$ & $C_{xyz}$ & $\varphi_1[T_1T_2T_3^{-1}I]$\\ 
8c&$23$&$T$& $(1/4,1/4,1/4)$, $(3/4,3/4,3/4)$ & $C_{\beta2}$ & $\varphi_2[T_3C_2, T_2C'_2]$\\ 
\hline
\multirow{3}{*}{24d} & \multirow{3}{*}{$2/m$} & \multirow{3}{*}{$C_{2h}$} & $(0,1/4,1/4)$, $(0,3/4,1/4)$, & \multirow{3}{*}{$A_i (B_\beta + B_{xy+xz+yz})$} & \multirow{3}{*}{$\varphi_1[T_1I]$}\\
& & & $(1/4,0,1/4)$, $(1/4,0,3/4)$, & & \\
& & & $(1/4,1/4,0)$, $(3/4,1/4,0)$ & & \\ \hline
\hline 
 \end{tabular} }
 \end{center}

\subsection*{No. 203: $Fd\overline3$}\label{subsub:sg203}

This group is generated by three translations $T_{1,2,3}$ as given in Eqs.~\eqref{TransBravaisF}, a two-fold rotation $C_2$, a two-fold rotation $C'_2$, a three-fold rotation $C_3$, and an inversion $I$:
\begin{subequations}
 \begin{align}
C_2 &\colon (x,y,z)\rightarrow (-x + 1/4, -y + 1/4, z),\\ 
C'_2 &\colon (x,y,z)\rightarrow (-x + 1/4, y, -z + 1/4),\\ 
C_3 &\colon (x,y,z)\rightarrow (z, x, y),\\ 
I &\colon (x,y,z)\rightarrow (-x, -y, -z).
\end{align}
\end{subequations}

The $\mathbb{Z}_2$ cohomology ring is given by

\begin{equation}
\mathbb{Z}_2[A_i,B_\alpha,B_{xy+xz+yz},C_{\alpha1},C_{\alpha2},C_\gamma]/\langle\mathcal{R}_3,\mathcal{R}_4,\mathcal{R}_5,\mathcal{R}_6\rangle
 \end{equation}
where the relations are 
\begin{subequations} 
 \begin{align}
\mathcal{R}_3\colon & ~~
A_i B_\alpha,\\
\mathcal{R}_4\colon & ~~
A_i C_{\alpha1},~~A_i C_{\alpha2},~~A_i C_\gamma,~~B_\alpha (B_\alpha + B_{xy+xz+yz}),~~B_\alpha^2 + A_i^2 B_{xy+xz+yz} + B_{xy+xz+yz}^2,\\
\mathcal{R}_5\colon & ~~
(B_\alpha + B_{xy+xz+yz}) C_{\alpha1},~~(B_\alpha + B_{xy+xz+yz}) C_{\alpha2},~~(B_\alpha + B_{xy+xz+yz}) C_\gamma,\\
\mathcal{R}_6\colon & ~~
B_\alpha^3 + C_{\alpha1}^2 + C_{\alpha1} C_{\alpha2} + C_{\alpha2}^2,~~C_\gamma (C_{\alpha1} + C_{\alpha2} + C_\gamma).
\end{align} 
 \end{subequations}
We have the following table regarding IWPs and group cohomology at degree 3.
\begin{center}
\begin{tabular}{c|cc|c|c|c}\hline\hline \multirow{3}{*}{\shortstack[l]{Wyckoff\\position}}&\multicolumn{2}{c|}{Little group}& {Coordinates}&\multirow{3}{*}{LSM anomaly class}&\multirow{3}{*}{Topo. inv.} \\ \cline{2-4} & \multirow{2}{*}{Intl.} & \multirow{2}{*}{Sch\"{o}nflies} & $(0,0,0) + ~(0,1/2,1/2) + $ & & \\ & & & $ (1/2,0,1/2) + ~(1/2,1/2,0) +$ & &\\ \hline
8a&$23$&$T$& $(1/8,1/8,1/8)$, $(7/8,7/8,7/8)$ & $C_\gamma$ & $\varphi_2[C_2, C'_2]$\\ 
8b&$23$&$T$& $(5/8,5/8,5/8)$, $(3/8,3/8,3/8)$ & $C_{\alpha1} + C_{\alpha2} + C_\gamma$ & $\varphi_2[T_3C_2, T_2C'_2]$\\ 
\hline
\multirow{2}{*}{16c} & \multirow{2}{*}{$\overline{3}$} & \multirow{2}{*}{$C_{3i}$} & $(0,0,0)$, $(3/4,3/4,0)$, & \multirow{2}{*}{$A_i (A_i^2 + B_{xy+xz+yz})$} & \multirow{2}{*}{$\varphi_1[I]$}\\
& & & $(3/4,0,3/4)$, $(0,3/4,3/4)$ & & \\ \hline
\multirow{2}{*}{16d} & \multirow{2}{*}{$\overline{3}$} & \multirow{2}{*}{$C_{3i}$} & $(1/2,1/2,1/2)$, $(1/4,1/4,1/2)$, & \multirow{2}{*}{$A_i B_{xy+xz+yz}$} & \multirow{2}{*}{$\varphi_1[T_1T_2T_3^{-1}I]$}\\
& & & $(1/4,1/2,1/4)$, $(1/2,1/4,1/4)$ & & \\ \hline
\hline 
 \end{tabular} 
 \end{center}

\subsection*{No. 204: $Im\overline3$}\label{subsub:sg204}

This group is generated by three translations $T_{1,2,3}$ as given in Eqs.~\eqref{TransBravaisI}, a two-fold rotation $C_2$, a two-fold rotation $C'_2$, a three-fold rotation $C_3$, and an inversion $I$:
\begin{subequations}
 \begin{align}
C_2 &\colon (x,y,z)\rightarrow (-x, -y, z),\\ 
C'_2 &\colon (x,y,z)\rightarrow (-x, y, -z),\\ 
C_3 &\colon (x,y,z)\rightarrow (z, x, y),\\ 
I &\colon (x,y,z)\rightarrow (-x, -y, -z).
\end{align}
\end{subequations}

The $\mathbb{Z}_2$ cohomology ring is given by

\begin{equation}
\mathbb{Z}_2[A_i,A_{x+y+z},B_\alpha,B_\beta,C_{\alpha1},C_{\alpha2},C_\gamma,C_{xyz},D_\delta]/\langle\mathcal{R}_2,\mathcal{R}_3,\mathcal{R}_4,\mathcal{R}_5,\mathcal{R}_6,\mathcal{R}_7,\mathcal{R}_8\rangle
 \end{equation}
where the relations are 
\begin{subequations} 
 \begin{align}
\mathcal{R}_2\colon & ~~
A_{x+y+z} (A_i + A_{x+y+z}),\\
\mathcal{R}_3\colon & ~~
A_{x+y+z} B_\alpha,~~A_{x+y+z} (A_i^2 + B_\beta),\\
\mathcal{R}_4\colon & ~~
A_{x+y+z} C_{\alpha1},~~A_{x+y+z} C_{\alpha2},~~A_{x+y+z} C_\gamma,~~A_{x+y+z} C_{xyz},\\
\mathcal{R}_5\colon & ~~
A_{x+y+z} D_\delta,~~A_i B_\alpha B_\beta + A_i^2 C_\gamma + B_\alpha C_\gamma + B_\beta C_\gamma + A_i^2 C_{xyz} + B_\alpha C_{xyz} + A_i D_\delta,\nonumber\\
&~~A_i^4 A_{x+y+z} + A_i^3 B_\beta + A_i B_\alpha B_\beta + B_\beta C_{\alpha1} + B_\beta C_{\alpha2} + B_\beta C_{xyz},\\
\mathcal{R}_6\colon & ~~
A_i^2 B_\alpha B_\beta + B_\alpha^2 B_\beta + B_\alpha B_\beta^2 + A_i B_\beta C_{\alpha1} + A_i B_\beta C_{\alpha2} + A_i^3 C_\gamma + A_i B_\beta C_\gamma + C_{\alpha1} C_\gamma + A_i^3 C_{xyz} + A_i^2 D_\delta + B_\alpha D_\delta,\nonumber\\
&~~A_i^4 B_\beta + A_i^2 B_\alpha B_\beta + B_\alpha^2 B_\beta + B_\beta^3 + A_i B_\beta C_{\alpha2} + A_i^3 C_\gamma + A_i B_\alpha C_\gamma + A_i^3 C_{xyz} + C_{\alpha1} C_{xyz} + A_i^2 D_\delta + B_\alpha D_\delta + B_\beta D_\delta,\nonumber\\
&~~B_\alpha^3 + C_{\alpha1}^2 + C_{\alpha1} C_{\alpha2} + C_{\alpha2}^2,~~A_i^2 B_\alpha B_\beta + A_i B_\beta C_{\alpha1} + A_i^3 C_\gamma + A_i B_\beta C_\gamma + C_{\alpha2} C_\gamma + A_i^3 C_{xyz} + A_i^2 D_\delta + B_\alpha D_\delta,\nonumber\\
&~~A_i^5 A_{x+y+z} + A_i^4 B_\beta + A_i B_\beta C_{\alpha1} + A_i^3 C_\gamma + A_i B_\alpha C_\gamma + A_i^3 C_{xyz} + C_{\alpha2} C_{xyz} + A_i^2 D_\delta + B_\alpha D_\delta + B_\beta D_\delta,\nonumber\\
&~~A_i^5 A_{x+y+z} + B_\alpha^2 B_\beta + A_i^2 B_\beta^2 + A_i^3 C_\gamma + A_i B_\alpha C_\gamma + C_\gamma^2 + A_i^3 C_{xyz} + A_i^2 D_\delta,\nonumber\\
&~~B_\alpha^2 B_\beta + B_\alpha B_\beta^2 + A_i B_\beta C_{\alpha2} + A_i^3 C_\gamma + A_i B_\alpha C_\gamma + C_\gamma C_{xyz},\nonumber\\
&~~A_i^5 A_{x+y+z} + B_\alpha^2 B_\beta + B_\beta^3 + A_i B_\beta C_{\alpha1} + A_i B_\beta C_{\alpha2} + A_i^3 C_\gamma + A_i B_\alpha C_\gamma + A_i B_\beta C_\gamma + C_{xyz}^2 + A_i^2 D_\delta,\\
\mathcal{R}_7\colon & ~~
A_i^5 B_\beta + A_i B_\alpha^2 B_\beta + A_i B_\alpha B_\beta^2 + A_i B_\beta^3 + A_i^2 B_\beta C_{\alpha1} + B_\alpha B_\beta C_{\alpha1} + B_\beta^2 C_{\alpha1} + B_\alpha B_\beta C_{\alpha2} + B_\beta^2 C_{\alpha2}\nonumber\\
&~~+ A_i^2 B_\alpha C_\gamma + B_\alpha^2 C_\gamma + A_i^2 B_\beta C_\gamma + A_i B_\beta D_\delta + C_{\alpha1} D_\delta,\nonumber\\
&~~A_i^6 A_{x+y+z} + A_i^5 B_\beta + A_i^3 B_\alpha B_\beta + B_\alpha B_\beta C_{\alpha1} + B_\beta^2 C_{\alpha1} + A_i^2 B_\alpha C_\gamma + B_\alpha^2 C_\gamma + A_i^2 B_\beta C_\gamma + A_i B_\beta D_\delta + C_{\alpha2} D_\delta,\nonumber\\
&~~A_i^6 A_{x+y+z} + A_i^5 B_\beta + A_i^3 B_\beta^2 + A_i B_\beta^3 + A_i^2 B_\beta C_{\alpha1} + B_\alpha B_\beta C_{\alpha2} + A_i^4 C_\gamma + A_i^2 B_\alpha C_\gamma + A_i B_\alpha D_\delta + A_i B_\beta D_\delta + C_\gamma D_\delta,\nonumber\\
&~~A_i^6 A_{x+y+z} + A_i^3 B_\alpha B_\beta + A_i B_\alpha^2 B_\beta + A_i B_\alpha B_\beta^2 + A_i B_\beta^3 + A_i^2 B_\beta C_{\alpha1} + B_\alpha B_\beta C_{\alpha2} + B_\beta^2 C_{\alpha2} \nonumber\\
&~~+ A_i^3 D_\delta + A_i B_\alpha D_\delta + C_{xyz} D_\delta,\\
\mathcal{R}_8\colon & ~~
A_i^7 A_{x+y+z} + A_i^6 B_\beta + B_\alpha^3 B_\beta + B_\alpha^2 B_\beta^2 + A_i^2 B_\beta^3 + B_\beta^4 + A_i B_\beta^2 C_{\alpha1} + A_i B_\beta^2 C_{\alpha2} + A_i^5 C_\gamma + A_i B_\alpha^2 C_\gamma + A_i^3 B_\beta C_\gamma \nonumber\\
&~~+ A_i B_\alpha B_\beta C_\gamma + A_i^4 D_\delta + A_i^2 B_\alpha D_\delta + A_i^2 B_\beta D_\delta + B_\alpha B_\beta D_\delta + B_\beta^2 D_\delta + D_\delta^2.
\end{align} 
 \end{subequations}
We have the following table regarding IWPs and group cohomology at degree 3.
\begin{center}
\resizebox{\columnwidth}{!}{
\begin{tabular}{c|cc|c|c|c}\hline\hline {Wyckoff}&\multicolumn{2}{c|}{Little group}& {Coordinates}&\multirow{2}{*}{LSM anomaly class}&\multirow{2}{*}{Topo. inv.} \\ \cline{2-4} position & Intl. & Sch\"{o}nflies & $(0,0,0) + ~(1/2,1/2,1/2) + $ & &\\ \hline
2a&$m\overline{3}$&$T_h$& $(0,0,0)$ & $A_i^3 + A_i^2 A_{x+y+z} + A_i B_\alpha + C_{\alpha1} + C_{\alpha2} + C_{xyz}$ & $\varphi_1[I]$\\ 
\hline
\multirow{2}{*}{6b} & \multirow{2}{*}{$mmm$} & \multirow{2}{*}{$D_{2h}$} & $(0,1/2,1/2)$, $(1/2,0,1/2)$, & \multirow{2}{*}{$C_{xyz}$} & \multirow{2}{*}{$\varphi_2[C_2, T_1T_2C'_2]$}\\
& & & $(1/2,1/2,0)$ & & \\ \hline
\multirow{2}{*}{8c} & \multirow{2}{*}{$\overline{3}$} & \multirow{2}{*}{$C_{3i}$} & $(1/4,1/4,1/4)$, $(3/4,3/4,1/4)$, & \multirow{2}{*}{$A_i^2 A_{x+y+z}$} & \multirow{2}{*}{$\varphi_1[T_1T_2T_3I]$}\\
& & & $(3/4,1/4,3/4)$, $(1/4,3/4,3/4)$ & & \\ \hline
\hline 
 \end{tabular} }
 \end{center}

\subsection*{No. 205: $Pa\overline3$}\label{subsub:sg205}

This group is generated by three translations $T_{1,2,3}$ as given in Eqs.~\eqref{TransBravaisP}, a two-fold screw $S_2$, a two-fold screw $S'_2$, a three-fold rotation $C_3$, and an inversion $I$:
\begin{subequations}
 \begin{align}
S_2 &\colon (x,y,z)\rightarrow (-x + 1/2, -y, z + 1/2),\\ 
S'_2 &\colon (x,y,z)\rightarrow (-x, y + 1/2, -z + 1/2),\\ 
C_3 &\colon (x,y,z)\rightarrow (z, x, y),\\ 
I &\colon (x,y,z)\rightarrow (-x, -y, -z).
\end{align}
\end{subequations}

The $\mathbb{Z}_2$ cohomology ring is given by

\begin{equation}
\mathbb{Z}_2[A_i,C_\beta]/\langle\mathcal{R}_6\rangle
 \end{equation}
where the relations are 
\begin{subequations} 
 \begin{align}
\mathcal{R}_6\colon & ~~
C_\beta (A_i^3 + C_\beta).
\end{align} 
 \end{subequations}
We have the following table regarding IWPs and group cohomology at degree 3.
\begin{center}
\begin{tabular}{c|cc|c|c|c}\hline\hline {Wyckoff}&\multicolumn{2}{c|}{Little group}& \multirow{2}{*}{Coordinates}&\multirow{2}{*}{LSM anomaly class}&\multirow{2}{*}{Topo. inv.} \\ \cline{2-3} position & Intl. & Sch\"{o}nflies & & & \\ \hline
\multirow{2}{*}{4a} & \multirow{2}{*}{$\overline{3}$} & \multirow{2}{*}{$C_{3i}$} & $(0,0,0)$, $(1/2,0,1/2)$, & \multirow{2}{*}{$C_\beta$} & \multirow{2}{*}{$\varphi_1[I]$}\\
& & & $(0,1/2,1/2)$, $(1/2,1/2,0)$ & & \\ \hline
\multirow{2}{*}{4b} & \multirow{2}{*}{$\overline{3}$} & \multirow{2}{*}{$C_{3i}$} & $(1/2,1/2,1/2)$, $(0,1/2,0)$, & \multirow{2}{*}{$A_i^3 + C_\beta$} & \multirow{2}{*}{$\varphi_1[T_3I]$}\\
& & & $(1/2,0,0)$, $(0,0,1/2)$ & & \\ \hline
\hline 
 \end{tabular} 
 \end{center}

\subsection*{No. 206: $Ia\overline3$}\label{subsub:sg206}

This group is generated by three translations $T_{1,2,3}$ as given in Eqs.~\eqref{TransBravaisI}, a two-fold rotation $C_2$, a two-fold rotation $C'_2$, a three-fold rotation $C_3$, and an inversion $I$:
\begin{subequations}
 \begin{align}
C_2 &\colon (x,y,z)\rightarrow (-x, -y + 1/2, z),\\ 
C'_2 &\colon (x,y,z)\rightarrow (-x + 1/2, y, -z),\\ 
C_3 &\colon (x,y,z)\rightarrow (z, x, y),\\ 
I &\colon (x,y,z)\rightarrow (-x, -y, -z).
\end{align}
\end{subequations}

The $\mathbb{Z}_2$ cohomology ring is given by

\begin{equation}
\mathbb{Z}_2[A_i,A_{x+y+z}]/\langle\mathcal{R}_4\rangle
 \end{equation}
where the relations are 
\begin{subequations} 
 \begin{align}
\mathcal{R}_4\colon & ~~
A_i^2 A_{x+y+z} (A_i + A_{x+y+z}).
\end{align} 
 \end{subequations}
We have the following table regarding IWPs and group cohomology at degree 3.
\begin{center}
\begin{tabular}{c|cc|c|c|c}\hline\hline {Wyckoff}&\multicolumn{2}{c|}{Little group}& {Coordinates}&\multirow{2}{*}{LSM anomaly class}&\multirow{2}{*}{Topo. inv.} \\ \cline{2-4} position & Intl. & Sch\"{o}nflies & $(0,0,0) + ~(1/2,1/2,1/2) + $ & &\\ \hline
\multirow{2}{*}{8a} & \multirow{2}{*}{$\overline{3}$} & \multirow{2}{*}{$C_{3i}$} & $(0,0,0)$, $(1/2,0,1/2)$, & \multirow{2}{*}{$A_i^2 (A_i + A_{x+y+z})$} & \multirow{2}{*}{$\varphi_1[I]$}\\
& & & $(0,1/2,1/2)$, $(1/2,1/2,0)$ & & \\ \hline
\multirow{2}{*}{8b} & \multirow{2}{*}{$\overline{3}$} & \multirow{2}{*}{$C_{3i}$} & $(1/4,1/4,1/4)$, $(1/4,3/4,3/4)$, & \multirow{2}{*}{$A_i^2 A_{x+y+z}$} & \multirow{2}{*}{$\varphi_1[T_1T_2T_3I]$}\\
& & & $(3/4,3/4,1/4)$, $(3/4,1/4,3/4)$ & & \\ \hline
\multirow{6}{*}{24d} & \multirow{6}{*}{$2$} & \multirow{6}{*}{$C_2$} & $(x,0,1/4)$, $(-x+1/2,0,3/4)$, & \multirow{6}{*}{$A_i A_{x+y+z} (A_i + A_{x+y+z})$} & \multirow{6}{*}{$\varphi_2[T_1C'_2I, C_2]$}\\
& & & $(1/4,x,0)$, $(3/4,-x+1/2,0)$, & & \\
& & & $(0,1/4,x)$, $(0,3/4,-x+1/2)$, & & \\
& & & $(-x,0,3/4)$, $(x+1/2,0,1/4)$, & & \\
& & & $(3/4,-x,0)$, $(1/4,x+1/2,0)$, & & \\
& & & $(0,3/4,-x)$, $(0,1/4,x+1/2)$ & & \\ \hline
\hline 
 \end{tabular} 
 \end{center}

\subsection*{No. 207: $P432$}\label{subsub:sg207}

This group is generated by three translations $T_{1,2,3}$ as given in Eqs.~\eqref{TransBravaisP}, a two-fold rotation $C_2$, a two-fold rotation $C'_2$, a three-fold rotation $C_3$, and a two-fold rotation $C''_2$:
\begin{subequations}
 \begin{align}
C_2 &\colon (x,y,z)\rightarrow (-x, -y, z),\\ 
C'_2 &\colon (x,y,z)\rightarrow (-x, y, -z),\\ 
C_3 &\colon (x,y,z)\rightarrow (z, x, y),\\ 
C''_2 &\colon (x,y,z)\rightarrow (-y, -x, -z).
\end{align}
\end{subequations}

The $\mathbb{Z}_2$ cohomology ring is given by

\begin{equation}
\mathbb{Z}_2[A_{c''},A_{x+y+z},B_\alpha,B_{xy+xz+yz},C_\alpha,C_{xyz}]/\langle\mathcal{R}_3,\mathcal{R}_4,\mathcal{R}_5,\mathcal{R}_6\rangle
 \end{equation}
where the relations are 
\begin{subequations} 
 \begin{align}
\mathcal{R}_3\colon & ~~
A_{c''} A_{x+y+z} (A_{c''} + A_{x+y+z}),\\
\mathcal{R}_4\colon & ~~
A_{c''} C_\alpha,~~A_{x+y+z} (A_{c''}^3 + A_{x+y+z}^3 + A_{c''} B_\alpha + A_{x+y+z} B_\alpha + C_\alpha),~~A_{c''} (A_{x+y+z} B_{xy+xz+yz} + C_{xyz}),\nonumber\\&~~A_{x+y+z} (A_{c''} B_{xy+xz+yz} + C_{xyz}),~~B_{xy+xz+yz} (A_{c''} A_{x+y+z} + A_{x+y+z}^2 + B_\alpha + B_{xy+xz+yz}),\\
\mathcal{R}_5\colon & ~~
A_{c''}^2 A_{x+y+z} B_{xy+xz+yz} + A_{x+y+z}^3 B_{xy+xz+yz} + A_{x+y+z} B_\alpha B_{xy+xz+yz} + B_{xy+xz+yz} C_\alpha + B_\alpha C_{xyz},\nonumber\\&~~B_{xy+xz+yz} (A_{c''}^2 A_{x+y+z} + A_{x+y+z}^3 + A_{x+y+z} B_\alpha + C_\alpha + C_{xyz}),\\
\mathcal{R}_6\colon & ~~
A_{c''} A_{x+y+z} B_\alpha B_{xy+xz+yz} + C_\alpha C_{xyz} + C_{xyz}^2.
\end{align} 
 \end{subequations}
We have the following table regarding IWPs and group cohomology at degree 3.
\begin{center}
\resizebox{\columnwidth}{!}{
\begin{tabular}{c|cc|c|c|c}\hline\hline {Wyckoff}&\multicolumn{2}{c|}{Little group}& \multirow{2}{*}{Coordinates}&\multirow{2}{*}{LSM anomaly class}&\multirow{2}{*}{Topo. inv.} \\ \cline{2-3} position & Intl. & Sch\"{o}nflies & & & \\ \hline
1a&$432$&$O$& $(0,0,0)$ & $A_{c''}^2 A_{x+y+z} + A_{x+y+z}^3 + A_{c''} B_\alpha + A_{x+y+z} B_\alpha + A_{c''} B_{xy+xz+yz} + C_\alpha + C_{xyz}$ & $\varphi_2[C_2, C'_2]$\\ 
1b&$432$&$O$& $(1/2,1/2,1/2)$ & $C_{xyz}$ & $\varphi_2[T_1T_2C_2, T_1T_3C'_2]$\\ 
\hline
\multirow{2}{*}{3c} & \multirow{2}{*}{$422$} & \multirow{2}{*}{$D_4$} & $(0,1/2,1/2)$, $(1/2,0,1/2)$, & \multirow{2}{*}{$(A_{c''} + A_{x+y+z}) B_{xy+xz+yz}$} & \multirow{2}{*}{$\varphi_2[T_1T_2C_2, T_1C'_2]$}\\
& & & $(1/2,1/2,0)$ & & \\ \hline
\multirow{2}{*}{3d} & \multirow{2}{*}{$422$} & \multirow{2}{*}{$D_4$} & $(1/2,0,0)$, $(0,1/2,0)$, & \multirow{2}{*}{$A_{x+y+z} (A_{c''}^2 + A_{x+y+z}^2 + B_\alpha + B_{xy+xz+yz})$} & \multirow{2}{*}{$\varphi_2[T_1C_2, T_1C'_2]$}\\
& & & $(0,0,1/2)$ & & \\ \hline
\hline 
 \end{tabular} }
 \end{center}

\subsection*{No. 208: $P4_232$}\label{subsub:sg208}

This group is generated by three translations $T_{1,2,3}$ as given in Eqs.~\eqref{TransBravaisP}, a two-fold rotation $C_2$, a two-fold rotation $C'_2$, a three-fold rotation $C_3$, and a two-fold rotation $C''_2$:
\begin{subequations}
 \begin{align}
C_2 &\colon (x,y,z)\rightarrow (-x, -y, z),\\ 
C'_2 &\colon (x,y,z)\rightarrow (-x, y, -z),\\ 
C_3 &\colon (x,y,z)\rightarrow (z, x, y),\\ 
C''_2 &\colon (x,y,z)\rightarrow (-y + 1/2, -x + 1/2, -z + 1/2).
\end{align}
\end{subequations}

The $\mathbb{Z}_2$ cohomology ring is given by

\begin{equation}
\mathbb{Z}_2[A_{c''},A_{x+y+z},B_\alpha,B_{xy+xz+yz},C_\alpha,C_\beta]/\langle\mathcal{R}_3,\mathcal{R}_4,\mathcal{R}_5,\mathcal{R}_6\rangle
 \end{equation}
where the relations are 
\begin{subequations} 
 \begin{align}
\mathcal{R}_3\colon & ~~
A_{c''} (A_{c''} A_{x+y+z} + A_{x+y+z}^2 + B_\alpha),\\
\mathcal{R}_4\colon & ~~
A_{c''} C_\alpha,~~A_{x+y+z} (A_{c''} A_{x+y+z}^2 + A_{x+y+z}^3 + A_{x+y+z} B_\alpha + C_\alpha),~~A_{c''} (A_{c''}^2 A_{x+y+z} + A_{x+y+z}^3 + C_\beta),\nonumber\\
&~~A_{c''}^2 A_{x+y+z}^2 + A_{x+y+z}^4 + B_\alpha^2 + A_{c''} A_{x+y+z} B_{xy+xz+yz} + A_{x+y+z}^2 B_{xy+xz+yz} + B_{xy+xz+yz}^2,\nonumber\\
&~~A_{c''} A_{x+y+z}^3 + A_{x+y+z}^4 + A_{x+y+z}^2 B_\alpha + B_\alpha^2 + A_{c''} A_{x+y+z} B_{xy+xz+yz} + A_{x+y+z}^2 B_{xy+xz+yz} + B_\alpha B_{xy+xz+yz} \nonumber\\
&~~+ A_{x+y+z} C_\beta,\\
\mathcal{R}_5\colon & ~~
A_{c''}^2 A_{x+y+z}^3 + A_{x+y+z}^5 + B_\alpha C_\alpha + B_{xy+xz+yz} C_\alpha + A_{x+y+z}^2 C_\beta,\nonumber\\&~~A_{c''}^3 A_{x+y+z}^2 + A_{c''}^2 A_{x+y+z}^3 + A_{x+y+z}^3 B_\alpha + A_{x+y+z} B_\alpha^2 + A_{c''}^2 A_{x+y+z} B_{xy+xz+yz} + A_{x+y+z}^3 B_{xy+xz+yz} + A_{x+y+z}^2 C_\beta \nonumber\\&+ B_\alpha C_\beta + B_{xy+xz+yz} C_\beta,\\
\mathcal{R}_6\colon & ~~
A_{c''}^4 A_{x+y+z}^2 + A_{c''}^3 A_{x+y+z}^3 + A_{c''}^2 A_{x+y+z}^4 + A_{c''} A_{x+y+z}^5 + B_\alpha^3 + C_\alpha^2 + C_\alpha C_\beta + C_\beta^2.
\end{align} 
 \end{subequations}
We have the following table regarding IWPs and group cohomology at degree 3.
\begin{center}
\resizebox{\columnwidth}{!}{
\begin{tabular}{c|cc|c|c|c}\hline\hline {Wyckoff}&\multicolumn{2}{c|}{Little group}& \multirow{2}{*}{Coordinates}&\multirow{2}{*}{LSM anomaly class}&\multirow{2}{*}{Topo. inv.} \\ \cline{2-3} position & Intl. & Sch\"{o}nflies & & & \\ \hline
2a&$23$&$T$& $(0,0,0)$, $(1/2,1/2,1/2)$ & $A_{c''} A_{x+y+z}^2 + A_{x+y+z}^3 + A_{x+y+z} B_\alpha + C_\alpha$ & $\varphi_2[C_2, C'_2]$\\ 
\hline
\multirow{2}{*}{4b} & \multirow{2}{*}{$32$} & \multirow{2}{*}{$D_3$} & $(1/4,1/4,1/4)$, $(3/4,3/4,1/4)$, & \multirow{2}{*}{N/A} & \multirow{2}{*}{N/A}\\
& & & $(3/4,1/4,3/4)$, $(1/4,3/4,3/4)$ & & \\ \hline
\multirow{2}{*}{4c} & \multirow{2}{*}{$32$} & \multirow{2}{*}{$D_3$} & $(3/4,3/4,3/4)$, $(1/4,1/4,3/4)$, & \multirow{2}{*}{N/A} & \multirow{2}{*}{N/A}\\
& & & $(1/4,3/4,1/4)$, $(3/4,1/4,1/4)$ & & \\ \hline
\multirow{3}{*}{6d} & \multirow{3}{*}{$222$} & \multirow{3}{*}{$D_2$} & $(0,1/2,1/2)$, $(1/2,0,1/2)$, & \multirow{3}{*}{$A_{x+y+z} (A_{c''} A_{x+y+z} + A_{x+y+z}^2 + B_\alpha)$} & \multirow{3}{*}{$\varphi_2[T_1C_2, T_1C'_2]$}\\
& & & $(1/2,1/2,0)$, $(0,1/2,0)$, & & \\
& & & $(1/2,0,0)$, $(0,0,1/2)$ & & \\ \hline
\multirow{3}{*}{6e} & \multirow{3}{*}{$222$} & \multirow{3}{*}{$D_2$} & $(1/4,0,1/2)$, $(3/4,0,1/2)$, & \multirow{3}{*}{$A_{c''} (A_{c''} A_{x+y+z} + A_{x+y+z}^2 + B_{xy+xz+yz})$} & \multirow{3}{*}{$\varphi_2[T_2C_2, C''_2]$}\\
& & & $(1/2,1/4,0)$, $(1/2,3/4,0)$, & & \\
& & & $(0,1/2,1/4)$, $(0,1/2,3/4)$ & & \\ \hline
\multirow{3}{*}{6f} & \multirow{3}{*}{$222$} & \multirow{3}{*}{$D_2$} & $(1/4,1/2,0)$, $(3/4,1/2,0)$, & \multirow{3}{*}{$A_{c''} B_{xy+xz+yz}$} & \multirow{3}{*}{$\varphi_2[T_1C_2, C''_2]$}\\
& & & $(0,1/4,1/2)$, $(0,3/4,1/2)$, & & \\
& & & $(1/2,0,1/4)$, $(1/2,0,3/4)$ & & \\ \hline
\hline 
 \end{tabular} }
 \end{center}

\subsection*{No. 209: $F432$}\label{subsub:sg209}

This group is generated by three translations $T_{1,2,3}$ as given in Eqs.~\eqref{TransBravaisF}, a two-fold rotation $C_2$, a two-fold rotation $C'_2$, a three-fold rotation $C_3$, and a two-fold rotation $C''_2$:
\begin{subequations}
 \begin{align}
C_2 &\colon (x,y,z)\rightarrow (-x, -y, z),\\ 
C'_2 &\colon (x,y,z)\rightarrow (-x, y, -z),\\ 
C_3 &\colon (x,y,z)\rightarrow (z, x, y),\\ 
C''_2 &\colon (x,y,z)\rightarrow (-y, -x, -z).
\end{align}
\end{subequations}

The $\mathbb{Z}_2$ cohomology ring is given by

\begin{equation}
\mathbb{Z}_2[A_{c''},B_\alpha,B_\beta,B_{xy+xz+yz},C_\alpha,C_{\gamma1},C_{\gamma2},C_{xyz}]/\langle\mathcal{R}_4,\mathcal{R}_5,\mathcal{R}_6\rangle
 \end{equation}
where the relations are 
\begin{subequations} 
 \begin{align}
\mathcal{R}_4\colon & ~~
A_{c''} C_\alpha,~~A_{c''} (C_{\gamma1} + C_{\gamma2}),~~A_{c''}^2 B_\beta + B_\alpha B_\beta + B_\alpha B_{xy+xz+yz} + A_{c''} C_{\gamma1},~~B_\alpha B_\beta + B_\beta^2 + A_{c''} C_{\gamma1},\nonumber\\&~~A_{c''}^2 B_\alpha + B_\alpha B_\beta + A_{c''}^2 B_{xy+xz+yz} + B_\beta B_{xy+xz+yz} + A_{c''} C_{\gamma1} + A_{c''} C_{xyz},~~A_{c''}^2 B_\beta + B_\alpha B_\beta + B_{xy+xz+yz}^2 + A_{c''} C_{\gamma1},\\
\mathcal{R}_5\colon & ~~
(B_\beta + B_{xy+xz+yz}) C_\alpha,~~A_{c''}^3 B_\alpha + A_{c''} B_\alpha B_\beta + A_{c''}^3 B_{xy+xz+yz} + B_\beta C_{\gamma1} + B_{xy+xz+yz} C_{\gamma1} + A_{c''}^2 C_{xyz},\nonumber\\&~~B_\beta C_\alpha + B_\alpha C_{\gamma1} + B_\alpha C_{\gamma2},~~B_\beta (C_\alpha + C_{\gamma1} + C_{\gamma2}),\nonumber\\&~~A_{c''}^3 B_\alpha + A_{c''} B_\alpha B_\beta + A_{c''}^3 B_{xy+xz+yz} + B_\beta C_\alpha + B_\beta C_{\gamma1} + B_{xy+xz+yz} C_{\gamma2} + A_{c''}^2 C_{xyz},\nonumber\\&~~A_{c''} B_\alpha^2 + A_{c''}^3 B_\beta + B_\beta C_{\gamma1} + B_\alpha C_{xyz},~~B_\beta (C_{\gamma1} + C_{xyz}),\nonumber\\&~~A_{c''}^3 B_\alpha + A_{c''} B_\alpha B_\beta + A_{c''}^3 B_{xy+xz+yz} + B_\beta C_{\gamma1} + A_{c''}^2 C_{xyz} + B_{xy+xz+yz} C_{xyz},\\
\mathcal{R}_6\colon & ~~
A_{c''}^2 B_\alpha B_\beta + B_\alpha^2 B_\beta + A_{c''}^3 C_{\gamma1} + C_{\gamma1}^2 + C_\alpha C_{\gamma2},~~A_{c''}^2 B_\alpha B_\beta + B_\alpha^2 B_\beta + A_{c''}^3 C_{\gamma1} + C_\alpha C_{\gamma2} + C_{\gamma1} C_{\gamma2} + C_\alpha C_{xyz},\nonumber\\&~~A_{c''}^2 B_\alpha B_\beta + B_\alpha^2 B_\beta + A_{c''}^3 C_{\gamma1} + C_\alpha C_{\gamma1} + C_\alpha C_{\gamma2} + C_\alpha C_{xyz} + C_{\gamma1} C_{xyz},\nonumber\\&~~A_{c''}^2 B_\alpha B_\beta + B_\alpha^2 B_\beta + A_{c''}^3 C_{\gamma1} + C_\alpha C_{\gamma1} + C_{\gamma2}^2,~~A_{c''}^2 B_\alpha B_\beta + B_\alpha^2 B_\beta + A_{c''}^3 C_{\gamma1} + C_\alpha C_{\gamma1} + C_\alpha C_{\gamma2} + C_{\gamma2} C_{xyz},\nonumber\\&~~A_{c''}^2 B_\alpha^2 + A_{c''}^4 B_\beta + A_{c''}^2 B_\alpha B_\beta + B_\alpha^2 B_\beta + A_{c''}^3 C_{\gamma1} + A_{c''} B_\alpha C_{\gamma1} + A_{c''} B_\beta C_{\gamma1} + C_\alpha C_{\gamma1} + C_\alpha C_{\gamma2} + C_\alpha C_{xyz} + C_{xyz}^2.
\end{align} 
 \end{subequations}
We have the following table regarding IWPs and group cohomology at degree 3.
\begin{center}
\resizebox{\columnwidth}{!}{
\begin{tabular}{c|cc|c|c|c}\hline\hline \multirow{3}{*}{\shortstack[l]{Wyckoff\\position}}&\multicolumn{2}{c|}{Little group}& {Coordinates}&\multirow{3}{*}{LSM anomaly class}&\multirow{3}{*}{Topo. inv.} \\ \cline{2-4} & \multirow{2}{*}{Intl.} & \multirow{2}{*}{Sch\"{o}nflies} & $(0,0,0) + ~(0,1/2,1/2) + $ & & \\ & & & $ (1/2,0,1/2) + ~(1/2,1/2,0) +$ & &\\ \hline
4a&$432$&$O$& $(0,0,0)$ & $C_\alpha + C_{\gamma2} + C_{xyz}$ & $\varphi_2[C_2, C'_2]$\\ 
4b&$432$&$O$& $(1/2,1/2,1/2)$ & $A_{c''} B_\alpha + A_{c''} B_{xy+xz+yz} + C_{\gamma1} + C_{xyz}$ & $\varphi_2[C_2, T_1T_2T_3^{-1}C'_2]$\\ 
8c&$23$&$T$& $(1/4,1/4,1/4)$, $(1/4,1/4,3/4)$ & $C_{\gamma1} + C_{\gamma2}$ & $\varphi_2[T_3C_2, T_2C'_2]$\\ 
\hline
\multirow{3}{*}{24d} & \multirow{3}{*}{$222$} & \multirow{3}{*}{$D_2$} & $(0,1/4,1/4)$, $(0,3/4,1/4)$, & \multirow{3}{*}{$A_{c''} B_{xy+xz+yz}$} & \multirow{3}{*}{$\varphi_2[T_3C_2, T_3C''_2]$}\\
& & & $(1/4,0,1/4)$, $(1/4,0,3/4)$, & & \\
& & & $(1/4,1/4,0)$, $(3/4,1/4,0)$ & & \\ \hline
\hline 
 \end{tabular} }
 \end{center}

\subsection*{No. 210: $F4_132$}\label{subsub:sg210}

This group is generated by three translations $T_{1,2,3}$ as given in Eqs.~\eqref{TransBravaisF}, a two-fold rotation $C_2$, a two-fold rotation $C'_2$, a three-fold rotation $C_3$, and a two-fold rotation $C''_2$:
\begin{subequations}
 \begin{align}
C_2 &\colon (x,y,z)\rightarrow (-x, -y, z),\\ 
C'_2 &\colon (x,y,z)\rightarrow (-x, y, -z),\\ 
C_3 &\colon (x,y,z)\rightarrow (z, x, y),\\ 
C''_2 &\colon (x,y,z)\rightarrow (-y + 1/4, -x + 1/4, -z + 1/4).
\end{align}
\end{subequations}

The $\mathbb{Z}_2$ cohomology ring is given by

\begin{equation}
\mathbb{Z}_2[A_{c''},B_\alpha,B_{xy+xz+yz},C_\alpha,C_\beta,C_\gamma]/\langle\mathcal{R}_3,\mathcal{R}_4,\mathcal{R}_5,\mathcal{R}_6\rangle
 \end{equation}
where the relations are 
\begin{subequations} 
 \begin{align}
\mathcal{R}_3\colon & ~~
A_{c''} B_\alpha,\\
\mathcal{R}_4\colon & ~~
A_{c''} C_\alpha,~~A_{c''} C_\beta,~~A_{c''} C_\gamma,~~B_\alpha B_{xy+xz+yz},~~B_{xy+xz+yz}^2,\\
\mathcal{R}_5\colon & ~~
B_{xy+xz+yz} C_\alpha,~~B_{xy+xz+yz} C_\beta,~~B_{xy+xz+yz} C_\gamma,\\
\mathcal{R}_6\colon & ~~
B_\alpha^3 + C_\alpha^2 + C_\alpha C_\beta + C_\beta^2,~~C_\gamma (C_\alpha + C_\gamma).
\end{align} 
 \end{subequations}
We have the following table regarding IWPs and group cohomology at degree 3.
\begin{center}
\begin{tabular}{c|cc|c|c|c}\hline\hline \multirow{3}{*}{\shortstack[l]{Wyckoff\\position}}&\multicolumn{2}{c|}{Little group}& {Coordinates}&\multirow{3}{*}{LSM anomaly class}&\multirow{3}{*}{Topo. inv.} \\ \cline{2-4} & \multirow{2}{*}{Intl.} & \multirow{2}{*}{Sch\"{o}nflies} & $(0,0,0) + ~(0,1/2,1/2) + $ & & \\ & & & $ (1/2,0,1/2) + ~(1/2,1/2,0) +$ & &\\ \hline
8a&$23$&$T$& $(0,0,0)$, $(3/4,1/4,3/4)$ & $C_\alpha + C_\gamma$ & $\varphi_2[C_2, C'_2]$\\ 
8b&$23$&$T$& $(1/2,1/2,1/2)$, $(1/4,3/4,1/4)$ & $C_\gamma$ & $\varphi_2[C_2, T_1T_2T_3^{-1}C'_2]$\\ 
\hline
\multirow{2}{*}{16c} & \multirow{2}{*}{$32$} & \multirow{2}{*}{$D_3$} & $(1/8,1/8,1/8)$, $(7/8,3/8,5/8)$, & \multirow{2}{*}{$A_{c''} B_{xy+xz+yz}$} & \multirow{2}{*}{$\varphi_2[T_1T_2^{-1}, C''_2]$}\\
& & & $(3/8,5/8,7/8)$, $(5/8,7/8,3/8)$ & & \\ \hline
\multirow{2}{*}{16d} & \multirow{2}{*}{$32$} & \multirow{2}{*}{$D_3$} & $(5/8,5/8,5/8)$, $(3/8,7/8,1/8)$, & \multirow{2}{*}{Same as 16c } & \multirow{2}{*}{Same as 16c}\\
& & & $(7/8,1/8,3/8)$, $(1/8,3/8,7/8)$ & & \\ \hline
\hline 
 \end{tabular} 
 \end{center}

\subsection*{No. 211: $I432$}\label{subsub:sg211}

This group is generated by three translations $T_{1,2,3}$ as given in Eqs.~\eqref{TransBravaisI}, a two-fold rotation $C_2$, a two-fold rotation $C'_2$, a three-fold rotation $C_3$, and a two-fold rotation $C''_2$:
\begin{subequations}
 \begin{align}
C_2 &\colon (x,y,z)\rightarrow (-x, -y, z),\\ 
C'_2 &\colon (x,y,z)\rightarrow (-x, y, -z),\\ 
C_3 &\colon (x,y,z)\rightarrow (z, x, y),\\ 
C''_2 &\colon (x,y,z)\rightarrow (-y, -x, -z).
\end{align}
\end{subequations}

The $\mathbb{Z}_2$ cohomology ring is given by

\begin{equation}
\mathbb{Z}_2[A_{c''},A_{x+y+z},B_\alpha,B_\beta,C_\alpha,C_\gamma,C_{xyz}]/\langle\mathcal{R}_2,\mathcal{R}_3,\mathcal{R}_4,\mathcal{R}_5,\mathcal{R}_6\rangle
 \end{equation}
where the relations are 
\begin{subequations} 
 \begin{align}
\mathcal{R}_2\colon & ~~
A_{x+y+z} (A_{c''} + A_{x+y+z}),\\
\mathcal{R}_3\colon & ~~
A_{c''}^2 A_{x+y+z} + A_{x+y+z} B_\alpha + A_{c''} B_\beta,~~A_{x+y+z} (A_{c''}^2 + B_\alpha + B_\beta),\\
\mathcal{R}_4\colon & ~~
A_{c''} C_\alpha,~~A_{x+y+z} C_\alpha,~~A_{x+y+z} C_\gamma + A_{c''} C_{xyz},~~A_{x+y+z} (C_\gamma + C_{xyz}),\\
\mathcal{R}_5\colon & ~~
A_{c''} A_{x+y+z} C_\gamma + B_\beta C_\gamma + B_\alpha C_{xyz},~~B_\beta (C_\alpha + C_\gamma + C_{xyz}),\\
\mathcal{R}_6\colon & ~~
A_{c''}^3 A_{x+y+z} B_\alpha + A_{c''} A_{x+y+z} B_\alpha^2 + B_\alpha^2 B_\beta + B_\alpha B_\beta^2 + C_\alpha C_\gamma,\nonumber\\&~~A_{c''}^5 A_{x+y+z} + A_{c''} A_{x+y+z} B_\alpha^2 + B_\alpha B_\beta^2 + B_\beta^3 + C_\alpha C_{xyz},~~A_{c''} A_{x+y+z} B_\alpha^2 + B_\alpha^2 B_\beta + A_{c''} B_\alpha C_\gamma + C_\gamma^2,\nonumber\\&~~A_{c''}^3 A_{x+y+z} B_\alpha + B_\alpha B_\beta^2 + A_{x+y+z} B_\alpha C_\gamma + C_\gamma C_{xyz},\nonumber\\&~~A_{c''}^5 A_{x+y+z} + A_{c''}^3 A_{x+y+z} B_\alpha + A_{c''} A_{x+y+z} B_\alpha^2 + B_\beta^3 + A_{x+y+z} B_\alpha C_\gamma + C_{xyz}^2.
\end{align} 
 \end{subequations}
We have the following table regarding IWPs and group cohomology at degree 3.
\begin{center}
\resizebox{\columnwidth}{!}{
\begin{tabular}{c|cc|c|c|c}\hline\hline {Wyckoff}&\multicolumn{2}{c|}{Little group}& {Coordinates}&\multirow{2}{*}{LSM anomaly class}&\multirow{2}{*}{Topo. inv.} \\ \cline{2-4} position & Intl. & Sch\"{o}nflies & $(0,0,0) + ~(1/2,1/2,1/2) + $ & &\\ \hline
2a&$432$&$O$& $(0,0,0)$ & $A_{c''} B_\alpha + A_{x+y+z} B_\alpha + C_\alpha + C_\gamma + C_{xyz}$ & $\varphi_2[C_2, C'_2]$\\ 
\hline
\multirow{2}{*}{6b} & \multirow{2}{*}{$422$} & \multirow{2}{*}{$D_4$} & $(0,1/2,1/2)$, $(1/2,0,1/2)$, & \multirow{2}{*}{$C_\gamma + C_{xyz}$} & \multirow{2}{*}{$\varphi_2[C_2, T_1T_2C'_2]$}\\
& & & $(1/2,1/2,0)$ & & \\ \hline
\multirow{2}{*}{8c} & \multirow{2}{*}{$32$} & \multirow{2}{*}{$D_3$} & $(1/4,1/4,1/4)$, $(3/4,3/4,1/4)$, & \multirow{2}{*}{$C_{xyz}$} & \multirow{2}{*}{$\varphi_2[T_1T_2^{-1}, T_1T_2T_3C''_2]$}\\
& & & $(3/4,1/4,3/4)$, $(1/4,3/4,3/4)$ & & \\ \hline
\multirow{3}{*}{12d} & \multirow{3}{*}{$222$} & \multirow{3}{*}{$D_2$} & $(1/4,1/2,0)$, $(3/4,1/2,0)$, & \multirow{3}{*}{$A_{x+y+z} B_\alpha$} & \multirow{3}{*}{$\varphi_2[T_2T_3C_2, T_1T_2T_3C''_2]$}\\
& & & $(0,1/4,1/2)$, $(0,3/4,1/2)$, & & \\
& & & $(1/2,0,1/4)$, $(1/2,0,3/4)$ & & \\ \hline
\hline 
 \end{tabular} }
 \end{center}

\subsection*{No. 212: $P4_332$}\label{subsub:sg212}

This group is generated by three translations $T_{1,2,3}$ as given in Eqs.~\eqref{TransBravaisP}, a two-fold screw $S_2$, a two-fold screw $S'_2$, a three-fold rotation $C_3$, and a two-fold rotation $C''_2$:
\begin{subequations}
 \begin{align}
S_2 &\colon (x,y,z)\rightarrow (-x + 1/2, -y, z + 1/2),\\ 
S'_2 &\colon (x,y,z)\rightarrow (-x, y + 1/2, -z + 1/2),\\ 
C_3 &\colon (x,y,z)\rightarrow (z, x, y),\\ 
C''_2 &\colon (x,y,z)\rightarrow (-y + 1/4, -x + 1/4, -z + 1/4).
\end{align}
\end{subequations}

The $\mathbb{Z}_2$ cohomology ring is given by

\begin{equation}
\mathbb{Z}_2[A_{c''},C_\gamma]/\langle\mathcal{R}_6\rangle
 \end{equation}
where the relations are 
\begin{subequations} 
 \begin{align}
\mathcal{R}_6\colon & ~~
C_\gamma^2.
\end{align} 
 \end{subequations}
We have the following table regarding IWPs and group cohomology at degree 3.
\begin{center}
\begin{tabular}{c|cc|c|c|c}\hline\hline {Wyckoff}&\multicolumn{2}{c|}{Little group}& \multirow{2}{*}{Coordinates}&\multirow{2}{*}{LSM anomaly class}&\multirow{2}{*}{Topo. inv.} \\ \cline{2-3} position & Intl. & Sch\"{o}nflies & & & \\ \hline
\multirow{2}{*}{4a} & \multirow{2}{*}{$32$} & \multirow{2}{*}{$D_3$} & $(1/8,1/8,1/8)$, $(3/8,7/8,5/8)$, & \multirow{2}{*}{$C_\gamma$} & \multirow{2}{*}{$\varphi_2[T_1T_2^{-1}, C''_2]$}\\
& & & $(7/8,5/8,3/8)$, $(5/8,3/8,7/8)$ & & \\ \hline
\multirow{2}{*}{4b} & \multirow{2}{*}{$32$} & \multirow{2}{*}{$D_3$} & $(5/8,5/8,5/8)$, $(7/8,3/8,1/8)$, & \multirow{2}{*}{Same as 4a } & \multirow{2}{*}{Same as 4a}\\
& & & $(3/8,1/8,7/8)$, $(1/8,7/8,3/8)$ & & \\ \hline
\hline 
 \end{tabular} 
 \end{center}

\subsection*{No. 213: $P4_132$}\label{subsub:sg213}

This group is generated by three translations $T_{1,2,3}$ as given in Eqs.~\eqref{TransBravaisP}, a two-fold screw $S_2$, a two-fold screw $S'_2$, a three-fold rotation $C_3$, and a two-fold rotation $C''_2$:
\begin{subequations}
 \begin{align}
S_2 &\colon (x,y,z)\rightarrow (-x + 1/2, -y, z + 1/2),\\ 
S'_2 &\colon (x,y,z)\rightarrow (-x, y + 1/2, -z + 1/2),\\ 
C_3 &\colon (x,y,z)\rightarrow (z, x, y),\\ 
C''_2 &\colon (x,y,z)\rightarrow (-y + 3/4, -x + 3/4, -z + 3/4).
\end{align}
\end{subequations}

The $\mathbb{Z}_2$ cohomology ring is given by

\begin{equation}
\mathbb{Z}_2[A_{c''},C_\gamma]/\langle\mathcal{R}_6\rangle
 \end{equation}
where the relations are 
\begin{subequations} 
 \begin{align}
\mathcal{R}_6\colon & ~~
C_\gamma^2.
\end{align} 
 \end{subequations}
We have the following table regarding IWPs and group cohomology at degree 3.
\begin{center}
\begin{tabular}{c|cc|c|c|c}\hline\hline {Wyckoff}&\multicolumn{2}{c|}{Little group}& \multirow{2}{*}{Coordinates}&\multirow{2}{*}{LSM anomaly class}&\multirow{2}{*}{Topo. inv.} \\ \cline{2-3} position & Intl. & Sch\"{o}nflies & & & \\ \hline
\multirow{2}{*}{4a} & \multirow{2}{*}{$32$} & \multirow{2}{*}{$D_3$} & $(3/8,3/8,3/8)$, $(1/8,5/8,7/8)$, & \multirow{2}{*}{$C_\gamma$} & \multirow{2}{*}{$\varphi_2[T_1T_2^{-1}, C''_2]$}\\
& & & $(5/8,7/8,1/8)$, $(7/8,1/8,5/8)$ & & \\ \hline
\multirow{2}{*}{4b} & \multirow{2}{*}{$32$} & \multirow{2}{*}{$D_3$} & $(7/8,7/8,7/8)$, $(5/8,1/8,3/8)$, & \multirow{2}{*}{Same as 4a } & \multirow{2}{*}{Same as 4a}\\
& & & $(1/8,3/8,5/8)$, $(3/8,5/8,1/8)$ & & \\ \hline
\hline 
 \end{tabular} 
 \end{center}

\subsection*{No. 214: $I4_132$}\label{subsub:sg214}

This group is generated by three translations $T_{1,2,3}$ as given in Eqs.~\eqref{TransBravaisI}, a two-fold rotation $C_2$, a two-fold rotation $C'_2$, a three-fold rotation $C_3$, and a two-fold rotation $C''_2$:
\begin{subequations}
 \begin{align}
C_2 &\colon (x,y,z)\rightarrow (-x, -y + 1/2, z),\\ 
C'_2 &\colon (x,y,z)\rightarrow (-x + 1/2, y, -z),\\ 
C_3 &\colon (x,y,z)\rightarrow (z, x, y),\\ 
C''_2 &\colon (x,y,z)\rightarrow (-y + 1/4, -x + 1/4, -z + 1/4).
\end{align}
\end{subequations}

The $\mathbb{Z}_2$ cohomology ring is given by

\begin{equation}
\mathbb{Z}_2[A_{c''},A_{x+y+z},C_\gamma]/\langle\mathcal{R}_6\rangle
 \end{equation}
where the relations are 
\begin{subequations} 
 \begin{align}
\mathcal{R}_6\colon & ~~
C_\gamma (A_{c''}^2 A_{x+y+z} + A_{c''} A_{x+y+z}^2 + C_\gamma).
\end{align} 
 \end{subequations}
We have the following table regarding IWPs and group cohomology at degree 3.
\begin{center}
\begin{tabular}{c|cc|c|c|c}\hline\hline {Wyckoff}&\multicolumn{2}{c|}{Little group}& {Coordinates}&\multirow{2}{*}{LSM anomaly class}&\multirow{2}{*}{Topo. inv.} \\ \cline{2-4} position & Intl. & Sch\"{o}nflies & $(0,0,0) + ~(1/2,1/2,1/2) + $ & &\\ \hline
\multirow{2}{*}{8a} & \multirow{2}{*}{$32$} & \multirow{2}{*}{$D_3$} & $(1/8,1/8,1/8)$, $(3/8,7/8,5/8)$, & \multirow{2}{*}{N/A} & \multirow{2}{*}{N/A}\\
& & & $(7/8,5/8,3/8)$, $(5/8,3/8,7/8)$ & & \\ \hline
\multirow{2}{*}{8b} & \multirow{2}{*}{$32$} & \multirow{2}{*}{$D_3$} & $(7/8,7/8,7/8)$, $(5/8,1/8,3/8)$, & \multirow{2}{*}{N/A} & \multirow{2}{*}{N/A}\\
& & & $(1/8,3/8,5/8)$, $(3/8,5/8,1/8)$ & & \\ \hline
\multirow{3}{*}{12c} & \multirow{3}{*}{$222$} & \multirow{3}{*}{$D_2$} & $(1/8,0,1/4)$, $(3/8,0,3/4)$, & \multirow{3}{*}{$A_{c''}^2 A_{x+y+z} + A_{c''} A_{x+y+z}^2 + C_\gamma$} & \multirow{3}{*}{$\varphi_2[C_2, C''_2]$}\\
& & & $(1/4,1/8,0)$, $(3/4,3/8,0)$, & & \\
& & & $(0,1/4,1/8)$, $(0,3/4,3/8)$ & & \\ \hline
\multirow{3}{*}{12d} & \multirow{3}{*}{$222$} & \multirow{3}{*}{$D_2$} & $(5/8,0,1/4)$, $(7/8,0,3/4)$, & \multirow{3}{*}{$C_\gamma$} & \multirow{3}{*}{$\varphi_2[C_2, T_1T_2C''_2]$}\\
& & & $(1/4,5/8,0)$, $(3/4,7/8,0)$, & & \\
& & & $(0,1/4,5/8)$, $(0,3/4,7/8)$ & & \\ \hline
\hline 
 \end{tabular} 
 \end{center}

\subsection*{No. 215: $P\overline43m$}\label{subsub:sg215}

This group is generated by three translations $T_{1,2,3}$ as given in Eqs.~\eqref{TransBravaisP}, a two-fold rotation $C_2$, a two-fold rotation $C'_2$, a three-fold rotation $C_3$, and a mirror $M$:
\begin{subequations}
 \begin{align}
C_2 &\colon (x,y,z)\rightarrow (-x, -y, z),\\ 
C'_2 &\colon (x,y,z)\rightarrow (-x, y, -z),\\ 
C_3 &\colon (x,y,z)\rightarrow (z, x, y),\\ 
M &\colon (x,y,z)\rightarrow (y, x, z).
\end{align}
\end{subequations}

The $\mathbb{Z}_2$ cohomology ring is given by

\begin{equation}
\mathbb{Z}_2[A_m,A_{x+y+z},B_\alpha,B_{xy+xz+yz},C_\alpha,C_{xyz}]/\langle\mathcal{R}_3,\mathcal{R}_4,\mathcal{R}_5,\mathcal{R}_6\rangle
 \end{equation}
where the relations are 
\begin{subequations} 
 \begin{align}
\mathcal{R}_3\colon & ~~
A_m A_{x+y+z}^2,\\
\mathcal{R}_4\colon & ~~
A_m C_\alpha,~~A_{x+y+z} (A_{x+y+z}^3 + A_{x+y+z} B_\alpha + C_\alpha),~~A_m (A_{x+y+z} B_{xy+xz+yz} + C_{xyz}),~~A_{x+y+z} C_{xyz},\nonumber\\&~~B_{xy+xz+yz} (A_{x+y+z}^2 + B_\alpha + B_{xy+xz+yz}),\\
\mathcal{R}_5\colon & ~~
A_{x+y+z}^3 B_{xy+xz+yz} + A_{x+y+z} B_\alpha B_{xy+xz+yz} + B_{xy+xz+yz} C_\alpha + B_\alpha C_{xyz},\nonumber\\&~~B_{xy+xz+yz} (A_{x+y+z}^3 + A_{x+y+z} B_\alpha + C_\alpha + C_{xyz}),\\
\mathcal{R}_6\colon & ~~
C_{xyz} (C_\alpha + C_{xyz}).
\end{align} 
 \end{subequations}
We have the following table regarding IWPs and group cohomology at degree 3.
\begin{center}
\begin{tabular}{c|cc|c|c|c}\hline\hline {Wyckoff}&\multicolumn{2}{c|}{Little group}& \multirow{2}{*}{Coordinates}&\multirow{2}{*}{LSM anomaly class}&\multirow{2}{*}{Topo. inv.} \\ \cline{2-3} position & Intl. & Sch\"{o}nflies & & & \\ \hline
1a&$\overline{4}3m$&$T_d$& $(0,0,0)$ & $A_{x+y+z}^3 + A_{x+y+z} B_\alpha + C_\alpha + C_{xyz}$ & $\varphi_2[C_2, C'_2]$\\ 
1b&$\overline{4}3m$&$T_d$& $(1/2,1/2,1/2)$ & $C_{xyz}$ & $\varphi_2[T_1T_2C_2, T_1T_3C'_2]$\\ 
\hline
\multirow{2}{*}{3c} & \multirow{2}{*}{$\overline{4}2m$} & \multirow{2}{*}{$D_{2d}$} & $(0,1/2,1/2)$, $(1/2,0,1/2)$, & \multirow{2}{*}{$A_{x+y+z} B_{xy+xz+yz}$} & \multirow{2}{*}{$\varphi_2[T_1T_2C_2, T_1C'_2]$}\\
& & & $(1/2,1/2,0)$ & & \\ \hline
\multirow{2}{*}{3d} & \multirow{2}{*}{$\overline{4}2m$} & \multirow{2}{*}{$D_{2d}$} & $(1/2,0,0)$, $(0,1/2,0)$, & \multirow{2}{*}{$A_{x+y+z} (A_{x+y+z}^2 + B_\alpha + B_{xy+xz+yz})$} & \multirow{2}{*}{$\varphi_2[T_1C_2, T_1C'_2]$}\\
& & & $(0,0,1/2)$ & & \\ \hline
\hline 
 \end{tabular} 
 \end{center}

\subsection*{No. 216: $F\overline43m$}\label{subsub:sg216}

This group is generated by three translations $T_{1,2,3}$ as given in Eqs.~\eqref{TransBravaisF}, a two-fold rotation $C_2$, a two-fold rotation $C'_2$, a three-fold rotation $C_3$, and a mirror $M$:
\begin{subequations}
 \begin{align}
C_2 &\colon (x,y,z)\rightarrow (-x, -y, z),\\ 
C'_2 &\colon (x,y,z)\rightarrow (-x, y, -z),\\ 
C_3 &\colon (x,y,z)\rightarrow (z, x, y),\\ 
M &\colon (x,y,z)\rightarrow (y, x, z).
\end{align}
\end{subequations}

The $\mathbb{Z}_2$ cohomology ring is given by

\begin{equation}
\mathbb{Z}_2[A_m,B_\alpha,B_\beta,B_{xy+xz+yz},C_\alpha,C_\beta,C_\gamma,C_{xyz}]/\langle\mathcal{R}_4,\mathcal{R}_5,\mathcal{R}_6\rangle
 \end{equation}
where the relations are 
\begin{subequations} 
 \begin{align}
\mathcal{R}_4\colon & ~~
A_m C_\alpha,~~A_m C_\beta,~~B_\alpha B_{xy+xz+yz} + A_m C_\gamma,~~B_\alpha B_\beta + B_\beta^2 + A_m C_\gamma,~~A_m^2 B_\beta + B_\beta B_{xy+xz+yz} + A_m C_{xyz},~~B_{xy+xz+yz}^2,\\
\mathcal{R}_5\colon & ~~
B_{xy+xz+yz} C_\alpha,~~B_\beta C_\alpha + B_\alpha C_\beta,~~B_\beta (C_\alpha + C_\beta),~~B_{xy+xz+yz} C_\beta,~~B_{xy+xz+yz} C_\gamma,~~A_m B_\alpha B_\beta + B_\beta C_\gamma + B_\alpha C_{xyz},\nonumber\\&~~A_m B_\alpha B_\beta + A_m^2 C_\gamma + B_\beta C_\gamma + B_\beta C_{xyz},~~A_m^3 B_\beta + A_m^2 C_{xyz} + B_{xy+xz+yz} C_{xyz},\\
\mathcal{R}_6\colon & ~~
C_\beta (C_\alpha + C_\beta),~~C_\beta C_\gamma + C_\alpha C_{xyz},~~(C_\alpha + C_\beta) C_{xyz},~~C_\gamma (C_\alpha + C_\gamma),~~A_m B_\beta C_\gamma + C_\alpha C_{xyz} + C_\gamma C_{xyz},\nonumber\\&~~A_m^2 B_\alpha B_\beta + A_m^3 C_\gamma + C_\alpha C_{xyz} + C_{xyz}^2.
\end{align} 
 \end{subequations}
We have the following table regarding IWPs and group cohomology at degree 3.
\begin{center}
\begin{tabular}{c|cc|c|c|c}\hline\hline \multirow{3}{*}{\shortstack[l]{Wyckoff\\position}}&\multicolumn{2}{c|}{Little group}& {Coordinates}&\multirow{3}{*}{LSM anomaly class}&\multirow{3}{*}{Topo. inv.} \\ \cline{2-4} & \multirow{2}{*}{Intl.} & \multirow{2}{*}{Sch\"{o}nflies} & $(0,0,0) + ~(0,1/2,1/2) + $ & & \\ & & & $ (1/2,0,1/2) + ~(1/2,1/2,0) +$ & &\\ \hline
4a&$\overline{4}3m$&$T_d$& $(0,0,0)$ & $C_\alpha + C_\gamma + C_{xyz}$ & $\varphi_2[C_2, C'_2]$\\ 
4b&$\overline{4}3m$&$T_d$& $(1/2,1/2,1/2)$ & $C_\beta + C_\gamma + C_{xyz}$ & $\varphi_2[C_2, T_1T_2T_3^{-1}C'_2]$\\ 
4c&$\overline{4}3m$&$T_d$& $(1/4,1/4,1/4)$ & $C_\beta + C_{xyz}$ & $\varphi_2[T_3C_2, T_2C'_2]$\\ 
4d&$\overline{4}3m$&$T_d$& $(3/4,3/4,3/4)$ & $C_{xyz}$ & $\varphi_2[T_3^{-1}C_2, T_2^{-1}C'_2]$\\ 
\hline
\hline 
 \end{tabular} 
 \end{center}

\subsection*{No. 217: $I\overline43m$}\label{subsub:sg217}

This group is generated by three translations $T_{1,2,3}$ as given in Eqs.~\eqref{TransBravaisI}, a two-fold rotation $C_2$, a two-fold rotation $C'_2$, a three-fold rotation $C_3$, and a mirror $M$:
\begin{subequations}
 \begin{align}
C_2 &\colon (x,y,z)\rightarrow (-x, -y, z),\\ 
C'_2 &\colon (x,y,z)\rightarrow (-x, y, -z),\\ 
C_3 &\colon (x,y,z)\rightarrow (z, x, y),\\ 
M &\colon (x,y,z)\rightarrow (y, x, z).
\end{align}
\end{subequations}

The $\mathbb{Z}_2$ cohomology ring is given by

\begin{equation}
\mathbb{Z}_2[A_m,A_{x+y+z},B_\alpha,B_\beta,C_\alpha,C_{xyz}]/\langle\mathcal{R}_2,\mathcal{R}_3,\mathcal{R}_4,\mathcal{R}_5,\mathcal{R}_6\rangle
 \end{equation}
where the relations are 
\begin{subequations} 
 \begin{align}
\mathcal{R}_2\colon & ~~
A_{x+y+z}^2,\\
\mathcal{R}_3\colon & ~~
A_{x+y+z} B_\alpha + A_m B_\beta,~~A_{x+y+z} (B_\alpha + B_\beta),\\
\mathcal{R}_4\colon & ~~
A_m C_\alpha,~~A_{x+y+z} C_\alpha,~~A_{x+y+z} C_{xyz},\\
\mathcal{R}_5\colon & ~~
A_{x+y+z} B_\alpha^2 + B_\beta C_\alpha + B_\beta C_{xyz},\\
\mathcal{R}_6\colon & ~~
B_\alpha^2 B_\beta + B_\beta^3 + C_\alpha C_{xyz},~~B_\alpha^2 B_\beta + B_\beta^3 + C_{xyz}^2.
\end{align} 
 \end{subequations}
We have the following table regarding IWPs and group cohomology at degree 3.
\begin{center}
\begin{tabular}{c|cc|c|c|c}\hline\hline {Wyckoff}&\multicolumn{2}{c|}{Little group}& {Coordinates}&\multirow{2}{*}{LSM anomaly class}&\multirow{2}{*}{Topo. inv.} \\ \cline{2-4} position & Intl. & Sch\"{o}nflies & $(0,0,0) + ~(1/2,1/2,1/2) + $ & &\\ \hline
2a&$\overline{4}3m$&$T_d$& $(0,0,0)$ & $A_{x+y+z} B_\alpha + C_\alpha + C_{xyz}$ & $\varphi_2[C_2, C'_2]$\\ 
\hline
\multirow{2}{*}{6b} & \multirow{2}{*}{$\overline{4}2m$} & \multirow{2}{*}{$D_{2d}$} & $(0,1/2,1/2)$, $(1/2,0,1/2)$, & \multirow{2}{*}{$C_{xyz}$} & \multirow{2}{*}{$\varphi_2[C_2, T_1T_2C'_2]$}\\
& & & $(1/2,1/2,0)$ & & \\ \hline
\multirow{3}{*}{12d} & \multirow{3}{*}{$\overline{4}$} & \multirow{3}{*}{$S_4$} & $(1/4,1/2,0)$, $(3/4,1/2,0)$, & \multirow{3}{*}{$A_{x+y+z} B_\alpha$} & \multirow{3}{*}{$\varphi_2[T_1T_2T_3C'_2M, T_1T_3C_2]$}\\
& & & $(0,1/4,1/2)$, $(0,3/4,1/2)$, & & \\
& & & $(1/2,0,1/4)$, $(1/2,0,3/4)$ & & \\ \hline
\hline 
 \end{tabular} 
 \end{center}

\subsection*{No. 218: $P\overline43n$}\label{subsub:sg218}

This group is generated by three translations $T_{1,2,3}$ as given in Eqs.~\eqref{TransBravaisP}, a two-fold rotation $C_2$, a two-fold rotation $C'_2$, a three-fold rotation $C_3$, and a glide $G$:
\begin{subequations}
 \begin{align}
C_2 &\colon (x,y,z)\rightarrow (-x, -y, z),\\ 
C'_2 &\colon (x,y,z)\rightarrow (-x, y, -z),\\ 
C_3 &\colon (x,y,z)\rightarrow (z, x, y),\\ 
G &\colon (x,y,z)\rightarrow (y + 1/2, x + 1/2, z + 1/2).
\end{align}
\end{subequations}

The $\mathbb{Z}_2$ cohomology ring is given by

\begin{equation}
\mathbb{Z}_2[A_m,B_\alpha,B_\beta,B_{xy+xz+yz},C_\alpha,C_\beta,C_\gamma]/\langle\mathcal{R}_2,\mathcal{R}_3,\mathcal{R}_4,\mathcal{R}_5,\mathcal{R}_6\rangle
 \end{equation}
where the relations are 
\begin{subequations} 
 \begin{align}
\mathcal{R}_2\colon & ~~
A_m^2,\\
\mathcal{R}_3\colon & ~~
A_m B_\beta,\\
\mathcal{R}_4\colon & ~~
A_m C_\alpha,~~A_m C_\beta,~~A_m C_\gamma,~~B_\beta^2 + B_\alpha B_{xy+xz+yz} + B_\beta B_{xy+xz+yz} + B_{xy+xz+yz}^2,\\
\mathcal{R}_5\colon & ~~
A_m B_\alpha^2 + B_\alpha C_\alpha + B_{xy+xz+yz} C_\alpha + B_\alpha C_\beta + B_\beta C_\beta,\nonumber\\&~~A_m B_\alpha B_{xy+xz+yz} + B_\beta C_\alpha + B_{xy+xz+yz} C_\alpha + B_\alpha C_\beta + B_{xy+xz+yz} C_\beta + B_\alpha C_\gamma,\nonumber\\&~~A_m B_\alpha^2 + A_m B_\alpha B_{xy+xz+yz} + B_\alpha C_\alpha + B_{xy+xz+yz} C_\alpha + B_\alpha C_\beta + B_{xy+xz+yz} C_\beta + B_\beta C_\gamma,\nonumber\\&~~A_m B_\alpha^2 + B_\alpha C_\alpha + B_\beta C_\alpha + B_\alpha C_\beta + B_{xy+xz+yz} C_\beta + B_{xy+xz+yz} C_\gamma,\\
\mathcal{R}_6\colon & ~~
B_\alpha B_\beta^2 + B_\beta^3 + C_\alpha C_\gamma,~~B_\alpha B_\beta^2 + C_\alpha^2 + C_\alpha C_\beta + C_\beta^2,~~B_\alpha B_\beta^2 + B_\beta^2 B_{xy+xz+yz} + C_\beta C_\gamma,~~B_\alpha B_\beta^2 + B_\beta^3 + C_\gamma^2.
\end{align} 
 \end{subequations}
We have the following table regarding IWPs and group cohomology at degree 3.
\begin{center}
\begin{tabular}{c|cc|c|c|c}\hline\hline {Wyckoff}&\multicolumn{2}{c|}{Little group}& \multirow{2}{*}{Coordinates}&\multirow{2}{*}{LSM anomaly class}&\multirow{2}{*}{Topo. inv.} \\ \cline{2-3} position & Intl. & Sch\"{o}nflies & & & \\ \hline
2a&$23$&$T$& $(0,0,0)$, $(1/2,1/2,1/2)$ & $A_m B_\alpha + C_\alpha + C_\gamma$ & $\varphi_2[C_2, C'_2]$\\ 
\hline
\multirow{3}{*}{6b} & \multirow{3}{*}{$222$} & \multirow{3}{*}{$D_2$} & $(0,1/2,1/2)$, $(1/2,0,1/2)$, & \multirow{3}{*}{$C_\gamma$} & \multirow{3}{*}{$\varphi_2[C_2, T_3C'_2]$}\\
& & & $(1/2,1/2,0)$, $(0,1/2,0)$, & & \\
& & & $(1/2,0,0)$, $(0,0,1/2)$ & & \\ \hline
\multirow{3}{*}{6c} & \multirow{3}{*}{$\overline{4}$} & \multirow{3}{*}{$S_4$} & $(1/4,1/2,0)$, $(3/4,1/2,0)$, & \multirow{3}{*}{$A_m (B_\alpha + B_{xy+xz+yz})$} & \multirow{3}{*}{$\varphi_2[C'_2G, T_1^{-1}C_2]$}\\
& & & $(0,1/4,1/2)$, $(0,3/4,1/2)$, & & \\
& & & $(1/2,0,1/4)$, $(1/2,0,3/4)$ & & \\ \hline
\multirow{3}{*}{6d} & \multirow{3}{*}{$\overline{4}$} & \multirow{3}{*}{$S_4$} & $(1/4,0,1/2)$, $(3/4,0,1/2)$, & \multirow{3}{*}{$A_m B_{xy+xz+yz}$} & \multirow{3}{*}{$\varphi_2[T_1C'_2G, T_2C_2]$}\\
& & & $(1/2,1/4,0)$, $(1/2,3/4,0)$, & & \\
& & & $(0,1/2,1/4)$, $(0,1/2,3/4)$ & & \\ \hline
\hline 
 \end{tabular} 
 \end{center}

\subsection*{No. 219: $F\overline43c$}\label{subsub:sg219}

This group is generated by three translations $T_{1,2,3}$ as given in Eqs.~\eqref{TransBravaisF}, a two-fold rotation $C_2$, a two-fold rotation $C'_2$, a three-fold rotation $C_3$, and a glide $G$:
\begin{subequations}
 \begin{align}
C_2 &\colon (x,y,z)\rightarrow (-x, -y, z),\\ 
C'_2 &\colon (x,y,z)\rightarrow (-x, y, -z),\\ 
C_3 &\colon (x,y,z)\rightarrow (z, x, y),\\ 
G &\colon (x,y,z)\rightarrow (y + 1/2, x + 1/2, z + 1/2).
\end{align}
\end{subequations}

The $\mathbb{Z}_2$ cohomology ring is given by

\begin{equation}
\mathbb{Z}_2[A_m,B_\alpha,B_{xy+xz+yz},C_\alpha,C_\beta,F_\gamma,F_\delta]/\langle\mathcal{R}_3,\mathcal{R}_4,\mathcal{R}_5,\mathcal{R}_6,\mathcal{R}_7,\mathcal{R}_8,\mathcal{R}_9,\mathcal{R}_{12}\rangle
 \end{equation}
where the relations are 
\begin{subequations} 
 \begin{align}
\mathcal{R}_3\colon & ~~
A_m^3,\\
\mathcal{R}_4\colon & ~~
A_m^2 B_\alpha,~~A_m^2 B_{xy+xz+yz},~~A_m C_\alpha,~~A_m C_\beta,~~B_{xy+xz+yz} (B_\alpha + B_{xy+xz+yz}),\\
\mathcal{R}_5\colon & ~~
B_{xy+xz+yz} C_\alpha + B_\alpha C_\beta,~~B_{xy+xz+yz} (C_\alpha + C_\beta),\\
\mathcal{R}_6\colon & ~~
C_\beta (C_\alpha + C_\beta),\\
\mathcal{R}_7\colon & ~~
A_mF_\gamma,~~A_mF_\delta,\\
\mathcal{R}_8\colon & ~~ B_{xy+xz+yz}F_\gamma+ B_\alpha F_\delta + B_\alpha C_\alpha C_\beta,~~
(B_{xy+xz+yz}+B_\alpha) F_\delta,\\
\mathcal{R}_9\colon & ~~ C_\beta F_\gamma +  C_\beta F_\delta+C_\alpha^2 C_\beta,~~
(C_\alpha+C_\beta) F_\delta,\\
\mathcal{R}_{12}\colon & ~~ F_\gamma^2+C_\alpha^2(F_\gamma+B_\alpha^3+C_\alpha^2),~~
F_\delta^2 +C_\alpha C_\beta(F_\delta+B_{xy+xz+yz}^3+C_\alpha C_\beta),~~
F_\gamma F_\delta
+
C_\alpha C_\beta(B_\alpha^3+C_\alpha^2).
\end{align} 
 \end{subequations}
We have the following table regarding IWPs and group cohomology at degree 3.
\begin{center}
\resizebox{\columnwidth}{!}{
\begin{tabular}{c|cc|c|c|c}\hline\hline \multirow{3}{*}{\shortstack[l]{Wyckoff\\position}}&\multicolumn{2}{c|}{Little group}& {Coordinates}&\multirow{3}{*}{LSM anomaly class}&\multirow{3}{*}{Topo. inv.} \\ \cline{2-4} & \multirow{2}{*}{Intl.} & \multirow{2}{*}{Sch\"{o}nflies} & $(0,0,0) + ~(0,1/2,1/2) + $ & & \\ & & & $ (1/2,0,1/2) + ~(1/2,1/2,0) +$ & &\\ \hline
8a&$23$&$T$& $(0,0,0)$, $(1/2,1/2,1/2)$ & $A_m B_\alpha + A_m B_{xy+xz+yz} + C_\alpha + C_\beta$ & $\varphi_2[C_2, C'_2]$\\ 
8b&$23$&$T$& $(1/4,1/4,1/4)$, $(3/4,3/4,3/4)$ & $A_m B_{xy+xz+yz} + C_\beta$ & $\varphi_2[T_3C_2, T_2C'_2]$\\ 
\hline
\multirow{3}{*}{24c} & \multirow{3}{*}{$\overline{4}$} & \multirow{3}{*}{$S_4$} & $(0,1/4,1/4)$, $(0,3/4,1/4)$, & \multirow{3}{*}{$A_m B_{xy+xz+yz}$} & \multirow{3}{*}{$\varphi_2[T_2C'_2G, T_1T_2^{-1}C_2]$}\\
& & & $(1/4,0,1/4)$, $(1/4,0,3/4)$, & & \\
& & & $(1/4,1/4,0)$, $(3/4,1/4,0)$ & & \\ \hline
\multirow{3}{*}{24d} & \multirow{3}{*}{$\overline{4}$} & \multirow{3}{*}{$S_4$} & $(1/4,0,0)$, $(3/4,0,0)$, & \multirow{3}{*}{$A_m (B_\alpha + B_{xy+xz+yz})$} & \multirow{3}{*}{$\varphi_2[T_2^{2}T_3^{-1}C'_2G, C_2]$}\\
& & & $(0,1/4,0)$, $(0,3/4,0)$, & & \\
& & & $(0,0,1/4)$, $(0,0,3/4)$ & & \\ \hline
\hline 
 \end{tabular} }
 \end{center}

\subsection*{No. 220: $I\overline43d$}\label{subsub:sg220}

This group is generated by three translations $T_{1,2,3}$ as given in Eqs.~\eqref{TransBravaisI}, a two-fold rotation $C_2$, a two-fold rotation $C'_2$, a three-fold rotation $C_3$, and a glide $G$:
\begin{subequations}
 \begin{align}
C_2 &\colon (x,y,z)\rightarrow (-x, -y + 1/2, z),\\ 
C'_2 &\colon (x,y,z)\rightarrow (-x + 1/2, y, -z),\\ 
C_3 &\colon (x,y,z)\rightarrow (z, x, y),\\ 
G &\colon (x,y,z)\rightarrow (y + 1/4, x + 1/4, z + 1/4).
\end{align}
\end{subequations}

The $\mathbb{Z}_2$ cohomology ring is given by

\begin{equation}
\mathbb{Z}_2[A_m,B_\alpha,C_\gamma]/\langle\mathcal{R}_2,\mathcal{R}_6\rangle
 \end{equation}
where the relations are 
\begin{subequations} 
 \begin{align}
\mathcal{R}_2\colon & ~~
A_m^2,\\
\mathcal{R}_6\colon & ~~
C_\gamma (A_m B_\alpha + C_\gamma).
\end{align} 
 \end{subequations}
We have the following table regarding IWPs and group cohomology at degree 3.
\begin{center}
\resizebox{\columnwidth}{!}{
\begin{tabular}{c|cc|c|c|c}\hline\hline {Wyckoff}&\multicolumn{2}{c|}{Little group}& {Coordinates}&\multirow{2}{*}{LSM anomaly class}&\multirow{2}{*}{Topo. inv.} \\ \cline{2-4} position & Intl. & Sch\"{o}nflies & $(0,0,0) + ~(1/2,1/2,1/2) + $ & &\\ \hline
\multirow{3}{*}{12a} & \multirow{3}{*}{$\overline{4}$} & \multirow{3}{*}{$S_4$} & $(3/8,0,1/4)$, $(1/8,0,3/4)$, & \multirow{3}{*}{$C_\gamma$} & \multirow{3}{*}{$\varphi_2[T_1T_2C'_2G, C_2]$}\\
& & & $(1/4,3/8,0)$, $(3/4,1/8,0)$, & & \\
& & & $(0,1/4,3/8)$, $(0,3/4,1/8)$ & & \\ \hline
\multirow{3}{*}{12b} & \multirow{3}{*}{$\overline{4}$} & \multirow{3}{*}{$S_4$} & $(7/8,0,1/4)$, $(5/8,0,3/4)$, & \multirow{3}{*}{$A_m B_\alpha + C_\gamma$} & \multirow{3}{*}{$\varphi_2[C'_2G, C_2]$}\\
& & & $(1/4,7/8,0)$, $(3/4,5/8,0)$, & & \\
& & & $(0,1/4,7/8)$, $(0,3/4,5/8)$ & & \\ \hline
\multirow{4}{*}{16c} & \multirow{4}{*}{$3$} & \multirow{4}{*}{$C_3$} & $(x,x,x)$, $(-x+1/2,-x,x+1/2)$, & \multirow{4}{*}{N/A} & \multirow{4}{*}{N/A}\\
& & & $(-x,x+1/2,-x+1/2)$, $(x+1/2,-x+1/2,-x)$, & & \\
& & & $(x+1/4,x+1/4,x+1/4)$, $(-x+1/4,-x+3/4,x+3/4)$, & & \\
& & & $(x+3/4,-x+1/4,-x+3/4)$, $(-x+3/4,x+3/4,-x+1/4)$ & & \\ \hline
\hline 
 \end{tabular} }
 \end{center}

\subsection*{No. 221: $Pm\overline3m$}\label{subsub:sg221}

This group is generated by three translations $T_{1,2,3}$ as given in Eqs.~\eqref{TransBravaisP}, a two-fold rotation $C_2$, a two-fold rotation $C'_2$, a three-fold rotation $C_3$, a mirror $M$, and an inversion $I$:
\begin{subequations}
 \begin{align}
C_2 &\colon (x,y,z)\rightarrow (-x, -y, z),\\ 
C'_2 &\colon (x,y,z)\rightarrow (-x, y, -z),\\ 
C_3 &\colon (x,y,z)\rightarrow (z, x, y),\\ 
M &\colon (x,y,z)\rightarrow (y, x, z),\\ 
I &\colon (x,y,z)\rightarrow (-x, -y, -z).
\end{align}
\end{subequations}

The $\mathbb{Z}_2$ cohomology ring is given by

\begin{equation}
\mathbb{Z}_2[A_i,A_m,A_{x+y+z},B_\alpha,B_{xy+xz+yz},C_\alpha,C_{xyz}]/\langle\mathcal{R}_3,\mathcal{R}_4,\mathcal{R}_5,\mathcal{R}_6\rangle
 \end{equation}
where the relations are 
\begin{subequations} 
 \begin{align}
\mathcal{R}_3\colon & ~~
A_m A_{x+y+z} (A_i + A_{x+y+z}),\\
\mathcal{R}_4\colon & ~~
A_i^2 A_{x+y+z}^2 + A_i A_{x+y+z}^3 + A_{x+y+z}^4 + A_i A_{x+y+z} B_\alpha + A_{x+y+z}^2 B_\alpha + A_i A_{x+y+z} B_{xy+xz+yz} + A_{x+y+z} C_\alpha + A_i C_{xyz},\nonumber\\
& ~~A_m C_\alpha,~~A_m (A_i^2 A_{x+y+z} + A_{x+y+z} B_{xy+xz+yz} + C_{xyz}),\nonumber\\
& ~~A_{x+y+z} (A_i^3 + A_i A_{x+y+z}^2 + A_{x+y+z}^3 + A_i B_\alpha + A_{x+y+z} B_\alpha + A_i B_{xy+xz+yz} + C_\alpha + C_{xyz}),\nonumber\\
& ~~B_{xy+xz+yz} (A_i^2 + A_i A_m + A_i A_{x+y+z} + A_{x+y+z}^2 + B_\alpha + B_{xy+xz+yz}),\\
\mathcal{R}_5\colon & ~~
A_i^4 A_{x+y+z} + A_i^3 A_{x+y+z}^2 + A_i^2 A_{x+y+z}^3 + A_i A_{x+y+z}^4 + A_i A_{x+y+z}^2 B_\alpha + A_i^2 A_{x+y+z} B_{xy+xz+yz} \nonumber\\
& ~~+ A_{x+y+z}^3 B_{xy+xz+yz} + A_{x+y+z} B_\alpha B_{xy+xz+yz} + A_i A_{x+y+z} C_\alpha + B_{xy+xz+yz} C_\alpha + B_\alpha C_{xyz},\nonumber\\
& ~~B_{xy+xz+yz} (A_i^2 A_{x+y+z} + A_i A_m A_{x+y+z} + A_{x+y+z}^3 + A_{x+y+z} B_\alpha + C_\alpha + C_{xyz}),\\
\mathcal{R}_6\colon & ~~
A_i^5 A_{x+y+z} + A_i^3 A_{x+y+z}^3 + A_i^2 A_{x+y+z}^4 + A_i A_{x+y+z}^3 B_\alpha + A_{x+y+z}^4 B_\alpha + A_i A_{x+y+z} B_\alpha^2 + A_{x+y+z}^2 B_\alpha^2 \nonumber\\
& ~~+ A_i^3 A_{x+y+z} B_{xy+xz+yz} + A_i^2 A_m A_{x+y+z} B_{xy+xz+yz} + A_i A_{x+y+z} B_\alpha B_{xy+xz+yz} + A_{x+y+z} B_\alpha C_\alpha + C_\alpha C_{xyz} + C_{xyz}^2.
\end{align} 
 \end{subequations}
We have the following table regarding IWPs and group cohomology at degree 3.
\begin{center}
\resizebox{\columnwidth}{!}{
\begin{tabular}{c|cc|c|c|c}\hline\hline {Wyckoff}&\multicolumn{2}{c|}{Little group}& \multirow{2}{*}{Coordinates}&\multirow{2}{*}{LSM anomaly class}&\multirow{2}{*}{Topo. inv.} \\ \cline{2-3} position & Intl. & Sch\"{o}nflies & & & \\ \hline
\multirow{2}{*}{1a}&\multirow{2}{*}{$m\overline{3}m$}&\multirow{2}{*}{$O_h$}& \multirow{2}{*}{$(0,0,0)$} & $A_i^3 + A_i^2 A_m + A_i A_m A_{x+y+z} + A_i A_{x+y+z}^2 + A_{x+y+z}^3 $ & \multirow{2}{*}{$\varphi_1[I]$}\\
&&&& $+ A_i B_\alpha + A_{x+y+z} B_\alpha + A_i B_{xy+xz+yz} + C_\alpha + C_{xyz}$ &\\\hline
1b&$m\overline{3}m$&$O_h$& $(1/2,1/2,1/2)$ & $A_i^2 A_{x+y+z} + C_{xyz}$ & $\varphi_1[T_1T_2T_3I]$\\ 
\hline
\multirow{2}{*}{3c} & \multirow{2}{*}{$4/mmm$} & \multirow{2}{*}{$D_{4h}$} & $(0,1/2,1/2)$, $(1/2,0,1/2)$, & \multirow{2}{*}{$(A_i + A_{x+y+z}) B_{xy+xz+yz}$} & \multirow{2}{*}{$\varphi_2[T_1T_2C_2, T_1C'_2]$}\\
& & & $(1/2,1/2,0)$ & & \\ \hline
\multirow{2}{*}{3d} & \multirow{2}{*}{$4/mmm$} & \multirow{2}{*}{$D_{4h}$} & $(1/2,0,0)$, $(0,1/2,0)$, & \multirow{2}{*}{$A_{x+y+z} (A_i^2 + A_i A_m + A_i A_{x+y+z} + A_{x+y+z}^2 + B_\alpha + B_{xy+xz+yz})$} & \multirow{2}{*}{$\varphi_2[C_2, T_3C'_2]$}\\
& & & $(0,0,1/2)$ & & \\ \hline
\hline 
 \end{tabular} }
 \end{center}

\subsection*{No. 222: $Pn\overline3n$}\label{subsub:sg222}

This group is generated by three translations $T_{1,2,3}$ as given in Eqs.~\eqref{TransBravaisP}, a two-fold rotation $C_2$, a two-fold rotation $C'_2$, a three-fold rotation $C_3$, a glide $G$, and an inversion $I$:
\begin{subequations}
 \begin{align}
C_2 &\colon (x,y,z)\rightarrow (-x + 1/2, -y + 1/2, z),\\ 
C'_2 &\colon (x,y,z)\rightarrow (-x + 1/2, y, -z + 1/2),\\ 
C_3 &\colon (x,y,z)\rightarrow (z, x, y),\\ 
G &\colon (x,y,z)\rightarrow (y + 1/2, x + 1/2, z + 1/2),\\ 
I &\colon (x,y,z)\rightarrow (-x, -y, -z).
\end{align}
\end{subequations}

The $\mathbb{Z}_2$ cohomology ring is given by

\begin{equation}
\mathbb{Z}_2[A_i,A_m,B_\alpha,B_\beta,C_\alpha,C_\gamma]/\langle\mathcal{R}_2,\mathcal{R}_3,\mathcal{R}_4,\mathcal{R}_5,\mathcal{R}_6\rangle
 \end{equation}
where the relations are 
\begin{subequations} 
 \begin{align}
\mathcal{R}_2\colon & ~~
A_m (A_i + A_m),\\
\mathcal{R}_3\colon & ~~
A_i (B_\alpha + B_\beta),~~A_i B_\alpha + A_m B_\beta,\\
\mathcal{R}_4\colon & ~~
A_i C_\alpha,~~A_m C_\alpha,~~(A_i + A_m) C_\gamma,\\
\mathcal{R}_5\colon & ~~
(B_\alpha + B_\beta) (C_\alpha + C_\gamma),\\
\mathcal{R}_6\colon & ~~
B_\alpha B_\beta^2 + B_\beta^3 + C_\alpha C_\gamma,~~A_i^5 A_m + A_i^4 B_\alpha + B_\alpha B_\beta^2 + B_\beta^3 + A_i B_\alpha C_\gamma + C_\gamma^2.
\end{align} 
 \end{subequations}
We have the following table regarding IWPs and group cohomology at degree 3.
\begin{center}
\begin{tabular}{c|cc|c|c|c}\hline\hline {Wyckoff}&\multicolumn{2}{c|}{Little group}& \multirow{2}{*}{Coordinates}&\multirow{2}{*}{LSM anomaly class}&\multirow{2}{*}{Topo. inv.} \\ \cline{2-3} position & Intl. & Sch\"{o}nflies & & & \\ \hline
2a&$432$&$O$& $(1/4,1/4,1/4)$, $(3/4,3/4,3/4)$ & $A_i^2 A_m + A_i B_\alpha + C_\alpha + C_\gamma$ & $\varphi_2[C_2, C'_2]$\\ 
\hline
\multirow{3}{*}{6b} & \multirow{3}{*}{$422$} & \multirow{3}{*}{$D_4$} & $(3/4,1/4,1/4)$, $(1/4,3/4,1/4)$, & \multirow{3}{*}{$A_i^2 A_m + A_i B_\alpha + A_m B_\alpha + C_\gamma$} & \multirow{3}{*}{$\varphi_2[C_2, T_3C'_2]$}\\
& & & $(1/4,1/4,3/4)$, $(1/4,3/4,3/4)$, & & \\
& & & $(3/4,1/4,3/4)$, $(3/4,3/4,1/4)$ & & \\ \hline
\multirow{4}{*}{8c} & \multirow{4}{*}{$\overline{3}$} & \multirow{4}{*}{$C_{3i}$} & $(0,0,0)$, $(1/2,1/2,0)$, & \multirow{4}{*}{$A_i^2 (A_i + A_m)$} & \multirow{4}{*}{$\varphi_1[I]$}\\
& & & $(1/2,0,1/2)$, $(0,1/2,1/2)$, & & \\
& & & $(0,0,1/2)$, $(1/2,1/2,1/2)$, & & \\
& & & $(0,1/2,0)$, $(1/2,0,0)$ & & \\ \hline
\multirow{6}{*}{12d} & \multirow{6}{*}{$\overline{4}$} & \multirow{6}{*}{$S_4$} & $(0,3/4,1/4)$, $(1/2,3/4,1/4)$, & \multirow{6}{*}{$(A_i + A_m) B_\alpha$} & \multirow{6}{*}{$\varphi_2[C'_2G, T_1^{-1}C_2]$}\\
& & & $(1/4,0,3/4)$, $(1/4,1/2,3/4)$, & & \\
& & & $(3/4,1/4,0)$, $(3/4,1/4,1/2)$, & & \\
& & & $(3/4,0,1/4)$, $(3/4,1/2,1/4)$, & & \\
& & & $(0,1/4,3/4)$, $(1/2,1/4,3/4)$, & & \\
& & & $(1/4,3/4,1/2)$, $(1/4,3/4,0)$ & & \\ \hline
\hline 
 \end{tabular} 
 \end{center}

\subsection*{No. 223: $Pm\overline3n$}\label{subsub:sg223}

This group is generated by three translations $T_{1,2,3}$ as given in Eqs.~\eqref{TransBravaisP}, a two-fold rotation $C_2$, a two-fold rotation $C'_2$, a three-fold rotation $C_3$, a glide $G$, and an inversion $I$:
\begin{subequations}
 \begin{align}
C_2 &\colon (x,y,z)\rightarrow (-x, -y, z),\\ 
C'_2 &\colon (x,y,z)\rightarrow (-x, y, -z),\\ 
C_3 &\colon (x,y,z)\rightarrow (z, x, y),\\ 
G &\colon (x,y,z)\rightarrow (y + 1/2, x + 1/2, z + 1/2),\\ 
I &\colon (x,y,z)\rightarrow (-x, -y, -z).
\end{align}
\end{subequations}

The $\mathbb{Z}_2$ cohomology ring is given by

\begin{equation}
\mathbb{Z}_2[A_i,A_m,B_\alpha,B_\beta,B_{xy+xz+yz},C_\alpha,C_\beta,C_\gamma]/\langle\mathcal{R}_2,\mathcal{R}_3,\mathcal{R}_4,\mathcal{R}_5,\mathcal{R}_6\rangle
 \end{equation}
where the relations are 
\begin{subequations} 
 \begin{align}
\mathcal{R}_2\colon & ~~
A_m (A_i + A_m),\\
\mathcal{R}_3\colon & ~~
A_m (A_i^2 + B_\beta),\\
\mathcal{R}_4\colon & ~~
A_m C_\alpha,~~A_m (A_i B_\alpha + C_\beta),~~A_m (A_i B_{xy+xz+yz} + C_\gamma),\nonumber\\
&~~A_i^3 A_m + A_i^2 B_\alpha + A_i A_m B_\alpha + B_\beta^2 + A_i^2 B_{xy+xz+yz} + A_i A_m B_{xy+xz+yz} \nonumber\\
&~~+ B_\alpha B_{xy+xz+yz} + B_\beta B_{xy+xz+yz} + B_{xy+xz+yz}^2 + A_i C_\gamma,\\
\mathcal{R}_5\colon & ~~
A_i^5 + A_i^3 B_\alpha + A_m B_\alpha^2 + A_i B_\alpha B_\beta + A_i B_\beta^2 + A_i^2 C_\alpha + B_\alpha C_\alpha + B_{xy+xz+yz} C_\alpha + A_i^2 C_\beta + B_\alpha C_\beta + B_\beta C_\beta,\nonumber\\
&~~A_i^3 B_\alpha + A_i^2 A_m B_\alpha + A_m B_\alpha^2 + A_i^3 B_\beta + A_i B_\alpha B_\beta + A_i B_\beta^2 + A_i^3 B_{xy+xz+yz} + A_i B_\beta B_{xy+xz+yz} \nonumber\\
&~~+ B_\beta C_\alpha + B_{xy+xz+yz} C_\alpha + A_i^2 C_\beta + B_\alpha C_\beta + B_{xy+xz+yz} C_\beta + B_\alpha C_\gamma,\nonumber\\
&~~A_i^5 + A_m B_\alpha^2 + A_i B_\alpha B_\beta + A_i B_\beta^2 + A_i^3 B_{xy+xz+yz} + A_m B_\alpha B_{xy+xz+yz} + A_i B_\beta B_{xy+xz+yz} + A_i^2 C_\alpha \nonumber\\
&~~+ B_\alpha C_\alpha + B_{xy+xz+yz} C_\alpha + A_i^2 C_\beta + B_\alpha C_\beta + B_{xy+xz+yz} C_\beta + A_i^2 C_\gamma + B_\beta C_\gamma,\nonumber\\
&~~A_i^5 + A_i^4 A_m + A_m B_\alpha^2 + A_i^3 B_\beta + A_i B_\alpha B_\beta + A_i B_\beta^2 + A_i B_\beta B_{xy+xz+yz} + A_i^2 C_\alpha + B_\alpha C_\alpha \nonumber\\
&~~+ B_\beta C_\alpha + A_i^2 C_\beta + B_\alpha C_\beta + B_{xy+xz+yz} C_\beta + A_i^2 C_\gamma + B_{xy+xz+yz} C_\gamma,\\
\mathcal{R}_6\colon & ~~
A_i^6 + A_i^2 B_\alpha^2 + A_i A_m B_\alpha^2 + A_i^4 B_\beta + A_i^2 B_\alpha B_\beta + A_i^2 B_\beta^2 + B_\alpha B_\beta^2 + B_\beta^3 + A_i^3 C_\alpha + A_i B_\beta C_\alpha + C_\alpha C_\gamma,\nonumber\\
&~~A_i^6 + A_i^4 B_\alpha + A_i^2 B_\alpha^2 + A_i^2 B_\beta^2 + B_\alpha B_\beta^2 + C_\alpha^2 + C_\alpha C_\beta + C_\beta^2,\nonumber\\
&~~A_i^4 B_\alpha + A_i^3 A_m B_\alpha + A_i^2 B_\alpha^2 + A_i A_m B_\alpha^2 + B_\alpha B_\beta^2 + A_i^2 B_\alpha B_{xy+xz+yz} + A_i A_m B_\alpha B_{xy+xz+yz} \nonumber\\
&~~+ A_i^2 B_\beta B_{xy+xz+yz} + B_\beta^2 B_{xy+xz+yz} + A_i^3 C_\beta + A_i B_{xy+xz+yz} C_\beta + C_\beta C_\gamma,\nonumber\\
&~~A_i^6 + A_i^2 B_\alpha^2 + B_\alpha B_\beta^2 + B_\beta^3 + A_i^4 B_{xy+xz+yz} + A_i^3 A_m B_{xy+xz+yz} + A_i^2 B_\beta B_{xy+xz+yz} + A_i^3 C_\alpha \nonumber\\
&~~+ A_i B_\alpha C_\alpha + A_i B_{xy+xz+yz} C_\alpha + A_i^3 C_\beta + A_i B_\alpha C_\beta  + A_i B_{xy+xz+yz} C_\beta + A_i^3 C_\gamma + C_\gamma^2.
\end{align} 
 \end{subequations}
We have the following table regarding IWPs and group cohomology at degree 3.
\begin{center}
\begin{tabular}{c|cc|c|c|c}\hline\hline {Wyckoff}&\multicolumn{2}{c|}{Little group}& \multirow{2}{*}{Coordinates}&\multirow{2}{*}{LSM anomaly class}&\multirow{2}{*}{Topo. inv.} \\ \cline{2-3} position & Intl. & Sch\"{o}nflies & & & \\ \hline
2a&$m\overline{3}$&$T_h$& $(0,0,0)$, $(1/2,1/2,1/2)$ & $A_i^3 + A_i^2 A_m + A_m B_{xy+xz+yz} + C_\alpha + C_\gamma$ & $\varphi_1[I]$\\ 
\hline
\multirow{3}{*}{6b} & \multirow{3}{*}{$mmm$} & \multirow{3}{*}{$D_{2h}$} & $(0,1/2,1/2)$, $(1/2,0,1/2)$, & \multirow{3}{*}{$A_i B_\alpha + A_m B_\alpha + A_m B_{xy+xz+yz} + C_\gamma$} & \multirow{3}{*}{$\varphi_2[C_2, T_3C'_2]$}\\
& & & $(1/2,1/2,0)$, $(0,1/2,0)$, & & \\
& & & $(1/2,0,0)$, $(0,0,1/2)$ & & \\ \hline
\multirow{3}{*}{6c} & \multirow{3}{*}{$\overline{4}m2$} & \multirow{3}{*}{$D_{2d}$} & $(1/4,0,1/2)$, $(3/4,0,1/2)$, & \multirow{3}{*}{$A_m B_{xy+xz+yz}$} & \multirow{3}{*}{$\varphi_2[T_2C_2, GI]$}\\
& & & $(1/2,1/4,0)$, $(1/2,3/4,0)$, & & \\
& & & $(0,1/2,1/4)$, $(0,1/2,3/4)$ & & \\ \hline
\multirow{3}{*}{6d} & \multirow{3}{*}{$\overline{4}m2$} & \multirow{3}{*}{$D_{2d}$} & $(1/4,1/2,0)$, $(3/4,1/2,0)$, & \multirow{3}{*}{$A_m (B_\alpha + B_{xy+xz+yz})$} & \multirow{3}{*}{$\varphi_2[T_1C_2, GI]$}\\
& & & $(0,1/4,1/2)$, $(0,3/4,1/2)$, & & \\
& & & $(1/2,0,1/4)$, $(1/2,0,3/4)$ & & \\ \hline
\multirow{4}{*}{8e} & \multirow{4}{*}{$32$} & \multirow{4}{*}{$D_3$} & $(1/4,1/4,1/4)$, $(3/4,3/4,1/4)$, & \multirow{4}{*}{N/A} & \multirow{4}{*}{N/A}\\
& & & $(3/4,1/4,3/4)$, $(1/4,3/4,3/4)$, & & \\
& & & $(3/4,3/4,3/4)$, $(1/4,1/4,3/4)$, & & \\
& & & $(1/4,3/4,1/4)$, $(3/4,1/4,1/4)$ & & \\ \hline
\hline 
 \end{tabular} 
 \end{center}

\subsection*{No. 224: $Pn\overline3m$}\label{subsub:sg224}

This group is generated by three translations $T_{1,2,3}$ as given in Eqs.~\eqref{TransBravaisP}, a two-fold rotation $C_2$, a two-fold rotation $C'_2$, a three-fold rotation $C_3$, a mirror $M$, and an inversion $I$:
\begin{subequations}
 \begin{align}
C_2 &\colon (x,y,z)\rightarrow (-x + 1/2, -y + 1/2, z),\\ 
C'_2 &\colon (x,y,z)\rightarrow (-x + 1/2, y, -z + 1/2),\\ 
C_3 &\colon (x,y,z)\rightarrow (z, x, y),\\ 
M &\colon (x,y,z)\rightarrow (y, x, z),\\ 
I &\colon (x,y,z)\rightarrow (-x, -y, -z).
\end{align}
\end{subequations}

The $\mathbb{Z}_2$ cohomology ring is given by

\begin{equation}
\mathbb{Z}_2[A_i,A_m,A_{x+y+z},B_\alpha,C_\alpha]/\langle\mathcal{R}_3,\mathcal{R}_4\rangle
 \end{equation}
where the relations are 
\begin{subequations} 
 \begin{align}
\mathcal{R}_3\colon & ~~
(A_i + A_m) A_{x+y+z} (A_i + A_{x+y+z}),~~A_i (A_i A_{x+y+z} + A_{x+y+z}^2 + B_\alpha),\\
\mathcal{R}_4\colon & ~~
A_i C_\alpha,~~A_m C_\alpha,~~A_{x+y+z} (A_i A_{x+y+z}^2 + A_{x+y+z}^3 + A_{x+y+z} B_\alpha + C_\alpha).
\end{align} 
 \end{subequations}
We have the following table regarding IWPs and group cohomology at degree 3.
\begin{center}
\resizebox{\columnwidth}{!}{
\begin{tabular}{c|cc|c|c|c}\hline\hline {Wyckoff}&\multicolumn{2}{c|}{Little group}& \multirow{2}{*}{Coordinates}&\multirow{2}{*}{LSM anomaly class}&\multirow{2}{*}{Topo. inv.} \\ \cline{2-3} position & Intl. & Sch\"{o}nflies & & & \\ \hline
2a&$\overline{4}3m$&$T_d$& $(1/4,1/4,1/4)$, $(3/4,3/4,3/4)$ & $A_i A_{x+y+z}^2 + A_{x+y+z}^3 + A_{x+y+z} B_\alpha + C_\alpha$ & $\varphi_2[C_2, C'_2]$\\ 
\hline
\multirow{2}{*}{4b} & \multirow{2}{*}{$\overline{3}m$} & \multirow{2}{*}{$D_{3d}$} & $(0,0,0)$, $(1/2,1/2,0)$, & \multirow{2}{*}{$A_i (A_i + A_m) (A_i + A_{x+y+z})$} & \multirow{2}{*}{$\varphi_1[I]$}\\
& & & $(1/2,0,1/2)$, $(0,1/2,1/2)$ & & \\ \hline
\multirow{2}{*}{4c} & \multirow{2}{*}{$\overline{3}m$} & \multirow{2}{*}{$D_{3d}$} & $(1/2,1/2,1/2)$, $(0,0,1/2)$, & \multirow{2}{*}{$A_i (A_i + A_m) A_{x+y+z}$} & \multirow{2}{*}{$\varphi_1[T_1I]$}\\
& & & $(0,1/2,0)$, $(1/2,0,0)$ & & \\ \hline
\multirow{3}{*}{6d} & \multirow{3}{*}{$\overline{4}2m$} & \multirow{3}{*}{$D_{2d}$} & $(1/4,3/4,3/4)$, $(3/4,1/4,3/4)$, & \multirow{3}{*}{$A_{x+y+z} (A_i A_{x+y+z} + A_{x+y+z}^2 + B_\alpha)$} & \multirow{3}{*}{$\varphi_2[C_2, T_3C'_2]$}\\
& & & $(3/4,3/4,1/4)$, $(1/4,3/4,1/4)$, & & \\
& & & $(3/4,1/4,1/4)$, $(1/4,1/4,3/4)$ & & \\ \hline
\multirow{6}{*}{12f} & \multirow{6}{*}{$222$} & \multirow{6}{*}{$D_2$} & $(1/2,1/4,3/4)$, $(0,1/4,3/4)$, & \multirow{6}{*}{$A_i A_{x+y+z} (A_i + A_{x+y+z})$} & \multirow{6}{*}{$\varphi_2[T_2C_2, T_1T_2MI]$}\\
& & & $(3/4,1/2,1/4)$, $(3/4,0,1/4)$, & & \\
& & & $(1/4,3/4,1/2)$, $(1/4,3/4,0)$, & & \\
& & & $(1/2,3/4,1/4)$, $(0,3/4,1/4)$, & & \\
& & & $(1/4,1/2,3/4)$, $(1/4,0,3/4)$, & & \\
& & & $(3/4,1/4,1/2)$, $(3/4,1/4,0)$ & & \\ \hline
\hline 
 \end{tabular} }
 \end{center}

\subsection*{No. 225: $Fm\overline3m$}\label{subsub:sg225}

This group is generated by three translations $T_{1,2,3}$ as given in Eqs.~\eqref{TransBravaisF}, a two-fold rotation $C_2$, a two-fold rotation $C'_2$, a three-fold rotation $C_3$, a mirror $M$, and an inversion $I$:
\begin{subequations}
 \begin{align}
C_2 &\colon (x,y,z)\rightarrow (-x, -y, z),\\ 
C'_2 &\colon (x,y,z)\rightarrow (-x, y, -z),\\ 
C_3 &\colon (x,y,z)\rightarrow (z, x, y),\\ 
M &\colon (x,y,z)\rightarrow (y, x, z),\\ 
I &\colon (x,y,z)\rightarrow (-x, -y, -z).
\end{align}
\end{subequations}

The $\mathbb{Z}_2$ cohomology ring is given by

\begin{equation}
\mathbb{Z}_2[A_i,A_m,B_\alpha,B_\beta,B_{xy+xz+yz},C_\alpha,C_{\gamma},C_{xyz}]/\langle\mathcal{R}_4,\mathcal{R}_5,\mathcal{R}_6\rangle
 \end{equation}
where the relations are 
\begin{subequations} 
 \begin{align}
\mathcal{R}_4\colon & ~~
A_m C_\alpha,~~A_i C_\gamma + A_i^3 A_m + A_i^4 + A_i^2 B_\alpha + A_i^2 B_{xy+xz+yz} + A_i C_\alpha,\nonumber\\&~~A_m C_\gamma + A_i^3 A_m + A_i^2 A_m^2 + A_i A_m B_\alpha + A_i A_m B_{xy+xz+yz},~~B_\beta^2 + A_i A_m B_\beta + B_\alpha B_{xy+xz+yz},\nonumber\\&~~B_\beta B_{xy+xz+yz} + A_i^2 B_\beta + A_i A_m B_\beta + A_m C_{xyz} + B_\alpha B_\beta,\nonumber\\&~~B_{xy+xz+yz}^2 + A_i^2 B_{xy+xz+yz} + A_i A_m B_{xy+xz+yz} + B_\alpha B_{xy+xz+yz},\\
\mathcal{R}_5\colon & ~~
B_{xy+xz+yz} C_\alpha + B_\beta C_\alpha,\nonumber\\&~~B_\alpha C_\gamma + A_i^3 B_\alpha + A_i^2 A_m B_\alpha + A_i B_\alpha^2 + A_i B_\alpha B_{xy+xz+yz} + B_\alpha C_\alpha + B_\beta C_\alpha,\nonumber\\&~~B_\beta C_\gamma + A_i A_m C_{xyz},~~B_{xy+xz+yz} C_\gamma,~~B_\beta C_{xyz} + A_i A_m C_{xyz},~~B_{xy+xz+yz} C_{xyz},\\
\mathcal{R}_6\colon & ~~
C_\gamma^2 + A_i^4 A_m^2 + A_i^6 + A_i^4 B_{xy+xz+yz} + A_i^3 A_m B_{xy+xz+yz} + A_i^3 C_\alpha + A_i^2 B_\alpha^2\nonumber\\&~~{}+ A_i^2 B_\alpha B_{xy+xz+yz} + A_i B_\alpha C_\alpha + C_\alpha C_\gamma,\nonumber\\&~~C_\gamma C_{xyz} + A_i^3 C_{xyz} + A_i^2 A_m C_{xyz} + A_i B_\alpha C_{xyz} + C_\alpha C_{xyz},\nonumber\\&~~C_{xyz}^2 + A_i^3 C_{xyz} + A_i^2 A_m C_{xyz} + A_i B_\alpha C_{xyz} + C_\alpha C_{xyz}.
 \end{align} 
\end{subequations}
We have the following table regarding IWPs and group cohomology at degree 3.
\begin{center}
\resizebox{\columnwidth}{!}{
\begin{tabular}{c|cc|c|c|c}\hline\hline \multirow{3}{*}{\shortstack[l]{Wyckoff\\position}}&\multicolumn{2}{c|}{Little group}& {Coordinates}&\multirow{3}{*}{LSM anomaly class}&\multirow{3}{*}{Topo. inv.} \\ \cline{2-4} & \multirow{2}{*}{Intl.} & \multirow{2}{*}{Sch\"{o}nflies} & $(0,0,0) + ~(0,1/2,1/2) + $ & & \\ & & & $ (1/2,0,1/2) + ~(1/2,1/2,0) +$ & &\\ \hline
4a&$m\overline{3}m$&$O_h$& $(0,0,0)$ & $C_{\gamma}+C_{xyz}$ & $\varphi_1[I]$\\
4b&$m\overline{3}m$&$O_h$& $(1/2,1/2,1/2)$ & $C_{xyz}$ & $\varphi_1[T_1T_2T_3I]$\\
8c&$\overline{4}3m$&$T_d$& $(1/4,1/4,1/4)$, $(1/4,1/4,3/4)$ & $A_i^2A_m+A_i^3+A_iB_\alpha+A_iB_{xy+xz+yz}+C_\alpha+C_{\gamma}$ & $\varphi_2[T_3C_2, T_2C'_2]$\\
\hline
\multirow{3}{*}{24d} & \multirow{3}{*}{$mmm$} & \multirow{3}{*}{$D_{2h}$} & $(0,1/4,1/4)$, $(0,3/4,1/4)$, & \multirow{3}{*}{$A_iB_{xy+xz+yz}$} & \multirow{3}{*}{$\varphi_2[T_3C_2, T_3MI]$}\\
& & & $(1/4,0,1/4)$, $(1/4,0,3/4)$, & & \\
& & & $(1/4,1/4,0)$, $(3/4,1/4,0)$ & & \\ \hline
\hline 
 \end{tabular} }
 \end{center}

\subsection*{No. 226: $Fm\overline3c$}\label{subsub:sg226}

This group is generated by three translations $T_{1,2,3}$ as given in Eqs.~\eqref{TransBravaisF}, a two-fold rotation $C_2$, a two-fold rotation $C'_2$, a three-fold rotation $C_3$, a glide $G$, and an inversion $I$:
\begin{subequations}
 \begin{align}
C_2 &\colon (x,y,z)\rightarrow (-x, -y, z),\\ 
C'_2 &\colon (x,y,z)\rightarrow (-x, y, -z),\\ 
C_3 &\colon (x,y,z)\rightarrow (z, x, y),\\ 
G &\colon (x,y,z)\rightarrow (y + 1/2, x + 1/2, z + 1/2),\\ 
I &\colon (x,y,z)\rightarrow (-x, -y, -z).
\end{align}
\end{subequations}

The $\mathbb{Z}_2$ cohomology ring is given by

\begin{equation}
\mathbb{Z}_2[A_i,A_m,B_\alpha,B_{xy+xz+yz},C_\alpha,C_\beta,F_\delta]/\langle\mathcal{R}_3,\mathcal{R}_4,\mathcal{R}_5,\mathcal{R}_6,\mathcal{R}_7,\mathcal{R}_8,\mathcal{R}_9,\mathcal{R}_{12}\rangle
 \end{equation}
where the relations are 
\begin{subequations} 
 \begin{align}
\mathcal{R}_3\colon & ~~
A_m^2 (A_i + A_m),\\
\mathcal{R}_4\colon & ~~
A_m (A_i + A_m) B_{xy+xz+yz},~~A_m C_\alpha,~~A_i A_m B_\alpha + A_m^2 B_\alpha + A_i C_\beta,~~A_m C_\beta,\nonumber\\&~~B_{xy+xz+yz} (A_i^2 + A_i A_m + B_\alpha + B_{xy+xz+yz}),\\
\mathcal{R}_5\colon & ~~
A_i^2 A_m B_\alpha + A_i A_m^2 B_\alpha + A_i^2 C_\alpha + B_\alpha C_\alpha + B_{xy+xz+yz} C_\alpha + B_\alpha C_\beta,~~B_{xy+xz+yz} C_\beta,\\
\mathcal{R}_6\colon & ~~
C_\beta (C_\alpha + C_\beta),\\
\mathcal{R}_7\colon & ~~
A_m F_\delta,\\
\mathcal{R}_8\colon & ~~
B_{xy+xz+yz}F_\delta + A_i^2F_\delta+B_\alpha F_\delta,\\
\mathcal{R}_9\colon & ~~
C_\beta F_\delta,\\
\mathcal{R}_{12}\colon & ~~
F_\delta^2 + A_i^9 A_m B_\alpha + A_i^8 A_m^2 B_\alpha + A_i^{10}B_{xy+xz+yz}+A_i^9A_mB_{xy+xz+yz}+A_i^9C_\alpha\nonumber\\
&~~+A_i^8B_\alpha B_{xy+xz+yz}+A_i^7A_mB_\alpha B_{xy+xz+yz}+A_i^5A_mB_\alpha^3+A_i^4A_m^2B_\alpha^3+A_i^5B_\alpha^2C_\alpha\nonumber\\
&~~+A_i^4B_\alpha^3B_{xy+xz+yz}+A_i^3A_mB_\alpha^3B_{xy+xz+yz}+A_i^3C_\alpha^3+A_i^3C_\alpha F_\delta+A_i^2B_\alpha^4B_{xy+xz+yz}\nonumber\\
&~~+A_iA_mB_\alpha^4B_{xy+xz+yz}+A_iB_\alpha C_\alpha F_\delta+B_\alpha^2B_{xy+xz+yz}C_\alpha^2+C_\alpha^4+C_\alpha^3C_\beta+C_\alpha^2F_\delta.
\end{align} 
 \end{subequations}
We have the following table regarding IWPs and group cohomology at degree 3.
\begin{center}
\resizebox{\columnwidth}{!}{\begin{tabular}{c|cc|c|c|c}\hline\hline \multirow{3}{*}{\shortstack[l]{Wyckoff\\position}}&\multicolumn{2}{c|}{Little group}& {Coordinates}&\multirow{3}{*}{LSM anomaly class}&\multirow{3}{*}{Topo. inv.} \\ \cline{2-4} & \multirow{2}{*}{Intl.} & \multirow{2}{*}{Sch\"{o}nflies} & $(0,0,0) + ~(0,1/2,1/2) + $ & & \\ & & & $ (1/2,0,1/2) + ~(1/2,1/2,0) +$ & &\\ \hline
8a&$432$&$O$& $(1/4,1/4,1/4)$, $(3/4,3/4,3/4)$ & $A_m B_\alpha + A_m B_{xy+xz+yz} + C_\beta$ & $\varphi_2[T_3C_2, T_2C'_2]$\\ 
8b&$m\overline{3}$&$T_h$& $(0,0,0)$, $(1/2,1/2,1/2)$ & $A_i B_{xy+xz+yz} + A_m B_{xy+xz+yz} + C_\alpha + C_\beta$ & $\varphi_1[I]$\\ 
\hline
\multirow{3}{*}{24c} & \multirow{3}{*}{$\overline{4}m2$} & \multirow{3}{*}{$D_{2d}$} & $(1/4,0,0)$, $(3/4,0,0)$, & \multirow{3}{*}{$A_m B_{xy+xz+yz}$} & \multirow{3}{*}{$\varphi_2[C_2, T_3C_2GI]$}\\
& & & $(0,1/4,0)$, $(0,3/4,0)$, & & \\
& & & $(0,0,1/4)$, $(0,0,3/4)$ & & \\ \hline
\multirow{3}{*}{24d} & \multirow{3}{*}{$4/m$} & \multirow{3}{*}{$C_{4h}$} & $(0,1/4,1/4)$, $(0,3/4,1/4)$, & \multirow{3}{*}{$(A_i + A_m) (A_i^2 + B_\alpha + B_{xy+xz+yz})$} & \multirow{3}{*}{$\varphi_1[T_3I]$}\\
& & & $(1/4,0,1/4)$, $(1/4,0,3/4)$, & & \\
& & & $(1/4,1/4,0)$, $(3/4,1/4,0)$ & & \\ \hline
\hline 
 \end{tabular} }
 \end{center}

\subsection*{No. 227: $Fd\overline3m$}\label{subsub:sg227}

This group is generated by three translations $T_{1,2,3}$ as given in Eqs.~\eqref{TransBravaisF}, a two-fold rotation $C_2$, a two-fold rotation $C'_2$, a three-fold rotation $C_3$, a mirror $M$, and an inversion $I$:
\begin{subequations}
 \begin{align}
C_2 &\colon (x,y,z)\rightarrow (-x + 1/4, -y + 1/4, z),\\ 
C'_2 &\colon (x,y,z)\rightarrow (-x + 1/4, y, -z + 1/4),\\ 
C_3 &\colon (x,y,z)\rightarrow (z, x, y),\\ 
M &\colon (x,y,z)\rightarrow (y, x, z),\\ 
I &\colon (x,y,z)\rightarrow (-x, -y, -z).
\end{align}
\end{subequations}

The $\mathbb{Z}_2$ cohomology ring is given by

\begin{equation}
\mathbb{Z}_2[A_i,A_m,B_\alpha,B_{xy+xz+yz},C_\alpha,C_{\gamma}]/\langle\mathcal{R}_3,\mathcal{R}_4,\mathcal{R}_5,\mathcal{R}_6\rangle
 \end{equation}
where the relations are 
\begin{subequations} 
 \begin{align}
\mathcal{R}_3\colon & ~~
A_i B_\alpha,\\
\mathcal{R}_4\colon & ~~
A_i C_\alpha,~~A_m C_\alpha,~~A_i C_{\gamma},~~B_\alpha B_{xy+xz+yz} + A_m C_{\gamma},~~B_{xy+xz+yz} (A_i^2 + A_i A_m + B_{xy+xz+yz}),\\
\mathcal{R}_5\colon & ~~
B_{xy+xz+yz} C_\alpha,~~B_{xy+xz+yz} C_{\gamma},\\
\mathcal{R}_6\colon & ~~
C_{\gamma} (C_\alpha + C_{\gamma}).
\end{align} 
 \end{subequations}
We have the following table regarding IWPs and group cohomology at degree 3.
\begin{center}
\begin{tabular}{c|cc|c|c|c}\hline\hline \multirow{3}{*}{\shortstack[l]{Wyckoff\\position}}&\multicolumn{2}{c|}{Little group}& {Coordinates}&\multirow{3}{*}{LSM anomaly class}&\multirow{3}{*}{Topo. inv.} \\ \cline{2-4} & \multirow{2}{*}{Intl.} & \multirow{2}{*}{Sch\"{o}nflies} & $(0,0,0) + ~(0,1/2,1/2) + $ & & \\ & & & $ (1/2,0,1/2) + ~(1/2,1/2,0) +$ & &\\ \hline
8a&$\overline{4}3m$&$T_d$& $(1/8,1/8,1/8)$, $(7/8,3/8,3/8)$ & $C_\alpha + C_{\gamma}$ & $\varphi_2[C_2, C'_2]$\\ 
8b&$\overline{4}3m$&$T_d$& $(3/8,3/8,3/8)$, $(1/8,5/8,1/8)$ & $C_{\gamma}$ & $\varphi_2[T_3C_2, T_2C'_2]$\\ 
\hline
\multirow{2}{*}{16c} & \multirow{2}{*}{$\overline{3}m$} & \multirow{2}{*}{$D_{3d}$} & $(0,0,0)$, $(3/4,1/4,1/2)$, & \multirow{2}{*}{$A_i (A_i^2 + A_i A_m + B_{xy+xz+yz})$} & \multirow{2}{*}{$\varphi_1[I]$}\\
& & & $(1/4,1/2,3/4)$, $(1/2,3/4,1/4)$ & & \\ \hline
\multirow{2}{*}{16d} & \multirow{2}{*}{$\overline{3}m$} & \multirow{2}{*}{$D_{3d}$} & $(1/2,1/2,1/2)$, $(1/4,3/4,0)$, & \multirow{2}{*}{$A_i B_{xy+xz+yz}$} & \multirow{2}{*}{$\varphi_1[T_1T_2I]$}\\
& & & $(3/4,0,1/4)$, $(0,1/4,3/4)$ & & \\ \hline
\hline 
 \end{tabular} 
 \end{center}

\subsection*{No. 228: $Fd\overline3c$}\label{subsub:sg228}

This group is generated by three translations $T_{1,2,3}$ as given in Eqs.~\eqref{TransBravaisF}, a two-fold rotation $C_2$, a two-fold rotation $C'_2$, a three-fold rotation $C_3$, a glide $G$, and an inversion $I$:
\begin{subequations}
 \begin{align}
C_2 &\colon (x,y,z)\rightarrow (-x + 1/4, -y + 1/4, z),\\ 
C'_2 &\colon (x,y,z)\rightarrow (-x + 1/4, y, -z + 1/4),\\ 
C_3 &\colon (x,y,z)\rightarrow (z, x, y),\\ 
G &\colon (x,y,z)\rightarrow (y + 1/2, x + 1/2, z + 1/2),\\ 
I &\colon (x,y,z)\rightarrow (-x, -y, -z).
\end{align}
\end{subequations}

The $\mathbb{Z}_2$ cohomology ring is given by

\begin{equation}
\mathbb{Z}_2[A_i,A_m,B_\alpha,C_\alpha,F_\gamma]/\langle\mathcal{R}_3,\mathcal{R}_4,\mathcal{R}_7,\mathcal{R}_{12}\rangle
 \end{equation}
where the relations are 
\begin{subequations} 
 \begin{align}
\mathcal{R}_3\colon & ~~
A_m (A_i^2 + A_m^2),~~A_i (A_i A_m + A_m^2 + B_\alpha),\\
\mathcal{R}_4\colon & ~~
A_m (A_i^3 + A_i^2 A_m + A_m B_\alpha),~~A_i C_\alpha,~~A_m C_\alpha,\\
\mathcal{R}_7\colon & ~~
A_m(F_\gamma +B_\alpha^3+A_i^6),
~~A_i (F_\gamma + A_m^6),\\
\mathcal{R}_{12}\colon & ~~ F_\gamma^2 + C_\alpha^2 F_\gamma + C_\alpha^4 +  B_\alpha^6 + A_m^{12} .
\end{align} 
 \end{subequations}
We have the following table regarding IWPs and group cohomology at degree 3.
\begin{center}
\begin{tabular}{c|cc|c|c|c}\hline\hline \multirow{3}{*}{\shortstack[l]{Wyckoff\\position}}&\multicolumn{2}{c|}{Little group}& {Coordinates}&\multirow{3}{*}{LSM anomaly class}&\multirow{3}{*}{Topo. inv.} \\ \cline{2-4} & \multirow{2}{*}{Intl.} & \multirow{2}{*}{Sch\"{o}nflies} & $(0,0,0) + ~(0,1/2,1/2) + $ & & \\ & & & $ (1/2,0,1/2) + ~(1/2,1/2,0) +$ & &\\ \hline
\multirow{2}{*}{16a} & \multirow{2}{*}{$23$} & \multirow{2}{*}{$T$} & $(1/8,1/8,1/8)$, $(7/8,3/8,7/8)$, & \multirow{2}{*}{$A_i^2 A_m + A_i A_m^2 + A_m B_\alpha + C_\alpha$} & \multirow{2}{*}{$\varphi_2[C_2, C'_2]$}\\
& & & $(7/8,7/8,7/8)$, $(1/8,5/8,1/8)$ & & \\ \hline
\multirow{4}{*}{32b} & \multirow{4}{*}{$32$} & \multirow{4}{*}{$D_3$} & $(1/4,1/4,1/4)$, $(0,1/2,3/4)$, & \multirow{4}{*}{$A_i A_m (A_i + A_m)$} & \multirow{4}{*}{$\varphi_2[T_2T_3^{-1}C_2I, GI]$}\\
& & & $(1/2,3/4,0)$, $(3/4,0,1/2)$, & & \\
& & & $(3/4,3/4,3/4)$, $(0,1/2,1/4)$, & & \\
& & & $(1/2,1/4,0)$, $(1/4,0,1/2)$ & & \\ \hline
\multirow{4}{*}{32c} & \multirow{4}{*}{$\overline{3}$} & \multirow{4}{*}{$C_{3i}$} & $(0,0,0)$, $(1/4,3/4,1/2)$, & \multirow{4}{*}{$A_i (A_i^2 + A_m^2)$} & \multirow{4}{*}{$\varphi_1[I]$}\\
& & & $(3/4,1/2,1/4)$, $(1/2,1/4,3/4)$, & & \\
& & & $(3/4,1/4,0)$, $(1/2,1/2,1/2)$, & & \\
& & & $(1/4,0,3/4)$, $(0,3/4,1/4)$ & & \\ \hline
\multirow{6}{*}{48d} & \multirow{6}{*}{$\overline{4}$} & \multirow{6}{*}{$S_4$} & $(7/8,1/8,1/8)$, $(3/8,5/8,5/8)$, & \multirow{6}{*}{$A_m (A_i^2 + A_i A_m + B_\alpha)$} & \multirow{6}{*}{$\varphi_2[T_1^{-1}T_2C'_2G, C_2]$}\\
& & & $(1/8,7/8,1/8)$, $(5/8,3/8,5/8)$, & & \\
& & & $(1/8,1/8,7/8)$, $(5/8,5/8,3/8)$, & & \\
& & & $(7/8,1/8,7/8)$, $(3/8,5/8,3/8)$, & & \\
& & & $(5/8,3/8,7/8)$, $(1/8,7/8,3/8)$, & & \\
& & & $(7/8,3/8,1/8)$, $(3/8,7/8,5/8)$ & & \\ \hline
\hline 
 \end{tabular} 
 \end{center}

\subsection*{No. 229: $Im\overline3m$}\label{subsub:sg229}

This group is generated by three translations $T_{1,2,3}$ as given in Eqs.~\eqref{TransBravaisI}, a two-fold rotation $C_2$, a two-fold rotation $C'_2$, a three-fold rotation $C_3$, a mirror $M$, and an inversion $I$:
\begin{subequations}
 \begin{align}
C_2 &\colon (x,y,z)\rightarrow (-x, -y, z),\\ 
C'_2 &\colon (x,y,z)\rightarrow (-x, y, -z),\\ 
C_3 &\colon (x,y,z)\rightarrow (z, x, y),\\ 
M &\colon (x,y,z)\rightarrow (y, x, z),\\ 
I &\colon (x,y,z)\rightarrow (-x, -y, -z).
\end{align}
\end{subequations}

The $\mathbb{Z}_2$ cohomology ring is given by

\begin{equation}
\mathbb{Z}_2[A_i,A_m,A_{x+y+z},B_\alpha,B_\beta,C_\alpha,C_{xyz}]/\langle\mathcal{R}_2,\mathcal{R}_3,\mathcal{R}_4,\mathcal{R}_5,\mathcal{R}_6\rangle
 \end{equation}
where the relations are 
\begin{subequations} 
 \begin{align}
\mathcal{R}_2\colon & ~~
A_{x+y+z} (A_i + A_{x+y+z}),\\
\mathcal{R}_3\colon & ~~
A_{x+y+z} B_\alpha + A_m B_\beta,~~A_{x+y+z} (B_\alpha + B_\beta),\\
\mathcal{R}_4\colon & ~~
A_m C_\alpha,~~A_{x+y+z} C_\alpha,~~A_{x+y+z} C_{xyz},\\
\mathcal{R}_5\colon & ~~
A_i^2 A_{x+y+z} B_\alpha + A_{x+y+z} B_\alpha^2 + A_i^3 B_\beta + A_i B_\alpha B_\beta + B_\beta C_\alpha + B_\beta C_{xyz},\\
\mathcal{R}_6\colon & ~~
A_i A_{x+y+z} B_\alpha^2 + A_i^2 B_\alpha B_\beta + B_\alpha^2 B_\beta + B_\beta^3 + A_i B_\beta C_\alpha + C_\alpha C_{xyz},\nonumber\\&~~A_i A_{x+y+z} B_\alpha^2 + A_i^2 B_\alpha B_\beta + B_\alpha^2 B_\beta + B_\beta^3 + A_i B_\beta C_\alpha + A_i^3 C_{xyz} + A_i^2 A_m C_{xyz} + A_i B_\alpha C_{xyz} + C_{xyz}^2.
\end{align} 
 \end{subequations}
We have the following table regarding IWPs and group cohomology at degree 3.
\begin{center}
\resizebox{\columnwidth}{!}{
\begin{tabular}{c|cc|c|c|c}\hline\hline {Wyckoff}&\multicolumn{2}{c|}{Little group}& {Coordinates}&\multirow{2}{*}{LSM anomaly class}&\multirow{2}{*}{Topo. inv.} \\ \cline{2-4} position & Intl. & Sch\"{o}nflies & $(0,0,0) + ~(1/2,1/2,1/2) + $ & &\\ \hline
\multirow{2}{*}{2a}&\multirow{2}{*}{$m\overline{3}m$}&\multirow{2}{*}{$O_h$}& \multirow{2}{*}{$(0,0,0)$} & $A_i(A_i + A_m)(A_i + A_{x+y+z})+$ & \multirow{2}{*}{$\varphi_1[I]$}\\ 
&&&&$ A_i B_\alpha + A_{x+y+z} B_\alpha + C_\alpha + C_{xyz}$ &\\
\hline
\multirow{2}{*}{6b} & \multirow{2}{*}{$4/mmm$} & \multirow{2}{*}{$D_{4h}$} & $(0,1/2,1/2)$, $(1/2,0,1/2)$, & \multirow{2}{*}{$C_{xyz}$} & \multirow{2}{*}{$\varphi_2[C_2, T_1T_2C'_2]$}\\
& & & $(1/2,1/2,0)$ & & \\ \hline
\multirow{2}{*}{8c} & \multirow{2}{*}{$\overline{3}m$} & \multirow{2}{*}{$D_{3d}$} & $(1/4,1/4,1/4)$, $(3/4,3/4,1/4)$, & \multirow{2}{*}{$A_i (A_i + A_m) A_{x+y+z}$} & \multirow{2}{*}{$\varphi_1[T_1T_2T_3I]$}\\
& & & $(3/4,1/4,3/4)$, $(1/4,3/4,3/4)$ & & \\ \hline
\multirow{3}{*}{12d} & \multirow{3}{*}{$\overline{4}m2$} & \multirow{3}{*}{$D_{2d}$} & $(1/4,0,1/2)$, $(3/4,0,1/2)$, & \multirow{3}{*}{$A_{x+y+z} B_\alpha$} & \multirow{3}{*}{$\varphi_2[T_2T_3C_2, T_1T_2T_3MI]$}\\
& & & $(1/2,1/4,0)$, $(1/2,3/4,0)$, & & \\
& & & $(0,1/2,1/4)$, $(0,1/2,3/4)$ & & \\ \hline
\hline 
 \end{tabular} }
 \end{center}

\subsection*{No. 230: $Ia\overline3d$}\label{subsub:sg230}

This group is generated by three translations $T_{1,2,3}$ as given in Eqs.~\eqref{TransBravaisI}, a two-fold rotation $C_2$, a two-fold rotation $C'_2$, a three-fold rotation $C_3$, a glide $G$, and an inversion $I$:
\begin{subequations}
 \begin{align}
C_2 &\colon (x,y,z)\rightarrow (-x, -y + 1/2, z),\\ 
C'_2 &\colon (x,y,z)\rightarrow (-x + 1/2, y, -z),\\ 
C_3 &\colon (x,y,z)\rightarrow (z, x, y),\\ 
G &\colon (x,y,z)\rightarrow (y + 1/4, x + 1/4, z + 1/4),\\ 
I &\colon (x,y,z)\rightarrow (-x, -y, -z).
\end{align}
\end{subequations}

The $\mathbb{Z}_2$ cohomology ring is given by

\begin{equation}
\mathbb{Z}_2[A_i,A_m,B_\alpha,D_\gamma]/\langle\mathcal{R}_2,\mathcal{R}_4,\mathcal{R}_5,\mathcal{R}_8\rangle
 \end{equation}
where the relations are 
\begin{subequations} 
 \begin{align}
\mathcal{R}_2\colon & ~~
A_m (A_i + A_m),\\
\mathcal{R}_4\colon & ~~
A_i (A_i + A_m) B_\alpha,\\
\mathcal{R}_5\colon & ~~
(A_i + A_m) D_\gamma,\\
\mathcal{R}_8\colon & ~~ A_i^2B_\alpha^3+A_i^2B_\alpha D_\gamma+D_\gamma^2.
\end{align} 
 \end{subequations}
We have the following table regarding IWPs and group cohomology at degree 3.
\begin{center}
\begin{tabular}{c|cc|c|c|c}\hline\hline {Wyckoff}&\multicolumn{2}{c|}{Little group}& {Coordinates}&\multirow{2}{*}{LSM anomaly class}&\multirow{2}{*}{Topo. inv.} \\ \cline{2-4} position & Intl. & Sch\"{o}nflies & $(0,0,0) + ~(1/2,1/2,1/2) + $ & &\\ \hline
\multirow{4}{*}{16a} & \multirow{4}{*}{$\overline{3}$} & \multirow{4}{*}{$C_{3i}$} & $(0,0,0)$, $(1/2,0,1/2)$, & \multirow{4}{*}{$A_i^2 (A_i + A_m)$} & \multirow{4}{*}{$\varphi_1[I]$}\\
& & & $(0,1/2,1/2)$, $(1/2,1/2,0)$, & & \\
& & & $(3/4,1/4,1/4)$, $(3/4,3/4,3/4)$, & & \\
& & & $(1/4,1/4,3/4)$, $(1/4,3/4,1/4)$ & & \\ \hline
\multirow{4}{*}{16b} & \multirow{4}{*}{$32$} & \multirow{4}{*}{$D_3$} & $(1/8,1/8,1/8)$, $(3/8,7/8,5/8)$, & \multirow{4}{*}{N/A} & \multirow{4}{*}{N/A}\\
& & & $(7/8,5/8,3/8)$, $(5/8,3/8,7/8)$, & & \\
& & & $(7/8,7/8,7/8)$, $(5/8,1/8,3/8)$, & & \\
& & & $(1/8,3/8,5/8)$, $(3/8,5/8,1/8)$ & & \\ \hline
\multirow{6}{*}{24c} & \multirow{6}{*}{$222$} & \multirow{6}{*}{$D_2$} & $(1/8,0,1/4)$, $(3/8,0,3/4)$, & \multirow{6}{*}{$A_i B_\alpha$} & \multirow{6}{*}{$\varphi_2[C_2, GI]$}\\
& & & $(1/4,1/8,0)$, $(3/4,3/8,0)$, & & \\
& & & $(0,1/4,1/8)$, $(0,3/4,3/8)$, & & \\
& & & $(7/8,0,3/4)$, $(5/8,0,1/4)$, & & \\
& & & $(3/4,7/8,0)$, $(1/4,5/8,0)$, & & \\
& & & $(0,3/4,7/8)$, $(0,1/4,5/8)$ & & \\ \hline
\multirow{6}{*}{24d} & \multirow{6}{*}{$\overline{4}$} & \multirow{6}{*}{$S_4$} & $(3/8,0,1/4)$, $(1/8,0,3/4)$, & \multirow{6}{*}{$(A_i + A_m) B_\alpha$} & \multirow{6}{*}{$\varphi_2[C'_2G, C_2]$}\\
& & & $(1/4,3/8,0)$, $(3/4,1/8,0)$, & & \\
& & & $(0,1/4,3/8)$, $(0,3/4,1/8)$, & & \\
& & & $(3/4,5/8,0)$, $(3/4,3/8,1/2)$, & & \\
& & & $(1/8,1/2,1/4)$, $(7/8,0,1/4)$, & & \\
& & & $(0,1/4,7/8)$, $(1/2,1/4,1/8)$ & & \\ \hline
\hline 
 \end{tabular} 
 \end{center}

\end{appendices}

\clearpage
\newpage
\onecolumngrid

\addtocontents{toc}{\protect\SGCRestoreSubsectionsInTOC}
\bibliography{ref}

\end{document}